\documentclass[10pt]{book}
\pdfoutput=1

\usepackage{amsmath,amssymb}

\DeclareSymbolFont{AMSb}{U}{msb}{m}{n}
\DeclareMathSymbol{\Zdouble}{\mathalpha}{AMSb}{"5A}

\usepackage{bbm,relsize}

\usepackage{appendix}
\usepackage{chngcntr}
\usepackage{etoolbox}
\usepackage{cancel}

\usepackage{slashed}

\usepackage{graphicx}
\usepackage[font={small}]{caption}   

\usepackage[linkcolor={blue},citecolor={red},urlcolor={blue},colorlinks=true,linktocpage=true]{hyperref}

\usepackage[numbers,compress,sort]{natbib}

\usepackage{multicol}
\usepackage{bbold}
\usepackage[b]{esvect}

\AtBeginEnvironment{subappendices}{%
\chapter*{Appendix}
\addtocontents{toc}{Appendices}
\counterwithin{figure}{section}
\counterwithin{table}{section}
}

\def\pslash{\hspace{3pt}\raisebox{-.5pt}{$\slash$} \hspace{-6pt} p}

\def\SO{\textrm{SO}}
\def\SU{\textrm{SU}}
\def\U{\textrm{U}}
\def\Tr{\textrm{Tr}}
\def\beq{\begin{equation}}
\def\eeq{\end{equation}}
\def\bea{\begin{eqnarray}}
\def\eea{\end{eqnarray}}
\def\mc{\mathcal}
\def\Th{\widehat{T}}
\def\Id{\mathbb{1}}

\renewcommand{\epsilon}{\varepsilon}

\def \dblarrow#1{\overset{\text{\raisebox{-3pt}{$\leftrightarrow$}}}{#1}}
\def \ve#1{\vv{#1}}
\def \vet#1{\vv{#1}^{T}}

\makeatletter

\newcommand{\Rmnum}[1]{\expandafter\@slowromancap\romannumeral #1@}
\makeatother

\usepackage{mathrsfs}
\DeclareMathAlphabet{\mathpzc}{OT1}{pzc}{m}{it}

\numberwithin{equation}{section}

\title{
\vspace{-2cm}
\begin{flushright}
\small{DFPD-2015TH9}
\end{flushright}
\vspace{3cm}
\huge
\textbf{The Composite \\ Nambu--Goldstone Higgs}
\vspace{.2cm}}
\date{}
\author{
\hspace{-2.75em} {\large Giuliano Panico$^{a}$ and Andrea Wulzer$^{b}$}\\
[10mm]
\normalsize\itshape \hspace{-2.75em} $^a$ IFAE, Universitat Aut\` onoma de Barcelona, E-08193 Bellaterra, Barcelona, Spain\\
\normalsize\itshape \hspace{-2.75em} $^b$ Dipartimento di Fisica e Astronomia and INFN, Sezione di Padova,\\
\normalsize\itshape \hspace{-2.75em} Via Marzolo 8, I-35131 Padova, Italy\\
[15mm]
}

\date{%
    \centering \textbf{Abstract}
    \\[1.25\baselineskip]%
    \normalfont\normalsize%
    \parbox{1.0\linewidth}{%
The composite Higgs scenario, in which the Higgs emerges as a composite pseudo-Nambu-Goldstone boson, is extensively reviewed in these Notes. The material is presented in a pedagogical fashion, with great emphasis on the conceptual and technical foundations of the construction. A comprehensive summary of the flavor, collider and electroweak precision phenomenology is also presented.
    }
}

\begin{document}

\frontmatter
\maketitle
\tableofcontents

\chapter*{Preface}
\addcontentsline{toc}{chapter}{Preface}

Half a century after its formulation, the Standard Model (SM) is by now the established theory of Electro-Weak (EW) and Strong interactions, the discovery of the Higgs boson being the most recent of an impressive series of experimental confirmations. Still the SM is not the fundamental theory of Nature, and not just because no theory can be regarded as ``fundamental'' in natural sciences. Concrete reasons to extend the SM are the existence of gravity, for which no complete high-energy description is available, and other incontrovertible experimental facts such as dark matter, neutrino masses and oscillations. Next, there are a number of theoretical issues based on ``Naturalness'' considerations, among which the flatness and homogeneity of the universe that calls for cosmological inflation (which is also supported by observations), the strong CP problem and, of course, the Naturalness problem associated with the Higgs boson mass. This latter problem is the main motivation for the composite Higgs scenario which we will describe in the present Notes.

Since it is not fundamental, the SM is an effective theory, {\it{i.e.}}~a partial description of Nature that emerges, under suitable conditions, as an approximation of a more fundamental theory. In this extended theory the operators in the SM Lagrangian should find their origin as an effective description of the more fundamental dynamics and their coefficients, which are just phenomenological input parameters within the SM, should become calculable providing the explanation of their observed value.

Unveiling the fundamental origin of the SM is the ultimate goal of ``Beyond the SM'' (BSM) physics. Actually in this spirit the letter ``B'' of the acronym should better be read as ``Behind'' rather than ``Beyond'', in the sense that we are interested in departures from the SM predictions only to the extent to which they will guide us towards the understanding of its fundamental origin. A lack of discovery, {\it{i.e.}}~the exclusion of some hypothetical alternative model, could be equally or even more helpful in this respect.

The ambitious aim of BSM physics should not obscure the important byproducts that emerge from this line of research in the long path towards its final goal. First, BSM is of great help in developing a deep understanding of the SM itself, of the surprising and non-generic features that underlie its current phenomenological success and even to appreciate the true measure of this success. Consider for instance the precise measurements of the EW bosons properties performed at LEP in the 90's. It is impossible to explain why they provided such an important confirmation of the SM without referring to the alternative constructions, perfectly plausible at that time, which were predicting deviations and were excluded by these measurements. In this respect, BSM physics is of great pedagogical value. Second, BSM is essential to design further experimental tests of the SM. It offers an assessment of which sectors of the theory are less accurately tested, outlining the experimental directions in which a new physics discovery is more likely to come or, equivalently, those in which further non-trivial confirmations of the SM could be found. By purely working within the SM, {\it{i.e.}}~without comparing it with alternative models, one could only measure its parameters with increasing accuracy and check the statistical compatibility of the overall fit. If the latter program succeeds we will have established that the SM is one possible viable description of the data, but this will not strengthen our belief that it is really the SM, and not something else, what we are seeing in Nature. Exploring possible alternatives is essential for the latter purpose. As alternatives one could consider uncontrolled and unmotivated modifications of the SM Feynman rules, which are unfortunately often employed in SM studies, or sensible hypotheses resulting from BSM speculations. The third byproduct of BSM physics is that it stimulates theoretical research in quantum field theory, in a direction that lies in between pure SM phenomenology and abstract theoretical speculations. Being neither narrowly directed to a single theory like the former, nor detached from phenomenology like the latter, BSM offers a complementary viewpoint.

In this spirit, we wrote the present Notes with a threefold aim. First, to describe the composite Higgs scenario in view of its possible relevance as the true extension of the SM. Namely we will assess, at the best of the present-day theoretical and experimental understanding, how likely it is that a model of this class might be actually realized in Nature. Second, we will identify the most promising possible experimental manifestations of the scenario, outlining relevant directions for BSM discoveries or SM confirmations. These directions include indirect studies of the Higgs and the top quark couplings and the direct production of new particles with specific features. Third, we will carefully explain the tools that underlie the formulation of the scenario and the study of its implications. Some of these are old concepts. Some others are recent ideas or modern rephrasing of old ones. We think that these will find other  applications in the future, inside but also outside the composite Higgs domain. The material is presented in a pedagogical fashion. Basic knowledge of quantum field theory and of the SM is the only prerequisite. 

These Notes are organized as follows. The Introduction is devoted to the Naturalness problem and to how it is addressed by a composite Higgs. The next three chapters provide a first characterization of the phenomenology in the EW, top and Higgs sectors by only relying on symmetries and power-counting estimates. This leads to robust but semiquantitative conclusions, which should be confirmed by concrete models. A class of such models, based on collective breaking, is introduced in Chap.~\ref{ch:Pheno_models}. They serve as benchmarks for the detailed study of the collider and EW precision phenomenology presented in Chaps.~\ref{ch:LHC_pheno} and \ref{ch:EWPT}, respectively.

\subsection*{Acknowledgments}
We thank J.~D.~Wells (Springer Particle Physics Editor) for suggesting us to write down a volume on composite Higgs. We learned most of what we know on the subject by discussing and collaborating with C.~Grojean, A.~Pomarol and especially R.~Rattazzi. We also thank K.~Agashe, B.~Bellazzini, G.~Dall'Agata, J.~Serra and F.~Zwirner for their comments on the manuscript and R.~Contino, L.~Merlo and S.~Rigolin for useful discussions on four-derivative bosonic operators. The work of A.~W.~was partly supported by the ERC Advanced Grant no.267985 ({\emph{DaMeSyFla}}) and by the MIUR-FIRB Grant RBFR12H1MW.
G.~P.~acknowledges support by the Spanish Ministry MEC under grants FPA2013-44773-P, FPA2012-32828, and FPA2011-25948, by the Generalitat de Catalunya grant 2014-SGR-1450 and by the Severo Ochoa excellence program of MINECO (grant SO-2012-0234).

\mainmatter
\chapter{Introduction}
\label{ch:introduction}

The Higgs is a scalar boson and all the other known particles of this sort are bound states of a strongly interacting sector (namely QCD) whose confinement scale is not far from the particle's mass. It is thus legitimate to ask if the same could be true for the Higgs. Clearly QCD cannot be responsible for the formation of the Higgs particle and a new strongly interacting sector, {\it{i.e.}}~the existence of a new strong force, needs to be postulated if we want to explore this possibility.

The argument above provides a first, heuristic but strong, motivation for the composite Higgs scenario. However it is not the main one. The main motivation has to do with the very special role played by the Higgs in the Standard Model (SM), where it is responsible for the breaking of the Electro-Weak symmetry (EWSB).\footnote{Exhaustive textbooks on the subject are Refs.~\cite{Weinberg:1995mt,Weinberg:1996kr} and  \cite{Peskin:1995ev}.} Understanding the origin of the Higgs boson is thus an essential step towards the microscopic comprehension of the EWSB phenomenon. As we will see in the following, according to the Naturalness argument this comprehension will either come from TeV-scale physics or it will never come. The Higgs being a composite object with a compositeness scale (or geometric size) of TeV order is one of the very few known options for ``Naturally'' generating its mass and in turn the EWSB scale. 

The present chapter consists of three sections. In the first two sections we will describe the salient features of the SM, outlining the main structural features that underlie its phenomenological success, but also its main structural limitation, {\it{i.e.}}~the problem of Naturalness. In the third section we will provide a first qualitative description of the composite Higgs scenario explaining how it addresses the Naturalness issue.

\section{The SM {\emph{is}} an effective field theory}

A complete description of gravity is missing in the SM and this requires the latter to be extended. The statement can be made more precise because a partial description of gravity, obtained by quantizing general relativity in a semiclassical expansion, is instead unavoidably present in the SM. This quantum theory of gravity is intrinsically based on perturbation theory, which is perfectly applicable at low energy but looses its validity for energies above around the Planck mass, $E\gtrsim 4\pi M_P \simeq 10^{19} {\textrm{GeV}}$, because the effective gravity coupling strength grows like $g_G\simeq E/M_P$.\footnote{The result is obtained by estimating the energy scale where the four-graviton vertex,  $g_G^2$, reaches the perturbativity bound of $16\pi^2$.} Some new physics must emerge at $M_P$ to replace the perturbative gravity theory, or much below $M_P$ to stop the growth of the coupling strength. If we denote as ``$\Lambda_{\textrm{SM}}$'' the SM cutoff, {\it{i.e.}}~the energy at which non-SM particles and interactions emerge, we have that $\Lambda_{\textrm{SM}}\lesssim M_P$.

\begin{figure}[t]
\centering
\includegraphics[width=0.75\textwidth]{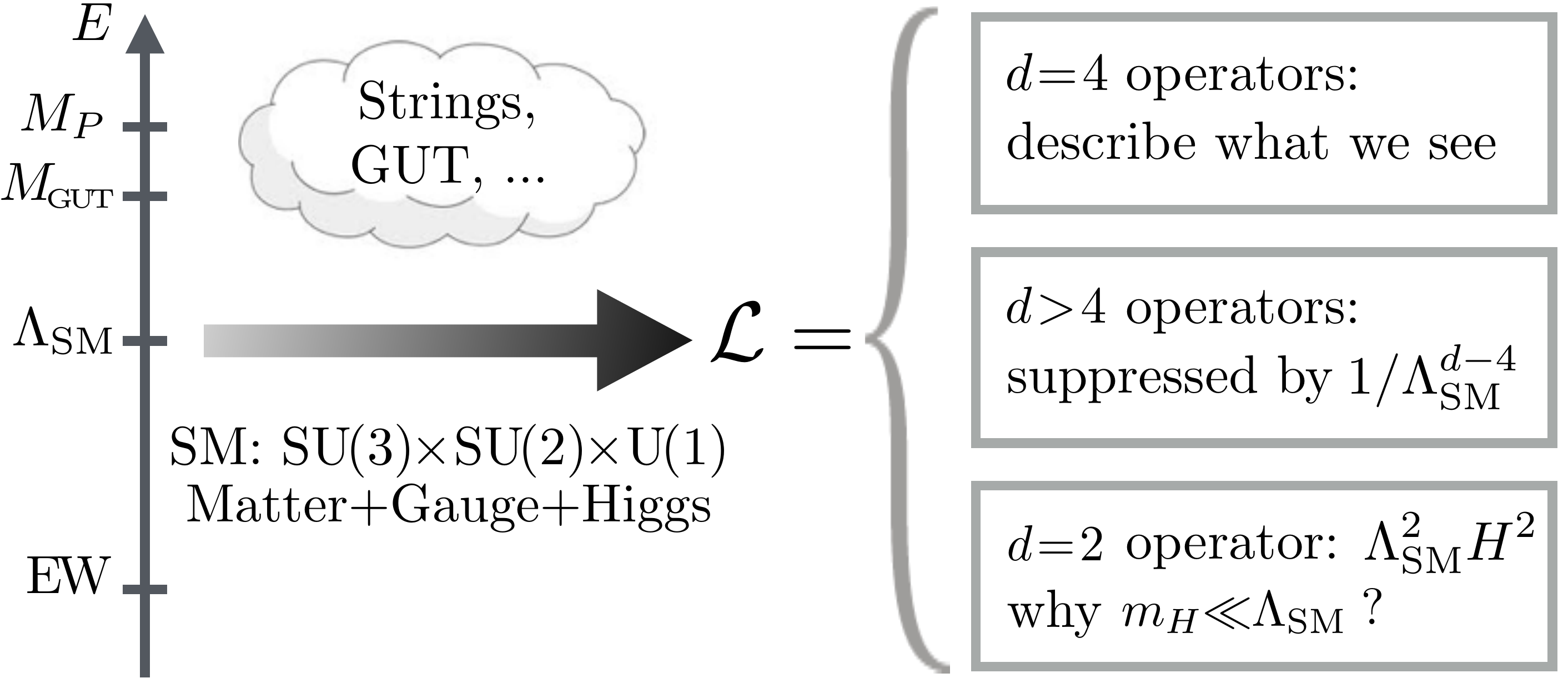}
\caption{Pictorial view of the SM as an effective field theory, with its Lagrangian generated at the scale $\Lambda_{\textrm{SM}}$.}
\label{SMEFT}
\end{figure}

The breakdown of perturbative quantum gravity serves to demonstrate that the SM, given that it has a finite cutoff, is for sure an effective field theory. This does not mean that the first layer of Beyond the SM (BSM) physics is necessarily the one that addresses the quantum gravity issue, nor that it must arise at the Planck scale. Instead, it might provide the microscopic explanation of other mysteries of the SM such as the origin of flavor, of neutrino masses or of EWSB. We will see below that EWSB plays the most important role in the discussion. This explanation will come to us in the form of predictions of the SM Lagrangian operators, and of their coefficients, in terms of the more fundamental parameters of the BSM theory. As depicted in Figure~\ref{SMEFT}, the SM Lagrangian will be computed at the scale $\Lambda_{\textrm{SM}}$ by integrating out the heavy dynamics and retaining in the theory only the light SM matter fermions, vector and Higgs boson degrees of freedom.\footnote{Other light degrees of freedom might well be present in the low-energy theory, provided they are coupled weakly enough to have escaped detection. Their presence would not affect the considerations that follow.} 

A technically consistent description of the vector bosons requires gauge invariance and phenomenological viability requires the gauge group of the SM \mbox{SU$(3)_c\times$SU$(2)_L\times$U$(1)_Y$}. But apart from being gauge (and Lorentz) invariant there is not much we can tell a priori on how the SM effective Lagrangian will look like. It will consist of an infinite series of local gauge-invariant operators with arbitrary energy dimension ``$d$'', with coefficients that on dimensional grounds (given that $[{\mathcal{L}}]=E^4$ and $\Lambda_{\textrm{SM}}$ is the only relevant scale) must be proportional to $1/\Lambda_{\textrm{SM}}^{d-4}$. We can then classify the operators by their energy dimension and discuss their implications.

First we consider the $d=4$ operators. They describe almost all what we have seen in Nature, namely EW and strong interactions, quarks and charged leptons masses. They define a renormalizable theory and thus, together with the $d=2$ operator we will discuss later, they are present in the textbook SM Lagrangian formulated in the old times when renormalizability was taken as a fundamental principle. 

Several books have been written (see for instance Refs.~\cite{OKUN,Cheng:1985bj,Schwartz:2013pla}) on the extraordinary phenomenological success of the renormalizable SM Lagrangian in describing the enormous set of experimental data \cite{Agashe:2014kda} collected in the past decades. In a nutshell, as emphasized in Ref.~\cite{10lectures}, most of this success is due to symmetries, namely to ``accidental'' symmetries. We call ``accidental'' a symmetry that arises by accident at a given order in the operator classification, without being imposed as a principle in the construction of the theory. The renormalizable ($d\leq4$) SM Lagrangian enjoys exact (or perturbatively exact) accidental symmetries, namely baryon and lepton family number, and approximate ones such as the flavor group and custodial symmetry. For brevity, we focus here on the former symmetries, which have the most striking implications. Baryon number makes the proton absolutely stable, in accordance with the experimental limit $\Gamma_p/m_p\lesssim10^{-64}$ on the proton width over mass ratio. It is hard to imagine how we could have accounted for the proton being such a narrow resonance in the absence of a symmetry. Similarly lepton family number forbids exotic lepton decays such as $\mu\rightarrow e\gamma$, whose branching ratio is experimentally bounded at the $10^{-12}$ level. From neutrino oscillations we know that the lepton family number is actually violated, in a way that however nicely fits in the SM picture as we will see below. Clearly this is connected with the neutrino masses, which exactly vanish at $d=4$ because of the absence, in what we call here ``the SM'', of right-handed neutrino fields.

We now turn to $d>4$ operators. Their coefficient is proportional to $1/\Lambda_{\textrm{SM}}^n$, with $n=d-4>0$, thus their effect on low-energy observables is suppressed by $(E/\Lambda_{\textrm{SM}})^n$ with respect to renormalizable terms. Current observations are at and below the EW scale, $E\lesssim m_{\textrm{EW}}\simeq100$~GeV, the suppression being effective thus requires $\Lambda_{\textrm{SM}}\gg 100$~GeV. This simple observation could explain why Nature is approximately well described by a renormalizable theory, without renormalizability being a principle. 

Non-renormalizable operators violate the $d=4$ accidental symmetries. Lepton family number stops being accidental already at $d=5$ because of the Weinberg operator \cite{Weinberg:1979sa}
\beq
\label{WOP}
\displaystyle
\frac1{\Lambda_{\textrm{SM}}}({\overline{\ell}}_L H^c)({{\ell}}_L^c H^c)\,,
\eeq
where $\ell_L$ denotes the lepton doublet, $\ell_L^c$ its charge conjugate, while $H$ is the Higgs doublet and $H^c = i \sigma^2 H^*$. The \mbox{SU$(2)_L$} indices are contracted within the parentheses and the spinor index between the two terms. A generic lepton flavor structure of the coefficient is understood. Surprisingly enough, the Weinberg operator is the unique $d=5$ term in the SM Lagrangian. For $\Lambda_{\textrm{SM}}\simeq10^{14}$~GeV and order one coefficient it generates neutrino masses of the correct magnitude ($m_\nu\sim0.1$~eV) and it can perfectly account for all observed neutrino oscillation phenomena. Baryon number is instead still accidental at $d=5$ and its violation is postponed to $d=6$. We thus perfectly understand, qualitatively, why lepton family violation effects are ``larger'', thus easier to discover, while baryon number violation like proton decay is still unobserved. At a more quantitative level we should actually remark that the bounds on proton decay from the $d=6$ operators, with order one numerical coefficients, set a limit $\Lambda_{\textrm{SM}}\gtrsim10^{16}$~GeV that is in slight tension with what required by neutrino masses. However few orders of magnitude are not a concern here, given that there is no reason why the operator coefficient should be of order one. A suppression of the proton decay operators is actually even expected  because they involve the first family quarks and leptons, whose couplings are reduced already at the renormalizable level. Namely, it is plausible that the same mechanism that makes the first-family Yukawa couplings small also reduces proton decay, while less suppression is expected in the third family entries of the Weinberg operator coefficient that might drive the generation of the heaviest neutrino mass.

The considerations above suggest an extremely plausible picture for high energy physics. Maybe the SM cutoff is extremely high; just to set a reference we might place it at $\Lambda_{\textrm{SM}}\sim10^{15}$~GeV~$=M_{\textrm{GUT}}$. This choice happens to coincide with the gauge coupling unification scale, but this doesn't mean that the new physics at the cutoff is necessarily a Grand Unified Theory (GUT) (see for instance Refs.~\cite{Langacker:1980js,Raby:2006sk}).
On the contrary, the physics at the cutoff can be very generic in this picture, precisely because the cutoff is high. Compatibility with low-energy observations is ensured by the large scale separation $\Lambda_{\textrm{SM}}\gg m_{\textrm{EW}}$. New physics at $M_{\textrm{GUT}}$ is not really one specific BSM scenario. Given that it makes no requirement on how the physics at the cutoff should look like and that it does not require new particles that we might be capable to observe in the foreseeable future\footnote{This doesn't make it completely untestable. Purely Majorana neutrino masses would be a strong indication of its validity while observing a large Dirac component would make it less appealing.}, it just coincides with what we typically call the ``SM-only'' option for high energy physics. We just rephrased it in the educated language of effective field theories. 

The effective field theory language is essential in order to properly discuss the limitations of the SM. One of those, which was already mentioned, is the hierarchy among the Yukawa couplings of the various quark and lepton flavors, which span few orders of magnitude. This tells us that the new physics at $\Lambda_{\textrm{SM}}$ cannot actually be completely generic, given that it must be capable of  generating such a hierarchy in its prediction for the Yukawa's. This limits the set of theories allowed at the cutoff but is definitely not a strong constraint. Whatever mechanism we might imagine to generate flavor hierarchies at $\Lambda_{\textrm{SM}}\sim M_{\textrm{GUT}}$, it will typically not be in contrast with observations given that the bounds on generic flavor-violating operators are  ``just'' at the $10^{8}$~GeV scale. Incorporating dark matter also requires some modification of the SM picture, but there are several ways in which this could be done without changing the situation dramatically. Perhaps the most appealing solution from the viewpoint of the SM is ``minimal dark matter'' \cite{Cirelli:2005uq}, a theory in which all the symmetries which are needed for phenomenological consistence are accidental. This includes not only the SM accidental symmetries, but also the additional $\Zdouble_2$ symmetry needed to keep the dark matter particle cosmologically stable. Similar considerations hold for the strong CP problem, for inflation and all other cosmological shortcomings of the SM. The latter could be addressed by light and extremely weakly-coupled new particles or by very heavy ones above the cutoff. In conclusion, none of the above-mentioned issues is powerful enough to put the basic idea of very heavy new physics scale in troubles. The only one that is capable to do so is the Naturalness (or Hierarchy) problem discussed below.\footnote{See Refs.~\cite{Barbieri:2013vca} and \cite{Giudice:2008bi} for recents essays on the Naturalness problem. The problem was first formulated in Refs.~\cite{thooftNat} and \cite{Dimopoulos:1979es,Susskind:1978ms}, however according to the latter references it was K.Wilson who first raised the issue.}

We have not yet encountered the Naturalness problem in our discussion merely because we voluntarily ignored, in our classification, the operators with $d<4$. The only such operator in the SM is the Higgs mass term, with $d=2$.\footnote{There is also the cosmological constant term, of $d=0$. It poses another Naturalness problem that we will mention later in this chapter.} When studying the $d>4$ operators we concluded that their coefficient is suppressed by $1/\Lambda_{\textrm{SM}}^{d-4}$. Now we have $d=2$ and we are obliged to conclude that the operator is {\emph{enhanced}} by $\Lambda_{\textrm{SM}}^2$, {\it{i.e.}}~that the Higgs mass term reads
\beq
\label{mhUV}
\displaystyle
c\,\Lambda_{\textrm{SM}}^2 H^\dagger H\,,
\eeq
with ``$c$'' a numerical coefficient. In the SM the Higgs mass term sets the scale of EWSB and it directly controls the Higgs boson mass. Today we know that $m_H=125$~GeV and thus the mass term is $\mu^2=m_H^2/2=(89\,{\textrm{GeV}})^2$. But if  $\Lambda_{\textrm{SM}}\sim M_{\textrm{GUT}}$, what is the reason for this enormous hierarchy? Namely
\beq
\displaystyle
{\textrm{why}}\;\;\frac{\mu^2}{\Lambda_{\textrm{SM}}^2}\sim10^{-28}\lll1\;\;{\textrm{{?}}}
\nonumber
\eeq
This is the essence of the Naturalness problem.

Further considerations on the Naturalness problem, its implications and the possible solution offered by the composite Higgs scenario are postponed to the next section. However, we can already appreciate here how radically it changes our expectations on high energy physics. The SM-only picture gets sharply contradicted by the Naturalness argument since the problem is based on the same logic ({\it{i.e.}}, dimensional analysis) by which its phenomenological virtues ({\it{i.e.}}, the suppression of $d>4$ operators) were established. The new picture is that $\Lambda_{\textrm{SM}}$ is low, in the $100$~GeV to few TeV range, such that a light enough Higgs is obtained ``Naturally'', {\it{i.e.}}~in accordance with the estimate in Eq.~(\ref{mhUV}). The new physics at the cutoff must now be highly non-generic, given that it cannot rely any longer on a large scale suppression of the BSM effects. To start with, baryon and lepton family number violating operators must come with a highly suppressed coefficient, which in turn requires baryon and lepton number being imposed as symmetries rather than emerging by accident. In concrete, the BSM sector must now respect these symmetries. This can occur either because it inherits them from an even more fundamental theory or because they are accidental in the BSM theory itself. Similarly, if  $\Lambda_{\textrm{SM}}\sim$~TeV flavor violation cannot be generic. Some special structure must be advocated on the BSM theory, Minimal Flavor Violation (MFV) \cite{Glashow:1976nt,D'Ambrosio:2002ex} being one popular and plausible option. The limits from EW Precision Tests (EWPT) come next; they also need to be carefully addressed for TeV scale new physics. On one hand this makes Natural new physics at the TeV scale very constrained. On the other hand it gives us plenty of indications on how it should, or it should not, look like.

\section{A Natural electroweak scale}
\label{natural_EWSB}

The reader might be unsatisfied with the formulation of the Naturalness problem we gave so far. All what Eq.~(\ref{mhUV}) tells us is that the numerical coefficient ``$c$'' that controls the actual value of the mass term beyond dimensional analysis should be extremely small, namely $c\sim10^{-28}$ for GUT scale new physics. Rather than pushing $\Lambda_{\textrm{SM}}$ down to the TeV scale, where all the above-mentioned constraints apply, one could consider keeping $\Lambda_{\textrm{SM}}$ high and try to invent some mechanism to explain why $c$ is small. After all, we saw that there are other coefficients that require a suppression in the SM Lagrangian, namely the light flavors Yukawa couplings. One might argue that it is hard to find a sensible theory where $c$ is small, while this is much simpler for the Yukawa's. Or that $28$ orders of magnitude are by far much more than the reduction needed in the Yukawa sector. But this would not be fully convincing and would not make full justice to the importance of the Naturalness problem.

In order to better understand Naturalness we go back to the essential message of the previous section. The SM is an effective field theory and thus the coefficients of its operators, which we regard today as fundamental input parameters, should actually be derived phenomenological parameters, to be computed one day in a more fundamental BSM theory. Things should work just like for the Fermi theory of weak interactions, where the Fermi constant $G_F$ is a fundamental input parameter that sets the strength of the weak force. We know however that the true microscopic description of the weak interactions is the EW theory. The reason why we are sure about this is that it allows us to predict $G_F$ in terms of its microscopic parameters $g_W$ and $m_W$, in a way that agrees with the low-energy determination. What we have in mind here is merely the standard textbook formula
\beq
\displaystyle
G_F=\frac{g_W^2}{4\sqrt{2}\,m_W^2}\,,
\eeq
that allows us to carry on, operatively, the following program. Measure the microscopic parameters $g_W$ and $m_W$ at high energy; compute $G_F$; compare it with low-energy observations.\footnote{Actually $G_F$ is taken as an input parameter in actual calculations because it is better measured than $g_W$ and $m_W$, but this doesn't affect the conceptual point we are making.} Since this program succeeds we can claim that the microscopic origin of weak interaction is well-understood in terms of the EW theory. We will now see that the Naturalness problem is an obstruction in repeating the same program for the Higgs mass and in turn for the EWSB scale.

Imagine knowing the fundamental, ``true'' theory of EWSB. It will predict the Higgs mass-term $\mu^2$ or, which is the same, the physical Higgs mass $m_H^2=2\mu^2$, in terms of its own input parameters ``$p_{\textrm{true}}$'', by a formula that in full generality reads
\beq
\label{mHtrue}
\displaystyle
m_H^2=\int_0^\infty\hspace{-10pt} dE\;\frac{d m_H^2}{dE}(E;p_{\textrm{true}})\,.
\eeq
The integral over energy stands for the contributions to $m_H^2$ from all the energy scales and it extends up to infinity, or up to the very high cutoff of the ``true'' theory itself. The integrand could be localized around some specific scale or even sharply localized by a delta-function at the mass of some specific particle, corresponding to a tree-level contribution to $m_H^2$. Examples of theories with tree-level contributions are GUT \cite{Langacker:1980js,Raby:2006sk} and Supersymmetric (SUSY)\footnote{Standard textbooks and reviews on supersymmetry are in Refs.~\cite{WessBagger,WeinbergSUSY,Sohnius:1985qm,Drees:1996ca,Martin:1997ns,Peskin:2008nw}.} models, where $m_H$ emerges from the mass terms of extended scalar sectors. The formula straightforwardly takes into account radiative contributions, which are the only ones present in the composite Higgs scenario. Also in SUSY, radiative terms have a significant impact given that the bounds on the scalar (SUSY and soft) masses that contribute at the tree-level are much milder than those on the colored stops and gluinos that contribute radiatively. In the language of old-fashioned perturbation theory \cite{Weinberg:1995mt}, ``$E$'' should be regarded as the energy of the virtual particles that run into the diagrams through which $m_H^2$ is computed.

\begin{figure}[t]
\centering
\includegraphics[width=1\textwidth]{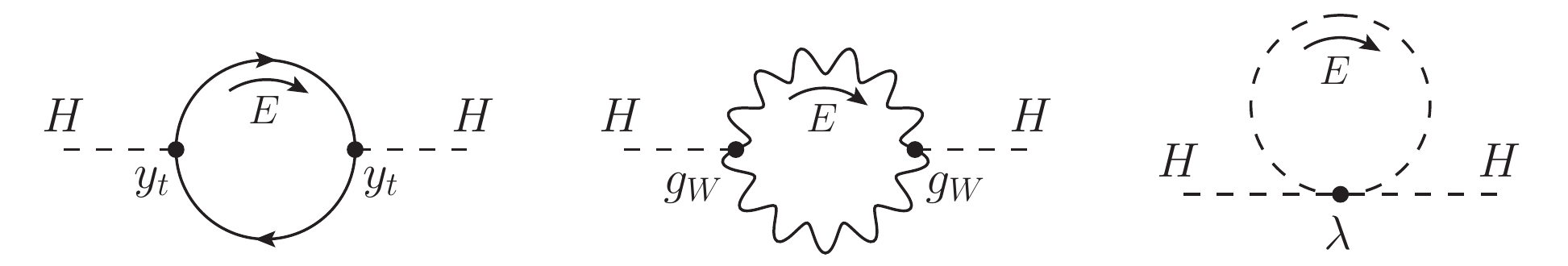}
\caption{Some representative top, gauge and Higgs boson loop diagrams that contribute to the Higgs mass.}
\label{HP}
\end{figure}

Consider now splitting the integral in two regions defined by an intermediate scale that we take just a bit below the SM cutoff. We have
\bea
\label{splitint}
\displaystyle
m_H^2&&=\int_0^{\lesssim{\Lambda_{\textrm{SM}}}}\hspace{-10pt} dE\;\frac{d m_H^2}{dE}(E;p_{\textrm{true}})+\int_{\lesssim{\Lambda_{\textrm{SM}}}}^\infty\hspace{-10pt} dE\;\frac{d m_H^2}{dE}(E;p_{\textrm{true}})\nonumber\\
&&=\delta_{\textrm{SM}}m_H^2+\delta_{\textrm{BSM}}m_H^2\,,
\eea
where $\delta_{\textrm{BSM}}m_H^2$ is a completely unknown contribution, resulting from energies at and above $\Lambda_{\textrm{SM}}$, while $\delta_{\textrm{SM}}m_H^2$ comes from virtual quanta below the cutoff, whose dynamics is by assumption well described by the SM. While there is nothing we can tell about $\delta_{\textrm{BSM}}m_H^2$ before we know what the BSM theory is, we can easily estimate $\delta_{\textrm{SM}}m_H^2$ by the diagrams in Figure~\ref{HP}, obtaining
\beq
\label{deltamh}
\delta_{\textrm{SM}}m_H^2=
\frac{3 y_t^2}{8\pi^2}\Lambda_{\textrm{SM}}^2
-
\frac{3 g_W^2}{8\pi^2}\left(\frac14+\frac1{8\cos^2\theta_W}\right)\Lambda_{\textrm{SM}}^2
-\frac{3 \lambda}{8\pi^2}\Lambda_{\textrm{SM}}^2\,,
\eeq
from, respectively, the top quark, EW bosons and Higgs loops. The idea is that we know that the BSM theory must reduce to the SM for $E<\Lambda_{\textrm{SM}}$. Therefore no matter what the physics at $\Lambda_{\textrm{SM}}$ is, its prediction for $m_H^2$ must contain the diagrams in Figure~\ref{HP} and thus the terms in Eq.~(\ref{deltamh}). These terms are obtained by computing $dm_H^2/dE$ from the SM diagrams and integrating it up to $\Lambda_{\textrm{SM}}$, which effectively acts as a hard momentum cutoff. The most relevant contributions come from the quadratic divergences of the diagrams, thus Eq.~(\ref{deltamh}) can be poorly viewed as the ``calculation'' of quadratic divergences. Obviously quadratic divergences are unphysical in quantum field theory. They are canceled by renormalization and they are even absent in certain regularizations schemes such as dimensional regularization. However the calculation makes sense, in the spirit above, as an estimate of the low-energy contributions to $m_H^2$.

The true nature of the Naturalness problem starts now to show up. The full finite formula for $m_H^2$ obtained in the ``true'' theory receives two contributions that are completely unrelated since they emerge from separate energy scales. At least one of those, $\delta_{\textrm{SM}}m_H^2$, is for sure very large if $\Lambda_{\textrm{SM}}$ is large. The other one is thus obliged to be large as well, almost equal and with opposite sign in order to reproduce the light Higgs mass we observe. A cancellation is taking place between the two terms, which we quantify by a fine-tuning $\Delta$ of at least
\beq
\label{deltatuning}
\Delta\geq\frac{\delta_{\textrm{SM}}m_H^2}{m_H^2}
=\frac{3\, y_t^2}{8\pi^2}\left(\frac{\Lambda_{\textrm{SM}}}{m_H}\right)^2\simeq \left(\frac{\Lambda_{\textrm{SM}}}{450\,{\textrm{GeV}}}\right)^2\,.
\eeq
Only the top loop term in Eq.~(\ref{deltamh}) has been retained for the estimate since the top dominates because of its large Yukawa coupling and because of color multiplicity. Notice that the one above is just a lower bound on the total amount of cancellation $\Delta$ needed to adjust $m_H$ in the true theory. The high energy contribution $\delta_{\textrm{BSM}}m_H^2$, on which we have no control, might itself be the result of a cancellation, needed to arrange for $\delta_{\textrm{BSM}}m_H^2\simeq -\delta_{\textrm{SM}}m_H^2$. Examples of this situation exist both in SUSY and in composite Higgs.

The problem is now clear. Even if we were able to write down a theory that formally predicts the Higgs mass, and even if this theory turned out to be correct we will never be able to really predict $m_H$ if $\Lambda_{\textrm{SM}}$ is much above the TeV scale, because of the cancellation. For $\Lambda_{\textrm{SM}} = M_{\textrm{GUT}}$, for instance, we have $\Delta\gtrsim10^{24}$. This means that in the ``true'' theory formula for $m_H$ a $24$ digits cancellation is taking place among two a priori unrelated terms. Each of these terms must thus be known with at least $24$ digits accuracy even if we content ourselves with an order one estimate of $m_H$. We will never achieve such an accuracy, neither in the experimental determination of the $p_{\textrm{true}}$ ``true'' theory parameters $m_H$ depends on, nor in the theoretical calculation of the Higgs mass formula. Therefore, we will never be able to repeat for $m_H$ the program we carried on for $G_F$ and we will never be able to claim we understand its microscopic origin and in turn the microscopic origin of the EWSB scale. A BSM theory with $\Lambda_{\textrm{SM}} = M_{\textrm{GUT}}$ has, in practice, the same predictive power on $m_H$ as the SM itself, where Eq.~(\ref{mHtrue}) is replaced by the much simpler formula
\beq
m_H^2=m_H^2\,.
\eeq
Namely if such an high-scale BSM theory was realized in Nature $m_H$ will remain forever an input parameter like in the SM. The microscopic origin of $m_H$, if any, must necessarily come from new physics at the TeV scale, for which the fine-tuning $\Delta$ in Eq.~(\ref{deltatuning}) can be reasonably small.

The Higgs mass term is the only parameter of the SM for which such an argument can be made. Consider for instance writing down the analog of Eq.~(\ref{mHtrue}) for the Yukawa couplings and splitting the integral as in Eq.~(\ref{splitint}). The SM contribution to the Yukawa's is small even for $\Lambda_{\textrm{SM}} = M_{\textrm{GUT}}$, because of two reasons. First, the Yukawa's are dimensionless and thus, given that there are no couplings in the SM with negative energy dimension, they do not receive quadratically divergent contributions. The quadratic divergence is replaced by a logarithmic one, with a much milder dependence on $\Lambda_{\textrm{SM}}$. Second, the Yukawa's break the flavor group of the SM. Therefore there exist selection rules (namely those of MFV) that make radiative corrections proportional to the Yukawa matrix itself. The Yukawa's, and the hierarchies among them, are thus ``radiatively stable'' in the SM. This marks the essential difference with the Higgs mass term and implies that their microscopic origin and the prediction of their values could come at any scale, even at a very high one. The same holds for all the other SM parameters apart from $m_H$.

The formulation in terms of fine-tuning (\ref{deltatuning}) turns the Naturalness problem from a vague aesthetic issue to a concrete semiquantitative question. Depending on the actual value of $\Delta$ the Higgs mass can be operatively harder or easier to predict, making the problem more or less severe. If for instance $\Delta\sim10$, we will not have much troubles in overcoming a one digit cancellation once we will know and we will have experimental access to the ``true'' theory. After some work, sufficiently accurate predictions and measurements will become available and the program of predicting $m_H$ will succeed. The occurrence of a one digit cancellation will at most be reported as a curiosity in next generation particle physics books and we will eventually forget about it. A larger tuning $\Delta=1000$ will instead be impossible to overcome. The experimental exploration of the high energy frontier will tell us, through Eq.~(\ref{deltatuning}), what to expect about $\Delta$. Either by discovering new physics that addresses the Naturalness problem or by pushing $\Lambda_{\textrm{SM}}$ higher and higher until no hope is left to understand the origin of the EWSB scale in the sense specified above. One way or another, a fundamental result will be obtained. 

The discovery of ``Unnaturalness'' by the non-observation of new physics at the TeV scale would prevent us from predicting $m_H$ as an effective phenomenological parameter, but this does not necessarily mean that we will never get some control on its value. The idea of anthropic vacuum selection, first applied by Weinberg to the cosmological constant problem \cite{Weinberg:1987dv}, might help in this respect. The cosmological constant operator suffers of exactly the same Naturalness problem as the Higgs mass. Provided we claim we understand gravity well enough to estimate them, radiative corrections push the cosmological constant to very high values, tens of orders of magnitude above what we knew it had to be (and was subsequently observed) in order for galaxies being able to form in the early universe. Weinberg pointed out that the most plausible value for the cosmological constant should thus be close to the maximal allowed value for the formation of galaxies because galaxies are essential for the development of intelligent life. The idea is that if many ground state configurations are possible in the fundamental theory, typically characterized by a very large cosmological constant but with a tail in the distribution that extends up to zero, the largest possible value compatible with galaxies formation, and thus with the very existence of the observer, will be actually observed. A similar argument can be made for the Higgs mass (see for instance Ref.~\cite{Hall:2007ja}), however it is harder in the SM to identify sharply the boundary of the anthropically allowed region of the parameter space. The anthropic argument basically says that we might be following a radically wrong path in our search for the fundamental laws of Nature. The Higgs mass might not be fundamental, but instead dictated by the environment, and in this case there could be no true mystery about its value. Poorly speaking, $m_H$ could be like the average temperature on the surface of Earth, of around $15\,^{\circ}\mathrm{C}$. It is the one we observe, rather than the one on the surface of the Sun, of around $5500\,^{\circ}\mathrm{C}$, because of anthropic selection. Plausible or not, the very existence of such speculative ideas demonstrates the relevance of the Naturalness problem and the importance of further investigating it. Once again, we see that even discovering Unnaturalness in $m_H$ by excluding Natural new physics at the TeV scale would be a fundamental result that would change our perspective on the physics of fundamental interactions.

\section{Dimensional transmutation}
\label{DT}

The composite Higgs scenario offers a simple solution to the problem of Naturalness. Suppose that the Higgs, rather than being a point-like particle as in the SM, is instead an extended object with a finite geometric size $l_H$. We will make it so by assuming that it is the bound state of a new strong force characterized by a confinement scale $m_*=1/l_H$ of TeV order. In this new theory the $dm_H^2/dE$ integrand in the Higgs mass formula (\ref{mHtrue}), which stands for the contribution of virtual quanta with a given energy, behaves as follows. Low energy quanta have too a large wavelength to resolve the Higgs size $l_H$. Therefore the Higgs behaves like an elementary particle and the integral grows linearly with $E$ like in the SM, resulting in a quadratic sensitivity to the upper integration limit. However this growth gets canceled by the finite size effects that start becoming visible when $E$ approaches and eventually overcomes $m_*$. Exactly like the proton when hit by a virtual photon of wavelength below the proton radius, the composite Higgs is transparent to high-energy quanta and the integrand decreases. The linear SM behavior is thus replaced by a peak at $E\sim m_*$ followed by a steep fall. The Higgs mass generation phenomenon gets localized at $m_*=1/l_H$ and $m_H$ is insensitive to much higher energies. This latter fact is also obvious from the fact that no Higgs particle is present much above $m_*$. Therefore there exist no Higgs field and no $d=2$ Higgs mass term to worry about.

Clearly it is essential for this to work that the Higgs size, or its inverse that we associated with a confinement scale, is itself Natural. Namely we must not encounter a new Naturalness problem when trying to put $m_*$ at the TeV scale. The complete composite Higgs picture, which by the way is common to all the non-SUSY attempts to address Naturalness, beginning with technicolor \cite{Weinberg:1975gm,Weinberg:1979bn,Susskind:1978ms}  (see Ref.~\cite{Lane:2002wv} for a review), is as follows. The main ingredient is a new ``composite sector'' that will eventually deliver the Higgs as a bound state. At least part of the composite sector dynamics will have to take place in a strongly-coupled non-perturbative regime, therefore it is appropriate to say that the new sector describes a new strong force. The composite sector emerges from an even more fundamental theory at a very high scale $\Lambda_{\textrm{UV}}\gg$~TeV, whose precise value will not matter for us given that the whole point of the construction is precisely to make the EW scale insensitive to it. We might think, just for definiteness, $\Lambda_{\textrm{UV}}\sim{M}_{\textrm{GUT}}$. 

At $\Lambda_{\textrm{UV}}$ the composite sector sits close to a fixed point of its Renormalization Group (RG) evolution and it is assumed that no strongly relevant deformation exists around this fixed point.
Namely, we assume that no operator in the composite sector Lagrangian has scaling dimension considerably below $4$, which corresponds to the absence of parameters with strongly positive dimensionality. This is the educated way to express the absence of unprotected energy scales in the UV theory, which would reintroduce the Naturalness problem. The SM, with its $d=2$ Higgs mass operator and the corresponding $d=2$ parameter $\mu^2$, is an example of how the composite sector should not look like. A sector with the features above, realized in Nature, is low-energy QCD. Low-energy QCD emerges, after integrating out the EW bosons and the heavy quarks, as a weakly-coupled theory close to the trivial (free) fixed point.\footnote{A theory with nearly massless $u$, $d$ and $s$ quarks, and all the others with masses at the EW scale $\sim m_Z$, at which $g_S\sim 4\pi/10$ well within the perturbative regime, is what we actually have in mind for our analogy.} No strongly relevant deformation is present because the only $d<4$ operators, the quark mass terms, are protected by the chiral symmetry and thus they evolve under the RG flow as if they had $d\simeq4$. 

The absence of strongly relevant deformations makes the RG flow towards the IR a ``slow'' process. Deviations from the fixed point are controlled by the RG ``time'' $t=\log[\Lambda_{\textrm{UV}}/E]$ so that the theory can significantly depart from the fixed point, confine and eventually develop the composite Higgs bound state, only at an exponentially suppressed scale $m_*$ defined by ${\overline{t}}=\log[\Lambda_{\textrm{UV}}/m_*]$. The time $\overline{t}$ by which the RG running ends can be arbitrarily long, depending on how close to the fixed point we started from at $\Lambda_{\textrm{UV}}$. It could be easily of order $10$, allowing for a Natural huge hierarchy between $\Lambda_{\textrm{UV}}$ and $m_*$. The analogy with QCD is once again extremely useful to clarify the situation. For $3$ light quark flavors and the running starting at $m_Z=\Lambda_{\textrm{UV}}$, the habitual formula for the QCD confinement scale $\Lambda_{\textrm{QCD}}=m_*$, can be expressed as 
\beq
\log[\Lambda_{\textrm{UV}}/m_*]=\frac{1}{18}\left(\frac{4\pi}{g_S}\right)^2\,.
\eeq
The loop expansion parameter $g_S^2/16\pi^2$, with $g_S$ evaluated at $\Lambda_{\textrm{UV}}=m_Z$ controls the departure of the UV theory from the free fixed point. This is why its inverse sets the total RG running time and hence the hierarchy between $\Lambda_{\textrm{UV}}$ and $m_*$. This mechanism, by which a scale $m_*$ is generated through the running without dimensionful parameters but only $d=0$ couplings being present in the UV theory is called ``dimensional transmutation'' in QCD textbooks. The name is appropriate also in the more general context we have in mind here, where the absence of relevant deformations corresponds to the absence of dimensionful parameters in the microscopic theory.

The second ingredient we need for a potentially realistic theory is one extra ``elementary'' sector that contains all those particles that we know, by phenomenology, can not be composite at the TeV scale unlike the Higgs.\footnote{Those particles might be ``partially composite'', a concept that we will introduce and discuss extensively in these Notes.} Those are basically all the other SM particles, the only possible exception being, as we will see, the right-handed component of the top quark. The elementary sector is a weakly-coupled gauge theory with the gauge group of the SM and its particle content aside from the Higgs. The most relevant operators in its Lagrangian, namely those that are not suppressed by $1/\Lambda_{\textrm{UV}}^n$, are thus just the ordinary $d=4$ SM gauge and fermion kinetic terms. Obviously no Yukawa couplings are present since there is no Higgs. The phenomenological need of describing the SM vector bosons as elementary gauge fields obliges the whole theory, including the composite sector, to respect the SM gauge symmetry. Namely, the composite sector will be characterized by an exact symmetry group ``${\mathpzc{G}}$'' containing one \mbox{SU$(2)_L\times$U$(1)_Y$} subgroup. The elementary $W_\mu$ and $B_\mu$ fields make the latter subgroup local by the habitual gauging procedure. This amounts to couple them to the global \mbox{SU$(2)_L\times$U$(1)_Y$} conserved current of the composite sector, giving rise to one sure communication channel between the elementary and the composite sectors.

In the analogy with QCD, the elementary sector consists of the photon field and the light leptons. The global group ${\mathpzc{G}}$ is the chiral symmetry \mbox{SU$(3)_L\times$SU$(3)_R$} and the photon gauges its U$(1)_{\textrm{e.m.}}$ subgroup. The major difference between our construction and  QCD is, in this respect, the fact that the chiral group is explicitly broken by the quark masses while our composite sector is exactly invariant under ${\mathpzc{G}}$. The latter assumption could actually be relaxed, allowing for some amount of breaking inside the composite sector. However this would not introduce radically new phenomena, therefore we will ignore this possibility in the following and assume that the explicit breaking of ${\mathpzc{G}}$ is entirely due to the presence of the elementary sector. This breaking gets transmitted to the composite sector by the elementary/composite interaction ${\mathcal{L}}_{\textrm{int}}$. The same would happen in QCD if we set the quark masses to zero. The breaking of the chiral group would solely come from the coupling of the elementary photon. 

In obvious analogy with QCD, the global group ${\mathpzc{G}}$ will generically be broken spontaneously to a subgroup ${\mathpzc{H}}$ at the confinement scale $m_*$, delivering exactly massless Nambu--Goldstone Bosons (NGB) in the ${\mathpzc{G}}/{\mathpzc{H}}$ coset. In the scenario described in these Notes, the Higgs is one of those Goldstone bosons. It acquires a mass and a potential, triggering EWSB, through the explicit breaking of the Goldstone symmetry group ${\mathpzc{G}}$ induced by the elementary sector. The Higgs being a NGB, or more precisely a pseudo NGB since the Goldstone symmetry is explicitly broken, is essential for the composite Higgs scenario having a chance to be realistic and marks the difference with the old composite Higgs constructions \cite{thooftNat,Dimopoulos:1979es} (see for instance Ref.~\cite{Terazawa:1976xx} for an even earlier attempt) where the Higgs was emerging as a generic bound state of the composite sector. In the latter case, the Higgs mass would be set directly by the confinement scale $m_*$, which we would thus be obliged to take in the $100$~GeV range rather than at the TeV or multi-TeV scale. But a large number of bound states, called ``resonances'', are expected at $m_*$, very much like plenty of hadrons emerge from the QCD confinement at $\Lambda_{\textrm{QCD}}$. None of such particles is observed at $100$~GeV, therefore $m_*$ must be at the TeV and $m_H$ much below that. This is Naturally the case for a NGB Higgs, whose mass is controlled by the explicit breaking of the Goldstone symmetry, provided of course the breaking effects are small. But ${\mathpzc{G}}$ breaking comes from the elementary sector and it is communicated to the composite one by the elementary/composite interactions ${\mathcal{L}}_{\textrm{int}}$. Keeping it small thus requires ${\mathcal{L}}_{\textrm{int}}$ being a weak perturbation of the composite sector dynamics. This condition is also crucial in order to ensure that the presence of the elementary sector does not destabilize the hierarchy among $m_*$ and $\Lambda_{\textrm{UV}}$ induced by the composite sector dynamics, invalidating our solution of the Naturalness problem. The elementary sector being a weak perturbation of the composite one at all scales between $\Lambda_{\textrm{UV}}$ and $m_*$ requires that no strongly relevant operator, compatible with symmetries and selection rules, should be present in ${\mathcal{L}}_{\textrm{int}}$. This is indeed the case in QCD and it is why the electromagnetic interactions are small corrections to the hadron dynamics. The approximate Goldstone symmetry can also explain, at the price of a mild tuning, why the composite Higgs particle not only is light, but also it couples in approximately the same way as if it were elementary as in the SM. This comes, as we will see in the next chapter, from the mechanism of ``vacuum misalignment'' discovered in Refs.~\cite{Kaplan:1983fs,Kaplan:1983sm,Dugan:1984hq}.

\begin{figure}[t]
\centering
\includegraphics[width=0.75\textwidth]{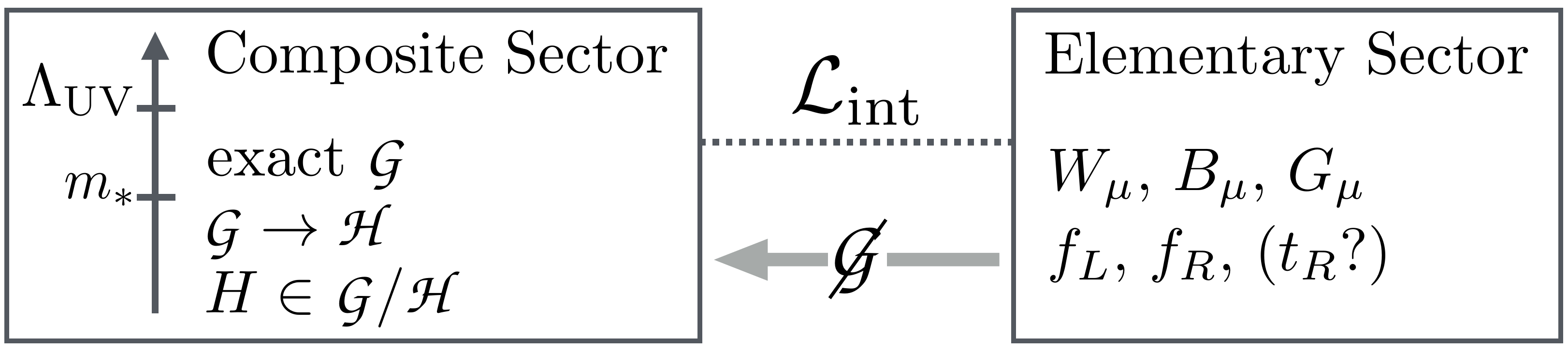}
\caption{The basic structure of the composite Higgs scenario.}
\label{elcomp}
\end{figure}

In summary, the composite Higgs setup is defined by the three basic elements depicted in Figure~\ref{elcomp}. We have a composite sector, an elementary one and an interaction ${\mathcal{L}}_{\textrm{int}}$ between the two, which transmits to the composite sector the explicit breaking of ${\mathpzc{G}}$. Characterizing the features they must possess in order to produce a potentially realistic theory is the purpose of the present Notes. The reader should be aware the she/he will not find here any attempt to formulate a microscopic UV realization of the composite Higgs scenario. It is relatively easy, and was done already in the original literature \cite{Kaplan:1983fs,Kaplan:1983sm,Dugan:1984hq}, to produce QCD-like models, weakly coupled at $\Lambda_{\textrm{UV}}$ and thus described by ``technigluons'' and ``techniquarks'' constituents at that scale, with a ${\mathpzc{G}}\rightarrow {\mathpzc{H}}$ symmetry breaking pattern suited to deliver a NGB Higgs. These models typically result in a phenomenologically satisfactory description of the EW bosons and Higgs dynamics. Very few attempts have been made \cite{Caracciolo:2012je,Marzocca:2013fza,Barnard:2013zea,Ferretti:2013kya,Ferretti:2014qta} to extend these constructions to the fermionic sector. Here the challenge is to find a microscopic realization of the ``partial compositeness'' hypothesis \cite{Kaplan:1991dc} that, as we will see, requires a peculiar structure of the elementary/composite fermionic interactions. The best examples we have of composite Higgs models with partial fermion compositeness are five-dimensional gauge theories on truncated anti-de Sitter space \cite{Contino:2003ve,Agashe:2004rs,Contino:2010rs}, which however are not UV-complete. Their completion might not be an asymptotically-free four-dimensional gauge theory, it could instead be some other kind of construction flowing towards a strongly-coupled UV fixed point. Given the extra dimensional nature of these model a string theory completion, possibly dual to a strongly-coupled theory in four dimensions, could be also envisaged. Our viewpoint on the UV completion of the composite Higgs scenario is the following. What we can definitely do is to identify and spell out as precisely as possible the required assumptions on the UV dynamics and, by the power of symmetries, selection rules and effective field theory techniques, study their compatibility with current observations and make prediction for future experimental searches. Which is by the way how we would proceed even if we knew the microscopic theory given the difficulty of performing explicit calculations in the strong coupling regime. If, as a result of this effort, something that resembles the composite Higgs is discovered, we would be sure that microscopic theory with the required features exists and finding one will become a priority. If conversely the composite Higgs scenario will be found to be experimentally excluded, or too much tuned to be relevant, finding a UV completion will still remain an interesting theoretical quantum field theory question, but with no direct phenomenological relevance.

\chapter{Goldstone boson Higgs}
\label{ch:Goldstone}

This chapter provides a first illustration of the composite Higgs scenario and a first characterization of its phenomenology. In particular of those aspects of the phenomenology that robustly follow from the Nambu--Goldstone Boson (NGB) nature of the Higgs in a model-independent way. Interestingly enough, this includes a specific pattern of Higgs coupling modifications with respect to the SM predictions. The basic concept behind the formulation of the composite Higgs scenarios is ``vacuum misalignment'', a mechanism by which the composite Higgs boson can effectively behave as an elementary one. This concept is explained in Sect.~\ref{sec:BI} and further illustrated in Sect.~\ref{sec:B2} with the help of two simple examples. Sect.~\ref{CCWZ} is more technical. It reviews the Callan--Coleman--Wess--Zumino (CCWZ) approach to spontaneously broken symmetries, an essential tool for the study of the composite Higgs scenario. Finally, in Sect.~\ref{sec:PC} we explain the concept of partial fermion compositeness, which is how the elementary SM fermions are assumed to couple with the composite sector in our construction.

\section{Vacuum misalignement}
\label{sec:BI}

We now consider, in addition to the SM fermions and gauge fields, a new sector endowed with a global Lie group of symmetries ${\mathpzc{G}}$. This is the ``composite sector'', in the language of the previous chapter. Since we want our setup to address the Naturalness problem by the mechanism of dimensional transmutation we imagine dealing with a strongly-interacting sector, conceptually similar to a QCD-like confining theory. However the forthcoming discussion applies to weakly-coupled theories as well and indeed illustrative weakly-coupled examples will be worked out in the following section. We assume that the vacuum state of the composite sector, when the latter is considered in isolation, is only invariant under a subgroup ${\mathpzc{H}}\subset {\mathpzc{G}}$ leading to ${\mathpzc{G}}\to {\mathpzc{H}}$ spontaneous breaking and thus to the appearance of massless NGB's in the coset ${\mathpzc{G}}/{\mathpzc{H}}$. The subgroup ${\mathpzc{H}}$ is assumed to contain the EW group $G_{\textrm{EW}}=\textrm{SU}(2)_L\times\textrm{U}(1)_Y\subseteq {\mathpzc{H}}$ and ${\mathpzc{G}}$ is assumed to be large enough for at least one Higgs doublet to be present in the coset. In order to study this system we must first of all introduce a reference system in the Lie algebra of ${\mathpzc{G}}$ by choosing a basis of linearly independent generators $T^A$ and splitting them into ``unbroken'' ($A=a=1,\ldots,{\textrm{dim}}[{\mathpzc{H}}]$) and ``broken'' ($A=\hat{a}=\hat{1},\ldots,{\textrm{dim}}[{\mathpzc{G}}/{\mathpzc{H}}]$) sets as
\beq
\{T^A\}\,=\,\{T^a,\,\Th^{\hat{a}}\}\,.
\label{gengen}
\eeq
The set $\{T^a\}$ generates the Lie algebra of the subgroup ${\mathpzc{H}}$. It is also convenient to introduce a 
reference vacuum field configuration $\ve{F}$, that describes one of the degenerate vacua of the composite sector. 
It is chosen to satisfy
\beq
T^a\ve{F}=0\,,\;\;\;\;\;\Th^{\hat{a}}\ve{F}\neq 0\,.
\label{vacgen}
\eeq
To be precise, what we mean with the second equation is that $\{\Th^{\hat{a}}\ve{F}\}$ forms a linearly independent (over the reals) set of vectors.

Notice that Eq.~(\ref{gengen}), and consequently Eq.~(\ref{vacgen}), is merely a conventional choice of the reference system in the ${\mathpzc{G}}$ algebra. From the viewpoint of the composite sector alone, for which ${\mathpzc{G}}$ is an exact symmetry, there is no preferred system. 
Any embedding of ${\mathpzc{H}}$ in ${\mathpzc{G}}$, obtained by acting on Eq.~(\ref{gengen}) with ${\mathpzc{G}}$ elements, is completely equivalent.\footnote{This is clearly not the case when ${\mathpzc{H}}$ is embeddable in multiple inequivalent ways in ${\mathpzc{G}}$, namely when different choices of the ${\mathpzc{H}}$ algebra generators are not all related by inner automorphisms. Which inequivalent embedding is selected is in this case a dynamical question and depends on the details of the underlying theory.} However, ${\mathpzc{G}}$ is eventually broken in our construction by identifying some of its generators with those of the EW group. In view of this breaking, it is convenient to choose the reference system in such a way that the embedding of ${\mathpzc{H}}$ contains all the $G_{\textrm{EW}}$ generators. Namely, the SM gauge fields $W_\mu^{1,2,3}$ and $B_\mu$, that gauge the $G_{\textrm{EW}}$ group, will couple to some of the global currents associated with the $\{T^a\}$'s and not to the $\{\Th^{\hat{a}}\}$'s. This is our definition of the $\{T^a\}$ set and of the reference vacuum $\ve{F}$, it does not entail any assumption on the ${\mathpzc{G}}\to {\mathpzc{H}}$ symmetry breaking pattern. 

As well known, and reviewed in Sect.~\ref{CCWZ}, the NGB fields are local transformations in the direction of the $\{\Th^{\hat{a}}\}$ generators and correspond to the ansatz
\beq
\displaystyle
\ve{\Phi}(x)=e^{i\,\theta^{\hat{a}}(x) \Th^{\hat{a}}}\ve{F}\,,
\label{gof}
\eeq
in the space of the field operators $\ve{\Phi}$ of the theory. Among the $\theta_{\hat{a}}$ fields we identify the four real components of one Higgs doublet, plus possibly other scalars of an enlarged Higgs sector. The Higgs field taking a Vacuum Expectation Value (VEV) eventually breaks $G_{\textrm{EW}}$ down to the electromagnetic group exactly like in the SM. To illustrate how this works, let us first consider the composite sector in isolation and ignore the ${\mathpzc{G}}$-breaking perturbations that arise from the coupling with the SM gauge fields.\footnote{The couplings with the SM fermions also break ${\mathpzc{G}}$ explicitly, as we will see in Sect.~\ref{sec:PC}.} In this case the $\theta$ fields are exact NGB's, therefore they have no potential and their VEV's $\langle\theta_{\hat{a}}\rangle$ are completely arbitrary. Moreover the VEV's are unobservable because any constant $\theta$ configuration merely corresponds to one equivalent vacuum obtained by acting on $\ve{F}$ with the ${\mathpzc{G}}$ transformation $\exp[-i \langle\theta_{\hat{a}}\rangle \Th^{\hat{a}}]$. Technically, we will be able to get rid of any $\langle\theta_{\hat{a}}\rangle$ by a suitable redefinition of the $\theta$ fields that induces the transformation $\ve{\Phi}\to\exp[-i \langle\theta_{\hat{a}}\rangle \Th^{\hat{a}} ] \ve{\Phi}$. In this way it is possible to set, in full generality, $\langle\theta^{\hat{a}}\rangle=0$. The concept that the composite Higgs VEV is unobservable in the absence of explicit breaking of ${\mathpzc{G}}$ is often useful in the study of composite Higgs theories.

\begin{figure}
\centering
\includegraphics[width=0.42\textwidth]{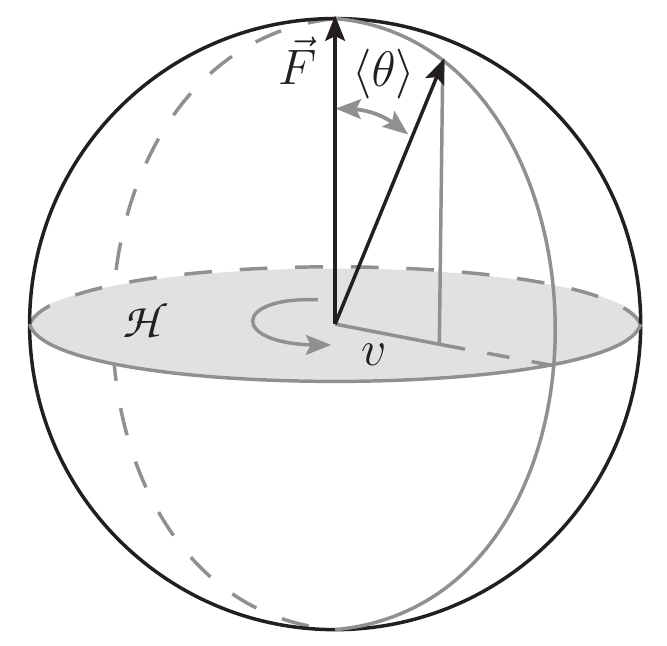}
\caption{A geometrical illustration of EWSB through vacuum misalignment, in the case of the spatial rotations group \mbox{${\mathpzc{G}}=\;$SO$(3)$} with \mbox{${\mathpzc{H}}=\;$SO$(2)$}. The \mbox{SO$(2)$} breaking from vacuum misalignment is proportional to the projection of $\vec{F}$ on the \mbox{SO$(2)$} plane, $v=f\sin\langle\theta\rangle$.}
\label{vmis}
\end{figure}

When we take ${\mathpzc{G}}$-breaking into account and $\theta$ becomes a pseudo NGB (pNGB) the situation changes. First of all, $\theta$ develops a potential and its VEV is not arbitrary anymore. Moreover, $\langle\theta\rangle$ becomes observable as it can not be set to zero by an exact symmetry transformation. Its physical effect is to break $G_{\textrm{EW}}$, embedded in ${\mathpzc{H}}$, giving rise to EWSB. Geometrically, as depicted in Fig.~\ref{vmis}, $\langle\theta\rangle$ measures the angle by which the vacuum is misaligned with respect to the reference vector $\ve{F}$, which we have chosen to be orthogonal to the plane of ${\mathpzc{H}}\supseteq G_{\textrm{EW}}$. The convenience of this choice should now be clear: the field $\theta$ defined by Eq.~(\ref{gof}) behaves exactly like the SM Higgs field in the sense that its non-vanishing VEV triggers EWSB. More precisely, we expect all the EWSB effects such as the SM particle masses to be controlled by the projection of $\ve{F}$ on the $G_{\textrm{EW}}$ plane, {\it{i.e.}}~we expect the EWSB scale to be set by $v=f\sin\langle\theta\rangle$ where $f=|\ve{F}|$ is the scale of ${\mathpzc{G}}\to {\mathpzc{H}}$ spontaneous breaking. This expectation is confirmed by the examples that follow.

The actual value of $\langle\theta\rangle$ depends on the details of the composite sector and on those of the symmetry-breaking perturbations. It can be obtained, in each given explicit model, by minimizing the pNGB potential. In the absence of some special mechanism or of an ad-hoc cancellation, we generically expect a minimum for $\langle\theta\rangle\sim1$. Namely, the vacuum does not generically point in a direction close to $\ve{F}$ and the EW symmetry is maximally broken, {\it{i.e.}}~$v\sim f$. If this is the case our setup is merely a non-minimal technicolor model  where we enlarged the group ${\mathpzc{H}}$ to contain also the broken SM generators and not only the electromagnetic \mbox{U$(1)$}. But if these additional generators are broken with maximal strength $v\sim f$ we will find no qualitative difference with the minimal technicolor case, aside from the presence of additional pNGB associated with the new broken generators. The composite Higgs construction becomes interesting, and different from technicolor, only if the misalignment angle is small, $\langle\theta\rangle\ll1$, such that a gap is generated among $f$ and the EWSB scale $v$. This condition is conveniently expressed as
\beq
\xi\,\equiv\,\frac{v^2}{f^2}=\sin^2\langle\theta\rangle\ll1\,,
\eeq
in terms of the important parameter $\xi$ which appears ubiquitously in the study of composite Higgs. The limit $\xi\to0$, at fixed $v$, corresponds to decoupling the composite sector from the low-energy physics by sending to infinity its typical scale $f$. In this limit, only the Goldstone boson Higgs remains in the spectrum while all the other bound states decouple. The theory, as shown explicitly in the examples that follow, systematically reduces to the SM for $\xi\to0$ and the composite Higgs becomes effectively elementary. Unlike technicolor, composite Higgs theories are endowed with one adjustable parameter $\xi$ that controls all the departures from the Standard Higgs model. The experimental confirmations of the SM, in particular its successful description of EW precision physics, can be systematically recovered by a small enough $\xi$. This mechanism is called ``vacuum misalignment'' \cite{Kaplan:1983fs,Kaplan:1983sm,Dugan:1984hq}.

However, we generically expect a large misalignment angle, $\langle\theta\rangle\sim1$, and therefore $\xi\sim1$. Two attitudes are possible towards the problem of obtaining a small enough $\xi$. The first one is to assume a certain degree of accidental cancellation, or fine-tuning, taking place in the scalar potential ensuring $\xi\ll1$. Though not completely satisfactory, this might well be the correct explanation as long as not too a small $\xi$, not much smaller than around $0.1$, is required. Moderate cancellations, of the order of one part into ten, are acceptable from the viewpoint of Naturalness, or at least of its formulation in terms of fine-tuning we insisted on in the previous chapter. A microscopic model of EWSB, which explains the huge Planck (or GUT) to weak hierarchy and complies with all the precise tests of the SM at the only price of a $10\%$ tuning of a single parameter would definitely be an extremely plausible possibility. The second approach to the small $\xi$ problem is to try to design some specific mechanism which leads to a small $\langle\theta\rangle$, not by tuning but rather through a structural cancellation taking place in the potential. One incarnation of this idea is provided by the so-called ``little Higgs'' program \cite{ArkaniHamed:2002qx,ArkaniHamed:2002qy} (for a review see Refs.~\cite{Schmaltz:2005ky,Perelstein:2005ka}), which foresees a parametric reduction of the Higgs mass term in the potential relative to the quartic, leading to a naturally small VEV. Unfortunately this is not achieved by reducing the Higgs mass term, but by enlarging the quartic, leading to a serious tension with the observed Higgs mass. Furthermore the proposed constructions are rather complicated and rely on seemingly artificial model-building. In comparison, the explanation based on some degree of unnatural tuning seems more plausible. We will not elaborate any further on these ideas, nevertheless they could be the right starting point towards the construction of a more Natural ``self-tuned'' version of the composite Higgs scenario.

\section{Two simple examples}
\label{sec:B2}

For a concrete illustration of the general idea we discuss two examples, formulated as renormalizable and weakly-coupled scalar theories or, as we will sometimes denote them in what follows, ``linear $\sigma$-models''. The first one is a toy model with \mbox{SO$(3)\to$ SO$(2)$} spontaneous breaking, which provides a composite realization of the Abelian Higgs model. The second example, based on the breaking \mbox{SO$(5)\to$ SO$(4)$}, is instead more realistic and leads to a SM-like Higgs theory of EWSB.

\subsection{The Abelian composite Higgs model}

We consider a triplet $\ve\Phi$ of real scalar fields, described by the Lagrangian
\beq
{\mc{L}}_C=\frac12 \partial_\mu{\ve\Phi}^T\partial^\mu\ve\Phi-\frac{g_*^2}8\left({\ve\Phi}^T\ve\Phi-f^2\right)^2\,.
\label{old}
\eeq
In spite of being just a simple scalar theory, which we will study in the perturbative weakly-coupled regime $g_*<4\pi$, we interpret it here as the strongly-interacting composite sector we described in the previous chapter. The theory is invariant under \mbox{SO$(3)$} transformations acting on $\ve\Phi$ as
\beq
\ve\Phi\;\to\;g\cdot\ve\Phi\,,\;\;\;\;\;g=e^{i\alpha_AT^A}\in\textrm{SO}(3)\,,
\eeq
where the \mbox{SO$(3)$} generators, normalized to $\Tr[T^AT^B]=\delta^{AB}$, can be conveniently chosen as $T^A=\{T,\,\Th^{i}\}$
\beq
\displaystyle
T=\frac1{\sqrt{2}}\left[\begin{array}{rrr}
\ 0 & -i & \ 0\\
i & 0 & 0\\
0& 0 & 0
\end{array}\right]\,,\;\;\;\;\;\Th^{i}=\left\{
\frac1{\sqrt{2}}\left[\begin{array}{rrr}
\ 0 & \ 0 & -i\\
0 & 0 & 0\\
i & 0 & 0
\end{array}\right]\,,\,\frac1{\sqrt{2}}\left[\begin{array}{rrr}
0 & 0 & 0\\
\ 0 & \ 0 & -i \\
0& i & 0
\end{array}\right]
\right\}\,,
\label{gso3}
\eeq
with $i={1,2}$. Geometrically, the three generators correspond to rotations in the \mbox{$1$-$2$}, \mbox{$1$-$3$} and \mbox{$2$-$3$} planes. 

The field $\ve\Phi$ acquires a non vanishing VEV breaking \mbox{SO$(3)$} to the \mbox{SO$(2)$} subgroup of rotations around $\langle\ve\Phi\rangle$. The tree-level minimization condition reads $\langle\vet\Phi\rangle \langle\ve\Phi\rangle=f^2$, so that the manifold of equivalent vacua is the two-sphere depicted in  Fig.~\ref{vmis}. Given the basis ($\ref{gso3}$) we adopted for the generators, the representative vacuum, selected by the condition in Eq.~(\ref{vacgen}), reads
\beq
\displaystyle
\ve{F}=\left[\begin{array}{ccc}
0\\0\\ f
\end{array}
\right]\,.
\eeq
In order to study the fluctuations around the vacuum it is convenient to perform a field redefinition and to trade the three $\vec\Phi$ components for one radial coordinate $\sigma$ plus two ``angular'' variables $\Pi_{1,2}$ (the Goldstone fields) describing the fluctuations around the broken generators as in Eq.~(\ref{gof}). We write
\beq
\displaystyle
\ve\Phi=e^{i\frac{\sqrt{2}}{f} 
\Pi^{i}(x) \Th_{i}
}
\left[
\begin{array}{c}
 0\\
 0\\
f+\sigma(x)
\end{array}
\right]\,,
\label{fred}
\eeq
where the normalization factor has been chosen (see below) to obtain a canonical kinetic term for the Goldstone boson fields $\Pi_i$. The exponential matrix in the above equation is a space-time dependent element of \mbox{SO$(3)$} which we call the ``Goldstone matrix'' $U\left[\Pi\right]$. It can be defined for any ${\mathpzc{G}}\rightarrow {\mathpzc{H}}$ breaking and it ubiquitously appears in composite Higgs. 

The Goldstone matrix can be computed explicitly in this simple case and it is given by
\beq
\displaystyle
U\left[\Pi\right]=e^{i\frac{\sqrt{2}}{f} \Pi_{i}(x) \Th^{i}}
=\left[
\begin{array}{cc}
\Id-\big(1-\cos{\frac{\Pi}{f}}\big)\frac{\ \ve\Pi\,\vet{\Pi}}{\Pi^2} & \ \ \sin{\frac{\Pi}{f}}\frac{\ve\Pi}{\Pi}\ \\
- \sin{\frac{\Pi}{f}}\frac{\ \vet\Pi}{\Pi} & \cos{\frac{\Pi}{f}}
\end{array}
\right]\,,
\label{uform}
\eeq
where $\Pi=\sqrt{\vet\Pi\ve\Pi}$. Actually, the expression above is more general and holds for any \mbox{SO$(N)\rightarrow\;\;$SO$(N-1)$} breaking provided the $N-1$ broken generators are chosen, in analogy with Eq.~(\ref{gso3}), to have one non-vanishing entry in the last line and column. The field redefinition (\ref{fred}) becomes
\beq
\ve{\Phi}=(f+\sigma)\left[
\begin{array}{c}
\sin{\frac{\Pi}{f}}\frac{\ve{\Pi}}{\Pi}\\
\cos{\frac{\Pi}{f}}
\end{array}
\right]\,.
\eeq
We see that the new variables furnish a full one-to-one parametrization of the field space, aside from the singular point $\ve\Phi=0$, provided $f+\sigma$ is taken to be positive and the Goldstones are restricted to the region $\Pi\in[0,\pi f)$. By substituting in the Lagrangian we straightforwardly obtain
\bea
\label{lagnl}
&&{\mathcal{L}}_C=\frac12\partial_\mu\sigma \partial^\mu\sigma-\frac{(g_*f)^2}{2}\sigma^2-\frac{g_*^2 f}2\sigma^3
-\frac{g_*^2}8\sigma^4\\
&&+\frac12\left(1+\frac\sigma{f}\right)^2\left[
\frac{f^2}{\Pi^2}\sin^2{\frac\Pi{f}}\partial_\mu\vet{\Pi}\partial^\mu\ve{\Pi}+\frac{f^2}{4\Pi^4}\left(
\frac{\Pi^2}{f^2}-\sin^2{\frac{\Pi}{f}}
\right)\partial_\mu\Pi^2\partial^\mu\Pi^2
\right]\,.\nonumber
\eea

Many interesting and generic properties of the composite Higgs scenario are well illustrated by the expression above. First, by Taylor-expanding around $\Pi=0$ (which is a perfectly regular point) we see that the Lagrangian contains an infinite set of local interactions involving an arbitrary number of Goldstone fields but only two derivatives. Each Goldstone leg insertion is weighted by the Goldstone symmetry breaking scale $f$. This is simply because $\ve{\Pi}$ enters in the Goldstone matrix as $\ve{\Pi}/f$. In analogy with the theory of QCD pions, where the role of $f$ is played by the pion decay constant $f_\pi$, we will sometimes refer to $f$ as the ``Higgs decay constant''. In agreement with the Goldstone theorem the $\Pi$'s describe two massless bosons associated with the two broken generators $\Th^{{1},{2}}$. 

The $\sigma$ field has instead a mass
\beq
m_*=g_* f\,.
\eeq
In analogy with a strongly coupled sector, which we would like to mimic by our example, the $\sigma$ particle is called a ``resonance''. We generically call a resonance any particle that emerges from the composite sector aside from the Goldstone bosons. In the analogy, the mass $m_*$ corresponds to the strong sector's confinement scale, conceptually similar to the QCD scale $\Lambda_{\textrm{QCD}}$. The parameter $g_*$ controls the interactions in our Lagrangian. It is thus interpreted as an effective low-energy coupling of the composite sector. In Chap.~\ref{ch:SILH} we will see that in a genuine strong theory $g_*$ could easily be of order $4\pi$, outside the perturbative regime, but it could also be parametrically reduced in the case of a confining gauge group with a large number of colors. 

We now inspect the symmetries of the Lagrangian in the non-linear form of Eq.~(\ref{lagnl}). We immediately recognize the presence of an $\SO(2)$ group under which $\ve\Pi$ forms a doublet and transforms as
\beq
\displaystyle
\ve\Pi\to e^{i\alpha\sigma_2}\ve\Pi\,.
\eeq
We call this a ``linearly realized'' symmetry as it acts in a linear and homogeneous way on the field variables. We can switch to the complex notation by defining
\beq
\label{cfi}
H=\frac{\Pi_1-i\,\Pi_2}{\sqrt{2}}\,,
\eeq
which we identify with the Higgs field, with unit charge under \mbox{U$(1)=\SO(2)$}, of the Abelian Higgs model we are constructing. Obviously, the linearly realized $\SO(2)$ invariance follows from one of the symmetries of the original Lagrangian (\ref{old}). It is indeed immediate to see that it induces an $\SO(3)$ rotation along the unbroken generator $T$
\beq
\displaystyle
\ve\Pi\to e^{i\alpha\sigma_2}\ve\Pi\;\;\;\mathlarger{\Leftrightarrow}\;\;\ve\Phi\to e^{i\sqrt{2}\alpha T}\ve\Phi\,.
\eeq
This correspondence guarantees the invariance of the Lagrangian in Eq.~(\ref{lagnl}), which is a mere rewriting of the original one. 

For the two broken generators $\Th^{i}$ identical considerations hold. Therefore, even if it would have been hard to tell at a first sight, Eq.~(\ref{lagnl}) must have other symmetries. It is not hard to work out, at the infinitesimal level, the Goldstone field transformations that induce rotations of $\ve\Phi$ along the broken generators. These read
\bea
\label{brtr}
&\displaystyle\ve\Pi\to\ve\Pi+\Pi \cot{\frac{\Pi}{f}}\, \ve\alpha+\left(
\frac{f}{\Pi} - {\cot{\frac{\Pi}{f}}}
\right)\left(\vet\alpha\ve\Pi\right)\frac{\ve\Pi}{\Pi}\,,&\\
&\displaystyle\mathlarger{\mathlarger{\Updownarrow}}&\nonumber\\
&\ve\Phi\to\ve\Phi+i\alpha_{i}\Th^{i}\vec\Phi\,.&\nonumber
\eea
As for the unbroken \mbox{U$(1)$}, the correspondence ensures that the transformations above are symmetries of the Lagrangian in Eq.~(\ref{lagnl}). This can also be directly verified by a lengthy but straightforward calculation. Differently from those associated with the unbroken \mbox{U$(1)$}, the broken transformations act non-linearly on the Goldstone field variables $\Pi$ and thus they are said to be ``non-linearly realized''. Moreover they also act non-homogeneously, in the sense that the zero field configuration is transformed into one with constant $\ve\Pi$ fields, {\it{i.e.}}~$\ve{0}\to f\ve{\alpha}$. Conversely, any constant field configuration, such as the one that defines a generic vacuum $\langle\ve\Pi\rangle$, can be transformed into the trivial vacuum $\langle\ve\Pi\rangle=0$. This implies that, as explained on general grounds in the previous section, the composite Higgs VEV has no physical effect in the absence of an explicit breaking of the Goldstone symmetry. Clearly for the argument above being conclusive we would need the finite form of the transformation, which however is too involved to be written explicitly. An implicit but compact form is reported in the following section.

Now that a NGB Higgs scalar has been obtained the last ingredient to construct the Abelian Higgs model is a \mbox{U$(1)$} gauge field. Rather intuitively it is introduced by gauging the unbroken \mbox{U$(1)$} subgroup, namely by replacing in the original Lagrangian
\beq
\partial_\mu\ve\Phi\;\;\mathlarger{\rightarrow}\;\;D_\mu\ve\Phi=\left(\partial_\mu-i\,\sqrt{2} e\, A_\mu T\right)\ve\Phi\,,
\label{covd}
\eeq
where $A_\mu$ is a \mbox{U$(1)$} gauge field with canonical kinetic term. In the language of Chap.~\ref{ch:introduction}, $A_\mu$ is an elementary sector field and its gauge couplings with $\Phi$ are elementary/composite interactions. The gauging, since it selects one generator among three, breaks \mbox{SO$(3)$} explicitly to \mbox{SO$(2)$}. The composite Higgs has now became a pNGB. One might wonder whether the choice of the embedding in \mbox{SO$(3)$} of the \mbox{SO$(2)$} gauge group, which we take exactly aligned with the generator $T$, hides some dynamical assumption. In view of this possible confusion we stressed in the previous section that the choice of the generators of the ${\mathpzc{G}}$ group, performed in such a way that the gauged directions are exclusively in ${\mathpzc{H}}$, is completely conventional and does not rely on any assumption. In the present example this is immediately verified. Suppose we had started from a generic embedding of the gauge group, defined by an arbitrary (but normalized) linear combination of generators $\overline{T}$. In full generality, $\overline{T}$ can be rotated to $T$ by an \mbox{SO$(3)$} transformation, namely it can be expressed as ${\overline{T}}={\overline{g}}T{\overline{g}}^T$. In this case our prescription for the choice of the reference system in the algebra would have been to work with rotated generators, obtained by acting with ${\overline{g}}$ on Eq~(\ref{gso3}). However this would not have changed the results because $\overline{g}$ can be eliminated by a field redefinition $\ve\Phi\rightarrow{\overline{g}}\,\ve\Phi$ and thus it has no physical effect. Therefore the choice $\overline{T}=T$ in Eq.~(\ref{covd}) is completely general.

We can finally write down our Abelian composite Higgs theory. The only effect of the gauging is to turn ordinary derivatives into covariant ones in Eq.~(\ref{lagnl}), with
\beq
D_\mu\ve\Pi=\left(\partial_\mu-i\,e\,A_\mu\sigma_2\right)\ve\Pi\,.
\eeq
By turning to the complex field notation the terms of the Lagrangian involving the Higgs become
\bea
\frac12\left(1+\frac\sigma{f}\right)^2&&\left[
\frac{f^2}{|H|^2}\sin^2{\frac{\sqrt{2} |H|}{f}} D_\mu H^\dagger D^\mu H\right.\\
&&\left.+\frac{f^2}{4|H|^4}\left(
2\frac{|H|^2}{f^2}-\sin^2{\frac{\sqrt{2}|H|}{f}}
\right)\left(\partial_\mu |H|^2\right)^2
\right]\,.\nonumber
\label{lagab}
\eea
while the $\sigma$ field Lagrangian remains unchanged. The covariant derivative in the previous equation is just the usual one
\beq
D_\mu H =\partial_\mu H -i\,e\,A_\mu H\,.
\eeq

Now that the Goldstone symmetry has been broken by the gauging two new important features emerge. First, the emergence of a Higgs potential is no longer forbidden, as it would be in the unbroken case because of the non-linearly realized symmetry of Eq.~(\ref{brtr}). Even if the potential still vanishes in our tree-level Lagrangian, it is radiatively generated by the gauge field loops, which transmit to the Higgs sector the Goldstone symmetry breaking. This potential, whose generation is not particularly enlightening and not worth discussing in this example, eventually gives a VEV to the composite Higgs field. Second, the Higgs VEV becomes observable and the breaking of the \mbox{U$(1)$} symmetry can take place. By setting the Higgs to its VEV  
\beq
\displaystyle
H=\langle H\rangle\equiv \frac{V}{\sqrt{2}}\,,
\eeq
the first term in the square bracket of Eq.~(\ref{lagab}) gives to the gauge field a mass 
\beq
m_A=e f \sin\frac{V }{f}\equiv e\, v\,.
\eeq
In the second equality of the above equation we have defined the scale $v$ of \mbox{U$(1)$} symmetry breaking in analogy with the ordinary elementary Abelian Higgs mass formula. In the latter case the scale $v$ is directly provided by the Higgs field VEV while in the composite case 
\beq
\label{xval}
\displaystyle
v=f \sin\frac{V }{f} \;\;\Rightarrow \;\; \xi =\frac{v^2}{f^2}=\sin^2\frac{V }{f}\,.
\eeq
The situation is thus precisely the one depicted in Fig.~\ref{vmis}, where the symmetry breaking scale is provided by the projection of the vacuum configuration on the plane associated with the unbroken generator.

\subsection{The minimal composite Higgs model}
\label{sec:MCHMlin}

It is not hard to turn our toy example into a model of EWSB. To this end we must enlarge the unbroken group in a way that it contains $G_{\textrm{EW}}$. Furthermore we need at least four spontaneously broken generators giving rise to one complex doublet of NGB Higgs fields. One possibility is to consider \mbox{SO$(5)$} spontaneously broken to an \mbox{SO$(4)$} subgroup. As explained in Appendix~\ref{so4}, \mbox{SO$(4)$} is locally isomorphic to the chiral group \mbox{SU$(2)_L\times$SU$(2)_R$}. We interpret the \mbox{SU$(2)_L$} factor as the SM one and we identify the hypercharge with the third \mbox{SU$(2)_R$} generator, $Y=t_R^3$. The $10$ \mbox{SO$(5)$} generators acting on the $\mathbf{5}$ representation, normalized as \mbox{Tr$[T^AT^B]=\delta^{AB}$}, are conveniently taken to be $T^A=\{T^a,\widehat{T}^i\}$ where 
\bea
\label{so5gen}
&&T^a=\left\{T_L^\alpha=\left[\begin{array}{cc} t_L^\alpha & 0 \\
0 & 0
\end{array}
\right]\,,\;
T_R^\alpha=\left[\begin{array}{cc} t_R^\alpha & 0 \\
0 & 0
\end{array}
\right]
\right\}\,,\nonumber\\
&&\displaystyle\left(\widehat{T}^i\right)_{IJ}=-\frac{i}{\sqrt{2}}\left(\delta_I^i\delta_J^5-\delta_J^i\delta_I^5\right)\,.
\eea
The $6$ first generators $T_a$ span the \mbox{SO$(4)$} sub-algebra and are written in \mbox{SU$(2)_L\times$SU$(2)_R$} notation in terms of the $4\times4$ generators $t_{L,R}^\alpha$ defined in Appendix~\ref{so4}. The remaining four, $\Th_i$, are instead broken and the associated NGB fields provide the two complex Higgs doublet components. As shown below the Goldstones transform in the ${\mathbf{4}}=\mathbf{(2,2)}$ of the unbroken \mbox{SO$(4)$}  and thus they have the correct SM quantum numbers to be identified with the Higgs field by applying Eq.~(\ref{hd}) in Appendix~\ref{so4}.

The composite sector Lagrangian is again the one in Eq.~(\ref{old}), where $\ve\Phi$ is now an \mbox{SO$(5)$} fiveplet. The manifold of equivalent vacua is the $4$-sphere and the representative vacuum configuration points along the fifth component. In complete analogy with the Abelian model example, the $5$ real components of $\ve\Phi$ are conveniently parametrized as 
\beq
\displaystyle
\ve\Phi=e^{i\frac{\sqrt{2}}{f} 
\Pi_{i}(x) \Th^{i}
}
\left[
\begin{array}{c}
 \ve 0\\
f+\sigma(x)
\end{array}
\right]=(f+\sigma)\left[
\begin{array}{c}
\sin{\frac{\Pi}{f}}\frac{\ve{\Pi}}{\Pi}\\
\cos{\frac{\Pi}{f}}
\end{array}
\right]\,,
\label{fred5}
\eeq
in terms of the resonance field $\sigma$ and of four NGB's $\ve\Pi$. The above expression was derived by employing Eq.~(\ref{uform}) which, as previously stated, applies in general to \mbox{SO$(N)\rightarrow$ SO$(N-1)$} and thus in particular to the present case. By substituting in the Lagrangian we obtain Eq.~(\ref{lagnl}) exactly like in the Abelian model. 

The symmetry content of the Lagrangian is also a trivial generalization of the Abelian case. The linearly-realized group consists now of \mbox{SO$(4)$} rotations of the $\ve\Pi$ fourplet and it corresponds to the action of the unbroken generators $T^a$ on $\ve\Phi$. This is immediately verified by noticing that a rotation of $\ve\Pi$ in Eq.~(\ref{fred5}) induces a rotation of $\ve\Phi$ in the \mbox{SO$(4)$} subgroup embedded in the first $4\times4$ block, {\it{i.e.}}
\beq
\displaystyle
\ve\Pi\to e^{i\alpha_a t^a}\ve\Pi\;\;\mathlarger{\Leftrightarrow}\;\;\ve\Phi\to e^{i \alpha_a T^a}\ve\Phi\,.
\eeq
As anticipated, the Goldstones live in the fourplet of \mbox{SO$(4)$} and thus they can be expressed, by inverting Eq.~(\ref{hd}) in terms of the two Higgs doublet components $H=(h_u,h_d)^T$ as
\begin{equation}
\label{dh}
\vec{\Pi}=\left[\begin{matrix}\Pi_1\\ \Pi_2\\ \Pi_3\\ \Pi_4\end{matrix}\right]=
\frac1{\sqrt{2}}\left[\begin{matrix}-i\,(h_u-h_u^\dagger)\\ 
h_u+h_u^\dagger
 \\ i\,(h_d-h_d^\dagger)\\ 
h_d+h_d^\dagger
 \end{matrix}\right]\,.
\end{equation}
The theory is of course also invariant under four non-linearly realized transformations associated with the broken generators $\widehat{T}^i$. Their infinitesimal action on the fields is the same as in the Abelian model, reported in Eq.~(\ref{brtr}). 

The electroweak interactions are introduced in the theory by gauging, with coupling strength $g$ and $g'$, the \mbox{SU$(2)_L\times$U$(1)_Y$} subgroup of \mbox{SO$(4)$}. The covariant derivative reads
\beq
D_\mu\ve\Phi=\left(\partial_\mu -i\, g W_{\mu}^{\alpha} T_L^\alpha - i\, g'B_\mu T_R^3\right)\ve\Phi\,.
\label{covdev4}
\eeq
Non-Abelian gauge kinetic terms are also introduced and collected in a purely elementary Lagrangian
\beq
\label{galag}
\displaystyle
{\mathcal{L}}_E=-\frac1{4}W_{\mu\nu}^\alpha W^{\mu\nu}_\alpha-\frac14 B_{\mu\nu} B^{\mu\nu}\,.
\eeq
This implies that at leading order in $\xi$ the EW boson propagators and self-interactions vertices are identical to the SM ones. Important subleading modifications will be discussed in Chap.~\ref{ch:SILH}.

Now that the model is fully specified we can discuss its phenomenology. It describes the bosonic sector of the SM, namely the Higgs and the EW bosons, plus the resonance $\sigma$ with a mass
\beq
\label{resmass}
m_*=g_* f\,.
\eeq
Given that no particles are observed beyond the SM ones we will be interested in a situation where the resonance is heavy, with $m_*$ in the TeV or multi-TeV range much above the EW scale.\footnote{Actually in the present example, in which the resonance is just an EW-neutral scalar singlet, there is not a concrete phenomenological need of taking it so heavy. We assume $m_*$ above the TeV in order to mimic the generic situation encountered in the genuine strongly-coupled models we eventually aim to describe where such a strong bound applies.} This separation of scales emerges from two combined effects. First, some gap among $v$ and $f$ is due to the condition $\xi\ll1$ which, as anticipated in Sect.~\ref{sec:BI} and discussed at length in the following chapters, is essential to ensure the viability of the composite Higgs scenario. However making $\xi$ small costs fine-tuning and our hope is to achieve a realistic theory for $\xi \gtrsim 0.1$, for which the separation among $v$ and $f$ is moderate. Second, a gap between $f$ and $m_*$ is naturally achieved by a strong enough composite sector coupling $g_*$. Thorough this review we will consider $g_*$ values in the range $g_*\in (1,4\pi)$, taking however into account that a moderately large values $g_*>1$ are preferred to keep the resonances far from the EW scale. Because of the scale separation, resonances are effectively decoupled from the SM particles and their presence can be simply ignored at a first approximation as we will do in the forthcoming discussion. Their effects on the SM particles phenomenology can be treated in a low-energy effective theory expansion as we will see in Chap.~\ref{ch:SILH}.

Ignoring the resonance, the Lagrangian (\ref{lagnl}) becomes
\begin{equation}
\label{CHlag0}
\frac{f^2}{2|H|^2}\sin^2{\frac{\sqrt{2} |H|}{f}} D_\mu H^\dagger D^\mu H+\frac{f^2}{8|H|^4}\left(
2\frac{|H|^2}{f^2}-\sin^2{\frac{\sqrt{2}|H|}{f}}
\right)\left(\partial_\mu |H|^2\right)^2,
\end{equation}
where we employed the standard Higgs covariant derivative
\beq
\displaystyle
D_\mu H =\left(\partial_\mu  -i\,g\,W_{\mu}^{\alpha} \frac{\sigma_\alpha}2-i\,g'\,B_\mu\frac{\Id}2 \right) H\,.
\eeq
The phenomenological implications of Eq.~(\ref{CHlag0}) can be illustrated in two ways, which we describe in turn. The first approach is to compute directly the physical couplings by going to the unitary gauge, defined as usual by
\beq
\label{ugauge}
\displaystyle
H=\left[\begin{array}{c} 0 \\ \frac{V+h(x)}{\sqrt{2}}\end{array}\right]\,,
\eeq
where $V$ denotes the Higgs VEV, which we take to be real without loss of generality, and $h(x)$ describes the physical Higgs fluctuations. In the unitary gauge the Lagrangian is surprisingly simple 
\beq
\frac12\left(\partial_\mu h\right)^2 + \frac{g^2}4 f^2 \sin^2{\frac{V+h}{f}} \left(|W|^2+\frac1{2c_w^2}Z^2\right)\,,
\label{LUN}
\eeq
where $W$ and $Z$ denote the ordinary SM mass and charge eigenstate fields, $c_w$ is the cosine of the weak mixing angle defined as usual by \mbox{$\tan\theta_w=g'/g$}. We immediately read the vector bosons masses
\beq
\displaystyle
m_W=c_w m_Z=\frac12 \, g f \sin\frac{V}{f}\equiv \frac12 \, g\, v\,,
\label{mwz}
\eeq
out of which we have extracted the definition of the physical EWSB scale $v\simeq 246$~GeV. As already discussed in the Abelian model, the latter is not directly provided by the composite Higgs VEV, but it is related to it and to the Higgs decay constant $f$ precisely as in Eq.~(\ref{xval}).

On top of the vector boson masses, the Lagrangian also contains an infinite set of local interactions involving two gauge and an arbitrary number of Higgs fields. By Taylor-expanding around $h=0$ we easily compute the first few terms
\beq
\displaystyle
\frac{g^2 v^2}{4} \left(|W|^2+\frac1{2c_w^2}Z^2\right)\left[2\sqrt{1-\xi}\,\frac{h}{v} +(1-2\xi)\,\frac{h^2}{v^2} -\frac43 \xi\sqrt{1-\xi}\,\frac{h^3}{v^3}+\ldots\right]\,,
\eeq
where we traded the parameters $V$ and $f$ for the physical EWSB scale $v$ and $\xi=v^2/f^2$. Exactly like in the SM we find single- and double-Higgs vertices, but with modified couplings
\beq
k_V\equiv \frac{g_{hVV}^{\textrm{CH}}}{g_{hVV}^{\textrm{SM}}}=\sqrt{1-\xi}<1\,,\;\;\;\;\;\frac{g_{hhVV}^{\textrm{CH}}}{g_{hhVV}^{\textrm{SM}}}=1-2\xi\,,
\label{hvc}
\eeq
Moreover, higher-dimensional vertices with more Higgs field insertions emerge and might trigger new interesting phenomena which are absent in the case of an elementary SM Higgs.  It is important to remark that in the limit $\xi\rightarrow0$, taken at fixed $v$ by sending $f\rightarrow\infty$, both these effects disappear. The couplings approach those of the elementary Higgs and the new interactions are suppressed, being weighted by inverse powers of $f$. The composite Higgs becomes effectively elementary in this limit.

An alternative way to inspect our Lagrangian in Eq.~(\ref{CHlag0}), which helps in clarifying why the composite Higgs reduces to the elementary SM one for small $\xi$, is to expand it for large $f$, obtaining a series of two-derivative operators with higher and higher energy dimension weighted by inverse powers of $f$. The first terms in this expansion, up to dimension $6$, are
\beq
\label{opexp}
D_\mu H^\dagger D^\mu H -\frac{2}{3 f^2} |H|^2 D_\mu H^\dagger D^\mu H+\frac{1}{6 f^2}\partial_\mu\left(H^\dagger H\right)\partial^\mu\left(H^\dagger H\right)+\ldots \,.
\eeq
The first term is just the SM Higgs kinetic Lagrangian and this clearly does not occur by accident. The reason is that the Higgs  enters in the Lagrangian only through the Goldstone boson matrix $U$ which in turn depends on the combination $H/f$. For $f\rightarrow\infty$ the only relevant term is the one with two Higgs field insertions, which corresponds to a $d=4$ renormalizable operator. Since the elementary Higgs theory is the most general renormalizable Lagrangian compatible with gauge invariance, it is the only model we might have ended up with. The situation would have been slightly different if we had considered a generic ${\mathpzc{G}}\rightarrow {\mathpzc{H}}$ symmetry breaking pattern, which delivers more Goldstone bosons than just one Higgs doublet. In that case the $f\to\infty$ limit would not correspond to the SM, but to a renormalizable theory of an extended Higgs sector. The presence of extra scalars, which potentially mix with the SM Higgs, might induce additional corrections to the couplings on top of those emerging from the higher-dimensional operators.

Going back to our model, and to the Lagrangian in Eq.~(\ref{opexp}), we see that the $d=6$ operators provide the leading corrections to the SM. Actually the two operators have the same effect on the Higgs coupling to the vector bosons because one can be transformed in the other by a field redefinition, up to operators of even higher dimension. Namely, by sending $H\rightarrow H + [|H|^2/(3 f^2)] H$ the first can be eliminated in favor of the second one with coefficient $1/2f^2$. In the notation of Ref.~\cite{Giudice:2007fh}, where the $d=6$ operators that emerge in theories with a pNGB Higgs are classified, the operator $(\partial |H|^2)^2$ is denoted by $O_H$. It is interesting to notice that $O_H$ does not correct the Higgs vertices with vector bosons directly but it induces, after EWSB, corrections to the physical Higgs kinetic term and new Higgs derivative self-interactions. When the the canonical kinetic term is restored by the appropriate rescaling this leads to a modification of the trilinear $hVV$ coupling, which can be readily checked to match with $k_V$ in Eq.~(\ref{hvc}) for small $\xi$. Computing the quadrilinear coupling $hhVV$ requires more care because the new trilinear $h$ interactions induced by $O_H$ also contribute to the physical $hhVV$ amplitude through virtual Higgs exchange. Therefore they must be eliminated by a further field redefinition before reading the physical coupling and reproducing the second equation in (\ref{hvc}). Details are reported in Appendix~B of Ref.~\cite{Giudice:2007fh}.

A crucial phenomenological virtue of our model, which we have not yet outlined, is that it respects the tree-level $\rho=1$ relation at all orders in $1/f$, where $\rho=m_W^2/(c_w m_Z)^2$, as apparent from Eq.~(\ref{mwz}). Correspondingly, the $d=6$ operator \footnote{We use again the notation of Ref.~\cite{Giudice:2007fh}, see also Ref.~\cite{Barbieri:2004qk} for the bounds on this operator from EW precision measurements.}
\beq
\displaystyle
O_T=\frac{1}{2 f^2}\left(H^\dagger \overleftrightarrow{D_\mu} H\right)\left(H^\dagger \overleftrightarrow{D^\mu} H\right)\,,
\eeq
which would induce $\rho-1\sim v^2/f^2= \xi$, is not present in the expansion of Eq.~(\ref{opexp}). Experimentally, $\rho=1$ is valid at the percent level and the deviations are well described by SM loop effects. The accuracy of the measurement bounds non--SM contributions to $\rho$ at the per-mille level. Therefore if $\rho-1$ was of order $\xi$ reconciling the model with observations would require $\xi\lesssim 10^{-3}$ and thus an unacceptable level of tuning. The reason why $\rho$ equals $1$ in our case is ``custodial symmetry'', namely the fact that the SM group generators are embedded in the global unbroken \mbox{SO$(4)$}, which is an exact symmetry of the composite sector and is only violated by the gauging. More precisely, since \mbox{SO$(4)\simeq\;$SU$(2)_L\times$SU$(2)_R$}, the gauging of \mbox{SU$(2)_L$} preserves $\SO(4)$ and the only breaking is due to the hypercharge. The Higgs is a fourplet and therefore its VEV breaks \mbox{SO$(4)$} down to the custodial \mbox{SO$(3)_c$} subgroup. The $W_\mu^\alpha$ fields transform as a triplet under $\SO(3)_c$. This symmetry is sufficient to fix the ratio among the $W$ and $Z$ bosons mass terms ensuring $\rho=1$. For a careful description of this mechanism the reader is referred to the original literature \cite{Sikivie:1980hm} and to Appendix~\ref{sec:custodial_symmetries}. The $\rho=1$ constraint is so strong that all the viable composite Higgs models must be endowed with custodial symmetry protection. Our \mbox{SO$(5)\rightarrow\;$ SO$(4)$} example is the ``Minimal Composite Higgs Model'' (MCHM) \cite{Agashe:2004rs}, in the sense that it delivers the minimal number of pNGB Higgs fields and relies on the minimal number of symmetry generators but still obeys custodial symmetry. An even more minimal possibility would be the \mbox{SU$(3)\rightarrow\;\;$SU$(2)\times$ U$(1)$} breaking. However it must be discarded because of the lack of custodial protection. See \cite{Gripaios:2009pe,Mrazek:2011iu} for non-minimal composite Higgs constructions where additional Higgs scalars emerge.

After reading this section, where we worked out in detail a model with a pNGB Higgs based on a weakly-coupled linear $\sigma$-model, the reader might be led to overestimate the importance of this kind of constructions. We thus stress that models of this sort are not interesting, the examples presented here have exclusively an illustrative purpose and the theories we eventually aim to discuss are very different from these. The first obvious limitations is that these models are unable to address the Naturalness problem since they are formulated in terms of an elementary scalar multiplet $\ve\Phi$. Its mass $m_*$ is sensitive to the UV physics in exactly the same way as the Higgs mass in the SM. Still, one might think that the Higgs boson being a pNGB might lead to some advantage, given that at least its mass is protected from large radiative corrections by the Goldstone symmetry. However this protection is insufficient, the Higgs mass receives quadratically divergent contributions from loop diagrams involving the SM fields, like the ones depicted in Fig.~\ref{HP}. In particular consider those from the gauge fields loops. They originate from the Higgs coupling to the vector bosons, which we saw above are not much different than in the ordinary SM. Therefore the result is expectedly similar, namely 
\beq
\label{hmdiv}
\delta m_H^2\simeq \frac{g^2}{16\pi^2}\Lambda^2\,.
\eeq
Notice that no special cancellation occurs because the gauge field couplings break the Goldstone symmetry explicitly so that the Higgs mass is not protected at the radiative level. The occurrence of a divergence in $m_H$, while no Higgs mass term was present in the original Lagrangian, also signals that the model is actually non-renormalizable. More precisely, the composite sector defined by Eq.~(\ref{old}) is renormalizable if considered in isolation, renormalizability gets spoiled by the coupling to gauge fields through the covariant derivative in Eq.~(\ref{covdev4}) which breaks the Goldstone symmetry. In other words, the breaking requires the introduction of Goldstone-breaking counterterms, among which the Higgs mass, which was not present in the Lagrangian in Eq.~(\ref{old}). By looking at Eq.~(\ref{hmdiv}), and taking also into account that a contribution from the top quark loop, identical to the SM one, would have arisen if we had tried to introduce the top quark in our model, we immediately realize that no progress has been made. The essential feature of composite Higgs being a solution of the Naturalness problem is the strongly-coupled nature of the underlying UV theory, by which the Higgs mass is stabilized through dimensional transmutation. The Goldstone symmetry is required to keep the Higgs naturally lighter than the other strong sector's resonances, it is in itself of no help in addressing the Naturalness problem.

\section{General CCWZ construction}
\label{CCWZ}

It is clear, also in light of the previous discussion, that we can not rely on the example of the previous section for the study of composite Higgs phenomenology. Some of the results we obtained, and in particular the pattern of Higgs coupling modifications in Eq.~(\ref{hvc}), are actually of general validity and do not rely on the specific model we considered, but in order to establish this fact we need a general treatment of theories with spontaneous symmetry breakdown: the famous Callan--Coleman--Wess--Zumino (CCWZ) construction \cite{Coleman:1969sm,Callan:1969sn}. This formalism allows to write general low-energy effective Lagrangians for strongly- or weakly-coupled theories characterised by a generic ${\mathpzc{G}}\rightarrow {\mathpzc{H}}$ symmetry breaking pattern, describing the Goldstone bosons associated with the breaking and the heavy resonances. It is also readily extended to incorporate explicit symmetry breaking, a property we will extensively make use of in Sect.~\ref{sec:PC}. Furthermore the generality of the method makes it an essential tool for the systematic study of non-minimal cosets, when trying to go beyond the canonical  \mbox{SO$(5)\rightarrow $\;\;SO$(4)$} example.

\subsection{The basic formalism}

The starting point is to identify the correct degrees of freedom that describe the massless NGB, one for each broken generator as predicted by the Goldstone theorem. Suitable candidates are obtained by considering, in the field space of the underlying theory, configurations that are related to the representative vacuum $\ve{F}$ by a local ${\mathpzc{G}}$ transformation, namely
\beq
\displaystyle
\ve{\Phi}(x)=e^{i\,\theta_{A}(x) T^{A}}\ve{F}\,,
\label{ansgen}
\eeq
where $T^A$ denotes, following the notation of Sect.~\ref{sec:BI}, the full set of generators of the group ${\mathpzc{G}}$. Each $\theta_A(x)$ is potentially a massless field because any constant $\theta_A$ configuration corresponds to a global symmetry transformation and thus it leads to one of the equivalent vacua of the theory, with the same energy as the original one. Therefore $\theta_A(x)$ has no potential and consequently zero mass, its energy is entirely kinetic and it originates from derivative terms in the Lagrangian. 

However not all these fields are physical, some of them are redundant and they can be dropped from the ansatz (\ref{ansgen}). In order to see how this works, and with the purpose of introducing the basic formula out of which CCWZ is constructed, we notice that a generic, global or local group element $g[\alpha_A]$, can be decomposed in a unique way as the product \footnote{It is trivial to verify the equation that follows for an infinitesimal group transformation and it is not hard to believe that it can be extended by continuity to finite group elements that are continuously connected to the identity.}
\beq
\displaystyle
g[\alpha_A]=e^{i\,\alpha_A T^A}=e^{i\,f_{\hat{a}} [\alpha]  \Th^{\hat{a}}} \cdot e^{i\,f_a[\alpha] T^a}\,,
\label{fund}
\eeq
where, as in Eq.~(\ref{gengen}), $T^a$ and $\Th^{\hat{a}}$ denote the unbroken and broken generators, respectively. The transformation on the right is an element of the subgroup ${\mathpzc{H}}$ and as such it leaves the representative vacuum invariant.\footnote{An analogous decomposition obviously holds with the ${\mathpzc{H}}$ element on the left.} By applying the above equation to the ansatz (\ref{ansgen}) we thus obtain
\beq
\displaystyle
\ve{\Phi}(x)=U\left[\Pi\right]\ve{F}\,,
\label{ansgol}
\eeq
where we defined the Goldstone matrix
\beq
\displaystyle 
U\left[\Pi\right]=e^{i\frac{\sqrt{2}}{f} \Pi_{\hat{a}}(x) \Th^{\hat{a}}}\,,
\eeq
in terms of canonically-normalized scalar NGB fields $\Pi_{\hat{a}}(x)$. In accordance with the Goldstone theorem we have one massless scalar for each broken generator. The fields associated with the unbroken ones drop out from the ansatz and thus do not lead to physical degrees of freedom. Because of Eq.~(\ref{fund}), the Goldstone bosons span the left coset space ${\mathpzc{G}}/{\mathpzc{H}}$, defined as the equivalence class of ${\mathpzc{G}}$ modulo ${\mathpzc{H}}$ elements multiplication. 

Symmetries are the central aspect of the CCWZ construction, let us then work out the action of the ${\mathpzc{G}}$ group on the Goldstone bosons. Namely, we seek for an operation on the field variables
\beq
\ve\Pi(x)\;\rightarrow\;{{\ve\Pi}^{(g)}}(x)\,,
\label{goltrcomp}
\eeq
associated with a generic element $g\in {\mathpzc{G}}$, which results in a symmetry transformation of the ansatz configuration defined by Eq.~(\ref{ansgol}). The first attempt would be to look for a transformation that induces $U\rightarrow g\cdot U$ on the Goldstone matrix, however this immediately fails because $ g\cdot U$ is a generic element of ${\mathpzc{G}}$ and as such it can not be expressed as the exponential of broken generators only. Therefore it is impossible to induce $U\rightarrow g\cdot U$ by acting on the Goldstone fields. However by Eq.~(\ref{fund}) we can decompose $g\cdot U$, in a unique way, as the product of one broken generator exponential and one ${\mathpzc{H}}$ element and define $\Pi^{(g)}$ implicitly by the relation
\beq
g\cdot U\left[\Pi\right]=U[\Pi^{(g)}]\cdot h\left[\Pi ;\,g\right]\,,
\label{gtrdef}
\eeq
where
\beq
\displaystyle 
h\left[\Pi ;\,g\right] = e^{i\,\zeta_a[\Pi ;\,g] T^a}\,.
\label{hmatrix}
\eeq
Or, equivalently
\beq
U\left[\Pi\right]\;\rightarrow\;U\left[\Pi^{(g)}\right]=g\cdot U\left[\Pi\right] \cdot h^{-1}\left[\Pi ;\,g\right]\,.
\label{goltr}
\eeq
This operation, given that $h$ leaves $\ve{F}$ invariant, induces a symmetry transformation on the the ansatz (\ref{ansgol}) 
\beq
\ve{\Phi}(x)\;\rightarrow\;g\,\ve{\Phi}(x)\,.
\eeq
Being the latter a symmetry of the underlying theory, it has to be respected also by the effective Lagrangian for the Goldstone bosons. Only a very peculiar set of terms, invariant under Eq.~(\ref{goltr}), are thus allowed.

Before discussing the implication of the symmetry, few comments are in order. First, it must be clarified that the above derivations do not rely on the explicit matrix representation adopted for the group generators. Indeed the coefficients $f_a$ and $f_{\hat{a}}$ appearing in the decomposition (\ref{fund}) are uniquely fixed by the generators commutation relations and not by their explicit form. Consequently, the same holds for the Goldstone transformation function $\Pi^{(g)}$ and for the coefficients $\zeta_a[\Pi ;\,g]$ of $h$, which are derived from Eq.~(\ref{fund}). This means that the Goldstone matrix $U[\Pi]$ can actually be defined for any representation of the $T^A$ generators. Nevertheless if not otherwise specified we will call Goldstone matrix the one in the fundamental representation. Similar considerations apply to the $h$ transformation. Second, it is important to stress that the relevant symmetry for the classification of the operators is provided by the full group ${\mathpzc{G}}$ and not only by the unbroken subgroup ${\mathpzc{H}}$. A spontaneously broken symmetry is thus not really broken, in the sense that it still implies powerful constraints on the allowed operators, not much differently from an unbroken one. This is expressed mathematically by the fact the Goldstone boson transformation defined by Eq.~(\ref{goltr}) provides a full-fledged representation of the whole ${\mathpzc{G}}$ because it respects the group multiplication rule, namely
\beq
\Pi^{(g_1\cdot g_2)}=\left(\Pi^{(g_2)}\right)^{(g_1)}\,.
\eeq
Of course $\Pi^{(g)}$ is rather different from the habitual group representations. The latter ones simply consist of constant transformation matrices acting linearly on the field variables while $\Pi^{(g)}$ carries a complicated non-linear dependence on $\Pi$. For this reason it is called a ``non-linear'' representation and the spontaneously broken group is sometimes said to be ``non-linearly realized'' rather than broken. We will soon encounter other non-linear representations, suitable for describing the transformation properties of the heavy resonance fields.

It is easy and instructive to derive the Goldstone bosons transformation explicitly in the particular case of an ${\mathpzc{H}}$ subgroup transformation, we just need to recall few elementary properties of Lie algebras.\footnote{See for instance Ref.~\cite{Feger:2012bs} for a concise review on Lie algebras and a computer package that could be useful in the study of composite Higgs models with non-minimal cosets.} Namely, we will use the fact that the ${\mathpzc{G}}$ generator algebra decomposes as
\bea
\label{deco}
&&\big[T^a,\,T^b\big]=i\, f^{ab}_{\;\;\;\;c}T^c+i \,\cancel{f^{ab}_{\;\;\;\;{\hat{c}}}} \; \Th^{\hat{c}}\equiv T^{{c}} \left(t_{\textrm{Ad}}^a\right)_{{c}}^{\;\;{b}}
\,,\nonumber\\
&&\big[T^a,\,\Th^{\hat{b}}\big]=i\, f^{a {\hat{b}}}_{\;\;\;\;{\hat{c}}} \Th_{{\hat{c}}}+i \, \cancel{f^{a {\hat{b}}}_{\;\;\;\;c}} \; T^{c}
\equiv \Th^{\hat{c}} \left({t_{\pi}}^a\right)_{\hat{c}}^{\;\;\hat{b}}
\,,\nonumber\\
&&\big[\Th^{\hat{a}},\,\Th^{\hat{b}}\big]=i\, f^{{\hat{a}} {\hat{b}}}_{\;\;\;\;{c}} T^c+i\, f^{{\hat{a}} {\hat{b}}}_{\;\;\;\;{\hat{c}}} \Th^{\hat{c}}\,,
\eea
where the first equality, {\it{i.e.}}~the fact that no broken generator appears in the commutator of unbroken ones, is due to ${\mathpzc{H}}$ being a subgroup and the second equality follows from the first one because the ${\mathpzc{G}}$ structure constants $f^{ABC}$ are completely antisymmetric.\footnote{We consider here a compact Lie group ${\mathpzc{G}}$, whose structure constants are indeed completely antisymmetric. The extension of Eq.~(\ref{deco}) to non-compact case is possible. For a composite Higgs model based on a non-compact coset, see Ref.~\cite{Urbano:2013aoa}.} In the commutator of broken generators, instead, both terms are present aside from the special cases called ``symmetric cosets'' where only unbroken generators appear because $f^{{\hat{a}} {\hat{b}}}_{\;\;\;\;{\hat{c}}}=0$. Symmetric cosets are endowed with an algebra automorphism under $\Th\rightarrow -\Th$ which can be associated with a $\Zdouble_2$ parity on the Goldstones, leading sometimes to interesting phenomenological consequences. In Eq.~(\ref{deco}), ${t_{\textrm{Ad}}}^a$ is the adjoint representation of the Lie algebra of the subgroup ${\mathpzc{H}}$ while the matrices ${t_{\pi}}^a$, which can also be shown to obey the ${\mathpzc{H}}$ algebra by the Jacobi identity, form a not yet specified ${\mathpzc{H}}$ representation denoted as ${\mathbf{r}}_\pi$. We will readily see that ${\mathbf{r}}_\pi$ is the representation in which the Goldstones transform under ${\mathpzc{H}}$. It can be identified, for any coset, by looking at the decomposition under ${\mathpzc{H}}$ of the adjoint of ${\mathpzc{G}}$, namely at
\beq
{\textrm{\bf{Ad}}}_{\mathpzc{G}}={\textrm{\bf{Ad}}}_{\mathpzc{H}}\oplus {\mathbf{r}}_\pi\,.
\eeq
For \mbox{SO$(N)/$SO$(N-1)$} cosets, ${\mathbf{r}}_\pi$ is the fundamental representation in accordance with the results of the previous section. 

When $g\in {\mathpzc{H}}$, namely for 
\beq
\displaystyle 
g=g_{\mathpzc{H}}=e^{i\,\alpha_a T^a}\in {\mathpzc{H}}\,,
\eeq
Eq.~(\ref{gtrdef}) is immediately worked out and reads
\bea
\displaystyle 
g_{\mathpzc{H}}\cdot U\left[\Pi\right]=&&\exp\left[{i\frac{\sqrt{2}}{f} \Pi_{\hat{a}}(x) \, g_{\mathpzc{H}}\cdot \Th^{\hat{a}}\cdot g_{\mathpzc{H}}^{-1}}\right]\cdot g_{\mathpzc{H}}
\nonumber\\
=&&
U\left[e^{i\,\alpha_a {t_\pi}^a}\ve\Pi\right]\cdot g_{\mathpzc{H}}\,,
\eea
having employed the exponentiated version of the commutation relation in Eq.~(\ref{deco}), {\it{i.e.}}
\beq
e^{i\,\alpha_a T^a}\cdot \Th^{\hat{a}}\cdot e^{-i\,\alpha_a T^a} = \Th^{\hat{b}}\left(e^{i\,\alpha_a {t_\pi}^a}\right)_{\hat{b}}^{\;\;\hat{a}}\,.
\label{expv}
\eeq
Therefore, as anticipated, $\Pi$ transforms in ${\mathbf{r}}_\pi$
\beq
\Pi_{\hat{a}}\;\rightarrow\;{{\Pi_{\hat{a}}}^{(g_{\mathpzc{H}})}} = \left(e^{i\,\alpha_a {t_\pi}^a}\right)_{\hat{a}}^{\;\;\hat{b}}\Pi_{\hat{b}}\,,
\eeq
and the ${\mathpzc{H}}$ transformation on the right side of the Goldstone matrix in Eq.~(\ref{gtrdef}) is $h[\Pi;\,g_{\mathpzc{H}}]=g_{\mathpzc{H}}$. Contrary to the general ones, ${\mathpzc{H}}$ subgroup transformations act linearly on the Goldstones. Verifying ${\mathpzc{H}}$ invariance can thus serve as a simple check of the consistency of our effective Lagrangian.

The situation is different for the transformations along the broken generators. There is no simple way to write them explicitly, not even at the infinitesimal level, aside from particular cases such as the \mbox{SO$(N)/$SO$(N-1)$} coset in Eq.~(\ref{brtr}). Their action is relatively simple only on the Goldstone matrix, as shown in Eq.~(\ref{goltr}), and this is why $U[\Pi]$ is the fundamental object to construct invariants in CCWZ. The broken transformations can however be worked out in a combined expansion on the transformation parameters $\alpha_{\hat{a}}$, defined by $g=g_{{\mathpzc{G}}/{\mathpzc{H}}}\simeq \Id+i\,\alpha_{\hat{a}}\Th^{\hat{a}}$, and on the field variables, leading to
\beq
\displaystyle
\Pi_{\hat{a}}\;\rightarrow\;{\Pi^{(g_{{\mathpzc{G}}/{\mathpzc{H}}})}}_{\hat{a}}=\Pi_{\hat{a}}+\frac{f}{\sqrt{2}}\alpha_{\hat{\alpha}}+
{\mathcal{O}}\left(\alpha\,\frac{\Pi^2}{f}+\alpha\,\frac{\Pi^3}{f^2}\ldots\right)\,.
\label{shift}
\eeq
This is the famous ``shift symmetry'', which forbids non-derivative potential terms in the Goldstone Lagrangian. The implications of the symmetry on the terms involving derivatives are instead more subtle, and harder to recognize in the Lagrangian because of the polynomial corrections to the shift. The latter implies that the transformation relates operators with different number of fields, leading to invariant Lagrangians with an infinite series of polynomial terms. For instance, out of the Goldstone kinetic terms a whole set of two-derivative interactions with any number of Goldstone legs would be generated by applying  Eq.~(\ref{shift}). By the CCWZ construction all these terms are automatically written at once.

A  particular case in which the transformation can be written explicitly is when the Goldstone fields are constant, namely when they are set to their VEV $\Pi=\langle\Pi\rangle$ and we perform a transformation $g=U^{-1}[\langle\Pi\rangle]$. In this case we trivially find
\beq
\label{gzero}
\langle\ve\Pi\rangle\;\rightarrow\;\ve0\,, 
\eeq
showing, as anticipated in the previous sections, that the Goldstone bosons VEV is unobservable because it can be set to zero by a symmetry transformation. This makes that in composite Higgs EWSB effects are mediated by the explicit breaking of the Goldstone symmetry. 

Let us finally turn to the classification of the operators allowed by the symmetry. The two fundamental objects employed in the construction are the $d[\Pi]_{\mu,\,{\hat{a}}}$ and $e[\Pi]_{\mu,\,a}$ symbols, which carry, respectively, one $\hat{a}$ index of the Goldstone representation ${\mathbf{r}}_\pi$ and one $a$ in the adjoint of ${\mathpzc{H}}$. They also contain one derivative, leading to the space-time index $\mu$, because we saw that derivatives are needed to construct non-trivial invariants. The $d$ and $e$ symbols are defined by decomposing on the ${\mathpzc{G}}$ algebra the Maurer--Cartan form constructed with $U$, namely
\beq
i\,{U[\Pi]}^{-1}\cdot\partial_\mu{U[\Pi]}=d_{\mu,\,{\hat{a}}}[\Pi]\Th^{\hat{a}}+e_{\mu,\,{{a}}}[\Pi] T^{{a}}\equiv d_\mu+e_\mu\,,
\label{dedef}
\eeq
where the shorthand notations $d_\mu$ and $e_\mu$ have been introduced. Under ${\mathpzc{G}}$, following Eq.~(\ref{goltr}), the Maurer--Cartan form transforms as
\bea
\label{mctr}
i\,{U[\Pi]}^{-1}\cdot\partial_\mu{U[\Pi]}\;\;\rightarrow && \;\; h[\Pi;\,g]\cdot \left(i\,{U[\Pi]}^{-1}\cdot\partial_\mu{U[\Pi]}\right)\cdot {h[\Pi;\,g]}^{-1}\nonumber\\
&&+\,i\,h[\Pi;\,g]\cdot\partial_\mu {h[\Pi;\,g]}^{-1}
\,.
\eea
We notice that the shift term on the second line is itself a Maurer--Cartan form, the one associated with the transformation $h[\Pi;\,g]$, which is an element of the subgroup ${\mathpzc{H}}$. Therefore it decomposes on the Lie algebra of ${\mathpzc{H}}$ and it does not have components along the broken generators. It follows that the shift is carried entirely by the $e$ symbol while $d$ transforms linearly with $h$ \footnote{Of course $h$ is itself highly non-linear, therefore $d$ still lives in a non-linear representation of ${\mathpzc{G}}$.}
\bea
\label{detransf}
d_\mu[\Pi]\;&\rightarrow&\; h[\Pi;\,g]\cdot d_\mu[\Pi] \cdot {h[\Pi;\,g]}^{-1}\,,\nonumber\\
e_\mu[\Pi]\;&\rightarrow&\; h[\Pi;\,g]\cdot \left(e_\mu[\Pi]+i\,\partial_\mu\right)\cdot {h[\Pi;\,g]}^{-1}\,.
\eea
When rewritten in components by Eq.~(\ref{expv}), the $d$ symbol transformation is a simple rotation of the $\hat{a}$ index in the representation ${\mathbf{r}}_\pi$
\beq
d_{\mu,\,{\hat{a}}}\;\rightarrow\;{{d^{(g)}}_{\mu,\,{\hat{a}}}} = \left(e^{i\,\zeta_a[\Pi;\,g] {t_\pi}^a}\right)_{\hat{a}}^{\;\;\hat{b}}d_{\mu,\,{\hat{b}}}\,.
\label{dscom}
\eeq 
We see that $d_{\mu,\,{\hat{a}}}$ transforms like the Goldstones and this is not surprising because the $d$ symbol is a sort of derivative of the Goldstone fields, in the sense that when expanding it we find
\beq
d_{\mu,\,{\hat{a}}}\simeq-\frac{\sqrt{2}}{f}\partial_\mu\Pi_{\hat{a}}+{\mathcal{O}}\left(\partial\Pi/f \cdot \Pi^2/f^2\right)\,.
\eeq
However $d_\mu$, differently from $\partial_\mu\Pi$, keeps transforming in ${\mathbf{r}}_\pi$ under the full ${\mathpzc{G}}$ and not just under the subgroup ${\mathpzc{H}}$, this is why it is useful to construct invariants. The $e$ symbol components $e_{\mu,\,a}$ have an index in the adjoint of ${\mathpzc{H}}$ and they transform, as the above equation clearly shows, as if they were gauge fields associated with a local ${\mathpzc{H}}$ invariance. As such they can be employed to construct covariant derivatives and field-strengths, as we will see later, but they can not be inserted directly in the operators.

The CCWZ prescription is to construct ${\mathpzc{G}}$-invariant operators by combining $d$ and $e$ symbols and derivatives. With the remarkable exception of the Wess--Zumino--Witten term \cite{Wess:1971yu,Witten:1983tw,Chu:1996fr} (which signals the presence of a global anomaly in ${\mathpzc{G}}$), all the invariant operators can obtained in this way. The reason why this results in a systematic and simple procedure is that the transformation rules are now entirely expressed in terms of the linear action of the matrix $h$. Therefore we just have to worry about building ${\mathpzc{H}}$ invariants with the standard group theory tools and the full ${\mathpzc{G}}$ invariance will follow automatically. In doing so one must however remember that ${\mathpzc{H}}$ is effectively a local group because $h$ depends on the Goldstone fields $\Pi(x)$. The simplest operators we can think to, which contain only two derivatives, are the ones constructed by two powers of the $d$ symbol, with the two ${\mathbf{r}}_\pi$ indices properly contracted to form an ${\mathpzc{H}}$ invariant. The most general combination of such operators defines the so-called $2$-derivative non-linear $\sigma$-model Lagrangian. If ${\mathpzc{H}}$ is compact, since ${\mathbf{r}}_\pi$ is real, one such an invariant always exists and it is given by \footnote{It is not worth considering the case of non-compact ${\mathpzc{H}}$ since it leads to negative-defined kinetic terms for the Goldstones.}
\beq
\label{nlsm}
{\mathcal{L}}^{(2)}=\frac{f^2}4 d_{\mu,\,{\hat{a}}} \delta^{\hat{a}{\hat{b}}} {d^\mu}_{{\hat{b}}}=\frac12\partial_\mu\Pi_{\hat{a}}\partial^\mu\Pi_{\hat{a}}+\sum_n{\mathcal{O}}\left((\partial\Pi)^2 \cdot \Pi^n/f^n\right)
\,.
\eeq
It provides the Goldstone bosons kinetic terms plus an infinite set of two-derivative interactions, which are all fixed by the symmetry and controlled by the unique parameter $f$. In general, other $2$-derivative operators might exist, provided there is more than one way to form invariants out of two ${\mathbf{r}}_\pi$ indices. One example which might be relevant for composite Higgs is the coset \mbox{SO$(6)/$SO$(4)$} which delivers $9$ Goldstones in the representation ${\mathbf{r}}_\pi={\mathbf{4}}\oplus {\mathbf{4}}\oplus {\mathbf{1}}$. Being ${\mathbf{r}}_\pi$ reducible, several invariants can be formed ($4$, in this case) and more free parameters appear in the non-linear $\sigma$-model Lagrangian. 

For the minimal composite Higgs coset \mbox{SO$(5)/$SO$(4)$}, instead, ${\mathbf{r}}_\pi=\mathbf{4}$ is irreducible and there is only one invariant. Therefore all the $2$-derivative Higgs interactions are predicted in terms of the Higgs decay constant $f$. Given that the CCWZ construction is completely general, this means that any explicit composite Higgs model, provided it is based on the minimal symmetry breaking pattern \mbox{SO$(5)\rightarrow\;\;$SO$(4)$}, leads to the exact same Lagrangian and physical predictions at the $2$-derivative level. We will see in the next section that this remains true when gauge fields are included. We can thus conclude, even before computing the Lagrangian explicitly, that all the results previously obtained in the linear $\sigma$-model example are completely general, in spite of the fact that the linear $\sigma$-model is just one possible realization of the composite Higgs idea and furthermore not a particularly motivated one. This first application of the CCWZ method should be already sufficient to illustrate the tremendous predictive power of the non-linearly realized symmetry on the physics of a pNGB Higgs.  

\subsection{Gauge sources and local invariance}
\label{sources_and_local}

Before going on with the discussion we need to take one step back and to add an additional bit of complication. Until now we have been considering a global ${\mathpzc{G}}$ invariance, now we want to extend the formalism to the case in which ${\mathpzc{G}}$ is made local by introducing the appropriate set of gauge fields $A_{\mu,\,A}$, transforming in the standard way
\beq
A_\mu\equiv A_{\mu,\,A}T^A\;\;\mathlarger{\rightarrow}\;\;A_\mu^{(g)}=g(x)\cdot\left(A_\mu+i\,\partial_\mu\right)\cdot {g(x)}^{-1}\,,
\label{gaugetr}
\eeq
under a local ${\mathpzc{G}}$ element $g$. 

This is almost the situation we have to deal with in the study of composite Higgs theories where a subgroup of the global ${\mathpzc{G}}$ is gauged to describe the EW vector bosons. On general grounds, the gauging procedure means modifying the composite sector Lagrangian by adding couplings with the currents $J^{\mu,\,A}$ associated with the global symmetry generators
\beq
{\mathcal{L}}_C\;\;\mathlarger{\rightarrow}\;\;{\mathcal{L}}_C+A_{\mu,\,A} J^{\mu,\,A}\,.
\label{gaugesources}
\eeq
The $A\cdot J$ term is of course precisely designed, by the definition of the Noether currents, to compensate for the variation of ${\mathcal{L}}_C$ under space-time dependent ${\mathpzc{G}}$ transformations leading to a locally invariant theory. If the full ${\mathpzc{G}}$ group has to be gauged, kinetic terms have to be introduced for all the $A_{\mu,\,A}$'s, but nothing forbids us to gauge a subgroup by giving a kinetic term only to a subset of the fields. Equivalently, all the fields can be formally gauged and the unwanted ones eventually decoupled by an infinite kinetic term, which corresponds to vanishing coupling strength. Any process which does not involve gauge bosons propagation is completely insensitive to the presence or to the absence of the kinetic terms and to the couplings, therefore all the $A_{\mu,\,A}$'s can be treated on the same footing as external sources. Some of them will be made dynamical by the kinetic terms and the others will be regarded as non-dynamical and eventually set to zero at the end of the calculation. The advantage of retaining the extra non-dynamical fields is of course the presence of the enlarged local ${\mathpzc{G}}$ group under which the sources transform as in Eq.~(\ref{gaugetr}). With this method one can study the effective composite Higgs Lagrangian for Goldstone, gauge and possibly resonance fields obtained by integrating out the composite sector dynamics. Propagating gauge fields effects, which of course are sensitive to the kinetic terms, are conveniently added at a second stage by working in perturbation theory. Notice that the latter effects break ${\mathpzc{G}}$ because the truly dynamical fields do not fill a complete ${\mathpzc{G}}$ representation. This structure complies with the picture outlined in the Introduction and with the examples of the previous section. In the minimal \mbox{SO$(5)/$SO$(4)$} model we turn on all the $10$ \mbox{SO$(5)$} gauge fields and eventually identify as physical only the $4$ ones associated with the SM \mbox{SU$(2)\times\;$U$(1)$} gauge group. In this case the physical value of the $A_\mu$ source reads
\beq
\label{physg}
A_\mu=A_{\mu,\,A} T^A=g W_{\mu}^{\alpha} T_L^\alpha + g'B_\mu T_R^3\,.
\eeq
in accordance with Eq.~(\ref{covdev4}). 

It is not hard to generalize the CCWZ construction to the local case. The Goldstone transformation property is still defined by Eq.~(\ref{goltr}) where $g$ is now a local group element. The Maurer--Cartan form, which we used to define the $d$ and $e$ symbols, generalizes to the object ${\overline{A}}_\mu$ constructed with the following logic: the Goldstone matrix $U[\Pi]$ is a local element of ${\mathpzc{G}}$ and as such it can be used to act on the gauge field $A_\mu$ following Eq.~(\ref{gaugetr}). The result of this operation is still an element of the ${\mathpzc{G}}$ algebra and as such it can be decomposed in terms of broken and unbroken generators similarly to what we did for the Maurer--Cartan form in Eq.~(\ref{dedef}). Actually it is worth acting with $U^{-1}$, and not with $U$, defining
\beq
\label{dedefgauge}
\displaystyle
{\overline{A}}_\mu={{A}}_\mu^{(U^{-1})}={U[\Pi]}^{-1}\cdot\left(A_\mu+i\,\partial_\mu\right)\cdot {U[\Pi]}\equiv d_\mu [\Pi,\,A]+e_\mu[\Pi,\,A]\,.
\eeq
The Maurer--Cartan form is immediately recovered in the ungauged limit $A_\mu=0$.\footnote{Notice that ${\overline{A}}_\mu$ can also be regarded as the trivial generalization of the Maurer--Cartan form obtained by replacing the ordinary derivative with the covariant one. Expressing it in terms of the gauge-transformed $A_\mu$ is not just a fancy but also a useful rewriting.} Furthermore, ${\overline{A}}_\mu$ transforms under ${\mathpzc{G}}$ exactly like the Maurer--Cartan form does in the global case, namely as in Eq.~(\ref{mctr}). By exploiting the group multiplication rules the latter property is shown by a one-line calculation
\beq
\displaystyle
{\overline{A}}_\mu={{A}}_\mu^{(U^{-1})}\rightarrow{{{A}}_\mu^{(g)}}^{(h\cdot U^{-1}\cdot g^{-1})}={{{A}}_\mu^{(U^{-1})}}^{(h)}=h\left({\overline{A}}_\mu+i\,\partial_\mu\right){h}^{-1}\,.
\eeq
The generalized $d$ and $e$ symbols thus transform precisely as in Eq.~(\ref{detransf})
\bea
\label{detrg}
d_\mu[\Pi,\,A]\;&\rightarrow&\; h[\Pi;\,g]\cdot d_\mu[\Pi,\,A] \cdot {h[\Pi;\,g]}^{-1}\,,\nonumber\\
e_\mu[\Pi,\,A]\;&\rightarrow&\; h[\Pi;\,g]\cdot \left(e_\mu[\Pi,\,A]+i\,\partial_\mu\right) {h[\Pi;\,g]}^{-1}\,,
\eea
though of course, now, under the full local ${\mathpzc{G}}$ group. According to the definition (\ref{dedefgauge}), the generalized $d$ and $e$ symbols are provided by those of the global case, which contain one derivative of the Goldstone field, plus non-derivative terms proportional to $A_\mu$. In order to maintain the homogeneity of the operators constructed out of $d$ and $e$ it is thus convenient to treat the gauge fields on the same footing as derivatives and to regard both $\partial$ and $A$ as ${\mathcal{O}}(\partial)$ or equivalently, by going to the momentum space, as ${\mathcal{O}}(p)$ objects. The reader can get an idea of how the $d$ and $e$ symbols look like in concrete by Eq.~(\ref{de54}) in Appendix~\ref{so54}, where these objects are explicitly computed for the  minimal Coset  \mbox{SO${(5)}/$SO${(4)}$}.

From this point on, since we encountered exactly the same transformation rules, the classification of locally invariant operators proceeds in parallel with the globally invariant ones. In both cases, local ${\mathpzc{H}}$ invariance is all what we have to worry about. The $2$-derivative non-linear $\sigma$-model Lagrangian, for instance, trivially generalizes to
\beq
\label{smg}
{\mathcal{L}}^{(2)}=\frac{f^2}4 d_{\mu,\,{\hat{a}}}[\Pi,\,A] \delta^{\hat{a}{\hat{b}}} {d^\mu}_{{\hat{b}}}[\Pi,\,A]\,,
\eeq
and contains now not only the Goldstone kinetic terms and self-couplings, but also interactions involving the gauge fields, which are all dictated by the local ${\mathpzc{G}}$ invariance and predicted in terms of the sigma-model scale $f$. All the previous considerations about the unicity of the $\sigma$-model Lagrangian in \mbox{SO$(5)/$SO$(4)$} remain the same, meaning that also the Higgs/gauge interactions are completely determined by the coset structure. In particular the modifications of the Higgs couplings to the gauge fields we obtained in the linear $\sigma$-model example in Eq.~(\ref{hvc}) must be regarded as robust model-independent predictions of the \mbox{SO$(5)/$SO$(4)$} composite Higgs. We verify in Appendix~\ref{so54} that ${\mathcal{L}}^{(2)}$ as given above coincides as expected with the Lagrangian (\ref{CHlag0}) we derived in the linear $\sigma$-model.

\subsection{Two derivative tensors and resonances}
\label{2dt}

The classification of  ${\mathcal{O}}(p^2)$ (two-derivatives) invariant operators, out of which we defined the non-linear $\sigma$-model Lagrangian, has been extremely simple because the $d$ symbol was the only object which could have appeared in these operators. Going to higher orders, as we will need to do in the next chapter, requires more care and some additional technicality. A simple way to proceed is to forget momentarily about invariants and classify instead all the possible two-derivative tensor operators which transforms homogeneously under $h$. 

The first ones we might think to are those constructed with two $d$-symbols, of the form
\beq
\displaystyle
d_{\mu,\,{\hat{a}}}d_{\nu,\,{\hat{b}}}\in {\mathbf{r}}_\pi\otimes\; {\mathbf{r}}_\pi = {\textrm{\bf{Ad}}}_{\mathpzc{H}}\oplus {\mathbf{1}}_{\mathpzc{H}} \oplus \ldots\,.
\eeq
They transform in the tensor product of two Goldstone representations, whose decomposition contains the adjoint, the singlet, plus eventually other representations depending on the nature of the group ${\mathpzc{H}}$. Notice that the decomposition in irreducible representations is what matters for the CCWZ method. Since ${\mathpzc{H}}$ invariance is all what is required, tensor product components belonging to different irreducible representations can be employed separately to construct invariant operators. The singlet and the adjoint components are immediately worked out for a generic coset
\beq
\label{dcomb}
\displaystyle
({d_{\mathbf{1}}^2)}_{\mu\nu}=\delta^{{\hat{a}}{\hat{b}}}d_{\mu,\,{\hat{a}}} d_{\nu,\,{\hat{b}}}\,,\;\;\;\;\;
({d_{\textrm{\bf{Ad}}}^2})_{\mu\nu,\,c}=f^{{\hat{a}} {\hat{b}}}_{\;\;\;\;{c}}\,d_{\mu,\,{\hat{a}}} d_{\nu,\,{\hat{b}}}\,,
\eeq
in terms of the structure constants $f$ appearing in Eq.~(\ref{deco}). The existence and the form of  other tensors depend instead on the coset. In \mbox{SO$(5)/$SO$(4)$} one extra $d^2$ tensor is found in the ${\mathbf{9}}$ of \mbox{SO$(4)$}.

A second class of operators is formed by acting on the $d$ symbol with one derivative, or more precisely with a suitable covariant derivative which takes care of the local nature of the $d$ symbol transformation rule in Eq.~(\ref{dscom}). The $e$ symbol transforms precisely like a gauge connection (\ref{detrg}) and thus the covariant derivative is given by
\beq
\label{Dd}
\displaystyle
(D\cdot d)_{\mu\nu,\,{\hat{a}}}\equiv\partial_\mu d_{\nu,\,{\hat{a}}} - i\, e_{\mu,\,{a}}\left({t_\pi}^a\right)_{\hat{a}}^{\;\;\hat{b}} d_{\nu,\,{\hat{b}}}\in {\mathbf{r}}_\pi\,.
\eeq
It transforms, by definition, in the representation ${\mathbf{r}}_\pi$. Whether $D\cdot d$ constitutes a single CCWZ tensor or not depends on the number of irreducible components of ${\mathbf{r}}_\pi$. 

Another object we can form, exploiting once again the fact that $e_\mu$ transforms like a gauge field, is the field strength tensor
\beq
\label{emunu}
E_{\mu\nu}\equiv E_{\mu\nu,\,a}T^a=i\,\left[D_\mu,\,D_\nu\right]=\partial_\mu e_{\nu}-\partial_\nu e_{\mu}-i\,[e_\mu,\,e_\nu]\,,
\eeq
which transforms homogeneously with $h$ in the adjoint representation, namely
\beq
E_{\mu\nu}[\Pi,\,A]\;\rightarrow\; h[\Pi;\,g]\cdot E_{\mu\nu}[\Pi,\,A] \cdot {h[\Pi;\,g]}^{-1}\,.
\eeq
In components, $E$ reads
\beq
\label{emuncomp}
E_{\mu\nu,\,a}=\partial_\mu e_{\nu,\,a}-\partial_\nu e_{\mu,\,a}+f^{bc}_{\;\;\;\;a} e_{\mu,\,b} e_{\nu,\,c}\in {\textrm{\bf{Ad}}}_{\mathpzc{H}}\,.
\eeq

It turns out that $d^2$, $D\cdot d$ and $E$ exhaust the most general ${\mathcal{O}}(p^2)$ tensor,
but this is not yet apparent because in our discussion we have ignored the possibility of constructing tensors directly from the gauge fields $A_\mu$ rather than starting from the $d$ and the $e$ symbols. Notice that $A_\mu$ is very different from the other CCWZ objects because it transforms directly with the ${\mathpzc{G}}$ transformation $g(x)$, as in Eq.~(\ref{gaugetr}), rather than with the ${\mathpzc{H}}$ matrix $h$. This is the reason why we had to ``dress'' it with the Goldstone matrix in Eq.~(\ref{dedefgauge}): the Goldstone matrix, since in transforms with $g$ on one side and with $h$ on the other, is precisely what is needed to change an index transforming with $g$ into a one transforming with $h$. We might consider dressing the $A_\mu$ field strength defining an object
\bea
\label{deff}
{\mathcal{F}}_{\mu\nu}&=&{U[\Pi]}^{-1}\cdot F_{\mu\nu}[A]\cdot U[\Pi]\nonumber\\
&=&{U[\Pi]}^{-1}\cdot \left(\partial_\mu A_{\nu}-\partial_\nu A_{\mu}-i\,[A_\mu,\,A_\nu]
\right)\cdot U[\Pi]\,,
\eea
that transforms homogeneously with $h$. Clearly, ${\mathcal{F}}$ belongs to the algebra of ${\mathpzc{G}}$ and therefore  decomposes  along the broken and the unbroken generators leading to two tensors
\beq
\label{deff1}
{\mathcal{F}}_{\mu\nu}=\left({\mathcal{F}}_{\textrm{\bf{Ad}}}\right)_{\mu\nu,\,a}T^a+\left({\mathcal{F}}_{{\mathbf{r}}_\pi}\right)_{\mu\nu,\,\hat{a}}\Th^{\hat{a}}\;\in\; {\textrm{\bf{Ad}}}_{\mathpzc{H}}\oplus {\mathbf{r}}_\pi\,,
\eeq
in the adjoint and in the ${\mathbf{r}}_\pi$ representation, respectively. However the ${\mathcal{F}}$ tensors are redundant, and thus they can be ignored in the operator classification because they can be expressed as linear combinations of the others. This is readily shown by noticing that the field strength $F_{\mu\nu}[A]$ reacts linearly to a gauge transformation of the argument so that 
\bea
\label{2trel}
{\mathcal{F}}_{\mu\nu}&=&{U[\Pi]}^{-1}\cdot F_{\mu\nu}[A]\cdot U[\Pi]=F_{\mu\nu}\left[{{A}}^{(U^{-1})}\right]\nonumber\\
&=&F_{\mu\nu}\left[ d +e  \right]\,,
\eea
where we made use of the definition of $d$ and $e$ in Eq.~(\ref{dedefgauge}). The object on the second line of the equation is an ${\mathcal{O}}(p^2)$ tensor constructed in terms of the $d$ and the $e$ symbols and thus it must be, according to the previous classification, a linear combination of $d^2$, $D\cdot d$ and $E$.

Notice that Eq.~(\ref{2trel}) could be used in two ways, either for eliminating ${\mathcal{F}}$ in favor of the others or for expressing two linear combinations of $d^2$, $D\cdot d$ and $E$ in terms of ${\mathcal{F}}_{\textrm{\bf{Ad}}}$ and ${\mathcal{F}}_{{\mathbf{r}}_\pi}$. This second option might be convenient for certain applications. For instance in composite Higgs one might want to separate the operators constructed with the ${\mathcal{F}}$'s, which are entirely induced by the SM gauging and vanish in the ungauged limit $A_\mu\rightarrow0$, from the ones which emerge from the composite sector alone. Moreover the ${\mathcal{F}}$ components are easy to compute because they contain no Goldstone boson derivatives and thus it might be useful to express the other tensors in terms of them. For those applications Eq.~(\ref{2trel}) needs to be explicitly worked out. By exploiting the commutation relations in Eq.~(\ref{deco}) we obtain 
\bea
\label{CCWZID}
&&\left({\mathcal{F}}_{\textrm{\bf{Ad}}}\right)_{\mu\nu,\,a} = E_{\mu\nu,\,a} + ({d_{\textrm{\bf{Ad}}}^2})_{\mu\nu,\,a}\,, \nonumber\\
&&\left({\mathcal{F}}_{{\mathbf{r}}_\pi}\right)_{\mu\nu,\,\hat{a}}=(D\cdot d)_{\mu\nu,\,{\hat{a}}}-(D\cdot d)_{\nu\mu,\,{\hat{a}}}+({d_{{\mathbf{r}}_\pi}^2})_{\mu\nu,\,\hat{a}}\,,
\eea
where ${d_{{\mathbf{r}}_\pi}^2}$ is a $d^2$ operator in the ${\mathbf{r}}_\pi$ representation defined as
\beq
 ({d_{{\mathbf{r}}_\pi}^2})_{\mu\nu,\,\hat{a}}=f^{{\hat{a}} {\hat{b}}}_{\;\;\;\;{\hat{c}}}\,d_{\mu,\,{\hat{a}}} d_{\nu,\,{\hat{b}}}\,.
\eeq
Notice that for symmetric cosets, where $f^{{\hat{a}} {\hat{b}}}_{\;\;\;\;{\hat{c}}}=0$, ${d_{{\mathbf{r}}_\pi}^2}$ vanishes.

Aside from being a derivation of Eq.~(\ref{CCWZID}), the discussion above illustrates how the gauge source fields can be ignored in the classification of CCWZ tensors or invariant operators. The $d_\mu$ symbol, the $E_{\mu\nu}$ field-strength and the covariant derivatives is all what is needed, any object directly formed with the sources can be systematically expressed in terms of the latter as we saw above for the ${\mathcal{F}}$ tensors. It is actually even simpler than that because $E_{\mu\nu}$ is the commutator of two covariant derivatives as in Eq.~(\ref{emunu}). Therefore $d_\mu$ and $D_\mu$ are, strictly speaking, the only needed objects. However for practical purposes derivative commutators are conveniently traded for the field-strength. In Sect.~\ref{p4bos} we will make use of those rules for the classification of the complete \mbox{SO$(5)/$SO$(4)$} Lagrangian at ${\mathcal{O}}(p^4)$.

In what follows we will occasionally need to include in our model, on top of the Goldstone and of the gauge bosons, also some of the composite sector resonances. The latter states are parametrically heavier than the former ones, but it might still be worth including some of them in the effective field theory if their mass happens to be smaller than the typical composite sector scale. CCWZ  is the ideal framework to discuss resonances. In full analogy with what we found for the $d$-symbol components in Eq.~(\ref{dscom}), we introduce resonance fields $\Psi_i$ transforming as 
\beq
\displaystyle
\Psi_{i}\;\rightarrow\;{{\Psi^{(g)}}}_{i} = \left(e^{i\,\zeta_a[\Pi;\,g] {t_\Psi}^a}\right)_{i}^{\;\;j}\Psi_{j}\equiv h_{{\mathbf{r}}_\Psi}[\Pi;\,g]_{i}^{\;\;j}\Psi_{j}\,,
\eeq
where $\zeta_a[\Pi;\,g]$ are the parameters of the ${\mathpzc{H}}$ transformation $h[\Pi;\,g]$  as in Eq.~(\ref{hmatrix}) and ${t_\Psi}^a$ are the generators of ${\mathpzc{H}}$ in a given representation ${\mathbf{r}}_\Psi$. The $d$-symbol transformation property corresponds to the particular case ${\mathbf{r}}_\Psi={\mathbf{r}}_\pi$. This is a consistent assignment because it respects the ${\mathpzc{G}}$ group multiplication rules
\beq
\label{mrres}
\Psi^{(g_1\cdot g_2)}=\left(\Psi^{(g_2)}\right)^{(g_1)}\,,
\eeq
thanks to the following property of $h$
\beq
\label{cpr}
h\left[\Pi ;\,g_1\cdot g_2\right]=h\left[\Pi^{(g_2)} ;\,g_1\right]\cdot h\left[\Pi;\,g_2\right]\,.
\eeq
The latter is easily shown from the definition in Eq.~(\ref{gtrdef}). Notice that all the structural properties of $h$, among which the one above, only depend on the commutators algebra of the generators employed in its definition and not on their explicit representation. Therefore Eq.~(\ref{cpr}) holds for $h_{{\mathbf{r}}_\Psi}$ as well, in spite the latter being defined as the exponential of ${\mathpzc{H}}$ generator matrices ${t_\Psi}^a$. 

We stress once again that the resonance transformation property is defined in terms of a representation of ${\mathpzc{H}}$, and not of the full ${\mathpzc{G}}$ group. Therefore the resonance fields organize themselves in ``short'' multiplets, corresponding to ${\mathpzc{H}}$ irreducible representations, each of which can be treated separately. If for instance ${\mathpzc{G}}=\textrm{SO}(5)$ and ${\mathpzc{H}}=\textrm{SO}(4)$, resonances in the ${\mathbf{4}}$ or in the ${\mathbf{1}}$ can be introduced individually rather than grouping  them in a complete ${\mathbf{5}}$ of ${\mathpzc{G}}$ as we would have been obliged to do if ${\mathpzc{G}}$ were unbroken. Indeed we know that in spontaneously broken theories no degenerate ${\mathpzc{G}}$ multiplets are expected and the spectrum is classified in terms of the irreducible representations of the unbroken group. It is thus reassuring that the formalism allows us to deal with ${\mathpzc{H}}$ multiplets individually. 

In CCWZ, effective Lagrangian are written in terms of the resonance fields and of the $d$ and the $e$ symbols, with the indices properly contracted to form ${\mathpzc{H}}$ invariants. Clearly when taking derivatives of the resonances the local nature of ${\mathpzc{H}}$ must be taken into account by using the covariant derivative
\beq
\label{gencd}
D_\mu\Psi_i=\partial_\mu \Psi_{i} - i\, e_{\mu,\,{a}}\left({t_{\mathbf{r}_\pi}}^a\right)_{i}^{\;\;j} \Psi_j\,.
\eeq
If for instance $\Psi$ is a fermionic resonance its kinetic Lagrangian is simply
\beq
\label{CCWZferres}
i\,\overline{\Psi}\gamma^\mu D_\mu \Psi - m_\Psi \overline{\Psi}\Psi\,.
\eeq
Notice that it contains, through the ``$e$'' term in the covariant derivative, a full set of interactions with the Goldstone and the gauge fields which are all dictated by the symmetry and controlled by the $\sigma$-model scale $f$.

\section{Partial fermion compositeness}
\label{sec:PC}

Nothing has been said up to now on how we plan to introduce in our construction the SM matter fermions, their interactions with the Higgs and eventually generate their mass. This is actually a delicate task, which is achieved in modern composite Higgs models by relying on the so-called ``partial compositeness'' hypothesis. partial compositeness was first introduced by D.B.~Kaplan in Ref.~\cite{Kaplan:1991dc} and more recently rediscovered in the context of extra-dimensional models of EWSB \cite{Contino:2004vy}. Interestingly enough, partial compositeness emerges naturally and automatically in those models. This provides a hint that the partial compositeness hypothesis might eventually find a microscopic realization. See Sect.~\ref{DT} for additional considerations on this aspect.

\subsection{The basic idea}
\label{sec:PCbasic}

It is convenient to illustrate partial compositeness in opposition to the ``standard'' approach to matter fermions in strongly-coupled EWSB models, namely the one which was originally adopted in technicolor models \cite{Weinberg:1975gm,Weinberg:1979bn,Susskind:1978ms,Dimopoulos:1979es,Lane:2002wv} and later in the composite Higgs context~\cite{Dugan:1984hq} (see also Ref.~\cite{Galloway:2010bp} for a recent example). In technicolor, and in composite Higgs as well as explained in Sect.~\ref{DT}, the SM fermions are introduced as elementary fields external to the composite sector and coupled to the latter by bilinear operators of the form
\beq
\label{inttc}
\displaystyle
{\mathcal{L}}_{\textrm{Int}}=\frac{\lambda_t}{\Lambda_{\textrm{UV}}^{d-1}}\overline{q}_L{\mathcal{O}}_S^c t_R+
\frac{\lambda_b}{\Lambda_{\textrm{UV}}^{d-1}}\overline{q}_L{\mathcal{O}}_S b_R+\textrm{h.c.}\,,
\eeq
where $q_L=(t_L,b_L)^T$, $t_R$ and $b_R$ denotes one of the SM quark fields families, even though for the present discussion we will be mainly interested in the third one as the notation suggests. Leptons could be included along similar lines but they play no role in what follows. All the fields are assumed to be canonically normalized, with SM-like kinetic terms
\beq
\label{elm_q_kin}
{\mathcal{L}}_{E}={\overline{q}}_Li\slashed{D} q_L + {\overline{t}}_Ri\slashed{D} t_R  + {\overline{b}}_R i\slashed{D} b_R  + \ldots \,,
\eeq
that originate, exactly like for the gauge fields, from the elementary sector of the theory. In equation (\ref{inttc}), ${\mathcal{O}}_S$ is meant to be one Lorentz scalar operator composed of strong sector fields. Its SM quantum numbers are precisely those of the Higgs field and as such it can form Yukawa-like couplings. 

\begin{figure}
\centering
\includegraphics[width=1.0\textwidth]{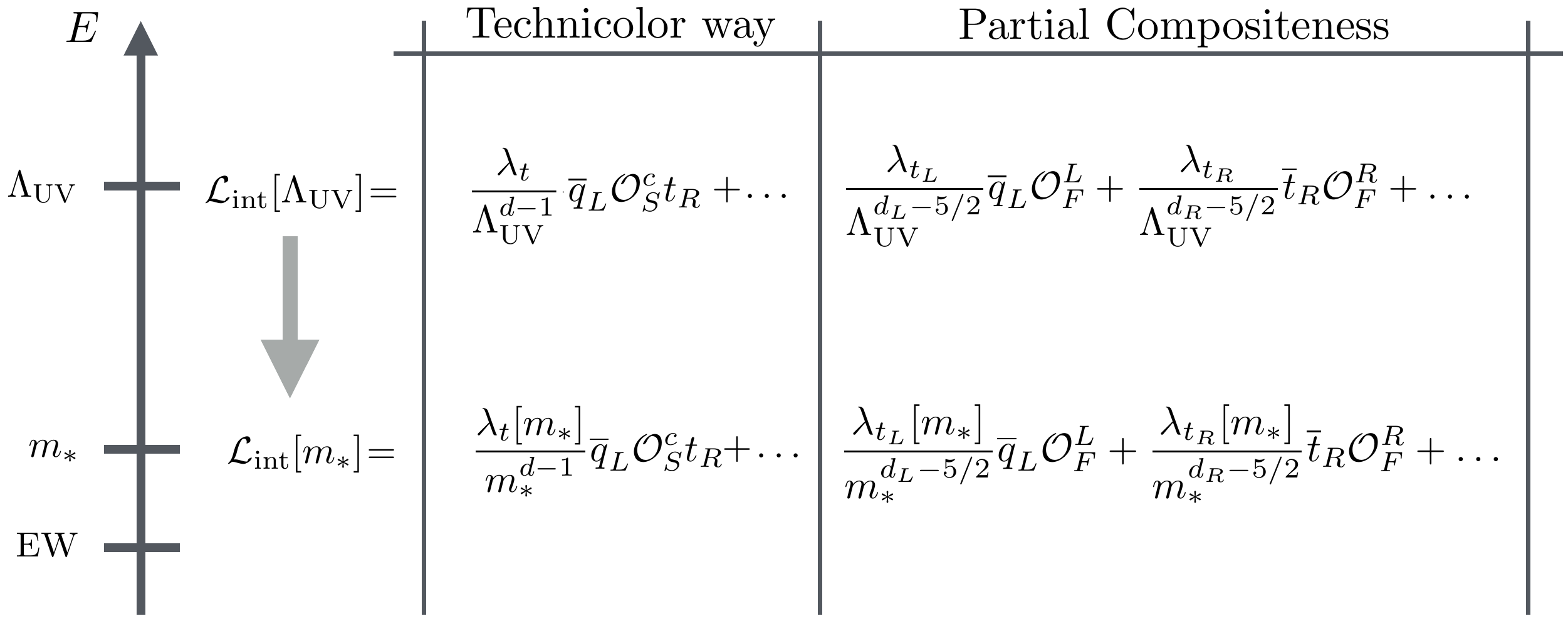}
\caption{A pictorial view of the partial compositeness approach to SM fermion couplings compared with the old-fashioned technicolor way.}
\label{scaleferm}
\end{figure}

Writing down ${\mathcal{O}}_S$ explicitly in one example helps to clarify what we have in mind. In minimal technicolor, which consists of a scaled version of two-flavor QCD, we have
\beq
\displaystyle
({\mathcal{O}}_S)_{i}^{\; j}=(\Id\,{\mathcal{O}}_S^4+i\,\sigma_\alpha {\mathcal{O}}_S^\alpha)_{i}^{\; j}=\overline{\Psi}_R^j\Psi_{L,\,i}\,,
\eeq
where $\Psi_{L,R}$ are the chiral techniquark fields, endowed with flavour indices $i,j=1,2$ in \mbox{SU$(2)_L\times$SU$(2)_R$}. The equation above provides the ${\mathcal{O}}_S$ components in the real fourplet notation, one could switch to the complex doublet notation trough Eq.~(\ref{hd}). Therefore in technicolor ${\mathcal{O}}_S$ is a techniquark bilinear with energy dimension $d=3$, the elementary/composite interactions are dimension-$6$ four-fermion operators and as such are suppressed by two powers of the high scale, $\Lambda_{\textrm{UV}}$, at which they are generated. Eq.~(\ref{inttc}) provides the obvious generalization, based on dimensional analysis, for an arbitrary ${\mathcal{O}}_S$ dimension $d$.

The physical origin of the suppression scale $\Lambda_{\textrm{UV}}$ is pictorially represented in Fig.~\ref{scaleferm} and it can be understood as follows, along the lines of Sect.~\ref{DT}. At around the TeV the strong sector confines and it dynamically generates the new physics scale $m_*$, which can be identified with the typical mass of the composite resonances. Above that scale the strong sector approaches a conformal fixed point around which the energy scaling is dictated by the operators dimensionality. In minimal technicolor models the conformal fixed point merely correspond to the free theory of techniquarks and technigluons, weakly perturbed by the technicolor interactions, but on general grounds strongly interacting fixed points might also be considered. See Ref.~\cite{Piai:2010ma} for a review and Ref.~\cite{Luty:2004ye} for a concise but clear discussion. An explicit realization is the so-called ``walking technicolor'' model \cite{Holdom:1984sk,Akiba:1985rr,Appelquist:1986an,Yamawaki:1985zg,Appelquist:1986tr,Appelquist:1987fc}. The fermion fields are not part of the strong sector, therefore their origin as physical particles and their interactions will emerge from a more complete theory at a scale $\Lambda_{\textrm{UV}}\gg m_*$. The scale $\Lambda_{\textrm{UV}}$ should be regarded as the cutoff scale of the BSM theory itself. For example $\Lambda_{\textrm{UV}}$ might be the extended technicolor scale (see Refs.~\cite{Eichten:1979ah,Lane:2002wv}) at which the four-fermion interactions among quarks and techniquarks are generated by the exchange of heavy gauge fields. 

The interaction strength is dictated by dimensional analysis, up to dimensionless coefficients $\lambda_t$ and $\lambda_b$ which depend on the couplings of the underlying microscopic theory. In the extended technicolor example, $\lambda_{t,b}\sim g_{\textrm{ETC}}^2$, where $g_{\textrm{ETC}}$ is the extended technicolor gauge coupling. On general grounds $\lambda_{t,b}$ could be small, if the underlying couplings are weak, but they can not be too much larger than unity if the underlying theory has to remain perturbative. This implies an upper bound $\lambda_{t,b}<\lambda_{\textrm{Max}}$, where for simplicity we treat $\lambda_{\textrm{Max}}$ as an order-one parameter even if its numerical value might be larger, for instance $\lambda_{\textrm{Max}}\sim 16\pi^2$ in extended technicolor. More formally, the bound on $\lambda$ comes from the fact that the elementary quark interaction must be a small perturbation of the strong sector dynamics in the full $[m_*,\,\Lambda_{\textrm{UV}}]$ range, otherwise our picture is not self-consistent and the scale separation among $m_*$ and $\Lambda_{\textrm{UV}}$ gets destabilized.

As depicted in Fig.~\ref{scaleferm}, the elementary quark interactions in Eq.~(\ref{inttc}) must be evolved down to $m_*$ before reading their low-energy implications. In our hypothesis the evolution is driven by the operator dimension  \footnote{The equation below trivially follows from the fact that the interaction operator stays unchanged during the evolution at a fixed point. The coupling runs just because of the different normalization of the operator in the IR, which is provided by the scale $m_*$ rather than by $\Lambda_{\textrm{UV}}$. The IR normalization is the appropriate one to read the low-energy effects of the interaction.}
\beq
\displaystyle
\lambda_{t,b}[m_*]\simeq  \lambda_{t,b}\left(\frac{m_*}{\Lambda_{\textrm{UV}}}\right)^{d-1}\,,
\eeq
up to corrections due to departures from the fixed point, either intrinsically present in the strong sector or induced by the elementary quark interaction itself. The quark masses, or equivalently the Yukawa couplings, are thus estimated to be
\beq
\label{etsytbil}
\displaystyle
y_{t,b}=\sqrt{2}\frac{m_{t,b}}{v}\simeq \lambda_{t,b}[m_*]\simeq  \lambda_{t,b}\left(\frac{m_*}{\Lambda_{\textrm{UV}}}\right)^{d-1}\,.
\eeq
By taking into account that $d-1$ can be shown to be necessarily positive by unitarity arguments, so that the couplings get power suppressed when running to the IR, we see that the above equation is problematic in two respects. First, the presence of the upper bound $\lambda<\lambda_{\textrm{Max}}$ makes hard to generate large Yukawas in our setup, which foresees a considerable scale separation $\Lambda_{\textrm{UV}}\gg m_*$. If for instance $\lambda_{\textrm{Max}}\sim 1$ and $d$ is significantly larger than $1$ no realistic top Yukawa coupling $y_t\simeq1$ can be obtained. Second, even if it was possible to get $y_t$ right by a large enough $\lambda_{\textrm{Max}}$ this could definitely not be achieved for an arbitrarily large scale separation. Since $m_*$ is tied to the TeV scale by the Naturalness problem we end up with an upper bound on $\Lambda_{\textrm{UV}}$, which reads
\beq
\displaystyle
{\Lambda_{\textrm{UV}}}\simeq \left( \frac{\lambda_t}{y_t}\right)^{\frac1{d-1}}{m_*}<\lambda_{\textrm{Max}}^{\frac1{d-1}}\,\textrm{TeV}\,.
\eeq
But $\Lambda_{\textrm{UV}}$ is where the Yukawa's are generated, therefore the full flavor structure of the SM must emerge at that scale. If it is not heavy enough, above around $10^5$~TeV, large and phenomenologically unacceptable extra flavor-violating interactions will also arise, at least in the absence of special mechanisms and selection rules in the underlying microscopic theory. In the technicolor case, where $d=3$ and $\lambda_{\textrm{Max}}=16\pi^2$, ${\Lambda_{\textrm{UV}}}$ can be quantitatively estimated to be
\beq
{\Lambda_{\textrm{UV}}} < 10\,{\textrm{TeV}}\,,
\eeq
far below what phenomenologically acceptable. 

As proposed in Ref.~\cite{Luty:2004ye}, one way out to this situation would be to consider theories where the scalar operator dimension $d$ is close to one, namely $d=1+\epsilon$. This would allow, in principle, to maintain a large scale separation while still obtaining the correct top quark Yukawa. The problem here is that $d=1$ is the dimension of the elementary Higgs field and furthermore it can be shown that the only theory where $d$ is exactly equal to $1$ is the one of a free scalar, which suffers from the Naturalness problem. This somehow suggests that by taking $d=1+\epsilon$ we might run into the risk of reintroducing the Naturalness problem in our construction. More concretely, the issue comes if we ask ourselves about the scaling dimension of the scalar operator squared, ${\mathcal{O}}_S^2$, which is the analog of the Higgs mass term in the SM. If $d[{\mathcal{O}}_S^2]<4$, its presence in the Lagrangian reintroduces the Naturalness problem like for a free scalar where  $d[{\mathcal{O}}_S^2]=2$. In Ref.~\cite{Rattazzi:2008pe}, the following bound was derived
\beq
d[{\mathcal{O}}_S^2]\leq f(d)\,,
\eeq
where $f(d)$ is a continuos function and $f(1)=2$. Therefore for $d=1+\epsilon$
\beq
d[{\mathcal{O}}_S^2]\leq 2+{\mathcal{O}}(\epsilon)\,,
\eeq
and we are pushed into the dangerous region $d[{\mathcal{O}}_S^2]<4$. Though qualitative and not completely accurate \footnote{For instance, potentially important numerical factors have been ignored in the estimate of Eq.~(\ref{etsytbil}) while they could emerge in concrete technicolor-like theories.}, the above discussion is sufficient to illustrate the difficulties with the standard technicolor-like approach to fermion mass generation and to motivate the study of alternative mechanisms.

We now turn to partial compositeness. Also in this case, matter fermions are introduced as elementary fields external to the composite sector and coupled to the latter at the high scale $\Lambda_{\textrm{UV}}$, with the only possible exception of the right-handed top quark which might instead be a completely composite state. Leaving aside this possibility, on which we will return later, the elementary/composite interactions now read
\beq
\label{intpc}
{\mathcal{L}}_{\textrm{Int}}=
\frac{\lambda_{t_L}}{\Lambda_{\textrm{UV}}^{d_L-5/2}}\overline{q}_L{\mathcal{O}}_F^L+\frac{\lambda_{t_R}}{\Lambda_{\textrm{UV}}^{d_R-5/2}}\overline{t}_R{\mathcal{O}}_F^R+\ldots\,,
\eeq
plus analogous terms for the bottom and the other quarks. The crucial difference with the technicolor way is that the interaction terms are linear in the elementary fields rather than bilinear and correspondingly the composite sector operators ${\mathcal{O}}_F^{L,R}$ are fermionic rather than scalar. The reason why this setup is called ``partial compositeness'' is that the linear couplings give rise, in the IR, to mixings of the elementary quarks with some composite resonances so that the physical mass eigenstates are linear combinations of elementary and composite degrees of freedom. The compositeness fraction, and eventually the Yukawa couplings which emerge from this mechanism as we will see below, is controlled by the couplings $\lambda_{t_{L,R}}$, duly evolved to the IR scale $m_*$. Assuming for simplicity $d_{L,R}> 5/2$ \footnote{The unitarity bound on fermionic operators is $d_{L,R}\geq 3/2$, therefore we might well consider also the case $3/2<d_{L,R}< 5/2$ \cite{Contino:2004vy}.} and ignoring self-induced contributions to the Renormalization Group evolution, which become numerically important for $d_{L,R}$ close to $5/2$ \cite{Contino:2004vy} but do not change the picture qualitatively, the low-energy couplings are
\beq
\label{coupev}
\displaystyle
\lambda_{t_L}[m_*]\simeq  \lambda_{t_L}\left(\frac{m_*}{\Lambda_{\textrm{UV}}}\right)^{d_L-5/2}\hspace{-4pt},\;\;\;\;\;
\lambda_{t_R}[m_*]\simeq  \lambda_{t_R}\left(\frac{m_*}{\Lambda_{\textrm{UV}}}\right)^{d_R-5/2}\hspace{-4pt}.
\eeq
We see that for a large scale separation  $\Lambda_{\textrm{UV}}\gg m_*$ they can remain sizable, thus generating a large enough top Yukawa, if the operator dimensions are taken to be close to the critical value $d_{L,R}\sim 5/2$.

Partial compositeness is superior to the technicolor way in two respects. First, there is no known obstruction in having fermionic operators of dimension nearly $5/2$ while we saw that $d\simeq 1$ for a scalar is problematic. There is no risk of reintroducing an Naturalness problem because already in the simplest cases where $d[\phi^2]=2\,d[\phi]$, like in the free theory or for gauge theories with a large number of colors, the operator square $|{\mathcal{O}}_F|^2$ has a safely irrelevant dimension equal to $5$. Furthermore the five-dimensional models in anti-de Sitter space do provide explicit examples of conformal field theories (in the sense that they obey the conformal group algebra and unitarity) where dimensions close to $5/2$ can be realized. Differently from the scalar operator of dimension one, there can thus not be any first principle obstruction against $d_{L,R}\sim 5/2$. 

The second point in favor of partial compositeness concerns the generation of the flavor hierarchies. With the technicolor approach the small masses of the bottom, the charm and the other quarks must find a justification in the underlying microscopic theory, where some selection rule must be at work ensuring an hierarchy among the $\lambda_t$ and $\lambda_b$ UV couplings of the different families. These hierarchies are maintained by the running since all the Yukawa-like operators have the same scaling dimension, which is dictated by the one of the scalar operator ${\mathcal{O}}_S$. In partial compositeness, instead, each quark flavor couples to the composite sector through its own set of fermionic operators ${{\mathcal{O}}_{F,f}^{L,R}}$, each characterized by its own dimensions $d_{L,R}^f$. The operators in top quark sector are required to have $d_{L,R}^t\sim 5/2$ while the others could have different dimensions. If $d_{L,R}^{b,\,c,\,\ldots}- 5/2={\mathcal{O}}(1)>0$ the couplings at $m_*$ are naturally reduced by the running as in Eq.~(\ref{coupev}) and light quark masses are obtained even if no hierarchy was present in the UV couplings $\lambda_{b,\,c,\,\ldots}\sim\lambda_{t}$. More details on this mechanism, and on how it could lead to a realistic $V_{CKM}$ matrix and to the suppression of extra flavor transition among the light quarks, will be reviewed in Chap.~\ref{ch:flavor}.

Partial compositeness means, as mentioned above, that the physical SM particles are linear superimpositions of elementary and composite degrees of freedom, namely
\beq
\label{parC}
|{\textrm{Phys.}}_i\rangle = \cos{\theta_i}|{\textrm{Elem.}}_i\rangle+\sin{\theta_i}|{\textrm{Comp.}}_i\rangle\,,
\eeq
let us see how this works in detail. At the scale $m_*$ the strong sector condenses and it generates, on top of the pNGB Higgs, a set of resonances with typical mass $m_*$. At least one resonance is expected for each gauge-invariant local operator, in the sense that each operator is expected to be capable to excite from the vacuum a single-particle state with the same Lorentz and internal quantum numbers. The fermionic operators  ${{\mathcal{O}}_{F}^{L,R}}$ are thus associated, for each family, to fermionic resonances $Q$ and $\tilde{T}$ for which
\beq
\langle0|{{\mathcal{O}}_{F}^{L}}|Q\rangle\neq0\,,\;\;\;\;\;
\langle0|{{\mathcal{O}}_{F}^{R}}|\tilde{T}\rangle\neq0
\,,
\eeq
and similarly for the down-type sector. The resonances are called ``partners'' and their basic properties are easily deduced from the equation above. First of all, they must carry the same \mbox{SU$(2)_L\times$U$(1)_Y$} quantum numbers of the corresponding SM fields, namely $Q$ and $\tilde{T}$ are one doublet with $1/6$ hypercharge and one singlet with $Y=2/3$, respectively. One hypercharge $-1/3$ singlet, $\tilde{B}$, will also emerge from the bottom sector. These representations are actually part of larger multiplets because in composite Higgs the SM group is embedded in an unbroken \mbox{SO$(4)$}. Second, the Partner's mass originates from the strong sector confinement irregardless of the breaking of the EW symmetry. The partners must thus be endowed with a Dirac mass term, as opposite to a Majorana one because they are charged, which means that both chiralities must be present with the same quantum numbers. Fermions of this kind are said to be ``vector-like''. 

Finally, and very importantly, the partners carry QCD color because the fermionic operators must come in color triplets in order to be consistently coupled with the quarks. This property marks another relevant conceptual difference with the technicolor approach to fermion masses. In the latter case the composite sector needs not to carry color and all its constituents might well be assumed to be color singlets. In partial compositeness this is not possible: the color group must be present as a global and unbroken \mbox{SU$(3)$} symmetry of the composite sector and the QCD interactions must be introduced by gauging this symmetry through elementary gluon fields similarly to what we saw in the previous section for the \mbox{SU$(2)_L\times$U$(1)_Y$} group. This has important phenomenological implications on which we will come back later.  

Since the partners are excited from the vacuum by ${\mathcal{O}}^{L,R}_{F,f}$, the partial compositeness interactions in Eq.~(\ref{intpc}) makes them mix with the quarks, with a strength proportional to the IR couplings. The energy dimensionality of the mixing is carried by the composite sector's confinement scale $m_*$ and the result is also weighted, as we will discuss in details in the next chapter, by one inverse power of the typical composite sector coupling $g_*$. The mass terms of the quark/quark-partners system are estimated to be 
\bea
\label{mixlag}
\displaystyle
&&{\mathcal{L}}_{\textrm{Mass}}^L=-m_*\overline{Q}Q-\frac{\lambda_{f_L}}{g_*} m_*\, (\overline{q}_LQ+h.c.)\,,\nonumber\\
&&{\mathcal{L}}_{\textrm{Mass}}^R=-m_*\overline{\tilde{T}}\tilde{T}-\frac{\lambda_{f_R}}{g_*} m_*\, (\overline{t}_R \tilde{T}+h.c.)\,.
\eea
Even if we did not indicate it explicitly, the $\lambda_{f_{L,R}}$ couplings in the above equation are clearly the ones evolved to $m_*$ according to Eq.~(\ref{coupev}). However from the low-energy viewpoint we can ignore their microscopic origin and regard them as free input parameters. The mass matrices are easily diagonalized leading to two massless Eigenstates, which we identify with the physical $q_L$ and $t_R$ quarks, plus heavy resonances. The light states are partially composite as in Eq.~(\ref{parC}) with compositeness fractions 
\beq
\label{angles}
\displaystyle
\sin\theta_L^f=\frac{\lambda_{f_L}}{\sqrt{g_*^2+(\lambda_{f_L} )^2}}\simeq\frac{\lambda_{f_L}}{g_*}\,,\;\;\;\;\;\sin\theta_R^f=\frac{\lambda_{f_R}}{\sqrt{g_*^2+(\lambda_{f_R} )^2}}\simeq\frac{\lambda_{f_R}}{g_*}\,,
\eeq
In the second set of equalities we took the limit $\lambda_{f_L}\ll g_*$, which is most likely appropriate for the light flavors but not necessarily so for the top quark.

\begin{figure}
\centering
\includegraphics[width=0.7\textwidth]{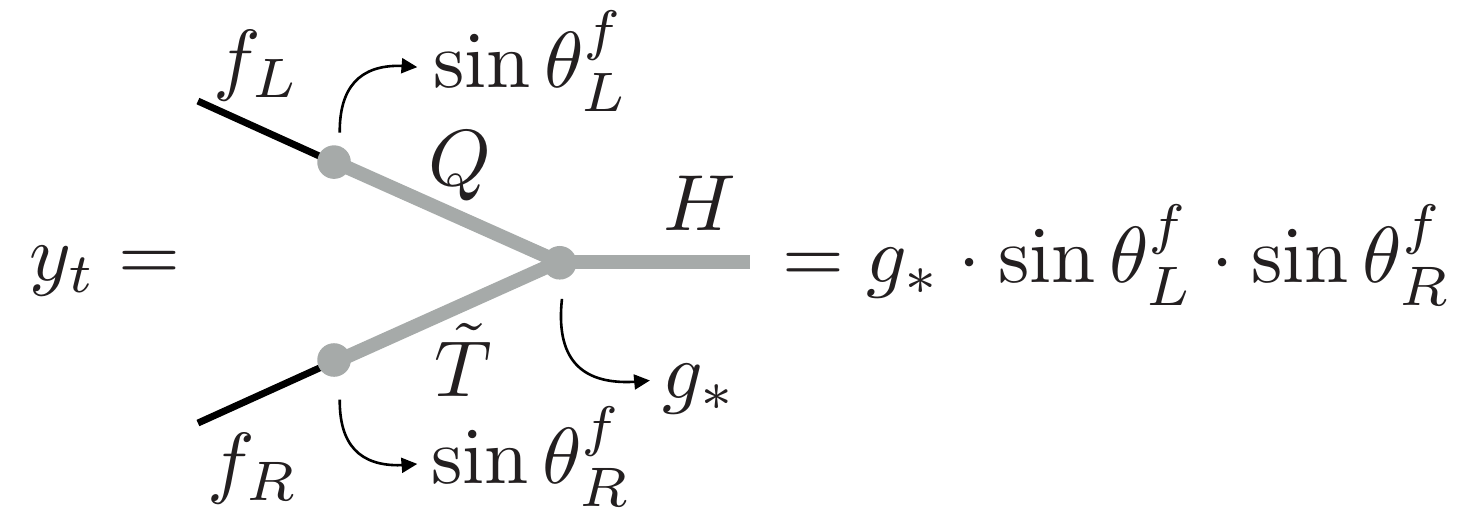}
\caption{Yukawa couplings generation in partial compositeness, under the supplementary hypothesis of VMD as explained in the text.}
\label{yukc}
\end{figure}

We are finally in the position to estimate the Yukawa couplings, out of which the SM particles eventually acquire their mass after EWSB. The $Q$ and $\tilde{T}$ partners couple to the Higgs with coupling strength $g_*$ and this gives rise, after the rotation to the mass basis, to Yukawa couplings of the massless eigenstates which are proportional to the left- and right-handed compositeness fractions. As depicted in Fig.~\ref{yukc} the Yukawas are given by
\beq
\label{yukpar}
\displaystyle
y_f= g_*\sin\theta_L^f \sin\theta_R^f \simeq \frac{\lambda_{f_L} \lambda_{f_R}}{g_*}\,.
\eeq
Light SM particles, with small Yukawas, are thus characterized by small $\lambda$'s and thus by a tiny compositeness fraction $\sin\theta_{L,R}^f\ll1$ while the top is obliged to be composite to a large extent in order to obtain its large Yukawa.

This concludes our first illustration of partial compositeness. Though qualitative, it should be sufficient to transmit the general idea. However it is important to stress that our discussion, and in particular the derivation of the Yukawa couplings in Eq.~(\ref{yukpar}), is not only qualitative but it also relies on one extra assumption which is not intrinsic of partial compositeness and could have been avoided. Namely, the estimate (\ref{yukpar}) is based on the idea that the only interactions of the elementary degrees of freedom with the Higgs are the ones mediated by the mixing with the composite resonances. Otherwise, extra contributions to the Yukawa would have been present in Fig.~\ref{yukc} from contact interactions involving the elementary states directly. This might be motivated by an analogy with hadron physics, where it is known that the interactions of the photon are mostly driven by the mixing with a resonance of appropriate quantum numbers, the $\rho$ meson. This is the so-called Vector Meson Dominance (VMD) hypothesis which we might generalize in the present context to the dominance of the partner's exchange in the interaction of the elementary fields with the composite sector. However, VMD is not a robust and theoretically well understood feature and we should not take it too seriously.\footnote{Nevertheless, as stressed in Ref.~\cite{Contino:2006nn}, it might still be a convenient simplifying assumption for the study of composite Higgs models.} It is thus important to remark that Eq.~\ref{yukpar} does not rely on VMD. The reader will easily realize this after reading Chap.~\ref{ch:SILH}.

\subsection{Higgs couplings to fermions}
\label{sec:HCsigmaM}

Now that the general idea is clear, let us see how to implement partial compositeness concretely in the composite Higgs framework. We will show how sharp leading-order predictions can be obtained for the physical Higgs couplings to fermions. However, differently from the gauge boson couplings in Eq.~(\ref{hvc}) which we saw to be model-independent and completely determined by the choice of the coset, the fermion interactions depend on one extra model-building assumption related with the detailed implementation of partial compositeness. The simplest and most common options will be discussed in turn.

In partial compositeness the quarks interact with the composite sector through fermionic operators ${\mathcal{O}}_F^{L,R}$  as in Eq.~(\ref{hvc}). In order to make quantitative predictions more details must be specified on the nature of those operators. In particular, we do need to specify their representation under the \mbox{SO$(5)$} global group. Notice that the full  \mbox{SO$(5)$}, irregardless of being spontaneously broken to \mbox{SO$(4)$}, is what matters here because it must be kept in mind that the elementary/composite interactions were originally written at the high $\Lambda_{\textrm{UV}}$ scale, far above the one where spontaneous breaking occurs. At that scale the operators are classified into full \mbox{SO$(5)$} multiplets. Stated differently, the UV operators are made of the strong sector constituents of the underlying microscopic theory, for which \mbox{SO$(5)$} is still an unbroken flavor symmetry and particles are grouped in \mbox{SO$(5)$} representations. Choosing the \mbox{SO$(5)$} representations of ${\mathcal{O}}_F^{L,R}$ is the model-building ambiguity we were referring to in the previous paragraph. A priori, any complicated and reducible representation might be considered. However the only mixing that  matters in the IR is the one with the operator of lowest dimension, given that the effect of the others is washed out by the running. This singles out a unique irreducible representation, barring the implausible possibility that the strong sector contains  several multiplets with accidentally comparable scaling dimensions. A case-by-case study of the irreducible representations, starting from the smallest multiplets which are more likely to have low energy dimension, is thus sufficient to cover all the plausible theoretical options. 

The operator multiplets must be such as to contain the SM quarks representations, $\mathbf{2}_{1/6}$, $\mathbf{1}_{2/3}$ and $\mathbf{1}_{-1/3}$, when decomposed under the EW group \mbox{$G_{\textrm{EW}}$}. These components of the multiplet are the only ones that actually participate in the elementary/composite interactions. However, it turns out that no such representation exists if $G_{\textrm{EW}}$ is entirely embedded in \mbox{SO$(5)$}. An extension of the global symmetry group of the composite sector is required in order to implement partial compositeness. The simplest possibility is to add a new unbroken \mbox{U$(1)_X$} factor, extending the original breaking pattern \mbox{SO$(5)\rightarrow\;$ SO$(4)$} to \footnote{Partial compositeness requires, as described in the previous section, one further extension by an unbroken color \mbox{SU$(3)_c$} group under which the fermionic operators are triplets. The complete group is thus  \mbox{SO$(5)\times$U$(1)_X\times$SU$(3)_c$}.\label{foot1}}
\beq
\textrm{SO}(5)\times{\textrm{U}}(1)_X\;\rightarrow\;\;\textrm{SO}(4)\times {\textrm{U}}(1)_X\,.
\eeq
In the scheme we had in mind until now, which was introduced in Sect.~\ref{sec:MCHMlin}, $G_{\textrm{EW}}$ was embedded in the unbroken \mbox{SO$(4)=\;$SU$(2)_L\times$SU$(2)_R$} and the hypercharge was provided by the third \mbox{SU$(2)_R$} generator, $Y=T^3_R$. We now instead give to the hypercharge also one component along the newly introduced \mbox{U$(1)_X$} generator, namely
\beq
\label{hyp}
Y=T^3_R+X\,.
\eeq
In practice, this means introducing a new term in the elementary/composite interactions of Eq.~(\ref{gaugesources}), namely to take \footnote{An analogous term $g_S \,G_a\cdot J_{c}^a$ has to be introduced for the gluon fields $G_{\mu,\,a}$, coupled to the currents of the \mbox{SU$(3)_c$} color group introduced in Footnote~\ref{foot1}.\label{foot2}}
\beq
\label{gscoup}
{\mathcal{L}}_{\textrm{int}}^{\textrm{gauge}}=A_{\mu,\,A} J^{\mu,\,A}+X_\mu J^\mu_X\,,
\eeq
where $J^\mu_X$ is the \mbox{U$(1)_X$} global current and $X_\mu$ is the associated source field. The physical value of the new source is 
\beq
\label{Xsv}
X_\mu=g'B_\mu\,,
\eeq
where $B_\mu$ is the hypercharge field. $B_\mu$ also enters in the $T_R^3$ term of $A_{\mu}$ (\ref{physg}) and therefore it couples with $Y$ as required by Eq.~(\ref{hyp}).

This might seem a radical deformation of our setup, however it is not because \mbox{U$(1)_X$} is unbroken and thus it does not lead to a new Goldstone. Furthermore it commutes with \mbox{SO$(5)$} and therefore the bosonic fields whose dynamics we have been studying so far, namely the Higgs and the gauge fields, are all neutral objects. All the derivations of the previous sections, the definition of the CCWZ $d$ and $e$ symbols and their properties are unaffected by the extension of the group and hold in exactly the same way. The presence of the \mbox{U$(1)_X$} must of course be taken into account when dealing with charged fields, in particular the covariant derivative of Eq.~(\ref{gencd}) should be supplemented by a term with the source $X_\mu$ in order to respect local \mbox{U$(1)_X$} invariance.

We now return to our problem of identifying suitable representations for the fermionic operators. The simplest one is the $\mathbf{5}$, let us thus start from the case in which the ${\mathcal{O}}_F^{L,R}$, namely those that mix with the elementary $q_L$ and $t_R$ as in Eq.~(\ref{intpc}), are part of a fiveplet with an appropriate choice, $X=2/3$, of the \mbox{U$(1)_X$} charge. When decomposed under \mbox{SO$(4)$}, the fiveplet splits into a singlet plus one fourplet  $\mathbf{4}={\mathbf{(2,2)}}$, which in turn leads to two \mbox{SU$(2)_L$} doublets of opposite $T^3_R$ charges as shown in Appendix~\ref{so4}.  Given our new definition of the hypercharge (\ref{hyp}), the $\mathbf{5}_{2/3}$ decomposes under $G_{\textrm{EW}}$ as
\beq
\label{5dec}
\mathbf{5}_{2/3}\rightarrow\; \mathbf{4}_{2/3}\oplus \mathbf{1}_{2/3}\rightarrow \;\mathbf{2}_{7/6}\oplus \mathbf{2}_{1/6}\oplus \mathbf{1}_{2/3}\,.
\eeq
The two last terms could couple to $q_L$ and to $t_R$ respectively. The easiest one is the $t_R$ coupling. The fermionic operator, $({\mathcal{O}}_F^{R})_I$, is endowed with a fiveplet index $I$ but the only component which couples to $t_R$ is the singlet embedded in the last entry $I=5$. Nevertheless, it is convenient to express the interaction as
\beq
\label{pcrew}
{\mathcal{L}}_{\textrm{int}}^{t_R}=\lambda_{t_R}\overline{t}_R\left({\mathcal{O}}_F^R\right)_5+h.c.=\lambda_{t_R}\left(\overline{T}_R\right)^I\left({\mathcal{O}}_F^R\right)_I+h.c.\,,
\eeq
in terms of an incomplete fiveplet
\beq
\label{tremb}
T_R=\{0,\,0,\,0,\,0,\,t_R\}^T\,,
\eeq
which we denote as the embedding of $t_R$ in the $\mathbf{5}$. Notice that the interaction has been written in a shorthand notation in which $\lambda_{t_R}$ represents the coupling strength at the IR and the powers of $m_*$ needed to match energy dimensionality have been reabsorbed in the operator normalization. The `` $^T$ '' symbols is just the transpose, needed because we want to  work with column vectors.

The rewriting in terms of $T_R$ is extremely useful to read the implications of the symmetries on the elementary fermions interactions. Suppose one is willing to compute the effective Lagrangian for the SM fermions, the gauge and the Higgs fields, obtained by integrating out the composite sector dynamics and ignoring the virtual effects from elementary fields exchange, which could be possibly added on top. The fact that $T_R$ is an incomplete multiplet is irrelevant for this calculation, we might formally uplift it to a complete multiplet of external source fields and eventually set it to its physical form by Eq.~(\ref{tremb}). The idea is exactly the one we introduced in Sect.~\ref{sources_and_local} to deal with the elementary gauge fields. But if $T_R$ is regarded as a complete multiplet we can consider transforming it under the global group and its effective Lagrangian must stay invariant. The transformation must clearly be the same one of ${\mathcal{O}}_F^R$ in order for the interaction to be formally invariant. Therefore under \mbox{SO$(5)$}
\beq
\left({T}_R\right)_I\rightarrow\;g_{I}^{\;\;J}\left({T}_R\right)_J\,,
\eeq
and the \mbox{U$(1)_X$} charge is equal to $2/3$. In order to write down invariant Lagrangians by employing the general CCWZ construction it is convenient to ``dress'' the source with the Goldstone matrix and to define the following objects
\beq
\label{so4mult}
\left\{T_R^{\mathbf{4}},\,T_R^{\mathbf{1}}\right\}^T=U[\Pi]^{-1}\cdot T_R\,.
\eeq
This dressing procedure is fully analog to the one we adopted for the gauge source in Eq.~(\ref{dedefgauge}): by multiplying with $U^{-1}$ we turn an index transforming with $g$ into one transforming with $h[\Pi;\,g]$. The latter can be eventually  contracted, together with all the other CCWZ objects defined in Sect.~\ref{CCWZ}, by respecting the local \mbox{SO$(4)$} symmetry and the result will be automatically invariant under the full \mbox{SO$(5)$}. Notice that $h[\Pi;\,g]$ is the exponential of unbroken \mbox{SO$(4)$} generators only, for which we took a block-diagonal form. Therefore $h$ itself is block diagonal, with the first $4\times4$ block made of an \mbox{SO$(4)$} rotation and ``$1$'' in the remaining entry. The two objects defined above, $T_R^{\mathbf{4}}$ and $T_R^{\mathbf{1}}$, thus belong to two distinct \mbox{SO$(4)$} representation, namely
\beq
T_R^{\mathbf{4}}\;\in\; \mathbf{4}_{2/3}\,,\;\;\;\;\;T_R^{\mathbf{1}}\;\in\; \mathbf{1}_{2/3}\,,
\eeq
with $2/3$ \mbox{U$(1)_X$} charge. They can be employed independently in the construction of invariants.

We now turn to the $q_L$ coupling. The corresponding operator, $({\mathcal{O}}_F^{L})_I$, is still an \mbox{SO$(5)$} fiveplet with $X=2/3$, but it is not necessarily related with $({\mathcal{O}}_F^{R})_I$. Two independent operators might well exist in the composite sector, characterized by different scaling dimensions $d_L\neq d_R$. Furthermore, the opposite chirality components of ${\mathcal{O}}_F^{L}$ and ${\mathcal{O}}_F^{R}$ participate to the mixing, namely the right-handed chirality for ${\mathcal{O}}_F^{L}$, which mixes with $q_L$, and the left-handed one for ${\mathcal{O}}_F^{R}$. If the composite sector does not respect the parity symmetry the two chiralities correspond to independent operators. If on the contrary the composite sector does respect parity we might be entitled to regard  ${\mathcal{O}}_F^{L}$ and ${\mathcal{O}}_F^{R}$ as the two components of one single Dirac operator and in this case $d_L=d_R$. The discussion which follows is independent of which of the two options is realized. 

In very much the same way as for the $t_R$ we write the $q_L$ interaction as 
\beq
\label{pcrew1}
{\mathcal{L}}_{\textrm{Int}}^{q_L}=\lambda_{t_R}\left(\overline{Q}_{t_L}\right)^I\left({\mathcal{O}}_F^L\right)_I+h.c.\,,
\eeq
where ${Q}_{t_L}$ is again an incomplete multiplet, this time given by
\beq
\label{qlemb}
Q_{t_L}=\frac1{\sqrt{2}}\{-i\,b_L,\,-b_L,\,-i\,t_L,\,t_L,\,0\}^T\,.
\eeq
The $q_L$ doublet is embedded in the \mbox{SO$(4)$} fourplet components of the fiveplet according to Eq.~(\ref{emb5}), it corresponds to the $T_R^3=-1/2$ doublet $\Psi_-$. Exactly like for $T_R$ (\ref{so4mult}), we can act on $Q_L$ with the Goldstone matrix and obtain two new \mbox{SO$(4)$} multiplets
\beq
\label{spurs4}
Q_{t_L}^{\mathbf{4}}\;\in\; \mathbf{4}_{2/3}\,,\;\;\;\;\;Q_{t_L}^{\mathbf{1}}\;\in\; \mathbf{1}_{2/3}\,,
\eeq
which we will employ in the construction of the invariants.

The leading order invariants are the ones made of two source fields and no derivatives. These are $\mathcal{O}(p^0)$ operators in the counting we introduced in Sect.~\ref{CCWZ}. The contraction of two $T_R$ or of two $Q_{t_L}$ sources vanishes because of chirality and one is left with mixed $Q_{t_L}$-$T_R$ terms. Two invariants might be formed, a priori, by contracting $Q_{t_L}^{\mathbf{4}}$ with $T_R^{\mathbf{4}}$ and $Q_{t_L}^{\mathbf{1}}$ with $T_R^{\mathbf{1}}$, however the two are not independent because of the following relation
\bea
\label{invrel}
\left(\overline{Q}_{t_L}^{\mathbf{4}}\right)^i\left({T}_R^{\mathbf{4}}\right)_i+\,\overline{Q}_{t_L}^{\mathbf{1}}{T}_R^{\mathbf{1}}&&=
\left({\overline{Q}_{t_L}}\right)^I\left[U_{I}^{\;i}{U^\dagger}_{i}^{\;J}+U_{I}^{\;5}{U^\dagger}_{5}^{\;J}\right]\left({T}_R\right)_J\nonumber\\
&&=\left(\overline{Q}_{t_L}\right)^I\left({T}_R\right)_I=0\,,
\eea
which vanishes because ${Q}_{t_L}$ and ${T}_R$ are orthogonal. We thus find a unique effective operator with two elementary fermions and no derivatives, which leads to a generalized top Yukawa Lagrangian
\bea
\label{yukup}
\displaystyle
{\mathcal{L}}_{\textrm{Yuk}}^{t}&&=-c^t\frac{\lambda_{t_L}\lambda_{t_R}}{g_*^2}m_*\overline{Q}_{t_L}^{\mathbf{1}}{T}_R^{\mathbf{1}}+h.c.\nonumber\\
&&=-c^t\frac{\lambda_{t_L}\lambda_{t_R}}{g_*^2}m_*\frac1{2\sqrt{2} |H|}\sin\frac{2\sqrt{2}|H|}{f}\overline{q}_L H^c t_R+h.c.
\eea
where $H^c=i\,\sigma^2 H^*$. The parametrization of the operator coefficient in the equation above is actually irrelevant for the results which follow, however it requires some explanation. The  $Q_L$ and $T_R$ sources interact with the composite sector only in combination with the coupling strength $\lambda_{t_L}$ and $\lambda_{t_R}$, respectively. Therefore each source insertion is necessarily accompanied by one power of the corresponding coupling. The composite sector scale $m_*$ provides the energy dimensionality of the operator and $c^t$ is an order one free parameter. The justification of the $1/g_*^2$ factor, which is similar to one we encountered in Eq.~(\ref{mixlag}), is instead more complicated and it is postponed to the next chapter. 

Eq.~(\ref{yukup}) is an infinite series of operators with more and more insertions of the Higgs field, each weighted by the Higgs decay constant $f$. The leading operator is just the dimension $4$ SM up-type Yukawa interaction. All the others, starting from $d=6$, provide corrections to the Higgs--top coupling with respect to the SM. This whole set of interactions is controlled, at fixed $f$, by a single multiplicative parameter which however is not free because we still have to impose the constraint of the top quark mass. When the Higgs is set to its VEV, Eq.~(\ref{yukup}) becomes the top mass term with
\beq
m_t=c^t\frac{\lambda_{t_L}\lambda_{t_R}}{g_*^2}m_*\frac{\sqrt{\xi(1-\xi)}}{\sqrt{2}}\,.
\eeq
By trading the prefactor for $m_t$ and going to the unitary gauge the generalized Yukawa Lagrangian in Eq.~(\ref{yukup}) becomes
\bea
\label{yukupug}
\displaystyle
{\mathcal{L}}_{\textrm{Yuk}}^{t}&&=-\frac{m_t}{2} \frac{1}{\sqrt{\xi(1-\xi)}}\sin\frac{2(V+h)}{f}\overline{t} t \nonumber\\
&&=-m_t \overline{t} t - k_t \frac{m_t}{v}  h \,  \overline{t} t - c_2 \frac{ m_t}{v^2}h^2  \overline{t} t+\ldots
\eea
It provides the top mass term, plus a set of top interactions with the physical Higgs. The first one is a SM-like coupling, but with a modified strength 
\beq
k_t^{\mathbf{5}}\equiv \frac{g_{htt}^{\textrm{comp}}}{g_{htt}^{\textrm{SM}}}=\frac{1-2\,\xi}{\sqrt{1-\xi}}\,.
\label{htc5}
\eeq
The result is labeled by the superscript ``${\ }^{\mathbf{5}}$'' because it relies on our choice of embedding the operators in the fiveplet. The second interaction is a dimension $5$ vertex with two Higgs bosons, obviously absent in the SM, with coefficient 
\beq
c_2^{\mathbf{5}}=-2\xi\,.
\eeq
As expected on general grounds, the couplings reduce to the SM ones in the limit $\xi\rightarrow0$. Namely $k_t^{\mathbf{5}}\rightarrow1$ and $c_2^{\mathbf{5}}\rightarrow0$.

The bottom quark sector, namely the interactions needed to generate the bottom mass and Yukawa couplings, are introduced in complete analogy with the top ones. We consider, on top of Eq.~(\ref{intpc}), two further elementary/composite couplings 
\beq
\label{intpcb}
{\mathcal{L}}_{\textrm{Int}}^b=
\frac{\lambda_{b_L}}{\Lambda_{\textrm{UV}}^{d_L^b-5/2}}\overline{q}_L{\mathcal{O}^b}_F^L+\frac{\lambda_{b_R}}{\Lambda_{\textrm{UV}}^{d_R^b-5/2}}\overline{b}_R{\mathcal{O}^b}_F^R\,,
\eeq
where ${\mathcal{O}^b}_F^L$ and ${\mathcal{O}^b}_F^R$ are, respectively, in the $\mathbf{2}_{1/6}$ and $\mathbf{1}_{-1/3}$ of the SM group. Notice that ${\mathcal{O}^b}_F^L$ has the same SM quantum numbers of the corresponding operator in the top sector. In spite of this, the two are independent objects, a priori. The bottom sector operators can be embedded in a $\mathbf{5}_{-1/3}$ of $\textrm{SO}(5)\times{\textrm{U}}(1)_X$, which decomposes as
\beq
\mathbf{5}_{-1/3}\rightarrow\; \mathbf{4}_{-1/3}\oplus \mathbf{1}_{-1/3}\rightarrow \;\mathbf{2}_{1/6}\oplus \mathbf{2}_{5/6}\oplus \mathbf{1}_{-1/3}\,,
\eeq
under $G_{\textrm{EW}}$. The interaction (\ref{intpcb}) is rewritten as
\beq
{\mathcal{L}}_{\textrm{int}}^{b}=\lambda_{b_L}\left(\overline{Q}_{b_L}\right)^I\left({\mathcal{O}^b}_F^L\right)_I+\lambda_{b_R}\left(\overline{B}_R\right)^I\left({\mathcal{O}^b}_F^R\right)_I\,,
\eeq
where, for shortness, the dimensionful normalization has been reabsorbed in the operators. The two new source fields $Q_{b_L}$ and $B_R$ are given by
\bea
\label{qlbemb}
&&Q_{b_L}=\frac1{\sqrt{2}}\{-i\,t_L,\,t_L,\,i\,b_L,\,b_L,\,0\}^T\,,\nonumber\\
&&B_R=\{0,\,0,\,0,\,0,\,b_R\}^T\,.
\eea
Differently from the one for the top in Eq.~(\ref{qlemb}), the $Q_{b_L}$ source is chosen to project on the $T_R^3=1/2$ doublet in the decomposition (\ref{emb5}).

By acting with the inverse of the Goldstone matrix we turn the sources into \mbox{SO$(4)$} multiplets
\bea
&&Q_{b_L}^{\mathbf{4}}\;\in\; \mathbf{4}_{-1/3}\,,\;\;\;\;\;Q_{b_L}^{\mathbf{1}}\;\in\; \mathbf{1}_{-1/3}\,,\nonumber\\
&&B_R^{\mathbf{4}}\;\in\; \mathbf{4}_{-1/3}\,,\;\;\;\;\;\;\;B_R^{\mathbf{1}}\;\in\; \mathbf{1}_{-1/3}\,,
\eea
out of which we can form invariants. Contractions with the top sector sources are forbidden by \mbox{U$(1)_X$} and the only invariant is found to be
\bea
\label{yukdown}
\displaystyle
{\mathcal{L}}_{\textrm{Yuk}}^{b}&&=-c^b\frac{\lambda_{b_L}\lambda_{b_R}}{g_*^2}m_*\overline{Q}_{b_L}^{\mathbf{1}}{B}_R^{\mathbf{1}}+h.c.\nonumber\\
&&=-c^b\frac{\lambda_{b_L}\lambda_{b_R}}{g_*^2}m_*\frac1{2\sqrt{2} |H|}\sin\frac{2\sqrt{2}|H|}{f}\overline{q}_L H b_R+h.c.
\eea
This generalized down-type Yukawa Lagrangian gives mass to the bottom quark. After trading the prefactor for $m_b$ and going to the unitary gauge
\bea
\label{yukdownug}
\displaystyle
{\mathcal{L}}_{\textrm{Yuk}}^{b}&&=-\frac{m_b}{2} \frac{1}{\sqrt{\xi(1-\xi)}}\sin\frac{2(V+h)}{f}\overline{b} b \nonumber\\
&&=-m_b \overline{b} b - k_b \frac{m_b}{v}  h \,  \overline{b} b +\ldots
\eea
where $k_b$, the modification of the bottom--Higgs coupling with respect to the SM is found to be
\beq
k_b^{\mathbf{5}}\equiv \frac{g_{hbb}^{\textrm{comp}}}{g_{hbb}^{\textrm{SM}}}=\frac{1-2\,\xi}{\sqrt{1-\xi}}\,.
\label{hbc5}
\eeq
Additional higher dimensional vertices, such as $h^2\overline{b}b$, are also present in Eq.~(\ref{yukdownug}). However they are suppressed by the small bottom mass and thus, differently from the $h^2\overline{t}t$ coupling in Eq.~(\ref{yukupug}), hardly play a relevant phenomenological role.

The discussion proceeds along similar lines for any representation in which we might take the fermionic operators to transform. An ``economical'' choice is the spinorial $\mathbf{4}$ of \mbox{SO$(5)$} defined in Appendix~\ref{so54}. From the explicit form of its generators (\ref{gen45}) we see that it decomposes as ${\mathbf{4}}={(\mathbf{2},\mathbf{1})}\oplus {(\mathbf{1},\mathbf{2})}$ under \mbox{SO$(4)$}, therefore if we assign to it a \mbox{U$(1)_X$} charge $X=1/6$ it will simultaneously contain all the SM representations of one complete quark family, namely
\beq
\mathbf{4}_{1/6}\rightarrow\; (\mathbf{2},\mathbf{1})_{1/6}\oplus (\mathbf{1},\mathbf{2})_{1/6}\rightarrow \;\mathbf{2}_{1/6}\oplus \mathbf{1}_{2/3}\oplus \mathbf{1}_{-1/3}\,.
\eeq
The elementary field embeddings become now 
\bea
\label{embspin}
&&Q_{t_L}=Q_{b_L}=\{t_L,\,b_L,\,0,\,0\}^T\,,\nonumber\\
&&T_R=\{0,\,0,\,t_R,\,0\}^T\,,\nonumber\\
&&B_R=\{0,\,0,\,0,\,b_R\}^T\,,
\eea
out of which a total of $6$ CCWZ tensors, $3$ of which in the $(\mathbf{2},\mathbf{1})_{1/6}$ and the others in the $(\mathbf{1},\mathbf{2})_{1/6}$, can be constructed by acting with the inverse of the Goldstone matrix. Clearly, the Goldstone matrix in the spinorial, reported in Eq.~(\ref{gm4}), has to be employed. Out of the $4$ possible invariants only two are independent because, similarly to what we saw happening in Eq.~(\ref{invrel}) for the fiveplet, the sum of the $(\mathbf{2},\mathbf{1})$ and of the $(\mathbf{1},\mathbf{2})$ self-contractions vanishes. This is actually a general rule, which has to do with the fact the sum of the contractions among the different \mbox{SO$(4)$} multiplets reconstructs one  \mbox{SO$(5)$} invariants from which the Goldstone matrix, which acts as an  \mbox{SO$(5)$} transformation, systematically cancels. The net result is that only two invariants exist, corresponding to up- and down-type generalized Yukawas
\bea
\displaystyle
{\mathcal{L}}_{\textrm{Yuk}}^{t}&&=-i\,c^t\frac{\lambda_{t_L}\lambda_{t_R}}{g_*^2}m_*\overline{Q}_{t_L}^{(\mathbf{2},\mathbf{1})}\cdot{T_R}^{(\mathbf{2},\mathbf{1})}+h.c.\nonumber\\
&&=-c^t\frac{\lambda_{t_L}\lambda_{t_R}}{g_*^2}m_*\frac1{2 |H|}\sin\frac{\sqrt{2}|H|}{f}\overline{q}_L H^c t_R+h.c.\nonumber\\
{\mathcal{L}}_{\textrm{Yuk}}^{b}&&=-i c^b\frac{\lambda_{b_L}\lambda_{b_R}}{g_*^2}m_*\overline{Q}_{b_L}^{(\mathbf{2},\mathbf{1})}\cdot{B_R}^{(\mathbf{2},\mathbf{1})}+h.c.\nonumber\\
&&=-c^t\frac{\lambda_{t_L}\lambda_{t_R}}{g_*^2}m_*\frac1{2 |H|}\sin\frac{\sqrt{2}|H|}{f}\overline{q}_L H b_R+h.c.
\eea
where the dot stands for the doublet indices contraction. By trading the pre-factor for the physical top and bottom masses, going to the unitary gauge and Taylor-expanding the Lagrangian we obtain the top and bottom Yukawa couplings modification and the coefficient of the $h^2{\overline{t}}t$ operator
\beq
\displaystyle
k_t^{\mathbf{4}}=k_b^{\mathbf{4}}=\sqrt{1-\,\xi}\,,\;\;\;\;\;c_2^{\mathbf{4}}=-\frac{\xi}2\,.
\eeq
As anticipated, the predicted pattern of fermion coupling modification is different than the one we encountered in the case of the ${\mathbf{5}}$.

The scenarios described above, with fermionic operators in the ${\mathbf{5}}$ and in the ${\mathbf{4}}$, correspond to the most popular Minimal Composite Higgs Models, denoted as \mbox{MCHM$_5$} and \mbox{MCHM$_4$}, respectively. The \mbox{MCHM$_4$} was first proposed in Ref.~\cite{Agashe:2004rs} but subsequently abandoned in favour of the \mbox{MCHM$_5$}~\cite{Contino:2006qr} because it was found, as reviewed in the following chapters, that it leads to a large and phenomenologically unacceptable correction to the $Z$ boson coupling to the left-handed bottom. The \mbox{MCHM$_4$} and the \mbox{MCHM$_5$}, aside from the fact that they predict different modified couplings, share two peculiar features which are not representative of the generic situation one might encounter for other representations. The first peculiarity is that the Higgs couplings are uniquely predicted in terms of $\xi$. This came because in both cases we found only one invariant operator for the top and one for the bottom sector, the theory was thus completely specified in terms of two parameters only which we traded for $m_t$ and $m_b$. There exist representations for which this is not the case. Suppose for instance embedding the top sector operators in an adjoint of \mbox{SO$(5)$} with $X=2/3$, the \mbox{SO$(4)\times$U$(1)_X$} decomposition of this representation is
\beq
{\mathbf{10}}_{2/3}\rightarrow\;(\mathbf{3},\mathbf{1})_{2/3}\oplus (\mathbf{1},\mathbf{3})_{2/3} \oplus (\mathbf{2},\mathbf{2})_{2/3}\,.
\eeq
This is a viable representation since $q_L$ can be embedded in the bidoublet and $t_R$ in one of the components of the $(\mathbf{1},\mathbf{3})_{2/3}$. After being dressed with the Goldstone matrix, the $Q_L$ and $T_R$ sources give rise, by the above decomposition, to three separate irreducible representations which we can use to write down the effective Lagrangian. Three singlets can thus be formed by contracting $Q_L$ with $T_R$ and only one combination of those is trivial because it corresponds to a full \mbox{SO$(5)$} invariant. We are thus left with two independent operators in the top sector and thus with two free parameters.\footnote{By making extra assumptions on the composite sector the number of parameters might be reduced back to one. For instance, if we postulate that the composite sector is invariant under a $P_{LR}$ parity which interchanges left and right \mbox{SO$(4)$} generators the operators constructed from the $(\mathbf{3},\mathbf{1})$ and the $(\mathbf{1},\mathbf{3})$  would be obliged to have the same coefficient.} One combination is fixed by the top mass but the other remains free and affects the Higgs couplings which are no longer predicted.

\begin{figure}
\centering
\includegraphics[width=.65\textwidth]{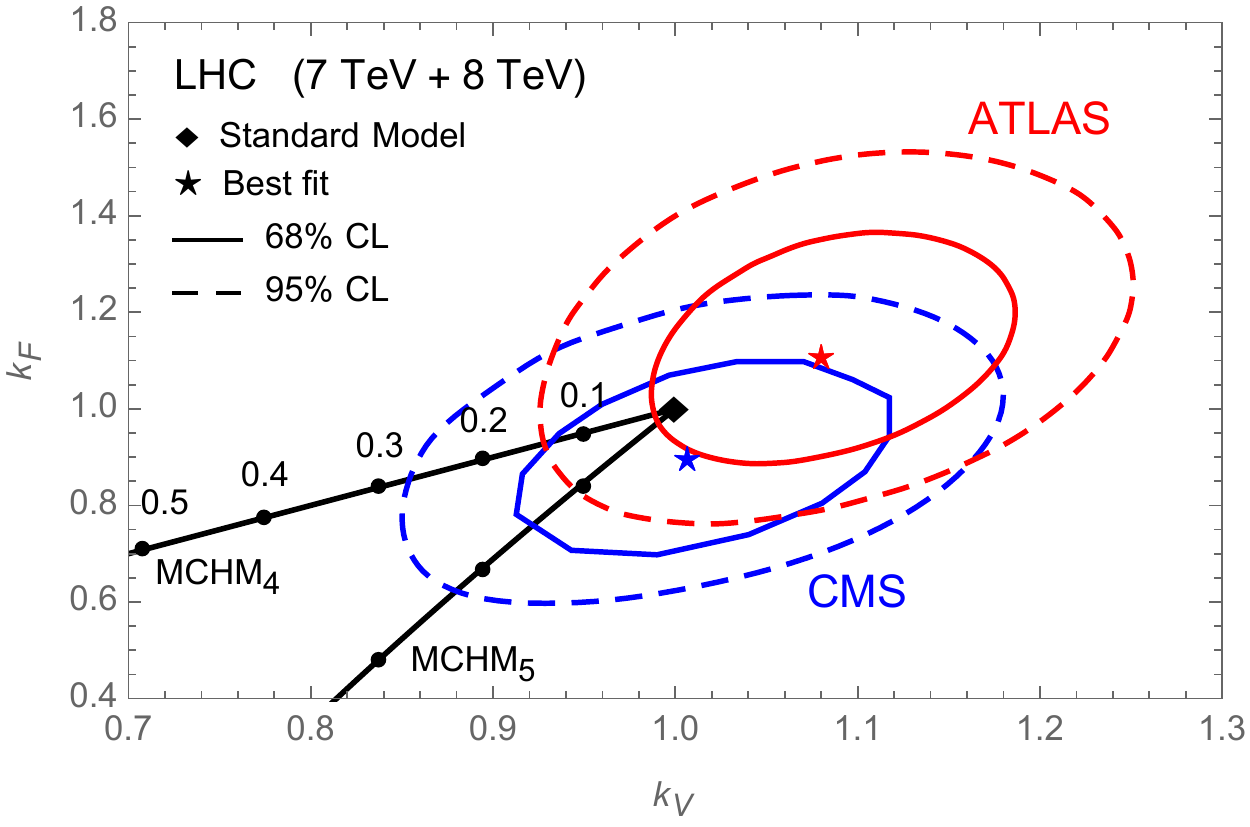}
\caption{Fit of the Higgs coupling strength to the gauge bosons ($k_V$) and fermions ($k_F$) obtained by the
ATLAS (red contours)~\cite{ATLASHiggsCouplings} and CMS Collaborations (blue contours)~\cite{Khachatryan:2014jba} from the combination of the $7$ and $8\ \mathrm{TeV}$ LHC data.
The solid black lines show the predictions in the \mbox{MCHM$_{5,4}$} models for different values of $\xi$.}
\label{fig:kvkf}
\end{figure}

The second peculiarity is that in both the \mbox{MCHM$_5$} and the \mbox{MCHM$_4$} scenarios the physical Higgs couplings to top and bottom are modified by the same amount with respect to the SM, namely $k_t=k_b$. This is obviously not a general feature, the simplest counterexample is to consider to different \mbox{SO$(5)$} representation for the top and the bottom sectors. If for instance we had taken the top sector operators are in the $\mathbf{5}$ but the bottom ones in the $\mathbf{4}$ we would have found $k_t=k_t^{\mathbf{5}}\neq k_b=k_b^{\mathbf{4}}$. Generic or not, the equality of top and bottom couplings modification is a robust feature of the \mbox{MCHM$_{4,5}$} models, which we can exploit for a fast and easy comparison with the experimental measurements of the Higgs boson couplings. One way in which those experimental results are presented are Confidence Level (CL) curves on the $k_V$-$k_F$ plane, where $k_V$ is the modification of the Higgs coupling to vectors and $k_F$ is a universal rescaling of the coupling to fermions with respect to the SM. Since the current experimental results are dominated by the top and bottom Yukawa couplings while the sensitivity to leptons and light quark couplings is extremely mild or absent, we can interpret $k_F$ as $k_F=k_t=k_b$, without worrying about the other fermions. By combining our predictions for $k_F$ as a function of $\xi$ with $k_V$ obtained in Eq.~(\ref{hvc}) we can draw a line on the $k_V$-$k_F$ plane for each of the two models and compare with the CL curves. The result, displayed in Fig.~\ref{fig:kvkf}, shows that relatively large values of $\xi$, above around $\xi\gtrsim0.2$ are excluded by the Higgs data. As we will see in the following chapters, these limits are barely competitive with other constraints from the LEP electroweak precision tests and from direct resonance searches at the LHC, but they are sufficient to give an idea of the allowed parameter space of the composite Higgs models.

For reasons related with the radiative generation of the Higgs potential which will become clear in the following chapter, one last case which is worth discussing is when the top sector operator ${\mathcal{O}^t}_F^L$ is in a ${\mathbf{14}}_{2/3}$ of \mbox{SO$(5)$}  and ${\mathcal{O}^t}_F^R$ in a singlet ${\mathbf{1}}_{2/3}$. The setup could be completed by different choice of representations in the bottom sector, which will turn out to have a mild impact on the Higgs potential. The ${\mathbf{14}}$ is the symmetric traceless tensor product of $2$ fiveplets, thus it decomposes in \mbox{SO$(4)$} in terms the corresponding $2$-index tensor, the ${\mathbf{9}}=(\mathbf{3},\mathbf{3})$, plus one fourplet sitting on the last line and column and one singlet in the $5$-$5$ entry, namely
\beq
{\mathbf{14}}_{2/3}\rightarrow\;(\mathbf{3},\mathbf{3})_{2/3} \oplus (\mathbf{2},\mathbf{2})_{2/3}\oplus (\mathbf{1},\mathbf{1})_{2/3}\,.
\eeq
The $q_L$ doublet mixes, as in previous cases, with the $(\mathbf{2},\mathbf{2})_{2/3}$, its embedding in the $(Q_{t_L})_{IJ}$ source tensor reads
\beq
\label{14source}
Q_{t_L}=\frac1{\sqrt{2}}\left[
\begin{array}{ccccc}
0 & 0 & 0 & 0 &-i\,b_L\\
0 & 0 & 0 & 0 &-\,b_L\\
0 & 0 & 0 & 0 &-i\,t_L\\
0 & 0 & 0 & 0 &t_L \\
-i\,b_L & -\,b_L & -i\,t_L & t_L & 0
\end{array}
\right]\,.
\eeq
When dressed with the Goldstone matrix, the ${\mathbf{14}}_{2/3}$ splits into three irreducible representations in accordance with the decomposition above, but the only one that matters is the singlet 
\beq
\displaystyle
{Q_{t_L}}^{\mathbf{1}}=\left(U[\Pi]^\dagger\right)_{5I}\left(U[\Pi]^\dagger\right)_{5J}{Q_{t_L}}^{IJ}=\frac1{\sqrt{2}|H|}\sin{\frac{2\sqrt{2} |H|}{f}}\left(H^c\right)^{\dagger}\cdot q_L\,.
\eeq
which we will combine with the $T_R$ to form an invariant operator.
\beq
\label{yuk141}
{\mathcal{L}}_{\textrm{Yuk}}^{t}=-c^t\frac{\lambda_{t_L}\lambda_{t_R}}{g_*^2}m_*\overline{Q}_{t_L}^{\mathbf{1}}{T_R}+h.c\,,
\eeq
where $T_R=t_R$ is the trivial embedding of $t_R$ in the singlet. With the usual procedure the modified Higgs couplings are easily found to be
\beq
\displaystyle
k_t^{\mathbf{14}}=\frac{1-2\,\xi}{\sqrt{1-\xi}}\,,\;\;\;\;\;c_2^{\mathbf{14}}=-2{\xi}\,.
\eeq
The results happens to coincide with the one obtained for the $\mathbf{5}$.

One further reason for being interested in the $\mathbf{14}\oplus\mathbf{1}$ pattern is that it can be also interpreted as describing the case of a completely composite $t_R$ field, emerging directly from the composite sector rather than originating in the elementary one. This is possible because the $t_R$ couples to an operator in the ${\mathbf{1}}_{\mathbf{2/3}}$ of \mbox{SO$(5)\times$U$(1)_X$}, therefore its interaction
\beq
{\mathcal{L}}_{\textrm{int}}^{t_R}= \lambda_{t_R}
\overline{t}_R{\mathcal{O}}_{\mathbf{1}}^R+{\textrm{h.c.}}\,,
\eeq
is perfectly invariant under the composite sector symmetry group, provided the $t_R$ field is interpreted as a composite sector bound state in the ${\mathbf{1}}_{\mathbf{2/3}}$ of the unbroken group \mbox{SO$(4)\times$U$(1)_X$}. Also, being an \mbox{SO$(4)$} singlet, the CCWZ kinetic term for such a resonance (see Eq.~(\ref{CCWZferres})) would contain no $e_\mu$ term and it would just coincide with the SM one we assumed in Eq.~(\ref{elm_q_kin}).\footnote{The gauging of \mbox{U$(1)_X$} leads to an additional term in the resonance covariant derivative in Eq.~(\ref{gencd}) which precisely reproduce the SM gauge interaction.} Therefore all the terms we wrote down for $t_R$ could have come from the composite sector, compatibly with all its symmetries. The difference between the elementary and composite $t_R$ interpretation of the $\mathbf{14}\oplus\mathbf{1}$ setup is purely quantitative and resides in the expected strength of $\lambda_{t_R}$. In the former case, $\lambda_{t_R}$ is an elementary sector coupling and thus we take it smaller than the composite sector one: $\lambda_{t_R}<g_*$. In the latter one, $\lambda_{t_R}$ is naturally of order $g_*$. By treating it as a free parameter and making no a priori assumption on its value we can smoothly interpolate among the two interpretations. In particular in Eq.~(\ref{yuk141}) we can take $\lambda_{t_R}\sim g_*$ showing (using the relation $m_*=g_* f$ which we will discuss in the next chapter) that in the composite $t_R$ case the left-handed coupling must be $\lambda_{t_L}\sim y_t$ in order to reproduce the top quark Yukawa coupling $y_t$. These considerations will be important when we will study the generation of the Higgs potential.

\begin{subappendices}

\section[The \mbox{SO$(4)$} algebra]{The \mbox{SO$\mathbf{(4)}$} algebra}
\label{so4}

The Lie algebra of \mbox{SO$(4)$} is the $6$-dimensional space of traceless Hermitian imaginary $4\times4$ matrices that define the fundamental (the ${\mathbf{4}}$) representation of the group. For applications to composite Higgs the most convenient choice of the Lie algebra basis is the one that makes explicit its connection with the algebra of the chiral group \mbox{SU$(2)_L\times$SU$(2)_R$}, which has also dimension $6$. The two groups are indeed locally isomorphic, {\it{i.e.}}
\beq
\label{iso}
{\textrm{SO}}(4)\simeq {\textrm{SU}}(2)_L\times {\textrm{SU}}(2)_R\,,
\eeq
which means that they have the same algebra. In order to prove the isomorphism and to derive the \mbox{SO$(4)$} basis we proceed as follows. Be $\ve\Pi$ a real vector in the ${\mathbf{4}}$ of \mbox{SO$(4)$}, its four components are in one-to-one correspondence with the elements of a $2\times2$ pseudo-real matrix 
\beq
\label{hmat}
\displaystyle
{\Sigma}=\frac1{\sqrt{2}}\left( i\,\sigma_\alpha\Pi^\alpha +\Id_2 \Pi^4  \right)=\frac1{\sqrt{2}}{\overline{\sigma}}_i\Pi^i\,,
\eeq
where $\alpha=1,2,3$, $\sigma_\alpha$ are Pauli matrices and
\beq
\label{sbardef}
{\overline{\sigma}}_i=\{i\,\sigma_\alpha,\Id_2 \}\,.
\eeq
The ${\overline{\sigma}}$'s obey the following normalization, completeness and reality conditions
\bea
\label{sbarcond}
&{\textrm{Tr}}[{\overline{\sigma}}_i^\dagger {\overline{\sigma}}_j]=2\,\delta_{ij}\,,\;\;\;\;\;
\sum\limits_{i=1}^{4}({\overline{\sigma}}_i^\dagger)_{a}^{\;\;b}\left({\overline{\sigma}}_i\right)_{c}^{\;\;d}=2\,\delta_a^d \delta_c^b\,,
&\nonumber\\
&({\overline{\sigma}}_i)^*=\sigma_2 {\overline{\sigma}}_i \sigma_2\,,\;\;\;\;\;
{\overline{\sigma}}_i {\overline{\sigma}}_j^\dagger-{\overline{\sigma}}_j {\overline{\sigma}}_i^\dagger=2\, {\overline{\sigma}}_i {\overline{\sigma}}_j^\dagger - 2\,\delta_{ij}\Id_2\,,
&
\eea
from which ${\Sigma}$ is immediately seen to be pseudo-real, {\it{i.e.}}
\beq
\label{prel}
{\Sigma}^*=\sigma_2{\Sigma}\sigma_2\,.
\eeq
The chiral group acts on ${\Sigma}$ by matrix multiplication, 
\beq
\label{ctr}
{\Sigma}\;\rightarrow\;g_L{\Sigma} g_R^\dagger.
\eeq
and it preserves the pseudo-reality condition (\ref{prel}). Therefore the matrix ${\Sigma}$ offers a consistent representation of the chiral group, which we call a pseudo-real bidoublet ${\mathbf{(2,2)}}$ with a self-explanatory notation. In order to demonstrate the local isomorphism among the two groups we consider an infinitesimal chiral transformation on ${\Sigma}$ and we show that it has the same effect as an \mbox{SO$(4)$} rotation on the $\ve\Pi$ vector. This is because 
\beq
{\textrm{Tr}}\left[{\Sigma}^\dagger {\Sigma}\right]= |\ve\Pi |^2\,.
\eeq
The trace is invariant under Eq.~(\ref{ctr}), which means that the norm of $\ve\Pi$ is unchanged by the chiral transformations. Since \mbox{SO$(4)$} contains the most general norm-preserving infinitesimal transformation of a four-component vector, this demonstrates that any chiral transformation is an element of \mbox{SO$(4)$} and therefore the chiral group algebra is contained in the \mbox{SO$(4)$} one. However no sub-algebra exists, aside from the full algebra itself, with the same dimensionality of the original one. The isomorphism (\ref{iso}) is thus proven. 

Let us turn to the determination of the \mbox{SO$(4)$} generators. In light of the discussion above we can split them into two sets $t^a=\{t_L^\alpha,t_R^\alpha\}$ with $\alpha=1,2,3$. Each set obeys \mbox{SU$(2)$} commutation relations and the two sets commute in accordance with the \mbox{SU$(2)_L\times$SU$(2)_R$} algebra, namely
\bea
\label{commso4}
&\left[t_L^\alpha,t_L^\beta\right]=i\epsilon^{\alpha\beta\gamma} t_L^\gamma\,,\;\;\;\;\;\left[t_R^\alpha,t_R^\beta\right]=i\epsilon^{\alpha\beta\gamma} t_R^\gamma\,,&\nonumber\\
&\left[t_L^\alpha,t_R^\beta\right]=0\,.&
\eea
Those generators, in the fundamental ${\mathbf{4}}$ representation, are easily extracted from the infinitesimal variations
\begin{eqnarray}
\label{vLR}
&&\delta_L {\Sigma} = i\,\delta_\alpha^L\frac{\sigma^\alpha}2 {\Sigma}\,,\nonumber\\
&&\delta_R {\Sigma} = -\,i\,\delta_\alpha^R  {\Sigma} \frac{\sigma^\alpha}2\,,
\end{eqnarray}
under chiral transformations $g_{L,R}\simeq \Id + i\delta_\alpha^{L,R}\sigma^\alpha/2$. The corresponding variations of $\ve\Pi$ have the form
\bea
\delta_L \ve\Pi = i\,\delta_\alpha^L t^\alpha_L\ve\Pi\,,\nonumber\\
\delta_R \ve\Pi = i\,\delta_\alpha^R t^\alpha_R\ve\Pi\,,
\eea
from which, by matching with Eq.~(\ref{vLR}), we obtain
\bea
(t^\alpha_{L})_{ij} =\frac14{\textrm{Tr}}[{\overline{\sigma}}_i^\dagger\sigma^\alpha {\overline{\sigma}}_j]= -\frac{i}{2}\left[\varepsilon_{\alpha\beta\gamma}
\delta_i^\beta \delta_j^\gamma +
\left(\delta_i^\alpha \delta_j^4 - \delta_j^\alpha \delta_i^4\right)\right]\,,\nonumber\\
(t^\alpha_{R})_{ij} =\frac14{\textrm{Tr}}[{\overline{\sigma}}_i\sigma^\alpha {\overline{\sigma}}_j^\dagger]= -\frac{i}{2}\left[\varepsilon_{\alpha\beta\gamma}
\delta_i^\beta \delta_j^\gamma  -
\left(\delta_i^\alpha \delta_j^4 - \delta_j^\alpha \delta_i^4\right)\right]\,.
\label{eq:SO4_gen}
\eea
The generators obey the commutation relations in Eq.~(\ref{commso4}) and they are subject to the normalization and completeness relations
\bea
\label{normcso4}
&{\textrm{Tr}}\left[t_L^\alpha t_L^\beta\right]={\textrm{Tr}}\left[t_R^\alpha t_R^\beta\right]=\delta^{ab}\,,
\;\;\;\;\;{\textrm{Tr}}\left[t_L^\alpha t_R^\beta\right]=0&\nonumber\\
&\sum\limits_{\alpha=1}^{3}\left[ \left(t_L^\alpha\right)_{ij} \left(t_L^\alpha\right)_{kl}+\left(t_R^\alpha\right)_{ij} \left(t_R^\alpha\right)_{kl}\right]=-\frac12\left(\delta_{ik}\delta_{jl}-\delta_{il}\delta_{jk}\right)\,,&\nonumber\\
&\sum\limits_{\alpha=1}^3\left[ \left(t_L^\alpha\right)_{ij} \left(t_L^\alpha\right)_{kl}-\left(t_R^\alpha\right)_{ij} \left(t_R^\alpha\right)_{kl}\right]=-\frac12\epsilon_{ijkl}\,,
\eea
where $\epsilon_{ijkl}$ is the anti-symmetric Levi-Civita tensor in $4$ dimensions.

In composite Higgs models the \mbox{SU$(2)_L$} group is identified with the SM left-handed group and the hypercharge \mbox{U$(1)_Y$} is the third \mbox{SU$(2)_R$} generator up to the \mbox{U$(1)_X$} charge (see Sect.~\ref{sec:HCsigmaM}),which however vanishes for the Higgs field. In this case the four real components of the ${\mathbf{(2,2)}}$ representation defined in Eq.~(\ref{hmat}) form one complex SM-like Higgs doublet with $1/2$ hypercharge. This is immediately verified by noticing that ${\Sigma}$, thanks to pseudo-reality, can be written as
\beq
\label{bdH}
{\Sigma}=\left(H^c,\,H\right)\,,
\eeq
in terms of the doublet $H$ and of its conjugate $H^c=i\sigma_2 H^*$. By remembering that $H^c$ is also a doublet but with $-1/2$ hypercharge it is immediate to verify that the action of the chiral group in Eq.~(\ref{ctr}) matches the expected Higgs transformation rules under the \mbox{SU$(2)_L\times$U$(1)_{Y}$}. By the definition (\ref{hmat}) the $H$ components are expressed as
\beq
H=\left[\begin{array}{c}h_u\\ h_d\end{array} \right]=\frac1{\sqrt{2}}\left[\begin{array}{c}\Pi^2+i\Pi^1\\ \Pi^4-i\Pi^3\end{array}\right]\,,
\label{hd}
\eeq
in terms of the fourplet fields $\Pi_i$. Conversely, one real \mbox{SO$(4)$} fourplet or, equivalently, one pseudo-real ${\mathbf{(2,2)}}$, can be rewritten in terms of one complex Higgs doublet as in Eq.~(\ref{dh}). This is to say that the  real \mbox{SO$(4)$} fourplet decomposes as
\beq
{\mathbf{4}}={\mathbf{(2,2)}}\;\rightarrow\;\mathbf{2}_{{1/2}}\,,
\eeq
under the \mbox{SU$(2)_L\times$U$(1)_{Y}$} subgroup.

Similar considerations hold for the complex \mbox{SO$(4)$} fourplet, which we will encounter in the main text when dealing with the SM matter fermions. Its complex components $\psi^i$ can be traded for the elements of a generic \mbox{$2\times2$ matrix}
\beq
\label{psimat}
\displaystyle
\Psi=\frac1{\sqrt{2}}\left(\psi^4 + i\,\sigma_\alpha\psi^\alpha \right)=\frac1{\sqrt{2}}{\overline{\sigma}}_i\psi^i\,,
\eeq
which transforms in the ${\mathbf{(2,2)}}$ representation as in Eq.~(\ref{ctr}). Since it does not obey the pseudo-reality condition we dub it a complex bidoublet ${\mathbf{(2,2)_c}}$. Under the \mbox{SU$(2)_L\times$U$(1)_{Y}$} subgroup the two columns of $\Psi$ form two doublets with opposite $\pm1/2$ $Y$ charge, namely
\beq
\label{f4dec}
\Psi=\frac1{\sqrt{2}}\left[
\begin{array}{cc}
\psi^4+i\,\psi^3 &\psi^2+i\,\psi^1 \\
-\psi^2+i\,\psi^1 &\psi^4-i\,\psi^3
\end{array}
\right]\equiv\left(\Psi_-,\,\Psi_+
\right)\,.
\eeq
This corresponds to the decomposition
\beq
{\mathbf{4_c}}={\mathbf{(2,2)_c}}\;\rightarrow\;\mathbf{2}_{{1/2}}\oplus \mathbf{2}_{{-1/2}}\,.
\eeq
From Eq.~(\ref{f4dec}) we can easily read the up and down components of the two doublets in terms of the fourplet fields. Conversely, the fourplet components are written in terms of $\Psi_\pm^{u,d}$ as
\beq
\label{emb5}
\ve\psi=\frac1{\sqrt{2}}\{-i\,\Psi_+^u -i\,\Psi_-^d,\;\Psi_+^u - \Psi_-^d,\;i\,\Psi_+^d -i\,\Psi_-^u,\;\Psi_+^d + \Psi_-^u\}^T\,.
\eeq
The above equation is often referred to as the embedding of the two doublets in the complex ${\mathbf{4}}$.

Other relevant representations are the ${\mathbf{(2,1)}}$ and the ${\mathbf{(1,2)}}$. As the notation suggests these are doublets under one of the chiral \mbox{SU$(2)$} factors and they are invariant under the other one. Their  \mbox{SU$(2)_L\times$U$(1)_{Y}$} decomposition is obviously
\bea
&&{\mathbf{(2,1)}}\;\rightarrow\;\mathbf{2}_{{0}}\,,\nonumber\\
&&{\mathbf{(1,2)}}\;\rightarrow\;\mathbf{1}_{{1/2}}\oplus \mathbf{1}_{{-1/2}} \,.
\eea
The adjoint of \mbox{SO$(4)$}, the ${\mathbf{6}}$, also deserves some comment. Given that the algebra splits into the tensor product of two \mbox{SU$(2)$}'s, the adjoint is a reducible representation and it is represented, in \mbox{SU$(2)_L\times$SU$(2)_{R}$} notation, as 
\beq
{\mathbf{6}}={\mathbf{(3,1)}}\oplus{\mathbf{(1,3)}}\,,
\eeq
where two terms correspond to the generators $t_L^\alpha$ and $t_R^\alpha$, respectively. The decomposition reads
\bea
&&{\mathbf{(3,1)}}\;\rightarrow\;\mathbf{3}_{{0}}\,,\nonumber\\
&&{\mathbf{(1,3)}}\;\rightarrow\;\mathbf{1}_{{0}}\oplus \mathbf{1}_{{1}}\oplus\mathbf{1}_{{-1}}  \,.
\eea
The last representation which is worth mentioning is the ${\mathbf{9}}={\mathbf{(3,3)}}$. It corresponds to a real $3\times3$ matrix with the chiral group acting in the spin one representation on the two sides, or to the symmetric traceless tensor product of two fourplets. It decomposes as 
\beq
{\mathbf{9}}={\mathbf{(3,3)}}\;\rightarrow\;\mathbf{3}_{{0}}\oplus\mathbf{3}_{{1}}\oplus\mathbf{3}_{{-1}}\,.
\eeq

\section[Explicit CCWZ for \mbox{SO$(5)/$SO$(4)$}]{Explicit CCWZ for \mbox{SO$\mathbf{(5)}/$SO$\mathbf{(4)}$}}
\label{so54}

The abstract definitions of Sect.~\ref{CCWZ}, where the CCWZ construction is illustrated for a generic ${\mathpzc{G}}/{\mathpzc{H}}$ coset, become concrete and fully explicit in the particular case of the minimal coset  \mbox{SO${(5)}/$SO${(4)}$}.

The \mbox{SO${(5)}$} generators, reported explicitly in Eq.~(\ref{so5gen}) for the fundamental ${\mathbf{5}}$ representation, can be split into an unbroken subset $T^a$ which represents the \mbox{SO${(4)}$} subgroup and obeys the commutation relations in Eq.~(\ref{commso4}) and a broken one $\Th^i$ associated to the four Goldstone bosons, with commutation relations
\bea
&\left[T^a,\Th^i\right]=i\,f^{ai}_{\;\;\;\;j}\Th^j=\Th^j\left(t^a\right)_{j}^{\;\;i}\,,&\nonumber\\
&\left[\Th^i,\Th^j\right]=i\,f^{ij}_{\;\;\;\;a}T^a=\left(t_a\right)^{ji} T^a\,,&
\eea
where $t^a=\{t_L^\alpha,t_R^\alpha\}$ are the \mbox{SO${(4)}$} generators in the ${\mathbf{4}}$ as in Eq.~(\ref{eq:SO4_gen}). The generators in the ${\mathbf{5}}$, defined in Eq.~(\ref{so5gen}), obey  normalization and completeness conditions
\bea
\displaystyle
&{\textrm{Tr}}\left[T^A T^B\right]=\delta^{AB}\,,&\nonumber\\
&
\displaystyle
\sum\limits_{A=1}^{10} \left(T^A\right)_{IJ} \left(T^A\right)_{KL}=-\frac12\left(\delta_{IK}\delta_{JL}-\delta_{IL}\delta_{JK}\right)\,.&
\eea

Given the generators, it is not hard to compute the Goldstone matrix in the fundamental representation
\beq
\label{gm5}
\displaystyle
U=e^{ i\frac{\sqrt{2}}{f} \Pi_{i}(x) \Th^{i}}
=\left[
\begin{array}{cc}
\displaystyle \Id-\Big(1-\cos{\frac{\Pi}{f}}\Big)\frac{\ \ve\Pi\,\vet{\Pi}}{\Pi^2} &\displaystyle \ \ \sin{\frac{\Pi}{f}}\frac{\ve\Pi}{\Pi}\ \\
\displaystyle - \sin{\frac{\Pi}{f}}\frac{\ \vet\Pi}{\Pi} &\displaystyle \cos{\frac{\Pi}{f}}
\end{array}
\right]\,,
\eeq
in terms of the four real Higgs field components. The complex Higgs doublet notation can be reached afterwards by substituting Eq.~(\ref{dh}). The Goldstone matrix considerably simplifies in the unitary gauge (\ref{ugauge}) and thus it is worth reporting it
\beq
\label{gmug}
\displaystyle
U\underset{\textrm{UG}}{=}\left[
\begin{array}{ccc}
\displaystyle \Id_3 & \displaystyle \vec0^T & \displaystyle 0 \\
\displaystyle \vec0 & \displaystyle \cos{\frac{V+h}f} &  \displaystyle \sin{\frac{V+h}f}\\
\displaystyle 0 &\displaystyle -\sin{\frac{V+h}f} & \displaystyle \cos{\frac{V+h}f}
\end{array}
\right]\,,
\eeq
where $\Id_3$ is the $3\times 3$ identity matrix and $\ve0$ is the $3$-dimensional null vector. The Goldstone matrix in the unitary gauge is a rotation in the $4$-$5$ plane of the five-dimensional space.

As explained in the main text, the Goldstone matrix can be defined in any representation of the group as the exponential of the appropriate generator matrices. Above we computed the one in the fundamental and one should worry of how to obtain the others. For all the representations constructed as tensor product of fundamentals, which can thus be expressed as tensors with fiveplet indices, this is completely straightforward and does not require any additional calculation: the Goldstone matrix acts by rotating each index with the $5\times 5$ matrix $U$. However not all the \mbox{SO${(5)}$} representations are tensor product of  fundamentals, the simplest counterexample is the spinorial, for which the Goldstone matrix needs to be recomputed. The spinorial has dimension $4$ and its generators are
\bea
\label{gen45}
&\displaystyle
{T_{\mathbf{4}}}_L^\alpha=\frac12\left[
\begin{array}{cc}
\sigma^\alpha & 0 \\
0 & 0
\end{array}
\right]\,,\;\;\;\;\;{T_{\mathbf{4}}}_R^\alpha=\frac12\left[
\begin{array}{cc}
0 & 0\\
0 & \sigma^\alpha
\end{array}
\right]\,,&\nonumber\\
&\displaystyle
{\Th_{\mathbf{4}}}^i=\frac{1}{2\sqrt{2}}\left[
\begin{array}{cc}
0 & {\overline\sigma}_i\\
{\overline\sigma}_i^\dagger & 0
\end{array}
\right]\,,&
\eea
where $\sigma^\alpha$ denotes the three Pauli matrices and ${\overline\sigma}$ is defined in Eq.~(\ref{sbardef}). The spinorial  can be also regarded as the fundamental of the symplectic group \mbox{Sp$(4)$}, which is isomorphic to \mbox{SO$(5)$}. The generators indeed obey the symplectic condition
\beq
\label{symp_cond}
\displaystyle
\Omega\cdot T_{\mathbf{4}}^A+\left(T_{\mathbf{4}}^A\right)^T\cdot\Omega=0\,,
\eeq
with the antisymmetric unitary matrix
\beq
\displaystyle
\Omega=e^{i\,\pi\left[{T_{\mathbf{4}}}_L^2-{T_{\mathbf{4}}}_R^2\right]}=\left[
\begin{array}{cc}
i\,\sigma^2 & 0 \\
0 & -i\,\sigma^2
\end{array}\right]\,.
\eeq
For completeness, we report normalization and completeness relations also for the spinorial
\bea
&
\displaystyle
{\textrm{Tr}}\left[T_{\mathbf{4}}^A T_{\mathbf{4}}^B\right]=\frac12\delta^{AB}\,,&\nonumber\\
&
\displaystyle
\sum\limits_{A=1}^{10} \left(T_{\mathbf{4}}^A\right)_{I}^{\;\;J} \left(T_{\mathbf{4}}^A\right)_{K}^{\;\;L}=\frac14\left(\delta_{I}^{L}\delta_{K}^{J}-\Omega_{IK}\Omega^{JL}\right)\,.&
\eea

The Goldstone matrix in the spinorial is straightforwardly obtained by exponentiating the broken generators and it turns out to be most easily expressed in the complex doublet Higgs notation rather than in terms of the real fourplet $\ve\Pi$. It reads
\beq
\label{gm4}
\displaystyle
U_{\mathbf{4}}=e^{i\frac{\sqrt{2}}{f} \Pi_{i}(x){\Th_{\mathbf{4}}}^i}
=\left[
\begin{array}{cc}
 \displaystyle \cos{\frac{|H|}{\sqrt{2} f}}\Id_2 & \displaystyle i\, \sin{\frac{|H|}{\sqrt{2} f}}\frac{\Sigma}{|H|}\\
\displaystyle i\, \sin{\frac{|H|}{\sqrt{2} f}}\frac{{\Sigma}^\dagger}{|H} & \displaystyle \cos{\frac{|H|}{\sqrt{2} f}}\Id_2
\end{array}
\right]\,,
\eeq
where ${\Sigma}$ is the pseudo-real bidoublet representation of the Higgs as defined in Eq.~(\ref{bdH}). The result further simplifies in the unitary gauge
\beq
\displaystyle
U_{\mathbf{4}}\underset{\textrm{UG}}{=}
\left[
\begin{array}{cc}
\displaystyle \cos{\frac{V+h}{2 f}}\Id_2 & \displaystyle i\, \sin{\frac{V+h}{2 f}}\Id_2\\
\displaystyle  i\, \sin{\frac{V+h}{2 f}} \Id_2 & \displaystyle \cos{\frac{V+h}{2 f}}\Id_2
\end{array}
\right]\,.
\eeq
Any \mbox{SO${(5)}$} representation, including the ${\mathbf{5}}$, is the tensor product of spinorials (the conjugate $\overline{\mathbf{4}}$ is equivalent to the spinorial itself and its Goldstone matrix is $U_{\overline{\mathbf{4}}}=U_{{\mathbf{4}}}^*=\Omega U_{{\mathbf{4}}} \Omega^\dagger$). The knowledge of $U_{{\mathbf{4}}}$ thus allows to derive the Goldstone matrix in any representation. 

Let us now turn to the determination of the $d_\mu$ and $e_\mu$ symbols. Those are defined in Eq.~(\ref{dedefgauge}) in the presence of non-dynamical source gauge fields $A_{\mu,\,A}$, one for each of the 10 \mbox{SO$(5)$} generators. However, only a subset of those sources will be made physical by giving them a kinetic term, all the others will be eventually set to zero. The physical sources are the ones in the SM subgroup, which is embedded in the unbroken \mbox{SO$(4)$}.\footnote{Actually the hypercharge has a \mbox{U$(1)_X$} component introduced in Sect.~\ref{sec:HCsigmaM}, which however plays nor role in the calculation of the $d$ and $e$ symbols.} We can thus split the  $A_{\mu,\,A}$'s in unbroken and broken components
\beq
A_{\mu,\,A}=\left\{A_{\mu,\,a}=\{A_{\mu,\,\alpha}^L,\,A_{\mu,\,\alpha}^R\},\, A_{\mu,\,i} =0\right\}\,,
\eeq
and already set the latter ones to zero while retaining, for the moment, all the unbroken generator sources. The unbroken sources have been further split in the two sets that correspond to the two \mbox{SU$(2)$} factors of ${\textrm{SO}}(4)\simeq {\textrm{SU}}(2)_L\times {\textrm{SU}}(2)_R$. The only truly dynamical sources are the ones associated with the four SM gauge fields, namely we will eventually set
\bea
\label{physs}
&&A_{\mu,\,\alpha}^L=\left\{g\,W_\mu^1,\,g\,W_\mu^2,\,g\,W_\mu^3\right\}\,,\nonumber\\
&&A_{\mu,\,\alpha}^R=\left\{0,\,0,\,g'B_\mu\right\}\,,
\eea
in accordance with Eq.~(\ref{physg}).

The $d$ and $e$ symbols can be straightforwardly computed from the definition (\ref{dedefgauge}), or obtained in a somewhat faster way by first classifying the possible structures which they can contain compatibly with the ${\textrm{SO}}(4)$ symmetry. The result is 
\bea
\label{de54}
\displaystyle
&&{d_{\mu}}^i=\sqrt{2}\left(\frac1{\Pi}\sin{\frac{\Pi}{f}}-\frac1{f}\right)\frac{\ve\Pi^TD_\mu\ve\Pi}{\Pi^2}\Pi^i-
\frac{\sqrt{2}}{\Pi}\sin{\frac{\Pi}{f}}D_\mu\Pi^i\,,\nonumber\\
&&{e^L_{\mu}}^\alpha={A^L}_{\mu}^\alpha-\frac{4 }{\Pi^2} \sin^2{\frac{\Pi}{2f}}\ve\Pi^T i\,t_L^\alpha D_\mu\ve\Pi\,,\nonumber\\
&&{e^R_{\mu}}^\alpha={A^R}_{\mu}^\alpha-\frac{4 }{\Pi^2} \sin^2{\frac{\Pi}{2f}}\ve\Pi^T i\, t_R^\alpha D_\mu\ve\Pi\,,
\eea
where $D_\mu\ve\Pi$ is the \mbox{SO$(4)$} covariant derivative
\beq
\label{picov}
D_\mu\ve\Pi=\left(\partial_\mu - i A_{\mu,\,\alpha}^L t_L^\alpha-  i A_{\mu,\,\alpha}^R t_R^\alpha\right)\ve\Pi\,,
\eeq
not to be confused with the CCWZ covariant derivative introduced in Sect.~\ref{2dt}. We have split the $e_{\mu,\,a}$ symbol in two components associated with the decomposition ${\mathbf{6}}={\mathbf{(3,1)}}\oplus{\mathbf{(1,3)}}$ of the adjoint in irreducible representations. In the absence of additional symmetries, the two objects can be employed separately in the construction of invariants. For instance, it is possible to define two independent field-strength tensors following Eq.~(\ref{emuncomp})
\bea
\label{emunudef}
&&{E^L_{\mu\nu}}^\alpha=\partial_\mu {e^L}_\nu^\alpha-\partial_\nu {e^L}_\mu^\alpha+\epsilon^{\alpha\beta\gamma} {e^L}_{\mu,\,\beta} {e^L}_{\nu,\,\gamma}\,,\nonumber\\
&&{E^R_{\mu\nu}}^\alpha=\partial_\mu {e^R}_\nu^\alpha-\partial_\nu {e^R}_\mu^\alpha+\epsilon^{\alpha\beta\gamma} {e^R}_{\mu,\,\beta} {e^R}_{\nu,\,\gamma}\,,
\eea
in the ${\mathbf{(3,1)}}$ and ${\mathbf{(1,3)}}$, respectively. In the following we will also occasionally employ a collective notation $E_{\mu\nu}^a=\{{E^L}_{\mu\nu}^\alpha,{E^R}_{\mu\nu}^\alpha\}$ for the six field-strength tensor components.

For some practical calculation, especially when willing to switch to the complex Higgs doublet notation, the $d$ and $e$ objects in Eq.~(\ref{de54}) are conveniently expressed in terms of $2\times2$ matrices obtained by contracting them with ${\overline\sigma}$ and $\sigma$, namely
\bea
\label{matrixnot}
&&
\displaystyle
d_\mu^{(2)}=d_\mu^i{\overline{\sigma}}_i=\left(\frac1{\sqrt{2} |H| }\sin{\frac{\sqrt{2}|H|}{f}}-\frac1{f}\right)\frac{\partial_\mu |H|^2}{|H|^2}{\Sigma}-\frac{\sqrt{2}}{|H|}\sin{\frac{\sqrt{2}|H|}{f}}D_\mu{\Sigma}\,,
\nonumber\\
&&
\displaystyle
{e_L^{(2)}}_\mu={e_\mu^L}^\alpha\frac{{{\sigma}}_\alpha}2=A^L_{\mu}+\frac{i}{2|H|^2}\sin^2{\frac{|H|}{\sqrt{2}f}}\left[{\Sigma}\, D_\mu {\Sigma}^\dagger-D_\mu {\Sigma}\, {\Sigma}^\dagger
\right]\,,
\nonumber\\
&&
\displaystyle
{e_R^{(2)}}_\mu={e_\mu^R}^\alpha\frac{{{\sigma}}_\alpha}2=A^R_{\mu}+\frac{i}{2|H|^2}\sin^2{\frac{|H|}{\sqrt{2}f}}\left[{\Sigma}^\dagger D_\mu {\Sigma}-D_\mu {\Sigma}^\dagger {\Sigma}
\right]\,,
\eea
where the Higgs matrix covariant derivative, in accordance with (\ref{picov}), is
\beq
D_\mu{\Sigma}=\partial_\mu{\Sigma} - \frac{i}{2} A_\mu^L\, {\Sigma} +\frac{i}{2} {\Sigma} \,A_\mu^R \,,
\eeq
with $A_\mu^{L,R}=A^{L,R}_{\mu,\,\alpha} \sigma^\alpha/2$.
Notice that the $d_\mu$ symbol matrix representation is pseudo-real and those of $e_\mu^{L,R}$ are Hermitian and traceless, as obvious from the definition. In the chiral notation, where the \mbox{SO$(4)$} rotation gets split into two \mbox{SU$(2)_L\times$SU$(2)_R$} transformations $g_L$ and $g_R$, $d^{(2)}$ and $e^{(2)}_{L,R}$ transform as
\bea
&&d_\mu^{(2)} \;\rightarrow\; g_L \cdot d_\mu^{(2)}\cdot g_R^\dagger\,,\nonumber\\
&&{e_L^{(2)}}_\mu \rightarrow g_L \cdot ( {e_L^{(2)}}_\mu+i\partial_\mu)\cdot g_L^\dagger\,,\nonumber\\
&&{e_R^{(2)}}_\mu \rightarrow g_R \cdot ( {e_R^{(2)}}_\mu+i\partial_\mu)\cdot g_R^\dagger\,,
\eea
{\it{i.e.}}~respectively like one bidoublet and two gauge fields.

Now that the basic objects are known we can straightforwardly apply the general CCWZ machinery and derive some useful formulas. First, we  compute the $2$-derivative non-linear $\sigma$-model Lagrangian of Eq.~(\ref{smg}) and verify that it agrees with the expression reported in the main text. After setting the gauge sources to their physical value (\ref{physs}) we obtain
\bea
\label{smgchk}
&&{\mathcal{L}}^{(2)}=\frac{f^2}4 d_{\mu,\,{i}} {d^{\mu,\,i}}= \frac{f^2}8 {\textrm{Tr}}[(d_\mu^{(2)})^\dagger d^{(2),\,\mu}]=\ \\
&&\frac{f^2}{2|H|^2}\sin^2{\frac{\sqrt{2} |H|}{f}} D_\mu H^\dagger D^\mu H+\frac{f^2}{8|H|^4}\left(
2\frac{|H|^2}{f^2}-\sin^2{\frac{\sqrt{2}|H|}{f}}
\right)\left(\partial_\mu |H|^2\right)^2\,,\nonumber
\eea
in accordance with the result in Eq.~(\ref{CHlag0}) obtained for the linear $\sigma$-model. 

We would also like to compute the $E_{\mu\nu}$ field-strength components and the antisymmetric part of the  two-derivative tensor $D\cdot d$ defined in Eq.~(\ref{Dd}), which we will need in the following chapter. These objects can be obtained directly  from their definitions in Eqs.~(\ref{emunudef}) and (\ref{Dd}), or derived in fast way by employing the identity (\ref{CCWZID}) proven in Sect.~(\ref{2dt}). In this second case we proceed by first computing the ``dressed'' field-strength tensors ${\mathcal{F}}$ defined in Eqs.~(\ref{deff}), (\ref{deff1}), which in our case consist of $3$ CCWZ multiplets in the $\mathbf{(3,1)}$, ${\mathbf{(1,3)}}$ and ${\mathbf{(2,2)}}$ representations. Those are rather simple because they contain no derivatives of the Goldstone fields and read
\bea
&&\displaystyle{{\mathcal{F}}_L}^\alpha_{\mu\nu}=\cos^2{\frac{\Pi}{2f}}{A^L}_{\mu\nu}^\alpha-\frac{4}{\Pi^2}\sin^2{\frac{\Pi}{2f}}\,\ve\Pi^Tt_L^\alpha ({A^R}_{\mu\nu,\,\beta }\, t_R^\beta) \ve\Pi\,,  \nonumber\\
&&\displaystyle{{\mathcal{F}}_R}^\alpha_{\mu\nu}=\cos^2{\frac{\Pi}{2f}}{A^R}_{\mu\nu}^\alpha-\frac{4}{\Pi^2}\sin^2{\frac{\Pi}{2f}}\,\ve\Pi^Tt_R^\alpha ({A^L}_{\mu\nu,\,\beta }\, t_L^\beta) \ve\Pi \,, \nonumber\\
&&\displaystyle{{\mathcal{F}}_{\mathbf{4}}}_{\mu\nu}^i=\frac{\sqrt{2}}{\Pi}\sin{\frac{\Pi}{f}}\left( 
{A^L}_{\mu\nu,\,\alpha }\, i\, t_L^\alpha+{A^R}_{\mu\nu,\,\alpha }\, i\, t_R^\alpha\right)^{i j} \Pi_j  \,,
\eea
where $A_{\mu\nu}^{L,R}$ denote the field-strengths associated with the gauge sources
\beq
\label{gauge_sources_field_st}
{A_{\mu\nu}^{L,R}}^\alpha=\partial_\mu {A_\nu^{L,R}}^\alpha-\partial_\nu {A_\mu^{L,R}}^\alpha
+\epsilon^{\alpha\beta\gamma} {A_\mu^{L,R}}_\beta {A_\nu^{L,R}}_\gamma\,.
\eeq
After setting the sources to their physical values in Eq.~(\ref{physs}), they reduce to the familiar $W_{\mu\nu}$ and $B_{\mu\nu}$ SM tensors. 

The last object we need in order to apply Eq.~(\ref{CCWZID}) (since ${d_{{\mathbf{r}}_\pi}^2}=0$ for a symmetric coset) is ${d_{\textrm{\bf{Ad}}}^2}$, the adjoint tensor formed out of two $d$-symbols defined in Eq.~(\ref{dcomb}). In our case it splits in two components
\beq
{d_L^2}_{\mu\nu}^\alpha=d_\mu^i(i\,t^\alpha_L)_{i j}d_\nu^j\,,\;\;\;\;\;{d_R^2}_{\mu\nu}^\alpha=d_\mu^i(i\,t^\alpha_R)_{i j}d_\nu^j\,.
\eeq
The explicit form of ${d_{L,R}^2}$ in terms of $\Pi$ can be easily worked out, however the expression in terms of the $d$-symbol provided by the equation above is already the simplest one for practical calculations. The field-strengths ${E^{L,R}}_{\mu\nu}$ and ${D\cdot d}_{[\mu,\nu]}$ are, finally
\bea
\label{eddcomp}
&&\displaystyle{E_L}_{\mu\nu}^\alpha={{\mathcal{F}}_L}^\alpha_{\mu\nu}\,-\,{d_L^2}_{\mu\nu}^\alpha\,,\nonumber\\
&&\displaystyle{E_R}_{\mu\nu}^\alpha={{\mathcal{F}}_R}^\alpha_{\mu\nu}\,-\,{d_R^2}_{\mu\nu}^\alpha\,,\nonumber\\
&&\displaystyle(D\cdot d)_{[\mu,\nu]}^i=D_\mu d_\nu^i-D_\nu d_\mu^i=\displaystyle{{\mathcal{F}}_{\mathbf{4}}}_{\mu\nu}^i\,.
\eea
The above formulas can be also obtained by computing $E$ and $D\cdot d$ directly from their definitions. This provides a non-trivial cross-check of Eq.~(\ref{CCWZID}). In the matrix notation, $E$ and $D\cdot d$ become
\bea
\displaystyle {E_L^{(2)}}_{\mu\nu}=&&\cos^2{\frac{|H|}{\sqrt{2}f}}A_{\mu\nu}^L+\frac{1}{|H|^2}\sin^2{\frac{|H|}{\sqrt{2}f}}
\left({\Sigma} A_{\mu\nu}^R{\Sigma}^\dagger-\frac12\textrm{Tr}[{\Sigma} A_{\mu\nu}^R{\Sigma}^\dagger] \right)\nonumber\\
&&\displaystyle+\frac{i}{8}\left(d_\mu^{(2)}{d_\nu^{(2)}}^\dagger-d_\nu^{(2)}{d_\mu^{(2)}}^\dagger\right)\,,\nonumber\\
\displaystyle {E_R^{(2)}}_{\mu\nu}=&&\cos^2{\frac{|H|}{\sqrt{2}f}}A_{\mu\nu}^R+\frac{1}{|H|^2}\sin^2{\frac{|H|}{\sqrt{2}f}}
\left({\Sigma}^\dagger A_{\mu\nu}^L{\Sigma}-\frac12\textrm{Tr}[{\Sigma}^\dagger A_{\mu\nu}^L{\Sigma}] \right)\nonumber\\
&&\displaystyle+\frac{i}{8}\left({d_\mu^{(2)}}^\dagger d_\nu^{(2)}-{d_\nu^{(2)}}^\dagger d_\mu^{(2)}\right)\,,\nonumber\\
\displaystyle {D\cdot d}^{(2)}_{[\mu,\nu]}=&&\frac{\sqrt{2}\,i}{ |H|}\sin{\frac{\sqrt{2}|H|}{f}}\left(A_{\mu\nu}^L{\Sigma}-{\Sigma}A_{\mu\nu}^R\right)\,,
\eea
where $A_{\mu\nu}^{L,R}={A_{\mu\nu}^{L,R}}^\alpha\sigma_\alpha/2$.

All the formulas above greatly simplify in the unitary gauge, in which
\beq
\displaystyle{\Sigma}=\frac{V+h}{\sqrt{2}}\,\Id_2\,.
\eeq
For the $d$ and $e$ symbols we have
\bea
\label{edUG}
&&
\displaystyle
d_\mu^{(2)}\underset{\textrm{UG}}{=}-\frac{\sqrt{2}}{f}\partial_\mu h\,\Id_2 + \sqrt{2}\, i \sin{\frac{V+h}{f}}(A^L_\mu-A^R_\mu)\,,
\nonumber\\
&&
\displaystyle
{e_L^{(2)}}_\mu\underset{\textrm{UG}}{=}A_\mu^L-\sin^2{\frac{V+h}{2\,f}}(A^L_\mu-A^R_\mu)\,,
\nonumber\\
&&
\displaystyle
{e_R^{(2)}}_\mu\underset{\textrm{UG}}{=}A_\mu^R+  \sin^2{\frac{V+h}{2\,f}}(A^L_\mu-A^R_\mu)\,,
\eea
while for $E$ and $D\cdot d$ one finds
\bea
\label{eddUG}
&&\displaystyle {E_L^{(2)}}_{\mu\nu}\underset{\textrm{UG}}{=}\cos^2{\frac{V+h}{2f}}A_{\mu\nu}^L+\sin^2{\frac{V+h}{2f}}A_{\mu\nu}^R
+\frac{i}{8}\left(d_\mu^{(2)}{d_\nu^{(2)}}^\dagger-d_\nu^{(2)}{d_\mu^{(2)}}^\dagger\right)\,,
\nonumber\\
&&\displaystyle {E_R^{(2)}}_{\mu\nu}\underset{\textrm{UG}}{=}\cos^2{\frac{V+h}{2f}}A_{\mu\nu}^R+\sin^2{\frac{V+h}{2f}}A_{\mu\nu}^L
+\frac{i}{8}\left({d_\mu^{(2)}}^\dagger{d_\nu^{(2)}}-{d_\nu^{(2)}}^\dagger{d_\mu^{(2)}}\right)\,,
\nonumber\\
&&\displaystyle {D\cdot d}^{(2)}_{[\mu,\nu]}\underset{\textrm{UG}}{=}\sqrt{2}\,i \sin{\frac{V+h}{f}}\left(A_{\mu\nu}^L-A_{\mu\nu}^R\right)\,.
\eea

\end{subappendices}

\chapter{Beyond the sigma-model}
\label{ch:SILH}

In the previous chapter we restricted our attention to a specific class of new physics effects, that we can classify as ``non-linear $\sigma$-model effects''. These are modifications of the SM driven by the pNGB nature of the Higgs and the associated non-linear $\sigma$-model structure of the effective Lagrangian. The non-linear $\sigma$-model effects are encapsulated in the generalized Higgs kinetic term and Yukawa Lagrangians that deliver the SM operators that couple the Higgs to vector bosons and fermions,  plus a series of terms with additional Higgs field insertions weighted by the inverse of the non-linear $\sigma$-model scale $f$. The non-linear operators can be regarded as the minimal extension of the SM required to enforce the Goldstone symmetry on the Lagrangian. Thus they define the minimal set of SM deformations which we will unmistakably encounter in any explicit model with pNGB Higgs, if based on the minimal coset \mbox{SO$(5)/$SO$(4)$}. The Goldstone symmetry relates operators with different number of Higgs insertions but with the same number of derivatives and gauge fields. For the uplift of the SM couplings to Goldstone symmetry invariants, therefore, only CCWZ operators of the minimal derivative order are considered. Those of ${\mathcal{O}}(p^2)$ in the bosonic sector and of ${\mathcal{O}}(p^0)$ in the fermionic one. Extending the operator analysis to higher orders is one of the goals of the present chapter.

Going beyond the $\sigma$-model will carry us one step ahead in the theoretical comprehension of the composite Higgs scenario, but it will also force us to make additional assumptions on the dynamics of the composite  sector besides the occurrence of spontaneous Goldstone symmetry breakdown. Namely, we will need to make a hypothesis on the typical size of the higher derivative operators induced by the composite dynamics in the low energy effective theory. Such operators estimate is what we call the ``power-counting rule'' of the effective field theory. Power counting did not play a major role in the previous chapter because the lower derivative order operators come in such a limited number that their coefficient needs not to be estimated but it can just be fixed by observations. This is what we saw happening in the fermion sector in Sect.~\ref{sec:HCsigmaM}, where the generalized Yukawa coefficients have been traded for the top and bottom masses. Up to few exceptions this led us to coupling predictions that are uniquely dictated by the $\sigma$-model scale $f$ or, equivalently, by the parameter $\xi=v^2/f^2$. The situation will be different with higher derivative operators, each coming with its own independent parameter. By power-counting we will estimate the operator coefficients and thus their contribution to the physical observables, even if we will still be unable to determine them completely. Their precise value depends on the microscopic details of the composite sector and it is not calculable within the effective theory. Other than estimating coefficients, and more importantly than that, power-counting serves as a criterion to identify the most relevant operators and to select only a finite number of them for practical calculations. Power-counting also tells us the size of the operators that we have neglected when truncating the effective field theory Lagrangian and thus it provides us with an estimate of the accuracy of our predictions. 

The second problem that we will address in the present chapter concerns the physics of the resonances, {\it{i.e.}} the additional non-Goldstone bound states that are present in the composite sector. While by powerful symmetry principles the pNGB low-energy dynamics is well under control, nothing has been said on the resonances up to now. But we need to characterize their phenomenology and thus we must find a way to at least estimate resonance couplings and masses.

\begin{figure}
\centering
{\includegraphics[height=0.15\textwidth]{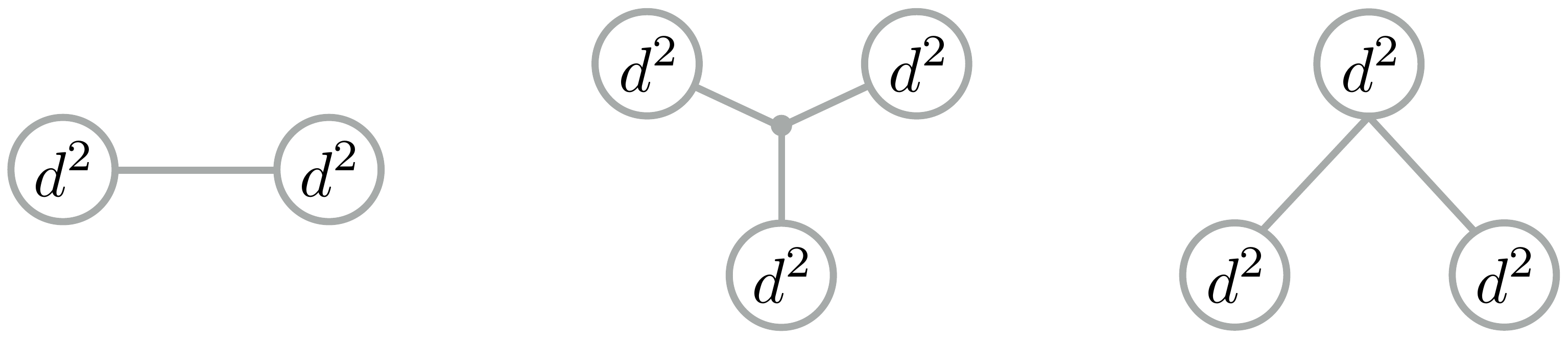}}
\caption{A schematic view of the Feynman diagrams that generate the effective field theory operators from the virtual exchange of the massive resonances. The blob denotes the resonance coupling to the light Higgs and gauge fields, given by Eq.~(\ref{sigmac}).}
\label{EOP}
\end{figure}

The two problems are actually related in the case of perturbative composite sector models because higher derivative operators originate from integrating out resonance fields. By an estimate of resonance couplings one can thus derive the effective operators power-counting. This is conveniently illustrated in the linear $\sigma$-model example of Sect.~\ref{sec:MCHMlin}, which describes one single scalar resonance $\sigma$ on top of the pNGB Higgs and the SM gauge fields. The Lagrangian contains the ${\mathcal{O}}(p^2)$ non-linear $\sigma$-model operator (\ref{CHlag0}) (or equivalently (\ref{smgchk})) plus $\sigma$ resonance self-interactions (\ref{lagnl}) and $\sigma$ couplings to the SM fields which are given by
\beq
\label{sigmac}
\displaystyle
\left(\frac{f}2\sigma+\frac14\sigma^2\right)d_{\mu,\,i}d^{\mu,\,i}\,.
\eeq
By integrating out $\sigma$ and Taylor-expanding for momenta below the resonance mass $m_*$, the interactions above generate operators of higher and higher derivative order. The first one is ${\mathcal{O}}(p^4)$ and emerges from the first diagram in Fig.~\ref{EOP} by the exchange of one virtual $\sigma$ particle, it is
\beq
{\mathcal{L}}^{(4)}_{\textrm{EFT}}=\frac12\left(\frac{f}2\right)^2\frac1{m_*^2}\left(d^2\right)^2=\frac{f^2}{8 m_*^2}\left(d^2\right)^2\,.
\eeq
Aside from the $1/2$ prefactor, which results from the explicit calculation, the emergence of the other factors is easily understood from the diagram in Fig.~(\ref{EOP}). The two powers of $f/2$ come from the $\sigma$ vertices and $1/m_*^2$ is the low-momentum propagator. If we compare the result with the ${\mathcal{O}}(p^2)$ operator with $f^2/4$ coefficient, we reach the  unsurprising conclusion that the ${\mathcal{O}}(p^4)$ term is weighted by one extra $1/m_*$ factor for each extra derivative. The simplest guess is that this pattern continues to higher orders and the ${\mathcal{O}}(p^n)$ operators scale as
\beq
\label{pc0}
{\mathcal{L}}^{(n)}_{\textrm{EFT}}=\frac{f^2}{m_*^{n-2}}O^{(n)}\,,
\eeq
where $O^{(n)}$ denotes a linear combination of ${\mathcal{O}}(p^n)$ CCWZ invariants with order-one numerical coefficients. It is not hard to verify that the guess is correct. Consider for instance the ${\mathcal{O}}(p^6)$ operators. Some of them come from the single resonance exchange diagram by taking the second order in the propagator expansion. This leads to one $p^2/m_*^2$ factor, which is $\partial^2/m_*^2$ in coordinate space, and Eq.~(\ref{pc0}) is respected. The second source of ${\mathcal{O}}(p^6)$ operators is the second diagram in Fig.~\ref{EOP}, which gives
\beq
{\mathcal{L}}^{(6)}\sim f^3 \frac1{m_*^6} (g_*^2f)\left(d^2\right)^3=f^2 \frac{(g_* f)^2}{m_*^6} \left(d^2\right)^3\,,
\eeq
where the three $f$ factors come from the $\sigma$ couplings to SM fields, $1/m_*^6$ is due to the three $\sigma$ propagators and $g_*^2 f$ is the trilinear $\sigma$ self-coupling. The result agrees with Eq.~(\ref{pc0}) because we saw that the resonance mass is related to $g_*$ and $f$ by
\beq
m_*=g_* f\,.
\eeq
The same is found for the third contribution depicted in Fig.~\ref{EOP} and it would not be hard to demonstrate by diagram inspection that the validity of the power-counting (\ref{pc0}) extends to all orders in the derivative expansion. This has a simple and structural origin as we will explain in the following section.

\newpage

\section{One Scale One Coupling}
\label{OSOC}

\stepcounter{figure}

\begin{figure}
\centering
\includegraphics[width=0.7\textwidth]{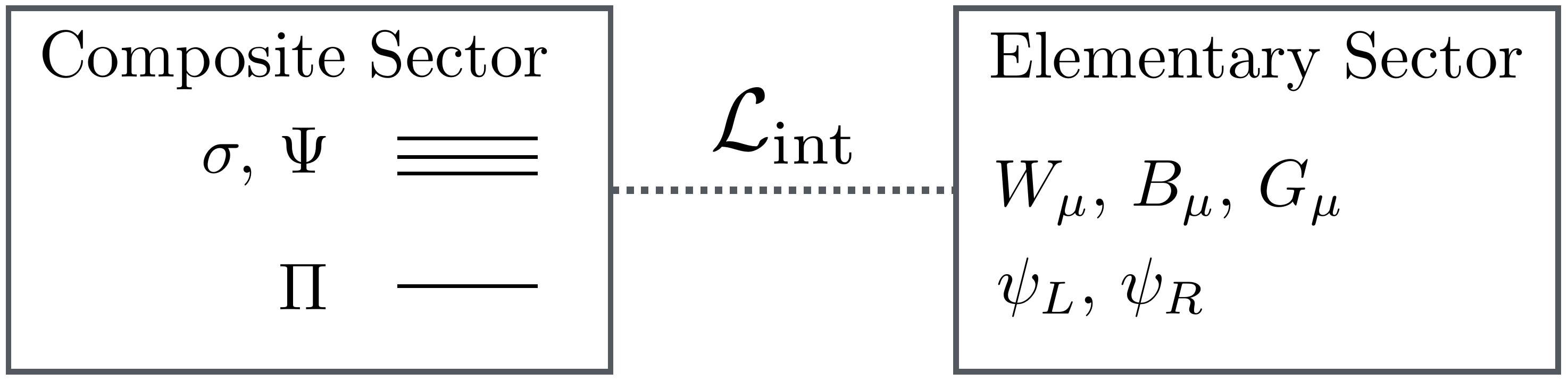}
\caption{The composite Higgs setup. The elementary SM gauge fields are the three $W$'s, the hypercharge boson $B$ and the eight QCD gluons. The elementary fermionic quark and lepton fields are collectively denoted as $\psi_L$ and $\psi_R$.}
\label{ecomp}
\end{figure}

In order to illustrate the assumptions that we are going to make on the composite sector dynamics we start by summarizing, with the help of Fig.~\ref{ecomp}, the physical setup that we have in mind. There exists a new sector that delivers the NGB Higgs plus a set of massive resonances. We will generically denote the latter states as $\sigma$, that stands for bosonic spin~$0$ or spin~$1$ particles, and $\Psi$, that represents spin~$1/2$ fermions. Ideally also higher spin resonances might emerge, however since a weakly-coupled Lagrangian description of this kind of objects is problematic we are not going to discuss them explicitly in what follows.

We consider a situation where the composite sector dynamics can be described perturbatively by a weakly-coupled Lagrangian ${\mathcal{L}}_C$. A simple example, the linear $\sigma$-model, was studied explicitly in Sect.~\ref{sec:MCHMlin}. One could imagine complicating it by adding new scalar, fermion or massive gauge fields. Provided they fulfill the condition described below this kind of generalized linear $\sigma$-models are within the scope of the present discussion. More interesting examples are the $5$-dimensional holographic composite Higgs models, many of which have been proposed in the literature \cite{Scrucca:2003ra,Agashe:2004rs,Contino:2006qr}. In this case what we call the composite sector is the $5$-d bulk dynamics of the theory, namely the $5$-d gauge theory with Dirichlet boundary conditions on the UV brane. After Kaluza--Klein reduction this theory describes the NGB Higgs plus an infinite set of weakly-coupled resonance fields. Details on the interpretation of $5$-d models in these terms can be found in the original literature and in a comprehensive review \cite{Contino:2010rs}. Another set of explicit constructions to which these ideas apply will be discussed in Chap.~\ref{ch:Pheno_models}. Clearly, the hope here is that those models might faithfully reproduce the low-energy manifestations of a genuine strongly-coupled confining sector, possibly described at the microscopic level by a $4$-dimensional gauge theory. We will discuss at the end of this section how a genuine strong sector might indeed fit in the present discussion and truly obey the assumption which we are going to make on the resonance dynamics. The composite Higgs models also contains, as shown in Fig.~\ref{ecomp}, elementary SM fields coupled to the composite sector in a peculiar manner explained in the previous chapter. We momentarily ignore the elementary fields interactions and we focus on the couplings among composite particles.

Our assumption is that  the composite sector Lagrangian involving space--time derivatives, Goldstones and resonance fields is of the form
\beq
\displaystyle
\label{pwc0}
{\mathcal{L}}_C=\frac{m_*^4}{g_*^2}\widehat{\mathcal{L}}\left[\frac{\partial}{m_*},\frac{g_*\Pi}{m_*},\frac{g_*\sigma}{m_*},\frac{g_*\Psi}{{m_*}^{3/2}}\right]\,,
\eeq
where $\widehat{\mathcal{L}}$ is a dimensionless polynomial functional with arbitrary order-one numerical coefficients. The fields that appear in the equation are all supposed to be canonically normalized and indeed when applied to the kinetic terms our formula consistently predicts order-one coefficients. Notice that Eq.~(\ref{pwc0}) provides an estimate of the composite particle interaction vertices in terms of two parameters only, the typical resonance scale $m_*$ and the typical resonance coupling $g_*$. Theories obeying Eq.~(\ref{pwc0}) are thus said to be ``One Scale One Coupling'' ($1$S$1$C) models. We will show below how to define the $1$S$1$C assumption rigorously by dimensional analysis. The scaling of the Goldstone field insertions in Eq.~(\ref{pwc0}) should be compared with what is predicted by the Goldstone symmetry, {\it{i.e.}}~the fact that the $\Pi$ field only enters in the Lagrangian through the Goldstone matrix and thus it is weighted by the non-linear $\sigma$-model scale $f$. We must then identify
\beq
\displaystyle
f=\frac{m_*}{g_*}\,.
\eeq
Notice that by this identification the parametric scaling of the ${\mathcal{O}}(p^2)$ Lagrangian (\ref{smgchk}) is immediately seen to match with Eq.~(\ref{pwc0}). Furthermore, it is not hard to check that Eq.~(\ref{pwc0}) is verified by the whole linear $\sigma$-model Lagrangian, including $\sigma$ self-interactions and its coupling with the Goldstone in Eq.~(\ref{sigmac}). This qualifies the linear $\sigma$-model as a $1$S$1$C theory.

The $1$S$1$C hypothesis can be phrased in a fancy but useful way in terms of dimensional analysis. The argument is extremely simple, however it cannot be made by just counting energy dimensions as we are used to do by working in Natural Units where \mbox{$\hbar=c=1$}. We must take care of all the three fundamental physical quantities of the MKS system, namely the  length $L$, the mass $M$ and the time $T$. For the sake of the present discussion we will actually trade these three quantities for the energy $E=M\,L^2/T^2$, length and time. In MKS units the quantum-mechanical action has dimension of $E\cdot T$, which is the one of the reduced Planck constant $\hbar$. Therefore the Lagrangian density must have dimension \footnote{We take $x^\mu=\{c\,t,\,\ve x\}$, therefore the space--time volume is $d^4x=dx^0d^3x=c\,dt\,d^3x$ and $\partial_\mu=\partial/\partial{x^\mu}$ has dimension of $L^{-1}$.}
\beq
\label{dimlag}
[{\mathcal{L}}]=E \, T/L^4=[\hbar]/L^4\,,
\eeq
and thus canonically normalized bosonic and fermionic fields have dimensions
\beq
\label{fdim}
[\Pi]=[\sigma]=[\hbar]^{1/2}/L\,,\;\;\;\;\;[\Psi]=[\hbar]^{1/2}/L^{3/2}\,.
\eeq
In order for the interactions in Eq.~(\ref{pwc0}) to have the correct dimension we must set
\beq
\label{pardim}
[m_*]=L^{-1}\,,\;\;\;\;\; [g_*]=[\hbar]^{-1/2}\,.
\eeq
This sharply defines $m_*$ as a scale, namely a length scale, and $g_*$ as a ``coupling'' $C$, {\it{i.e.}}~a parameter that carries a dimension
\beq
C=[\hbar]^{-1/2}=(ET)^{-1/2}\,.
\eeq
In Natural Units, $C$ dimensions are lost and one might be tempted to treat $g_*$ on the same footing as a dimensionless numerical parameter. However it is enough to reintroduce $\hbar$ to appreciate the fundamental difference between $g_*$ and a pure number. Similarly, we can appreciate the difference between a ``pure scale'' like $m_*$ and an object like the Goldstone decay constant $f$, with dimension
\beq
[f]=C^{-1}L^{-1}\,.
\eeq
Dimensional analysis is extremely useful in many contexts and not just in the study of composite Higgs theory. Table~\ref{dimt} reports the dimensions of some important SM parameters. Those allow us to better understand a number of known results. For instance from the fact that the Higgs quartic coupling $\lambda_H$ has dimension $C^2$ we understand how it can appear at the first power in a tree-level \mbox{$2\rightarrow 2$} scattering amplitude while the electric charge $e$, of dimension $C$, must come with the square.\footnote{We can take creation/annihilation operators to have dimension $L$ (and not to have $\hbar$ in their canonical commutators) by factoring out a $\sqrt{\hbar}$ in the Fourier decomposition of the fields. The $n$-particles states thus have dimension $[|n\rangle]= L^n$. The $2\rightarrow n$ Feynman amplitude is conveniently defined as \mbox{$\langle n,{\textrm{out}}| 2,{\textrm{in}}\rangle\hspace{-1pt} =\hspace{-1pt} (2\pi)^4\delta^4(p_{\textrm{out}}-p_{\textrm{in}})\,\hbar^{n/2}{\mathcal{M}}_n$}, so that $[{\mathcal{M}}_n]=C^{n} L^{n-2}$ and in particular $[{\mathcal{M}}_2]=C^{2} $. Having stripped out $\hbar^{n/2}$ from the definition, no further powers of $\hbar$ appears in ${\mathcal{M}}_n$ at tree-level while a factor of $\hbar^L$ emerges at $L$ loops.} We also understand why supersymmetric relations among $\lambda_H$ and the weak coupling are of the form $\lambda_H\sim g^2$. Similarly we understand in which sense the Fermi constant is a coupling, or actually a coupling squared. This is why it can mediate \mbox{$2\rightarrow 2$} scattering.

\begin{table}
\centering
\begin{tabular}{ r  | c | c | c | c | c | c }
Parameter & $e\;$ & $\lambda_H$ & $G_F$ & $m_{W,Z}$ & $v$ & $f_\pi$  \\
\hline
\ & \ & \ & \ & \ & \\[-10pt]
{Dimension} & $C$   & $C^2$ & $C^2 L^2$ &  $L^{-1}$ & $C^{-1}L^{-1}$ & $C^{-1}L^{-1}$
\end{tabular}
\caption{The dimension of some important SM parameter. The results agree with the ones in Sect.~6-2-1 of the classical book in Ref.~\cite{IZ}. Comparing them with the PDG ones, namely those reported in Table~1 ``Physical Constants'' of Ref.~\cite{Agashe:2014kda} requires more care since the latter are given in an old-fashioned conventions (see for instance Ref.~\cite{Sakurai}) where $\hbar$ and $c$ appear explicitly in the Lagrangian.}
\label{dimt}
\end{table}

By turning around the dimension counting we can reformulate the \mbox{$1$S$1$C} hypothesis (\ref{pwc0}) in an equivalent and simpler way. The \mbox{$1$S$1$C}  theories are those in which $m_*$ and $g_*$ are the only dimensionful parameters, with the dimensions in Eq.~(\ref{pardim}). The composite particle couplings are thus obliged to be as in Eq.~(\ref{pwc0}) because this is the only form they can have by dimensional analysis. This reformulation is useful in at least two respects. First, it allows us to tell if the \mbox{$1$S$1$C} hypothesis is verified without having to work out the Goldstone and resonance couplings explicitly and to check that they agree with Eq.~(\ref{pwc0}). For instance we might have told that the linear $\sigma$-model fits in this category by just looking at its original Lagrangian in Eq.~(\ref{old}), with no need of performing the field redefinition which isolated the Goldstones from the resonance $\sigma$. A more interesting example is the one of $5$-d holographic models. In that case the only two dimensionful parameters are the $5$-d gauge coupling $g_5$ with $[g_5]=C\,L^{1/2}$ and the length of the extra dimension in conformal coordinates $1/\mu_{IR}$ with $[1/\mu_{IR}]=L$. This qualifies the  $5$-d model as a \mbox{$1$S$1$C} theory with
\beq
m_*=\mu_{IR}\,,\;\;\;\;\;g_*=g_5\,\mu_{IR}^{1/2}.
\eeq
We can thus conclude that the resonance couplings obey Eq.~(\ref{pwc0}) without computing them explicitly through a complicated Kaluza--Klein reduction.

The second way in which the reformulation is useful is that it makes very simple to draw the implications of the \mbox{$1$S$1$C} hypothesis. Let us consider the low-energy effective theory for the Goldstones obtained by integrating out the heavy resonance fields. From our estimate of the resonance couplings we should be able to derive the effective field theory power-counting. If the resonances are integrated out at tree level, no dimensionful parameter appears in the calculation aside from $m_*$ and $g_*$ and therefore dimensional analysis tells us that the operator scaling in Eq.~(\ref{pwc0}) must be respected also by the effective field theory Lagrangian. This indeed coincides with the power-counting formula (\ref{pc0}), which we derived explicitly in the linear $\sigma$-model example. One can immediately check this by noticing that in the CCWZ invariants $O^{(n)}$ the Goldstones enter as $\Pi/f$. The situation is different for radiative corrections, where one extra dimensionful parameter, $\hbar$, appears.
Indeed it is well known that $\hbar$ controls the semiclassical perturbative expansion and one power of $\hbar$ is associated to each loop. By dimensional analysis the complete effective field theory Lagrangian must thus scale like
\bea
\label{POWCStrong}
{\mathcal{L}}_{\textrm{EFT}}&&=\;\;\frac{m_*^4}{g_*^2}\widehat{\mathcal{L}}_{\textrm{tree}}\left[\frac{\partial}{m_*},\frac{g_*\Pi}{m_*},\frac{g_*\sigma}{m_*},\frac{g_*\Psi}{{m_*}^{3/2}}\right]\nonumber\\
&&+\,\frac{g_*^2 \hbar}{16\pi^2}\frac{m_*^4}{g_*^2}\widehat{\mathcal{L}}_{1\textrm{-loop}}\left[\frac{\partial}{m_*},\frac{g_*\Pi}{m_*},\frac{g_*\sigma}{m_*},\frac{g_*\Psi}{{m_*}^{3/2}}\right]+\ldots\,,
\eea
where $1/16\pi^2$ is the habitual factor from the loop integral. In the above equation we retained the dependence on the resonance fields $\sigma$ and $\Psi$ because we might sometimes be interested in integrating out only some of the resonances and retaining the others in the effective field theory. Obviously this does not make sense if all the resonance are equally heavy, but it can have a justification if some of them are somewhat lighter than the others for some numerical coincidence or for a structural reason.

The above formula goes under the name of ``SILH power-counting'', from the title of Ref.~\cite{Giudice:2007fh} where it was introduced in the study of composite Higgs theories.\footnote{The derivation of the SILH power-counting based on dimensional analysis was known to the experts of the field. See for instance Ref.~\cite{Thamm:2014fba} and Ref.~\cite{Pomarol:2014dya}.}
It was first discussed in a different context by Ref.~\cite{Georgi:1989xy} as a generalization of the NDA counting \cite{Manohar:1983md}. The NDA formula is recovered from Eq.~(\ref{POWCStrong}) in the limit \mbox{$g_*\hspace{-2pt}\rightarrow 4\pi/\sqrt{\hbar}$}, which is the maximum value of the coupling for which the modelling of resonances of Eq.~(\ref{pwc0}) makes sense before entering the non-perturbative regime. The perturbativity bound \mbox{$g_*<4\pi/\sqrt{\hbar}$} follows from the fact that the dimensionless parameter combination which controls the loop expansion is $g_*^2\hbar/16\pi^2$. Below that value the theory of resonances is perturbative and consistent with unitarity, and so is the effective field theory constructed out of it.\footnote{A not commonly appreciated puzzle is that the maximal coupling estimate based on unitarity of $2\rightarrow2$ processes, see for instance \cite{Lee:1977eg}, is actually $\sqrt{2\pi}$ times lower, meaning that there exist perturbative theories which are formally non-unitary. This discrepancy comes from the $2\pi$ enhancement of the imaginary part of the one-loop amplitude, which makes the latter comparable with the tree-level real part at smaller coupling. Given that the imaginary part is actually a tree-level process we consider this fact as a signal that the conventional unitarity argument, based on the habitual but artificial separation among tree and loop, should be reconsidered.}

One important comment to be made on Eq.~(\ref{POWCStrong}) concerns the distinction among tree- and loop-generated effective operators, which clearly depends on the details on the composite sector theory. For instance in the linear $\sigma$-model we found that the only ${\mathcal{O}}(p^4)$ operator that is generated at the tree-level is $(d^2)^2$, whereas many others would arise in a more complicated theory such as a $5$-d model. Not having a strong enough argument to distinguish loop from tree operators, in what follows we will adopt the agnostic attitude that all the operators compatible with symmetries might arise at tree-level, so that the estimate of their coefficients is provided by the first line of Eq.~(\ref{POWCStrong}). It must however be kept in mind that loop factor reductions of the coefficients are actually possible in  specific classes of composite sector models~\cite{Giudice:2007fh}. 

We now consider the elementary sector. The \mbox{$1$S$1$C} hypothesis is sufficient to generalize the power-counting in Eq.~(\ref{POWCStrong}) to include elementary fields insertions. This will not cost us any new assumption, aside of course from the one we made already in the previous chapter on the nature of the elementary/composite interactions ${\mathcal{L}}_{\textrm{int}}$. We start from the gauge fields (see Eq.~(\ref{gscoup}) and Footnotes~{\ref{foot1},\ref{foot2}}), which are introduced with the gauging procedure and thus talk with the composite sector through
\beq
{\mathcal{L}}_{\textrm{int}}^{\textrm{gauge}}=g\,A_\mu J^\mu\,.
\eeq
In the above equation, $g$ collectively denotes the gauge couplings and $J_\mu$ are the composite sector global symmetry currents that are gauged by the fields  $A_\mu$. Among the gauge fields we have the four \mbox{{\textrm{SU}}$(2)_L\times{{\textrm{U}}(1)}_Y$} ones plus the eight QCD gluons $G_\mu^a$. We indeed showed in Sect.~\ref{sec:PC} that partial compositeness obliges us to make the composite sector colored and thus coupled to gluons. Now that we know the structure of the gauge field couplings to the composite sector, which by the way is the same through which the QED photon couples to QCD, we should be capable to estimate their interactions with the composite Particles. More precisely, we will now focus on the leading contribution to the interactions, namely the one that comes purely from the composite sector dynamics, {\it{i.e.}}~in the limit of vanishing $g$ coupling. Technically, what we want to do is to estimate the effective field theory operators obtained by integrating out only the resonance fields (possibly retaining some of them) while treating the gauge fields as external non-dynamical sources. In those operators the gauge field necessarily comes accompanied by one power of $g$ and no other powers of the elementary couplings can arise since the only dynamics that is integrated out is the composite sector one. Elementary fields and elementary couplings thus only emerge in the combination
\beq
g\cdot A_\mu\,,
\eeq
where of course the appropriate coupling $g=\{g,\,g',\,g_S\}$ has to be picked out for each field.

The dependence on the composite sector parameters $g_*$ and $m_*$ is now uniquely fixed by dimensional analysis. In order to work it out we first of all need to find the dimension of the elementary couplings $g$. Not surprisingly it is \footnote{This is because the global current operator, as extracted from the Noether formula, has dimensions
\beq
[J]=[{\mathcal{L}}]\cdot L=[\hbar]/L^3=C^{-2}/L^3\,.
\eeq
Therefore, given that $[A_\mu]=C^{-1}L^{-1}$, $[g]=C$ is required for the interaction Lagrangian to have the correct dimension.}
\beq
\label{gdim}
[g]=C\,.
\eeq
The $g\cdot A$ combination, given the canonical fields dimension in Eq.~(\ref{fdim}), has dimension of an inverse length. The dimensionless object which will appear in the effective Lagrangian is thus
\beq
\frac{g\cdot A_\mu}{m_*}\,.
\eeq
We see that $g\cdot A$ weights as much as a derivative in the power counting. This result is compatible with the structure of the gauge covariant derivative and with the dependence on the gauge fields of the CCWZ operators described in the previous chapter.

The elementary fermion fields can be discussed in a similar way. The only conceptual difference with the gauge fields is that the fermionic composite operators are not as sharply defined as the global currents are. In concrete, when we write the elementary/composite interactions as
\beq
{\mathcal{L}}_{\textrm{int}}=\lambda\,\overline{\psi} {\mathcal{O}}\,,
\eeq
we have not yet specified ${\mathcal{O}}$ but only its quantum numbers. The normalization is still arbitrary and indeed in Sect.~\ref{sec:HCsigmaM} we already exploited the ambiguity in the definition of ${\mathcal{O}}$ to reabsorb in it the powers of $m_*$ needed to make the elementary/composite couplings $\lambda$ dimensionless in Natural Units. Now we have to deal with length and coupling dimensions but we can still reabsorb powers of $m_*$ and $g_*$ in ${\mathcal{O}}$ giving to $\lambda$ any conventional dimension. All choices are equivalent but the convenient one is to take $\lambda$ to be a coupling
\beq
[\lambda]=C\,.
\eeq
With this choice the dimensionless object is
\beq
\frac{\lambda\cdot \psi}{m_*^{3/2}}\,,
\eeq
where the appropriate $\lambda$ has to be chosen for a given fermion species and chirality. We saw in Sect.~\ref{sec:HCsigmaM} that four $\lambda$ couplings exist already in the third family sector, namely $\lambda_{t_{L,R}}$ for the top and $\lambda_{b_{L,R}}$ for the bottom. Given that those couplings generate fermion masses after EWSB and the top mass is larger than the one of the bottom, the top sector couplings are typically larger than the bottom ones. Similarly, we expect the light fermion couplings to be very small even though this is not necessarily the case as we will discuss in Chap.~\ref{ch:flavor}.

In summary, our power-counting formula reads
\beq
\label{POWCFull}
{\mathcal{L}}_{\textrm{EFT}}=\;\;\frac{m_*^4}{g_*^2}\widehat{\mathcal{L}}\left[\frac{\partial}{m_*},\frac{g_*\Pi}{m_*},\frac{g_*\sigma}{m_*},\frac{g_*\Psi}{{m_*}^{3/2}},\frac{g\cdot A_\mu}{m_*},\frac{\lambda\cdot \psi}{m_*^{3/2}}
\right]\,,
\eeq
up to higher orders in the $g_*$ loop expansion. The one above is the power-counting rule for purely composite sector contributions to the operators, we will see in Sect.~\ref{CHPOT} how to generalize it in order include the radiative effects from elementary field propagation, in particular we will consider those responsible for the generation of the Higgs potential.

\subsubsection*{Large-${\mathbf{N}}$ power counting}

Until now we have considered the case of a weakly-coupled composite sector and we have derived our power-counting rules based on the \mbox{$1$S$1$C} hypothesis. We have also mentioned concrete examples, the most interesting being the one of $5$-d holographic models, where the \mbox{$1$S$1$C} hypothesis holds and thus the power counting rule applies. It is important to remark that our argument, though based on such simple and robust considerations like dimensional analysis, is intrinsically based on semiclassical perturbation theory and there is no hope of extending it to the non-perturbative case. The reason is that in a non-perturbative theory, where all the loop orders are equally important, the Planck constant can appear in all places in a completely uncontrolled manner, while for our reasoning it was crucial to keep track of the powers of $\hbar$ that arise at each order of the loop expansion. Coupling dimensional analysis becomes not predictive in the strongly-coupled limit because $C$ dimensions can always be compensated by $\hbar$, without paying the price of the small loop expansion parameter. The QCD theory with vanishing quark masses is an example of this situation. It is definitely a \mbox{$1$S$1$C} theory because it is defined in the UV in terms of just one coupling $g_S$ measured at one scale $\mu$, however it is clear that QCD hadrons do not obey the power counting (\ref{POWCFull}), or more precisely they do not obey it with $m_*$ identified with $\mu$ and $g_*$ identified with $g_S$. Low-energy QCD physics is controlled by the confinement scale  $\Lambda_{\textrm{QCD}}$ (which is a highly non-trivial combination of $\mu$, $g_S$ and $\hbar$) and there is no obvious candidate for the resonance coupling $g_*$.

Nevertheless, Eq.~(\ref{POWCFull}) still holds, at least in the mesonic sector of the theory. In order to understand how this works we must consider the large-$N$ expansion of QCD invented by G.~'t~Hooft in Ref.~\cite{'tHooft:1973jz}. It consists in generalizing the \mbox{SU$(3)$} color group to \mbox{SU$(N)$}, allowing $N$ to be large and asymptotically infinite. An expansion is thus set up in terms of the small parameter $1/N\ll1$. Notice that the expansion must be performed with care in order to get sensible results. In particular, the strong coupling constant must be taken to scale with $N$ as $g_S=g_{\textrm{'t}}/\sqrt{N}$ with a constant 't~Hooft coupling $g_{\textrm{'t}}$. At high energies the 't~Hooft coupling is small and the canonical semiclassical expansion can also be set up together with the one in $1/N$. In the IR, which is the relevant regime for our discussion, the 't~Hooft coupling becomes non perturbative and $1/N$ remains as the only expansion parameter.  The topic of large-$N$ expansion is extremely broad, however for our argument we will only need few basic results. The reader is referred to the Coleman lectures \cite{CL} and to the introduction of Ref.~\cite{Witten:1979kh} for their derivation.

We consider here the mesonic sector, {\it{i.e.}}~only those bound states that can be excited from the vacuum by a quark bilinear operator
\beq
{\mathcal{M}}\sim {\overline{q}}_i(\ldots)q^i\,,
\eeq
where $i=1,\,\ldots\,N$ is the color index contracted to form a color singlet. The dots in the parentheses represent an arbitrary combination of Gamma matrices and derivatives so that ${\mathcal{M}}$ could be a scalar, a vector, or any other Lorentz tensor. The associated particles could thus be the Goldstone bosons, {\it{i.e.}}~the pions, or spin-$1$ resonances like the $\rho$ or the $a_1$ mesons or higher-spin bosonic particles. Quark chiralities and flavor indices are not reported in our schematic notation. The four combinations of spinor chiralities ${\overline{q}}_{L,R}(\ldots) {{q}}_{L,R}$ can be considered, with the flavor indices contracted in all the possible ways to form irreducible representations of the chiral group. Clearly, a limited set of representations can be formed in this way and correspondingly only particles with certain isospin (or Gell-Mann \mbox{SU$(3)$}) quantum numbers can be described by this class of operators. The others are outside the mesonic sector and the considerations which follow do not hold for them.

\begin{figure}
\vspace{15pt}
\centering
\includegraphics[width=1\textwidth]{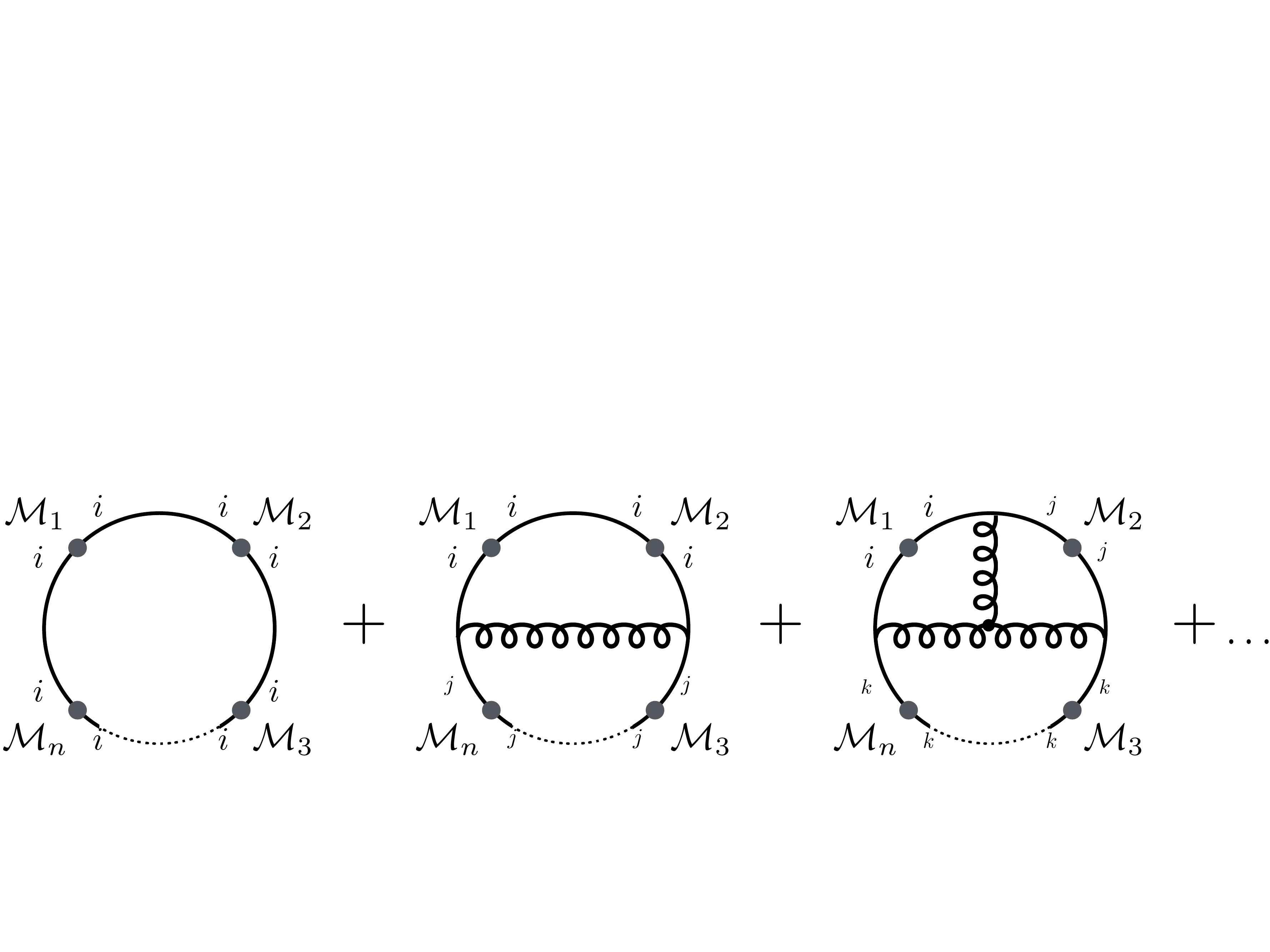}
\vspace{5pt}
\caption{Some planar diagrams contributing to the mesonic correlators. Summation over the free color indices $i$, $j$, $k$ is understood. }
\label{Correlators}
\end{figure}

We will now estimate the scattering amplitudes among mesonic particles and for this we first need to discuss the connected correlators of the mesonic operators. Diagrammatically those are shown in Fig.~\ref{Correlators} where each operator is represented as a $q$--${\overline{q}}$ vertex with the same color index on the two legs. Only few representative diagrams are reported in the figure but it is clear that an infinite set of them should be considered since the theory is in a non-perturbative regime and any loop order gives a comparable contribution. However we can still rely on the large-$N$ expansion and apply the classical result \cite{'tHooft:1973jz,CL,Witten:1979kh} that the leading diagrams are the ``planar'' ones, namely those that can be drawn on a plane without line intersections aside from the interaction vertices. Actually the true definition of planarity is a bit more refined than that and the only truly planar diagrams in this case are the ones which involve a single quark loop. It turns out that all the planar diagrams give a contribution to the correlator which is proportional to $N$ while the others, depending on their ``degree of non-planarity'', scale like $N^0$, $1/N$ and so on, providing subleading corrections. We will not demonstrate this here, however the reader can easily verify that at least the first three planar diagrams in Fig.~\ref{Correlators} really scale like $N$. This is obvious for the first one because there is only one free color index to be summed over. For the second one there are two free indices, which gives $N^2$, but also two powers of the coupling, $g_S^2$, which scales like $1/N$ in the \mbox{'t Hooft} limit. In the third one there are $3$ indices but $4$ couplings and so on for higher loop orders. In summary, the correlators scale as
\beq
\langle{\mathcal{M}}_1\ldots{\mathcal{M}}_n\rangle\propto N\,,
\eeq
for an arbitrary number $n$ of mesonic operator insertions. 

We can get a better parametric control on the correlator by dimensional analysis. Since coupling dimension does not help in the case of a strongly-coupled theory we can go back to Natural Units and just count energy dimensions. Low-energy QCD is well-know (or at least widely believed) to be a one-scale theory, with a single energy scale $\Lambda_{\textrm{QCD}}$ from confinement. But if it is so, all the infinite series of diagrams in Fig.~\ref{Correlators} must conspire to give a dimensionally correct result which only depends on the external momenta $p_i$ and on \mbox{$m_*=\Lambda_{\textrm{QCD}}$}, namely 
\beq
\label{corrsc}
\langle{\mathcal{M}}_1\ldots{\mathcal{M}}_n\rangle\sim \frac{N}{16\pi^2}m_*^\Delta F[p_i/m_*]\,,
\eeq
where the energy dimensionality of the correlator, $\Delta$, could be easily computed in terms of those of the ${\mathcal{M}}_i$'s. The actual value of $\Delta$ does not matter much. Since $m_*$ is the only dimensionful object in the problem aside from the external momenta the overall dependence on it is trivially fixed. It could be safely ignored and restored at the end of the calculation by dimensional analysis. What is less trivial in the above equation is the $1/16\pi^2$ loop factor, which has been introduced with the following logic. The theory being non-perturbative means that loop diagrams of all orders give comparable contributions, therefore it should be sufficient to pick up one in order to obtain an estimate of the final result. The $1/16\pi^2$ factor is found by looking at the simplest (one loop) diagram in Fig.~\ref{Correlators}.

In order to obtain the scattering amplitudes we must compute the residual of the correlators at the on-shell poles (which amounts to multiply by $p_i^2-m_i^2$ for each external leg and to take the on-shell limit) and eventually divide by the wave function renormalization factors $\sqrt{Z_i}$. Schematically, wave functions are extracted from the two-point correlators as
\bea
&\displaystyle
\langle {\mathcal{M}}_i {\mathcal{M}}_i\rangle\underset{p_i^2\rightarrow m_i^2}{=}\frac{Z_i}{p_i^2-m_i^2}\sim\frac{N}{16\pi^2}m_*^{\Delta_i} F[p_i/m_*]&\nonumber\\
&\displaystyle{\mathlarger{\mathlarger{\Downarrow}}}&\nonumber\\
&\displaystyle\sqrt{Z_i}\sim\frac{\sqrt{N}}{4\pi}m_*^{\frac{\Delta_i}2+1}&
\eea
where $\Delta_i$ is the dimension of the two-point correlator. Dividing by $\sqrt{Z_i}$ leads to one $4\pi/\sqrt{N}$ factor for each of the external legs. The powers of $m_*$ combine to give the correct dimension of the $n$-point Feynman amplitude, which is $[{\mathcal{A}}_n]=E^{4-n}$, and we find
\beq
\label{amplitudesm}
{\mathcal{A}}_n\sim m_*^4 \frac{N}{16\pi^2}\left(\frac{4\pi}{\sqrt{N}}\right)^n\left(\frac1{m_*}\right)^n  F[p_i/m_*]\,.
\eeq

The result is remarkable. It shows that any sensible amplitude (with $n>2$) is suppressed at large-$N$ and eventually vanishes in the strict limit $1/N=0$. Therefore mesons asymptotically become free particles and this is surprising since they originate as bound states of a strong dynamics. At finite $N$ they behave like weakly-interacting particles with a coupling factor
\beq
\displaystyle
g_*=\frac{4\pi}{\sqrt{N}}\,,
\eeq
which suppresses the amplitude for each external leg. The behavior of the amplitudes in Eq.~(\ref{amplitudesm}) must be reproduced, in the low-energy effective description of the mesonic particles, by a peculiar dependence of the effective operator coefficients on $g_*$ and $m_*$. Namely,  for each field insertion one needs one $g_*$ and an appropriate power of $m_*$ ($m_*^{-1}$ for bosons and $m_*^{-3/2}$ for fermions) fixed by the energy dimension of the low-energy meson fields.  Derivatives correspond to momenta in the Fourier space and thus they come with $1/m_*$. An overall additional factor of $m_*^4/g_*^2$ must also be present to match with Eq.~(\ref{amplitudesm}). We end up with the \mbox{$1$S$1$C} power-counting in Eq.~(\ref{POWCFull}). As anticipated it holds in the mesonic sector of low-energy QCD even though for a radically different reason than for perturbative theories. The actual value of $g_*$ depends on $N$, it can range from $0$ to $4\pi$. For real-world QCD $N=3$ and one is entitled to approximate $g_*=4\pi/\sqrt{3}\simeq 4\pi$, in which case the power-counting formula reduces to the NDA counting \cite{Manohar:1983md}. The latter is known to give phenomenologically valid estimates of the chiral Lagrangian operators, providing a partial experimental confirmation of  Eq.~(\ref{POWCFull}). 

The interactions involving elementary fields are easily estimated by proceeding along similar lines. In this case one has to compute scattering amplitudes in the presence of external elementary field sources, at a fixed order in the elementary/composite couplings $g$ and $\lambda$. For this we need correlators with insertions of the composite operators ($J$ and ${\mathcal{O}}$) the elementary fields are coupled to. Under the assumption that also those operators are mesonic, the correlator estimate is still the one in Eq.~(\ref{corrsc}) and the derivation proceeds as before, with the only difference that  no wave-function factor has to be taken out for the elementary field legs. Therefore, there is no $1/\sqrt{Z}\sim g_*$ for the elementary field insertions, but instead we have the elementary coupling $g$ or $\lambda$ from the source interaction coefficient. Dimensional analysis does the rest and we recover Eq.~(\ref{POWCFull}). All this holds if the operators $J$ and ${\mathcal{O}}$ the sources couple to are  mesonic operators. This is definitely the case in QCD where the relevant source is the electromagnetic field coupled to the quark current. Photon interactions with the hadrons can thus be estimated by Eq.~(\ref{POWCFull}), once again with good phenomenological success.

The fact that the \mbox{$1$S$1$C} power-counting applies to the mesonic sector of QCD gives a strong support to its validity. It shows that it could hold not just for ``fake'' weakly-coupled composite Higgs models but also for genuine strongly-coupled theories. The two things might actually coincide, for instance the $5$-d holographic models might be dual to some strong $4$-d dynamics, but we do not  need to assume an exact duality to apply the \mbox{$1$S$1$C} power-counting. Nevertheless, those are just indications and no conclusive statement can be made on which one is the true power-counting of the true composite sector (if any) the pNGB Higgs emerges from. This is because of several reasons, the most obvious one being that we do not know anything about the underlying microscopic theory. For sure it is not QCD, nor a rescaled version of it. If it is a radically different theory its hadron scattering amplitudes might obey a different power-counting. Actually even if the theory was QCD-like the power-counting could be different if the relevant composite particles and operators were not mesons but other color singlets formed by more complicated contractions than ${\overline{q}}_i\cdot q^i$. For instance the glueball amplitudes in QCD are not controlled by $g_*\sim 1/\sqrt{N}$ but by $g_*\sim 1/N$  \cite{'tHooft:1973jz,CL,Witten:1979kh}. The composite Higgs fermionic operators $\mathcal{O}$ deserve a special mention in this context. We assumed them to be mesons, but of course in QCD there are no mesonic operators with Fermi statistics. The only fermionic candidates are baryons which however behave at large-$N$ in a radically different manner. They cannot even be described by a Lagrangian because they are solitons  \cite{Witten:1979kh}. The femionic operators we are talking about and the associated particles should emerge from some different theory, for instance from objects like $\phi_i^\dagger q^i$ in some supersymmetric model with scalar quark partners $\phi^i$. Non-supersymmetric proposals for the microscopic origin of these operators have been made in Refs.~\cite{Barnard:2013zea,Ferretti:2013kya,Ferretti:2014qta}. 

In conclusion, the \mbox{$1$S$1$C} power-counting should be regarded as a plausible guess on the unknown composite sector dynamics, which can be supported and motivated by different viewpoints, but it should not be taken as absolute truth. Also the fact that it has no known alternative does not mean that it is necessarily unique, other options could be explored.\footnote{Several confusing statements about power-counting have appeared in the recent literature. For instance that power-counting is a convention and any guess is equally plausible. Or the converse one, that power-counting should be inferred from the effective field theory itself by some ``consistency'' requirement. Both those statements are false. Power-counting is the result of a set of assumptions on the UV theory, thus it is not unique but any sensible one, possibly alternative to \mbox{$1$S$1$C}, must be founded on alternative physics hypotheses.}
With this caution remarks in mind, we will make extensive use of it in what follows.

\section{Higher derivative operators}

We now have all the ingredients to improve our low-energy description of the composite Higgs, the SM vectors and fermions, by going to the next order in the derivative expansion. This will induce new effects which were absent in the leading-order Lagrangian. In particular, the only leading-order corrections to the SM have been found to be the modified interactions of the Higgs particle (see for instance Eqs.~(\ref{hvc}), (\ref{htc5})), induced by additional non-renormalizable operators with extra powers of the Higgs field but without extra derivatives or gauge field insertions. The leading terms are $d=6$ operators suppressed by $1/f^2$ with respect to their $d=4$ SM counterparts. The higher order CCWZ operators discussed in what follows will instead induce, in accordance with the power-counting in Eq.~(\ref{POWCFull}), $d=6$ corrections with extra derivatives and gauge fields, suppressed by $1/m_*^2=1/g_*^2\,1/f^2$. Their effect is thus generically subdominant if $g_*>1$ and this is precisely why it makes sense to neglect them in a first approximation. Nevertheless they are not at all irrelevant. First of all, because $g_*$ might be dangerously close to one. Of course we would like to have it large to make the resonances heavy, however we will see in Sect.~\ref{CHPOT} that it should better not be too large in order for the Higgs being naturally light. Second, higher derivative operators induce corrections to the SM gauge field propagators and interactions which were absent at the leading order. Though suppressed, those effects are strongly bounded by Electro-Weak Precision Tests (EWPT) and thus they give rise, as we will see below, to important phenomenological constraints on the composite Higgs scenario. Finally, higher derivative operators give enhanced contributions to high momentum reactions. Therefore they might become  relevant for a program of high-energy measurements to be performed at the LHC and at future colliders.

\subsection[Order $p^4$ bosonic]{Order ${\mathbf{p^4}}$ bosonic}
\label{p4bos}

The most general CCWZ invariant operator can be written, as discussed in Sect.~\ref{2dt}, in terms of the $d_\mu$ symbol and the covariant derivative $D_\mu$, with the unbroken group and the Lorentz indices properly contracted to obtain an invariant Lorentz scalar. The ${\mathcal{O}}(p^4)$ operators can be classified in three groups: those obtained by four $d$-symbol insertions and no derivative, those with two $d$'s and two $D$'s and the ones with four $D$'s and no $d$-symbols. No \mbox{SO${(4)}$} invariant can be constructed with an odd number of $d$ symbol insertions so the three classes above exhaust all possibilities. It is convenient to start the classification from the operators in the last category, which are obtained by taking covariant derivative commutators, forming the $E_{\mu\nu}$ field-strength following Eq.~(\ref{emunu}) and contracting indices to form a scalar. Given that the $E$ tensor is a reducible representation of \mbox{SO${(4)}$}  (see Appendix~\ref{so54} ) and taking care that the Lorentz indices can be contracted either with  $\eta_{\mu\nu}$'s or with the Levi--Civita antisymmetric tensor $\epsilon^{\mu\nu\rho\sigma}$, four independent operators exist
\bea
\displaystyle
&&O_{++,\,1}={E^L}_{\mu\nu}^\alpha {E^L}^{\mu\nu}_\alpha + {E^R}_{\mu\nu}^\alpha {E^R}^{\mu\nu}_\alpha\,,\nonumber\\
&&O_{+-,\,1}={E^L}_{\mu\nu}^\alpha {E^L}^{\mu\nu}_\alpha - {E^R}_{\mu\nu}^\alpha {E^R}^{\mu\nu}_\alpha\,,\nonumber\\
&&R_{-+}=\epsilon^{\mu\nu\rho\sigma}\left({E^L}_{\mu\nu}^\alpha {E^L}_{\rho\sigma,\,\alpha} + {E^R}_{\mu\nu}^\alpha {E^R}_{\rho\sigma,\,\alpha}\right)\,,\nonumber\\
&&R_{--}=\epsilon^{\mu\nu\rho\sigma}\left({E^L}_{\mu\nu}^\alpha {E^L}_{\rho\sigma,\,\alpha} -{E^R}_{\mu\nu}^\alpha {E^R}_{\rho\sigma,\,\alpha}\right)\,,
\eea
where the operators labelling reflects the quantum numbers under the CP and $P_{LR}$ $\Zdouble_2$ symmetries defined and discussed in Appendix~\ref{discrete}. As explained in the Appendix, all the operators are even under charge conjugation and their CP quantum number is only dictated by parity. The ones constructed with the $\epsilon$ tensor are odd while the others are even. The $P_{LR}$ parity assignment is also evident since $P_{LR}$ interchanges $L$ with $R$. We will show below that the operators labeled as ``$R$'' are redundant.

The second class of operators to be discussed are those with four $d$-symbols and no covariant derivatives. The four fourplet indices can be contracted with two $\delta_{ij}$'s or with the completely antisymmetric \mbox{SO${(4)}$}-invariant tensor $\epsilon_{ijkl}$. In the first case, $\eta_{\mu\nu}$'s have to be employed for the Lorentz contractions while the Levi--Civita $\epsilon_{\mu\nu\rho\sigma}$ is needed in the second one. We have a total of three new operators
\bea
\displaystyle
&&O_{++,\,2}=(d_\mu^id^\mu_i)^2 \,,\nonumber\\
&&O_{++,\,3}=d_\mu^id^\mu_j d_\nu^jd^\nu_i \,,\nonumber\\
&&O_{--,\,1}=\epsilon_{ijkl}\epsilon^{\mu\nu\rho\sigma}d_\mu^i d_\nu^j d_\rho^k d_\sigma^l\,,
\eea
which are again classified according to their CP and $P_{LR}$ properties. Notice that the $\epsilon$ tensor in \mbox{SO${(4)}$}  produces a $P_{LR}$-odd object because $P_{LR}$, as discussed in Appendix~\ref{discrete}, is just parity in \mbox{O${(4)}$} and it is represented by a matrix with determinant $-1$.

The operators in the last class, made of two $d$'s and two $D$'s, can be further divided into two subclasses. The first ones are  the commutator of two $D$'s, {\it{i.e.}}~the $E_{\mu\nu}^{L,R}$ field-strengths, multiplied by two $d$'s. There are three of them which are relevant 
\bea
\label{thridrc}
\displaystyle
&&R_{++,\,1}={E^L}_{\mu\nu}^\alpha d^\mu_i\left(i\,t_L^\alpha\right)^{ij}d^\nu_j + {E^R}_{\mu\nu}^\alpha d^\mu_i\left(i\,t_R^\alpha\right)^{ij}d^\nu_j  \,,\\
&&O_{+-,\,2}={E^L}_{\mu\nu}^\alpha d^\mu_i\left(i\,t_L^\alpha\right)^{ij}d^\nu_j - {E^R}_{\mu\nu}^\alpha d^\mu_i\left(i\,t_R^\alpha\right)^{ij}d^\nu_j  \,,\nonumber\\
&&O_{--,\,2}=\epsilon^{\mu\nu\rho\sigma}\left(
{E^L}_{\mu\nu}^\alpha d_{\rho,\,i}\left(i\,t_L^\alpha\right)^{ij}d_{\sigma,\,j} - 
{E^R}_{\mu\nu}^\alpha d_{\rho,\,i}\left(i\,t_R^\alpha\right)^{ij}d_{\sigma,\,j}\right)\,.\nonumber
\eea
A fourth one, like $O_{--,\,2}$ but with ``$+$'' instead of ``$-$'' in the parenthesis, has been voluntarily ignored and will be discussed below. The second set of operators are those constructed with two $d$-symbols and two covariant derivatives acting on them. In the classification of the operators in this subclass the covariant derivatives must be taken to commute as if they were ordinary derivatives because covariant derivative commutators produce $E_{\mu\nu}$ field-strengths and lead to operators we already counted in the first subclass. Actually, since we missed one operator in the previous subclass, one exception to this rule is needed and three operators must be considered
\bea
\displaystyle
&&O_{++,\,4}=\left(D\cdot d\right)_{[\mu,\nu]}^i \left(D\cdot d\right)^{[\mu,\nu]}_i\,,\nonumber \\
&&O_{-+}=\epsilon^{\mu\nu\rho\sigma} \left(D\cdot d\right)_{[\mu,\nu]}^i \left(D\cdot d\right)_{[\rho,\sigma],\,i}\,,\nonumber \\
&&R_{++,\,2}=\left(D_\mu d^\mu\right)_i \left(D_\nu d^\nu\right)^i\,,
\eea
where $D\cdot d$ denotes the covariant derivative of the $d$-symbol. We have chosen to express the operators in terms of $D\cdot d$ anti-symmetrized on the Lorentz indices because this has a rather simple explicit form, provided by Eq.~(\ref{eddcomp}). In writing $O_{-+}$ we made the above-mentioned exception, indeed $O_{-+}$ would vanish if the covariant derivatives had to commute. By integrating by parts $O_{-+}$ can be rewritten in terms of the covariant derivatives commutator, {\it{i.e.}}~of the field-strength tensors, reproducing the operator which we were missing in Eq.~(\ref{thridrc}). 

We end up with a total of $10$ operators (distributed as $4$--$2$--$2$--$1$ in the $++$, $--$, $+-$ and $-+$ categories) and we claim that this is all what is needed to describe the composite Higgs dynamics at ${\mathcal{O}}(p^4)$ in the bosonic sector. However we still have to deal with the four operators $R_{-+}$, $R_{--}$, $R_{++,\,1}$ and $R_{++,\,2}$ and to show that they are redundant as anticipated. We start from $R_{-+}$ and $R_{--}$, which are redundant simply because they are total derivatives. In order to show this one has to remember (see for instance Ref.~\cite{Nakahara}) that out of a generic gauge field ${\mathcal{A}}_\mu$ one can construct an object, called the ``Chern--Pontryagin'' density, which is a total derivative. Namely
\beq
\label{cpden}
{\textrm{Tr}}\big[{\mathcal{A}}_{\mu\nu} {\tilde{\mathcal{A}}}^{\mu\nu} \big]=
\epsilon^{\mu\nu\rho\sigma}{\textrm{Tr}}\left[{\mathcal{A}}_{\mu\nu} {\mathcal{A}}_{\rho\sigma} \right]=\partial_\mu\Omega^\mu[\mathcal{A}]\,,
\eeq
where ${\mathcal{A}}_{\mu\nu}$ is the field-strength and the $\Omega[\mathcal{A}]$ is the (Hodge dual) Chern--Simons form constructed with $\mathcal{A}$. But ${E^L}$ and $E^R$ are precisely field-strength tensors, constructed with $e^L$ and $e^R$ as if they were \mbox{SU$(2)$} gauge fields, and the operators $R_{-+}$ and $R_{--}$ are linear combinations of the left and right Chern--Pontryagin densities. Therefore they are total derivatives.

We now turn to discuss $R_{++,\,1}$. The basic and generic observation is that there exist linear combinations of CCWZ invariants that are ``too symmetric'', in the sense that they reconstruct an invariant of a linearly realized ${\mathpzc{G}}$ group (\mbox{SO$(5)$}, in our case) and not just of the unbroken ${\mathpzc{H}}$. These objects are typically trivial. In Sect.~\ref{sec:HCsigmaM} we already encountered some examples of this situation when we proved (see for instance Eq.~(\ref{invrel})) the existence of a relation among the two ${\mathcal{O}}(p^0)$ fermionic operators obtained from the classification. This happened because the linear combination of CCWZ tensors on the left-hand-side of Eq.~(\ref{invrel}) reconstructs a singlet under a fictitious linearly realized \mbox{SO$(5)$} that rotates the dressed sources in the $\mathbf{4}$ and $\mathbf{1}$ as if they formed a single fiveplet.\footnote{As explained at length in the previous chapter this is not the way in which the genuine non-linearly realized \mbox{SO$(5)$} acts. Each CCWZ operator is automatically invariant under the latter symmetry.} The Goldstone matrix used to define the dressed sources drops because it corresponds to a transformation under this fictitious group and therefore the result is independent of the Goldstone fields. In particular in Eq.~(\ref{invrel}) it reduced to the product of two orthogonal sources and thus it vanished. Coming back to the present case, it is not hard to identify trivial ${\mathcal{O}}(p^4)$ operators starting from the ${\mathcal{F}}$ tensors defined in Eqs.~(\ref{deff}), (\ref{deff1}) and worked out explicitly in Appendix~\ref{so54}. Let us consider
\beq
{\textrm{Tr}}[F_{\mu\nu}F_{\rho\sigma}]={\textrm{Tr}}[{\mathcal{F}}_{\mu\nu}{\mathcal{F}}_{\rho\sigma}]={{\mathcal{F}}_L}_{\mu\nu}^\alpha {{\mathcal{F}}_L}_{\rho\sigma}^\alpha+{{\mathcal{F}}_R}_{\mu\nu}^\alpha {{\mathcal{F}}_R}_{\rho\sigma}^\alpha+{{\mathcal{F}}_{\mathbf{4}}}_{\mu\nu}^i {{\mathcal{F}}_{\mathbf{4}}}_{\rho\sigma}^i\,,
\eeq
where in the first equality we exploited the fact that the Goldstone matrix we used to dress $F$ drops when we form, by taking the trace, a linearly realized \mbox{SO$(5)$} singlet. Clearly the Lorentz scalars obtained by the above tensors are not new operators because the ${\mathcal{F}}$'s can be expressed in terms of $E$ and $D\cdot d$ by Eq.~(\ref{eddcomp}). By contracting the Lorentz indices with $\epsilon$ or with two $\eta$'s and making use of Eq.~(\ref{normcso4}) we obtain
\bea
&&\epsilon^{\mu\nu\rho\sigma}{\textrm{Tr}}[F_{\mu\nu}F_{\rho\sigma}] = R_{-+}\,,\\
&&{\textrm{Tr}}[F_{\mu\nu}F^{\mu\nu}]=2\,R_{++,\,1}+ O_{++,\,1}+\frac12 O_{++,\,2}-\frac12 O_{++,\,3}+O_{++,\,4}\,.\hspace{15pt}\nonumber
\eea

From the first line of the above equation we see once again that $R_{-+}$ is a total derivative, given that Eq.~(\ref{cpden}) can be also applied to the Chern--Pontryagin density constructed with the gauge source fields. The second equation allows us to relate $R_{++,\,1}$ to ${\textrm{Tr}}[F^2]$, which has the form of a  kinetic term for the source fields. When the sources will be set to their physical value by Eq.~(\ref{physs}), this term will just provide one additional contribution to the elementary fields kinetic term we introduced in Eq.~(\ref{galag}) and therefore its only effect will be an unobservable redefinition of the gauge coupling parameters $g$ and $g'$.\footnote{If we denote as $c/(4g_*^2)$ the operator coefficient, the coupling redefinition that eliminates it is $1/g^2\rightarrow 1/g^2+c/g_*^2$ and $1/g^{\prime \,2}\rightarrow 1/{g}^{\prime \,2}+c/g_*^2$. This is best seen by first performing the field redefinition $W\rightarrow W/g$ and $B\rightarrow B/g'$ by which the coupling strength is moved to the kinetic term normalization.} The operator $R_{++,\,1}$ can thus be eliminated in favor of the other ones in the $++$ category up to an unobservable parameter redefinition. Identical considerations allow us to get rid of the operators we might have constructed with the \mbox{U$(1)_X$} source $X_\mu$ introduced in Sect.~\ref{sec:HCsigmaM}. We voluntarily ignored $X_\mu$ until now, but clearly it can be used to construct the CCWZ invariants, where it will appear as an \mbox{SO$(4)$} singlet endowed with his own unbroken local \mbox{U$(1)_X$} symmetry.  The only two ${\mathcal{O}}(p^4)$ operators with $X_\mu$ are $F\tilde{F}$ and the $FF$ kinetic term, which are respectively a total derivative and equivalent to one further redefinition of $g'$.

Let us finally turn to $R_{++,\,2}$, it is redundant like the other two, but for a different reason which has to do with the equations of motion of the Goldstone bosons at ${\mathcal{O}}(p^2)$. A generic property of perturbative theories, whose Lagrangian terms are classified by some expansion parameter $\varepsilon$, is that all the operators of order $\varepsilon^n$ that vanish on the ${\mathcal{O}}(\epsilon^{n-1})$ equations of motion can be safely ignored because they are actually equivalent to ${\mathcal{O}}(\epsilon^{n+1})$ terms. This follows from the possibility of performing redefinitions of the field variables, by which the operators proportional to the lower order equations of motion can be systematically traded for higher order terms. The expansion parameter is $\epsilon=p^2$ in our case and we are classifying up to ${\mathcal{O}}(p^4)$. We can thus remove from the classification all the operator which vanish on the ${\mathcal{O}}(p^2)$ equations of motion, namely on the equations of motion obtained by the ${\mathcal{O}}(p^2)$ Lagrangian in Eq.~(\ref{smgchk}). Deriving these equations directly from the Lagrangian is not completely straightforward, however we can guess the result on symmetry basis. The equations of motions associated to the four Goldstone boson fields must be a CCWZ tensor in the $\mathbf{4}$ of \mbox{SO$(4)$},  and of course a Lorentz scalar. But from the classification of two derivative tensors performed in Sect.~\ref{2dt} and Appendix~\ref{so54} one easily shows that only one such term exists and therefore the  ${\mathcal{O}}(p^2)$ equations of motion are necessarily
\beq
D^\mu d_\mu^i=0\,.
\eeq
We can thus eliminate the operator $R_{++,\,2}$ from our basis.

This concludes the operator classification, which would be the right starting point for a detailed phenomenological analysis of the indirect effects associated with the composite nature of the Higgs. However this topic is currently under development (see \cite{Contino:2011np,Azatov:2013ura,Alonso:2014wta,Hierro:2015_inprep} for recent discussions) and it goes beyond the scope of the present Notes. Here we will limit ourselves to illustrate two specific effects associated with ${\mathcal{O}}(p^4)$ operators, namely a correction to the $S$ parameter of EWPT and the possible modification of the Higgs boson coupling to two photons. The first effect is important because it provides a powerful phenomenological constraint on the composite Higgs scenario. The second one is important because it is actually absent, or more precisely it is reduced with respect to the naive expectation thanks to the pNGB nature of the Higgs. Even if we will not discus it here, identical considerations could be made for the Higgs coupling to gluons.

Let us start from the $S$ parameter. It measures possible deviations from the SM of the neutral gauge bosons propagator matrix and it is defined, with the conventions of Ref.~\cite{Barbieri:2004qk}, as
\beq
\displaystyle
\widehat{S}=g^2 \frac{d\;}{dq^2}\Pi_{W_3B}|_{q^2=0}\,,
\eeq
in terms of the transverse components of the two-point vacuum polarization amplitudes at the tree-level order, denoted as $\Pi_{VV'}(q^2)$, with two gauge bosons \mbox{$V,V'=\{W^\alpha, B\}$} on the external legs. The external fields $V$ and $V'$ are taken to be the ones that couple to the light SM quarks and leptons with unit strength, which means that the EW couplings $g$ and $g'$ are reabsorbed by a field redefinition and are carried by the kinetic terms in this field basis.
In the definition of $\widehat{S}$ it is implicitly assumed that the couplings to vector bosons of the light matter fermions are identical the SM ones. This qualifies $\widehat{S}$ as an ``oblique'' correction to the SM dynamics, namely a correction which occurs in the $2\rightarrow2$ reactions among light particles (precisely measured at LEP) only through a modification of the vector boson propagators and not of the vector bosons/matter interaction vertexes. This is typically a good approximation in composite Higgs because we saw in Sect.~\ref{sec:PCbasic} that the light matter fermions are characterized by a tiny compositeness fraction. Therefore their coupling to vector bosons is dominantly the one from the covariant derivative in the canonical kinetic term coming from the elementary sector in Eq.~(\ref{elm_q_kin}), with small corrections from the composite sector. We will elaborate more on this aspect in the following section and in Chaps.~\ref{ch:flavor} and \ref{ch:EWPT}, but for the moment we just ignore non-oblique corrections and we identify the unit-coupling EW boson fields in the definition of the vacuum polarization with the rescaled gauge sources $gW^\alpha$ and $g'B$.

No direct correction to $\widehat{S}$ came from the non-linear $\sigma$-model terms we considered in the previous chapter because their only effect is to modify the Higgs boson couplings while leaving the EW boson propagators identical to the SM ones.\footnote{We will discuss in Chap.~\ref{ch:EWPT} the oblique corrections which are radiatively induced by the modified Higgs couplings.} Such corrections do instead arise at ${\mathcal{O}}(p^4)$, in particular from the operator $O_{++,\,1}$. By setting the Higgs to its VEV and using Eq.~(\ref{eddUG}) it gives quadratic terms
\beq
\label{Shat0}
-\frac{c_{++,\,1}}{4\,g_*^2}\left[\sin^2{\frac{V}{f}}{A^L_{\mu\nu}}^\alpha {A^{R,\,\mu\nu}}_\alpha +\frac14\left(3+\cos{\frac{2\,V}{f}}\right)\left({A^L_{\mu\nu}}^2+{A^R_{\mu\nu}}^2\right)\right]\,,
\eeq
with $A^{L,R,\,\mu\nu}$ as in Eq.~(\ref{gauge_sources_field_st}), having estimated the operator coefficient as
\beq
-\frac{c_{++,\,1}}{4\,g_*^2}\,,
\eeq
in accordance with the \mbox{$1$S$1$C} power-counting in Eq.~(\ref{POWCFull}). The presence of the minus sign is purely conventional: $c_{++,\,1}$ is an unknown ${\mathcal{O}}(1)$ coefficient and it  could have any sign from the viewpoint of our low-energy effective field theory. This convention is adopted because $c_{++,\,1}$ is typically positive when computed in explicit models, eventually leading to a positive $\widehat{S}$. Similar considerations hold for the $1/4$ normalization. From the mixed $L$--$R$ term in Eq.~(\ref{Shat0}) we obtain a mixed $W$--$B$ vacuum polarization, and thus 
\beq
\label{Shatest}
\displaystyle
\widehat{S}=\frac{c_{++,\,1}g^2}{2\,g_*^2}\xi = 2\, c_{++,\,1} \frac{m_W^2}{m_*^2}\,.
\eeq

Both the expressions for $\widehat{S}$ reported above, in terms of $g_*$ and $f$ or of $m_*$, are interesting and deserve a comment. Clearly, the second one is more compact and it shows that the current experimental constraint on $\widehat{S}$, which we can take to be approximately $\widehat{S}\lesssim10^{-3}$ (for references and more details, see Chap.~\ref{ch:EWPT}), implies a $g_*$-independent limit
\beq
m_*\gtrsim \sqrt{2\, c_{++,\,1} }\cdot 2.5~{\textrm{TeV}}\,,
\eeq
on the composite sector scale $m_*$. While this is obviously not a sharp constraint because it is subject to ${\mathcal{O}}(1)$ uncertainties in the coefficient, it constitutes a rather robust starting point for the construction of any realistic composite Higgs model. It is also worth noticing that the limit could be made weaker, or stronger, by other contributions to EWPT originating from other sectors of the theory as we will exhaustively discuss in Chap.~\ref{ch:EWPT}. The first $\widehat{S}$ formula in Eq.~(\ref{Shatest}) is interesting because we can recognize that it coincides with the well-known Technicolor result (see for instance \cite{Peskin-Takeuchi} and remember that $1/g_*^2\sim N/16\pi^2$ in our formalism) aside from the $\xi$ factor. This provides one further illustration of how the composite Higgs is superior to Technicolor, leading to smaller new physics effects, only for small misalignment angle $\xi\ll1$ while it reduces to the latter for $\xi=1$. A small $\xi$ gives not just a naturally SM-like Higgs boson as discussed in the previous chapter, but also to an improved agreement with the EWPT.

We now turn to the Higgs coupling to two photons. No such coupling is induced by any of the operators listed above and this is not hard to verify by proceeding as follows. Given that we are interested in operators involving photons and no other EW bosons we can set the $W$ and $Z$ fields to zero obtaining, in the notation of Appendix~\ref{so54}, gauge sources of the form
\beq
\label{photononly}
A^L_\mu=A^R_\mu=e\gamma_\mu\frac{\sigma_3}2\,,
\eeq
where $\gamma$ denotes the photon field and $e$ is the electric charge, defined as $1/e^2=1/g^2+1/g^{\prime 2}$ like in the SM. If we now look, in Eqs.~(\ref{edUG}), (\ref{eddUG}), at the unitary gauge version of the CCWZ objects $d$, $D\cdot d$ and $E$ out of which our operators are constructed, we immediately see that the dependence on the photon fields drops in all of them, aside from the field-strengths $E_{L,R}$, which however become trivial 
\beq
\displaystyle
{E_{L}^{(2)}}_{\mu\nu}={E_{R}^{(2)}}_{\mu\nu}=e\,\gamma_{\mu\nu}\frac{\sigma_3}2\,,
\eeq
and independent of the physical Higgs field $h$. The only non-redundant operator with two photons thus comes from $O^{(4)}_{++,\,1}$, which however reduces to an harmless correction to the photon kinetic term. The $h\gamma\gamma$ vertex is thus not induced by the ${\mathcal{O}}(p^4)$ operators. This fact is an important phenomenological virtue of our construction. If it was not the case, we might have obtained a large correction relative to the SM, where $h\gamma\gamma$ is radiatively generated and thus suppressed. Such a correction would be already within the sensitivity of current Higgs coupling measurements.

Being the absence of $h\gamma\gamma$ such an important phenomenological property, let us further investigate its origin, which has to do with the pNGB nature of the Higgs. In order to understand how this comes we have to go back to Eq.~(\ref{gscoup}), where we declared how the EW boson sources interact with the composite sector. Since we are only interested in photon couplings we can restrict the sources to the photon field by Eq.~(\ref{photononly}), obtaining
\beq
{\mathcal{L}}_{\textrm{int}}^{\gamma}=e \gamma^\mu\left( J_\mu^{L,\,3}+J_\mu^{R,\,3}+J_\mu^{X} \right)=e\gamma^\mu J_\mu^{\textrm{e.m.}}\,,
\eeq
where we also made use of Eq.~(\ref{Xsv}). The full set of EW boson couplings to the currents breaks the composite sector global group completely. However the photon interactions displayed above preserve some symmetry transformations, namely all and only those associated with electrically neutral generators, which commute with the electric charge $Q=T_L^3+T_R^3+X$ and thus leave $J_\mu^{\textrm{e.m.}}$ invariant. This defines an \mbox{SU$(2)\times$U$(1)$} subgroup of \mbox{SO$(5)$}, whose detailed structure is however not very important for the present discussion. What matters for us is that the generator $\Th^4$, which is the one associated with the physical Higgs boson, is obviously neutral and belongs to this unbroken subgroup. Being $\Th^4$ an exact symmetry, only  spontaneously broken by the composite sector, the physical Higgs component is an exact Goldstone boson, endowed with an exact shift symmetry induced by $\Th^4$ whose explicit action on $\Pi^4$ could be extracted from Eq.~(\ref{brtr}). It is this symmetry that forbids the $h\gamma\gamma$ vertex. 

Indeed, $h\gamma\gamma$ at the two-derivative order might only come (up to a CP-odd term discussed below) from the interaction
\beq
\label{hgg}
\frac{c_\gamma e^2 v}{m_*^2} h\,\gamma_{\mu\nu}\gamma^{\mu\nu}\,.
\eeq
The two derivatives need to be carried by two photon field strength $\gamma_{\mu\nu}$ tensors, which is the only object we can use because of the residual QED gauge invariance. Given that no derivative is acting on $h$, this term is incompatible with the shift symmetry and thus it is forbidden.\footnote{An even sharper argument would be to imagine putting the theory in an electromagnetic field background. The operator above would induce a potential (a tadpole term) for the Higgs, which is definitely incompatible with its Goldstone nature.}  If instead the Higgs was not a Goldstone, but another bound state of the composite sector, there would be no reason why the operator (\ref{hgg}) should not be present with the coefficient predicted (taking into account the need of one insertion of the EWSB scale) by the power-counting formula (\ref{POWCFull}). This would give $c_\gamma\sim 1$ and thus too a strong effect in the $h\gamma\gamma$ partial width to be compatible with observations, providing further phenomenological support to the idea that the Higgs, if composite, must be a Goldstone boson.

Notice that above argument does not imply that the operator (\ref{hgg}) is exactly absent to all orders in perturbation theory because the Higgs shift symmetry is broken by the coupling of the other EW bosons and that an even stronger breaking comes from the couplings $\lambda_{t_{L,R}}$ with the third family quarks. This breaking will show up in the radiative corrections induced by loops of the elementary gauge and fermionic fields, eventually leading to a loop-suppressed effective operator. We have not yet discussed this kind of effects, however by dimensional analysis it is not hard to understand that they are further suppressed with respect to Eq.~(\ref{POWCFull}) by an additional loop factor $\hbar/16\pi^2$ times the square of the relevant coupling. We thus expect a $c_\gamma$ in Eq.~(\ref{hgg}) which, in Natural Units, is of order
\beq
\displaystyle
c_\gamma\sim \frac{{\lambda_{t_{L,R}}^2}}{16\pi^2}\,.
\eeq 
The same loop suppression factor that is present in the SM also arises in composite Higgs, on top of the $1/m_*^2$ reduction, making the relative correction to $h\gamma\gamma$ well under control.

The other possible $h\gamma\gamma$ vertex, which is the CP-odd operator
\beq
\label{hggcpo}
\epsilon^{\mu\nu\rho\sigma}\frac{{c^\prime}_\gamma e^2 v}{m_*^2} h\,\gamma_{\mu\nu}\gamma_{\rho\sigma}\,,
\eeq
deserves some additional comment. At a first sight one might conclude that it is forbidden by the Higgs shift symmetry exactly like the CP-even one. However thinking more carefully one finds an interesting subtlety. Under a constant shift of $h$, which corresponds to a global \mbox{SO$(5)$} transformation, the CP-odd operator is not invariant but its variation is proportional to $\gamma\tilde{\gamma}$, which is a total derivative. Therefore, at least at the global level and for field configurations with appropriate boundary conditions, the operator leads to an invariant action and is thus allowed in principle. Nevertheless, we did not encounter it in our classification, but this might just be because we are only writing down strictly invariant Lagrangian terms, constructed out of the $d$ and $e$ CCWZ symbols, which respect also the local version of the \mbox{SO$(5)$} group. There is only one term which might not have this structure and generate the CP-odd $h\gamma\gamma$  vertex in Eq.~(\ref{hggcpo}): the gauged Wess--Zumino--Witten term \cite{Wess:1971yu,Witten:1983tw,Chu:1996fr}. This is definitely irrelevant for the minimal composite Higgs model because it turns out that the Wess--Zumino--Witten term does not exist for the minimal coset \mbox{SO$(5)/$SO$(4)$}, but it could lead to interesting phenomenological consequences for non-minimal cosets. Notice that the Wess--Zumino--Witten term is intrinsically linked to the presence of anomalies in the global group. Therefore the theorem according to which $h\gamma\gamma$ is suppressed by the Goldstone symmetry is not violated. The point is that the anomaly actually corresponds to a sizable breaking of the symmetry.

\subsection[Order $p$ fermionic]{Order ${\mathbf{p}}$ fermionic}
\label{pfer}

Let us now come to the fermionic sector, with the purpose of extending to  ${\mathcal{O}}(p)$ the operator classification of Sect.~\ref{sec:HCsigmaM}. The result and its physical implications strongly depend on the  \mbox{SO$(5)\times$U$(1)_X$} quantum numbers of  the fermionic operators that realize partial compositeness; the two options of a spinorial and of a fundamental representation will be discussed in turn.

In order to properly read the implications of the symmetries, the elementary SM fields must be embedded into fermionic source multiplets with the same quantum number of the corresponding operators. In the case of the spinorial $\mathbf{4_{1/6}}$ representation the four embeddings $Q_{t_L}$, $T_R$, $Q_{b_L}$ and $B_R$ are defined in Eq.~(\ref{embspin}). However the bottom sector sources $Q_{b_L}$ and $B_R$ will not be relevant for the present discussion because their couplings $\lambda_{b_{L,R}}$ to the composite sector are typically  small, much smaller than those of the top sector $\lambda_{t_{L,R}}$, as needed to reproduce the small bottom quark mass. The effective operators constructed with the bottom sources are thus power-counting suppressed (\ref{POWCFull}) and negligible in comparison with the ones from the top sources. They can still be relevant, but only if they trigger qualitatively new effects as we will see happening in the case of the fundamental representation. For the spinorial, the relevant objects are thus $Q_{t_L}$ and $T_R$, out of which a total of $4$ CCWZ multiplets can be obtained by dressing the sources with the Goldstone matrix as explained in Sect.~\ref{sec:HCsigmaM}. These are four \mbox{SO$(4)\hspace{-1pt}\simeq$\,SU$(2)_L\times$SU$(2)_R$} doublets \footnote{The definition of these objects is not explicitly reported in Sect.~\ref{sec:HCsigmaM}, it is however completely analogous to the one given in Eq.~(\ref{so4mult}) for the case of the fundamental representation.}
\beq
Q_{t_L}^{\mathbf{2_L}}\,,\;\;\;\;\;Q_{t_L}^{\mathbf{2_R}}\,,\;\;\;\;\;T_R^{\mathbf{2_L}}\,,\;\;\;\;\;T_R^{\mathbf{2_R}}\,,
\eeq
with a common \mbox{U$(1)_X$} charge of $1/6$. In the above equation we denoted, for shortness, the $\mathbf{(2,1)}$ and $\mathbf{(1,2)}$ representations as $\mathbf{2_L}$ and $\mathbf{2_R}$, respectively. We will now classify the invariant operators constructed with two of those sources and one derivative, which can be either carried by the $d_\mu$ symbol or by the covariant derivative $D_\mu$. 

There are four Hermitian operators with $d_\mu$
 \bea
\displaystyle
&&O^{Q}_{+}={\overline{Q}}_{t_L}^{\mathbf{2_L}}\gamma^\mu d_\mu^{(2)}Q_{t_L}^{\mathbf{2_R}}+{\overline{Q}}_{t_L}^{\mathbf{2_R}}\gamma^\mu {d_\mu^{(2)}}^\dagger Q_{t_L}^{\mathbf{2_L}}\,,\nonumber\\
&&O^{Q}_{-,\,1}=i\,{\overline{Q}}_{t_L}^{\mathbf{2_L}}\gamma^\mu d_\mu^{(2)}Q_{t_L}^{\mathbf{2_R}}-i\,{\overline{Q}}_{t_L}^{\mathbf{2_R}}\gamma^\mu {d_\mu^{(2)}}^\dagger Q_{t_L}^{\mathbf{2_L}}\,,\nonumber\\
&&O^{T}_{+}={\overline{T}}_R^{\mathbf{2_L}}\gamma^\mu d_\mu^{(2)}{{T}}_R^{\mathbf{2_R}}+{\overline{T}}_R^{\mathbf{2_R}}\gamma^\mu {d_\mu^{(2)}}^\dagger {{T}}_R^{\mathbf{2_L}}\,,\nonumber\\
&&O^{T}_{-,\,1}=i\,{\overline{T}}_R^{\mathbf{2_L}}\gamma^\mu d_\mu^{(2)}{{T}}_R^{\mathbf{2_R}}-i\,{\overline{T}}_R^{\mathbf{2_R}}\gamma^\mu {d_\mu^{(2)}}^\dagger {{T}}_R^{\mathbf{2_L}}\,,
\eea
expressed in terms of the $d_\mu$ symbol in the $2\times 2$ matrix representation of Appendix~(\ref{so54}) (see Eq.~(\ref{matrixnot})), which is typically the most convenient one when dealing with the spinorial representation. The operator labeling refers to their intrinsic $P_{LR}$ parity, which is easily worked out by noticing that $d_\mu^{(2)}$ transforms into its conjugate while the two doublets get interchanged by $P_{LR}$ as shown Appendix~\ref{discrete}. The CP quantum number will instead not be specified for the fermionic operators. 

By acting with a covariant derivative, four more operators can be formed 
 \bea
\displaystyle
&&R^{Q}_{+}=i\,{\overline{Q}}_{t_L}^{\mathbf{2_L}}\gamma^\mu D_\mu Q_{t_L}^{\mathbf{2_L}}+i\,{\overline{Q}}_{t_L}^{\mathbf{2_R}}\gamma^\mu D_\mu Q_{t_L}^{\mathbf{2_R}}\,,\nonumber\\
&&O^{Q}_{-,\,2}=i\,{\overline{Q}}_{t_L}^{\mathbf{2_L}}\gamma^\mu D_\mu Q_{t_L}^{\mathbf{2_L}}-i\,{\overline{Q}}_{t_L}^{\mathbf{2_R}}\gamma^\mu D_\mu Q_{t_L}^{\mathbf{2_R}}\,,\nonumber\\
&&R^{T}_{+}=i\, {\overline{T}}_R^{\mathbf{2_L}}\gamma^\mu D_\mu {{T}}_R^{\mathbf{2_L}}+i\,{\overline{T}}_R^{\mathbf{2_R}}\gamma^\mu D_\mu {{T}}_R^{\mathbf{2_R}}\,,\nonumber\\
&&O^{Q}_{-,\,2}=i\,{\overline{T}}_R^{\mathbf{2_L}}\gamma^\mu D_\mu {{T}}_R^{\mathbf{2_L}}-i\,{\overline{T}}_R^{\mathbf{2_R}}\gamma^\mu D_\mu {{T}}_R^{\mathbf{2_R}}\,.
\eea
Notice that the covariant derivative $D_\mu$ acting on the fermionic fields is not only given by the CCWZ one we introduced in Sect.~\ref{CCWZ} (see for instance Eq.~(\ref{gencd})). The latter takes care of the non-linearly realized \mbox{SO$(5)$} but other terms are needed to account for the unbroken \mbox{U$(1)_X$} and \mbox{SU$(3)_c$} color local groups, involving the corresponding sources, under which the fermionic fields are charged. The complete covariant derivatives are
\bea
&&\displaystyle D_\mu Q_{t_L}^{\mathbf{2_L}} = \left[\partial_\mu -i\, {e_L^{(2)}}_\mu -i\, \frac16 X_\mu -i\, g_S\,G_\mu\right]Q_{t_L}^{\mathbf{2_L}}\,,\nonumber\\
&& \displaystyle D_\mu Q_{t_L}^{\mathbf{2_R}} = \left[\partial_\mu -i\, {e_R^{(2)}}_\mu -i\, \frac16 X_\mu -i\, g_S\,G_\mu\right]Q_{t_L}^{\mathbf{2_R}}\,,
\eea
and analogously for $T_R$. The $1/6$ factor reflects the  \mbox{U$(1)_X$} charge of the fields and $G_\mu=G_\mu^a \lambda_a/2$ are the gluon fields in the matrix notation, acting on the triplet color index of the source. The $2\times2$ matrix representation for the $e_\mu$ symbol, decomposed in left and right components as in Appendix~\ref{so54}, has been used to express the CCWZ part of the covariant derivative. 

The two operators $R^{Q}_{+}$ and $R^{T}_{+}$ are redundant, as the notation suggests, and by following the same logic of the previous section it is easy to understand why. The point is that there exist two operators, constructed by working directly with the source fields rather than with the dressed ones, which have a trivial effect on the theory. These two operators are \footnote{The covariant derivative here is acting on the sources, which transform linearly under \mbox{SO${(5)}$} and not like the CCWZ objects do. Therefore $D_\mu=\partial_\mu-i\,A_\mu-i\,Q_X X_\mu-i\,G_\mu$, with no $e_\mu$ symbol appearing.}
\bea
\label{kin_term}
&&\displaystyle i\,{\overline{Q}}_{t_L}\gamma^\mu D_\mu Q_{t_L} = i\, {\overline{q}}_L\gamma^\mu D_\mu q_L\,,\nonumber\\
&&\displaystyle i\,{\overline{T}}_R\gamma^\mu D_\mu T_R = i\, {\overline{t}}_R\gamma^\mu D_\mu t_R\,,
\eea
and they just reduce, when the fermionic and gauge sources are restricted to their physical values, to corrections of the elementary quark fields kinetic terms. Such corrections can be  eliminated by a rescaling and reabsorbed in an unobservable redefinition of the elementary/composite interaction couplings $\lambda_{t_L}$ and $\lambda_{t_R}$. The ones above are not new operators, but instead they are linear combinations of $R^{Q,T}_{+}$ and $O^{Q,T}_{+}$. This is immediately verified by inverting the definition of the dressed sources obtaining 
\bea
&&Q_{t_L}=U_{\mathbf{4}}[\Pi]\cdot \{Q_{t_L}^{\mathbf{2_L}} ,\,Q_{t_L}^{\mathbf{2_R}} \}^T\,,\nonumber\\
&&T_R=U_{\mathbf{4}}[\Pi]\cdot \{T_R^{\mathbf{2_L}} ,\,T_R^{\mathbf{2_R}} \}^T\,,
\eea
where $U_{\mathbf{4}}$ is the Goldstone matrix in the spinorial we derived in Appendix~\ref{so54}. By plugging into Eq.~(\ref{kin_term}), remembering that the definition of $d$ and $e$ (\ref{dedefgauge}) applies to any representation including the spinorial, we  obtain 
\bea
&&\displaystyle i\,{\overline{Q}}_{t_L}\gamma^\mu D_\mu Q_{t_L} = R_+^Q+\frac1{2\sqrt{2}} O_+^Q\,,\nonumber\\
&&\displaystyle i\,{\overline{T}}_R\gamma^\mu D_\mu T_R = R_+^T+\frac1{2\sqrt{2}} O_+^T\,,
\eea
showing that $R^{Q,T}_+$ are redundant as anticipated.

The operators above trigger a number of interesting physical effects, among which further modifications of the Higgs coupling to fermions on top of those outlined in Sect.~\ref{sec:HCsigmaM} and corrections to the $W$-$t$-$b$ vertex $V_{tb}$. Here we will limit ourselves to discuss their effect on the $Z$ boson coupling to the left-handed bottom, $g_{b_L}$, which was precisely measured at LEP. We focus  on $O_{+}^{Q}$, which is the simpler to deal with. Other contributions of the same order are expected from $O_{-,2}^{Q}$ but not from  $O_{-,1}^{Q}$ which is CP-odd. The corresponding $T_R$ source operators modify the top coupling to the $Z$, which however is not yet measured well enough to be relevant. After going to the unitary gauge and setting the sources to their physical values we find
\beq
O_{+}^{Q}=\frac{g}{\sqrt{2}c_w}\xi\,\overline{b}_L\gamma^\mu Z_\mu b_L\,+\,\ldots \,,
\eeq
which sums up with the SM vertex which originates from the elementary kinetic term in Eq.~(\ref{elm_q_kin}), leading to corrections.

In order to quantify these corrections we rely on the \mbox{$1$S$1$C} power counting (\ref{POWCFull}) estimate of the operator coefficient, which in this case leads to two powers of the $\lambda_{t_L}$ coupling from the two source insertions, a prefactor of $1/g_*^2$ and no  $m_*$ since the operator has dimension four. Therefore we have 
\beq
\label{glcorr}
\delta g_{b_L}\simeq\left(\frac{{\lambda_{t_L}}}{g_*}\right)^2\xi\,,
\eeq
having stripped out a $g/c_w$ factor in the definition of the coupling as customary in the literature (for more details, see Sect.~\ref{sec:Zbb}). This correction is experimentally bounded at the per-mille level, implying a constraint 
\beq
\xi\lesssim \left(\frac{g_*}{{\lambda_{t_L}}}\right)^2\cdot 10^{-3}\,.
\eeq

This limit can be more or less stringent, depending on the size of $\lambda_{t_L}$. However $\lambda_{t_L}$ controls the generation of the top quark Yukawa as in Eq.~(\ref{yukpar}) (in accordance with the more quantitative results of Sect.~\ref{sec:HCsigmaM}), namely
\beq
\label{y_trep}
y_t\simeq\frac{\lambda_{t_L} \lambda_{t_R}}{g_*}\,,
\eeq
therefore it is not a completely free parameter. The two options which is worth considering are
\beq
\displaystyle
\label{cases_comp}
{\textrm{(\Rmnum{1})}}\;\;\lambda_{t_R}\simeq \lambda_{t_L} \simeq \sqrt{y_t g_*}\,,\;\;\;\;\;\;\;\;\;\;\;\;{\textrm{(\Rmnum{2})}}\;\;\lambda_{t_R}\simeq g_*,\, \lambda_{t_L} \simeq y_t \,,
\eeq
for which the constraint, using $g_*<4\pi$ and $y_t\simeq1$, becomes
\beq
{\textrm{(\Rmnum{1})}}\;\;\xi\lesssim \frac{g_*}{y_t} \cdot 10^{-3}< 1.3 \cdot 10^{-2} \,,\;\;\;\;\;\;\;\;\;\;\;\;{\textrm{(\Rmnum{2})}}\;\;\xi\lesssim \frac{g_*^2}{y_t^2} \cdot 10^{-3}< 0.16\,.
\label{zbblimit}
\eeq
Further lowering $\lambda_{t_L}$ unfortunately is not an option, so that the coupling pattern $({\textrm{\Rmnum{2}}})$ is already the most favorable one, because of two reasons. First, by applying Eq.~(\ref{y_trep}) we would conclude that $\lambda_{t_L}<y_t$ requires $\lambda_{t_R}>g_*$, {\it{i.e.}}~that an elementary coupling becomes larger than the composite one. This goes against our basic philosophy, outlined already in the Introduction, according to which the elementary sector must be a mild deformation of the composite one, characterized by a weaker interaction strength. Second, and more concretely, we notice that Eq.~(\ref{y_trep}) is an approximate formula that looses its validity when $\lambda_{t_{L,R}}>g_*$. In this case the more complete expression for $y_t$ in Eqs.~(\ref{yukpar}) and (\ref{angles}) should be considered, which displays how keeping raising $\lambda_{t_R}$ above $g_*$ doesn't help because the compositeness fraction $\sin\theta^t_R$ saturates to one. From the complete formula one can derive the upper limit $\lambda_{t_L}<y_t$, which is already saturated with the choice $({\textrm{\Rmnum{2}}})$ showing that no improvement is possible.

The bound from $\delta{g}_{b_L}$ is extremely strong in case $({\textrm{\Rmnum{1}}})$ of Eq.~(\ref{zbblimit}), it requires two digits of unnatural cancellation in $\xi$, but it seems tolerable in case $({\textrm{\Rmnum{2}}})$, not far from the one obtainable from $\widehat{S}$ (\ref{Shatest}) for maximal $g_*$. However the estimate $({\textrm{\Rmnum{2}}})$ is too optimistic, for two reasons. First, the favored coupling pattern is not $({\textrm{\Rmnum{2}}})$ but $({\textrm{\Rmnum{1}}})$, because of the structure of the Higgs potential, which will be discussed in the following section, and the need of obtaining a realistic VEV for EWSB. Second, the study of the Higgs potential will also reveal that $g_*$ should not be maximal for the Higgs to be naturally light. The limits in Eq.~(\ref{zbblimit}) rapidly deteriorate for $g_*<4\pi$, carrying the model into the unnatural regime.

The strong bound on $\delta{g}_{b_L}$, which we have seen emerging for fermionic operators in the spinorial representation, leads us to discard this possibility and to search for alternatives where $\delta{g}_{b_L}$ corrections are smaller. Interestingly enough, it is sufficient to go to the next-to-minimal representation, the fundamental, to dramatically improve the situation.\footnote{The concept of minimality based on the dimension of the representations is rather questionable. From the viewpoint of a strongly-coupled microscopic theory it is hard to tell what is ``minimal'' or ``easier'' to be realized. In the case at hand, it is enough to have composite sector constituents not living in the spinorial, but only in representations with  congruency class (see e.g.~Ref.~\cite{Feger:2012bs}) equal to zero, for not being capable to form composite operators in the spinorial representation and being forced to consider alternatives.} In this case, two sources $Q_{t_L}$ and $T_R$ are introduced in the top sector, both in the $\mathbf{5}_{2/3}$, and used to form four dressed source in the $\mathbf{4}_{2/3}$ and in the $\mathbf{1}_{2/3}$. The complete set of ${\mathcal{O}}(p)$ operators reads
 \bea
\displaystyle
&&O^{Q}_{+,\,1}=({\overline{Q}}_{t_L}^{\mathbf{4}})_i\gamma^\mu d_\mu^i Q_{t_L}^{\mathbf{1}}+{\overline{Q}}_{t_L}^{\mathbf{1}}\gamma^\mu d_\mu^i (Q_{t_L}^{\mathbf{4}})_i\,,\nonumber\\
&&O^{Q}_{+,\,2}=i\,({\overline{Q}}_{t_L}^{\mathbf{4}})_i\gamma^\mu d_\mu^i Q_{t_L}^{\mathbf{1}}-i\,{\overline{Q}}_{t_L}^{\mathbf{1}}\gamma^\mu d_\mu^i (Q_{t_L}^{\mathbf{4}})_i\,,\nonumber\\
&&O^{Q}_{+,\,3}=i\,{\overline{Q}}_{t_L}^{\mathbf{1}}\gamma^\mu D_\mu Q_{t_L}^{\mathbf{1}}\,,
\label{op5}
\eea
plus the corresponding terms with $T_R$. In the above equation we have already dropped two redundant operators, associated with the source kinetic terms.

It is easy to verify explicitly that none of the operators in Eq.~(\ref{op5}) contributes to $\delta{g}_{b_L}$, neither directly by a \mbox{$Z$-$b$-$\overline{b}$} vertex nor indirectly through a correction of the $b_L$ kinetic term. Large effects like the ones in Eq.~(\ref{glcorr}) are thus absent in the case of the fundamental representation. Corrections to ${g}_{b_L}$ will emerge at higher orders in the derivative or loop expansion and, as we will see below, from operators involving the bottom sector sources. The cancellation of $\delta{g}_{b_L}$  appears miraculous at this point, however it can be nicely understood by symmetries. This has to do with the fact that all the operators we found are accidentally $P_{LR}$ even, and to the fact that $P_{LR}$ is not broken by the $b_L$ embedding into the sources. Namely, if we look at Eq.~(\ref{qlemb}) and compare it with Eq.~(\ref{plrfund}) we see that $P_{LR}$ is preserved if $b_L$ is regarded as an odd field. The $t_L$ embedding does instead break $P_{LR}$ but this will not show up in the \mbox{$Z$-$b$-$\overline{b}$} vertex at this order. $P_{LR}$, as we will see in Appendix~\ref{sec:custodial_symmetries}, is not even broken by the Higgs VEV, it survives in the low-energy bottom Lagrangian and it ensures $\delta{g}_{b_L}$ cancellation in a way that is similar to the cancellation of the $\widehat{T}$ oblique parameter due to custodial \mbox{SO$(3)_c$}. For this reason, $P_{LR}$ is called a custodial symmetry for \mbox{$Z$-$b$-$\overline{b}$} \cite{Agashe:2006at}. The situation is different for the operators constructed with the bottom sector sources $Q_{b_L}$ in which, as Eq.~(\ref{qlbemb}) shows, the $b_L$ embedding breaks $P_{LR}$. One operator that induces $\delta{g}_{b_L}$ is, for instance 
\beq
({\overline{Q}}_{b_L}^{\mathbf{4}})_i\gamma^\mu d_\mu^i Q_{b_L}^{\mathbf{1}}+{\overline{Q}}_{b_L}^{\mathbf{1}}\gamma^\mu d_\mu^i (Q_{b_L}^{\mathbf{4}})_i= \frac{g}{\sqrt{2}c_w}\xi\,\overline{b}_L\gamma^\mu Z_\mu b_L\,+\ldots
\label{bsex}
\eeq
Notice that the correction emerges in spite of the fact that the operator is $P_{LR}$ even, the reason being that the embedding breaks the symmetry as previously explained.

Corrections coming from the operators constructed with the bottom sources like the one in Eq.~(\ref{bsex}) are much less dangerous than those from the top sector because, according to the power counting in Eq.~(\ref{POWCFull}), they are suppressed by the bottom coupling. The correction is now estimated to be
\beq
\delta g_{b_L}\simeq\left(\frac{{\lambda_{b_L}}}{g_*}\right)^2\xi\,,
\eeq
where the parameter $\lambda_{b_L}$ can easily be much smaller than $\lambda_{t_L}$ as it serves to give the mass to the bottom quark and not to the top. Assuming for simplicity $\lambda_{b_L}\sim \lambda_{b_R}$, the estimate in Eq.~(\ref{yukpar}) gives now $\lambda_{b_L}\sim\sqrt{y_b g_*}$ so that
\beq
\delta g_{b_L}\simeq\frac{{y_b}}{g_*}\xi\simeq  \frac{\xi}{g_*} \, 3 \cdot 10^{-2} \,.
\eeq
Even for a rather small $g_*\sim 2$ and mild tuning $\xi\sim 0.1$ the correction is close to the experimental bound of $10^{-3}$. The effect is thus small enough to be under control, but still potentially relevant for a quantitative compatibility with EWPT. A careful discussion of this and other contributions to $\delta g_{b_L}$, in the context of explicit models where the effects can be calculated and not just estimated as we did here, is postponed to Sect.~\ref{sec:Zbb}.

This concludes our description of the ${\mathcal{O}}(p)$ fermionic Lagrangian in the case of the spinorial and of the fundamental representations. This is representative of the generic situation and straightforwardly generalizable to other representations. On top of illustrating the operator classification technique, the discussion led us to a phenomenological criterion to identify the most suitable representations for the fermionic operators. They have to be such that large corrections to $g_{b_L}$ from the top sector sources operators are forbidden, like for the fundamental representation and unlike the spinorial. Even if we postponed the discussion of this point to Appendix~\ref{sec:custodial_symmetries}, we have anticipated that the cancellation relies on the $P_{LR}$ accidental symmetry of the ${\mathcal{O}}(p)$ operators. Concerning the other representations we considered in Sect.~\ref{sec:HCsigmaM}, it turns out that the case of a ${\mathbf{14}}_{2/3}\oplus{\mathbf{1}}_{2/3}$ behaves like the fundamental and $\delta g_{b_L}$ is suppressed in exactly the same way. The case of a ${\mathbf{10}}$ is slightly different because $P_{LR}$ does not emerge automatically as an accidental symmetry, but it needs to be imposed as a symmetry of the composite sector. If this is done the cancellation holds, at the price of an additional assumption which instead is not needed in the other cases.

\section{The composite Higgs potential}
\label{CHPOT}

The origin and the structure of the Higgs potential, out of which the composite Higgs field acquires a VEV producing EWSB and the Higgs particle acquires a mass, is the last subject to be discussed in the present chapter. The generation of the potential is definitely a phenomenon that goes ``beyond the sigma-model'', though for a slightly different reason than the effects previously discussed in this chapter. Rather than from higher-derivative operators, it is due to lower (zero) derivative ones which were forbidden in the $\sigma$-model by the Goldstone symmetry. Rather than studying the implications of the Goldstone symmetry group ${\mathpzc{G}}$ on higher derivative operators we must now study the implications of the explicit breaking of the symmetry on the operators of lowest possible derivative order. Notice that the seeds of such breaking have been already introduced in the theory, in the form of the couplings of the elementary sector fields to the composite sector operators. Given that the former fields do not come in multiplets of the ${\mathpzc{G}}$ group, their couplings (with few remarkable exceptions) break the symmetry and thus are capable to trigger the generation of the Higgs potential.

The discussion is organized in two steps. First, in Sects.~\ref{CHPCH} and \ref{pwcpot_sec}, we will outline the technical and conceptual tools through which the operators contributing to the potential can be classified and estimated within our assumptions, specified in Sect.~\ref{OSOC}, on the dynamics of the composite sector. Provided the elementary sector couplings are weak, much weaker than the composite sector interaction strength $g_*$, the ${\mathpzc{G}}$ group is a valid and predictive approximate symmetry, with powerful implications on the allowed form the potential can assume. At the second step, in Sect.~\ref{HVMT}, we will discuss how and at what price, in terms of fine-tuning, a phenomenologically viable potential can be obtained, giving rise to realistic EWSB and to the correct mass for the Higgs boson. All these aspects will be illustrated in the case of the minimal coset where \mbox{${\mathpzc{G}}=\textrm{SO}(5)$} and \mbox{${\mathpzc{H}}=\textrm{SO}(4)$}, however they are of general validity and straightforwardly generalizable to non-minimal cosets.

\subsection{Higgs potential characterized}
\label{CHPCH}
Given that the Higgs potential can only come from the explicit breaking of the Goldstone symmetry, the right starting point for the discussion is to identify the interactions that realize this breaking in our construction. The composite sector is perfectly invariant: as discussed at length, the spontaneous symmetry breakdown merely realizes the symmetry in a non-linear fashion, it doesn't really break it. Explicit breaking only comes from the elementary sector and the way in which this breaking is communicated to the composite sector, where the Higgs field originates, is from the various elementary/composite interactions we have in our theory. In particular, we first focus on the gauge interactions specified in Eqs.~(\ref{gscoup}), (\ref{physg}) and (\ref{Xsv}). These read
\beq
\label{lint}
{\mathcal{L}}_{\textrm{int}}^{\textrm{gauge}}=g\,W_{\mu,\,\alpha}J_L^{\mu\,,\alpha}+g'B_\mu J_R^{\mu,\,3}+g'B_\mu J_X^{\mu}.
\eeq
Until now we found convenient to collect the physical $W$ and $B$ fields into the sources $A_\mu^A$ and $X_\mu$ and to carry all their components, including the non-physical ones, in the operator classification. We will now instead restrict to the physical fields from the very beginning, finding another way, namely the method of spurions, to study the implications of the Goldstone symmetry.

Let us first consider the terms in Eq.~(\ref{lint}) that involve the hypercharge field $B_\mu$ and the corresponding coupling $g'$. The last term is clearly harmless because the \mbox{U$(1)_X$} current $J_X^\mu$ is an \mbox{SO$(5)$} singlet. Coupling it to $B_\mu$ does not break the Goldstone symmetry and therefore that term is not capable to generate the potential and it can be ignored in the present discussion. The other term instead breaks the symmetry, but we can rewrite it in the formally invariant fashion 
\beq
\label{spuriongprime}
{\mathcal{L}}_{g'}^{\slash\hspace{-5pt}{\mathpzc{G}}}=g'B^\mu J^{3}_{R,\,\mu}\equiv B^\mu {{\mathcal{G}}'}_AJ_\mu^A\,,
\eeq
where $A$ runs over the $10$ components of the \mbox{SO$(5)$} global current multiplet and ${{\mathcal{G}}'}_A$ is what we call a ``spurion''. The one above is a mere rewriting of the interaction Lagrangian. Therefore the only non-vanishing entry of the ${{\mathcal{G}}'}$ vector is the one associated with the generator $T_R^3$, and it is equal to $g'$. By switching to a matrix notation, the physical value assumed by the spurion is expressed as
\beq
\label{spurgprimephys}
{{\mathcal{G}}'}\equiv {{\mathcal{G}}'}_AT^A=g'T_R^3\,.
\eeq

The rewriting in Eq.~(\ref{spuriongprime}) becomes useful if we imagine for a moment to promote all the $10$ spurion components to independent parameters, making ${{\mathcal{G}}'}$ a generic adjoint matrix of couplings. Any result we might obtain in this extended theory will reproduce the ones we are interested in once the coupling matrix is restricted to its physical value in Eq.~(\ref{spurgprimephys}). The advantage of this approach is that drawing the implications of the \mbox{SO$(5)$} symmetry is much simpler in the extended theory than in the original one, because the former is \mbox{SO$(5)$} invariant if we assign to the spurion a transformation property \footnote{Here and in what follows we only deal with the global version of the group, differently from the previous analyses where we considered its uplift to a local invariance by making it act also on the gauge sources.}
\beq
\label{spuriongprimetrans}
{{\mathcal{G}}'}\;\rightarrow\; g\cdot {{\mathcal{G}}'}\cdot g^\dagger\,,\;\;\;\;\;g\in{\textrm{SO}}(5)\,.
\eeq
The spurion transformation, in the adjoint representation, compensates for the current operators transformation making Eq.~(\ref{spuriongprime}) invariant. Specifically, this means that by acting with a symmetry transformation on any correlator computed within the extended theory, with a certain value ${{\mathcal{G}}'}$ of the coupling matrix, we will end up with the result we would have obtained within the same theory but with a coupling matrix rotated according to Eq.~(\ref{spuriongprimetrans}).\footnote{More precisely, if we act with $g$ on the correlators the result we get is the one obtained with the spurion matrix rotated by the inverse transformation $g^{-1}$.} Therefore all the correlators, and in turn all the physical quantities we might extract out of them, must depend on the spurion matrix in a special way, such as to stay invariant under the simultaneous action of \mbox{SO$(5)$} on the field variables and on the spurion. This is to say that we can still use \mbox{SO$(5)$} as a symmetry to classify the operators. The symmetry of course is broken, but only at the very end of the calculation when the spurion is set to its physical value in Eq.~(\ref{spurgprimephys}), which is not \mbox{SO$(5)$}-invariant.

An alternative, pedagogically valid but potentially misleading interpretation of the method of spurions goes as follows. We can imagine ${{\mathcal{G}}'}$  being a collection of scalar fields in the adjoint representation, artificially introduced in the theory and coupled to $B_\mu$ and $J_\mu$ as in Eq.~(\ref{spuriongprime}). Through a suitably designed scalar potential we might give it a VEV that equals the physical value of the spurion in Eq.~(\ref{spurgprimephys}). On this vacuum, the original $B$--$J$ interaction is reproduced. In view of this interpretation, the physical value of the spurion is sometimes called the ``spurion VEV''. Notice that this interpretation should be taken with care, keeping clear in mind that the validity of the spurion method relies in no way on the physical existence of new extra scalar degrees of freedom, neither of light ones which appear explicitly in the low-energy effective theory nor of heavy ones above the cutoff. 

Now that the foundations of the spurion method are clear, we come to its application. The goal is to classify the operators constructed with the spurions and the Goldstone bosons (plus possibly other physical fields of the theory) that are invariant under the non-linearly realized \mbox{SO$(5)$} symmetry. We can make this classification systematic by introducing the concept of ``dressed spurion'', in full analogy with the one of dressed sources we discussed in Chap.~\ref{ch:Goldstone}. Just as in the latter case, the basic observation is that the Goldstone matrix transforms with $g\in{\textrm{SO}}(5)$ on the left and with $h\in{\textrm{SO}}(4)$ on the right, namely
\beq
U\;\rightarrow\; g\cdot U\cdot h^{-1}\,.
\eeq
Therefore by acting with $U^{-1}$ we can turn an object with ${\textrm{SO}}(5)$ indices into a dressed one, which transforms with the CCWZ ${\textrm{SO}}(4)$ matrix $h$. The dressed object can be split into several ${\textrm{SO}}(4)$ irreducible representations, according to the decomposition in ${\textrm{SO}}(4)$ of the original ${\textrm{SO}}(5)$ multiplet. The dressed ${{\mathcal{G}}'}$ spurion is 
\beq
\label{dressedSP}
{{\mathcal{G}}'}_D\equiv U^\dagger \cdot {{\mathcal{G}}'}\cdot U\,,
\eeq
and it decomposes as 
\beq
\label{gpspdec}
{\mathbf{10}}={\mathbf{(3,1)}}\oplus{\mathbf{(1,3)}}\oplus{\mathbf{(2,2)}}\,,
\eeq
into three separate ${\textrm{SO}}(4)$ multiplets. Explicitly, these are
\bea
\label{gpspmult}
&&({{\mathcal{G}}'}_D^{\mathbf{3_L}})_{\alpha_L}\equiv {\textrm{Tr}}[{{\mathcal{G}}'}\cdot T_{L,\,{\alpha_L}}]\,,\nonumber\\
&&
({{\mathcal{G}}'}_D^{\mathbf{3_R}})_{\alpha_R}\equiv {\textrm{Tr}}[{{\mathcal{G}}'}\cdot T_{R,\,{\alpha_R}}]\,,\nonumber\\
&&
({{\mathcal{G}}'}_D^{\mathbf{4}})_i\equiv {\textrm{Tr}}[{{\mathcal{G}}'}\cdot \Th_{i}]\,.
\eea
In the absence of extra symmetries, each multiplet can be employed independently in the construction of the invariant operators.

Let us now specialize our discussion to the Higgs potential. It is made of operators involving the Goldstone matrix, no derivatives, plus of course some insertion of the spurions that carry the explicit breaking of the Goldstone symmetry. The physical value of the spurion ${{\mathcal{G}}'}$ is proportional to the hypercharge coupling $g'$, which is a ``weak'' coupling. We thus intuitively expect that we might be allowed to work in a $g'$ expansion, with the largest contribution to the potential coming from the operators with the smallest number of spurions. This intuition will be confirmed, later in the present section, by the power-counting estimate of the operator coefficients. There are no operators of ${\mathcal{O}}(g')$, because of two independent reasons. The first one is that no singlet is present in the decomposition (\ref{gpspdec}) of the dressed spurion under \mbox{SO$(4)$}. If a singlet appeared, as it would in non-minimal cosets with \mbox{U$(1)$} factors in the unbroken group, we might have taken it as a candidate operator. In the current situation instead two or more spurion multiplets have to be multiplied to form and invariant. The second reason has to do with a $\Zdouble_2$ symmetry of the elementary sector, the operation $B_\mu\rightarrow - B_\mu$ on the hypercharge gauge field.\footnote{The breaking of this symmetry due to the coupling with the elementary quarks plays no role in this discussion. It would become relevant only if we had to discuss mixed contributions to the potential from both the quark and the gauge field spurions.} This can be taken to act on the spurion as
\beq
{{\mathcal{G}}'}\;\rightarrow\;-\,{{\mathcal{G}}'}\,,
\eeq
while leaving invariant the whole composite sector and thus in particular the $J_\mu$ current. Being a symmetry of the interaction (\ref{spuriongprime}), this ``elementary'' $\Zdouble_2$ parity must be respected by the operators, very much like the composite sector ${\textrm{SO}}(5)$ group. No operator with and odd number of ${{\mathcal{G}}'}$'s should thus be considered, not even if ${\textrm{SO}}(5)$ invariant. The general idea is that the spurion is an object that ensures the communication among two sectors: the elementary and the composite one. As such it can inherit symmetries from both sides, leading to extra selection rules. Other examples of elementary symmetries will emerge in the forthcoming discussion.

Three operators can be written at ${\mathcal{O}}({g}^{\prime\,2})$, corresponding to the ${\textrm{SO}}(4)$ singlets one can form with the three irreducible components of the dressed spurion in Eq.~(\ref{gpspmult}). These are conveniently expressed as
\bea
&& O_{{{g}^{\prime\,2}} +}=({{\mathcal{G}}'}_D^{\mathbf{4}})_i ({{\mathcal{G}}'}_D^{\mathbf{4}})^i\,,\nonumber\\
&& O_{{{g}^{\prime\,2}} -}=({{\mathcal{G}}'}_D^{\mathbf{3_L}})_{\alpha_L} ({{\mathcal{G}}'}_D^{\mathbf{3_L}})^{\alpha_L}-
({{\mathcal{G}}'}_D^{\mathbf{3_R}})_{\alpha_R} ({{\mathcal{G}}'}_D^{\mathbf{3_R}})^{\alpha_R}\,,\nonumber\\
&& R_{{{g}^{\prime\,2}} +}=({{\mathcal{G}}'}_D^{\mathbf{3_L}})_{\alpha_L} ({{\mathcal{G}}'}_D^{\mathbf{3_L}})^{\alpha_L}+
({{\mathcal{G}}'}_D^{\mathbf{3_R}})_{\alpha_R} ({{\mathcal{G}}'}_D^{\mathbf{3_R}})^{\alpha_R}\,,
\eea
with $O_{{{g}^{\prime\,2}} -}$ being odd and the others even under $P_{LR}$. As usual, when dealing with dressed objects attention must be paid to possible redundancies, associated with combinations of operators which are ``too symmetric'' to be relevant. In the present case $R_{{{g}^{\prime\,2}} +}$  is redundant, because of the identity
\beq
\sum\limits_{A=1}^{10}\left\{
{\textrm{Tr}}[{{\mathcal{G}}'}\cdot T^A]
\right\}^2=
\sum\limits_{A=1}^{10}\left\{
{\textrm{Tr}}[{{\mathcal{G}}'}_D\cdot T^A]
\right\}^2=R_{{{g}^{\prime\,2}} +}+O_{{{g}^{\prime\,2}} +}\,.
\eeq
The first term of the equality is independent of the Higgs, thus it does not contribute to the potential but only to the vacuum energy. It is equals to the second one (which in turn is immediately rewritten as the sum of the two even operators) because it is invariant under a fictitious linearly-realized ${\textrm{SO}}(5)$ and the dressing procedure precisely amounts to a rotation in this group. Being proportional to $O_{{{g}^{\prime\,2}}+}$ up to a constant, $R_{{{g}^{\prime\,2}}+}$ is redundant and can be eliminated from the classification. Notice that the redundancy is associated with the ${\textrm{SO}}(5)$ singlet one can form by employing directly the spurion, rather than its dressed version. Given that a unique singlet can be formed by two powers of ${{\mathcal{G}}'}\in{\mathbf{10}}$, only one redundancy is expected at ${\mathcal{O}}({g}^{\prime\,2})$, the one outlined above.

The power and the simplicity of our result is better appreciated when we set the spurion to its physical value in Eq.~(\ref{spurgprimephys}) and we go to the unitary gauge in Eq.~(\ref{gmug}), obtaining
\beq
\label{gaugeinv1}
O_{{{g}^{\prime\,2}} +}=\frac{g^{\prime\,2}}{2}\sin^2\frac{H}{f}\,,\;\;\;\;\;
O_{{{g}^{\prime\,2}} -}={g^{\prime\,2}}\cos\frac{H}{f}\,,
\eeq
with $H=V+h$ denoting the real neutral Higgs field component (times $\sqrt{2}$). We have found that the ${\mathcal{O}}({g}^{\prime\,2})$ potential is a linear combination of these two operators, therefore its functional form is fixed, up to numerical coefficients, to be
\beq
\label{gprimepot}
V_{{{g}^{\prime\,2}}}\propto c_{+} \cdot \frac{g^{\prime\,2}}{2}\sin^2\frac{H}{f}+ c_{ -}\cdot {g^{\prime\,2}}\cos\frac{H}{f}\,.
\eeq
If the composite sector happens to be invariant under $P_{LR}$, the result would further simplifies since $c_{ -}=0$.

The classification is straightforwardly extended to other sources of explicit breaking, starting from the one associated with the gauging of the SM ${\textrm{SU}}(2)_L$ subgroup, {\it{i.e.}}~from the $W$--$J_L$ term in Eq.~(\ref{lint}). We rewrite the latter as 
\beq
\label{spuriong}
{\mathcal{L}}_{g}^{\slash\hspace{-5pt}{\mathpzc{G}}}=g W_{\mu,\,\alpha} J_{L}^{{\mu,\,\alpha}}\equiv W_{\mu,\,\alpha} {\mathcal{G}}^\alpha_A J^{\mu,\,A}\,,
\eeq
in terms of three spurions in the adjoint, with physical value
\beq
\label{spurgphys}
{{\mathcal{G}}^\alpha}\equiv {{\mathcal{G}}}_A^\alpha T^A=g\, T_L^\alpha\,,
\eeq
where $\alpha=1,2,3$. Just like we did for ${{\mathcal{G}}'}$, dressed spurions can be defined as in Eq.~(\ref{dressedSP}), decomposed in \mbox{SO$(4)$} multiplets as in Eq.~(\ref{gpspmult}) and eventually used to construct invariants. The only difference with the previous case is that now we have three spurions rather than one, all of them are in the same representation. One might thus construct singlets by pairing them in all possible combinations, leading to a proliferation of the number of invariants. However one must also take into account the existence of an \mbox{SU$(2)$} global symmetry of the elementary sector, call it \mbox{SU$(2)_E$}, under which the $W^\alpha$'s form a triplet while the composite sector is invariant. This is nothing but the global version of the local \mbox{SU$(2)_L$} restricted to act on the elementary fields only, under which the elementary sector Lagrangian in isolation is exactly invariant. Analogously to the elementary $\Zdouble_2$ discussed above, \mbox{SU$(2)_E$} can be formally uplifted to a symmetry of the interactions by assigning suitable transformation rules to the spurions, namely by taking the index $\alpha$ in the triplet.\footnote{The spurion VEV breaks \mbox{SU$(2)_E$} to its diagonal combination with the \mbox{SU$(2)_L$} subgroup of \mbox{SO$(5)$} and this latter unbroken symmetry is the SM \mbox{SU$(2)_L$}. Imposing the spurionic \mbox{SU$(2)_E$} automatically ensures the invariance of the potential under the SM group even after the spurions are set to their physical values.} This symmetry must be also respected, on top of the \mbox{SO$(5)$} group, therefore the only allowed operators are 
\bea
&& O_{{{g}^{2}} +}=\sum\limits_{\alpha=1}^3({{\mathcal{G}}^\alpha}_D^{\mathbf{4}})_i ({{\mathcal{G}}^\alpha}_D^{\mathbf{4}})^i\,,\nonumber\\
&& O_{{{g}^{2}} -}=\sum\limits_{\alpha=1}^3({{\mathcal{G}}^\alpha}_D^{\mathbf{3_L}})_{\alpha_L} ({{\mathcal{G}}^\alpha}_D^{\mathbf{3_L}})^{\alpha_L}-\sum\limits_{\alpha=1}^3
({{\mathcal{G}}^\alpha}_D^{\mathbf{3_R}})_{\alpha_R} ({{\mathcal{G}}^\alpha}_D^{\mathbf{3_R}})^{\alpha_R}\,.
\eea
A third one, which has not been reported, is fully analogous to $R_{{{g}^{\prime\,2}} +}$ and it is redundant for a very similar reason.

After setting the spurions to their VEV and going to the unitary gauge we obtain
\beq
\label{gaugeinv2}
O_{{{g}^{2}} +}=\frac{3\,g^{2}}{2}\sin^2\frac{H}{f}\,,\;\;\;\;\;
O_{{{g}^{2}} -}=-3\, {g^{2}}\cos\frac{H}{f}\,.
\eeq
The ${\mathcal{O}}(g^2)$ contribution to the potential is thus found to be 
\beq
\label{gpot}
V_{{{g}^{2}}}\propto c_{+} \cdot \frac{3\,g^{2}}{2}\sin^2\frac{H}{f}- c_{ -}\cdot {3\,g^{2}}\cos\frac{H}{f}\,,
\eeq
up to the unknown constants $c_+$ and $c_{-}$. Notice that the latter constants are not new parameters, they are just the same ones that appear in the ${\mathcal{O}}({g}^{\prime\,2})$ potential in Eq.~(\ref{gprimepot}). This is because each of the three ${\mathcal{G}}^\alpha$ components is completely identical to ${{\mathcal{G}}'}$, from the viewpoint of the composite sector. Namely they all couple to the same operator, {\it{i.e.}}~the global current multiplet. Of course they differ for their physical value but this difference is washed out when the spurion is regarded as a matrix of independent couplings. Thus the operators written in terms of  ${\mathcal{G}}^\alpha$ and ${{\mathcal{G}}'}$ must have the same coefficient, from which the equality of $c_\pm$ in Eqs.~(\ref{gprimepot}) and (\ref{gpot}). Actually there is a difference between ${{\mathcal{G}}}$ and ${{\mathcal{G}}'}$: the fact that they are coupled to the Abelian $B$ and to the non-Abelian $W^\alpha$ fields, respectively. The $W^\alpha$ self-interactions can differentiate the two contributions to the potential, but definitely not at ${\mathcal{O}}({g}^{2})$ as the self-interaction would carry further powers of $g$. Furthermore it will become clear in the following that self-interactions can play a role only at high order in the loop expansion.

While interesting and illustrative of the spurion method, the gauge contributions to the potential derived above are of limited phenomenological relevance as they are typicallly overwhelmed by other terms, in particular by those emerging from the top quark sector. The elementary/composite couplings $\lambda_{t_{L,R}}$ of the third generation $q_L$ and $t_R$ are indeed, usually, the largest sources of Goldstone symmetry breaking in our theory and thus they give the dominant contribution to the potential. The physical reason to have them large is of course the need of reproducing the large top Yukawa (see Eq.~(\ref{y_trep})), as discussed at length in Sect.~\ref{pfer}. The top contribution to the potential can be easily worked out by spurion analysis, however differently from the gauge one it is not universal: it depends on the quantum numbers of the composite sector operators $q_L$ and $t_R$ mix with. The two cases of mixing with two ${\mathbf{5}}$'s or with a ${\mathbf{14}}$ and a ${\mathbf{1}}$ will be discussed in turn.

In the case of the ${\mathbf{5}}\oplus{\mathbf{5}}$, see Eqs.~(\ref{pcrew}) and (\ref{pcrew1}), the ${\mathpzc{G}}$-breaking interactions read
\beq
\label{spurionlambda}
{\mathcal{L}}_{\lambda_t}^{\slash\hspace{-5pt}{\mathpzc{G}}}={\overline{q}}_L^\alpha(\Lambda_L)_\alpha^I ({\mathcal{O}}_F^L)_I
+{\overline{t}}_R(\Lambda_R)^I ({\mathcal{O}}_F^R)_I+{\textrm{h.c.}}
\,,
\eeq
where the spurions $\Lambda_{L,R}$ are in the ${\mathbf{5_{-2/3}}}$ of \mbox{SO$(5)\times$U$(1)_X$}\footnote{The \mbox{U$(1)_X$} charge must be opposite to the one of the operators for the interaction to be invariant.} and their physical values are
\bea
&&\displaystyle  (\Lambda_L)_\alpha^I = \frac{\lambda_{t_L}}{\sqrt{2}}\left[
\begin{array}{ccccc}
0 & 0 & +i  & +1& 0\\
+i  & -1 & 0 & 0&\,0\,
\end{array}
\right]_\alpha^{\;\;I}\,,\nonumber\\
&&\displaystyle  (\Lambda_R)^I = \lambda_{t_R}[
\begin{array}{ccccc}
0 & 0 & 0 & 0 & 1
\end{array}
]^{I}\,.\label{eq:spurion_55_pot}
\eea
As in the gauge sector, elementary spurionic symmetries must be also considered in the operator classification. These are a \mbox{U$(2)_E^L$} group under which $q_L^\alpha$ form a doublet of, say, unit charge, and consequently $(\Lambda_L)_\alpha$ is a conjugate doublet with charge $-1$, and a \mbox{U$(1)_E^R$} phase shift of $t_R$ and $\Lambda_R$. Taking these symmetries into account, only two invariants are found at ${\mathcal{O}}(\lambda_{t_{L,R}}^2)$
\beq
\label{t5op}
O_{\lambda_L^{2}}=({\Lambda_{L,\,D}^{\mathbf{1}}}^*)^\alpha (\Lambda_{L,\,D}^{\mathbf{1}})_\alpha\,,
\;\;\;\;\;
O_{\lambda_R^{2}}={\Lambda_{R,\,D}^{\mathbf{1}}}^*\, \Lambda_{L,\,D}^{\mathbf{1}}\,,
\eeq
where $\Lambda_{L,\,D}^{\mathbf{1}}$ and $\Lambda_{L,\,D}^{\mathbf{1}}$ are the singlet dressed spurions, emerging from the decomposition
\beq
{\mathbf{5}}={\mathbf{4}}\oplus {\mathbf{1}}\,.
\eeq
Explicitly, the dressed spurion multiplets in the ${\mathbf{4}}$ and in the ${\mathbf{1}}$, for both chiralities, are defined as
\beq
\displaystyle
\left[\begin{array}{c}
\Lambda_D^{\mathbf{4}}\\
\Lambda_D^{\mathbf{1}}
\end{array}
\right]\equiv U^\dagger\cdot\Lambda\,,
\eeq
Two invariants could have been written, for each chirality, by paring two ${\mathbf{4}}$ or two ${\mathbf{1}}$ components. However one combination of the two, which corresponds to the \mbox{SO$(5)$} singlet one can form out to two ${\mathbf{5}}$'s, only contributes to the vacuum energy and can be dropped. This is why we are left with a total of two operators in Eq.~(\ref{t5op}). On the spurion's VEV, and in the unitary gauge, we find \footnote{The phases of $\lambda_{t_{L,R}}$ can be reabsorbed by a redefinition of the elementary quark fields, we thus take these parameters real.}
\beq
\displaystyle 
O_{\lambda_L^{2}}=\frac{{\lambda_{t_L}}^2}{{2}}\sin^2\frac{H}{f}\,,
\;\;\;\;\;
\displaystyle 
O_{\lambda_R^{2}}={\lambda_{t_R}}^2\cos^2\frac{H}{f}\,,
\eeq
and thus the potential is expressed as
\beq
\label{toppot2}
\displaystyle 
V_{{{\lambda}^{2}}}^{{\mathbf{5}}\oplus{\mathbf{5}}}\propto \left(\frac{c_{L}}2\lambda_{t_L}^2-c_R\lambda_{t_R}^2\right) \sin^2\frac{H}{f}+{\textrm{const.}}\,,
\eeq
in terms of two unknown constants $c_L$ and $c_R$.\footnote{A priori, $c_L$ and $c_R$ are completely unrelated because the two chiral  fermionic operators ${\mathcal{O}}_F^L$ and ${\mathcal{O}}_F^R$ the spurions couple to are distinct operators, in spite of having the same quantum numbers under the global group. If they were related by some other symmetry, for instance by spatial parity, we would have $c_L=c_R$, but in general this is not the case.}

The above result is surprising in one aspect, which is relevant for phenomenology as we will see in the following section. Namely, it shows that the ${\mathcal{O}}(\lambda^2)$ potential has a fixed functional dependence on the Higgs field. In particular it is proportional to $\sin^2H/f$ and the only freedom we have is to choose, by varying $\lambda_{t_{L,R}}$ or by acting on the composite sector that controls $c_{L,R}$, the size and the sign of the overall coefficient. Therefore this contribution to the potential, which we claimed to be the dominant one, can only have a minimum at $H=0$ or $H=\pi f/2$ (modulus $\pi f$), and thus in no case it gives a realistic theory for which we need $\xi=\sin^2\langle H\rangle/f\ll1$. This implies that the ${\mathcal{O}}(\lambda^2)$ term, though formally dominant, must be made accidentally small such as to compete with other contributions allowing for a tunable $\xi$ parameter. In view of this fact, which will be discussed in details in Sect.~\ref{HVMT}, we should extend our classification to the first subleading order, {\it{i.e.}}~to ${\mathcal{O}}(\lambda_L^4)$, ${\mathcal{O}}(\lambda_R^4)$ and ${\mathcal{O}}(\lambda_L^2\lambda_R^2)$ operators.\footnote{Symmetries forbid ${\mathcal{O}}(\lambda^3)$ terms.} A total of $6$  ${\mathcal{O}}(\lambda_L^4)$ invariants exist, and the same amount of  ${\mathcal{O}}(\lambda_R^4)$. However $2$ of them are redundant and can be removed, in each category, by exploiting the existence of two full \mbox{SO$(5)$} singlets in the tensor product of two $\Lambda$ times two $\Lambda^*$ spurions. Similarly at ${\mathcal{O}}(\lambda_L^2\lambda_R^2)$ we count $10$ singlets and $3$ redundancies, for a total of $7$ operators. Fortunately listing all these operators is not necessary because many of them end up having the same trigonometrical structure up to constant vacuum energy terms. A complete coverage of the possible structures is provided by the following terms
\bea
\label{op4sp5}
&& \displaystyle
O_{\lambda_L^4,\,1}=\left[({\Lambda_{L,\,D}^{\mathbf{1}}}^*)^\alpha (\Lambda_{L,\,D}^{\mathbf{1}})_\alpha\right]^2
=\frac{\lambda_{t_L}^4}{{4}}\sin^4\frac{H}{f}
\,,\\
&& \displaystyle
O_{\lambda_L^4,\,2}=\left[({\Lambda_{L,\,D}^{\mathbf{1}}}^*)^\alpha (\Lambda_{L,\,D}^{\mathbf{1}})_\alpha\right]\,
\left[({\Lambda_{L,\,D}^{\mathbf{4}}}^*)^\alpha_i (\Lambda_{L,\,D}^{\mathbf{4}})_\alpha^i\right]\nonumber\\
&&\;\;\;\;\;\;\;\;\;\; =\frac{\lambda_{t_L}^4}{{4}}\left[4\,\sin^2\frac{H}{f}-\sin^4\frac{H}{f}\right]\,,
\nonumber\\
&&O_{\lambda_R^4,\,1}=\left[{\Lambda_{R,\,D}^{\mathbf{1}}}^* \Lambda_{R,\,D}^{\mathbf{1}}\right]^2
=\lambda_{t_R}^4\cos^4\frac{H}{f}=\lambda_{t_R}^4\left[1-2\,\sin^2\frac{H}{f}+\sin^4\frac{H}{f}\right]
\,,\nonumber\\
&& \displaystyle
O_{\lambda_R^4,\,2}=\left[{\Lambda_{R,\,D}^{\mathbf{1}}}^* \Lambda_{R,\,D}^{\mathbf{1}}\right]\,
\left[({\Lambda_{R,\,D}^{\mathbf{4}}}^*)_i (\Lambda_{R,\,D}^{\mathbf{4}})^i\right]={\lambda_{t_R}^4}\left[\sin^2\frac{H}{f}-\sin^4\frac{H}{f}\right]\,,
\nonumber\\
&&O_{\lambda_L^2 \lambda_R^2,\,1}=\left[({\Lambda_{L,\,D}^{\mathbf{1}}}^*)^\alpha (\Lambda_{L,\,D}^{\mathbf{1}})_\alpha\right]\,
\left[{\Lambda_{R,\,D}^{\mathbf{1}}}^* \Lambda_{R,\,D}^{\mathbf{1}}\right]\nonumber\\
&&\;\;\;\;\;\;\;\;\;\;\;\;\;\;
=\frac{\lambda_{t_L}^2\lambda_{t_R}^2}{2}\left[\sin^2\frac{H}{f}-\sin^4\frac{H}{f}\right]\,,\nonumber\\
&& \displaystyle
O_{\lambda_L^2 \lambda_R^2,\,2}=\left[({\Lambda_{L,\,D}^{\mathbf{1}}}^*)^\alpha (\Lambda_{L,\,D}^{\mathbf{1}})_\alpha\right]\,
\left[({\Lambda_{R,\,D}^{\mathbf{4}}}^*)_i (\Lambda_{R,\,D}^{\mathbf{4}})^i\right]=\frac{\lambda_{t_L}^2 \lambda_{t_R}^2}{2}\sin^4\frac{H}{f}\,.\nonumber
\eea
Up to constants, all the invariants are linear combinations of $\sin^2$ and $\sin^4$ and thus the complete ${\mathcal{O}}(\lambda^4)$ potential can be expressed as
\bea
\label{toppot4}
\displaystyle 
V_{{{\lambda}^{4}}}^{{\mathbf{5}}\oplus{\mathbf{5}}}\propto&& ({c_{LL}}\lambda_{t_L}^4+c_{RR}\lambda_{t_R}^4+c_{LR}\lambda_{t_R}^2\lambda_{t_L}^2) \sin^2\frac{H}{f}\nonumber\\
+&&({c'_{LL}}\lambda_{t_L}^4+c'_{RR}\lambda_{t_R}^4+c'_{LR}\lambda_{t_R}^2\lambda_{t_L}^2) \sin^4\frac{H}{f}\,.
\eea
By properly adjusting, or tuning, the two terms, taking of course also into account the ${\mathcal{O}}(\lambda^2)$ contributions in Eq.~(\ref{toppot2}), realistically small values of $\xi$ can now be obtained.

The case of elementary quarks mixing to a ${\mathbf{14}}\oplus{\mathbf{1}}$ is considerably different. The elementary/composite interaction takes the form
\beq
{\mathcal{L}}_{\textrm{Int}}=
\lambda_{t_L}\,\overline{q}_L{\mathcal{O}}_{\mathbf{14}}^L+{\lambda_{t_R}}\,\overline{t}_R{\mathcal{O}}_{\mathbf{1}}^R+{\textrm{h.c.}}\,,
\eeq
where the fermionic operator coupled to the $q_L$ doublet transforms in the ${\mathbf{14}}_{{2/3}}$ of \mbox{SO$(5)\times$U$(1)_X$}, while the one coupled to $t_R$ is in the ${\mathbf{1}}_{{2/3}}$. Being the latter a singlet, the $t_R$ mixing does not break the Goldstone symmetry and therefore the $\lambda_{t_R}$ coupling is not capable, in this case, to trigger the generation of the Higgs potential. The only source of breaking is $\lambda_{t_L}$, which we embed into a spurion in the ${\mathbf{14}}$ by rewriting the interaction as 
\beq
\label{spurionlambda14}
{\mathcal{L}}_{\lambda_t}^{\slash\hspace{-5pt}{\mathpzc{G}}}={\overline{q}}_L^\alpha(\Lambda_L)_\alpha^{IJ} ({\mathcal{O}}_F^L)_{IJ}
\,,
\eeq
where $\Lambda_L$ is a symmetric traceless $2$-tensor. The spurion VEV is (see Eq.~(\ref{14source}))
\beq
\displaystyle
(\Lambda_L)_{IJ}^\alpha=\frac{{\lambda_{t_L}}}{\sqrt{2}}
\left\{
\left[
\begin{array}{ccccc}
0 & 0 & 0 & 0 &0\\
0 & 0 & 0 & 0 &0\\
0 & 0 & 0 & 0 &+i\\
0 & 0 & 0 & 0 &+1 \\
0 & 0 & +i & +1 & 0
\end{array}
\right]_{IJ},\,
\left[
\begin{array}{ccccc}
0 & 0 & 0 & 0 &+i\\
0 & 0 & 0 & 0 &-1\\
0 & 0 & 0 & 0 &0\\
0 & 0 & 0 & 0 &0 \\
+i & -1 & 0 & 0 & 0
\end{array}
\right]_{IJ}
\right\}^\alpha\,.
\eeq

Notice that the $t_R$ coupling being invariant under the Goldstone symmetry nicely fits with the alternative interpretation of the ${\mathbf{14}}\oplus{{1}}$ setup, outlined at the end of Sect.~\ref{sec:HCsigmaM}, in terms of a completely composite $t_R$ field in the ${\mathbf{1}}_{{2/3}}$ of \mbox{SO$(4)\times$U$(1)_X$}. This is possible only because $t_R$ mixes with a singlet. Indeed if the $t_R$ has to be part of the composite sector, its interactions must respect the composite sector symmetry group and thus they cannot generate the Higgs potential. In this interpretation, the coupling strength is of order $\lambda_{t_R}\sim g_*$, while in the elementary case $\lambda_{t_R}< g_*$. In the following section we will consider both cases by keeping the mixing as a free parameter. We will see that the completely composite option is favored from the viewpoint of the generation of a viable Higgs VEV and mass.

Notice that the completely composite $t_R$ interpretation only relies on ${\mathcal{O}}_F^R$ being a singlet, it does not require ${\mathcal{O}}_F^L$ in the ${\mathbf{14}}$. The reason for making the latter choice, rather than for instance ${\mathcal{O}}_F^L\in{\mathbf{5}}_{2/3}$, resides in the structure of the potential generated by the spurion in the ${\mathbf{14}}$. This representation decomposes as
\beq
{\mathbf{14}}_{2/3}\rightarrow\;(\mathbf{3},\mathbf{3})_{2/3} \oplus (\mathbf{2},\mathbf{2})_{2/3}\oplus (\mathbf{1},\mathbf{1})_{2/3}\,,
\eeq
in terms of three rather than two irreducible \mbox{SO$(4)$} multiplets.  Three invariant operators can thus be written, but only one of them is redundant given that only one \mbox{SO$(5)$} singlet is present in the tensor product of two ${\mathbf{14}}$'s. We now end up with two independent operators, which we can take to be
\bea
\label{141_ops}
&& \displaystyle
O_{\lambda_L^2,\,1}=({\Lambda_{L,\,D}^{\mathbf{4}}}^*)^\alpha_i (\Lambda_{L,\,D}^{\mathbf{4}})_\alpha^i
=\lambda_{t_L}^2 \left[2-\frac72 \sin^2\frac{H}{f}+2 \sin^4\frac{H}{f}\right]\,,\nonumber\\
&& \displaystyle
O_{\lambda_L^2,\,2}=({\Lambda_{L,\,D}^{\mathbf{1}}}^*)^\alpha (\Lambda_{L,\,D}^{\mathbf{1}})_\alpha=
2\, \lambda_{t_L}^2\left[4\,\sin^2\frac{H}{f}-\sin^4\frac{H}{f}\right]\,,
\eea
where the dressed spurion multiplets are obtained, in full analogy with the previous cases, by acting with $U^{-1}$ on the two $\Lambda$ indices and splitting the indices of the resulting matrix in \mbox{SO$(4)$} components. The ${\mathcal{O}}(\lambda^2)$ potential is thus a combination of $\sin^2$ and $\sin^4$ functions, of the form
\beq
\label{toppot214}
\displaystyle 
V_{{{\lambda}^{2}}}^{{\mathbf{14}}\oplus{\mathbf{1}}}\propto \lambda_{t_L}^2\left(c_1 \sin^2\frac{H}{f}+c_2 \sin^4\frac{H}{f}\right)\,.
\eeq
Differently from the ${{\mathbf{5}}\oplus{\mathbf{5}}}$ one, this potential can be tuned to achieve realistic EWSB without relying on additional higher order terms.

\subsection{Higgs potential estimated}
\label{pwcpot_sec}
A qualitative but sharp assumption was made in Sect.~\ref{OSOC} on the nature of the composite sector, which we characterized as a  \mbox{$1$S$1$C} (One Scale One Coupling) model. By this assumption the power-counting formula in Eq.~(\ref{POWCFull}) was derived to estimate the expected size of operators in the low-energy effective field theory describing the pNGB Higgs plus the other SM particles. We got the same result for large-$N$ QCD-like strongly-coupled theories, showing that they might effectively behave as \mbox{$1$S$1$C} models. These results were valid exclusively for effective operators generated by the composite sector dynamics alone, {\it{i.e.}}~by the sole exchange of composite sector virtual states with no elementary fields propagating in the internal lines. Extending the analysis to the Higgs potential operators, which do not belong to the latter category, is the purpose of the present section. We do this with a twofold aim. First, we want to estimate the overall magnitude of the potential that controls such an important observable like the Higgs mass. Second, we want to check if and to what extent the intuitive idea that we can expand in the elementary couplings, and thus in the number of spurion insertions, is actually valid or not. If it was not, the predictive power of the spurion method would get completely washed out since we would be forced to consider an infinite series of operators with arbitrary powers of $\sin^2H/f$, leading to a potential of completely generic form.

\begin{figure}
\vspace{15pt}
\centering
\includegraphics[height=0.2\textwidth]{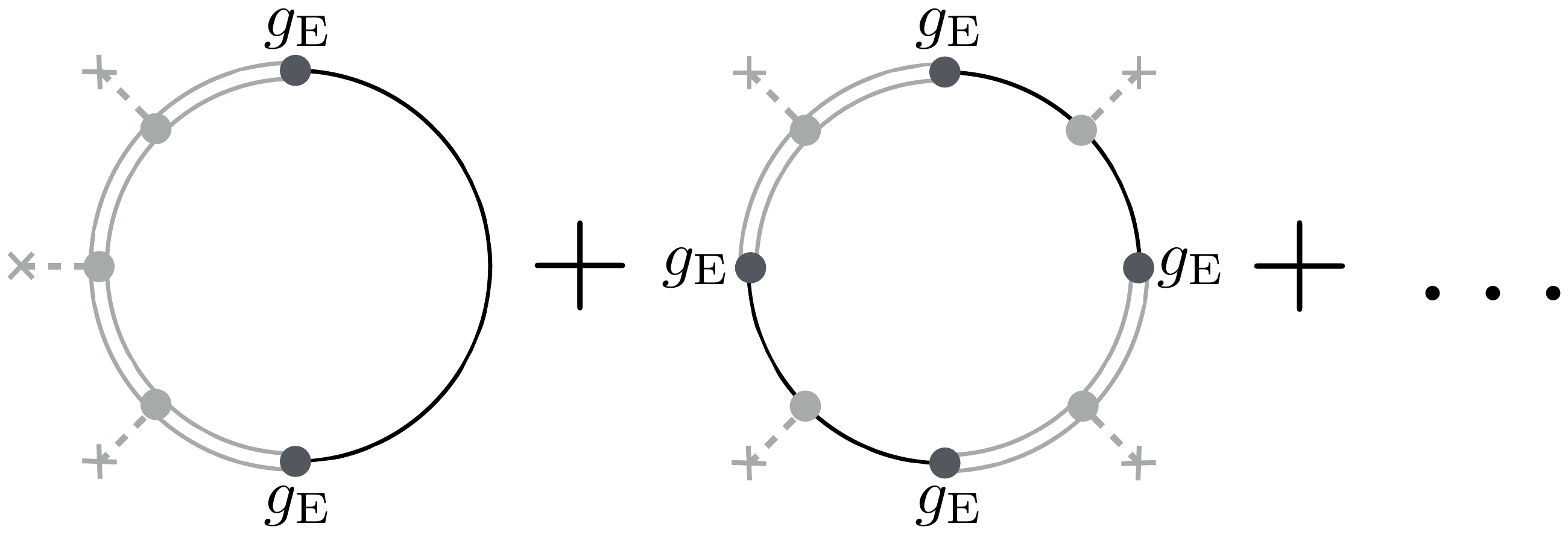}
\vspace{-5pt}
\caption{The structure of the leading diagrams contributing to the potential.}
\label{potential}
\end{figure}

Deriving the power-counting estimate for the potential starts from outlining its origin in terms of Feynman diagrams. The potential is, almost by definition, the sum of \mbox{$1$PI} (one particle irreducible) diagrams with zero-momentum external Higgs lines. Since the Higgs is part of the composite sector, it does not couple directly to the elementary sector fields. Therefore no diagram should be considered with only elementary internal lines. Furthermore the Higgs is a NGB and thus it gets no potential from the purely composite sector diagrams because they respect the Goldstone symmetry. Mixed diagrams need to be considered, where at least one elementary internal line is present. In order to make them \mbox{$1$PI} the elementary line must close into a loop, therefore the potential gets generated only at the radiative level. The structure of the leading diagrams is reported in Fig.~\ref{potential}, where the dashed lines ending on crosses denote Higgs field insertions, the black single lines are elementary sector gauge or fermionic fields and $g_{\textrm{E}}$ collectively denotes the elementary/composite couplings. Depending on which elementary sector state is exchanged, $g_{\textrm{E}}=\{g,g',\lambda_{t_L},\lambda_{t_R}\}$. The double lines represent portions of the graph made of purely composite sector propagators and vertices. We denoted them as lines because in weakly-coupled models (such as the ones we will deal with in Chap.~\ref{ch:Pheno_models}) they are indeed single particle propagators, making the ones in Fig.~\ref{potential} one-loop diagrams. However in a generic strongly-interacting composite sector we should think to them as two-point correlators 

\begin{figure}[t]
\vspace{15pt}
\centering
\includegraphics[height=0.2\textwidth]{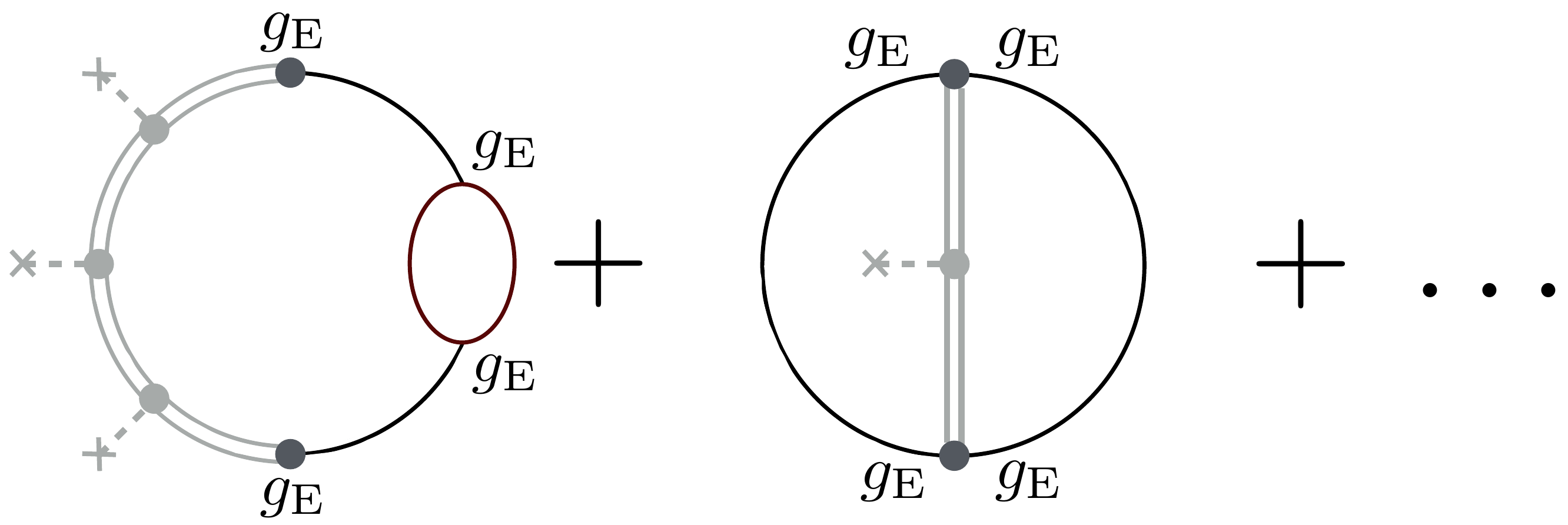}
\vspace{-5pt}
\caption{Subleading contributions to the potential.}
\label{potential1}
\end{figure}

\begin{figure}[h]
\vspace{-5pt}
\centering
\includegraphics[width=0.22\textwidth]{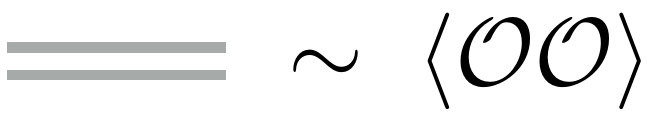}
\vspace{-8pt}
\end{figure}
\noindent
of the composite sector operators to which the corresponding elementary line is coupled. Namely, ${\mathcal{O}}$ is the global current $J$ or one of the fermionic operators ${\mathcal{O}}_{L,R}$.

The contributions to the potential from the diagrams in Fig.~\ref{potential}, for each number of $g_{\textrm{E}}$ insertions, is immediately worked out in the \mbox{$1$S$1$C} hypothesis. The potential is one term in the Lagrangian density, thus it has the dimension (see Eq.~(\ref{dimlag})) of $C^{-2}\cdot L^{-4}$, where $C$ is the coupling dimension we defined as $C=[\hbar]^{-1/2}$. The correct coupling dimension is already saturated by the one-loop factor $\hbar/16\pi^2$ and thus the dimension carried by elementary $g_{\textrm{E}}$ coupling insertions must be canceled by other dimensionful objects. In the \mbox{$1$S$1$C} hypothesis the only such object that is present in the composite sector is $g_*$ and similarly $m_*$ is the only one that carries $L$ dimension. Thus the potential, setting $\hbar=1$, is
\beq
\label{pwcpot}
\displaystyle
V=\frac{N_c m_*^4}{16\pi^2}\left[\left(\frac{g_{\textrm{E}}}{g_*}\right)^2{\mathcal{V}}_{(2)}[H/f]+\left(\frac{g_{\textrm{E}}}{g_*}\right)^4{\mathcal{V}}_{(4)}[H/f]+\ldots\right]\,,
\eeq
where we took into account that the Higgs, being a Goldstone, must appear as $H/f$ as also apparent from the spurion analysis in the previous section. An overall factor of $N_c$ that counts QCD color multiplicity and equals $1$ for the gauge and $3$ for the fermionic diagrams has been included in the estimate.

This was derived for perturbative \mbox{$1$S$1$C} models, but the same result holds for strongly-coupled large-$N$ theories. Indeed we saw in Sect.~\ref{OSOC} that the $\langle\mathcal{O}\mathcal{O}\rangle$ two-point correlators, under the assumption of mesonic ${\mathcal{O}}$, scale with $N$ like
\beq
\displaystyle
\langle\mathcal{O}\mathcal{O}\rangle\propto\frac{N}{16\pi^2}\equiv\frac1{g_*^2}\,.
\eeq
By taking this estimate for the double lines in Fig.~\ref{potential}, we end up once again with the power-counting in Eq.~(\ref{pwcpot}).

Other diagram topologies, such as the ones in Fig.~\ref{potential1}, also contribute to the potential. They contain extra insertions of the elementary sector fields self-couplings (namely, the gauge couplings) or composite sector correlators with more operator insertions. In both cases this makes an additional loop suppression with respect to the ones in Fig.~\ref{potential}
\beq
V_{2\textrm{-loop}}\sim\frac{g_{\textrm{E}}^2}{16\pi^2}V_{1\textrm{-loop}}\,.
\eeq
Provided the elementary sector couplings are perturbative, these contributions are negligible and we are left with the estimate in  Eq.~(\ref{pwcpot}).

The power-counting formula provides the desired justification of the spurion expansion we performed in the previous section. Each spurion carries one $g_{\textrm{E}}$ and each $g_{\textrm{E}}$ is weighted by $1/g_*$. This allows to set up a perturbative expansion if 
\beq
g_{\textrm{E}}<g_*\,.
\eeq
The elementary sector couplings being weaker than the composite sector one is the obvious criterion by which the elementary sector behaves as a weak perturbation of the composite one and thus in particular the composite sector group is a good approximate symmetry. This fact is explicitly verified here, showing that the spurion classification is predictive if and only if $g_{\textrm{E}}<g_*$.

\subsubsection*{The IR potential}
There are contributions to the potential that do not need to be estimated but can just be computed, up to a mild logarithmic sensitivity to the details of the composite sector dynamics. The calculation is interesting under several respects, however it lies somewhat outside the main line of development of these Notes since the result will not introduce qualitatively new effects and thus it will be ignored in the phenomenological analysis that follows. 

The basic observation is that there exists a regime where we do have complete control of the theory so that a real calculation is possible. This is the energy range below the resonance scale $m_*$ where the effective field theory, describing the pNGB Higgs and the other SM particles, is perturbative and accurately describes the dynamics. Below (or much below) $m_*$ the dominant operators are the leading ones in the derivative expansion, namely the ``non-linear $\sigma$-model'' terms we worked out in Chap.~\ref{ch:Goldstone}. Within the non-linear $\sigma$-model we can take one loop of the elementary gauge and fermionic fields and compute the potential, ending up with the standard Coleman--Weinberg formula
\bea
\label{potIR}
\displaystyle
&&V_{\textrm{gauge}}[H]=\frac{3}{32\pi^2}\Lambda^2\sum\limits_i M_{i}^2 +\frac{3}{64\pi^2}\sum\limits_i[M_{i}^4(\log{M_{i}^2/\Lambda^2-1/2)}]\,,\nonumber\\
\displaystyle
&&V_{\textrm{top}}[H]=-\frac{N_c}{8\pi^2} \Lambda^2 M_t^2-\frac{N_c}{16\pi^2} M_t^4(\log{M_t^2/\Lambda^2-1/2)}\,,
\eea
where $\Lambda$ is the hard momentum cutoff of the loop integral.\footnote{The calculation could equally well be performed in dimensional regularization, leading eventually to the same physical result. Working in a scheme where the quadratic divergence appears explicitly is however more interesting for the present discussion.} In the equation, $M_{i}$ denotes the field-dependent gauge field masses, namely those of the $W$ and $Z$ bosons in the presence of a constant background for the $H$ field. Analogously,  $M_t$ is the $H$-dependent top mass. Clearly the bottom and the other quarks and leptons contribute to the potential as well, however they can be safely ignored because of their small masses. No contribution comes instead at one loop from diagrams with propagating Higgs boson lines. Indeed the Higgs self-interactions respect the Goldstone symmetry and thus they cannot generate the potential. Furthermore the Higgs does not mix with the elementary fields so that mixed loops need not to be considered.

The masses in Eq.~(\ref{potIR}) can be written in a form that, albeit somewhat involved, allows to make contact with the spurion notation. We start from the top mass and we consider the case of fermion embedding in the ${\mathbf{5}}\oplus{\mathbf{5}}$ for illustration. The ${\mathbf{14}}\oplus{\mathbf{1}}$ or other cases could be similarly worked out. The mass comes from Eq.~(\ref{yukup})
\beq
\displaystyle
-c^t\frac{\lambda_{t_L}\lambda_{t_R}}{g_*}m_*{\overline{Q}_L^t}^{\mathbf{1}}{T}_R^{\mathbf{1}}\,.
\eeq
expressed in terms of the dressed sources ${{Q}_L^t}^{\mathbf{1}}$ and ${T}_R^{\mathbf{1}}$ in the singlet of \mbox{SO$(4)$}. By writing this explicitly as in the second line of Eq.~(\ref{yukup}) one could immediately obtain
\beq
\label{mtH0}
M_t^2=\frac18 {c^t}^2\lambda_{t_L}^2\lambda_{t_R}^2 f^2 \sin^2\frac{2H}{f}=\frac{y_t^2}{4}\frac{f^2}{1-\xi} \sin^2\frac{2H}{f}\,.
\eeq
where we expressed the prefactor in terms of the top Yukawa coupling and of $\xi=v^2/f^2$. However it is interesting to take an intermediate step, noticing that the dressed sources are related to the dressed spurions as
\beq
\lambda_{t_L}{{Q}_L^t}^{\mathbf{1}}=\left({\Lambda_{L,\,D}^{\mathbf{1}}}^*\right)^\alpha q_{L,\alpha}\,,\;\;\;\;\;
\lambda_{t_R}{{T}_R}^{\mathbf{1}}={\Lambda_{R,\,D}^{\mathbf{1}}}^*\,t_R\,.
\eeq
This is evident from the definitions in Eq.~(\ref{qlemb}) and (\ref{spurs4}), and similarly for $T_R$. The mass term in the spurion notation is thus
\beq
\displaystyle
-c^t f \,{\overline{q}_L^\alpha}\left({\Lambda_{L,\,D}^{\mathbf{1}}}\right)_\alpha {\Lambda_{R,\,D}^{\mathbf{1}}}^* t_R+h.c.\;,
\eeq
and the squared top mass  reads \footnote{Since no $b_R$ is present, the fermionic mass matrix is $2\times1$. The top mass is computed as $M_t^2={M_F}^\dagger\cdot M_F$.}
\bea
M_t^2&&={c^t}^2 f^2\, \left[({\Lambda_{L,\,D}^{\mathbf{1}}}^*)^\alpha \left({\Lambda_{L,\,D}^{\mathbf{1}}}\right)_\alpha \right]
\left[
{\Lambda_{R,\,D}^{\mathbf{1}}}^* \,{\Lambda_{R,\,D}^{\mathbf{1}}}\right]
\nonumber\\
&&
={c^t}^2 f^2\,O_{\lambda_L^2 \lambda_R^2,\,1}\,.
\eea
After setting the spurions to their VEV, this correctly reproduces Eq.~(\ref{mtH0}). The rewriting of $M_t^2$ in terms of spurions shows that it is proportional to one of the ${\mathcal{O}}(\lambda^4)$ operators listed in Eq.~(\ref{op4sp5}), as one could have also demonstrated by inspection, comparing Eq.~(\ref{mtH0}) with the list of operators expressed in the unitary gauge. This shows explicitly that $M_t$ is invariant under the spurionic symmetries. Similarly, by using Eq.~(\ref{mwz}), we can express the $W$ and $Z$ boson masses in terms of the gauge sector invariants in Eq.~(\ref{gaugeinv1}) and (\ref{gaugeinv2}):
\bea
\displaystyle
&&M_W^2=\frac16 f^2O_{g^2 +}=\frac{g^2 f^2}4\sin^2\frac{H}{f}\,,\nonumber\\
&&M_Z^2=\frac16 f^2O_{g^2 +}+\frac12 f^2O_{{{g}^{\prime\,2}} +}=\frac1{c_w^2}M_W^2\,.
\eea

Now we have all the elements to discuss the structure of the potential in Eq.~(\ref{potIR}). The first comment is that the potential is a function of the field-dependent masses, which we saw above being invariant under the spurionic symmetries. Therefore it is itself invariant, in accordance with the general argument of Sect.~\ref{CHPCH}. Second, some of the terms it contains are not new contributions to the potential. In particular the quadratically divergent ones, proportional to the squared masses, are linear combinations of the ${\mathcal{O}}(g_{\textrm{E}}^2)$ operators we discussed in Sect.~\ref{CHPCH}. Notice that not only their functional dependence on the Higgs is the one we expected, but also their size. Indeed when the momentum cutoff is identified with the physical cutoff of the effective field theory, {\it{i.e.}}~$\Lambda\simeq m_*$, their coefficient obeys the power-counting in Eq.~(\ref{pwcpot}). Similar considerations hold for the finite terms. They are ${\mathcal{O}}(g_{\textrm{E}}^4)$ operators and their coefficients are the expected ones. The only terms in Eq.~(\ref{potIR}) that are new and interesting are the logarithmically divergent ones. Furthermore these are the only terms we can actually compute since the others are respectively unphysical divergences and scheme-dependent finite contributions. The logarithms, instead, even after the divergence is canceled by renormalization, result in scheme-independent terms
\bea
\displaystyle
&&V_{\textrm{gauge}}^{\textrm{IR}}[H]= \frac{3f^4}{64\pi^2}\frac{3g^4+2g^2g'^2+g'^4}{16}\sin^4\frac{H}{f}\left(\log\left[\frac{g^2}{g_*^2}\sin^2\frac{H}{f}\right]+{\textrm{const.}}\right),\nonumber\\
\displaystyle
&&V_{\textrm{top}}^{\textrm{IR}}[H]=-\frac{N_c}{256\pi^2}\frac{y_t^4f^4}{(1-\xi)^2}\sin^4\frac{2H}{f}\left(\log\left[\frac{y_t^2}{g_*^2}\sin^2\frac{2H}{f}\right]+{\textrm{const.}}\right)\hspace{-1pt},
\eea 
where we used $\Lambda=m_*=g_*f$ as the divergence subtraction point and we reabsorbed ${\mathcal{O}}(1)$ factor in the incalculable constant. We interpret this IR contribution to the potential as the Renormalization Group running of the potential operators from the resonance scale $m_*$ down to the energies that are relevant for EWSB. The separation between these two scales makes it log enhanced.

All in all, the complete structure of the potential is as follows. The terms that are polynomial in the spurions, with coefficient dictated by the power counting in Eq.~(\ref{pwcpot}), are operators of UV origin. Namely they emerge from the composite sector dynamics at the scale $m_*$ and  they are computed at that scale by integrating out the composite sector degrees of freedom. They come from Feynman diagrams with finite external momenta and thus they can be Taylor expanded in the $g_{\textrm{E}}$ couplings. Of course the true potential is computed at zero external momentum and not at $m_*$. But how the potential changes from $m_*$ to zero can be computed within the realm of validity of the effective field theory, leading to the calculable IR contribution we described above. This IR term is not polynomial in the couplings and not even expandable in Taylor series around $g_{\textrm{E}}=0$. This is why it escaped the spurion classification in spite of being perfectly invariant under the spurionic symmetries.

\subsection{Higgs VEV, mass and tuning}
\label{HVMT}
We are finally in the position to understand under what conditions and at what price, in terms of fine-tuning, a realistic EWSB scale and Higgs mass can be obtained in our framework. Apart from the IR term, which we ignore for simplicity, the potential takes the generic form 
\beq
\label{pot_form}
\displaystyle
V[H]=-\alpha f^2 \sin^2\frac{H}{f}+\beta f^2 \sin^4\frac{H}{f}\,,
\eeq
where the signs and the $f^2$ normalization are chosen for future convenience. The parameter $\alpha$ receives contributions from both the gauge and the fermionic sectors at order $g^2$ (\ref{gprimepot}), $g^{\prime 2}$ (\ref{gpot}) and $\lambda_t^2$ (\ref{toppot2}), (\ref{toppot214}), while $\beta$ is of ${\mathcal{O}}(\lambda_t^2)$ in the case of the ${\mathbf{14}}\oplus{\mathbf{1}}$ and of ${\mathcal{O}}(\lambda_t^4)$ (\ref{toppot4}) for the ${\mathbf{5}}\oplus{\mathbf{5}}$ choice. The gauge contribution is typically subleading and it will be ignored in what follows. The functional form of Eq.~(\ref{pot_form}) is also obtained for fermions in the ${\mathbf{10}}\oplus{\mathbf{10}}$ while for the spinorial ${\mathbf{4}}\oplus{\mathbf{4}}$ sines of half the frequency ($H/f\rightarrow H/2f$) should be considered. We will not treat the latter case here because it is phenomenologically disfavored as we saw in Sect.~\ref{pfer}. It turns out that it behaves exactly like the ${\mathbf{5}}\oplus{\mathbf{5}}$ as far as the issues of VEV and mass generation are concerned. 

Irregardless of their origin, the parameters $\alpha$ and $\beta$ must be such to give realistic EWSB, {\it{i.e.}}~$\sin^2\langle H\rangle/f=\xi\ll1$ and $m_H=126$~GeV. These requirements correspond to the conditions
\bea
\label{pot_tun_cond1}
&&\alpha=2\,\beta\,\xi\,,\\
\label{pot_tun_cond2}
&&m_H^2=8\,\xi(1-\xi)\,\beta\,.
\eea
Based on the above equations, two kinds of considerations can be made. On one hand we can pragmatically observe that by inverting for $\alpha$ and $\beta$ the potential is completely specified up to the unknown parameter $\xi=v^2/f^2$
\beq
\displaystyle
V[H]=\frac{m_H^2 f^2}{8\,\xi\,(1-\xi)}\left[\sin^2\frac{H}{f} - \xi \right]^2+{\textrm{const}}\,.
\eeq
This allows us to compute the Higgs boson self-interactions, starting from the trilinear coupling. The latter is modified with respect to the SM by the relative amount 
\beq
\displaystyle
k_{3\,h}=\frac{1-2\,\xi}{\sqrt{1-\xi}}\,.
\eeq
Unfortunately a precise enough measurement of this coupling is far to come, therefore the prediction above is of limited phenomenological relevance for the time being. On the other hand, more thoughtfully, we can ask ourselves how easy or difficult it is in our framework to get $\alpha$ and $\beta$ such as to obey Eqs.~(\ref{pot_tun_cond1}), (\ref{pot_tun_cond2}) leading to realistic EWSB and Higgs mass. The rest of the present section is devoted to this issue.

The VEV condition (\ref{pot_tun_cond1}) obliges us to take $\alpha/\beta=2\,\xi\ll1$, a situation which is definitely unnatural in the ${\mathbf{14}}\oplus{\mathbf{1}}$  case (and even more so for the ${\mathbf{5}}\oplus{\mathbf{5}}$, as we will see) where $\alpha$ and $\beta$ emerge at the same order and thus are expected to be of comparable size. The power counting in Eq.~(\ref{pwcpot}) tells us
\beq
\displaystyle
\alpha_{{\mathbf{14}}\oplus{\mathbf{1}}}=a\,\frac{N_c}{16\pi^2}\lambda_{t_L}^2 m_*^2\,,\;\;\;\;\;\beta_{{\mathbf{14}}\oplus{\mathbf{1}}}=b\,\frac{N_c}{16\pi^2}\lambda_{t_L}^2 m_*^2\,,
\eeq
with $a$ and $b$\  ${\mathcal{O}}(1)$ coefficients. Furthermore the $\sin^2$ and $\sin^4$ contributions to the potential normally arise together from the composite sector microscopic dynamics. Namely, by integrating out the composite sector resonances we expect to obtain operators like the ones in Eq.~(\ref{141_ops}) that contain both functions. The parameters $\alpha$ and $\beta$ are thus linear combinations of unrelated operator coefficients so that suppressing one with respect to the other requires cancellation. This cancellation, or fine-tuning, can be quantified as
\beq
\label{tuning_vev_14}
\displaystyle
\Delta_\xi^{{\mathbf{14}}\oplus{\mathbf{1}}}=\frac{\left(\alpha/\beta\right)_{\textrm{expected}}}{\left(\alpha/\beta\right)_{\textrm{needed}}}=\frac1{2\,\xi}\,,
\eeq
in terms of the naturally expected size of the parameter over the value we actually need. Such an amount of tuning, needed to achieve a small enough $\xi$, is the one we advocated since the very beginning of these Notes to be required for a potentially realistic setup. Concrete constructions can have more tuning than $\sim1/\xi$, as we will see below, but none happens to have less even though no robust argument actually forbids this possibility. A priori one might imagine a model where $\alpha$ emerges at a subleading order in the spurions with respect to $\beta$, making $\alpha\ll\beta$ and thus $\xi$ naturally small. No example of this ``self-tuned'' configuration exists, while we do have examples of the opposite situation, $\alpha\gg\beta$, and the ${\mathbf{5}}\oplus{\mathbf{5}}$ model is one of those. Indeed 
\beq
\displaystyle
\alpha_{{\mathbf{5}}\oplus{\mathbf{5}}}=a\,\frac{N_c}{16\pi^2}\lambda_t^2 m_*^2\,,\;\;\;\;\;\beta_{{\mathbf{5}}\oplus{\mathbf{5}}}=b\,\frac{N_c}{16\pi^2}\lambda_t^4 f^2\,,
\eeq
having denoted as $\lambda_t$, for shortness, any of the $\lambda_{t_{L,R}}$. Actually for the sake of the present discussion $\lambda_t$ should be regarded as the largest of the two couplings. Achieving small $\xi$ is thus more difficult in the ${\mathbf{5}}\oplus{\mathbf{5}}$, {\it{i.e.}}
\beq
\label{delta_xi_5p5}
\displaystyle
\Delta_\xi^{{\mathbf{5}}\oplus{\mathbf{5}}}=\frac{\left(\alpha/\beta\right)_{\textrm{expected}}}{\left(\alpha/\beta\right)_{\textrm{needed}}}=\left(\frac{g_*}{\lambda_t}\right)^2\cdot\frac1{2\,\xi}=\left(\frac{g_*}{\lambda_t}\right)^2\cdot\Delta_\xi^{{\mathbf{14}}\oplus{\mathbf{1}}}\,.
\eeq
This comes out, in a sense, from a ``double tuning'': a first tuning needed to make $\alpha$ comparable with $\beta$ overcoming the different orders at which they are generated, times a second cancellation needed for $\alpha/\beta\sim\xi\ll1$. The first step of the cancellation if not needed for the ${\mathbf{14}}\oplus{\mathbf{1}}$, which thus provides a ``minimally tuned'' scenario. 

From the viewpoint of achieving a small enough $\xi$, {\it{i.e.}}~of satisfying Eq.~(\ref{pot_tun_cond1}), ${\mathbf{14}}\oplus{\mathbf{1}}$ is favored over ${\mathbf{5}}\oplus{\mathbf{5}}$, let us now see what happens with the second condition (\ref{pot_tun_cond2}) needed to obtain a realistic Higgs mass. With the ${\mathbf{14}}\oplus{\mathbf{1}}$ estimate of $\beta$ we obtain
\beq
\label{mh14p1}
\displaystyle
(m_H^2)_{{\mathbf{14}}\oplus{\mathbf{1}}}=(1-\xi)\,b\,\frac{N_c}{2\pi^2}\left(\lambda_{t_L}g_*\right)^2v^2\simeq b\,(100\;{\textrm{GeV}})^2\left(\lambda_{t_L}g_*\right)^2\,.
\eeq
The result is of the right order of magnitude if $\lambda_{t_L}$ and $g_*$ are of order one, but it rapidly becomes too large for larger values. However we saw in the previous sections that large $g_*$ is phenomenologically welcome in order to push $m_*$ high for moderate $\xi=v^2/f^2$. For instance in Sect.~\ref{p4bos} we concluded that $m_*$ at least above $2.5$~TeV is required by EWPT, which means $g_*\gtrsim 3$ already for quite a small $\xi=0.1$. The $g_*^2$ enhancement in Eq.~(\ref{mh14p1}) costs fine-tuning, to be performed by an unnatural reduction of $b$. Obviously the tuning gets less and less severe the smaller the $\lambda_{t_L}^2$ prefactor is. The most favorable situation is thus case (\Rmnum{2}) in Eq.~(\ref{cases_comp}) where, compatibly with the generation of the top Yukawa, $\lambda_{t_L}=y_t$ and $\lambda_{t_R}=g_*$. This corresponds to the completely composite $t_R$ limit, which, as anticipated, is the most favorable one from the viewpoint of the Higgs mass generation. But even in this favorable situation some extra tuning is needed
\beq
\displaystyle
\Delta_{m_h}^{{\mathbf{14}}\oplus{\mathbf{1}}}=\frac{\left(b\right)_{\textrm{expected}}}{\left(b\right)_{\textrm{needed}}}=\left(\frac{100\;{\textrm{GeV}}}{126\;{\textrm{GeV}}}\right)^2y_t^2 g_*^2 \simeq 6\,\left(\frac{g_*}3\right)^2\,.
\eeq
This second cancellation is completely unrelated to the one needed for the VEV in Eq.~(\ref{tuning_vev_14}). We leave to the reader the choice of how to combine the two tunings to quantify the total degree of unnaturalness of the scenario. One might consider multiplying them, which basically means measuring the area of the allowed region in the $a$--$b$ plane, or summing them in quadrature by a logarithmic derivative definition of the tuning in the spirit of Ref.~\cite{Barbieri:1987fn}. The second option is more close to the interpretation of the tuning as degree of cancellation we emphasized in the Introduction.

The situation is different in the ${\mathbf{5}}\oplus{\mathbf{5}}$ case. A priori, it seems better since $\beta$ comes at higher order and thus
\beq
\label{mh5p5}
\displaystyle
(m_H^2)_{{\mathbf{5}}\oplus{\mathbf{5}}}=(1-\xi)\,b\,\frac{N_c}{2\pi^2}\lambda_t^4 v^2\simeq b\,(100\;{\textrm{GeV}})^2\lambda_t^4\,,
\eeq
which is insensitive to $g_*$. However now the elementary coupling that controls $m_H$ is not just $\lambda_{t_L}$ as it was for the ${\mathbf{14}}\oplus{\mathbf{1}}$. Both chiralities contribute to the potential and $\lambda_t$ in the above equation represents the maximum between $\lambda_{t_L}$ and $\lambda_{t_R}$. Decreasing $\lambda_{t_L}$ while increasing $\lambda_{t_R}$ does not help now and the most favorable situation is case (\Rmnum{1}) in Eq.~(\ref{cases_comp}), {\it{i.e.}}~$\lambda_{t_L}\simeq\lambda_{t_R}\simeq\sqrt{y_t g_*}$.\footnote{One should also take into account that the VEV tuning in Eq.~(\ref{delta_xi_5p5}) gets worse for smaller $\lambda_t$ so that the truly optimal situation comes from the balance among the two sources of tuning and depends on how these are combined in the total degree of unnaturalness. If we multiply them, minimal $\lambda_t$ ({\it{i.e.}}~case (\Rmnum{1}) ) is favored, if we sum them in quadrature a somewhat  larger value could be better.} In this configuration the tuning on $b$ is just identical to the one in the ${\mathbf{14}}\oplus{\mathbf{1}}$ with completely composite $t_R$
\beq
\displaystyle
\Delta_{m_h}^{{\mathbf{5}}\oplus{\mathbf{5}}}=\frac{\left(b\right)_{\textrm{expected}}}{\left(b\right)_{\textrm{needed}}}=\left(\frac{100\;{\textrm{GeV}}}{126\;{\textrm{GeV}}}\right)^2y_t^2 g_*^2 \simeq 6\,\left(\frac{g_*}3\right)^2\,.
\eeq
Given that a larger tuning for the VEV has to be paid in the  ${\mathbf{5}}\oplus{\mathbf{5}}$ case, the  ${\mathbf{14}}\oplus{\mathbf{1}}$ option emerges from this analysis as the more natural configuration. Further details on this point will be given in Sect.~\ref{sec:5+5_model}, where explicit models will be analyzed. It will turn out that the mechanism by which a correct $m_H$ is obtained in concrete models is not a cancellation of the ``$b$'' coefficient, but rather a further reduction of $\lambda_t$ due to anomalously light particles. However the qualitative conclusions on the tuning will be unaffected.

We now take one step back from explicit models and we outline the general message that emerges from this discussion. We saw a preference for small $g_*$ in order for a light enough Higgs boson being generated. This is in tension with phenomenology which prefers a large $g_*$ in order to increase $m_*=g_*f$. This result is robust and can be also established by the following argument. Rather than writing $m_H$ in terms of $\beta$ as in Eq.~(\ref{pot_tun_cond2}), we can express it in terms of $\alpha$ given that the condition $\alpha/\beta=2\xi$ will anyhow have to be imposed at some point. Namely
\beq
m_H^2=4(1-\xi)\,\alpha\,.
\eeq
In both models we considered, and actually in all known models based on the minimal coset, $\alpha$ is of ${\mathcal{O}}(\lambda^2)$ and it is universally estimated to be
\beq
\label{alpha_univ}
\displaystyle
\alpha=a\,\frac{N_c}{16\pi^2}\lambda_t^2 m_*^2\,,
\eeq
so that 
\beq
\displaystyle
m_H^2\simeq a\,\frac{N_c}{8\pi^2}\lambda_t^2 m_*^2\,.
\eeq
Irregardless of how much tuning will be needed to adjust the VEV, an amount of cancellation of at least
\beq
\displaystyle
\Delta\geq\frac{N_c}{8\pi^2}\lambda_t^2\left(\frac{m_*}{m_H}\right)^2\,,
\eeq
will come from the Higgs mass condition. If furthermore we notice that $\lambda_t$ is in no case below $y_t$, according to the general rule that weak couplings can not produce a strong one and to the discussion in Sect.~\ref{pfer}, we conclude that $\Delta$ is at least
\beq
\label{tuning_gen}
\displaystyle
\Delta\geq\frac{N_c}{8\pi^2}y_t^2\left(\frac{m_*}{m_H}\right)^2=\left(\frac{m_*}{450\,{\textrm{GeV}}}\right)^2\,.
\eeq
We recognize here the general tuning formula we discussed back in the Introduction in Eq.~(\ref{deltatuning}), with $\Lambda_{\textrm{SM}}$ identified with the resonance scale $m_*$. The existence of a tension among large $m_*$ and moderate tuning could thus have been guessed from the very beginning of these Notes.

This tension is definitely an issue, however its importance should not be overrated.  First of all it is based on an oder of magnitude estimate and on the intrinsically semiquantitative concept of tuning. Second, the tension is not that sharp because the tuning argument places limits on resonances of different nature than those that drive the phenomenological constraints. Namely, the potential is dominantly generated by the fermionic sector of the theory while the corrections to EWPT observable (to ${\widehat{S}}$ in particular) emerge from the bosonic sector. Clearly in the strict \mbox{$1$S$1$C} hypothesis fermionic and bosonic resonances are characterized by the same typical scale, however ${\mathcal{O}}(1)$ departures are possible or even expected. We cannot exclude a situation where the fermionic resonance scale $m_*$ that appears in the tuning formula in Eq.~(\ref{tuning_gen}) is below $1$ or $1.5$~TeV, ensuring a moderate tuning, while the bosonic $m_*$ that controls ${\widehat{S}}$ in Eq.~(\ref{Shatest}) is above $2$ or $2.5$~TeV. A completely analogous situation is encountered in supersymmetric models, where a tuning formula similar to Eq.~(\ref{tuning_gen})  controls the mass of some specific supersymmetric particles, namely stops and gluinos. This requires the latter states to be light, not the full superparticle spectrum. In explicit composite Higgs models the particles controlling the tuning are the ``top partners'', whose nature and phenomenology will be carefully described in Chaps.~\ref{ch:Pheno_models} and \ref{ch:LHC_pheno}. The top partners, exactly like the corresponding supersymmetric particles, happen to carry QCD color so that they can be copiously produced at hadron colliders by QCD interactions. If they are light as dictated by the tuning formula they should be discovered at the $14$~TeV LHC. Otherwise excluding their existence will push stronger and stronger bonds on the degree of unnaturalness of the model, and ultimately on its plausibility.

Also in view of possible exclusions, it is interesting to ask ourselves if and at what price, in terms of model-building complication, the tension could be structurally avoided. Namely we would like to find a model with low tuning, below around $10\%$, with colored resonances in the multi-TeV region, violating of Eq.~(\ref{tuning_gen}). This can be achieved if $\alpha$ is structurally smaller than its estimate in Eq.~(\ref{alpha_univ}), which for instance might occur if it emerged at ${\mathcal{O}}(\lambda_t^4)$, with the ${\mathcal{O}}(\lambda_t^2)$ contribution having been canceled by some selection rule. If it was so we would estimate
\beq
\displaystyle
m_H^2\simeq a\,\frac{N_c}{8\pi^2}\lambda_t^4 f^2=\frac{a}{\xi}\left(\frac{\lambda_t}{y_t}\right)^4(50\,{\textrm{GeV}})^2\,,
\eeq
and the sensitivity of $m_H$ to $m_*$ would be avoided. Concrete composite Higgs scenarios of this sort have recently been proposed \cite{Geller:2014kta,Barbieri:2015lqa,Low:2015nqa}, based on the so-called ``twin Higgs'' mechanism \cite{Chacko:2005pe}. In these constructions a light Higgs can be Naturally obtained even for maximal $g_*=4\pi$, {\it{i.e.}}~$m_*\sim10$~TeV at $\xi\simeq 0.1$. The only tuning, of order $1/\xi$, is the one associated with the VEV. Total $t_R$ compositeness is favored also in those models. Twin composite Higgs models require non-minimal cosets and extra model-building ingredients such as a doubling of the SM spectrum and an approximate twin parity symmetry. Describing these constructions would carry us faraway from the pedagogical purpose of these Notes. They constitute an active model-building direction on which progress might come from future investigations.

\begin{subappendices}

\section{Discrete symmetries}
\label{discrete}

Discrete symmetries are often useful in the study of composite Higgs theory. In the case of the minimal coset \mbox{SO${(5)}/$SO${(4)}$} the relevant ones are space--time parity P, charge conjugation C (often combined with P to form CP) and a $\Zdouble_2$ external automorphism of the algebra called $P_{{LR}}$.

Concerning parity, there is not much to say. It corresponds to ordinary spatial coordinate reflection under which the Goldstone boson Higgs transforms like a scalar and the gauge fields like vectors. Notice that the action of parity does not flip $L$ and $R$ \mbox{SO${(4)}$} generators given that the $L$--$R$ labeling does not refer here to fermion chirality. The CCWZ $d$ and $e$ symbols are vectors under parity and thus the ${\mathcal{O}}(p^2)$ bosonic Lagrangian (\ref{smgchk}) is accidentally P-invariant even if parity is not imposed as a symmetry of the composite sector. Composite sector breaking of P can emerge at ${\mathcal{O}}(p^4)$ through the operators discussed in Sect.~\ref{p4bos}. Parity is obviously broken by the elementary fermion couplings to the SM gauge fields, and the same holds for charge conjugation. Nevertheless, one might still want to impose them as  symmetries of the composite sector. Even if we will not consider this possibility here, we mention that in this case the chiral fermionic operators ${\mathcal{O}}_F^{L,R}$ that realize partial compositeness (see Sect.~\ref{sec:HCsigmaM}) would be supplemented by their opposite chirality P- and C-conjugate counterparts, with the same scaling dimensions.

Charge conjugation is less trivial. It acts as $H\rightarrow H^*$ on the complex Higgs field, which in the real fourplet notation (\ref{dh}) means
\beq
\ve\Pi\;\rightarrow\;{\mathcal{C}}_{\mathbf{4}}\,\ve\Pi\,,\;\;\;\;\textrm{where}\;\; {\mathcal{C}}_{\mathbf{4}}=\textrm{diag}(-1,+1,-1,+1)\,.
\eeq
Notice that ${\mathcal{C}}_{\mathbf{4}}$ is a unit-determinant orthogonal matrix and as such it is a proper element of the unbroken group \mbox{SO${(4)}$}. Namely, it is
\beq
\label{c_exp}
\displaystyle
{\mathcal{C}}=e^{i\,\pi[T_L^2+T_R^2]}\,,
\eeq
which, with the suitable generator matrices, can be expressed in any representation of the complete group \mbox{SO${(5)}$} or of the unbroken subgroup \mbox{SO${(4)}$}. Given that the charge conjugation operation happens to act on the Goldstone fields like an element of the unbroken symmetry group, we can simply use the results of Sect.~\ref{CCWZ} to derive its action on the Goldstone matrix, which is
\beq
\label{gmc}
U_{\mathbf{r}}[\Pi]\;\rightarrow\;{\mathcal{C}}_{\mathbf{r}}\,U_{\mathbf{r}}[\Pi]\,{\mathcal{C}}_{\mathbf{r}}^{-1}\,,
\eeq
for a generic representation ${\mathbf{r}}$.

On the SM gauge fields, C acts as $W^\alpha\rightarrow  (-)^{1-\delta_{\alpha,2}}W^\alpha$ and $B_\mu\rightarrow -B_\mu$. This can be uplifted to the transformation rule
\beq
A_\mu\hspace{-2pt}=\hspace{-2pt}A_{\mu,\,A}T^A \rightarrow \;{\mathcal{C}}\cdot A_{\mu} \cdot {\mathcal{C}}^{-1}\,,
\eeq
which we assign to the whole set of dynamical and non-dynamical sources that gauge \mbox{SO$(5)$}. Therefore C coincides with \mbox{$\mathcal{C}\in$\;SO${(4)}$} even when acting on the $A_\mu$ sources. This makes very easy to work out the transformation rules of the $d$ and the $e$ symbols. They are just a fourplet and an adjoint of \mbox{SO$(4)$} and thus
\beq
d_\mu^i\rightarrow ({\mathcal{C}}_{\mathbf{4}})^i_{\;j}d_\mu^j\,,\;\;\;\;\; {e_{L,R}^\alpha}_\mu\rightarrow (-)^{1-\delta_{\alpha}^2} {e_{L,R}^\alpha}_\mu\,.
\eeq
Furthermore, the ${\mathcal{C}}$ operation is automatically a symmetry of our Lagrangian and thus charge conjugation invariance is guaranteed for all the composite sector operators involving $d$ and $e$ only. This includes the ${\mathcal{O}}(p^2)$ Lagrangian and the ${\mathcal{O}}(p^4)$ operators of Sect.~\ref{p4bos}. Notice that charge conjugation coincides with $\mathcal{C}$ only for the Goldstones and for the $A_\mu$ sources, not for the \mbox{U$(1)_X$} source $X_\mu$. Given that we embed the hypercharge gauge boson $B_\mu$ in it, it must transform with a minus sign
\beq
X_\mu\rightarrow -X_\mu\,.
\eeq
This sign flip needs not to be a symmetry of the theory, therefore C can be broken, but only through terms with odd powers of the \mbox{U$(1)_X$} source. Given that the latter can only enter through its field-strength tensor $\partial_{[\mu}X_{\nu]}$ because of local invariance, C breaking is postponed to high orders in the derivative expansion and it does not emerge at ${\mathcal{O}}(p^4)$ in the bosonic sector. 

Let us now turn to the fermionic sector. Given that both P and C are broken by the SM couplings it is not worth trying to define their actions on the fermionic source fields. This makes sense instead for the product of the two symmetries, CP, which is preserved by the SM matter quantum number assignment. We take, as is normally done in the SM, the CP action to be \mbox{$\chi(\ve{x},t)\rightarrow\chi^c(\ve{x},t)\hspace{-2pt}=\hspace{-2pt}-i\gamma^2\gamma^0 \chi^*(-\ve{x},t)$}, where $\chi$ denotes any of the elementary SM fields. Given the definition, the action on the fermionic sources in the various representations introduced in Sect.~\ref{sec:HCsigmaM} is immediately worked out. In all cases where the elementary SM fermions are embedded in a real \mbox{SO${(5)}$} representation such as the $\mathbf{5}$, the $\mathbf{10}$ or the $\mathbf{14}$, it is easy to verify that CP acts as the global $\mathcal{C}$ rotation in the appropriate representation times the $\chi\rightarrow\chi^c$ operation. For instance, in the case of the  $\mathbf{5}$ we have
\beq
(F)_I\rightarrow ({\mathcal{C}}_{\mathbf{5}})_I^{\;\;J}  (F^c)_I\,,\;\textrm{with}\;\;\;{\mathcal{C}}_{\mathbf{5}}=\textrm{diag}(-1,+1,-1,+1,+1)\,,
\eeq
where $F$ denotes in general the top ($Q_{t_L}$ and $T_R$) or bottom ($Q_{b_L}$ and $B_R$) sector fermionic sources. In order to construct the CCWZ invariants, as explained in Sect.~\ref{sec:HCsigmaM}, it is useful to define dressed sources by acting with the inverse of the Goldstone matrix. Given how the latter transforms, as in Eq.~(\ref{gmc}), their CP transformation reads
\beq
F_{\mathbf{r}}\rightarrow {\mathcal{C}}_{\mathbf{r}}F_{\mathbf{r}}^c\,,
\eeq
where $\mathbf{r}$ is the  \mbox{SO$(4)$} representation where the dressed source lives. We see that CP acts as the \mbox{SO$(4)$} transformation $\mathcal{C}$, under which all the operators are automatically invariant, times the ``intrinsic'' CP operation $\chi\rightarrow\chi^c$. Since the same holds for the bosonic fields, with the only exception of $X_\mu$, this makes very easy to establish the CP quantum numbers of the operators. For instance all the ${\mathcal{O}}(p^0)$ operators in Sect.~\ref{sec:HCsigmaM} are CP-even, once their coefficients are set to a real value to obtain a real mass, while some of those of ${\mathcal{O}}(p)$ in Sect.~\ref{pfer} break CP.

This was for real representations. When the elementary SM fermions are in the complex spinorial $\mathbf{4}$, instead, no imaginary phase is introduced in the embeddings, see Eq.~(\ref{embspin}) and therefore CP is just
\beq
(F)_i \rightarrow  (F^c)_i\,,
\eeq
when acting on the sources. We actually need the transformation property of the dressed sources, obtained by acting with the inverse Goldstone matrix and splitting the fourplet into two doublets, namely
\beq
\label{dressed_sp}
\left[\begin{array}{c}F^{\mathbf{2_L}}\\
F^{\mathbf{2_R}}\end{array}
\right] = U_{\mathbf{4}}^{-1} F\,.
\eeq
The Goldstone matrix transformation is immediately obtained from Eq.~(\ref{c_exp})
\beq
U_{\mathbf{4}}\;\rightarrow\;\hat{\mathcal{C}}_{\mathbf{4}}\cdot U_{\mathbf{4}}\cdot\hat{\mathcal{C}}_{\mathbf{4}}^{-1}\,,\;{\textrm{where}}\;\;\;\hat{\mathcal{C}}_{\mathbf{4}}=\left[\begin{array}{cc}
i\sigma_2 & 0\\
0& i\sigma_2
\end{array}\right]\,.
\eeq
This can be rewritten, using the symplectic condition in Eq.~(\ref{symp_cond}), in a seemingly more complicated way
\beq
\label{ccmat}
U_{\mathbf{4}}\;\rightarrow\;\hat{\mathcal{C}}_{\mathbf{4}}\cdot\Omega^{-1}\cdot U_{\mathbf{4}}^*\cdot\Omega\cdot \,
\hat{\mathcal{C}}_{\mathbf{4}}^{-1}\,,\;{\textrm{with}}\;\;\;\Omega\cdot\hat{\mathcal{C}}_{\mathbf{4}}^{-1}=\left[\begin{array}{cc}
\Id_2& 0\\
0& -\Id_2
\end{array}\right]\,,
\eeq
in terms of the complex conjugate of the Goldstone matrix. This becomes useful if we take into account that the physical fields are always embedded in the source $F$ in either the first two components of the fourplet or in one of the two last (we denote by $F_\pm$ the two cases) but never in both at the same time. Therefore the matrix $\Omega\cdot\hat{\mathcal{C}}_{\mathbf{4}}$ reduces to either an overall plus or minus sign when acting on them, leading eventually to the following result
\beq
\left[\begin{array}{c}F^{\mathbf{2_L}}\\
F^{\mathbf{2_R}}\end{array}
\right]\;\rightarrow\;\left[\begin{array}{c}\pm (F_\pm^{\mathbf{2_L}})^c\\
\mp(F_\pm^{\mathbf{2_R}})^c\end{array}
\right]\,.
\eeq
Notice that the `` $^c$ '' operation acts now on the dressed sources and thus it entails taking the complex conjugate of $U_{\mathbf{4}}$, which is where Eq.~(\ref{ccmat}) comes into play.

We now discuss $P_{LR}$. As the name suggests, it is a $\Zdouble_2$ transformation that interchanges $L$ and $R$ generators of the \mbox{SO${(4)}$} group in \mbox{SO${(5)}$}. It corresponds to parity in the \mbox{SO${(5)}$} space and it is represented, in the fundamental, by the matrix
\beq
\label{plr5}
P_{LR}^{\mathbf{5}}=\textrm{diag}(-1,-1,-1,+1,+1)\,.
\eeq
It acts on the generators as
\bea
\label{plralgebra}
&P_{LR}^{\mathbf{5}} T_L^\alpha P_{LR}^{\mathbf{5}}=T_R^\alpha\,,\;\;\;\;
P_{LR}^{\mathbf{5}} T_R^\alpha P_{LR}^{\mathbf{5}}=T_L^\alpha\,,&\nonumber\\
&P_{LR}^{\mathbf{5}} \Th^i P_{LR}^{\mathbf{5}}=\left({{P}}_{LR}^{\mathbf{4}}\right)^{i}_{\;\;j} \Th^j\,,&
\eea
where ${{P}}_{LR}^{\mathbf{4}}$ is
\beq
\label{plrfund}
{{P}}_{LR}^{\mathbf{4}}=\textrm{diag}(-1,-1,-1,+1)\,.
\eeq
The $P_{LR}$ operation belongs to \mbox{O${(4)}\subset\;$O${(5)}$}, therefore it is not an element of the symmetry group and thus it is not automatically a symmetry of the composite sector. It could be imposed or more interestingly, as in the case encountered in Sect.~\ref{pfer}, emerge as an accidental symmetry. Notice that the fourth real Higgs component, $\Pi^4$, is $P_{LR}$-even. Therefore $P_{LR}$, provided it was a symmetry of some sector of the theory, will not be broken spontaneously by the Higgs VEV. On the Goldstone fourplet and on the Goldstone matrix in the fundamental, $P_{LR}$ acts, respectively, as
\beq
\ve\Pi\rightarrow {{P}}_{LR}^{\mathbf{4}}\ve\Pi\,,\;\;\;\;\;\; U[\Pi]\rightarrow P_{LR}^{\mathbf{5}}\cdot U[\Pi]\cdot P_{LR}^{\mathbf{5}}\,,
\eeq
out of which the $d$ and $e$ symbols transformation rules (including the terms with the gauge sources, whose transformation rule is defined below) are found to be
\beq
d_{\mu,\,i}\rightarrow \left({{P}}_{LR}^{\mathbf{4}}\right)_i^{\;\;j}d_{\mu,\,j}\,,\;\;\;\;\;e_{\mu,\,\alpha}^L\leftrightarrow e_{\mu,\,\alpha}^R\,.
\eeq

By following this logic, $P_{LR}$ can be defined also on the elementary gauge and fermionic source fields. Of course $P_{LR}$, differently from CP discussed above, is not a symmetry of the elementary sectors, therefore the SM field embedding into the sources will normally break it completely. Nevertheless we can assign transformation properties, for instance
\beq
A_\mu \rightarrow \;P_{LR}\cdot A_{\mu} \cdot P_{LR}^{-1}\,,\;\;\;\;\; F_I\rightarrow (P_{LR}^{\mathbf{5}})_I^{\;\;J}F_J\,,
\eeq
to gauge and to fermions in the fundamental, respectively. The dressed fermion sources transformation rules, given how the Goldstone matrix transforms, are at this point completely obvious. 

The action of $P_{LR}$ on all the representations obtainable as tensor products of \mbox{SO${(5)}$} fiveplets are immediately inferred from Eq.~(\ref{plr5}): it will be sufficient to act with the $P_{LR}^{\mathbf{5}}$ parity on each index. Some work is instead needed to obtain the representation on the spinorial, which turns out to be
\beq
\widehat{P}_{LR}^{\mathbf{4}}=\left[
\begin{array}{cc}
0 & \Id\\
\Id & 0
\end{array}
\right]\,,
\eeq
not to be confused with ${P}_{LR}^{\mathbf{4}}$ acting on the \mbox{SO${(4)}$} fourplets. It is easy to verify, given the generators of the spinorial reported in Eq.~(\ref{gen45}), that $\widehat{P}_{LR}^{\mathbf{4}}$ correctly acts on them like for the ones in the fundamental representation in Eq.~(\ref{plralgebra}). Not surprisingly, being a $L$--$R$ interchange, $P_{LR}$ flips the $(\mathbf{2},\mathbf{1})$ and $(\mathbf{1},\mathbf{2})$ components of the fourplet. On the Goldstone matrix we obviously have
\beq
U_{\mathbf{4}}[\Pi]\rightarrow P_{LR}^{\mathbf{4}} \cdot U_{\mathbf{4}}[\Pi]\cdot P_{LR}^{\mathbf{4}}\,,
\eeq
and thus the dressed sources in Eq.~(\ref{dressed_sp}) simply get interchanged 
\beq
F^{\mathbf{2_L}}\;\leftrightarrow\;F^{\mathbf{2_R}}\,.
\eeq

\end{subappendices}

\chapter{Flavor}
\label{ch:flavor}

As we saw in Chap.~\ref{ch:Goldstone}, a fundamental ingredient of the composite Higgs scenarios
is the partial compositeness hypothesis, which provides a general framework to describe the
Standard Model (SM) fermions and to generate their masses and couplings. In most of the previous discussions we
focused our attention on the third-generations quarks, and in particular on the top.
In fact, due to its large mass, the top is usually the elementary state with the largest mixing
with the composite sector and is the one that almost completely determines the dynamics of
Electro-Weak Symmetry Breaking (EWSB).

Constructing a complete model, however, also requires a description of the light
fermions and the implementation of the three-families structure of the SM.
This is a non trivial task due to the highly non-generic features of the SM flavor structure.
The first peculiar aspect is the presence of a hierarchy of masses among
the different quark and lepton generations. A second important feature is the suppression of flavor-violating effects
due to a set of accidental flavor symmetries. In particular, flavor-changing transitions mediated
by the $Z$-boson and by the Higgs are extremely suppressed.
The only sizable flavor-violating effects are due to the $W$-boson couplings
and are controlled by two mixing matrices, $V_{\textrm{CKM}}$ in the quark sector and $V_{\textrm{PMNS}}$ in the lepton sector.
It is also remarkable the fact that flavor violation in the quark and lepton sectors seems to
follow two very different patterns. In the quark sector the $V_{\textrm{CKM}}$ matrix
is close to the identity with a strong hierarchical structure that suppresses off-diagonal couplings.
On the contrary, in the lepton sector a completely
anarchic $V_{\textrm{PMNS}}$ matrix seems preferred by the experimental data.
The expected order-one flavor-violation effects,
however, are present only in the neutrino sector, whereas
they are strongly suppressed for charged leptons due to the smallness of the neutrino masses.

Another important feature of the SM flavor structure is related to CP violation. The flavor symmetries of the
three-generation structure, broken only by the Yukawa couplings, allow to remove from each mixing matrix
all complex phases except one. Moreover, if only two (or one) quark generations are considered all the
complex phases can be removed, thus completely forbidding CP violation. An important consequence of this structure
is the strong suppression of Electric Dipole Moments (EDM's), in accordance with the strong experimental bounds.

As it can be easily understood, reproducing the SM flavor structure in a Beyond the SM (BSM) scenario can be quite challenging.
In fact the presence of additional dynamics usually breaks the accidental flavor symmetries of the SM
and leads to large flavor-violating effects. In the composite Higgs scenarios, in particular, dangerous effects can
come from the non-linear Higgs dynamics and from the presence of extra particles. For instance, spin-$1$ composite-sector (vector) resonances can mediate Flavor Changing Neutral Currents (FCNC's), while fermionic resonances (or fermionic partners) can introduce new large CP violating phases or modify the $Z$ and $W$ couplings. 
In this chapter we will see how, by a judicious extension of the partial compositeness framework,
these problems can be kept under control and the composite Higgs scenario can be endowed with a realistic flavor structure.

Several alternative flavor constructions can be conceived.
The ``classical'' implementation is the so-called ``anarchic'' scenario~\cite{Kaplan:1991dc,Grossman:1999ra,Huber:2000ie,Gherghetta:2000qt,Huber:2003tu,Agashe:2004cp},
which is probably the one that most directly follows and most fully exploits the partial compositeness hypothesis.
As we will see in Sect.~\ref{sec:anarchic_pc}, this scenario gives a dynamical origin to the hierarchies
of the fermion masses. Remarkably, the same mechanism automatically generates a hierarchical structure for the
$V_{\textrm{CKM}}$ matrix and a suppression of the flavor-violating effects involving the light SM fermions.
In spite of the successes of the anarchic scenario, some residual tension with the experimental data remains.
This led to the exploration of alternative constructions. These constructions usually do not offer an explanation
for the mass hierarchies, but can more efficiently suppress flavor-violating effects thanks to
the introduction of suitable flavor symmetries. We will discuss this class of models in Sect.~\ref{sec:flavor_symm}.

\section{Anarchic partial compositeness}\label{sec:anarchic_pc}

\index{partial compositeness}
We start our discussion of the flavor structure by considering the anarchic scenario.
In this and in the following two sections we will focus exclusively on the
quark sector, which is the one that so far has been most thoroughly scrutinized in the literature.
A discussion of the flavor structure of the lepton sector, for which instead only a limited
literature is available, is postponed to Sect.~\ref{sec:leptonic_sector}.

As we saw in Chap.~\ref{ch:Goldstone}, the partial compositeness framework is based
on the assumption that the matter fermions, realized as elementary fields, are linearly mixed
with some composite operators (see Eq.~(\ref{intpc})). The elementary/composite mixing
is external with respect to the composite sector dynamics and is generated at a very high energy scale $\Lambda_{\textrm{UV}}$.
By a simple generalization of Eq.~(\ref{intpc}), we can write the structure of the mixing
in the presence of multiple generations of elementary fermions. In the case of the up-type elementary singlets $u_R^i$,
for instance, we can write
\begin{equation}\label{eq:flavor_mixing}
{\mathcal L}_{\textrm{int}}[\Lambda_{\textrm{UV}}] = \frac{\lambda^{ij}_{u_R}}{\Lambda_{\textrm{UV}}^{d^j_{u_R} - 5/2}}\,
\overline u_R^i {\cal O}^{u_R\,j}_F + \mathrm{h.c.}\,,
\end{equation}
where $i,j = 1,2,3$ are family indices. 
In writing Eq.~(\ref{eq:flavor_mixing}), we considered arbitrary scaling dimensions  $d^j_{u_R}$ for the composite fermionic operators ${\cal O}^{u_R\,j}_F$. The $\lambda_{u_R}^{ij}$ matrix controls the strength of the couplings at the scale $\Lambda_{\textrm{UV}}$
and has, in general, an anarchic structure, {\it{i.e.}}~all its elements (including the off-diagonal ones)
are of the same order. This kind of structure is expected if all the couplings
are generated by a generic UV theory that does not possess any flavor symmetry. The fermionic operators ${\cal O}^{u_R\,1,2,3}_F$ are defined in the UV, where the global symmetry group of the composite sector is linearly realized. Therefore, as explained in Sect.~\ref{sec:HCsigmaM}, each of them must be part of some complete representation of the group. The same representation is assumed for all the three operators.

\index{RG running}
The elementary quarks interactions at low energy are related to the ones in Eq.~(\ref{eq:flavor_mixing})
by the Renormalization Group evolution (see Eq.~(\ref{coupev})):
\begin{equation}\label{eq:low-en_couplings_flavor}
\lambda^{ij}_{u_R}[m_*] \simeq \lambda^{ij}_{u_R} \left(\frac{m_*}{\Lambda_{\textrm{UV}}}\right)^{d^j_{u_R} - 5/2}
\equiv \widetilde \lambda\, c^{ij}_{u_R}\, \zeta^j\,,
\end{equation}
where we split the coupling matrix $\lambda^{ij}_{u_R}$ into an overall normalization $\widetilde \lambda$
and an anarchic matrix with order one entries, $c^{ij}_{u_R}$. The components of the vector 
\beq\zeta^j=(m_*/\Lambda_{\textrm{UV}})^{d^j_{u_R} - 5/2}\,,\eeq
are instead Naturally hierarchical, given that $m_*\ll\Lambda_{\textrm{UV}}$, if the composite operators have different scaling dimensions. We order them such that $\zeta^i \ll \zeta^j$, {\it{i.e.}}  $d^i_{u_R}>d^j_{u_R}$,  for $i < j$. The elementary/composite mixing Lagrangian in the IR, namely at the scale $m_*\sim$~TeV, can be rewritten as
\begin{equation}\label{eq:Lint_low-en_flavor}
{\mathcal L}_{\textrm{int}}[m_*] = \widetilde \lambda\, c^{ij}_{u_R}\, \zeta^j\, \overline u^i_R {\cal O}^{u_R\,j}_F + \mathrm{h.c.}\,,
\end{equation}
where we absorbed into the composite operator normalization the powers of $m_*$ needed to match the energy dimension.
At low energy the composite operators ${\cal O}^{u_R\,j}_F$ are not distinguished by any quantum number, we can thus
redefine them in order to put the elementary/composite mixing in a convenient form. In particular, we can perform a rotation of the operators, accompanied by a rotation of the elementary $u_R^i$ in the flavor space, and put the mixing in a diagonal form.\footnote{In the UV, where the composite sector is close to the fixed point, the operators are characterized by their different scaling dimensions $d^j_{u_R}$, which can be regarded as their eigenvalue under dilatation. The presence of this additional quantum number makes them distinguishable and does not allow to rotate them.}

To understand the structure of the result of this diagonalization we first need to state a little theorem on the singular value decomposition of a matrix. We consider a matrix of the form
\begin{equation}\label{eq:M_ij_th}
M^{ij} = \zeta_L^i c^{ij} \zeta_R^j\,,
\end{equation}
where $c^{ij}$ is an anarchic matrix with ${\cal O}(1)$ entries and $\zeta_{L,R}$ are vectors with generic entries, which we order so that
\begin{equation}
\zeta_L^i \leq \zeta_L^j\,, \qquad \zeta_R^i \leq \zeta_R^j\,,
\qquad \textrm{for} \quad i < j\,.
\end{equation}
The theorem will provide interesting information only when one or both the $\zeta_{L,R}$ vectors have strongly hierarchical components, {\it{i.e.}}~when a strong ordering $\zeta_{L,R}^i \ll \zeta_{L,R}^j$ for $i < j$ is present. The theorem states that the singular value decomposition of $M$
\begin{equation}
M^{ij} = U_L^{ik} m^{kl} (U_R^\dagger)^{lj}\,,
\end{equation}
involves a real diagonal matrix $m$ with entries of order
\begin{equation}
\label{eq:diag_th}
m^{ii} \sim \zeta_L^i \zeta_R^i\,,
\end{equation}
and that the elements of the $U_{L}$ unitary transformation are
\begin{equation}\label{eq:UL_rot}
U_L^{ij} \sim
\left\{
\begin{array}{l@{\qquad\textrm{for}\quad}l}
{\zeta_L^i}/{\zeta_L^j} & i < j\,,\\
\rule{0pt}{1.15em}1 & i = j\,,\\
\rule{0pt}{1.15em}{\zeta_L^j}/{\zeta_L^i} & i > j\,,
\end{array}
\right.
\end{equation}
and analogously for $U_R$. Clearly, the result is non-trivial only for hierarchical $\zeta_L$ or $\zeta_R$. In this case the singular values of the $M^{ij}$ matrix are hierarchical,
\begin{equation}
m^{ii} \ll m^{jj} \qquad \textrm{for} \quad i<j\,,
\end{equation}
and the unitary transformations corresponding to the hierarchical $\zeta_{L,R}$ vectors are close to the
identity.

We now apply the theorem to the diagonalization of Eq.~(\ref{eq:Lint_low-en_flavor}), where $\zeta_L^i$ is of order one and $\zeta_R=\zeta$ is strongly ordered. After an anarchic $U_R$ rotation on the elementary fields and a hierarchical $U_L$ rotation on the composite operators, the elementary/composite Lagrangian takes the diagonal form 
 \begin{equation}
{\mathcal L}_{\textrm{int}} = \widetilde \lambda^{i}\, \overline u^i_R {\cal O}^{u_R\,i}_F + \mathrm{h.c.}\,,
\end{equation}
with real hierarchical mixing coefficients of order
\begin{equation}
\widetilde \lambda^i \sim \widetilde \lambda\, \zeta^i\,.
\end{equation}

A similar analysis can be performed for the other SM quark multiplets \mbox{$q_L^i = \{ u^i_L, d^i_L\}$} and $d^i_R$. Provided each of them mixes in the UV with only one set of composite operators as we assumed to be the case for $u^i_R$ in Eq.~(\ref{eq:flavor_mixing}), the elementary/composite Lagrangian can be put in a fully diagonal form
\begin{equation}\label{eq:single_mixing_lagr}
{\mathcal L}_{\textrm{int}} = \lambda_{q_L}^{i}\, \overline q^i_L {\cal O}^{q_L\,i}_F
+ \lambda_{u_R}^{i}\, \overline u^i_R {\cal O}^{u_R\,i}_F
+ \lambda_{d_R}^{i}\, \overline d^i_R {\cal O}^{d_R\,i}_F + \mathrm{h.c.}
\end{equation}
When instead one elementary field mixes with more sets of composite operators with different quantum numbers, the situation changes. More terms will be present in the interaction Lagrangian and they will not, in general, assume a diagonal form. Indeed, the diagonalization procedure outlined above requires one rotation being performed on the elementary fields, therefore only the mixings with one set of composite operators can be diagonalized. The scenarios where only one set of composite operators mixes with each SM quark representation, and Eq.~(\ref{eq:single_mixing_lagr}) holds, are denoted as  ``single mixing'' scenarios. In some cases, as we will see below, the presence of multiple mixings is however unavoidable to obtain a realistic model.

Several options exist for the choice of the $\SO(5) \times \U(1)_X$ global group representation the composite operators belong to, some of which are described in Sect.~\ref{sec:HCsigmaM}. Valid representations must fulfill two basic requirements. First of all, they must contain the appropriate SM multiplets in their decomposition. Second, the representation of the operators associated to the left and the right chiralities must be ``compatible'' with each other, in the sense that they must allow for the generation of the Yukawa couplings from the composite dynamics. The up-type Yukawa's come in the low-energy theory from composite sector two point functions $\langle {\mathcal O}_F^{q_L\,i} {\mathcal O}_F^{u_R\,j} \rangle$, and similarly for the down-type ones. These correlators can be non-vanishing only if they respect the unbroken group, {\it{i.e.}}~if the left and right composite operators contain the same representation of $\SO(4) \times \U(1)_X$.\footnote{The \mbox{SU$(3)_c$} color group is automatically respected, since all the operators are color triplets in order to mix with the elementary quarks.}

The simplest example is the ${\bf 4}_{1/6}$ spinorial representation we discussed in Sect.~\ref{sec:HCsigmaM}. It decomposes under $\SO(4) \times \U(1)_X \simeq \SU(2)_L \times \SU(2)_R \times \U(1)_X$ and eventually under the SM gauge group as
\begin{equation}
{\bf 4}_{1/6} \rightarrow ({\bf 2}, {\bf 1})_{1/6} \oplus ({\bf 1}, {\bf 2})_{1/6}
\rightarrow {\bf 2}_{1/6} \oplus {\bf 1}_{2/3} \oplus {\bf 1}_{-1/3}\,.
\label{eq:4decomposition}
\end{equation}
Given that it contains the SM representations of one complete quark family, the ${\bf 4}_{1/6}$ could be taken as the representation of all the composite operators ${\mathcal O}_F^{q_L}$, ${\mathcal O}_F^{u_R}$ and ${\mathcal O}_F^{d_R}$. This does not mean that the three sets of operators are identified as the components of one single multiplet. Three different multiplets, all in the ${\bf 4}_{1/6}$ representation of $\SO(5) \times \U(1)_X$ but characterized by different scaling dimensions $d_{q_L}$, $d_{u_R}$ and $d_{d_R}$, are introduced. Otherwise, all the elementary/composite mixings would have the same hierarchies and we would not be able to reproduce the quark masses and the $V_{\textrm{CKM}}$ matrix. The ${\bf 4}_{1/6}$, where three sets of operators are introduced as described above, is the simplest example of a ``single mixing'' scenario. Identical considerations hold for fermionic operators in the ${\bf 10}_{2/3}$ representation, which also contains one full SM quark family in its decomposition
\begin{eqnarray}
{\bf 10}_{2/3} & \rightarrow & ({\bf 2}, {\bf 2})_{2/3} \oplus ({\bf 3}, {\bf 1})_{2/3}
\oplus ({\bf 1}, {\bf 3})_{2/3} \nonumber\\
\rule{0pt}{1.1em} & \rightarrow & {\bf 2}_{7/6} \oplus {\bf 2}_{1/6} \oplus {\bf 3}_{2/3} \oplus {\bf 1}_{5/3} \oplus {\bf 1}_{2/3}
\oplus {\bf 1}_{-1/3}\,.
\label{eq:10_23decomposition}
\end{eqnarray}

The situation is different if we embed the operators in the fundamental $\SO(5)$ representation.
In order to find a component with the quantum numbers of the $u_R$ singlets we need to consider
the ${\bf 5}_{2/3}$ multiplet:
\begin{equation}
{\bf 5}_{2/3} \rightarrow ({\bf 2}, {\bf 2})_{2/3} \oplus ({\bf 1}, {\bf 1})_{2/3}
\rightarrow {\bf 2}_{7/6} \oplus {\bf 2}_{1/6} \oplus {\bf 1}_{2/3}\,.
\label{eq:5_23decomposition}
\end{equation}
This representation contains a doublet with the quantum numbers of the $q_L$
elementary fields, thus it can also be used to embed the ${\mathcal O}_F^{q_L}$
operators on top of ${\mathcal O}_F^{u_R}$. Having one ${\bf 5}_{2/3}$ multiplet mixing with the $u_R$ and another one mixing with the $q_L$ is already sufficient to generate the up-type Yukawa couplings that emerge, as previously explained, from the correlators of the left and the right composite operators. The ${\bf 5}_{2/3}$ representation, however, does not contain any state with the quantum numbers of the down-type singlets. The operator ${\mathcal O}_F^{d_R}$, mixing with the $d_R$, must then be taken to be in a different representation. One possibility is to consider the ${\bf 5}_{-1/3}$, which decomposes as
\begin{equation}
{\bf 5}_{-1/3} \rightarrow ({\bf 2}, {\bf 2})_{-1/3} \oplus ({\bf 1}, {\bf 1})_{-1/3}
\rightarrow {\bf 2}_{1/6} \oplus {\bf 2}_{-5/6} \oplus {\bf 1}_{-1/3}\,.
\label{eq:5_13decomposition}
\end{equation}
Introducing only this mixing is however not sufficient to generate the down-type Yukawa because ${\mathcal O}_F^{d_R}\in{\bf 5}_{-1/3}$ is ``incompatible'' with the left chirality operator ${\mathcal O}_F^{q_L}\in{\bf 5}_{2/3}$. Namely, they can not have a non-vanishing two-point function because they have different charges under the unbroken $\U(1)_X$ group. A second mixing, with a second set of operators ${{\mathcal O}'}_F^{q_L}$, must then be introduced for the $q_L$ elementary fields. Given that the ${\bf 5}_{-1/3}$ contains the $q_L$ doublet in its decomposition, the simplest choice that allows for the generation of down-type Yukawa's is to take ${{\mathcal O}'}_F^{q_L}$ in the ${\bf 5}_{-1/3}$ like ${\mathcal O}_F^{d_R}$. This scenario, where an additional set of operators needs to be introduced, does not respect the ``single mixing'' hypothesis and the mixing Lagrangian contains now an additional term
\begin{equation}\label{eq:double_mixing_lagr}
{\mathcal L}_{\textrm{int}} = \lambda_{q_L}^{i}\, \overline q^i_L {\cal O}^{q_L\,i}_F
+ \lambda_{u_R}^{i}\, \overline u^i_R {\cal O}^{u_R\,i}_F
+ \lambda'^i_{q_L}\, \overline q^i_L {{\cal O}'}^{q_L\,i}_F
+ \lambda_{d_R}^{i}\, \overline d^i_R {\cal O}^{d_R\,i}_F + \mathrm{h.c.}\,.
\end{equation}
In the above equation, a diagonal form has been given to the extra mixing $\lambda'$. However, as explained above, this is not the most general situation given that the extra mixing can not be always diagonalized. A diagonal form for the extra mixing is however phenomenologically required to avoid large flavor violating effects, therefore it will be assumed to be so in what follows. Speculations on how  some symmetry of the UV theory might force the two $q_L$ mixings to be aligned and thus simultaneously diagonalizable are discussed in Ref.~\cite{Csaki:2008eh}.

The last example we mention is the scenario with operators embedded in the symmetric ${\bf 14}$
representation. The ${\bf 14}_{2/3}$ multiplet, whose decomposition reads
\begin{eqnarray}
{\bf 14}_{2/3} & \rightarrow & ({\bf 2}, {\bf 2})_{2/3} \oplus ({\bf 3}, {\bf 3})_{2/3} \oplus ({\bf 1}, {\bf 1})_{2/3} \nonumber\\
& \rightarrow & {\bf 2}_{7/6} \oplus {\bf 2}_{1/6} \oplus {\bf 3}_{5/3} \oplus {\bf 3}_{2/3} \oplus {\bf 3}_{-1/3}
\oplus {\bf 1}_{2/3}\,,
\label{eq:14_23decomposition}
\end{eqnarray}
can serve as the representation of ${\cal O}^{q_L}_F$ and of ${\cal O}^{u_R}_F$. Analogously to the case of the fiveplet, a different multiplet (as, for instance, the ${\bf 14}_{-1/3}$) is needed for ${\mathcal O}_F^{d_R}$ and ${{\mathcal O}'}_F^{q_L}$ in order to generate the down-type Yukawa's. The ``single mixing'' hypothesis is violated also in this case.

\subsection*{Quark masses and mixings}

We can now discuss the generation of the quark masses and of the $V_{\textrm{CKM}}$ matrix. We saw in the previous chapters how to estimate the low-energy effective operators involving elementary fields that are linearly coupled to the composite sector. By applying those results we find that the
Yukawa couplings\footnote{For shortness we will not make an explicit distinction between the
couplings to the Higgs field responsible to generate the mass matrices and the linear couplings of the Higgs
fluctuations to the quarks. Obviously, due to the non-linear Higgs dynamics, the two things are in general different
and only coincide at leading order in the $v/f$ expansion.
In the following we will denote both couplings by ``Yukawa's'' and leave the exact interpretation of the
concept to the context.}
of the quarks have the structure \footnote{Below we report the structure of the Yukawa matrices at  leading-order in the $\lambda/g_*$ expansion. Subleading effects from of order $\lambda^2$ modifications of the kinetic terms induced by the composite sector will be discussed in Sect.~\ref{sec:h_couplings_flavor}.}
\begin{equation}\label{eq:Yukawas_1}
\displaystyle
y_u^{ij} = \frac{\lambda_{q_L}^i \lambda_{u_R}^{j}}{g_*} c_{ij}\,,
\qquad
\displaystyle
y_d^{ij} = \frac{\lambda'^i_{q_L} \lambda_{d_R}^{j}}{g_*} c'_{ij}\,,
\end{equation}
where $g_*$ is the typical coupling strength of the composite sector. The above formulae are valid in the scenarios in which the $q_L$ doublet is mixed with two sets
of composite operators (as for instance in the models based on the fundamental $\SO(5)$ representation).
In the ``single mixing'' scenarios we have the identification $\lambda'_{q_L} = \lambda_{q_L}$.
The $c_{ij}$ and $c'_{ij}$ parameters in Eq.~(\ref{eq:Yukawas_1}) are matrices in flavor space
and their structure might depend on the details of the composite sector.
In the anarchic partial compositeness scenario the composite sector is assumed to have
no flavor structure and to generate all possible flavor-violating couplings with similar strengths.
This translates in the assumption that the $c$ and $c'$ matrices are anarchic with all the elements
of the same order
\begin{equation}
c_{ij} \sim c'_{ij} \sim 1\,.
\end{equation}

Obviously some hierarchy is needed to generate the observed quark masses and mixings. In the
anarchic scenario these come from the structure of the elementary/composite couplings.
As we anticipated in the previous discussion, we assume that the elementary/composite mixings
are hierarchical
\begin{equation}
\lambda_{q_L}^1 \ll \lambda_{q_L}^2 \ll \lambda_{q_L}^3\,,
\qquad
\lambda_{u_R}^1 \ll \lambda_{u_R}^2 \ll \lambda_{u_R}^3\,,
\end{equation}
and analogously for the down sector couplings $\lambda'_{q_L}$ and $\lambda_{d_R}$.

The Yukawa matrices have the form of Eq.~(\ref{eq:M_ij_th}), with hierarchical $\zeta_L$ and $\zeta_R$. We can thus apply the theorem stated in Eqs.~(\ref{eq:diag_th}) and (\ref{eq:UL_rot}) to their diagonalization formulae
\begin{equation}
y_u = U_L y_u^D U_R^\dagger\,,
\qquad
y_d = D_L y_d^D D_R^\dagger\,.
\end{equation}
We find that the diagonalized
Yukawa's, $y^D_{u,d}$, are hierarchical and are of order
\begin{equation}\label{eq:diag_yukawa}
y_u^{D\,ii} \sim \lambda_{q_L}^i \lambda_{u_R}^i /g_*\,,
\qquad
y_d^{D\,ii} \sim {\lambda'}_{q_L}^i \lambda_{d_R}^i /g_*\,,
\end{equation}
and that the four chiral rotation matrices are close to the identity. Their entries can be estimated as
\begin{equation}
(U_L)_{ij} \sim
\left\{
\begin{array}{l@{\qquad\textrm{for}\quad}l}
\lambda_{q_L}^i/\lambda_{q_L}^{j} & i < j\\
\rule{0pt}{1.1em} 1 & i = j\\
\rule{0pt}{1.1em} \lambda_{q_L}^{j}/\lambda_{q_L}^{i} & i > j
\end{array}
\right.\,,
\end{equation}
and similarly for the $U_R$ and $D_{L,R}$ rotations.\index{V@$V_{\textrm{CKM}}$ matrix}
The $V_{\textrm{CKM}}$ matrix, which is the product of the rotations of the left-handed fields
\begin{equation}
V_{\textrm{CKM}} = U_L^\dagger D_L\,,
\end{equation}
is therefore also close to the identity. Its off-diagonal elements can be estimated as
\begin{equation}
V_{\textrm{CKM}}^{ij} \sim V_{\textrm{CKM}}^{ji} \sim
\lambda_{q_L}^i/\lambda_{q_L}^{j} \pm \lambda'^i_{q_L}/\lambda'^j_{q_L}
\qquad \textrm{for} \quad i < j\,,
\label{eq:CKM_estimate}
\end{equation}
where we inserted a $\pm$ sign to denote that the up and down contributions are determined up to order one coefficients which can also have arbitrary sign.

The elementary/composite mixings, and the hierarchies among them, are free parameters of the theory. We will now see how these hierarchies can be chosen in such a way that the size of the quark masses and of the $V_{\textrm{CKM}}$ elements are reproduced in a ``natural'' way, that is with order one $c_{ij}$ and $c'_{ij}$ parameters and no special relations among them. The quark masses give the first set of conditions
\begin{equation}\label{eq:mass_hierarchies}
\begin{array}{l@{\qquad}l}
\displaystyle \frac{m_u}{m_c} \sim \frac{\lambda_{q_L}^1}{\lambda_{q_L}^2} \frac{\lambda_{u_R}^1}{\lambda_{u_R}^2}\,,
& \displaystyle \frac{m_c}{m_t} \sim \frac{\lambda_{q_L}^2}{\lambda_{q_L}^3} \frac{\lambda_{u_R}^2}{\lambda_{u_R}^3}\,,\\
\rule{0pt}{2.25em}\displaystyle \frac{m_d}{m_s} \sim \frac{{\lambda'}_{q_L}^1}{{\lambda'}_{q_L}^2} \frac{\lambda_{d_R}^1}{\lambda_{d_R}^2}\,,
& \displaystyle \frac{m_s}{m_b} \sim \frac{{\lambda'}_{q_L}^2}{{\lambda'}_{q_L}^3} \frac{\lambda_{d_R}^2}{\lambda_{d_R}^3}\,.
\end{array}
\end{equation}
The second set of conditions comes from the non-diagonal elements of the $V_{\textrm{CKM}}$ matrix we estimated in Eq.~(\ref{eq:CKM_estimate}). By the latter formula we would like to reproduce the observed size of the $V_{\textrm{CKM}}$ elements, which is well described by the structure
\begin{equation}
V_{\textrm{CKM}} \sim
\left[
\begin{array}{c@{\hspace{.75em}}c@{\hspace{.75em}}c}
1 - \lambda_C^2/2 & \lambda_C & \lambda_C^3\\
\rule{0pt}{1.15em} \lambda_C & 1 - \lambda_C^2/2 & \lambda_C^2\\
\rule{0pt}{1.15em} \lambda_C^3 & \lambda_C^2 & 1
\end{array}
\right]\,,
\end{equation}
where $\lambda_C$ is of the order of the sine of the Cabibbo angle, $\lambda_C \sim \sin \theta_C \simeq 0.22$. It can be shown \cite{Barbieri:2012tu} that simultaneously reproducing mass and $V_{\textrm{CKM}}$ hierarchies requires that the second term in Eq.~(\ref{eq:CKM_estimate}) is either dominant or comparable with the first one. Namely, $\lambda_{q_L}^i/\lambda_{q_L}^{j} \lesssim \lambda'^i_{q_L}/\lambda'^j_{q_L}$. This means that the hierarchical structure is more pronounced in the up sector than in the down one and that the $V_{\textrm{CKM}}$ elements are thus mainly determined by the down-sector rotations. Therefore reproducing the structure of the $V_{\textrm{CKM}}$ fixes the hierarchy among the $\lambda'_{q_L}$ parameters~\cite{Huber:2003tu} to be
\begin{equation}
\lambda'^1_{q_L}/\lambda'^2_{q_L} \sim \lambda_C\,,
\qquad
\lambda'^2_{q_L}/\lambda'^3_{q_L} \sim \lambda_C^2\,.
\label{eq:mixing_estimate}
\end{equation}
In the ``single mixing'' models, where $\lambda'=\lambda$, this conditions obviously fix all the mixings of the left-handed $q_L^i$ doublets. Instead, if multiple mixings are present, the hierarchies among the up mixings $\lambda^i_{q_L}$ are to a large extent arbitrary and only restricted by mild constraints.
Once the left mixings are chosen, the sizes of the right-handed mixings are then determined by the requirement
of reproducing the hierarchies of the quark masses (see Eq.~(\ref{eq:mass_hierarchies})) and the value of
the top and bottom Yukawa's (see Eq.~(\ref{eq:diag_yukawa})).

The eight  conditions listed above allow to fix the size of almost all the $\lambda$ couplings. Only four quantities
remain undetermined and can be conveniently identified with
\begin{equation}
x_t \equiv \frac{\lambda_{q_L}^3}{\lambda_{u_R}^3}\,,
\qquad \textrm{and} \qquad
z_i \equiv \frac{\lambda_{q_L}^i}{{\lambda'}_{q_L}^i}\,,
\quad \textrm{i=1,\, 2,\, 3}\,.
\end{equation}
The $x_t$ parameter is related to the ratio between the amount of compositeness in the
left- and right-handed top components and it is restricted to vary in a limited range around one. This can be seen by
considering the estimate of the top Yukawa, which can be rewritten as
\begin{equation}\label{eq:yt_flavor}
y_t \simeq g_* \frac{\lambda_{q_L}^3}{g_*} \frac{\lambda_{u_R}^3}{g_*}\,.
\end{equation}
As we saw in the previous chapters, and in particular in Sect.~\ref{pfer}, the elementary/composite mixings can not overcome $g_*$. This gives two conditions, $\lambda_{q_L}^3\leq g_*$ and $\lambda_{u_R}^3\leq g_*$, which make $x_t$ live in the range
\begin{equation}
y_t/g_* \lesssim x_t \lesssim g_*/y_t\,.
\end{equation}
The $z_i$ parameters correspond to the ratio between
mixings of the left-handed up and down fields and parametrize the arbitrariness in the choice of the
$\lambda^i_{q_L}$ parameters in the models with multiple mixings. The values of the $z_i$ parameters
are subject to some mild restrictions.
Due to the large mass difference, it is natural to assume that the bottom compositeness is smaller
or at most equal to the top one. The condition $z_3 \gtrsim 1$ is thus usually verified
in the explicit models. As we discussed before, in order to obtain the quark masses and the hierarchies in
the $V_{\textrm{CKM}}$ matrix, the condition $\lambda_{q_L}^i/\lambda_{q_L}^{j} \lesssim {\lambda'}_{q_L}^i/{\lambda'}_{q_L}^{j}$
is required. This condition translates into the relations $z_1 \lesssim z_2 \lesssim z_3$.
A choice often encountered in the literature is to assume that the hierarchy in the up and down mixings are equal
so that
\begin{equation}\label{eq:lambda_sim_lambda_prime}
{\lambda}_{q_L}^1/{\lambda}_{q_L}^2 = \lambda'^1_{q_L}/\lambda'^2_{q_L} \sim \lambda_C\,,
\qquad
{\lambda}_{q_L}^1/{\lambda}_{q_L}^2 = \lambda'^1_{q_L}/\lambda'^2_{q_L} \sim\lambda_C^2\,.
\end{equation}
In this case the $z_i$ parameters are all equal, $z_1 = z_2 = z_3 \equiv z$.
This pattern obviously describes also the models with ``single mixing'', which are recovered for $z = 1$.
For simplicity, in our derivation of the flavor constraints in Sect.~\ref{sec:flavor_anarchy_constraints}
we will assume that the condition $z_1 = z_2 = z_3 \equiv z$ is realized.
Relaxing this assumption does not significantly modify the results, hence our estimates remain approximately valid for a generic pattern of up-type mixings.

\subsection{Higgs couplings and higher-order effects}\label{sec:h_couplings_flavor}

In the first part of this section we discussed the framework of anarchic partial compositeness scenarios
and how the quark masses
and the structure of the $V_{\textrm{CKM}}$ matrix are naturally generated through the hierarchies in the elementary/composite
mixings. We now want to extend the previous discussion and analyze the structure of the Higgs couplings and
their role in mediating flavor-violating effects. As we saw in Chap.~\ref{ch:Goldstone}, the Higgs interactions
induced by the leading ${\mathcal O}(p^0)$ operators are fixed by the Goldstone symmetry
and are fully determined by the $\SO(5)$ quantum numbers of the composite operators
mixed to the elementary quarks. For definiteness, in the following we will consider the scenario
with composite operators transforming in the fundamental representation of $\SO(5)$ and
we will only briefly discuss possible differences that arise in other set-ups.

The effective Lagrangian describing the Yukawa interactions of the up-type quarks can be obtained by a
simple generalization of Eq.~(\ref{yukup}). By taking into account the result in Eq.~(\ref{eq:Yukawas_1}) we get
\begin{eqnarray}
{\mathcal L}_{\textrm{Yuk}}^u &=&
- y_u^{ij} \frac{f}{2 \sqrt{2} | H |} \sin \frac{2 \sqrt{2} | H |}{f} \overline q_L^i H^c u_R^{j} + \textrm{h.c.}\,.\label{eq:Yuk_up_eff}
\end{eqnarray}
Analogously, from Eq.~(\ref{yukdown}) we get the down-type Yukawa operators
\begin{eqnarray}
{\mathcal L}_{\textrm{Yuk}}^d &=&
- y_d^{ij} \frac{f}{2 \sqrt{2} | H |} \sin \frac{2 \sqrt{2} | H |}{f} \overline q_L^i H d_R^{j} + \textrm{h.c.}\,.
\label{eq:Yuk_down_eff}
\end{eqnarray}
By going to the unitary gauge (see Eq.~(\ref{ugauge})),
\begin{eqnarray}
{\mathcal L}_{\textrm{Yuk}}^u &=& - \frac{f}{2 \sqrt{2}} \sin \frac{2 ( V + h )}{f}
y_u^{ij} \overline u_L^i u_R^{j} + \textrm{h.c.}\,,\\
{\mathcal L}_{\textrm{Yuk}}^d &=& - \frac{f}{2 \sqrt{2}} \sin \frac{2 ( V + h )}{f}
y_d^{ij} \overline d_L^i d_R^{j} + \textrm{h.c.}\,,
\end{eqnarray}
we discover the remarkable feature that the dependence on the Higgs field is completely factorized and is
disconnected from the flavor structure. As a consequence, in the mass eigenstate basis the leading Higgs
interactions are automatically diagonalized and FCNC's mediated by the Higgs are not present~\cite{Agashe:2009di}.

The absence of Higgs-mediated FCNC can also be understood as follows. In the low-energy Lagrangian in
Eqs.~(\ref{eq:Yuk_up_eff}) and (\ref{eq:Yuk_down_eff}) there are only two sources of flavor violation, namely
the $y_u^{ij}$ and the $y_d^{ij}$ matrices, which have the same flavor quantum numbers as the SM Yukawa matrices.
This means that an accidental
Minimal Flavor Violation (MFV)~\cite{D'Ambrosio:2002ex} is present, which suppresses FCNC's mediated by the Higgs.
Notice that for this result
to be valid it is essential that only two invariant operators are present in the effective theory at ${\mathcal O}(p^0)$.
Whether this structure is realized or not depends on the $\SO(5)$ quantum numbers of the composite
operators mixed with the elementary fields or, equivalently, on the representations we use to embed the
elementary quarks (see Sect.~\ref{sec:HCsigmaM}). For instance, a similar structure with only two invariants is realized in the
models based on the spinorial $\SO(5)$ representation. On the other hand, if more than two invariants are present,
the MFV structure is violated and the extra invariants would in general give rise
to flavor changing effects mediated by the Higgs. An example of scenarios of this type are the models
based on the representation ${\bf 10}$, which gives rise to two independent operators in the up sector
and two in the down sector.\footnote{This is true in the absence of additional symmetries in the composite
sector. For instance by imposing a $P_{LR}$ symmetry the number of invariants can be reduced to two and Higgs-mediated
FCNC's can be avoided.}

So far, in the analysis of the quark mass generation and of the Higgs couplings, we restricted our attention
to the leading effects coming from ${\mathcal O}(p^0)$ effective operators. As we discussed in Chap.~\ref{ch:SILH},
however, higher-order effects that modify the features of the low-energy dynamics are usually present.
These can lead to quantitative modifications of the estimates (in particular for effects related to fields
with a large amount of compositeness),
but can also produce important qualitative changes, if a symmetry
or selection rule is there at leading order as in the present situation.

The first corrections to the leading-order fermion dynamics come from operators of order $p$.
Restricting our attention to the $u_R$ singlets
two independent $P_{LR}$-invariant operators are found (see Sect.~\ref{pfer}), which can be expressed in the following form
\begin{eqnarray}
{\mathcal O}_u &=& \frac{1}{g_*^2} \kappa^u_{ij} \lambda_{u_R}^i \lambda_{u_R}^j\, \overline U_R^{{\mathbf{4}}\, i} i \gamma^\mu d_\mu U_R^{{\mathbf{1}}\, j} + \mathrm{h.c.}\,,\nonumber\\
{\mathcal O}'_u &=& \frac{1}{g_*^2} \kappa'^u_{ij} \lambda_{u_R}^i \lambda_{u_R}^j\, \overline U_R^{{\mathbf{4}}\, i} i \gamma^\mu D_\mu U_R^{{\mathbf{4}}\, j} + \mathrm{h.c.}\,,\label{eq:der_op_flavor}
\end{eqnarray}
where $\kappa_{ij}$ are generic complex matrices in flavor-space with order-one elements.
In the above formulae $D_\mu$ is the CCWZ covariant derivative and $d_\mu$ is the CCWZ $d$-symbol.
Finally $U_R^{{\mathbf{4}},{\mathbf{1}}}$ denote the elementary fields ``dressed'' with the Goldstone matrix (see Sect.~\ref{sec:HCsigmaM}) in, respectively, the ${\mathbf{4}}$ and the ${\mathbf{1}}$ \mbox{SO$(4)$} representations.
The two operators in Eq.~(\ref{eq:der_op_flavor}) have a non-trivial dependence on the Higgs field
and induce corrections to the Higgs couplings and, after EWSB, to the gauge bosons couplings.

The ${\mathcal O}(p)$ operators induce additional flavor-breaking interactions that break the MFV
structure in the Higgs couplings. We can easily understand this feature through a simple example. The operator
${\mathcal O}_u$ gives rise to derivative interactions of the Higgs:
\begin{equation}\label{eq:deriv_Higgs_int}
{\mathcal O}_u \supset 2 \sqrt{2}\,i\, \frac{v}{f^2} \widetilde \kappa^u_{ij}\, (\partial_\mu h)\,
\overline u_R^i \gamma^\mu u_R^j + \mathrm{h.c.}\,,
\end{equation}
where we absorbed the $\lambda_u$ factors and the $1/g_*^2$ normalization into the $\widetilde \kappa^u$ matrix
and we only considered the first term in the $\xi$ expansion (thus using the identification $V \simeq v$).
Notice that the $\widetilde \kappa^u$ matrix can be assumed to be Hermitian, given that the anti-Hermitian
part does not contribute to Eq.~(\ref{eq:deriv_Higgs_int}).
By integrating by parts and using the equations of motion for the $u_R$ fields
(or, equivalently, by performing a field redefinition), we can rewrite the operator in Eq.~(\ref{eq:deriv_Higgs_int})
in the following form
\begin{equation}
{\mathcal O}_u \supset 2\, \frac{v^2}{f^2} h\,
(y_u^\dagger \widetilde \kappa^u)^{i j}\, \overline u_L^i u_R^j + \mathrm{h.c.}\,.
\end{equation}
This operator induces a correction to the Higgs couplings to the up-type quarks.
Notice that, in general, the $\widetilde \kappa^u$ matrix is not aligned with the
quark mass matrix $y_u$, thus the Higgs interactions due to the operator ${\mathcal O}_u$
can mediate flavor-changing effects. To get an estimate of these effects it is convenient
to rewrite the coupling matrix in the mass-eigenstate basis:
\begin{equation}
y_u^\dagger \widetilde \kappa^u \rightarrow (U_L y_u^\dagger U_R^\dagger) (U_R \widetilde \kappa^u U_R^\dagger)
= y_u^D (U_R \widetilde \kappa^u U_R^\dagger)\,.
\end{equation}
By using the estimates for the quark masses and for the elements of the $U_R$ rotation matrix we finally get
\begin{equation}
{\mathcal O}_u \sim \frac{2}{g_*^2} \frac{v^2}{f^2}\frac{\lambda_{q_L}^i \lambda_{u_R}^i}{g_*}
\lambda_{u_R}^i \lambda_{u_R}^j\, h\, \overline u_L^i u_R^j + \mathrm{h.c.}\,.
\end{equation}
This result shows that the flavor-changing couplings of the Higgs arise at order $\xi$ and are weighted by four
powers of the elementary/composite mixings.

The second operator in Eq.~(\ref{eq:der_op_flavor}) induces similar effects although
the analysis is slightly more involved. As we saw, the $d$-symbol operators gives rise only to interaction
terms. On the contrary, the ${\mathcal O}'_u$ operator also induces a contribution to the kinetic terms of the elementary fields.
This correction can be removed by a field redefinition which does
not significantly alter the estimates we obtained for the $V_{\textrm{CKM}}$ matrix and for the quark masses.\footnote{The field redefinition
can be numerically relevant for the top quark given its sizable degree of compositeness. It is instead
typically negligible for all the other quarks.}
After the field redefinition an analysis similar to the one we described above
can be used to derive the induced Higgs flavor-violating couplings. Derivative operators analogous to the ones in Eq.~(\ref{eq:der_op_flavor}) can also be written for the $q_L^i$ and the $d_R^i$
elementary fields. They can be analyzed along the lines of the previous discussion and give rise to
similar Higgs flavor-changing interactions.

It is interesting to notice that, in addition to the Higgs couplings, the ${\mathcal O}(p)$ effective operators also
induce modifications of the EW bosons interactions which can generate additional flavor effects.
Important corrections can arise for the $W$ boson couplings. For instance couplings involving the right-handed quarks,
which are absent in the SM, can be generated (see Sect.~\ref{sec:DeltaF=1} for an analysis of these effects).
Moreover the couplings to the left-handed quarks can be modified
so that the unitarity of the CKM matrix is violated. This effect can be relevant for the quarks with a sizable amount
of compositeness, as the top in the anarchic scenarios. We postpone a discussion of this effect to Sect.~\ref{sec:top_couplings}.

\section{Constraints on the anarchic scenario}\label{sec:flavor_anarchy_constraints}

\begin{figure}
\centering
\includegraphics[width=.45\textwidth]{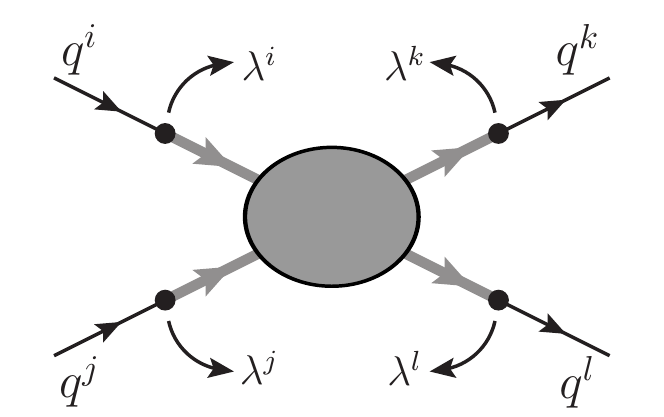}
\caption{Generation of $4$-fermion interactions in partial compositeness.}
\label{fig:4-fermion_flavor}
\end{figure}

\index{RS-GIM mechanism}
We are now ready to analyze the main flavor-violating effects in the anarchic scenario.
Before entering into the specific details, it is possible to derive an
important qualitative feature that stems from the general structure of partial compositeness.
As we discussed at length in Chap.~\ref{ch:Goldstone}, any insertion of an elementary field
in a low-energy effective operator is necessarily weighted by the corresponding amount of compositeness
or, in other words, is accompanied by the related elementary/composite mixing.
We exploited this feature in the anarchic construction
to generate the quark masses and the $V_{\textrm{CKM}}$ structure.
In this way we were naturally driven to the assumption
that the mixings have a hierarchical structure such that
the light quarks (essentially the ones belonging to the first and second generations)
have only tiny mixings with the composite dynamics.
The same structure also determines the size of the flavor-violating effective operators.
A typical example are the $4$-fermion contact interactions,
which are generated by the composite dynamics as shown in Fig.~\ref{fig:4-fermion_flavor}.
It is easy to understand that operators involving the light SM quarks are highly suppressed by
the tiny mixings and only induce small flavor-violating effects. This feature of the anarchic
scenario was first noticed in the holographic realizations of the composite Higgs idea
and is usually referred to as the ``RS-GIM mechanism''~\cite{Gherghetta:2000qt,Huber:2003tu,Agashe:2004cp}.

It is important to mention that, in scenarios that do not respect the ``single mixing''  hypothesis
multiple contributions to the effective flavor-violating operators can arise,
corresponding to the different $\lambda_i$ mixings. Usually the most important new physics contributions
are mediated by the largest mixings.
However, depending on the quantum numbers of the composite operators some selection rules can be present.
For instance this happens if the elementary states are mixed with composite operators
with different $\SO(4) \times \U(1)_X$ charges. In this case, analogously to what we discussed in Sect.~\ref{sec:anarchic_pc}
for the generation of the Yukawa couplings, the representations of the composite operators that mediate
the flavor-violating effects must be ``compatible'' with each other.

In the following we will discuss quantitatively the flavor-violating effects and
we will derive some estimates of the constraints on the anarchic scenario coming from
flavor measurements. For this purpose, as we did in the rest of this chapter, we will
adopt the One Scale One Coupling power counting described in Chap.~\ref{ch:SILH}.
The implications of relaxing this assumption will be discussed in Sect.~\ref{sec:VMD_interpretation}.

\index{FCNC}
Some of the most ``dangerous'' flavor-violating effects are related to the presence of FCNC's.
In the SM, FCNC's are absolutely absent at tree-level
thanks to the fact that the Higgs and $Z$-boson couplings are flavor-diagonal.
This is in general not true in BSM models. We already saw in Sect.~\ref{sec:h_couplings_flavor} that
in the composite scenario the Higgs field can mediate flavor changing effects, although,
in a large class of models, these effects are only generated at subleading order in
the elementary/composite mixings. As we will discuss in the following, the $Z$ couplings can also mediate
flavor-changing currents at tree-level. Furthermore, the composite dynamics can give rise
to contact interactions that do not preserve the flavor quantum numbers, as for instance
$4$-fermion operators involving fields from different generations. As we already mentioned, a typical mechanism
generating these operators is the exchange of vector resonances, as for instance ``heavy gluons''
or states with EW quantum numbers. These effects, however, can also arise as purely contact interactions
at the cut-off scale where the composite dynamics becomes completely non-perturbative.

The relevant flavor-violating observables belong to two broad categories, namely the
processes that involve $\Delta F = 1$ or $\Delta F = 2$ transitions. The new physics effects in each
class are determined by a different set of effective operators. In particular the $\Delta F = 1$ effects
are mainly mediated by flavor-changing distortions of the gauge field interactions, whereas the
leading contribution to $\Delta F = 2$ processes is due to $4$-fermion contact interactions.

In addition to the flavor-violating processes, it is also worth mentioning the constraints coming from
the bounds on the neutron EDM. These measurements can be used to
derive strong bounds on CP-violating effects induced by flavor-conserving operators with a structure
similar to the ones contributing to $\Delta F = 1$ processes.

Before analyzing in details the bounds it is important to stress that several new-physics
effects can simultaneously contribute to each observable.
For simplicity, in deriving the bounds we consider each new-physics operator separately. This is somewhat equivalent to
assume that the various contributions are uncorrelated and accidental cancellations do not happen.
However it should be kept in mind that, when more operators are active at the same time, the bounds
can get somewhat weaker (see for instance Fig.~6 of Ref.~\cite{Altmannshofer:2011gn}).

\subsection[$\Delta F = 1$ transitions]{${\mathbf{\Delta F = 1}}$ transitions}\label{sec:DeltaF=1}

We begin the discussion by considering the $\Delta F = 1$ observables. In the composite Higgs scenarios the new physics contributions
to these processes are mainly due to three classes of operators. The first one includes the dipole operators
\begin{equation}
{\mathcal L}_{\Delta F = 1} \sim \frac{\lambda_i \lambda_j}{g_*} \frac{v}{m_*^2} \overline f_i \sigma_{\mu\nu}
g_{\mathrm{SM}} F_{\mathrm{SM}}^{\mu\nu} f_j\,,
\label{eq:dipole_op}
\end{equation}
where $F_{\mathrm{SM}}^{\mu\nu}$ collectively denotes the field strength of the SM gauge fields
and $g_{\mathrm{SM}}$ is the corresponding coupling.
We used the $f_{i,j}$ symbols to denote any SM quark ($i,j$ must be interpreted as ``condensed''
indices collecting the flavor quantum numbers as well as the chirality), while $\lambda_{i,j}$ are the relevant
elementary/composite mixings.
To be as general as possible,
we estimated the size of the coefficient in Eq.~(\ref{eq:dipole_op}) by assuming that
the dipole operators are generated at tree level by the composite dynamics. In the literature,
however, the dipole operators are usually considered as one-loop effects, which implies an
additional suppression factor $g_*^2/(16 \pi^2)$. This choice comes from the observation that, in
the known explicit models, the dipole operators can not be generated through
the tree-level exchange of composite resonances and only arise at the radiative level.

The second class of operators contributing to the $\Delta F = 1$ processes contains the penguin
operators that lead to modifications of the $Z$ boson couplings. The leading operators in this set
arise at dimension 6 and contain two powers of the Higgs field:
\begin{equation}
{\mathcal L}_{\Delta F = 1} \sim \frac{\lambda_i \lambda_j}{m_*^2} \overline f_i \gamma^{\mu} f_j
i H^\dagger \dblarrow{D}_\mu H\,,
\label{eq:deltaZ_op}
\end{equation}
where $H^\dagger  \dblarrow{D}_\mu H \equiv H^\dagger D_\mu H - (D_\mu H)^\dagger H$.

The last class of $\Delta F = 1$ effects are related to operators that modify the $W$ boson couplings. Among this set, the
most relevant operators are those that induce interactions involving the right-handed quarks, whose
structure is given by
\begin{equation}
{\mathcal L}_{\Delta F = 1} \sim \frac{\lambda_i \lambda_j}{m_*^2} \overline u_{R}^i \gamma^{\mu} d_{R}^j
i H^{c\dagger} D_\mu H\,,
\label{eq:deltaW_op}
\end{equation}
where $H^c \equiv i \sigma^2 H^*$ is the charge-conjugated Higgs doublet.

Additional subleading contributions to the gauge boson couplings can come from
dimension-6 operators containing multiple derivatives.
For instance, the $Z$ boson couplings can receive extra corrections from operators of the form
\begin{equation}
\frac{\lambda_i \lambda_j}{m_*^2} \frac{1}{g_*^2}
\overline f_i \gamma_\mu f_j  g_{\textrm{SM}} D_\nu F^{\mu\nu}_{\textrm{SM}}\,.
\end{equation}
These operators lead to effects that are suppressed by a factor $\sim (g_{\textrm{SM}}/g_*)^2$ compared those
induced by the penguin operators in Eq.~(\ref{eq:deltaZ_op}) and can be usually neglected.
Similarly, the $W$ boson couplings can be modified by operators containing derivatives of the elementary quarks:
\begin{equation}
\frac{\lambda_i \lambda_j}{g_* m_*^2} \overline u_{R}^i i H^{c\dagger} \slashed D
\slashed D q_{L}^j\,,
\qquad
\frac{\lambda_i \lambda_j}{g_* m_*^2} \overline d_{R}^i i H^{\dagger} \slashed D
\slashed D q_{L}^j\,.
\label{eq:deltaW_op_2}
\end{equation}
By using the equations of motion for the elementary fermions, these operators can be put
in the same form of the operators in Eq.~(\ref{eq:deltaW_op}). They are however characterized
by a completely different power counting.
The operators in Eq.~(\ref{eq:deltaW_op}) arise at quadratic order in the elementary/composite mixings.
On the contrary, the operators in Eq.~(\ref{eq:deltaW_op_2}) generate a similar contribution but only at
order $\lambda^4$,\footnote{The additional factor $\lambda^2$ comes from the quark mass factor that
comes from the equations of motion.} and thus induce subleading corrections to the $W$ couplings.
Although formally subleading, the derivative operators in Eq.~(\ref{eq:deltaW_op_2}) can become relevant
for the analysis of the flavor bounds because in several cases the leading operators vanish as
a consequence of some discrete symmetries of the composite dynamics. We will discuss this aspect with more details
later on when we will analyze the experimental constraints.

Finally, $\Delta F = 1$ flavor-changing effects can also be mediated by the Higgs field.
As we showed in the previous section, in many minimal models, these effects are suppressed thanks
to the Goldstone nature of the Higgs and only arise at subleading order in the elementary/composite mixings.
In the following we assume that this mechanism is at work and we neglect these kind of effects.

We will now discuss the experimental bounds on each class of $\Delta F = 1$ operators. We will start by
analyzing the EW and QCD dipole operators and then we will consider the modifications of the $Z$ and of the $W$ couplings.

\subsubsection{The EW dipole operators}

One of  the strongest bounds on flavor-violating EW dipole operators comes from the $b \rightarrow s \gamma$
transitions. Following the standard notation we encode the new-physics effects in the effective Hamiltonian~\cite{Buras:1998raa}
\begin{equation}
{\mathcal H}_{b \rightarrow s\gamma} = \frac{G_F}{\sqrt{2}} V_{tb} V^*_{ts} \frac{e\, m_b}{4 \pi^2}
\left[C_{7\gamma} \overline s_L \sigma^{\mu\nu} b_R F_{\mu\nu}
+ C'_{7\gamma} \overline s_R \sigma^{\mu\nu} b_L F_{\mu\nu}\right]\,,
\label{eq:eff_Ham_C7}
\end{equation}
where $G_F = 1/(2 v^2)$ denotes the Fermi constant, $F_{\mu\nu}$ is the photon field strength
and $e$ is the electric charge. By matching the above operators with Eq.~(\ref{eq:dipole_op})
and using the estimates of the elementary/composite mixings derived in Section~\ref{sec:anarchic_pc},
we find that the composite Higgs contributions to the $C_{7\gamma}^{(\prime)}$ coefficients are of the order \footnote{In order
to properly match the coefficients of the effective operators with the general estimates of
partial compositeness, all the quantities must be evaluated at the $m_*$ scale. To keep our discussion as
simple as possible we avoid to explicitly include the running effects in our equations.
We however include them in the numerical results (for this we assume $m_* \sim 1\ \mathrm{TeV}$).
We refer the interested reader to the original literature~\cite{Agashe:2008uz,Gedalia:2009ws,Vignaroli:2012si,KerenZur:2012fr}.}
\begin{eqnarray}
&&C_{7\gamma} \sim \frac{\sqrt{2}}{G_F}\frac{4\pi^2}{m_b} \frac{1}{V_{tb}V^*_{ts}}
\frac{\lambda'^2_{q_L}\lambda_{d_R}^3}{g_*} \frac{v}{m_*^2}
\sim \frac{8 \pi^2}{G_F}\frac{1}{m_*^2} \frac{\lambda_C^2}{V_{tb} V_{ts}^*}\,,\label{eq:C7_dipole}\\
\rule{0pt}{1.75em}&&C'_{7\gamma} \sim \frac{\sqrt{2}}{G_F}\frac{4\pi^2}{m_b} \frac{1}{V_{tb}V^*_{ts}}
\frac{\lambda'^3_{q_L}\lambda_{d_R}^2}{g_*} \frac{v}{m_*^2}
\sim \frac{8 \pi^2}{G_F}\frac{1}{m_*^2} \frac{\lambda_C^2}{V_{tb} V_{ts}^*}
\left(\frac{m_s}{m_b} \frac{1}{\lambda_C^4}\right)\,.\hspace{3em}\label{eq:C7prime_dipole}
\end{eqnarray}

We can now compare these results with the experimental constraints.
The bounds on the new-physics contributions to the $C_{7\gamma}^{(\prime)}$
coefficients are approximately given by~\cite{Altmannshofer:2012az}
\begin{equation}
|\textrm{Re}\, C_{7\gamma}| \lesssim 0.2\,,
\qquad |\textrm{Im}\, C_{7\gamma}| \lesssim 0.6\,,
\qquad | C'_{7\gamma}| \lesssim 0.5\,.
\label{eq:C7_bounds}
\end{equation}
Due to the significant asymmetry, we decided to list separately the bounds on the real and imaginary
part of the $C_{7\gamma}$ coefficient. For the $C'_{7\gamma}$ coefficient, on the contrary,
the real and imaginary parts have comparable constraints and we only give the bound on the absolute value.
The results in Eq.~(\ref{eq:C7_bounds}) can be translated into lower bounds on the composite dynamics scale $m_*$.
In particular the bounds on $C_{7\gamma}$ correspond to
\begin{equation}
m_* \gtrsim 5\ \mathrm{TeV}
\hspace{.75em} \textrm{(from Re\,$C_{7\gamma}$)}\,,
\qquad
m_* \gtrsim 3\ \mathrm{TeV}
\hspace{.75em} \textrm{(from Im\,$C_{7\gamma}$)}\,,
\label{eq:bound_m*_C7}
\end{equation}
while the bound on $C'_{7\gamma}$ gives
\begin{equation}
m_* \gtrsim 11\ \mathrm{TeV}
\hspace{.75em} \textrm{(from $C'_{7\gamma}$)}\,,
\label{eq:bound_m*_C7prime}
\end{equation}
Notice that the new-physics contributions to the $C^{(\prime)}_{7\gamma}$ coefficients are in general complex
and their complex phase is expected to be of order one.
For this reason, to derive the bounds in Eq.~(\ref{eq:bound_m*_C7}) we assumed that
the effective operators are generated with maximal complex phases. We will adopt this assumption
also in the following sections to derive all the flavor bounds.

If the dipole operators do not arise at tree level and are only generated at the radiative level,
the new-physics effects are strongly reduced
and the bounds are significantly relaxed. For instance, if we assume that the
dipole operators arise at one loop, the experimental constraints can be expressed as lower bounds on $m_*/g_* \simeq f$
and can be estimated as
\begin{eqnarray}
& f \gtrsim 0.4\ \mathrm{TeV} \hspace{.75em} \textrm{(from Re\,$C_{7\gamma}$)}\,,
\qquad
f \gtrsim 0.25\ \mathrm{TeV} \hspace{.75em} \textrm{(from Im\,$C_{7\gamma}$)}\,.&\hspace{2.5em}\\
\rule{0pt}{1.35em}& f \gtrsim 0.9\ \mathrm{TeV} \hspace{.75em} \textrm{(from $C'_{7\gamma}$)}\,,&\hspace{2.5em}
\end{eqnarray}

Additional constraints on the flavor-violating EW dipole operators come from the $b \rightarrow d \gamma$ transitions.
The analysis of these effects is completely analogous to the one we used above for the
$b \rightarrow s \gamma$ processes, so we skip all the details.
The bounds on the composite dynamics coming from the $b \rightarrow d \gamma$ transitions
are roughly comparable to the ones given in Eqs.~(\ref{eq:bound_m*_C7}) and (\ref{eq:bound_m*_C7prime})~\cite{Konig:2014iqa}.

\subsubsection{The QCD dipole operators}

Another observable that can give strong constraints on the dipole operators is the direct CP violation in the
$K^0 \rightarrow 2 \pi$ decay, usually encoded in $\mathrm{Re}(\epsilon'_K/\epsilon_K)$~\cite{Gedalia:2009ws,Vignaroli:2012si,KerenZur:2012fr}.
This observable receives large contributions from the chromomagnetic operators
\begin{equation}
Q_G = C_G \frac{v}{\sqrt{2}} \overline s_R \sigma^{\mu \nu} g_s G_{\mu\nu} d_L\,,
\qquad
Q'_G = C'_G \frac{v}{\sqrt{2}} \overline s_L \sigma^{\mu \nu} g_s G_{\mu\nu} d_R\,,
\end{equation}
whose coefficients can be estimated as
\begin{equation}
C_G \sim \frac{{2} m_s \lambda_C}{v} \frac{1}{m_*^2}\,,
\qquad
C'_G \sim \frac{{2} m_d}{v \lambda_C} \frac{1}{m_*^2}\,.
\end{equation}
The present measurements lead to a bound on the imaginary part of the $C^{(\prime )}_G$
coefficients
\begin{equation}
\left|\frac{\mathrm{Im}\, C^{(\prime )}_G}{\sqrt{2} m_s/v}\right|
\lesssim \frac{1}{(34\ \mathrm{TeV})^2}\,,
\end{equation}
where $C^{(\prime )}_G$ and $m_s$ should be interpreted as running quantities computed
at an energy scale $E = 1\ \mathrm{TeV}$.
This result implies the following constraint on the $m_*$ scale
\begin{equation}
m_* \gtrsim 15\ \mathrm{TeV}\,,
\end{equation}
which is roughly valid for both effective operators.
If the dipole operators are generated at one loop, the above bound becomes
\begin{equation}
f \gtrsim 1.2\ \mathrm{TeV}\,.
\end{equation}

Constraints roughly of the same order can be obtained for the QCD dipole operators that mediate
$b \rightarrow s$, $b \rightarrow d$ and $c \rightarrow u$ transitions~\cite{Isidori:2011qw,Konig:2014iqa}.

\subsubsection{${\mathbf{Z}}$ couplings modifications}

Let us now consider the penguin operators whose general structure is given in Eq.~(\ref{eq:deltaZ_op}).
After EWSB these operators generate new gauge interactions involving the $Z$ boson.
In particular they can give rise to flavor-changing interactions which are absent at tree-level in
the SM. To compare the $Z$-mediated flavor-violating effects with the experimental data it is customary to encode
the new physics contributions into $4$-fermion effective operators.
By integrating out the $Z$, the powers of the Higgs VEV $v$
and the gauge couplings cancel against the $Z$ boson mass and the following effective
operator is obtained at leading order in the elementary/composite mixings
\begin{equation}
\frac{\lambda_i \lambda_j}{m_*^2} \overline f_i \gamma^\mu f_j J^{(Z)}_\mu\,.
\label{eq:penguin_correction}
\end{equation}
In the above equation $J^{(Z)}_\mu$ represents the usual SM current
\begin{equation}
J_\mu^{(Z)} = \sum \overline f_i \gamma_\mu \left[
(t^3_{L} - 2 q \sin^2 \theta_w) - t^3_{L} \gamma^5\right] f_i\,,
\end{equation}
where $t^3_{L}$ denotes the charge of the $f_i$ fermion with respect to the Abelian subgroup of $\SU(2)_L$
generated by $t^3_L$, while $q$ is the electric charge.

Before considering the implications of the experimental measurements, it is useful to discuss
how the presence of discrete symmetries can protect the $Z$-boson couplings by forbidding the generation of some of the operators in Eq.~(\ref{eq:deltaZ_op}).
In the class of composite Higgs scenarios based on the minimal coset $\SO(5)/\SO(4)$,
two discrete $\Zdouble_2$ symmetries can be used to forbid the generation
of penguin operators~\cite{Agashe:2006at}. The first one is the $P_{LR}$ symmetry
which acts by exchanging the $\SU(2)_L$ and $\SU(2)_R$ generators inside $\SO(4)$ (see Sect.~\ref{discrete} for more details). It will be shown in Appendix~\ref{sec:custodial_symmetries} that this symmetry protects the couplings of the $Z$ boson to any current built from elementary states whose
mixing with the composite dynamics preserves the $P_{LR}$ invariance.
The second symmetry that can protect the $Z$ couplings is a discrete $\Zdouble_2$ subgroup, called $P_C$, of the
custodial group $\SO(3)_c$.\footnote{See Sect.~\ref{sec:MCHMlin} for a first description of the custodial group and Appendix~\ref{sec:custodial_symmetries} for a complete discussion.} It is defined in such way that its action on the $\SU(2)_L \times \SU(2)_R$ eigenstates
exchanges the sign of the $t^3_{L,R}$ charges, namely
$| t_L, t_R; t_L^3, t_R^3 \rangle \rightarrow | t_L, t_R; -t_L^3, -t_R^3 \rangle$.
Analogously to the $P_{LR}$ case, a current which is even under $P_C$ has a protected coupling to the $Z$ boson. Typical cases are the currents built from fields with $t^3_L = t^3_R = 0$ which are eigenstates of $P_C$.

\begin{table}
\centering
\begin{tabular}{c|ccccc}
\rule[-.65em]{0pt}{1.em} & $({\bf 2}, {\bf 1})_{1/6}$ & $({\bf 2}, {\bf 2})_{2/3}$ & $({\bf 2}, {\bf 2})_{-1/3}$\\
\hline
\rule{0pt}{1.25em} $u_L$ & $\times$ & $\times$ & $P_{LR}$ \\
\rule[-.75em]{0pt}{1.85em} $d_L$ & $\times$ & $P_{LR}$ & $\times$\\
\hline
\hline
\rule[-.65em]{0pt}{1.75em} & $({\bf 1}, {\bf 1})_{2/3}$ & $({\bf 1}, {\bf 1})_{-1/3}$ & $({\bf 1}, {\bf 2})_{1/6}$ & $({\bf 1}, {\bf 3})_{2/3}$ & $({\bf 1}, {\bf 3})_{-1/3}$\\
\hline
\rule{0pt}{1.25em} $u_R$ & $P_{LR}\,, P_C$ &  & $\times$ & $P_C$ & $\times$\\
\rule[-.65em]{0pt}{1.85em} $d_R$ &  & $P_{LR}\,, P_C$  & $\times$ & $\times$ & $P_C$
\end{tabular}
\caption{List of the discrete $\Zdouble_2$ symmetries preserved by the mixing of the elementary
fermions with composite operators in different $\SU(2)_L \times \SU(2)_R \times \U(1)_X$ representations.
The $\times$ symbols indicates that both $P_{LR}$ and $P_C$ are broken.}
\label{tab:PLR_and_PC}
\end{table}

\index{P@$P_{LR}$ symmetry}
\index{P@$P_C$ symmetry}
\index{discrete symmetries}
The discrete $\Zdouble_2$ symmetries preserved by the mixing of the elementary quarks with the composite fermionic operators, depending on the \mbox{SO$(4)$} quantum numbers of the latter operators, are given in Table~\ref{tab:PLR_and_PC}. From this table it is straightforward to derive which couplings are
protected in the minimal models.
We will list a few common scenarios in the following. If the composite operators belong to the spinorial
$\SO(5)$ representation (see Eq.~(\ref{eq:4decomposition})) all the mixings break the
discrete symmetries and the $Z$ couplings are not protected. The situation is different for the models based on
the fundamental, the adjoint and the ${\bf 14}$ representations. In the case of the adjoint ${\bf 10}_{2/3}$
(see Eq.~(\ref{eq:10_23decomposition})) the couplings of the left-handed down-type quarks are protected
as well as the couplings of the right-handed up-type singlets. Finally in the case of the ${\bf 5}$ (Eqs.~(\ref{eq:5_23decomposition})
and (\ref{eq:5_13decomposition})) and of the ${\bf 14}$ (Eq.~(\ref{eq:14_23decomposition}))
a more complex pattern appears. The mixing of the right-handed quarks respects both discrete symmetries.
The left-handed $q_L$ doublet, instead, must be necessarily mixed with (at least) two operators in order to generate
all quark masses. The mixings with the $({\bf 2}, {\bf 2})_{2/3}$ operators, which induces the up-type quark masses,
respects the $P_{LR}$ symmetry for the $d^i_L$ fields but not for the $u_L^i$'s.
On the other hand, the mixing that leads to the down-type quark masses (with the $({\bf 2}, {\bf 2})_{-1/3}$ states)
preserves the $P_{LR}$ invariance only for the $u^i_L$ states. Following these patterns, in deriving the bounds we will
assume that the flavor-violating $Z$ couplings involving the $d_L^i$ fields are only generated through
the $\lambda'^i_{q_L}$ mixings and not through the up-type ones.

We can now discuss the implications of the flavor measurements.
The flavor-violating penguin operators more strongly constrained from the experimental data are
the ones involving the down-type quarks and, in particular, the ones  that lead
to $b \rightarrow s$ and $s \rightarrow d$ transitions.

The leading new-physics corrections to the $b \rightarrow s$ transitions can be encoded in the following effective
Hamiltonian~\cite{Altmannshofer:2011gn}
\begin{equation}
{\mathcal H}_{b \rightarrow s} =  \frac{G_F}{\sqrt{2}} V_{tb} V^*_{ts} \frac{e^2}{4 \pi^2}
\left[C_{10} (\overline s_L \gamma^\mu b_L)(\overline \ell \gamma_\mu \gamma^5 \ell)
+ C'_{10} (\overline s_R \gamma^\mu b_R)(\overline \ell \gamma_\mu \gamma^5 \ell)\right]\,.
\end{equation}
Additional operators involving the vector lepton current $\overline \ell \gamma^\mu \ell$ are
present, but they are suppressed with respect to the above ones by the small factor
$1 - 4 \sin^2 \theta_w = 0.08$ that appears in the SM $Z$-boson current.
By using the result in Eq.~(\ref{eq:penguin_correction}) the coefficients of the effective operators can
be estimated as
\begin{eqnarray}
C_{10} & \sim & \frac{4 \pi^2}{G_F e^2} \frac{\lambda_C^2}{V_{tb} V^*_{ts}}
\frac{1}{m_*^2} \frac{m_t}{v} \frac{g_* x_t}{z^2}\,,\label{eq:contr_c10}\\
\rule{0pt}{1.75em}C'_{10} & \sim & \frac{4 \pi^2}{G_F e^2} \frac{\lambda_C^2}{V_{tb} V^*_{ts}}
\frac{1}{m_*^2} \frac{m_t}{v} \frac{g_* z^2}{x_t} \left(\frac{m_s m_b}{\lambda_C^4 m_t^2}\right)\,.
\label{eq:contr_c10prime}
\end{eqnarray}
As we explained before, these estimates apply only to the models in which the elementary states are mixed with composite
operators in the ${\bf 4}$, ${\bf 5}$ or ${\bf 14}$ representations of $\SO(5)$. In the last two cases
the flavor-violating effects come only from the mixing with the $({\bf 2}, {\bf 2})_{-1/3}$ operators
needed to give mass to the down-type quarks. In the models based on the ${\bf 10}_{2/3}$ $\SO(5)$ representation,
on the contrary, the contributions to $C_{10}$ vanish at leading order in the elementary/composite
mixings.

The bounds on the effective operators can be derived from a global analysis of inclusive and exclusive
$b \rightarrow s \ell^+ \ell^-$ decays~\cite{Altmannshofer:2011gn,Altmannshofer:2012az}:
\begin{equation}
| C_{10} | \lesssim 2.6\,,
\qquad\quad
| C'_{10} | \lesssim 3.1\,.
\end{equation}
The bound on the left-handed operator translates into a relatively strong constraint on $m_*$:
\begin{equation}\label{eq:bound_sbL}
m_* \gtrsim 3 \frac{\sqrt{g_* x_t}}{z}\ \mathrm{TeV}\,.
\end{equation}
On the contrary, the bound on $C'_{10}$ does not give any significant constraint as a consequence of the
suppression factor, $m_s m_b/(\lambda_C^4 m_t^2) \simeq 2 \times 10^{-3}$, that appears in Eq.~(\ref{eq:contr_c10prime}).

Let us now consider the $s \rightarrow d$ transitions. One of the strongest constraint on these processes
comes from the $K_L \rightarrow \mu^+ \mu^-$ decay. The new-physics contributions come from the
flavor-changing $Z$ interactions and can be encoded in the following effective operators
\begin{equation}
{\mathcal L}_{s \rightarrow d} = -\frac{g}{c_w} Z_\mu \left(
\delta g_L^{ds} \overline d_L \gamma^\mu s_L + \delta g_R^{ds} \overline d_R \gamma^\mu s_R
+ \mathrm{h.c.}\right)\,.
\label{eq:lagr_s->d}
\end{equation}
The coefficients in the above formula can be estimated as
\begin{equation}
\delta g_L^{ds} \sim \frac{m_t v}{\sqrt{2}} \frac{1}{m_*^2} \frac{g_* x_t}{z^2} \lambda_C^5\,,
\qquad
\delta g_R^{ds} \sim \frac{m_t v}{\sqrt{2}} \frac{1}{m_*^2} \frac{g_* z^2}{x_t} \lambda_C^5
\left(\frac{m_d m_s}{\lambda_C^{10} m_t^2}\right)\,.
\end{equation}
Notice that the right-handed transitions are suppressed with respect to the left-handed ones
by a factor $m_d m_s/(\lambda_C^{10} m_t^2) \simeq 0.02$.
The experimental measurements correspond to a bound~\cite{Buras:2011ph}
\begin{equation}
\left| \delta g_{L,R}^{ds} \right| \lesssim 6 \times 10^{-7}\,,
\end{equation}
from which the following bounds can be derived
\begin{eqnarray}
m_* &\gtrsim& 4.7 \frac{\sqrt{g_* x_t}}{z}\ \mathrm{TeV} \qquad \textrm{(from $\delta g_{L}^{ds}$)}\,,\label{eq:bound_deltagLds}\\
\rule{0pt}{1.75em} m_* &\gtrsim& 0.7 \sqrt{\frac{g_*}{x_t}} z\ \mathrm{TeV} \qquad \textrm{(from $\delta g_{R}^{ds}$)}\,.
\label{eq:bound_deltagRds}
\end{eqnarray}
As a consequence of the $\Zdouble_2$ symmetries, in the ``single mixing'' models based on the ${\bf 10}_{2/3}$ representation,
the only relevant bound is the one coming from the right-handed current in Eq.~(\ref{eq:bound_deltagRds}).
The bound in Eq.~(\ref{eq:bound_deltagLds}), instead, applies to the scenarios in which the elementary quarks
are embedded in the spinorial or fundamental representation.

Constraints on the effective operators in Eq.~(\ref{eq:lagr_s->d}) can also be obtained from the rare Kaon decay
$K^+ \rightarrow \pi^+ \nu \overline \nu$ and from the measurement of the $\mathrm{Re}(\epsilon'_K/\epsilon_K)$
parameter.
We already showed how the latter observable can be used to derive strong constraints on the QCD dipole operators.
When considered in the context of penguin operators, the measurement of $\mathrm{Re}(\epsilon'_K/\epsilon_K)$
leads to a bound on the left-handed flavor-violating
$Z$ currents comparable to the one in Eq.~(\ref{eq:bound_deltagLds})~\cite{Bauer:2009cf,KerenZur:2012fr}.

To conclude the discussion we briefly mention another class of constraints on the penguin operators
related to flavor-conserving observables. In the anarchic scenarios the most relevant
effect of this type is the modification of the $Z$ interactions with the third-generation quarks. Among these observables
the coupling to the left-handed bottom quark is the one that has been measured with the best precision
(at the $0.1\%$ level) and can be used to set the most stringent constraints. Due to the high precision,
the measurement of the $Z b_L \overline b_L$ vertex is usually included among the EW precision tests
together with the oblique EW parameters~\cite{Altarelli:1993sz,Cacciapaglia:2006pk}. We postpone a detailed
discussion of the $Z b_L \overline b_L$ observable to Chapter~\ref{ch:EWPT}, where we will analyze it in the
more general context of the EW precision data. Here we only derive some rough estimates along the
lines of Sect~\ref{pfer} and compare the resulting bounds with the ones coming from flavor-violating processes.

From Eq.~(\ref{eq:deltaZ_op}) we find that the deviations in the bottom couplings to the $Z$ boson are
of order \footnote{See Eq.~(\ref{eq:Lagr_Z_b}) for the exact definition of the bottom couplings.}
\begin{equation}
\begin{array}{l}
\delta g_{b_L} \simeq \displaystyle \frac{(\lambda'^3_{q_L})^2}{2 m_*^2} v^2
\simeq \frac{1}{m_*^2} \frac{m_t v}{\sqrt{2}} \frac{g_* x_t}{z^2}\,,\\
\rule{0pt}{2.em}\delta g_{b_R} \simeq \displaystyle \frac{(\lambda^3_{d_R})^2}{2 m_*^2} v^2
\simeq \frac{1}{m_*^2} \frac{m_t v}{\sqrt{2}} \frac{g_* z^2}{x_t} \left(\frac{m_b}{m_t}\right)^2\,.
\end{array}
\end{equation}
The deviation in the right-handed coupling is usually much smaller than the experimental precision
on $g_{b_R}$ (see Fig.~\ref{fig:deltag_plane}). For the left-handed coupling the situation is very different. The present
data show a mild tension with the SM prediction at the $2.5\sigma$ level (see Fig.~\ref{fig:deltag_plane}).
The precise bound on $\delta g_{b_L}$ thus crucially depends on the sign of the new-physics contributions.
To get a rough model-independent estimate of the constraints we assume that the new-physics contributions should be
smaller than the overall precision on $g_{b_L}$ and we impose the condition $|\delta g_{b_L}| \lesssim 10^{-3}$.
In this way a strong lower bound on $m_*$ is obtained
\begin{equation}
m_* \gtrsim 5 \frac{\sqrt{g_* x_t}}{z}\ \mathrm{TeV}\,.
\end{equation}
This bound is comparable to the ones we derived from flavor-violating processes
(see Eqs.~(\ref{eq:bound_sbL}) and (\ref{eq:bound_deltagLds})).

\subsubsection{${\mathbf{W}}$ couplings modification}

\begin{figure}
\centering
\includegraphics[width=.425\textwidth]{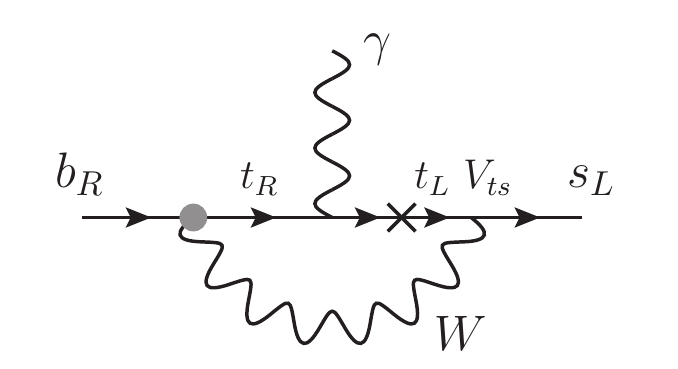}
\hspace{0.5em}
\includegraphics[width=.425\textwidth]{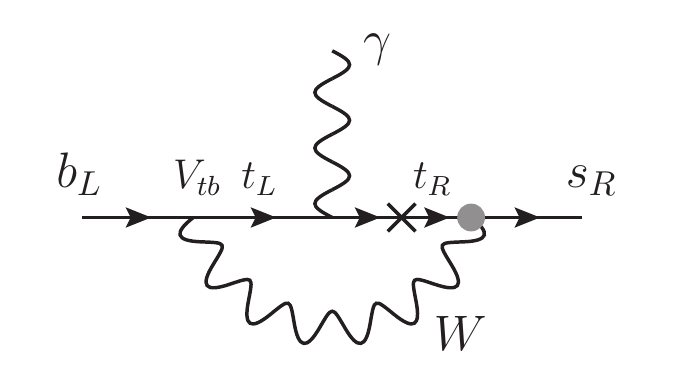}
\caption{Examples of one-loop diagrams contributing to the $C_{7\gamma}$ (on the left) and
$C'_{7\gamma}$ (on the right) coefficients of the dipole operators in Eq.~(\ref{eq:eff_Ham_C7}). The gray circles
represent the insertions of the effective $W t_R b_R$ and $W t_R s_R$ vertices.}
\label{fig:C7}
\end{figure}

The last class of operators relevant for the $\Delta F = 1$ processes includes the ones that lead to
distortions of the $W$-boson couplings. The general structure of these operators is shown in Eq.~(\ref{eq:deltaW_op}).
An effective way to constrain such operators is to focus on their contributions to $b \rightarrow s \gamma$
transitions~\cite{Vignaroli:2012si,Barbieri:2012tu}.
In composite Higgs scenarios the largest new-physics effects
are due to new $W$ interactions involving the right-handed top component, which
generate one-loop contributions to the $C_{7\gamma}$ and $C'_{7\gamma}$ coefficients
(see Eq.~(\ref{eq:eff_Ham_C7})). The structure of the relevant diagrams is shown in Fig.~\ref{fig:C7}.
Notice that similar contributions are also present in the SM. In this case, however, the purely left-handed
structure of the $W$ couplings implies a strong helicity suppression of the $b \rightarrow s \gamma$
transitions. This suppression is instead lifted if right-handed couplings exist.

In anarchic composite Higgs scenarios, the most relevant effects are due to the $W t_R b_R$ vertex,
which generates a correction to $C_{7\gamma}$,
and to the $W t_R s_R$ coupling, which contributes to $C'_{7\gamma}$.
The size of the $W t_R b_R$ coupling coming from the operators in Eq.~(\ref{eq:deltaW_op}) can be estimated as
\begin{equation}
\frac{g}{\sqrt{2}} \frac{v^2}{2} \frac{\lambda_{u_R}^3 \lambda_{d_R}^3}{m_*^2} (\overline t_R \slashed W b_R)
+ \mathrm{h.c.}
\simeq \frac{g}{2} \frac{z}{x_t} \frac{g_* v}{m_*^2} m_b (\overline t_R \slashed W b_R)
+ \mathrm{h.c.}\,,
\label{eq:Wtb_1}
\end{equation}
which leads to the following contribution to $C_{7\gamma}$~\cite{Vignaroli:2012si}:
\begin{equation}
C_{7\gamma} \sim \frac{1}{\sqrt{2}} m_t g_* v \frac{z}{x_t} \frac{1}{m_*^2} A_7(m_t^2/m_W^2)\,,
\label{eq:C7_Wtb}
\end{equation}
where $A_7(m_t^2/m_W^2) \simeq -0.8$.
Analogously we can derive an estimate of the size of the $W t_R s_R$ interaction
\begin{equation}
\frac{g}{\sqrt{2}} \frac{v^2}{2} \frac{\lambda_{u_R}^3 \lambda_{d_R}^2}{m_*^2} (\overline t_R \slashed W s_R)
+ \mathrm{h.c.}
\simeq \frac{g}{2} \frac{z}{x_t} \frac{g_* v}{m_*^2} \frac{m_s}{\lambda_C^2} (\overline t_R \slashed W s_R)
+ \mathrm{h.c.}\,,
\label{eq:Wts_1}
\end{equation}
and the corresponding contribution to $C'_{7\gamma}$:
\begin{equation}
C'_{7\gamma} \sim \frac{1}{\sqrt{2}} m_t g_* v \frac{z}{x_t} \frac{1}{m_*^2} \left(\frac{m_s}{m_b \lambda_C^4}\right)
A_7(m_t^2/m_W^2)\,.
\label{eq:C7prime_Wts}
\end{equation}
It is interesting to notice that the above correction to $C'_{7\gamma}$ is enhanced with respect
to the one to $C_{7\gamma}$ in Eq.~(\ref{eq:C7_Wtb}) by a factor $m_s/(m_b \lambda_C^4) \sim 8$.
This enhancement exactly matches the one that we found for the contributions to $C_{7\gamma}$
and $C'_{7\gamma}$ coming from the dipole
operators (see Eqs.~(\ref{eq:C7_dipole}) and (\ref{eq:C7prime_dipole})).

The bounds in Eq.~(\ref{eq:C7_bounds}) can be translated in the following constraints on the
composite mass scale $m_*$:
\begin{equation}
m_* \gtrsim 0.3 \sqrt{\frac{g_* z}{x_t}}\ \mathrm{TeV}\hspace{.75em} \textrm{(from $C_{7\gamma}$)}\,,
\quad
m_* \gtrsim 0.5 \sqrt{\frac{g_* z}{x_t}}\ \mathrm{TeV}\hspace{.75em} \textrm{(from $C'_{7\gamma}$)}\,.
\label{eq:bound_WRcoupl}
\end{equation}
These bounds are much weaker than the ones coming from the dipole operators, especially if
we assume that the latter are generated at tree-level (see Eqs.~(\ref{eq:bound_m*_C7}) and (\ref{eq:bound_m*_C7prime})).

It is interesting to notice that the operators in Eq.~(\ref{eq:deltaW_op}), which we used to
estimate the corrections to the $W$ couplings, are forbidden in many models as a consequence of the custodial invariance.
This happens if the mixing of the right-handed quarks $f_{Ri}$ to the composite dynamics preserves the $\SO(4)$ symmetry.
In this case the composite operators coupled to the $f_{Ri}$ fields are singlets under the custodial symmetry
and can not generate a current with the appropriate quantum numbers to couple to the $W$ bosons (which
transform as a triplet under the $\SO(3)_c$ custodial group).
An example of such situation are the models in which the composite operators belong to the fundamental
or to the ${\bf 14}$ representations of $\SO(5)$.

Notice that the custodial symmetry protects the $W$ couplings only at zero momentum.
Indeed, $W$ interactions with the right-handed quarks can still be generated by operators
involving derivatives of the quarks. These operators are non-vanishing because they are ``sensitive''
to the breaking of the custodial invariance induced by the mixing of the left-handed quarks.
Operators of this kind are shown in Eq.~(\ref{eq:deltaW_op_2}) and give rise to the relevant interactions
once we put the quarks on-shell.
The operators that generate the largest contribution to $W t_R b_R$ are
\begin{equation}
\frac{\lambda_{q_L}^3 \lambda_{u_R}^3}{g_* m_*^2} (\overline t_R i H^{c\dagger} \slashed D \slashed D q_L^3)\,,
\qquad
\frac{\lambda'^3_{q_L} \lambda_{d_R}^3}{g_* m_*^2} (\overline b_R i H^\dagger \slashed D \slashed D q_L^3)\,.
\end{equation}
By using the equations of motion for the elementary fermions, $i \slashed D f_{Li} = m_i f_{Ri}$,
we can see that both operators give rise to an effective interaction with a strength of order
\begin{equation}
2 \frac{g}{\sqrt{2}} \frac{m_t m_b}{m_*^2} (\overline t_R \slashed W b_R) + \mathrm{h.c.}\,.
\label{eq:Wtb_2}
\end{equation}
Similar considerations apply to the $W t_R s_R$ interaction, which is mainly due to the operator
\begin{equation}
\frac{\lambda'^3_{q_L} \lambda_{d_R}^2}{g_* m_*^2} (\overline s_R i H^\dagger \slashed D \slashed D q_L^3)\,,
\end{equation}
and can be estimated as
\begin{equation}
2 \frac{g}{\sqrt{2}} \frac{m_t m_s}{m_*^2} \frac{1}{\lambda_C^2} (\overline t_R \slashed W s_R) + \mathrm{h.c.}\,.
\label{eq:Wts_2}
\end{equation}
As expected, in the custodially-invariant scenarios, the right-handed $W$ couplings arise only at fourth order
in the elementary/composite mixings. Moreover they are necessarily weighted by the left-handed
mixings $\lambda^{(\prime )}_{q_L}$, which are the only ones that break the custodial symmetry.
As a consequence the estimates in Eqs.~(\ref{eq:Wtb_2}) and (\ref{eq:Wts_2}) differ
with respect to the general ones (see Eqs.~(\ref{eq:Wtb_1}) and (\ref{eq:Wts_1})) by a factor
\begin{equation}
\sim \frac{\lambda_{q_L}^3 \lambda'^3_{q_L}}{g_*^2}  \simeq \frac{\sqrt{2} m_t}{v} \frac{x_t}{g_* z}\,.
\end{equation}
This factor implies a mild suppression, which is significant only for large values of $g_*$. 
By using the bounds in Eq.~(\ref{eq:C7_bounds}) the following constraints on $m_*$ can be derived
\begin{equation}
m_* \gtrsim 0.3\ \mathrm{TeV}\hspace{.75em} \textrm{(from $C_{7\gamma}$)}\,,
\qquad
m_* \gtrsim 0.5\ \mathrm{TeV}\hspace{.75em} \textrm{(from $C'_{7\gamma}$)}\,.
\label{eq:bound_WRcoupl_2}
\end{equation}

\subsection[$\Delta F = 2$ transitions]{${\mathbf{\Delta F = 2}}$ transitions}

Another set of constraints on the anarchic composite Higgs scenarios can be derived from the $\Delta F = 2$
flavor-violating transitions. The main short-distance sources contributing to these processes
are the contact operators of the type
\begin{equation}
{\mathcal L}_{\Delta F = 2} \sim \frac{\lambda_i \lambda_j \lambda_k \lambda_l}{g_*^2} \frac{1}{m_*^2}
(\overline f_i \gamma^\mu f_j) (\overline f_k \gamma_\mu f_l)\,.
\label{eq:DeltaF2_short}
\end{equation}
Additional long-distance contributions can be mediated by the exchange of the $Z$ boson or the Higgs via a pair of
$\Delta F = 1$ flavor-violating vertices. By using the estimate in Eq.~(\ref{eq:deltaZ_op}), it is easy to show that the
$Z$-mediated effects are suppressed by a factor $v^2/f^2$ with respect to the short-distance ones in
Eq.~(\ref{eq:DeltaF2_short}). The Higgs-mediated effects, on the other hand, can lead to sizable effects
if they are generated at leading order in the elementary/composite mixings:
\begin{equation}
\frac{\lambda_i \lambda_j}{m_* f} \overline f_i H f_j H^\dagger H\,.
\end{equation}
By integrating out the Higgs, the ratio between the long-distance effects and the short-distance ones
in Eq.~(\ref{eq:DeltaF2_short}) is found to be of order
\begin{equation}
\sim g_*^2 \frac{v^2}{f^2} \frac{v^2}{m_h^2} \simeq g_*^2 \left(\frac{500\ \mathrm{GeV}}{f}\right)^2\,.
\end{equation}
This shows that, on general grounds, the Higgs-mediated effects could be large and could even dominate
over the short-distance ones. However, as we discussed
in the previous section, in a large class of minimal scenarios, the Higgs interactions are protected by the
Goldstone structure and flavor-violating vertices are strongly suppressed.
In this case the Higgs mediated contributions to $\Delta F = 2$ transitions are reduced
by a factor $\sim y_t^2/(4\pi)^2$ and are under control. In the following we assume that the protection mechanism is at work
and we only focus on the flavor-violating effects coming from short-distance operators.

By the use of Fierz identities the $4$-fermion contact operators relevant for $\Delta F = 2$ flavor-violating
transitions can be reduced to $8$ independent Lorentz structures. Their explicit form is
\begin{eqnarray}
{\mathcal Q}^{ij}_1 &=& (\overline f_{iL}^\alpha \gamma^\mu f_{jL}^\alpha) (\overline f_{iL}^\beta \gamma_\mu f_{jL}^\beta)\,,\\
{\mathcal Q}^{ij}_2 &=& (\overline f_{iR}^\alpha f_{jL}^\alpha) (\overline f_{iR}^\beta f_{jL}^\beta)\,,\\
{\mathcal Q}^{ij}_3 &=& (\overline f_{iR}^\alpha f_{jL}^\beta) (\overline f_{iR}^\beta f_{jL}^\alpha)\,,\\
{\mathcal Q}^{ij}_4 &=& (\overline f_{iR}^\alpha f_{jL}^\alpha) (\overline f_{iL}^\beta f_{jR}^\beta)\,,\\
{\mathcal Q}^{ij}_5 &=& (\overline f_{iR}^\alpha f_{jL}^\beta) (\overline f_{iL}^\beta f_{jR}^\alpha)\,,
\end{eqnarray}
and the remaining $3$ operators $\widetilde{\mathcal Q}^{ij}_{1,2,3}$ are obtained from the ${\mathcal Q}^{ij}_{1,2,3}$
operators by flipping the quark chiralities. In the above operators $i,j$ denote the flavor indices, while $\alpha, \beta$
are color indices. Within the One Scale One Coupling power counting, the coefficients of these operators only depend on the fermion species and chiralities that are involved. However extra suppressions are possible in explicit models where the operators are generated by the exchange of vector resonances. In this case, for instance, ${\mathcal Q}_{2,3}$ and $\widetilde {\mathcal Q}_{2,3}$ can not be  generated at the tree-level and thus their coefficient displays a loop suppression. Moreover, different operators are associated to the exchange of different kinds of resonances, so that their coefficient is controlled by the masses and the couplings of different particles, which can be numerically different in the specific model. For instance it is easy to see, using Fierz identities, that ${\mathcal Q}_4$ can only arise from the exchange of vector resonances with non-trivial QCD quantum numbers. On the other hand, vector states charged only under $\SO(5) \times \U(1)_X$
can only give rise to the ${\mathcal Q}_1$, $\widetilde {\mathcal Q}_1$ and ${\mathcal Q}_5$
operators. In what follows we will estimate the constraint by using the generic One Scale One Coupling power counting estimate. However since some of the strongest constraints come from ${\mathcal Q}_2$, $\widetilde {\mathcal Q}_2$ and ${\mathcal Q}_4$ operators, the selection rule described above can have an important impact on the interpretation of the experimental bounds in explicit models. Also numerical suppression or enhancement factors due to the multiplicities and the charge of the exchanged resonances can have an impact. We refer the interested reader to the Refs.~\cite{Csaki:2008zd,Barbieri:2012tu} for details. 

\subsubsection{The Kaon system}

The first set of experimental measurements we consider is the one related to the Kaon system.
The main flavor-violating effects in this system are related to the mixing of the
$K^0$ resonance with the corresponding anti-particle $\overline K^0$ and are encoded in the
observables $\Delta m_K$ and $\epsilon_K$.
The short-distance new-physics contributions to this processes are due to $\Delta S = 2$ contact interactions
involving two $s$ and two $d$ quarks.

For illustrative purposes we will focus only on the effective operators ${\mathcal Q}^{sd}_1$, ${\mathcal Q}^{sd}_2$
and ${\mathcal Q}^{sd}_4$, whose bounds lead to the strongest constraints on the new-physics
dynamics.\footnote{In the cases we consider the $\widetilde {\mathcal Q}_1$ operator
is generated with a smaller coefficient than ${\mathcal Q}_1$, while the bounds on the two operators are comparable.
The ${\mathcal Q}_2$ and ${\mathcal Q}_3$ operators are expected to have similar size, but the
bounds on the former are always tighter. The same happens for ${\mathcal Q}_4$
and ${\mathcal Q}_5$~\cite{Bona:2007vi}.} The experimental results are usually expressed as constraints on the
couplings $C$ that appear in the effective Lagrangian
\begin{equation}
{\mathcal L}_{\Delta F = 2} = \sum \frac{C({\mathcal Q}_i)}{\Lambda^2} {\mathcal Q}_i\,.
\end{equation}
The mass scale $\Lambda$ is conveniently chosen to be $\Lambda = 1\ \mathrm{TeV}$.
As we will see, this mass scale is of the same order of the typical flavor bounds on the new physics scale
in the composite Higgs scenarios, thus helping to keep running effects under control.
The current constraints can be expressed as follows~\cite{Bona:2007vi,Isidori:2013ez}
\begin{align}
&|\textrm{Re}\, C({\mathcal Q}^{sd}_1)| \lesssim 9.0 \times 10^{-7}\,,
\quad&
&|\textrm{Im}\, C({\mathcal Q}^{sd}_1)| \lesssim 3.4 \times 10^{-9}\,,\\
\rule{0pt}{1.35em}&|\textrm{Re}\, C({\mathcal Q}^{sd}_2)| \lesssim 1.9 \times 10^{-8}\,,
\quad&
&|\textrm{Im}\, C({\mathcal Q}^{sd}_2)| \lesssim 1.0 \times 10^{-10}\,,\\
\rule{0pt}{1.35em}&|\textrm{Re}\, C({\mathcal Q}^{sd}_4)| \lesssim 6.9 \times 10^{-9}\,,
\quad&
&|\textrm{Im}\, C({\mathcal Q}^{sd}_4)| \lesssim 2.6 \times 10^{-11}\,,
\end{align}
where, due to the significant asymmetry in the bounds, we reported separately the ones on the real and imaginary parts
of the coefficients. The bounds on $C({\mathcal Q}^{sd}_2)$ also apply to the $C(\widetilde {\mathcal Q}^{sd}_2)$
coefficient.

The size of the new-physics contributions to the effective operators can be estimated as
\begin{eqnarray}
C({\mathcal Q}^{sd}_1) &\sim& \left(\frac{\sqrt{2} m_t}{v}\right)^2 x_t^2 \lambda_C^{10}
\frac{1}{m_*^2}\,,\label{eq:K_q1}\\
\rule{0pt}{1.75em} C({\mathcal Q}^{sd}_2) &\sim& \left(\frac{\sqrt{2} m_s}{v}\right)^2 z^2 \lambda_C^2
\frac{1}{m_*^2}\,,\label{eq:K_q2}\\
\rule{0pt}{1.75em} C(\widetilde {\mathcal Q}^{sd}_2) &\sim& \left(\frac{\sqrt{2} m_d}{v}\right)^2 \frac{z^2}{\lambda_C^2}
\frac{1}{m_*^2}\,,\label{eq:K_q2t}\\
\rule{0pt}{1.75em} C({\mathcal Q}^{sd}_4) &\sim& \frac{\sqrt{2} m_s}{v}\frac{\sqrt{2} m_d}{v} z^2
\frac{1}{m_*^2}\,.\label{eq:K_q4}
\end{eqnarray}
Notice that, in deriving the above expressions, we took into account the fact that, if the left-handed quarks are mixed with
different composite operators, multiple independent contributions to the contact interactions can be generated.
Typical examples are the models in which the composite operators belong to the fundamental $\SO(5)$ representation.
As we discussed in Sect.~\ref{sec:anarchic_pc}, in these set-ups the left-handed quarks
are mixed with two composite operators belonging to the ${\bf 5}_{2/3}$
and ${\bf 5}_{-1/3}$ representations of $\SO(5) \times \U(1)_X$. In Eqs.~(\ref{eq:K_q1})-(\ref{eq:K_q4}) we reported the contributions coming from the mixing that generates the up-type quark masses, which
gives the leading new-physics effect. To obtain the results in the scenarios with only one mixing,
it is enough to set $z = 1$.

The experimental constraints can be easily translated into lower bounds on the scale of the composite dynamics $m_*$.
The numerical values of the bounds coming from the various operators are roughly comparable, although each
one has a different parametric dependence on the free parameters $x_t$ and $z$.
By assuming a maximal complex phase for the new-physics contributions, the constraints on the
${\mathcal Q}_1^{sd}$ operator give the following results
\begin{equation}
\begin{array}{ll}
m_* \displaystyle\gtrsim 0.4\, x_t\ \mathrm{TeV}
\qquad \qquad & \textrm{from $\mathrm{Re}\, C({\mathcal{Q}}^{sd}_1)$}\,,\\
\rule{0pt}{1.75em} m_* \displaystyle\gtrsim 6\, x_t\ \mathrm{TeV}
\qquad \quad & \textrm{from $\mathrm{Im}\, C({\mathcal{Q}}^{sd}_1)$}\,.
\end{array}
\end{equation}
The bounds on ${\mathcal Q}_2^{sd}$ and $\widetilde {\mathcal Q}_2^{sd}$ correspond respectively to
\begin{equation}
\begin{array}{ll}
m_* \displaystyle\gtrsim 0.4\, z\ \mathrm{TeV}
\qquad \hspace{2em} & \textrm{from $\mathrm{Re}\, C({\mathcal{Q}}^{sd}_2)$}\,,\\
\rule{0pt}{1.75em} m_* \displaystyle\gtrsim 5\, z\ \mathrm{TeV}
\qquad & \textrm{from $\mathrm{Im}\, C({\mathcal{Q}}^{sd}_2)$}\,,
\end{array}
\end{equation}
and
\begin{equation}
\begin{array}{ll}
m_* \displaystyle\gtrsim 0.4\, z\ \mathrm{TeV}
\qquad \hspace{2em} & \textrm{from $\mathrm{Re}\, C(\widetilde {\mathcal{Q}}^{sd}_2)$}\,,\\
\rule{0pt}{1.75em} m_* \displaystyle\gtrsim 6\, z\ \mathrm{TeV}
\qquad & \textrm{from $\mathrm{Im}\, C(\widetilde {\mathcal{Q}}^{sd}_2)$}\,.
\end{array}
\end{equation}
Finally, for the ${\mathcal Q}_4^{sd}$ operator we get
\begin{equation}
\begin{array}{ll}
m_* \displaystyle\gtrsim 0.6\, z\ \mathrm{TeV}
\qquad \qquad & \textrm{from $\mathrm{Re}\, C({\mathcal{Q}}^{sd}_4)$}\,,\\
\rule{0pt}{1.75em} m_* \displaystyle\gtrsim 10\, z\ \mathrm{TeV}
\qquad & \textrm{from $\mathrm{Im}\, C({\mathcal{Q}}^{sd}_4)$}\,.
\end{array}
\end{equation}
As can be easily seen from the above results, the constraints on $m_*$ coming from the bounds on the
imaginary part of the coefficients of the effective operators are quite stringent and favor a
compositeness scale around $10\ \mathrm{TeV}$. The bound can become even more severe in the models with
multiple mixings if $z > 1$, that is if the mixing to the up-type operators is
larger than the one to the down operators. In order to minimize the constraints one needs to ensure that
$z \simeq 1$, which requires the up-type mixings $\lambda^i_{q_L}$ to be of the same size
of the down-type ones $\lambda'^i_{q_L}$ (see discussion in Sect.~\ref{sec:anarchic_pc}).
Notice also that the bounds coming from the ${\mathcal Q}_1^{sd}$ operator can be reduced if $x_t < 1$.

\subsubsection{The ${\mathbf{B_d}}$ and ${\mathbf{B_s}}$ systems}

Other important constraints on the composite dynamics come from the flavor-violating processes involving
the $B_d$ and $B_s$ mesons. These observables can be used to put some bounds on the class of effective
$4$-fermion interactions leading to $\Delta B = 2$ transitions, namely the ${\mathcal Q}^{bd}$ and ${\mathcal Q}^{bs}$
operators. The analysis of these effects is completely analogous to the one we described for the Kaon system,
we thus skip all the details and we only list all the constraints on the $m_*$ scale in Table~\ref{tab:bounds_DeltaF=2}.

An interesting difference with respect to the Kaon bounds, is the fact that the constraints from the $B_{d,s}$ systems
are nearly independent of the complex phase in the new-physics operators. In particular the strongest constraints
come from the ${\mathcal Q}_1$ operators and are given by
\begin{equation}
\begin{array}{ll}
m_* \displaystyle\gtrsim 5\, x_t\ \mathrm{TeV}
\qquad & \textrm{from $\mathrm{Re}\, C({\mathcal{Q}}^{bd}_1)$}\,,\\
\rule{0pt}{1.75em} m_* \displaystyle\gtrsim 7\, x_t\ \mathrm{TeV}
\qquad & \textrm{from $\mathrm{Im}\, C({\mathcal{Q}}^{bd}_1)$}\,,
\end{array}
\label{eq:bound_Q1bd}
\end{equation}
for the $B_d$ system and
\begin{equation}
\begin{array}{ll}
m_* \displaystyle\gtrsim 5\, x_t\ \mathrm{TeV}
\qquad & \textrm{from $\mathrm{Re}\, C({\mathcal{Q}}^{bs}_1)$}\,,\\
\rule{0pt}{1.75em} m_* \displaystyle\gtrsim 8\, x_t\ \mathrm{TeV}
\qquad & \textrm{from $\mathrm{Im}\, C({\mathcal{Q}}^{bs}_1)$}\,,
\end{array}
\label{eq:bound_Q1bs}
\end{equation}
for the $B_s$ system. Analogously to what we found for the Kaon observables, the flavor bounds
in Eqs.~(\ref{eq:bound_Q1bd}) and (\ref{eq:bound_Q1bs}) can be relaxed if $x_t < 1$.

\begin{table}[t]
	\centering
	\begin{tabular}{c|cc|c}
		\rule[-.5em]{0pt}{1.25em} Operator & $\mathrm{Re}\, C$ & $\mathrm{Im}\, C$ & Observables\\
		\hline
		\hline
		\rule{0pt}{1.25em} ${\mathcal Q}_1^{sd}$ & $0.4\, x_t$ & $6\, x_t$ & $\Delta m_K; \epsilon_K$ \\
		\rule{0pt}{1.25em} ${\mathcal Q}_2^{sd}$ & $0.4\, z$ & $5\, z$ & ''\\
		\rule{0pt}{1.25em} $\widetilde {\mathcal Q}_2^{sd}$ & $0.4\, z$ & $6\, z$ & '' \\
		\rule[-.65em]{0pt}{1.9em} ${\mathcal Q}_4^{sd}$ & $0.6\, z$ & $10\, z$ & '' \\
		\hline
		\rule{0pt}{1.25em} ${\mathcal Q}_1^{bd}$ & $5\, x_t$ & $7\, x_t$ & $\Delta m_{B_d}; S_{\psi K_S}$ \\
		\rule{0pt}{1.25em} ${\mathcal Q}_2^{bd}$ & $1.4\, z$ & $2\, z$ & ''\\
		\rule[-.65em]{0pt}{1.9em} ${\mathcal Q}_4^{bd}$ & $0.6\, z$ & $0.8\, z$ & '' \\
		\hline
		\rule{0pt}{1.25em} ${\mathcal Q}_1^{bs}$ & $5\, x_t$ & $8\, x_t$ & $\Delta m_{B_s}$\\
		\rule{0pt}{1.25em} ${\mathcal Q}_2^{bs}$ & $0.6\, z$ & $1\, z$ & ''\\
		\rule[-.65em]{0pt}{1.9em} ${\mathcal Q}_4^{bs}$ & $0.5\, z$ & $1\, z$ & '' \\
		\hline
		\rule{0pt}{1.25em} ${\mathcal Q}_1^{cu}$ & $0.5\, x_t$ & $1.2\, x_t$ & $\Delta m_{D}; |q/p|, \phi_D$ \\
		\rule{0pt}{1.25em} ${\mathcal Q}_2^{cu}$ & $1.4$ & $3$ & ''\\
		\rule{0pt}{1.25em} ${\mathcal Q}_4^{cu}$ & $0.5$ & $1.1$ & '' \\
	\end{tabular}
	\caption{Lower bounds on the composite dynamics scale $m_*$ (expressed in $\mathrm{TeV}$) obtained from the
		$\Delta F = 2$ flavor-violating processes. The bounds coming from the constraints on the real and imaginary parts
		of the coefficients of the effective operators are listed separately and are derived by assuming maximal complex phases in the
		new-physics contributions. The constraints on the effective operators are taken from Refs.~\cite{Bona:2007vi,Isidori:2013ez}.
		The last column lists the observables used to derive the bounds.}
	\label{tab:bounds_DeltaF=2}
\end{table}

\subsubsection{The ${\mathbf{D}}$ system}

Another set of constraints on the $4$-fermion flavor-violating interactions comes from the $D$ system.
Differently from the $K$ and $B$ mesons, which are described by operators involving
the down-type quarks, the $D$ observables are influenced by operators involving the up-type quarks.
In particular the new-physics contact interactions that mediate $\Delta C = 2$ transitions are parametrized
by the ${\mathcal Q}^{cu}$ operators. The constraints coming from the present measurements
are listed in Table~\ref{tab:bounds_DeltaF=2}.

Notice that, as a consequence of the $\SU(2)_L$ symmetry, the effective operators parametrizing
the flavor-violating observables in the $D$ system are correlated to the ones related to the $K$ system.
In particular the ${\mathcal Q}_1^{cu}$ operator belongs to the same multiplet as the ${\mathcal Q}_1^{sd}$
operator, thus the constraints coming from the two operators are not independent. Due to the better determination of the
Kaon observables, the bounds coming from the ${\mathcal Q}_1^{sd}$ interaction are currently stronger
than the ones from ${\mathcal Q}_1^{cu}$. This is particularly noticeable on the bounds on the imaginary
part of the ${\mathcal Q}_1$ coefficient, while the bounds on the real part are of the same order.

\subsection{The neutron EDM}\label{sec:neutron_EDM}

As a last set of constraints on the flavor-anarchic scenarios we consider the ones coming from the
measurement of the neutron EDM $d_n$. To estimate the bounds we will closely follow the analysis
of Refs.~\cite{KerenZur:2012fr,Konig:2014iqa}. The largest new-physics contributions to $d_n$ arise from
the flavor-conserving EW and QCD dipole operators
\begin{equation}
{\mathcal O}_{f\!f\gamma} = C_{f\!f\gamma} \frac{e\, m_f}{16 \pi^2} \overline f \sigma^{\mu\nu} F_{\mu\nu} \gamma^5 f\,,
\qquad
{\mathcal O}_{f\!fg} = C_{f\!fg} \frac{g_s m_f}{16 \pi^2} \overline f \sigma^{\mu\nu} G_{\mu\nu} \gamma^5 f\,,
\end{equation}
where $f = u,d$ denotes the first generation quarks. The calculation of the contributions of the
quark dipole operators to the neutron EDM suffers from large hadronic uncertainties.
By using the QCD sum rules the following estimate can be derived~\cite{Pospelov:2000bw}
\begin{equation}
d_n = (1 \pm 0.5) \left[1.4\left(d_d - \frac{1}{4} d_u\right) + 1.1 e \left(\widetilde d_d + \frac{1}{2} \widetilde d_u\right)\right]\,,
\end{equation}
where $d_f$ and $\widetilde d_f$ denote the electric and chromoelectric dipole moments of the quarks defined as
\begin{equation}
d_f \equiv \frac{e\, m_f}{8 \pi^2} \mathrm{Im}\, C_{f\!f\gamma}\,,
\qquad
\widetilde d_f \equiv \frac{m_f}{8 \pi^2} \mathrm{Im}\, C_{f\!fg}\,.
\end{equation}

The current experimental constraint on the neutron EDM is~\cite{Agashe:2014kda}
\begin{equation}
d_n < 2.9 \times 10^{-26} e\, \mathrm{cm}
\qquad
\mathrm{at\ 90\%\ C.L.}\,.
\label{eq:bound_nEDM}
\end{equation}
Banning accidental cancellations among the various contributions, the following constraints on the composite
scale $m_*$ can be derived
\begin{eqnarray}
&& m_* \gtrsim 60\ \mathrm{TeV} \qquad \mathrm{from}\ \ {\mathcal O}_{dd\gamma}\,,\\
\rule{0pt}{1.15em} && m_* \gtrsim 20\ \mathrm{TeV} \qquad \mathrm{from}\ \ {\mathcal O}_{uu\gamma}\,,\\
\rule{0pt}{1.15em} && m_* \gtrsim 55\ \mathrm{TeV} \qquad \mathrm{from}\ \ {\mathcal O}_{ddg}\,,\\
\rule{0pt}{1.15em} && m_* \gtrsim 25\ \mathrm{TeV} \qquad \mathrm{from}\ \ {\mathcal O}_{uug}\,.
\end{eqnarray}
It is important to stress that the neutron EDM is only generated by the CP-violating effects induced by the
imaginary part of the coefficients of the dipole operators. The real part of $C_{f\!f\gamma}$ and $C_{f\!fg}$, instead,
does not contribute to $d_n$. The presence of the above bounds is thus crucially related to the fact that,
in the anarchic scenario, order-one complex phases are generically expected in all new-physics operators.

Indirectly, the neutron EDM is also sensitive to the QCD dipole operators involving the second and third generation quarks.
This effect is due to the QCD running of the dipole operators at low energy. When a heavy quark is integrated out a
threshold correction is generated that contributes to the three-gluon Weinberg operator. This operator, in turn,
directly contributes to the neutron EDM and mixes under renormalization with the light quark
dipole operators~\cite{Braaten:1990gq}.
Taking into account the experimental constraint in Eq.~(\ref{eq:bound_nEDM}) the following bounds on the charm,
bottom and top chromoelectric dipole operators can be derived~\cite{Chang:1990jv,Kamenik:2011dk,Sala:2013osa}
\begin{equation}
|\widetilde d_c| < 1.0 \times 10^{-22}\ \mathrm{cm}\,,
\qquad
|\widetilde d_b| < 1.1 \times 10^{-21}\ \mathrm{cm}\,,
\qquad
|\widetilde d_t| < 2.1 \times 10^{-19}\ \mathrm{cm}\,.
\end{equation}
Although weaker than the bounds from the operators involving the first-generation quarks,
the bounds on the $m_*$ scale coming from the heavier quarks are still quite stringent:
\begin{eqnarray}
&& m_* \gtrsim 14\ \mathrm{TeV} \qquad \mathrm{from}\ \ {\mathcal O}_{ccg}\,,\\
\rule{0pt}{1.15em} && m_* \gtrsim 9\ \mathrm{TeV} \qquad \ \, \mathrm{from}\ \ {\mathcal O}_{bbg}\,,\\
\rule{0pt}{1.15em} && m_* \gtrsim 5\ \mathrm{TeV} \qquad \ \, \mathrm{from}\ \ {\mathcal O}_{ttg}\,.
\end{eqnarray}

To conclude we remind that, as we did in the whole analysis, to derive the above bounds we assumed that the
dipole operators are generated at tree level. If they can only be induced by loop effects
the above constraints must be interpreted as bounds on the $f$ scale rather than on $m_*$.
To derive the corresponding bound it is sufficient to use the replacement $m_* \rightarrow 4 \pi f$.
Notice that, even assuming the extra suppression the bounds remain quite strong. For instance, from the
constraints on the operators with first-generation quarks we get
\begin{equation}
f \gtrsim 4.5\ \mathrm{TeV} \quad \textrm{($d$\ quark)}\,,
\qquad \quad
f \gtrsim 2\ \mathrm{TeV} \quad \textrm{($u$\ quark)}\,.
\end{equation}

\subsection{Beyond One Scale One Coupling}\label{sec:VMD_interpretation}

So far we analyzed the experimental constraints on the anarchic scenarios by adopting the
One Scale One Coupling assumption. However this can be a very crude parametrization of the composite dynamics. We will now discuss how to obtain some refinements.
Doing this is particularly important because of the tension, which we established in Chap.~\ref{ch:SILH}
(in particular, see the discussion at the end of Sect.~\ref{HVMT}), among the scale of the bosonic vector resonances
and the one of the composite fermionic states. The former controls the corrections to EWPT observables and is thus
preferentially heavy, whereas the latter determines
the generation of the Higgs potential and is preferentially light in order to avoid fine-tuning in the Higgs mass.
For this reason it is worth introducing a new mass scale $m_\psi$ (with its associated coupling $g_\psi=m_\psi/f$) that characterizes the fermionic states~\cite{Matsedonskyi:2012ym,KerenZur:2012fr,Panico:2012uw}. This scale is taken to be smaller than the one of the vector resonances, which we keep calling $m_*$. The mass $m_\psi$, and in turn the coupling $g_\psi$, controls the breaking of the chiral symmetry that is present in the limit of zero resonances mass. Any operator that breaks this symmetry is thus weighted by $g_\psi$. These operators include the coupling of the fermionic resonances
to the Higgs, which now is proportional to $g_\psi$ and not $g_*$. As a consequence the elementary/composite mixings are now connected to the quark Yukawa's by the relation $y_i \sim \lambda_{Li} \lambda_{Ri}/g_\psi$.

One further generalization of the One Scale One Coupling assumption that is worth considering is to allow for multiple mass scales and couplings for the vector resonances, distinguishing those that carry QCD charge from the ones that are color neutral. In what follows we will denote as $m_*^{\textrm{QCD}}$ and $g_*^{\textrm{QCD}}$ the QCD-colored particles mass and coupling, reserving $m_*$ and $g_*$ to characterize the color-neutral ones. Notice that the quantity $m_*^{\textrm{QCD}}/g_*^{\textrm{QCD}}$ does not necessarily coincide with the Goldstone boson decay constant $f$. On the other hand, for the fermionic resonances $m_\psi/g_\psi=f$, given that the latter relation is taken as the definition of the effective fermionic coupling $g_\psi$.

Going beyond the One Scale One Coupling assumption also requires one further assumption on the composite sector dynamics. In particular in what follows we will rely on the Vector Meson Dominance (VMD) hypothesis that we already introduced and discussed at the end of Sect.~\ref{sec:PCbasic}. This is the idea that the communication between the elementary and the composite sectors originates from mass-mixings of the elementary fields with composite resonances of appropriate quantum numbers. In this hypothesis the generation of the effective operators can be associate to Feynman diagrams and their coefficients estimated in terms of the relevant masses and couplings.

\begin{figure}
	\centering
	\includegraphics[width=.5\textwidth]{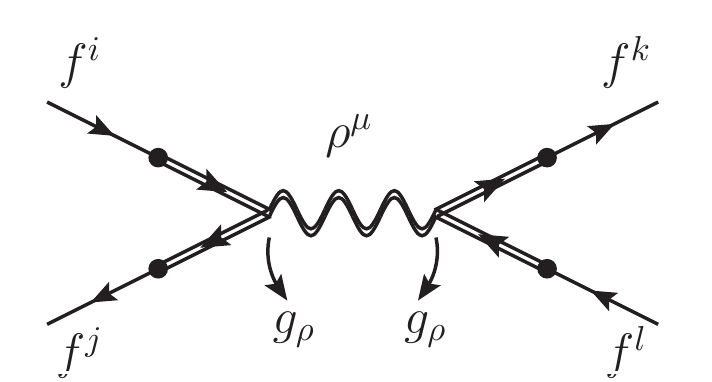}
	\caption{Schematic structure of a diagram giving rise to a $4$-fermion contact interaction through the exchange
		of a heavy vector resonance. The double-line propagators denote the composite states.}
	\label{fig:4-fermion_VMD}
\end{figure}

We start by considering the $4$-fermion effective interactions, associated with the diagram in Fig.~\ref{fig:4-fermion_VMD}. Each insertion of an elementary fermion $f_i$ is weighted by its amount of compositeness $\lambda_i/g_\psi$ that comes from the mixing of the elementary fermions with their fermionic partners, which has strength $\lambda f$ (see Eq.~(\ref{mixlag})), times the $1/m_\psi$ from the fermionic partner propagator. An additional weight $g_*^2/m_*^2$  comes from the vector resonance propagator and from its couplings to the composite fermionic resonances. By putting the factors together we find
\begin{equation}
\mathcal{L}_{\Delta F=2} \sim \frac{\lambda_i \lambda_j \lambda_k \lambda_l}{g_\psi^4} \frac{g_*^2}{m_*^2}
(\overline f_i \gamma^\mu f_j)(\overline f_k \gamma_\mu f_l)\,.
\end{equation}
The above derivation  holds for color-neutral vector resonances, if the effective operator is instead generated by the exchange of a colored vector resonance the $g_*^2/m_*^2$ factor must, obviously, be replaced by $(g_*^{\textrm{QCD}}/m_*^{\textrm{QCD}})^2$.

The modified power counting has some important consequences on the interpretation of the flavor bounds. Keeping the Yukawa's of the quarks fixed, the estimate of the $4$-fermion effective operators is rescaled
by a factor $(g_*/g_\psi)^2$ with respect to the One Scale One Coupling result in Eq.~(\ref{eq:DeltaF2_short}).
The presence of light fermionic partners thus worsens the flavor-violating effects.
It is easy to check that, once the Yukawa's are fixed, the coefficient of the effective operators
mediated by $\SO(5)$ vector resonances depends only on the fermionic partners mass.
This means that the constraints must be interpreted as lower bounds on $m_\psi$, and not on
the mass of the vector states.
The operators mediated by QCD states, instead, can be used to derive constraints on the quantity
$m_*^{\textrm{QCD}} (g_\psi/g_*^{\textrm{QCD}})$. These bounds favor heavy and relatively weakly coupled QCD resonances.

Similar considerations apply to the penguin operators whose estimate is modified as
\begin{equation}
{\mathcal L}_{\Delta F = 1} \sim \frac{\lambda_i \lambda_j}{g_\psi^2} \frac{g_*^2}{m_*^2}
\overline f_i \gamma^\mu f_j i H^\dagger \dblarrow{D}_\mu H\,.
\end{equation}
With respect to the estimate in Eq.~(\ref{eq:deltaZ_op}) the presence of light partners induces an
additional factor $(g_*/g_\psi)^2$. One power of $g_*/g_\psi$ is removed when we fix the Yukawa couplings, thus
leaving an enhancement $g_*/g_\psi$ of the effects of the penguin operators.

The situation is instead different for the dipole operators. In this case one has to take into account the need of a chirality flip in the fermionic resonance line, {\it{i.e.}}~of a chiral symmetry protection that  requires the
presence of an extra $m_\psi$ factor in the coefficient of the operator. We thus find
\begin{equation}\label{eq:bnd}
{\mathcal L}_{\Delta F = 1} \sim \frac{\lambda_i \lambda_j}{g_\psi} \frac{v}{m_*^2} \overline f_i \sigma_{\mu\nu}
g_{SM} F^{\mu\nu}_{SM} f_j\,.
\end{equation}
Once the Yukawa's are fixed this estimate coincides with the One Scale One coupling result in Eq.~(\ref{eq:dipole_op}),
thus the bounds are unchanged. Obviously in Eq.~(\ref{eq:bnd}) $m_*$ corresponds to the mass of the relevant vector resonances
contributing to the effective operator. For instance, in the case of the QCD dipole operators the $m_*$ mass scale
must be interpreted as $m_*^{\textrm{QCD}}$.

\section{Flavor symmetric scenarios}\label{sec:flavor_symm}

The comparison of the anarchic flavor scenario with the experimental data highlighted some
interesting features. The straightforward extension of the partial compositeness assumption
to all the quark generations automatically helps in reducing the flavor-violating effects in comparison with generic flavor-violating new physics at the TeV scale.\footnote{For instance, the experimental bounds
on the $\Delta F = 2$ transitions in the Kaon system require a suppression scale
$\Lambda \gtrsim 10^5\ \mathrm{TeV}$. The RS-GIM mechanism lowers this scale by four orders
of magnitude (see Table~\ref{tab:bounds_DeltaF=2}).} Thus it significantly lowers the suppression scale needed in flavor breaking operators and alleviates the tension between this scale and the energy scale of the EWSB dynamics.
Although the suppression is quite efficient,
the constraints on many flavor observables are so
strong that a residual tension is still present, pushing the natural scale of the
composite dynamics in the $10\ \mathrm{TeV}$ range. This bound is particularly
problematic since, for many observable, it applies directly to the mass of the fermionic partners
that control the amount of tuning in the Higgs potential. Partners with a mass above $10\ \mathrm{TeV}$
imply a minimal amount of tuning of order $0.2\%$ (see Eq.~(\ref{tuning_gen})).

Of course, order one corrections to the estimates and accidental cancellations may be present,
so that the $10\ \mathrm{TeV}$ bound and the related amount of fine-tuning
can be (slightly) relaxed in specific models. Lowering the compositeness scale to the
completely Natural $1\ \mathrm{TeV}$ level, however, seems not realistic. It is thus natural to ask if alternative
implementations of the flavor structure can be conceived that could lead to less constrained
scenarios. To answer this question it is useful to take a step back and reconsider the
broad features of the anarchic scenario in comparison with the flavor structure of the SM.

The fundamental feature characterizing the anarchic scenario is the RS-GIM mechanism,
which implies that new-physics flavor effects are weighted by the amount of compositeness
of the SM quarks. Due to the hierarchies in the elementary/composite mixings
(see Eqs.~(\ref{eq:CKM_estimate})-(\ref{eq:mixing_estimate})),
the amount of suppression in the flavor-violating observables is roughly equal to the corresponding
off-diagonal elements of the CKM matrix. The suppression is thus comparable to the one present in
MFV scenarios and, in particular, in the SM.

The anarchic scenario, however, differs from the SM in two fundamental aspects, namely the
generation of FCNC's and the violation of CP. Let us consider these issues more closely.

In the SM any FCNC can only be generated at loop level. In particular the couplings
of the $Z$ boson and of the Higgs are completely flavor-diagonal. As we saw in the previous sections,
this is not the case in anarchic partial compositeness. In this scenario the Higgs couplings can mediate flavor-changing transitions,
although in many models these effects can be suppressed due to a partial MFV in the Higgs sector.
The situation is slightly more complicated for the $Z$ couplings. In general the tree-level
$Z$ interactions are not protected and can mediate flavor-changing currents arising at leading
order in the elementary/composite mixings. Some protection can be present
in specific scenarios thanks to the existence of discrete $\Zdouble_2$ symmetries
(see Table~\ref{tab:PLR_and_PC}). In all the models we considered, however, a full protection
is never achieved and flavor-violation in the $Z$ sector always leads to significant bounds.
The Higgs and $Z$ interactions, however, are not the most dangerous sources of FCNC's in the anarchic scenarios.
In fact more problematic new-physics effects are due to $4$-fermion contact interactions
mediating $\Delta F = 2$ transitions. In models with VMD, these operators are due to the presence of
heavy vector resonances, which, in the absence of any flavor symmetry, generate arbitrary
$4$-point interactions among the fermionic partners. In the SM, instead, analogous effective interactions are only generated at loop level and are further suppressed by the GIM mechanism. These features emerge from the fact that the SM is a weakly-coupled perturbative theory and from the fact that it automatically  obeys the MFV hypothesis, namely that the only sources of breaking of the flavor symmetry group \mbox{U$(3)_{q_L}\times$U$(3)_{u_R}\times$U$(3)_{d_R}$} are the two Yukawa matrices. Both conditions are violated in the anarchic composite Higgs scenario.

The second main difference with respect to the SM is related to CP violation. In the SM, again due to MFV, CP is broken only in the presence of three quark families, otherwise the $V_{\textrm{CKM}}$ matrix could be made real by flavor symmetry rotations. Therefore CP-violating effect are  only possible in the SM if all the three generations are ``active'', {\it{i.e.}}~if they circulate in the loops. This implies a huge suppression in the quark dipole moments, in particular for the first generation quarks that determine the nucleons EDM's. In the anarchic composite Higgs case, instead, many additional
flavor-breaking sources are present besides the Yukawa matrices and the flavor symmetries are not enough to reduce all flavor-violating effects to the CKM matrix.
Moreover the flavor structures contain several new complex phases that can not be removed
and lead to new CP-violating effects. This structure has a direct impact on the bounds on the
composite dynamics. A first consequence is the presence of order-one complex phases in the flavor-violating operators.
This implies that the more stringent constraints coming from CP-violating observables apply (see for instance
the constraints from $\epsilon_K$ in $\Delta F = 2$ transitions).
Moreover, in general the CP violating phases can not be removed even if only one generation
of elementary quarks is considered. This leads to large contributions to the imaginary coefficients
of the quark dipole operators and, consequently, to the strong constraints from the neutron
EDM discussed in Sect.~\ref{sec:neutron_EDM}.

In the view of the shortcomings of the anarchic scenario it is natural to look for alternative implementations
of the flavor structure. The above discussion suggests a set of ``ideal'' features that allow a flavor
theory to pass the stringent experimental constraints. First of all it should include a suppression of the
flavor-breaking effects roughly comparable to the one provided by the RS-GIM mechanism or by MFV.
Second, FCNC's should not be generated at tree-level or, at least, should be suppressed by some symmetry
mechanism or selection rule. Finally, extra sources of CP violation should be kept under control.

An interesting way to reduce the flavor-violating effects is to assume that a set of global flavor symmetries is
present. As we will see in the following section, different symmetry patterns can be adopted for both the elementary
states and the composite dynamics.
Although the minimal constructions are still based on the partial compositeness hypothesis, the flavor symmetries
seem not compatible with any natural explanation of the hierarchies in the quark masses and in the CKM matrix. This partially undermines the original motivation for partial compositeness.

A more radical solution, which is not yet fully explored in the literature (see Refs.~\cite{Matsedonskyi:2014iha,Cacciapaglia:2015dsa}) and for this reason will not be described here, is to abandon the partial compositeness paradigm for the generation of the light quark Yukawa's. After all, we saw in Sect.~\ref{sec:PCbasic} that the Yukawa couplings could also be generated by bilinear elementary/composite interactions and that the obstruction to this possibility has to do with the generation of the heavy quark masses, not with the light ones. Therefore partial compositeness is the only known viable option for the top (and possibly bottom) Yukawa generation, but we are not obliged to extend it to the light quarks, whose Yukawa's might well be generated by bilinear couplings of the elementary fermions to composite scalar operators. Making this viable most likely requires flavor symmetries and a complete explanation of the mass and $V_{\textrm{CKM}}$ hierarchies is most likely impossible to achieve in this framework, however we encounter the same shortcomings in the flavor-symmetric partial compositeness scenario. Furthermore scalar operators with suitable quantum numbers to couple to the elementary fermion are necessarily present in the composite sector. By ignoring them, as we do in the partial compositeness scenario, we are effectively assuming that they are not coupled, or that their scaling dimension is so high that their effects are completely negligible in the IR. These are not necessarily plausible hypotheses. For this reason a ``mixed'' mechanism for Yukawa generation, with partial compositeness for the third family and bilinear couplings for the others appears a plausible option which is worth investigating. A similar approach will be taken for the study of lepton flavor in Sect.~\ref{sec:leptonic_sector}.

\subsection[The $\U(3)^3$ models]{The $\U{\mathbf{(3)^3}}$ models}
\label{U33mod}

The first class of models we consider is based on the assumption that the composite sector is invariant under
a global $\U(3)_{\textrm{cs}}$ flavor group~\cite{Barbieri:2008zt,Redi:2011zi,Barbieri:2012tu}.\footnote{The first
proposals of flavor symmetric composite Higgs scenarios were developed in the extra-dimensional
framework in Refs.~\cite{Rattazzi:2000hs,Cacciapaglia:2007fw,Santiago:2008vq,Delaunay:2010dw,Delaunay:2011vv}.}
The elementary/composite mixings are the only sources of breaking of the flavor symmetry of the composite dynamics
and of the $\U(3)_{q_L} \times \U(3)_{u_R} \times \U(3)_{d_R}$ symmetry of the elementary sector, thus they
encode the whole flavor structure. In order to suppress the new-physics flavor effects the MFV hypothesis is
realized by some ad hoc assumptions on the structure of the elementary/composite mixings.

One possibility
is to assume that the mixings of the $q_L^i$ doublets are trivial, while the mixings of the right-handed
singlets are proportional to the SM Yukawas:\footnote{This scenario can be easily realized also in the models with
a single left-handed mixing with the choice $\lambda'_{q_L} = \lambda_{q_L} \propto \Id$.}
\begin{equation}
\begin{array}{ll}
\lambda_{q_L} \propto \Id\,, \hspace{2.em}& \lambda'_{q_L} \propto \Id\,,\\
\rule{0pt}{1.em}\lambda_{u_R} \propto y_u\,, & \lambda_{d_R} \propto y_d\,.
\end{array}
\label{eq:U3_left}
\end{equation}
This choice has a direct implication for the compositeness of the left-handed quarks.
Due to the flavor symmetries and the conditions in Eq.~(\ref{eq:U3_left}), all the left-handed quark components $q_L^{i}$
share the same amount of compositeness. In order to reproduce the top quark mass
the $\lambda_{q_L}$ mixing must be relatively large, $\lambda_{q_L} \gtrsim y_t$, and the $q_L^3$ compositeness
must be sizable. As a consequence the light quarks of the first and second generations must also be quite composite.
Because of this feature, the above scenario is usually dubbed ``left-handed compositeness''.

When the right-handed mixings are set to zero, namely for vanishing Yukawa matrices $y_{u,d} = 0$, the theory has a global $\U(3)^3$ flavor symmetry. This acts as $\U(3)_{u_R} \times \U(3)_{d_R}$ transformations on the elementary right-handed fields, times one $\U(3)_{q_L}$ that rotates simultaneously the composite sector, by the $\U(3)_{\textrm{cs}}$ group, and the elementary left-handed doublets. The right-handed mixings, {\it{i.e.}}~the Yukawa matrices $y_{u,d}$, break the $\U(3)^3$ flavor symmetry down to the baryon number $\U(1)_B$. The MFV hypothesis, in which the Yukawa's are the only sources of flavor breaking, is thus realized. This implies in particular that flavor breaking can be entirely ascribed, after a field redefinition, to the $V_{\textrm{CKM}}$ matrix. In order to verify this we can consider the singular value decomposition for the right-handed mixings, which reduces the flavor structure to $4$ unitary rotations. The two ``right'' rotations can be removed by a
redefinition of the singlets $u_R^i$ and $d_R^i$. On the other hand, only one ``left'' rotation can be removed by
a rotation of the left-handed elementary fields accompanied by a corresponding flavor transformation
of the composite sector that leaves the left mixings unchanged. In this way only one flavor-violating structure remains,
which must coincide with the CKM matrix.

A similar construction, the so called ``right-handed compositeness'' can be realized by exchanging
the mixing structure of the left and right quarks:
\begin{equation}
\begin{array}{ll}
\lambda_{q_L} \propto y_u\,, \hspace{2.em}& \lambda'_{q_L} \propto y_d\,,\\
\rule{0pt}{1.em}\lambda_{u_R} \propto \Id\,, & \lambda_{d_R} \propto \Id\,.
\end{array}
\label{eq:U3_right}
\end{equation}
Obviously this scenario can only be realized if the left-handed $q_L$ doublets are mixed to two or more composite operators.
In the scenarios with single mixing the up and down quark masses as well as the CKM structure can not be accommodated. An extension of the flavor symmetry of the composite sector is needed in order to realize MFV in the right-handed compositeness scenario. Two composite sector flavor symmetries  $\U(3)_{\textrm{cs}}^u$ and $\U(3)_{\textrm{cs}}^d$ are needed, rather than one. The first group, $\U(3)_{\textrm{cs}}^u$ acts on the operators ${\mathcal O}_F^{u_R}$ that mix with $u_R$ and on ${\mathcal O}_F^{q_L}$ that mix with the $q_L$. The second group acts on ${\mathcal O}_F^{u_R}$ and on ${{\mathcal O}'}_F^{q_L}$.  Analogously to the models with left-handed compositeness, the requirement of obtaining the correct top mass forces
the $\lambda_{u_R}$ mixings to be large, thus implying a sizable amount of compositeness for all the right-handed
up-type quarks. To a lesser degree this is also true for the down-type quarks, although in this case the minimal amount
of compositeness can be significantly smaller being determined by the bottom mass.

The MFV structure is helpful to reduce some of the flavor constraints with respect to the anarchic case, below we will briefly describe the situation, skipping all details, which can be found in the original literature. The conclusion will be that the extra flavor suppression of this kind of constructions is not enough to completely remove the
tension with the experiments. 

Differently from the anarchic scenarios, the large flavor symmetry of the $\U(3)^3$ models allows to keep under
control the CP violating effects. In particular extra CP violation is not present if we assume that the
composite sector preserves CP. In this case the strong bounds from the neutron EDM disappear.

The flavor symmetry also ensures some cancellation in the FCNC's. In the left compositeness scenarios the
tree-level FCNC's are completely absent~\cite{Redi:2011zi,Barbieri:2012tu}, thus removing the bounds from $\Delta F = 2$
transitions. This is not the case in the right-compositeness models, in which minimally-flavor-violating
FCNC's are generated at tree level~\cite{Barbieri:2012uh,Redi:2012uj}. Under the VMD hypothesis, however,
the $4$-fermion interactions mediating $\Delta F = 2$ transitions contain only the left-handed quarks.
The relevant bounds are thus the ones coming from the ${\mathcal Q}_1$ operators, which still push the compositeness
scale in the multi-TeV range ($m_* \gtrsim 4\ \mathrm{TeV}$).

The large amount of compositeness of the light quarks leads to some additional tension with the
experiments~\cite{Barbieri:2012tu}.
An important effect is the modification of the gauge boson couplings that affects the EW precision measurements.
Significant constraints on the left-compositeness scenarios come form the measurement of the partial width
of the $Z$ boson into hadrons and from the violation of the quark-lepton universality. These observables
imply a lower bound on the compositeness scale of order $m_* \gtrsim 6\sqrt{g_*}\ \mathrm{TeV}$.
Stringent bounds also come from the angular distribution of dijet events at the LHC, which is sensitive to the
compositeness of the first-generation quarks. The corresponding bound is of the order $m_* \gtrsim 4\ \mathrm{TeV}$
and applies to the left-compositeness models as well as the right-compositeness ones~\cite{Domenech:2012ai}.

Another consequence of the large compositeness of the light quarks is an increased direct production cross section for the
fermionic partners. The current LHC bounds already push the mass scale of these resonances to
the few-TeV range~\cite{Redi:2013eaa,Delaunay:2013pwa}. Additional bounds can be derived for the vector resonances,
which can be easily produced from the light quarks. Significant bounds can be derived for the heavy gluons whose
present exclusion, in specific scenarios, can reach the $3\ \mathrm{TeV}$ scale~\cite{Barbieri:2012tu,Redi:2013eaa}.

Finally it is important to notice that in generic strongly coupled theories we expect resonances associated to
all the global symmetries. In particular in the left and right compositeness models we expect extra vector resonances
corresponding to the global flavor symmetry $\U(3)$ and $\U(3)^2$ respectively. These resonances are expected to
have a mass roughly comparable to the one of the usual EW or QCD resonances and could be an additional target for
collider experiments.

\subsection[The $\U(2)^3$ models]{The $\U{\mathbf{(2)^3}}$ models}

From the above discussion it emerges that the $\U(3)$ models can offer a partial improvement with respect to the
anarchic scenarios. The improvement is however not enough to bring down the composite dynamics scale around $1\ \mathrm{TeV}$.
Some of the most problematic aspects of the flavor symmetric models come from the fact that the $\U(3)$ invariance
necessarily connects the amount of compositeness of the light fermions with the one of the third-generation quarks.
This creates a big tension with the high-precision EW measurements involving the light quarks and significantly enhances
the direct collider bounds on the production of resonances (either fermionic partners or heavy gauge fields).

A possible way to modify the above scenarios is to reduce the flavor symmetry by excluding the heavy quarks.
In other words, we can assume only a $\U(2)_{q_L} \times \U(2)_{u_R} \times \U(2)_{d_R}$ symmetry, under which the first
two generation quarks transform as doublets, while the third generation fields
are singlets~\cite{Barbieri:2012uh,Redi:2012uj,Barbieri:2012tu}.
Compared to the $\U(3)^3$ models, a larger number of free parameters is present, however the flavor symmetry
can be broken more weakly since the top Yukawa is now invariant under $\U(2)^3$.
In complete analogy to the previous scenarios, models with left or with right compositeness can be
constructed. In the left-compositeness case the composite dynamics is invariant under a $\U(2)$ flavor symmetry,
while right compositeness requires an extended $\U(2)^2$ symmetry.\footnote{Mixed scenarios considering
a reduced $\U(2)$ flavor symmetry only for one quark chirality and full $\U(3)$ symmetry for the others can also
be constructed~\cite{Redi:2012uj}.}

The full MFV structure can not be reproduced in $\U(2)^3$ models, however a reduced version involving the
first two generations is still at work, suppressing flavor violation for processes involving the light quarks.
FCNC's are generically present both for left and for right compositeness. The largest contributions involve
only the left-handed quarks, thus the strongest constraints come from the ${\mathcal Q}_1$ operators, while
the bounds from ${\mathcal Q}_4$ are not significant~\cite{Barbieri:2012tu}. Although the constraints on
$m_*$ are roughly similar to the ones in the right-compositeness $\U(3)^3$ models,
the additional freedom in the choice of parameters allows a reduction of the bounds to the $m_* \sim 1\ \mathrm{TeV}$ scale.

Flavor changing interactions involving the $Z$ boson are present at tree-level in the models with
left compositeness, while they are absent in the right-compositeness case. The relevant effects only involve the
left-handed currents and the corresponding bounds can be lowered to the $1\ \mathrm{TeV}$ scale
with the same choice of parameters that reduces the tension from $\Delta F = 2$ transitions.

If the composite sector preserves CP, the corrections to the neutron EDM are only generated at higher order in the
parameters that break the $\U(2)^3$ symmetry and are well below the current experimental bounds.

Finally, the constraints from direct production of composite resonances and the ones from the EW precision
measurements on the light quarks become irrelevant due to the small amount of compositeness of these states.
The only bounds of this type come from the dynamics related to the third generation quarks,
in particular from the direct searches for top partners, from the oblique parameters
and from the corrections to the gauge couplings involving the $b$ quark.
These constraints are clearly common to all scenarios based on
partial compositeness and are comparable to the corresponding ones in the anarchic models.
Given their relevance for generic composite Higgs scenarios we discuss them in dedicated chapters,
namely Chaps.~\ref{ch:LHC_pheno} and \ref{ch:EWPT}.

\section{The lepton sector}\label{sec:leptonic_sector}

So far we focused our discussion on the flavor structure of the quark sector.
The construction of a complete model, however, also requires a full description of the lepton fields.
We will show in this section how a minimal implementation of the lepton sector can be obtained as a straightforward generalization
of the anarchic scenario~\cite{Agashe:2006iy,Agashe:2009tu,Csaki:2010aj,KerenZur:2012fr}.
As we will see, this minimal scenario is far from being satisfactory
since it suffers from extremely strong experimental constraints. It is however a useful toy example
to discuss the general features of the lepton sector and present the most important experimental constraints.
More refined models that improve the compatibility with the experiments through the introduction of
flavor symmetries can be constructed in analogy to what we discussed in Sect.~\ref{sec:flavor_symm} for the quark sector
or by advocating the presence of discrete symmetries.
For brevity, we will not enter into the details of such constructions and we refer the interested reader
to the original literature~\cite{Csaki:2008qq,delAguila:2010vg,Hagedorn:2011un,Hagedorn:2011pw,Redi:2013pga}.

The lepton sector presents two peculiar aspects that make it quite different from the quark sector.
The first peculiarity is related to the neutrino masses which are much smaller than the masses of the charged leptons.
This property makes it plausible that the neutrino masses come from a different source, thus allowing more freedom
in the flavor structure. An interesting possibility is to abandon the usual partial compositeness structure,
and generate the neutrino masses through a technicolor-like construction. We will discuss this scenario,
as well as the standard partial-compositeness one, in the following.

\index{V@$V_{\textrm{PMNS}}$ matrix}
The second peculiar aspect of the lepton sector is the fact that the lepton mixing matrix $V_{\textrm{PMNS}}$, contrary to the
$V_{\textrm{CKM}}$ matrix, seems to have a non-hierarchical structure. To understand the consequences of this structure
it is useful to recall the expression of the mixing matrix in terms of the chiral rotations that diagonalize the
Yukawa's:
\begin{equation}
V_{\textrm{PMNS}} = E_L^\dagger N_L\,,
\end{equation}
where $E_L$ and $N_L$ refer to the rotation of the left-handed fields in the charged lepton and neutrino sectors
respectively. A non-hierarchical structure for $V_{\textrm{PMNS}}$ generically occurs whenever $N_L$ is anarchic, irrespectively
of the charged lepton rotation matrix $E_L$. To obtain the correct flavor structure it is thus sufficient to generate
hierarchical Yukawa's for the charged leptons
\begin{equation}
y_e^i \sim \lambda^i_{\ell_L} \lambda^i_{e_R}/g_*\,,
\label{eq:lepton_mass}
\end{equation}
and to assume that the rotation matrix $N_L$ is non-hierarchical.

As a first scenario we consider the usual partial compositeness structure. In complete analogy to the quark sector we
introduce right-handed neutrinos and consider Dirac neutrino masses. As for the quarks, in order to generate
the lepton masses, the left-handed doublets $\ell_L^i$ should be mixed to one or more composite operators depending on
the $\SO(5)$ quantum numbers. For instance if the composite operators belong to the fundamental representation
at least two mixings are necessary: one with a ${\bf 5}_{-1}$ operator giving mass to the charged leptons and
one with a ${\bf 5}_0$ operator giving mass to the neutrinos. In this scenario we are left with a large freedom
in the choice of the mixings. Apart from the condition in Eq.~(\ref{eq:lepton_mass}), we need to assume that
the left neutrino mixing is non hierarchical, $\lambda'^i_{\ell_L}/\lambda'^j_{\ell_L} \sim 1$.
As we explained in Section~\ref{sec:anarchic_pc}, this condition ensures that the $N_L$ matrix is anarchic.

If the $\ell^i_L$ fields are mixed with only one composite operator the overall picture changes. In this case,
in order to obtain a non-hierarchical $N_L$, we need to assume $\lambda^i_{\ell_L}/\lambda^j_{\ell_L} \sim 1$.
This condition fixes the ratios between the $\lambda^i_{e_R}$ mixings, which must be chosen to reproduce
the charged lepton masses, $\lambda^i_{e_R}/\lambda_{e_R}^j \sim m_e^i/m_e^j$ for $i<j$.

As we mentioned before, an alternative scenario can be advocated for the generation of the neutrino masses.
Instead of coming from the usual partial compositeness mixing, the neutrino masses could arise from
couplings involving a bilinear of the elementary fields. This can be naturally realized if the neutrinos are
Majorana. In this case the leading operator responsible for generating neutrino mass terms is of the form
\begin{equation}
y_{ij}^\nu (\overline \ell_L^i)^c \ell_L^j {\mathcal O}\,.
\end{equation}
In the above equation $y_{ij}^\nu$ is an anarchic matrix and ${\mathcal O}$ is a composite operator
transforming as an $\SU(2)_L$ triplet and with unit hypercharge. If the scaling dimension of the composite operator is larger
than $2$ the Renormalization-Group running naturally generates a suppression of the neutrino masses.

A similar construction can also be realized if the neutrinos are Dirac by assuming that the mixing
of the right-handed fields $\lambda_{\nu_R}$ are negligibly small. In this case the dominant contribution
to the neutrino masses comes from higher dimensional operators involving the $\overline \ell_L^i \nu_R^j$
bilinear at the UV scale.

\subsection{Constraints}

The minimal implementations of the lepton sector we described before are in deep trouble when
compared with the stringent experimental data. The most problematic observables are the
electron EDM and the lepton flavor violating decay $\mu \rightarrow e \gamma$. As we will discuss
in the following these observables lead to bounds on the compositeness scale that are nearly one order
of magnitude stronger than the ones we found in the quark sector.

\subsubsection{The electron EDM}

We start by considering the electron EDM, which is induced by the dipole operator
\begin{equation}
{\mathcal O}_{ee\gamma} = C_{ee\gamma} \frac{e m_e}{16 \pi^2} \overline e \sigma^{\mu\nu} F_{\mu\nu} \gamma^5 e\,.
\end{equation}
If ${\mathcal O}_{ee\gamma}$ is induced a tree level, its contribution to the electron EDM can be estimated as
\begin{equation}
d_e = \frac{e m_e}{8 \pi^2} \mathrm{Im}\, C_{ee\gamma} \sim 2 \frac{m_e\, e}{m_*^2}\,.
\end{equation}
The current experimental bound is given by~\cite{Agashe:2014kda}
\begin{equation}
|d_e| < 0.87 \times 10^{-28}  e\, \mathrm{cm} \qquad \quad \textrm{at\ $90\%$\ C.L.}\,.
\end{equation}
This translates into the stringent constraint
\begin{equation}
m_* \gtrsim 480\ \mathrm{TeV}\,.
\end{equation}
Even if we assume that the dipole operator is generated only at loop
level, the bound on the composite dynamics remains quite strong,
$f \gtrsim 38\ \mathrm{TeV}$.

The corrections to the muon EDM can be analyzed in a similar way. The new-physics contributions,
however, are way below the present bounds and do not impose any significant constraint.

\subsubsection{Lepton flavor violation}

The second process that leads to a strong constraint on the minimal implementations of the lepton
sector is the $\mu \rightarrow e \gamma$ decay. The effective Lagrangian
mediating this process can be written as~\cite{KerenZur:2012fr}
\begin{equation}
{\mathcal L}_{\mu \rightarrow e \gamma} = m_\mu e F_{\mu\nu}
\left(\frac{\overline \mu_L \sigma^{\mu\nu} e_R}{\Lambda_L^2}
+ \frac{\overline \mu_R \sigma^{\mu\nu} e_L}{\Lambda_L^2}\right)\,.
\end{equation}
From this expression one can derive the branching ratio for the
flavor-violating decay
\begin{equation}
\mathrm{BR}(\mu \rightarrow e \gamma) = 96 \pi^2 e^2 \left(
\left|\frac{v}{\Lambda_L}\right|^4 + \left|\frac{v}{\Lambda_R}\right|^4\right)\,.
\end{equation}
By imposing the current experimental bound
$\mathrm{BR}(\mu \rightarrow e \gamma) < 5.7 \times 10^{-13}$~\cite{Agashe:2014kda},
one finds $\Lambda_{L,R} \gtrsim 900\ \mathrm{TeV}$.

The coefficients of the effective operators can be estimated as
\begin{eqnarray}
\frac{m_\mu}{\Lambda_L^2} &\sim& \sqrt{2} \frac{\sqrt{m_e m_\mu}}{m_*^2}
\left(\frac{\lambda^1_{\ell_L}}{\lambda^2_{\ell_L}} \sqrt{\frac{m_\mu}{m_e}}\right)^{-1}\,,\\
\frac{m_\mu}{\Lambda_R^2} &\sim& \sqrt{2} \frac{\sqrt{m_e m_\mu}}{m_*^2}
\left(\frac{\lambda^1_{\ell_L}}{\lambda^2_{\ell_L}} \sqrt{\frac{m_\mu}{m_e}}\right)\,.
\end{eqnarray}
The above expressions have been written in such a way to emphasize that, in order to minimize the
constraints, the optimal choice for the $\lambda^i_{\ell_L}$ parameters is
\begin{equation}
\frac{\lambda^1_{\ell_L}}{\lambda^2_{\ell_L}} \sim \sqrt{\frac{m_e}{m_\mu}}\,.
\end{equation}
With this choice we find that the bound on $m_*$ is given by
\begin{equation}
m_* \gtrsim 300\ \mathrm{TeV}\,.
\end{equation}
Also in this case the bound remains quite strong even if the dipole
operators are not generated at tree level ($f \gtrsim 25\ \mathrm{TeV}$).

Other minor constraints can be derived from the muon conversion inside
nuclei and from the $\mu \rightarrow 3e$ decay~\cite{KerenZur:2012fr}. These processes
are mediated by the penguin operators that lead to flavor-violating
$Z$ interactions.

\chapter{Phenomenological models}
\label{ch:Pheno_models}

In the previous chapters we focused on the broad qualitative features of the composite Higgs
scenarios that follow directly from the Nambu-Goldstone boson nature of the Higgs and from partial fermion compositeness.
As we discussed at length, the Goldstone structure determines many important
aspects of the elementary and composite dynamics and, when supplemented by a power-counting rule,
can be exploited to obtain a semi-quantitative understanding of the new-physics effects.
The full generality of this approach is at the same time the source of its advantages
and of its main limitations.
Most of the results we derived are indeed valid only as order of magnitude estimates and important numerical
corrections could be present in explicit models. Moreover, so far we mainly focused on the dynamics
of the Standard Model (SM) fields, but we did not consider in details the properties of the composite
resonances that unavoidably arise from the composite sector and constitute one of the most distinctive
features of the composite Higgs scenarios.

In this chapter we will change perspective and we will show how more complete descriptions of the
composite Higgs theories can be constructed. The main aim of our constructions will not be to make progresses on the microscopic origin of the composite Higgs scenario, but instead to obtain some phenomenological  model in which the relevant physical observables, such as the Higgs potential and the Electro-Weak (EW) parameters,
can be reliably predicted. We will see that, as a byproduct of the calculability requirement, a parametrization
of the dynamics of the composite resonances is also automatically introduced.

The construction of explicit models is important for several reasons. First of all, explicit realizations of
the composite scenarios allow to check in details the validity of the model-independent results derived in the
previous chapters. In particular they are useful to understand up to which extent the general estimates can be
modified and the various constraints coming from the experimental data can be relaxed
or tightened. Specific implementations
of the composite Higgs idea, indeed, often predict several correlations among the physical observables
or display some accidental cancellations. All these effects are very hard to be guessed in a fully model-independent approach.
For this reason explicit models can be used to get a more reliable determination of the impact of the experimental constraints
on the composite Higgs scenario (these topics will be the subjects of Chaps.~\ref{ch:LHC_pheno} and \ref{ch:EWPT}).

It is also important to stress that the explicit models also have a historical and practical relevance.
In fact, the first holographic implementations of the composite Higgs scenarios \cite{Agashe:2004rs,Contino:2006qr} provided a fundamental proof that this
kind of framework can be used to achieve realistic models of EW Symmetry Breaking
(EWSB). 
Explicit constructions, moreover, have often been the playground for the discovery of new features of the
composite scenarios, which then have been confirmed and more fully analyzed in a model-independent way.

In order to build explicit realizations of the composite Higgs scenarios different alternative constructions can be used.
One possibility is to rely on the holographic models formulated as gauge theories in five space-time
dimensions. These models provide calculable implementations
of the composite Higgs idea and, at the same time, include a full description of the composite sector resonances. This latter feature is however also at the origin of the main drawback of the
holographic approach, namely the fact that extracting predictions for the physical observables
is usually technically challenging. This is particularly so if we are interested in the collider phenomenology
of the composite resonances. The extra-dimensional models, indeed, necessarily include infinite towers of
composite states with increasing mass and formally describe the dynamics of each of them.
Obviously in a collider experiment only a few light resonances are accessible, thus retaining the heavier states
is an unnecessary complication. Furthermore only the light resonances are well-described by the theory, the heavy states are only formally present since their mass goes above the cutoff. 
On top of these reasons, a full description of the holographic models requires the introduction
of many technical tools and would drive us too far from the path we followed
in the previous chapters.\footnote{Good reviews explaining the extra-dimensional implementations of the
composite Higgs idea can be found in Refs.~\cite{Sundrum:2005jf,Serone:2009kf,Cheng:2010pt,Contino:2010rs}.
The holographic correspondence linking
these scenarios to the four-dimensional picture has been discussed in Refs.~\cite{Rattazzi:2000hs,PerezVictoria:2001pa,Contino:2004vy,Gherghetta:2006ha,Panico:2007qd,Serone:2009kf}.
}

In the following we will thus choose a different approach and we will focus on a class of explicit
implementations of the composite Higgs scenario based on fully four-dimensional constructions,
the so called ``multi-site'' models~\cite{Panico:2011pw,DeCurtis:2011yx}.
This approach is inspired by the holographic models and by the idea
of dimensional deconstruction~\cite{ArkaniHamed:2001ca,Hill:2000mu}, which consists in discretizing the extra space coordinate
by replacing it with a one-dimensional lattice with a finite number of points or ``sites''.
Each site is associated to a set of degrees of freedom that roughly correspond to one level of
Kaluza-Klein (KK) states. In this way the multi-site models provide a simplified version of the holographic theories in which only
few KK levels, {\it{i.e.}}~a limited set of composite resonances, are included.\footnote{It is important to stress
that the deconstructed models are deeply different from a naive truncation of the KK tower (for an effective model
based on this approach see Ref.~\cite{Contino:2006nn}). A naive truncation, indeed, implies a breaking
of the symmetries that protect the Higgs dynamics in the holographic models, thus not allowing to implement
calculability in the effective model.}

From the technical point of view, the construction of these models does not require any additional tool since it
relies only on the non-linear $\sigma$-model structure and on the CCWZ formalism introduced in Chap.~\ref{ch:Goldstone}.
The detailed structure of the models is dictated by the requirement of calculability, which is implemented
via a ``collective breaking'' mechanism.
As we will see, this mechanism has its roots in the holographic theories and naturally blends with the multi-site structure.


\section{Multi-site models: collective breaking}
\label{sec:multi-site}

In order to introduce the phenomenological multi-site models we proceed
in a constructive manner, following Ref.~\cite{Panico:2011pw}. The starting point is the simplest and most general
description of the composite Higgs dynamics, the non linear $\sigma$-model.
The additional ingredients needed to
include the dynamics of the composite resonances and to ensure the calculability
of the Higgs potential are then progressively introduced. This leads to
a simpler exposition, which allows to emphasize the key assumptions and
gradually describe the required technical tools.


\subsection[The non-linear $\sigma$-model]{The non-linear ${\mathbf{\sigma}}$-model}\label{sec:pheno_models_sigma-model}
\index{non-linear $\sigma$-model}

For definiteness we focus on the non-linear $\sigma$-model corresponding to the $\SO(5)/\SO(4)$
coset, which constitutes the ``minimal'' implementation of a composite Higgs.
This model only contains the pseudo Nambu-Goldstone Boson (pNGB) Higgs
and the SM gauge fields, whereas the composite sector resonances are
absent. The general structure of the non-linear $\sigma$-model has already been
discussed in Chap.~\ref{ch:Goldstone} and we rewrite here only the expressions relevant for
the present discussion.

\index{Goldstone matrix}
The operators appearing in the Lagrangian can be written in terms of
the Goldstone matrix $U$ defined as
\begin{equation}\label{eq:gold_matrix}
U[\Pi] = \exp\left(i \frac{\sqrt{2}}{f} \Pi_i \widehat T^i\right)\,,
\end{equation}
where $\Pi_i$ are the four NGB fields corresponding
to the components of the usual Higgs doublet and transform as a $\mathbf{4}$ of $\SO(4)$.
We denote collectively by $T^A = \{T^a, \widehat T^i\}$ the $\SO(5)$ generators in the
fundamental representation, corresponding to the generators of the unbroken
$\SO(4)$ subgroup ($T^a$ with $a = 1, \ldots, 6$) and to the generators of the
$\SO(5)/\SO(4)$ coset ($\widehat T^i$ with $i = 1, \ldots, 4$). For all the generators we choose
the normalization $\Tr [T^A T^B] = \delta^{AB}$. The explicit form of the
generators is reported in Eq.~(\ref{so5gen}).

The leading operators in the Lagrangian appear at the two-derivatives order and read
\begin{equation}\label{eq:lagr_sigma_model}
{\mathcal L}^\pi = \frac{f^2}{4} d_\mu^i d^\mu_i\,,
\end{equation}
where the $d$-symbol is defined, as in Eq.~(\ref{dedefgauge}), as the projection on the broken generators of the Maurer--Cartan form, namely
\begin{equation}\label{eq:dsymbol4}
d_\mu^i=i\,{\textrm{Tr}}\left[(U^T D_\mu U)\cdot \Th^i\right]\,.
\end{equation}
The covariant derivative of the Goldstone matrix
is defined as
\begin{equation}\label{eq:cov_der_and_elem_gauge}
D_\mu U = \partial_\mu U - i A_\mu U\,,
\qquad \textrm{with}
\qquad
A_\mu = g W^\alpha T^\alpha_L + g' B_\mu T^3_R\,,
\end{equation}
and gives rise to the interaction of the NGB Higgs with the elementary gauge
fields $W^\alpha_\mu$ ($\alpha = 1,2,3$) and $B_\mu$ corresponding to the
$\SU(2)_L \times \U(1)_Y$ SM gauge group embedded in $\SO(4)$.\footnote{We momentarily neglect
the presence of the extra $\U(1)_X$ charge in the definition of the hypercharge
and we set $Y = T^3_R$. The $\U(1)_X$ subgroup does not play any role until the
matter fermion fields are introduced.} In the above formula $g$ and $g'$ denote the
couplings of the $\SU(2)_L$ and $\U(1)_Y$ gauge subgroups.
The $T^\alpha_{L,R}$ generators correspond to
the $\SO(4)$ generators written in a basis that shows explicitly the equivalence
$\SO(4) \simeq \SU(2)_L \times \SU(2)_R$. To complete the leading order Lagrangian,
canonical kinetic terms for the elementary gauge fields must be introduced as well
\begin{equation}\label{eq:elem_gauge_lagr}
{\mathcal L}^g = -\frac{1}{4} \textrm{Tr}\left[W_{\mu\nu} W^{\mu\nu}\right]
- \frac{1}{4} B_{\mu\nu} B^{\mu\nu}\,,
\end{equation}
where we defined 
$W_{\mu\nu} = \partial_\mu W_\nu - \partial_\nu W_\mu - i g [W_\mu, W_\nu]$ with $W_\mu = W^\alpha_\mu T^\alpha_L$.

In the previous chapters we regarded the non-linear $\sigma$-model Lagrangian, ${\mathcal L}_0 = {\mathcal L}^\pi + {\mathcal L}^g$, as the leading-order low-energy description of the composite Higgs dynamics, valid below a physical cutoff that is provided by the composite resonances scale $m_*=g_*f$. We also saw how to estimate, by the  One-Coupling-One-Scale power-counting introduced  in Chap.~\ref{ch:SILH}, the infinite set of higher-order operators that appear in the 
effective field theory. Here instead we discuss the non-linear $\sigma$-model with a rather different purpose. Namely we want to see which physical observables are formally predictable within the leading-order version of the $\sigma$-model. More precisely, we ask ourselves which observables can be computed by only employing the leading-order Lagrangian ${\mathcal L}_0$, obtaining predictions for them in terms of its three parameters $f$, $g$ and $g'$. The calculability of the Higgs potential will be our main concern in the present section, however the considerations that follow will later find other applications in Chap.~\ref{ch:EWPT}.

All tree-level diagrams can be computed, obtaining formally valid predictions. Obstructions to calculability instead come from loop diagrams, which is where the non-renormalizable nature of the theory shows up. Order by order in the loop expansion new divergences will appear, associated to operators that were not present in ${\mathcal L}_0$. Canceling those divergences requires additional counterterms, which introduce new parameters in the calculation. If the contribution to the latter parameters is not suppressed and the final result is strongly sensitive to them, calculability is spoiled. In order to establish whether this occurs or not we need to count the degree of divergence of the Feynman diagrams with $L$ loops and an arbitrary set of external legs computed with the leading-order Lagrangian ${\mathcal L}_0$. This is provided by the following formula \footnote{The derivation is not particularly enlightening and thus it will not be reported here. The reader is referred to Ref.~\cite{Panico:2011pw}.}
\begin{equation}\label{eq:divergence_counting}
 \Lambda^2 f^2 \left(\frac{\Lambda}{4 \pi f}\right)^{2L}
\left(\frac{\Pi}{f}\right)^{E_\pi} \left(\frac{g V}{\Lambda}\right)^{E_V}
\left(\frac{\partial}{\Lambda}\right)^{d} \left(\frac{g f}{\Lambda}\right)^{2 \eta}\,,
\end{equation}
where $\Lambda$ is the hard momentum cutoff of the loop integrals, $g$ and $V$ collectively denote the $g$ and $g'$ gauge couplings and the corresponding gauge fields $W_\mu$ and $B_\mu$. The degree of divergence associated to a given $L$-loops diagram, with ``$E_\pi$'' Goldstone and ``$E_V$'' vector external legs is simply given by the power of $\Lambda$ predicted by the counting rule. A number ``$d$'' of external momenta insertions, which correspond to derivatives in the equivalent effective operators, is also taken into account. The $(g f/\Lambda)^{2\eta}$ factor, which reduces the degree of divergence ($\eta$ is necessarily positive) counts the number of elementary gauge field vertices that are present in the diagram. For each given observable, the leading divergence is the one of the diagram with the smallest number of gauge vertices insertions, {\it{i.e.}}~with the smallest $\eta$. However because of selection rules, as we will see for instance in the case of the Higgs potential, there might be an obstruction to reach the absolute minimum $\eta = 0$. One could be obliged to consider diagrams with $\eta \geq \eta_{min} > 0$, leading to a reduction of the degree of divergence. This mechanism will be the key ingredient for the construction of multi-site phenomenological models with a calculable Higgs potential.

The reader should be careful not to confuse the superficial degree of divergence counting in Eq.~(\ref{eq:divergence_counting}) with the power-counting estimate we derived in Chap.~\ref{ch:SILH} and in particular in Eq.~(\ref{POWCFull}). The two formulas provide answers to two distinct questions, therefore they are, a priori, completely unrelated. The divergence counting is an intrinsic property of the effective theory, or better of the leading-order Lagrangian ${\mathcal L}_0$. Indeed only the leading-order parameters appear in Eq.~(\ref{eq:divergence_counting}), the scale $\Lambda$ being just the unphysical and formally infinite loop momentum cutoff. Assumptions on the UV completion of the effective theory are instead needed for the power-counting estimate, and indeed the physical cutoff scale $m_*$, or equivalently the typical resonance coupling $g_*=m_*/f$, is present in Eq.~(\ref{POWCFull}). However, there exists a relation among the two formulas, which emerges if we identify, for a moment, the loop momentum cutoff $\Lambda$ with the physical cutoff $m_*$. ``Naturalness'' considerations, completely analogous to the ones we made in Sect.~\ref{natural_EWSB} for our estimate of the low-energy contributions to the Higgs mass in the SM, suggest that Eq.~(\ref{eq:divergence_counting}) should provide an approximate lower bound to the size of the effective operator coefficients that are generated from the UV theory. The idea is, precisely like for the Higgs mass-term in the SM, that the contribution of virtual quanta below the cutoff, which is well described by the effective-field-theory loops truncated at the cutoff scale, should be part of the complete result for the operator coefficient computed in the UV theory. The operator coefficient being ``Natural'', {\it{i.e.}}~not resulting from a finely-tuned cancellation, thus requires it to be larger or at most equal to the one estimated from the effective field theory loops. It is easy to verify that the power-counting estimate in Eq.~(\ref{POWCFull}) indeed obeys this condition, provided the coupling $g_*$ stays below the maximal value of $4\pi$ allowed by perturbativity in the UV theory.\footnote{Thinking backwards, this could have been a way to establish the bound $g_*\leq 4\pi$ in the operator estimate.} The lower bound on the operator coefficients is saturated, and the two estimates coincide, only in the ``NDA limit'' $g_*=4\pi$ (and consequently $\Lambda=m_*=4\pi f$), {\it{i.e.}}~when we consider a completely strongly-coupled UV completion for the effective field theory. More precisely, given that the effect of elementary gauge field loops was not taken into account in Eq.~(\ref{POWCFull}) the comparison with Eq.~(\ref{eq:divergence_counting}) is only possible for $\eta=0$. The $(g f/\Lambda)^{2 \eta}$ factor reduces to $(g^2/16\pi^2)^{\eta}$ for maximal cutoff $\Lambda=4\pi f$  and  and it corresponds to the loop suppression of the operators that are radiatively induced by the elementary fields, like the ones we encountered in Sect.~\ref{pwcpot_sec} in the estimate of the Higgs effective potential.

We now return to the issue of calculability of the Higgs potential. A naive usage of Eq.~(\ref{eq:divergence_counting}) would predict a very high degree of divergence, of the quartic order $\Lambda^4$, already at one loop, which is the minimal order at which the Goldstone boson potential can be generated. However the divergence is reduced because only the diagrams that are sensitive to the breaking of the Goldstone symmetry can contribute to the potential. Given that the gauge field couplings are the only source of such breaking in the Lagrangian, the only relevant diagrams are the ones that involve internal elementary gauge field lines and thus some insertion of the elementary gauge field couplings. The minimal number of coupling insertions is equal to two, therefore the lower bound $\eta\geq \eta_{min}=1$ must be considered in Eq.~(\ref{eq:divergence_counting}). From quartic, the divergence is thus reduced to quadratic because of a selection rule, namely because of the presence of the Goldstone symmetry and of its explicit breaking, which is entirely due to the gauge couplings.
\index{spurions}
Even if reduced, the divergence of the Higgs potential is still present in the non-linear $\sigma$-model. Since it emerges at the first order in the loop expansion at which the potential starts being generated, its presence clearly forbids us to compute the Higgs potential (and in turn its mass and self-couplings and the EWSB scale) in this theory. The Higgs potential is determined by a counterterm that cancels the divergence and leaves behind it a completely arbitrary finite part, which is merely one new free parameter of the theory with no connection with the ones that appear in the leading-order Lagrangian. Therefore it is not a calculable quantity.

The reduction of the divergence from quartic to quadratic is not sufficient to make the Higgs potential calculable, however it indicates the right direction to design calculable extensions of the non-linear $\sigma$-model. For this reason it is worth studying the cancellation of the leading quartic divergence in a more rigorous and systematic way by the method of spurions, which will be very useful in the rest of this chapter. The concept of spurions has already been introduced in Chap.~\ref{ch:SILH}
and we refer the reader to that discussion for more in-depth explanations.
Here we only focus on the minimal ingredients needed for the present analysis.
The concept of spurions is based on the possibility of keeping track of the consequences of
a symmetry when it is broken by a small explicit perturbation. This can be done by
formally restoring the original invariance by promoting the symmetry-breaking couplings to
spurions with definite transformation properties under the symmetry.
The only allowed operators in the effective theory are then the ones that respect the
whole original invariance and can be build from the usual fields and the spurions.

In the non-linear $\sigma$-model we restore the full $\SO(5)$ invariance by introducing
two spurions ${\mathcal G}$ and ${\mathcal G}'$ and rewriting the elementary gauge field
in Eq.~(\ref{eq:cov_der_and_elem_gauge}) as
\begin{equation}\label{eq:s_model_spurions}
A_\mu = {\mathcal G}_\alpha W^\alpha_\mu + {\mathcal G}' B_\mu = {\mathcal G}_{\alpha A} T^A W^\alpha_\mu
+ {\mathcal G}'_A T^A B_\mu\,,
\end{equation}
where the index $A$ is in the adjoint representation of $\SO(5)$. In this way the covariant derivative
transforms homogeneously under $\SO(5)$. The elementary gauge fields are now associated to
a new ``elementary'' group $\SU(2)_L^0$, which is a symmetry of the gauge Lagrangian in Eq.~(\ref{eq:elem_gauge_lagr}), and do not transform under $\SO(5)$.
In particular the three $W^\alpha_\mu$ belong to the adjoint representation of $\SU(2)_L^0$,
while the $B_\mu$ field and the Goldstone boson Higgs are invariant. For consistency, the index $\alpha$ of the spurion forms a triplet under the elementary group. The SM $\SU(2)_L$ gauge group, under which both the $W$ fields and the Higgs transform simultaneously, is given by the vector combination of the elementary $\SU(2)_L^0$ and the $\SU(2)_L$ subgroup of $\SO(5)$. One can also notice an additional symmetry, a $\Zdouble_2$ parity,
that acts by changing the sign of the spurion ${\mathcal G}'$ and of the $B_\mu$ field.

Clearly Eq.~(\ref{eq:s_model_spurions}) is just a rewriting of $A_\mu$, the physical values
of the spurions are indeed
\begin{equation}
{\mathcal G}_\alpha = g T^\alpha_L\,,
\qquad\quad
{\mathcal G}' = g' T^3_R\,,
\end{equation}
which give back the expression in Eq.~(\ref{eq:cov_der_and_elem_gauge}). The physical value
of the spurion ${\mathcal G}$ breaks the total $\SU(2)_L^0 \times \SO(5)$ symmetry
down to $\SO(4) \simeq \SU(2)_L \times \SU(2)_R$. The ${\mathcal G}'$ spurion then
produces the further breaking to the SM group. The important point about the spurions
is that before setting them to their physical values they have well-defined transformation properties under the total symmetry and they must enter in the allowed operators in
symmetry-preserving combinations with the usual fields. By using the symmetry we can
classify the local operators in terms of the number of spurions they contain.

Let us now go back to the issue of the calculability of the Higgs potential.
We said that the potential can only be generated due to the breaking of the $\SO(5)$
symmetry induced by the elementary gauging. In the language of spurions, this means that
the operators contributing to the potential must contain powers of ${\mathcal G}$ or ${\mathcal G}'$,
which carry powers of $g$ and $g'$ and thus lower the degree of divergence from quartic to quadratic. At the leading order
in the gauge couplings it is easy to classify the operators. There are only two of them
contributing to the Higgs potential \footnote{The classification of invariant operators constructed with the gauge spurions was already carried on in Sect.~\ref{CHPCH} with a different and more general technique based on dressed spurions. Two operators were found for each of the two ${\mathcal G}$ and ${\mathcal G}'$ spurions, however only the ones that are even under the $P_{LR}$ symmetry are generated by radiative corrections and are reported in the equations that follow. This is because the $2$-derivative non-linear $\sigma$-model Lagrangian is accidentally $P_{LR}$-invariant, thus it can not generate odd operators.}
\begin{equation}\label{eq:s_model_pot_1}
c_g f^4 \left(U^T {\mathcal G}_\alpha {\mathcal G}_\alpha U\right)_{55}
= \frac{3}{4} c_g f^4 g^2 \sin^2(H/f)
\end{equation}
and
\begin{equation}\label{eq:s_model_pot_2}
c_{g'} f^4 \left(U^T {\mathcal G}' {\mathcal G}' U\right)_{55}
= \frac{1}{4} c_{g'} f^4 g'^2 \sin^2(H/f)\,,
\end{equation}
where we denoted by $H$ the real neutral Higgs component (normalized in such way that
on the EW vacuum $\langle H\rangle\simeq v = 246\ \mathrm{GeV}$). In the notation used in Eq.~(\ref{eq:gold_matrix})
the physical Higgs corresponds to the fourth component of the $\Pi_i$ multiplet, $H = \Pi_4$.
Notice that to construct an $\SO(5)$ invariant operator we took the $(5,5)$ component
of the matrices in the parentheses.
This can be understood by recalling the transformation properties of
the Goldstone matrix $U_{IJ}$. The right index $I$ transforms linearly in the fundamental
representation of $\SO(5)$. On the contrary, the left index $J$ corresponds to a
non-linear realization of $\SO(5)$ obtained through $\SO(4)$ transformations.
As we saw in Chap.~\ref{ch:Goldstone} (compare Eq.~(\ref{goltr})) this corresponds to
\begin{equation}
\label{goltr1}
U[\Pi] \rightarrow  U(\Pi') = g \cdot U[\Pi] \cdot h^{-1}[\Pi,g]\,,
\end{equation}
where $g$ denotes a generic $\SO(5)$ transformation, while
$h[\Pi,g]$ is the element of $\SO(4)$ corresponding to the non-linear
realization of $\SO(5)$. Within our conventions the $\SO(4)$ generators are embedded in the right-top $4 \times 4$ block of
the fundamental $\SO(5)$ representation. The fifth component of a vector in the fundamental
representation is thus invariant under $\SO(4)$. It is easy to see that the
$U^T {\mathcal G}_\alpha {\mathcal G}_\alpha U$ operator in Eq.~(\ref{eq:s_model_pot_1}) transforms
with $h[\Pi,g]$ on both sides and its $(5,5)$ component
is $\SO(5)$ invariant. Similar considerations apply to the operator in Eq.~(\ref{eq:s_model_pot_2}).


\subsection{The two-site model}\label{sec:2-site_gauge}

\index{collective breaking}
We have seen that the Higgs potential is divergent in the non-linear $\sigma$-model
and therefore not predictable. The analysis of the degree of divergence, however,
led to an interesting insight on the dynamics responsible of the breaking of
the Goldstone symmetry.
In particular it highlighted the fact that the $\SO(5)$ invariance provides a partial
protection of the Higgs dynamics which lowers the degree of divergence associated
to the Higgs potential.
This result suggests a natural way to go beyond the non-linear $\sigma$-model description
and construct a predictive implementation of the composite Higgs scenario:
we must promote the $\SO(5)$ invariance to a larger set of independent symmetries
under which the Higgs behaves like a Goldstone boson. The Higgs potential is then
generated only if all the symmetries are broken and this reduces its degree of divergence.
This kind of mechanism is usually denoted as collective breaking~\cite{Georgi:1989xy,ArkaniHamed:2001nc}
and has been extensively used in the context of little Higgs theories
(for a review see Refs.~\cite{Schmaltz:2005ky,Perelstein:2005ka}).
As we will see in the following, an interesting byproduct of the collective-breaking structure
is the automatic connection with the dynamics of the composite sector resonances.
The additional symmetries are indeed a basic ingredient to introduce the composite
states in the effective models.

Before moving to the complete model in which the Higgs potential is fully calculable,
we focus on a simpler implementation of the idea of collective breaking,
the two-site model.
Although the two-site model has only a limited predictive power, it includes most of
the ingredients of the complete model and allows an easier presentation of the new technical
aspects.

The starting point of the two-site model is again a non-linear $\sigma$-model. In this case,
however, it is not based on the usual $\SO(5)/\SO(4)$ coset. Instead we add a second
$\SO(5)$ subgroup and consider the chiral group $\SO(5)_L \times \SO(5)_R$ spontaneously
broken to the vector subgroup $\SO(5)_V$. The new coset, $\SO(5)_L \times \SO(5)_R/\SO(5)_V$,
is parametrized by the $\SO(5)$ Goldstone matrix
\begin{equation}
{\mathcal U}[\Pi] = \exp\left(i\frac{\sqrt{2}}{f} \Pi_A T^A\right)\,,
\end{equation}
which transforms linearly under $\SO(5)_L \times \SO(5)_R$
\begin{equation}\label{eq:global_transf_2-site}
{\mathcal U}[\Pi] \rightarrow {\mathcal U}[\Pi'] = \gamma_L\, {\mathcal U}[\Pi]\, \gamma_R^T\,.
\end{equation}
This structure gives rise to ten Goldstones $\Pi_A$, which transform in the adjoint representation
of $\SO(5)_V$. Under $\SO(4) \subset \SO(5)_V$, four of these, the $\Pi_i$ corresponding to the
$\SO(5)/\SO(4)$ coset, form a fourplet
and are identified with the Higgs field. The remaining six, $\Pi_a$ in the adjoint of $\SO(4)$,
will be removed by gauging and will not appear in the spectrum as physical scalars. Notice that this notation, in which the Goldstone matrix transforms linearly under the group rather than with the non-linear $h\in \SO(4)$ as in Eq.~(\ref{goltr1}), is rather different from the one we used until now. This  formulation of the Goldstone boson theory, alternative to the general CCWZ one, is only possible for chiral groups broken to the vector combination and it is related to the latter by
\beq
U=\left[\begin{array}{cc}
{\mathcal U}[\Pi/2]& \\ 
0 & {\mathcal U}^\dagger[\Pi/2]
\end{array}\right]\,.
\eeq

With the three-fold purpose of removing the $\Pi_a$ scalars, of breaking the extra $\SO(5)_R$
invariance and of adding to the model a description of the vector resonances, we gauge
the $\SO(4)$ subgroup of $\SO(5)_R$ by introducing six gauge fields $\widetilde \rho^a_\mu$.
Given that $\SO(5)_R$ is spontaneously broken, the new gauge bosons become massive and acquire
their longitudinal components by eating the $\Pi_a$ Goldstones.
The $\widetilde \rho$ states are then interpreted as resonances of the composite 
sector and we assign them a coupling $\widetilde g_\rho$ of the order of the typical
composite sector coupling $g_*$. Their mass is given by
$\widetilde m_\rho \sim \widetilde g_\rho f$ (see Eq.~(\ref{eq:gauge_res_masses_2-site}))
and is of the order of the typical composite sector mass $m_*$.
For phenomenological reasons $\widetilde m_\rho$ is expected to be of TeV size and the coupling $\widetilde g_\rho$
is ``large'' though not maximal, $1 \lesssim \widetilde g_\rho < 4 \pi$. The latter assumption
is essential to ensure that the dynamics of the vector resonances can be described perturbatively.

The EW bosons are introduced by gauging the $\SU(2)_L \times \U(1)_Y$ subgroup
of $\SO(5)_L$ with the gauge fields $W^\alpha_\mu$ and $B_\mu$. The $W$ and $B$ fields
are interpreted as elementary fields. Their couplings $g_0$ and $g_0'$
almost coincide with the SM $g$ and $g'$ couplings and are typically much smaller than $\widetilde g_\rho$ (see Eq.~(\ref{eq:gauge_coupl_2-site})). Notice that the $\SU(2)_L \times \U(1)_Y$ subgroup
of $\SO(5)_L$ that is gauged by the elementary gauge fields does not coincide exactly with the SM gauge group. Indeed the SM group must be unbroken before EWSB, whereas $\SO(5)_L$ is spontaneously broken to its vector combination with $\SO(5)_R$. The correct identification of the SM group is thus with the vector combination of the elementary group $\SU(2)_L \times \U(1)_Y$ inside $\SO(5)_L$ and the analogous subgroup inside $\SO(5)_R$. This combination belongs to the $\SO(5)_V$ symmetry group and is clearly unbroken before the Higgs takes a VEV. A direct consequence of this construction is the fact that the SM gauge fields do not correspond just to the elementary fields $W$ and $B$, but they are a combination of the latter with the composite vector resonances $\widetilde \rho$. This structure clearly corresponds to the assumption of partial compositeness discussed in Chap.~\ref{ch:Goldstone}.
\index{partial compositeness}

\begin{figure}
\centering
\includegraphics[width=.355\textwidth]{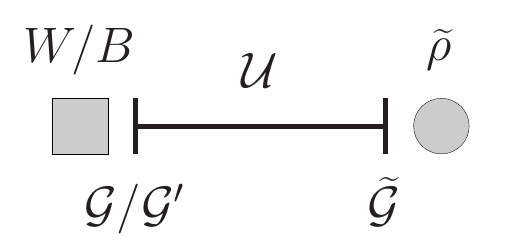}
\caption{Pictorial representation of the two-site model. The Goldstone matrix
${\mathcal U}$ is represented as a ``link'', i.e. a segment with verical lines at the
ends corresponding to the global $\SO(5)_L$ and $\SO(5)_R$ groups.
The elementary and composite states are associated to different ``sites'',
represented by the gray square and circle. The left site can be interpreted as the
elementary group $\SU(2)_L^0 \times \U(1)_Y^0$ under which $W$ and $B$
transform, the right one is the analogous group for $\widetilde \rho$,
${\widetilde \SO}(4)$. The corresponding spurions ${\mathcal G}$, ${\mathcal G}'$
and $\widetilde{\mathcal G}$ are also indicated. Their location reminds the symmetry groups
under which they transform.}\label{fig:2-site_moose}
\end{figure}

The structure of the model can also be represented in a schematic way by using
moose-like diagrams as shown in Fig.~\ref{fig:2-site_moose}.
The pictorial representation clarifies the two-site interpretation of the model.
This structure corresponds to a leading order Lagrangian
\begin{equation}
{\mathcal L}_0 = {\mathcal L}^\pi + {\mathcal L}^g_{cs} + {\mathcal L}^g_{el}\,,
\end{equation}
where we separated the $\sigma$-model kinetic term
\begin{equation}\label{eq:Goldstone_Lagr_2-site}
{\mathcal L}^\pi = \frac{f^2}{4} \Tr [(D_\mu {\mathcal U})^T D^\mu {\mathcal U}]\,,
\end{equation}
from the ``gauge'' terms ${\mathcal L}^g = {\mathcal L}^g_{cs} + {\mathcal L}^g_{el}$ which
only contain the kinetic terms of the elementary and the composite gauge bosons. The covariant derivative of
the Goldstone fields, which gives rise to all the interactions between the Goldstones
and the gauge fields, is defined as
\begin{equation}\label{eq:cov_deriv_2-site}
D_\mu {\mathcal U} = \partial_\mu {\mathcal U} - i A_\mu {\mathcal U} + i\, {\mathcal U} \widetilde R_\mu\,,
\end{equation}
with the elementary field $A_\mu$ given, in analogy to Eq.~(\ref{eq:cov_der_and_elem_gauge}), by
\begin{equation}\label{eq:elem_gauge_2-site}
A_\mu = g_0 W^\alpha_\mu T^\alpha_L + g'_0 B_\mu T^3_R\,,
\end{equation}
and with
\begin{equation}\label{eq:R_mu_2-site}
\widetilde R_\mu = \widetilde g_\rho \widetilde \rho^a_\mu T^a\,.
\end{equation}
The Lagrangian for the composite sector vector resonances is given by
\begin{equation}\label{eq:kin_tilderho}
{\mathcal L}^g_{cs} = -\frac{1}{4} \Tr[\widetilde \rho_{\mu\nu} \widetilde \rho^{\mu\nu}]\,,
\end{equation}
where $\widetilde \rho_{\mu\nu} = \partial_\mu \widetilde \rho_\nu - \partial_\nu \widetilde \rho_\mu - i \widetilde g_\rho [\widetilde \rho_\mu, \widetilde \rho_\nu]$. Finally, the Lagrangian for
the elementary gauge fields, ${\mathcal L}^g_{el}$, coincides with the one we introduced in the
non-linear $\sigma$-model in Eq.~(\ref{eq:elem_gauge_lagr}).

Let us now discuss briefly the mass spectrum of the gauge fields. The Goldstone
Lagrangian in Eq.~(\ref{eq:Goldstone_Lagr_2-site}) contains a mass term that mixes the elementary
gauge fields $W$ and $B$ with the composite fields $\widetilde \rho$. As can be inferred from the
previous discussion about the embedding of the SM gauge group, the massless gauge states
correspond to the unbroken $\SU(2)_L \times \U(1)_Y$ gauge invariance and are easily obtained by
diagonalizing the mass matrix. In this way one also finds the masses of the heavy vector resonances
\begin{equation}\label{eq:gauge_res_masses_2-site}
m_L^2 = \frac{g_0^2 + \widetilde g_\rho^2}{2} f^2 \simeq \frac{\widetilde g_\rho^2}{2} f^2\,,
\quad
m_{3R}^2 = \frac{g'^2_0 + \widetilde g_\rho^2}{2} f^2 \simeq \frac{\widetilde g_\rho^2}{2} f^2\,,
\quad
m_X^2 = \frac{\widetilde g_\rho^2}{2} f^2\,,
\end{equation}
which correspond respectively to the resonances associated to the $T^\alpha_L$, to the $T^3_R$
and to the $\SO(5)/\SO(4)$ coset generators. The SM gauge couplings are given by the expressions
\begin{equation}\label{eq:gauge_coupl_2-site}
\frac{1}{g^2} = \frac{1}{g_0^2} + \frac{1}{\widetilde g_\rho^2} \simeq \frac{1}{g_0^2}\,,
\qquad
\frac{1}{g'^2} = \frac{1}{g'^2_0} + \frac{1}{\widetilde g_\rho^2} \simeq \frac{1}{g'^2_0}\,,
\end{equation}
while the couplings among the heavy resonances are of order $\widetilde g_\rho$.

Similarly to the non-linear $\sigma$-model described in the Sect.~\ref{sec:pheno_models_sigma-model},
the two-site model is a non-renormalizable theory with a cut-off $\Lambda$ that can at most reach
the scale $4 \pi f$, where the $\sigma$-model
interactions in Eq.~(\ref{eq:Goldstone_Lagr_2-site}) become non-perturbative.
The divergence counting rule for the two-site model is similar
to the one given in Eq.~(\ref{eq:divergence_counting}).
The only new ingredients are the heavy vector field and the $\widetilde g_\rho$ coupling which enter
in a way similar to the elementary gauge bosons and couplings, namely
\begin{equation}\label{eq:divergence_counting2}
 \Lambda^2 f^2 \left(\frac{\Lambda}{4 \pi f}\right)^{2L}
\left(\frac{\Pi}{f}\right)^{E_\pi} \left(\frac{g V}{\Lambda}\right)^{E_V}
\left(\frac{\widetilde g_\rho \rho}{\Lambda}\right)^{E_\rho}
\left(\frac{\partial}{\Lambda}\right)^{d} \left(\frac{g f}{\Lambda}\right)^{2 \eta}
\left(\frac{\widetilde g_\rho f}{\Lambda}\right)^{2 \widetilde \eta}\,.
\end{equation}

The important novelty of the two-site construction, with respect to the minimal non-linear
$\sigma$-model of Sect.~\ref{sec:pheno_models_sigma-model}, is the fact that the Higgs is now
a NGB with respect to two independent symmetry groups instead of just one. This means that its dynamics and in particular its effective potential is ``doubly protected'' through a collective breaking mechanism. Let us explain with more details
how this mechanism works. Imagine setting the Higgs to its VEV, which corresponds to $\langle\Pi_4\rangle$ constant and the other components of $\Pi$ vanishing. This produces a constant matrix ${\mathcal U}$ that can be eliminated from the Lagrangian by either performing an $\SO(5)_L$ or an $\SO(5)_R$ transformation in Eq.~(\ref{eq:global_transf_2-site}) with either $\gamma_L={\mathcal U}^T$ or $\gamma_R={\mathcal U}$. This means that the Higgs VEV always cancels if any of the two groups is an exact symmetry of the theory. The Higgs potential, and also all the effects triggered by EWSB can only originate from the breaking of both symmetries.

In order to better exploit the implications of the symmetries we introduce the spurions ${\mathcal G}$,
${\mathcal G}'$ and $\widetilde {\mathcal G}$ associated to the gauging of the $\SU(2)_L$ and $\U(1)_Y$
subgroups of $\SO(5)$ and of the $\SO(4)$ subgroup of $\SO(5)_R$.
The ${\mathcal G}$ and ${\mathcal G}'$ spurions appear in the rewriting of the elementary gauge fields
\begin{equation}\label{eq:elem_spurions_2-site}
A_\mu = {\mathcal G}_\alpha W^\alpha_\mu + {\mathcal G}' B_\mu = {\mathcal G}_{\alpha\, A_L} T^{A_L}
W^\alpha_\mu + {\mathcal G}'_{A_L} T^{A_L} B_\mu\,,
\end{equation}
with the $A_L$ index in the adjoint of $\SO(5)_L$ and $\alpha$ in the triplet of the
elementary $\SU(2)^0_L$ group. The physical values of these spurions are
\begin{equation}
{\mathcal G}_\alpha = g_0 T^\alpha_L\,,
\qquad
{\mathcal G}' = g'_0 T^3_R\,.
\end{equation}
The new spurion $\widetilde G$ is introduced by replacing $\widetilde R_\mu$ in
Eq.~(\ref{eq:R_mu_2-site}) with
\begin{equation}\label{eq:rho_tilde_spurions_2-site}
\widetilde R_\mu = \widetilde {\mathcal G}_a \widetilde \rho^a_\mu
= \widetilde {\mathcal G}_{A_R\, a} T^{A_R} \widetilde \rho^a_\mu\,,
\end{equation}
where the $A_R$ index is in the adjoint of $\SO(5)_R$. The other index, $a$, is in the adjoint
of the group ${\widetilde \SO}(4)$, which we define as the group under which only the
$\widetilde \rho^a_\mu$ fields transform. The physical value of $\widetilde {\mathcal G}$ is
\begin{equation}
\widetilde {\mathcal G}_a = \widetilde g_\rho T^a\,,
\end{equation}
and it breaks the $\SO(5)_R \times {\widetilde \SO}(4)$ group to the diagonal $\SO(4)$ subgroup.

In order to generate an operator contributing to the Higgs potential we now need to insert, in addition to the
spurions related to the elementary couplings, also the ones that correspond to the composite $\widetilde \rho$ resonances.
It can be easily checked that the leading contributions to the potential come from operators
containing $4$ spurions. Examples of such operators are
\begin{equation}
\frac{c_g}{16 \pi^2} f^4 \Tr\left[{\mathcal G}_\alpha {\mathcal G}_\alpha {\mathcal U}
\widetilde {\mathcal G}_\alpha \widetilde {\mathcal G}_\alpha {\mathcal U}^T\right]
\quad
\textrm{and}
\quad
\frac{c_{g'}}{16 \pi^2} f^4 \Tr\left[{\mathcal G}' {\mathcal G}' {\mathcal U}
\widetilde {\mathcal G}_\alpha \widetilde {\mathcal G}_\alpha {\mathcal U}^T\right]\,.
\end{equation}
The presence of two additional powers of the gauge couplings
with respect to the case of the non-linear $\sigma$-model in Eqs.~(\ref{eq:s_model_pot_1}) and (\ref{eq:s_model_pot_2})
reduces the degree of divergence from quadratic to logarithmic.
To further reduce the divergence and make the potential finite at one loop we need
to introduce one additional symmetry under which the Higgs is a Goldstone. This is
achieved in the three-site model as we will discuss in the following section.


\subsection{The three-site model}\label{sec:3-site_gauge}

\begin{figure}
\centering
\includegraphics[width=.6\textwidth]{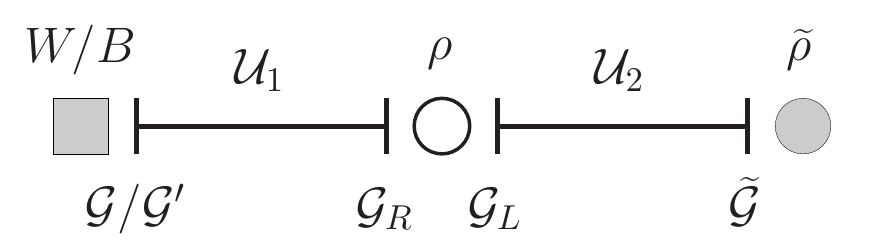}
\caption{Pictorial representation of the three-site model.}\label{fig:3-site_moose}
\end{figure}

The key ingredient for the construction of the three-site model,
schematically depicted in Fig.~\ref{fig:3-site_moose}, is a pair of identical $\sigma$-models
based on the coset $\SO(5)_L \times \SO(5)_R/\SO(5)_V$. The Goldstones are parametrized
by two $\SO(5)$ matrices ${\mathcal U}_1$ and ${\mathcal U}_2$ for a total of $20$ states,
$\Pi^A_1$ and $\Pi^A_2$. The leading-order Goldstone Lagrangian is given by
\begin{equation}\label{eq:Goldstone_Lagr_3-site}
{\mathcal L}^\pi = \frac{f_1^2}{4} \Tr \left[(D_\mu {\mathcal U}_1)^T D^\mu {\mathcal U}_1\right]
+ \frac{f_2^2}{4} \Tr \left[(D_\mu {\mathcal U}_2)^T D^\mu {\mathcal U}_2\right]\,.
\end{equation}
Notice that in Eq.~(\ref{eq:Goldstone_Lagr_3-site}) we kept the decay constants
of the two $\sigma$-models as independent parameters. The number of free parameters could be reduced
by imposing a $1 \leftrightarrow 2$ discrete symmetry, which enforces the relation $f_1 = f_2$.

The symmetries of the two $\sigma$-models, $\SO(5)_L^1 \times \SO(5)_R^1$ and
$\SO(5)_L^2 \times \SO(5)_R^2$ , are broken by gauging. As in the two-site case,
the ``first'' subgroup $\SO(5)_L^1$ is broken by the couplings with the elementary
gauge bosons and the ``last'' one, $\SO(5)_R^2$, by the couplings with $\widetilde \rho$.
The remaining subgroups, $\SO(5)^1_R$ and $\SO(5)^2_L$, are broken by gauging their vector
combination. The ten associated gauge fields, $\rho^A_\mu$, whose coupling is denoted by $g_\rho$,
become massive by eating ten Goldstones and are interpreted as resonances of the
composite sector.

The gauge structure, summarized in Fig.~\ref{fig:3-site_moose}, corresponds to the
covariant derivatives
\begin{eqnarray}
D_\mu {\mathcal U}_1 &=& \partial_\mu {\mathcal U}_1 - i A_\mu {\mathcal U}_1
+ i\, {\mathcal U}_1 R_\mu\,,\nonumber\\
D_\mu {\mathcal U}_2 &=& \partial_\mu {\mathcal U}_2 - i L_\mu {\mathcal U}_2
+ i\, {\mathcal U}_2 \widetilde R_\mu\,,\label{eq:cov_der_3-site}
\end{eqnarray}
where $R_\mu$ and $L_\mu$ are actually identical, $R_\mu = L_\mu = g_\rho \rho^A_\mu T_A$,
while $A_\mu$ and $\widetilde R_\mu$ are defined in Eqs.~(\ref{eq:elem_gauge_2-site})
and (\ref{eq:R_mu_2-site}).
After introducing the spurions, $R_\mu$ and $L_\mu$ are rewritten as
\begin{eqnarray}
R_\mu &=& {{\mathcal G}_R}^A \rho^A_\mu = {{\mathcal G}_R}^{A_R^1\, A}\, T^{A^1_R} \rho^A_\mu\,,\nonumber\\
L_\mu &=& {{\mathcal G}_L}^A \rho^A_\mu = {{\mathcal G}_R}^{A_L^2\, A}\, T^{A^2_L} \rho^A_\mu\,,
\end{eqnarray}
and transform under different symmetries, respectively $\SO(5)_R^1$ and $\SO(5)^2_L$.
The index $A$ carried by the two spurions is associated to an $\SO(5)_\rho$ group under
which the $\rho^A_\mu$ field transform in the adjoint representation while all the
other fields are invariant. The physical values of the spurions are
\begin{equation}
{{\mathcal G}_R}^A = {{\mathcal G}_L}^A = g_\rho\, T^A\,,
\end{equation}
and break the $\SO(5)_R^1 \times \SO(5)_L^2 \times \SO(5)_\rho$ to the vector combination.
As in the two-site model, additional spurions are associated to the SM couplings, ${\mathcal G}$
and ${\mathcal G}'$, and to the $\widetilde \rho$, $\widetilde {\mathcal G}$
(see Eqs.~(\ref{eq:elem_spurions_2-site}) and (\ref{eq:rho_tilde_spurions_2-site})).

In the three-site model the Higgs dynamics is triply protected by the Goldstone symmetries.
To understand this better let us set the Higgs to its VEV. This corresponds to constant
configurations $\langle \Pi^{\widehat 4}_{1,2} \rangle$ of the Goldstone fields
$\Pi^{\widehat 4}_{1,2}$. These constant configurations can be eliminated by a symmetry
transformation provided that at least one of the broken symmetries is restored by setting
the corresponding gauge coupling to zero. For instance if we set $g_\rho$ to zero the
$\SO(5)^1_R$ and $\SO(5)^2_L$ symmetries are restored and they can be used to get rid
of $\langle \Pi^{\widehat 4}_{1} \rangle$ and $\langle \Pi^{\widehat 4}_{2} \rangle$.
The situation is slightly more involved in the case of $\widetilde g_\rho$. To understand the
collective breaking mechanism in this case, it is important to remember that even in the presence
of a non-vanishing $g_\rho$ the vector combination of $\SO(5)^1_R$ and $\SO(5)^2_L$ is still
unbroken and it can be used to set $\langle \Pi^{\widehat 4}_{1} \rangle$ to zero.
The VEV is thus moved entirely onto $\Pi^{\widehat 4}_{2}$ and, if $\widetilde g_\rho$ is zero,
can be eliminated by an $\SO(5)^2_R$ transformation. The same holds for the
elementary couplings, $g_0$ and $g'_0$. Therefore, any physical effect of the Higgs VEV is
necessarily mediated by the three couplings $g_\rho$, $\widetilde g_\rho$ and $g_0$ (or $g'_0$).

The presence of the unbroken vector combination of $\SO(5)^1_R$ and $\SO(5)^2_L$ has also another
implication: any Higgs configuration, including a VEV, chosen for ${\mathcal U}_1$
can always be rotated away and moved to ${\mathcal U}_2$ and vice versa. To represent the
Higgs it is thus better to choose the invariant combination ${\mathcal U}_1 {\mathcal U}_2$.
Notice also that in the three-site model the Goldstone Higgs decay constant does not
coincide with any of the $\sigma$-models decay constants, instead it is linked to them by the relation
\begin{equation}
1/f^2 = 1/f_1^2 + 1/f_2^2\,.
\end{equation}
This relation can be easily
derived from the Higgs effective Lagrangian in which the gauge resonance have been
integrated out. We will explicitly perform this computation at the end of the section.

The triple collective breaking of the shift symmetry provides a further reduction of the
degree of divergence of the Higgs potential. In particular the Higgs potential becomes finite
because it must contain at least two additional powers of $g_\rho$. Performing the spurion analysis
one actually finds no contributions of order $g_\rho^2$. The leading operators, like
\begin{equation}\label{eq:gauge_pot_3-site}
\frac{c_{g'}}{(16 \pi^2)^3} \Tr[{\mathcal G}'\, {\mathcal U}_1\, {{\mathcal G}_R}^A\,
{\mathcal U}_1^T\, {\mathcal G}'\, {\mathcal U}_1\, {{\mathcal G}_R}^B {\mathcal U}_1^T]
\,\Tr[{{\mathcal G}_L}^A {\mathcal U}_2\,
\widetilde {\mathcal G}^a\, {\mathcal U}_2^T\, {{\mathcal G}_L}^B {\mathcal U}_2\,
\widetilde {\mathcal G}^a\, {\mathcal U}_2^T]\,,
\end{equation}
contain four powers of $g_\rho$. This further lowers the degree of divergence.
Indeed the rule in Eq.~(\ref{eq:divergence_counting2}) shows that the gauge contribution
to the potential is not only finite at one loop, but it starts diverging at the three-loop
order.

\begin{figure}
\centering
\includegraphics[width=.375\textwidth]{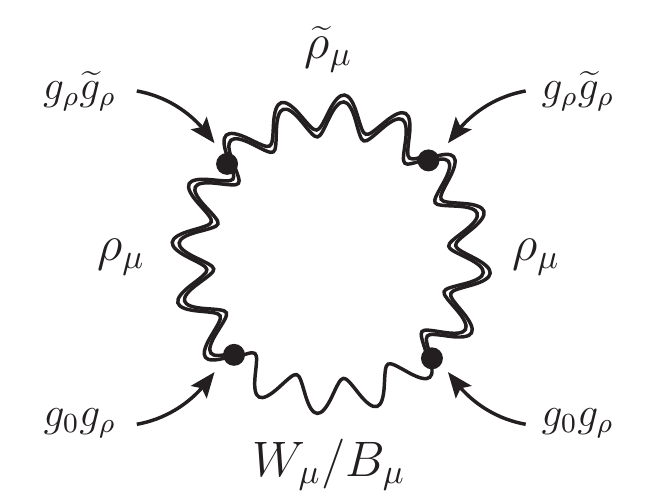}
\caption{Schematic representation of a one-loop contribution to the Higgs
potential induced by gauge fields. The double-line propagators represent the
heavy gauge fields, while the single-line one represents the elementary states.
The black dots denote the interaction vertices, which also include the Higgs field,
coming from the gauge interactions in Eq.~(\ref{eq:Goldstone_Lagr_3-site}).}
\label{fig:Higgs_pot_3-site}
\end{figure}

The reduction of the degree of divergence in the Higgs potential due to the collective breaking
can also be easily understood in a diagrammatic way. In order to be sensitive to the
breaking of all the Goldstone symmetries any diagram contributing to the Higgs potential
must include simultaneously the elementary gauge fields, the $\rho_\mu$ and the
$\widetilde \rho_\mu$ fields. The schematic structure of a one-loop contribution
is shown in Fig.~\ref{fig:Higgs_pot_3-site}. It is easy to see that at least four
vertices coming from the gauge interactions in the Goldstone
Lagrangian in Eq.~(\ref{eq:Goldstone_Lagr_3-site}) are needed.
As can be seen from the explicit form of the Lagrangian, each vertex carries two powers of the gauge couplings.
This explains why in Eq.~(\ref{eq:gauge_pot_3-site}) we found four powers of $g_\rho$ along with two powers
of the elementary coupling $g_0$ and of $\widetilde g_\rho$.

\subsubsection{Gauge fixing}
\index{gauge fixing}

Before concluding the discussion of the gauge sector of the three-site model it is
useful to briefly discuss the issue of gauge fixing. A common choice is to adopt
a ``partial'' unitary gauge, that is to remove the Goldstones eaten by the $\rho_\mu$
and $\widetilde \rho_\mu$ fields, while keeping in the Lagrangian the complete Higgs fourplet.
Due to a remnant of the symmetries at the middle and right-most site, the ``partial'' unitary gauge
choice does not completely fix the form of the Lagrangian: we can still choose how
the Higgs fourplet is embedded in the ${\mathcal U}_1$ and ${\mathcal U}_2$ matrices.

A convenient choice is to remove the Higgs from ${\mathcal U}_2$ by setting it to
the identity while keeping the full dependence on ${\mathcal U}_1$
\begin{equation}\label{eq:holo_gauge_3-site}
{\mathcal U}_1 = \overline{U} = \exp\left(i \frac{\sqrt{2}}{f} H_i T^i\right)\,,
\end{equation}
where $H_i$ correspond to the four Higgs components.
With this gauge choice the only dependence on the Goldstones appears at the
left-most site, resulting in a particularly simple form for the Lagrangian of the composite
sector.

Other simple gauge fixing choices are possible.
For instance one can move the full Higgs dependence on the ${\mathcal U}_2$ matrix while
setting ${\mathcal U}_1$ equal to the identity. In this way one recovers the usual
picture in which the Higgs interacts directly only with the composite states and not with the
elementary sector. A third possibility is to ``split'' the Higgs into the two
Goldstone matrices as
\begin{equation}\label{eq:uniform_gauge_3-site}
{\mathcal U}_{1,2} = \exp\left(\sqrt{2} i \frac{f}{f_{1,2}^2} H_i T^i\right)\,.
\end{equation}
The one in Eq.~(\ref{eq:uniform_gauge_3-site}) defines the ``unitary gauge'' of the theory, indeed it ensures that the Goldstones do not have a quadratic mixing with the gauge fields~\cite{Son:2003et}. 

As an example, we now compute the Goldstone Higgs decay constant as a function
of the original $\sigma$-models parameters $f_{1,2}$. For this purpose we use
the gauge choice in Eq.~(\ref{eq:holo_gauge_3-site}). In this gauge the only dependence
of the Lagrangian on the Goldstone matrix comes from the covariant derivative
$D_\mu \overline{U}$ (see Eq.~(\ref{eq:cov_der_3-site})). This quantity can be
rewritten by using the identity
\begin{equation}
\overline {U}^T D_\mu \overline {U} =
\overline {U}^T \partial_\mu \overline {U}
- i\, \overline {U}^T A_\mu \overline {U} + i R_\mu
= -i A^{\left(\overline {U}^T\right)} + i R_\mu\,,
\end{equation}
where $A^{(\overline {U}^T)}$ corresponds to a gauge transformation of the elementary
fields given by 
\beq
A_\mu^{(\overline {U}^T)} = \overline {U}^T (A_\mu + i\, \partial_\mu )\overline {U}\,.
\eeq
With the above definitions the Lagrangian in Eq.~(\ref{eq:Goldstone_Lagr_3-site}) can be written as
\begin{equation}\label{eq:Lagr_holo_gauge_3-site}
{\mathcal L}^\pi = \frac{f_1^2}{4} \Tr\left[\left(A_\mu^{(\overline {U}^T)} - g_\rho \rho_\mu\right)^2\right]
+ \frac{f_2^2}{4} \Tr\left[\left(g_\rho \rho_\mu - \widetilde g_\rho \widetilde \rho_\mu\right)^2\right]\,.
\end{equation}
From this expression we can derive an effective action for the Goldstone Higgs by integrating
out the massive vector resonances. The calculation is straightforward if we are only
interested in the two-derivative terms. The equations of motion at zero momentum for the $\widetilde \rho$ fields
and for the $\SO(4)$ components of $\rho$ imply
\begin{equation}\label{eq:EOM1}
\widetilde g_\rho \widetilde \rho_\mu^a = g_\rho \rho_\mu^a = \left(A_\mu^{(\overline {U}^T)}\right)^a\,,
\end{equation}
thus the $\SO(4)$ components in the Lagrangian (\ref{eq:Lagr_holo_gauge_3-site}) exactly cancel.
The equations of motion for the $\rho^i$ fields in the coset $\SO(5)/\SO(4)$, instead, give
\begin{equation}\label{eq:EOM2}
g_\rho \rho^i_\mu = \frac{f_1^2}{f_1^2 + f_2^2} \left(A_\mu^{(\overline {U}^T)}\right)^i\,.
\end{equation}
By substituting Eqs.~(\ref{eq:EOM1}) and (\ref{eq:EOM2}) back into the Lagrangian we obtain
\begin{equation}
{\cal L}^\pi_{eff} = \frac{f_1^2 f_2^2}{4 (f_1^2 + f_2^2)} \sum_i \left|\left(
\overline {U}^T \partial_\mu \overline U\right)^i\right|^2\,.
\end{equation}
From this expression we can immediately read the Goldstone decay constant
\begin{equation}
1/f^2 = 1/f_1^2 + 1/f_2^2\,.
\end{equation}
In the case $f_1 = f_2 = f$ the Higgs decay constant simply becomes $f = f_1/\sqrt{2} = f_2/\sqrt{2}$.


\subsection{The matter sector}\label{sec:metter_sector}

\index{partial compositeness}
So far we included in our effective model only the gauge degrees of freedom. To complete
the construction we also need to introduce the SM fermions and couple them to the Higgs.
Following the hypothesis of partial compositeness,
we assume that the SM fermions originate from elementary degrees of freedom, external
with respect to the composite sector. These states are then coupled linearly to some
composite sector operator.

The construction is analogous to the one of the SM vector bosons that we discussed in the
previous sections. The elementary vectors arise from gauging the SM subgroup embedded in
the $\SO(4)$ symmetry of the composite sector. This kind of construction gives rise to linear couplings
of the form $g_0 W^\alpha_\mu J^\mu_\alpha$, where $J^\mu_\alpha$  denotes a
current operator coming from the composite dynamics.
The linear coupling results in a mixing of the elementary fields with the
composite resonances. For the vector bosons, indeed, we find in our Lagrangian terms of the form
${\mathcal L}_{mix} \simeq (g_0/g_\rho) m_\rho^2\, W^\alpha_\mu\,\rho^\mu_\alpha$.\footnote{Notice
that, due to gauge invariance, the elementary/composite mixing arises from terms of the
form $f^2 (g_0 W^\alpha_\mu - g_\rho \rho_\mu^\alpha)^2$, see for instance Eq.~(\ref{eq:Goldstone_Lagr_2-site}).}
Analogously, for the elementary fermions, $q$, we assume a mixing of the form
${\mathcal L}_{mix} \simeq (y^q/g_\rho) m_\rho \overline q\, \psi$, where $y^q$ is the coupling
of the elementary field with the corresponding fermionic operator and $\psi$ generically
denotes the fermionic composite resonances. Because of the mixing, the light states,
which eventually describe the SM particles, are linear combinations of the elementary fields,
$q$, and the composite ones, $\psi$.

Before starting the explicit construction it is important to add a further comment.
In the case of the gauge and Goldstone Lagrangian constructed in the previous sections
the Goldstone symmetry almost completely determined the whole structure of the model,
leaving very narrow space for alternative constructions (on this point see
Sect.~\ref{sec:alternative_constructions}). The situation, instead, is considerably different
for the fermionic sector. First of all, the elementary fields can mix with composite
operators in many possible representations of the $\SO(5)$ symmetry group, thus leading
to different quantum numbers for the corresponding fermionic composite states.
On top of this, even the number of composite resonances included in the model
is to a large extent arbitrary and can give rise to a large number of alternatives.
In the following we will focus on a ``minimal'' scenario in which only one multiplet
of composite fermions is associated to each site. Moreover we will assume that the
elementary states are mixed with operators in the fundamental, the $\mathbf 5$,
representation of $\SO(5)$. For simplicity we will introduce in the model only the set of
fermionic resonances that are mixed with the top quark and are responsible for generating its mass,
the so called ``top partners''.
This minimal construction, however, contain all the main ingredients
of generic multi-site models and can be straightforwardly adapted to different cases.

\subsubsection{The two-site model}

In the case of two sites, focusing for simplicity on the top quark sector, we introduce
only one Dirac fiveplet of fermionic resonances $\widetilde \psi$. It transforms under
the $\SO(5)_R$ subgroup and is mixed with the elementary doublet $q_L=(t_L, b_L)$
and with the singlet $t_R$. The mixing term is
\begin{equation}\label{eq:Lagr_ferm_2-site}
{\mathcal L}_{mix} = y_L f {\overline Q_L}^I {\mathcal U}_{IJ} \widetilde \psi^J
+ y_R f {\overline T_R}^I {\mathcal U}_{IJ} \widetilde \psi^J + \textrm{h.c.}\,,
\end{equation}
where $Q_L$ and $T_R$ are the embeddings of the $q_L$ and $t_R$ elementary fermions
into incomplete $\SO(5)_L$ fiveplets. In Eq.~(\ref{eq:Lagr_ferm_2-site}), as customary,
we expressed the elementary/composite mixing in units of $f$. The $\mathbf 5$ representation decomposes as
$\mathbf{5} = (\mathbf{2}, \mathbf{2}) \oplus (\mathbf{1}, \mathbf{1})$ under
$\SO(4) \simeq \SU(2)_L \times \SU(2)_R$. The $(\mathbf{2}, \mathbf{2})$ part consists of
two $\SU(2)_L$ doublets with opposite $\U(1)^3_R$ charge $T^3_R = \pm 1/2$. We choose
to embed $q_L$ in the negative-charge doublet and $t_R$ in the singlet. The explicit
form of the embedding was given in Eqs.~(\ref{qlemb}) and (\ref{tremb})
and we report it here for completeness
\begin{equation}\label{eq:elem_ferm_embedding_5}
Q_L = \frac{1}{\sqrt{2}}
\left[
\begin{array}{c}
-i b_L\\
- b_L\\
-i t_L\\
t_L\\
0
\end{array}
\right]\,,
\qquad
T_R =
\left[
\begin{array}{c}
0\\
0\\
0\\
0\\
t_R
\end{array}
\right]\,.
\end{equation}

If we identify the $\U(1)_Y$ symmetry with the subgroup of $\SO(5)_L$ generated by $T^3_R$,
as we did in the previous sections, we do not obtain the correct hypercharges for the fermions.
We saw in the previous chapters that solving this problem requires an extra $\U(1)_X$ global symmetry, which
acts on the matter fields $q_L$, $t_R$ and $\widetilde \psi$ as a phase rotation with charge
$X = 2/3$. The hypercharge gauge field $B_\mu$ is now introduced by gauging the subgroup
of $\SO(5)_L \times \U(1)_X$ corresponding to the combination
\begin{equation}\label{eq:hypercharge_gen}
Y = T^3_R + X\,.
\end{equation}
Given that the Goldstones are not charged under the extra $\U(1)_X$, this change in the definition
of $Y$ does not affect their couplings to $B_\mu$ and the construction discussed in the previous
sections can be left unchanged. The $W^\alpha_\mu$ and the massive resonances $\widetilde \rho_\mu$
are included, as before, by gauging the $\SU(2)_L$ subgroup of $\SO(5)_L$ and the $\SO(4)$
subgroup of $\SO(5)_R$, whereas the $\U(1)_X$ symmetry acts on all sites and there is no
new composite resonance associated to it.

To keep the discussion as simple as possible we also assume that the QCD gauge group
$\SU(3)_c$ is ``external'' with respect to the site structure, similarly to the $\U(1)_X$ one. This means that it acts on all sites and, in particular, all the fermionic states we described before belong
to the fundamental $\SU(3)_c$ representation. Alternatively we could have extended the multi-site  construction presented in the previous sections for the EW gauge fields to the case of the $\U(1)_X$ and $\SU(3)_c$ groups.
In this way one would get additional vector resonances associated to these gauge
groups analogous to the $\widetilde \rho_\mu$ states. An important difference with respect to the
EW case is the fact that the replicas of the $\U(1)_Y$ and $\SU(3)_c$ groups at the various sites
are fully gauged, thus no extra physical Goldsone boson remains in the spectrum.
Moreover each new group factor is associated to its own coupling $f$, which a priori
is independent of the coupling $f$ related to the $\SO(5)$ invariance.

In order to make clear the symmetry pattern in the fermionic sector, we now discuss how to
introduce the spurions that we will use to analyze the fermionic contribution to
the Higgs potential. Due to the presence of the extra $\U(1)_X$ symmetry, the group under which
the elementary fields transform is enlarged to $\SU(2)_L^0 \times \U(1)_R^0 \times \U(1)_X^0$.
The charge of the elementary fields under $\U(1)^0_R$ corresponds to the charge under the
original subgroup of $\SO(5)_L$ generated by $T^3_R$, hence $q_L$ has charge $-1/2$
while $t_R$ is neutral. The spurions are two vectors $\Delta_L$ and $\Delta_R$ in the
$\mathbf 5$ representation of $\SO(5)_L$ and also transform under the elementary
$\SU(2)_L^0 \times \U(1)_R^0 \times \U(1)^0_X$ group. The $\Delta_L$ spurion is in the $\overline{\bf 2}$ 
representation of $\textrm{SU}(2)_L^0$ with $\U(1)_R^0$ charge $1/2$ and $X^0 = -2/3$ (the conjugate of the
$q_L$ representation), while $\Delta_R$ is a singlet with $X^0 = -2/3$ and is neutral under
$\U(1)_R^0$. The physical values of the $\Delta_{L,R}$ spurions are (compare Sect.~\ref{CHPCH})
\begin{equation}\label{eq:spurions_55}
(\Delta_L)_{\alpha}^{I} = \frac{y_L f}{\sqrt{2}}
\left[
\begin{array}{ccccc}
0 & 0 & +i & +1 & 0\\
+i & -1 & 0 & 0 & 0
\end{array}
\right]^I_\alpha\,,
\qquad
(\Delta_R)^I =
y_R f
\left[
\begin{array}{ccccc}
0 & 0 & 0 & 0 & 1
\end{array}
\right]^I\,.
\end{equation}
By using the previously defined objects we can rewrite the Lagrangian
in Eq.~(\ref{eq:Lagr_ferm_2-site}) as
\begin{equation}
{\mathcal L}_{mix} = {\overline q_L}^i \Delta_L^{iI} {\mathcal U}_{IJ} \widetilde \psi^J
+ {\overline t_R} \Delta_R^{I} {\mathcal U}_{IJ} \widetilde \psi^J + \mathrm{h.c.}\,,
\end{equation}
where $i$ denotes the $\SU(2)_L^0$ index, while $I$ and $J$ are the usual $\SO(5)$ indices.

The physical value of the spurions $\Delta_{L,R}$ breaks the composite-sector group
$\SO(5)_L \times \U(1)_X$ and the elementary one $\SU(2)_L^0 \times \U(1)_R^0 \times \U(1)^0_X$,
preserving the $\SU(2)_L \times \U(1)_Y$ subgroup that is gauged by the $W$ and $B$ elementary fields.
The unbroken $\SU(2)_L$ is the vectorial combination of $\SU(2)_L^0$ and the $\SU(2)_L$ subgroup
of $\SO(5)_L$, while the hypercharge is the combination of the $\U(1)_Y^0$ subgroup
of the elementary $\U(1)_R^0 \times \U(1)_X^0$, whose generator is specified in
Eq.~(\ref{eq:hypercharge_gen}), and the analogous combination coming from $\SO(5)_L \times \U(1)_X$.

As already discussed, in addition to the gauging at the left-most site, we also gauge the
$\SO(4)$ subgroup of $\SO(5)_R$ by introducing the $\widetilde \rho$ resonances. This leads to the
covariant derivatives
\begin{eqnarray}
D_\mu q_L &=& \left(\partial_\mu - i \frac{g_0}{2} W^\alpha_\mu \sigma_\alpha
- i \frac{g'_0}{6} B_\mu\right) q_L\,,\nonumber\\
D_\mu t_R &=& \left(\partial_\mu - i \frac{2 g'_0}{3} B_\mu \right) t_R\,,\nonumber\\
D_\mu \widetilde \psi &=& \left(\partial_\mu - i \frac{2 g'_0}{3} B_\mu
- i \widetilde g_\rho \widetilde \rho_\mu\right) \widetilde \psi\,.\label{eq:cov_der_ferm_2-site}
\end{eqnarray}
Notice that $\widetilde \psi$ is neutral under $\SO(5)_L$, therefore its covariant derivative
does not contain the $W^\alpha_\mu$ gauge fields. It contains instead $B_\mu$ because
$\widetilde \psi$ is charged under $\U(1)_X$ and the hypercharge is defined as in
Eq.~(\ref{eq:hypercharge_gen}). In terms of the covariant derivatives the elementary and
composite sector kinetic Lagrangians read
\begin{eqnarray}
{\mathcal L}^f_{el} &=& i \overline q_L \gamma^\mu D_\mu q_L
+ i \overline t_R \gamma^\mu D_\mu t_R\,,\nonumber\\
{\mathcal L}^f_{cs} &=& i \overline{\widetilde \psi} \gamma^\mu D_\mu \widetilde \psi
+ \widetilde m^{IJ} \overline{\widetilde \psi}_I \widetilde \psi_J\,,
\end{eqnarray}
where a mass term $\widetilde m = \mathrm{diag}(\widetilde m_Q, \widetilde m_T)$, different for the
fourplet and singlet components of $\widetilde \psi$, has also been introduced. This mass
matrix $\widetilde m^{IJ}$ is a spurion with two indices in the fundamental of $\SO(5)_R$
and breaks this symmetry to its $\SO(4)$ subgroup.

\begin{figure}
\centering
\includegraphics[width=.355\textwidth]{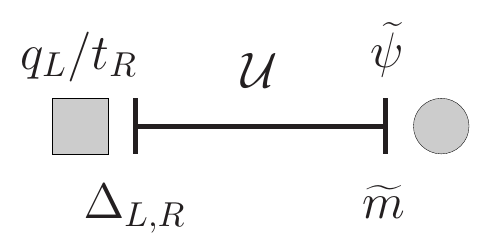}
\caption{Pictorial representation of the matter sector of the two-site model.}
\label{fig:2-site_ferm_moose}
\end{figure}

The fermionic sector of the two-site model is summarized in Fig.~\ref{fig:2-site_ferm_moose}.
With respect to the gauge sector presented in Sect.~\ref{sec:2-site_gauge}, the only new
parameters that we introduced are $\Delta_{L,R}$ and $\widetilde m$, which are masses,
not new couplings. This makes particularly easy to generalize the divergence counting rule of
Eq.~(\ref{eq:divergence_counting2}). The result is
\begin{equation}\label{eq:divergence_counting_ferm}
\Lambda^2 f^2 \left(\frac{\Lambda}{4 \pi f}\right)^{2L}
\left(\frac{\Pi}{f}\right)^{E_\pi} \left(\frac{g V}{\Lambda}\right)^{E_V}
\left(\frac{\psi}{\sqrt{\Lambda} f}\right)^{E_\psi}
\left(\frac{\partial}{\Lambda}\right)^{d} \left(\frac{g f}{\Lambda}\right)^{2 \eta}
\left(\frac{\mu}{\Lambda}\right)^\chi\,,
\end{equation}
where $\psi$ generically denotes the fermions $q_L$, $t_R$ or $\widetilde \psi$, while
$\mu$ is any of the masses $\Delta_{L,R}$ or $\widetilde m$. In the above formula
$V$ and $g$ collectively denote all the vector fields and the gauge couplings.
The positive integer $\chi$ counts the number of mass-term insertions and it is forced
by the chiral symmetry ($\psi_L \rightarrow -\psi_L$, $\psi_R \rightarrow \psi_R$
and $\mu \rightarrow -\mu$) to be even or odd depending on the chirality of the operator.

Now that we introduced all the necessary tools it is easy to generalize the discussion on the
calculability. The fermionic contribution to the Higgs potential is logarithmically divergent.
The local operators associated to the divergence are
\begin{equation}\label{eq:ferm_div_2-site_1}
\frac{c_R}{16\pi^2} \Delta_R^\dagger\, {\mathcal U}\, \widetilde m^2\, {\mathcal U}^T \Delta_R
= \frac{c_R}{16 \pi^2} y_R^2 f^2\left[\left(\widetilde m_Q^2 - \widetilde m_T^2\right)
\sin^2\left(\frac{H}{f}\right) + \widetilde m_T^2\right]
\end{equation}
and
\begin{equation}\label{eq:ferm_div_2-site_2}
\frac{c_L}{16\pi^2} \sum_\alpha (\Delta_L^\alpha)^\dagger\, {\mathcal U}\, \widetilde m^2\,
{\mathcal U}^T \Delta_L^\alpha
= \frac{c_L}{32 \pi^2} y_L^2 f^2\left[\left(\widetilde m_T^2 - \widetilde m_Q^2\right)
\sin^2\left(\frac{H}{f}\right) + 4 \widetilde m_Q^2\right]\,,
\end{equation}
and originate, respectively, from loops of the elementary $t_R$ and $q_L$. Notice that
the functional dependence on the Higgs is completely determined by the spurion analysis,
in agreement with the results of Chap.~\ref{ch:SILH}.

\subsubsection{The three-site model}

\begin{figure}
\centering
\includegraphics[width=.6\textwidth]{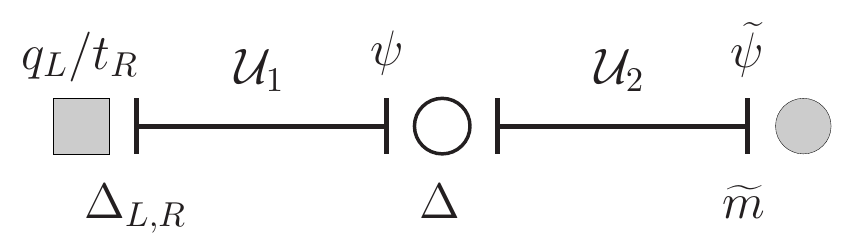}
\caption{Pictorial representation of the matter sector of the three-site model.}
\label{fig:3-site_ferm_moose}
\end{figure}

To obtain a calulable Higgs potential we need to consider the three-site model. As shown in
Fig.~\ref{fig:3-site_ferm_moose}, this is constructed by introducing two fiveplets of
fermionic Dirac resonances $\psi$ and $\widetilde \psi$. The former transforms under
the $\SO(5)_R^1$ group,\footnote{We could equivalently assume that the $\psi$ field
transforms under the vector combination of $\SO(5)_R^1$ and $\SO(5)_L^2$, in analogy with
the $\rho_\mu$ gauge fields. The choice made in the main text, however, helps in clarifying
the pattern of symmetry breaking induced by the fermions and makes simpler the introduction
of spurions.} while $\widetilde \psi$ transforms in the fundamental of $\SO(5)^2_R$.
The Lagrangian is similar to the one of the two-site case, with the difference that the
elementary fields mix now with $\psi$ and not with $\widetilde \psi$:
\begin{equation}\label{eq:mix_Lagr_ferm_3-site}
{\cal L}_{mix} = y_L f\, \overline Q_L\, {\mathcal U}_1 \psi + y_R f\, \overline T_R\, {\mathcal U}_2 \psi
+ \Delta\, \overline \psi\, {\mathcal U}_2 \widetilde \psi + \textrm{h.c.}\,.
\end{equation}
The other terms present in the leading-order Lagrangian are
\begin{eqnarray}
{\mathcal L}^f_{el} &=& i \overline q_L \gamma^\mu D_\mu q_L
+ i \overline t_R \gamma^\mu D_\mu t_R\,,\nonumber\\
{\mathcal L}^f_{cs} &=& i \overline{\widetilde \psi} \gamma^\mu D_\mu \widetilde \psi
+ \widetilde m^{IJ} \overline{\widetilde \psi}_I \widetilde \psi_J
+ i \overline \psi \gamma^\mu D_\mu \psi + m \overline \psi \psi\,,\label{eq:kin_Lagr_ferm_3-site}
\end{eqnarray}
where the covariant derivatives for the elementary fields and for $\widetilde \psi$ are defined
in Eq.~(\ref{eq:cov_der_ferm_2-site}) and
\begin{equation}
D_\mu \psi = \left(\partial_\mu - i \frac{2 g'_0}{3} B_\mu - i g_\rho T^A \rho_\mu^A\right)\psi\,.
\end{equation}
Notice that the $\psi$ mass term $m$ does not break any symmetry, differently from $\widetilde m$
which breaks $\SO(5)_R^2$ to its $\SO(4)$ subgroup.

To simplify the Lagrangian in Eqs.~(\ref{eq:mix_Lagr_ferm_3-site}) and (\ref{eq:kin_Lagr_ferm_3-site})
we assumed that the composite sector is invariant under parity ($\vec x \rightarrow -\vec x$) and
the only terms that break this symmetry are the mixings of the elementary fields with the
composite states. If we do not impose this invariance a different mixing is allowed between
the left- and right-handed components of $\psi$ and $\widetilde \psi$ in Eq.~(\ref{eq:kin_Lagr_ferm_3-site}).

We can now introduce spurions and analyze the fermion contributions to the Higgs potential.
The mixing Lagrangian in Eq.~(\ref{eq:mix_Lagr_ferm_3-site}) can be rewritten as
\begin{equation}
{\cal L}_{mix} = \overline q_L^i\, \Delta_L^{iI} ({\mathcal U}_1)_{IJ} \psi^J
+ \overline t_R\, \Delta_R^I ({\mathcal U}_2)_{IJ} \psi^J
+ \overline \psi^I {\Delta_I}^J ({\mathcal U}_2)_{IJ} \widetilde \psi^J + \textrm{h.c.}\,.
\end{equation}
The associated spurions, $\Delta_L$ and $\Delta_R$, transform under both the elementary
$\SU(2)_L^0 \times \U(1)_R^0 \times \U(1)_X^0$ and the $\SO(5)_L^1 \times \U(1)_X$ group and
break the global symmetry to the $\SU(2)_L \times \U(1)_Y$ subgroup as explained
in the two-site case. The new spurion $\Delta$ has indices in $\SO(5)_R^1$ and in $\SO(5)_L^2$.
Its physical value ${\Delta_I}^J = \Delta {\delta_i}^J$ is proportional to the identity
and therefore breaks $\SO(5)_R^1 \times \SO(5)_L^2$ to the vector subgroup.

The leading local contribution to the Higgs potential comes from operators like
\begin{equation}
\frac{c_R}{(16 \pi^2)^2} \frac{1}{f^2} \Delta_R^\dagger\, {\mathcal U}_1 \Delta\, {\mathcal U}_2\,
\widetilde m^2 {\mathcal U}_2^T\, \Delta^T\, {\mathcal U}_1^T \Delta_R\,,
\end{equation}
and similarly with $\Delta_L$. These fermionic contributions are finite at one loop and start
diverging only at two loops, differently from the gauge contributions
of Eq.~(\ref{eq:gauge_pot_3-site}) for which the divergence was postponed to three-loop order.

Before concluding the discussion of the three-site model it is useful to add a comment on the
choice of the $\widetilde \psi$ fermion representation. In our construction we assumed that
$\widetilde \psi$ is a complete $\SO(5)^2_R$ representation although we allowed,
in analogy to the gauge sector, an explicit soft breaking due to the mass
matrix $\widetilde m$ which only respect an $\SO(4)$ invariance. This choice was necessary
in order to obtain the protection due to the collective breaking mechanism. For instance
the assumption that the $\widetilde \psi$ fermions fill incomplete $\SO(5)_R^2$ representations
(of course always respecting the $\SO(4)$ invariance) would have induced a hard breaking
of the symmetry and no further reduction in the degree of divergence.
As a consequence the fermion contribution to the Higgs potential would have been logarithmically
divergent as in the two-site model. Introducing incomplete representations of fermions at the
last site can be compatible with the calculability of the Higgs potential only in models
with more than three sites, where the additional symmetries ensure enough protection to the
Higgs dynamics.


\subsection{Alternative constructions}
\label{sec:alternative_constructions}

As already discussed, the structure of the multi-site phenomenological models,
although quite constrained by the Goldstone symmetry and the collective-breaking principle,
admits some variations, especially in the fermionic sector. To give an idea of the
possible alternative constructions we present here the model proposed in Ref.~\cite{DeCurtis:2011yx}.

\begin{figure}
\centering
\includegraphics[width=.47\textwidth]{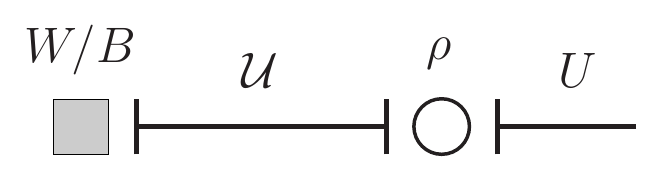}
\caption{Pictorial representation of the gauge sector of the alternative two-site model.
The ``half'' link on the right represents the non-linear $\sigma$-model
$\SO(5)^2_L/\SO(4)$.}
\label{fig:2-site_altern_moose}
\end{figure}

The structure of the gauge and Goldstone sector of the model resembles closely the one of the
three-site model presented in Sect.~\ref{sec:3-site_gauge}. It is based on two non-linear
$\sigma$-models: one corresponding to the $\SO(5)_L^1 \times \SO(5)_R^1/\SO(5)_V$
structure and the second given by $\SO(5)^2_L/\SO(4)$. Schematically it can be represented
as shown in Fig.~\ref{fig:2-site_altern_moose}.
The symmetry structure gives rise to 14 Goldstones, ten of them coming from the
$\SO(5)_L^1 \times \SO(5)_R^1$ $\sigma$-model and the remaining from the $\SO(5)^2_L/\SO(4)$
one. We encode the two sets of Goldstones in the matrices ${\mathcal U}$ and $U$ respectively.
Ten Goldstones are then eaten by the introduction of the composite gauge resonances $\rho$
that gauge the vector combination of the $\SO(5)_L^1$ and $\SO(5)^2_L$ groups.
The elementary gauge fields, $W$ and $B$, are instead associated to the gauging of the
$\SU(2)_L \times \U(1)_L$ subgroup of $\SO(5)_L^1$.
The leading-order Lagrangian of the Goldstone boson sector is given by
\begin{equation}
{\mathcal L}^\pi = \frac{f_1}{4} \Tr[(D_\mu {\mathcal U})^T D^\mu {\mathcal U}]
+ \frac{f^2_2}{4} d_\mu^i d^\mu_i\,,
\end{equation}
where the definition of the covariant derivative $D_\mu {\mathcal U}$ is similar to the
expression for ${\mathcal U}_1$ given in Eq.~(\ref{eq:cov_der_3-site})
and the $d_\mu^i$ symbol is defined in analogy to Eq.~(\ref{eq:dsymbol4}).
As in the three-site model the Goldstone Higgs decay constant $f$ is given by
$1/f^2 = 1/f_1^2 + 1/f_2^2$ and the Higgs field is represented by the
${\mathcal U}\cdot U$ combination which is invariant under the
unbroken vector subgroup of $\SO(5)_L^1$ and $\SO(5)^2_L$.
In the present set-up the gauge contribution to the Higgs potential is finite
at one-loop order, although it diverges at two loops.

Notice that the above construction is in some way ``intermediate'' between
the two-site and three-site models of Sects.~\ref{sec:2-site_gauge} and \ref{sec:3-site_gauge}.
For instance it can be formally obtained from the three-site construction by decoupling
(or integrating out) the $\widetilde \rho$ resonances. On the other hand the two-site
models can be recovered from the construction presented in this section
in the $f_2 \rightarrow \infty$ limit.

We can now describe the fermionic sector.
As in the two-site and three-site models, the $q_L$ and $t_R$ elementary fermions
are embedded in incomplete $\SO(5)_L^1$ representations. In the present construction they mix with two
composite multiplets $\psi_q$ and $\psi_t$, both in the fundamental
representation of $\SO(5)_R^1$. The kinetic terms for the fermions and the gauge interactions
are analogous to the ones described in the two-site model, with the only difference that now
a complete $\SO(5)$ group is gauged by the $\rho_\mu$ resonances.
The Lagrangian containing the mass terms can be split
into the part containing the elementary fields and the part that only involves the composite
states, ${\mathcal L}_{mix} = {\mathcal L}_{mix}^{el} + {\mathcal L}_{mix}^{cs}$.
The mixing part for the elementary states is
\begin{equation}
{\mathcal L}_{mix}^{el} = y_L f\, \overline Q_L\, {\mathcal U}\, \psi_q
+ y_R f\, \overline T_R\, {\mathcal U}\, \psi_t + \mathrm{h.c.}\,.
\end{equation}
The composite fermions mass terms contain, in addition to Dirac masses for the $\psi_q$ and
$\psi_t$ fields, also mixing mass terms that only couple the left-handed component of
$\psi_q$ and the right-handed component of $\psi_t$, namely
\begin{eqnarray}
{\mathcal L}_{mix}^{cs} &=& m_q \overline \psi_q \psi_q + m_t \overline \psi_t \psi_t\\
&& +\; m_1 \left(\overline \psi_{qL} U\right)_5 \left(U^T \psi_{tR}\right)_5
+ m_2\, \overline \psi_{qL} \psi_{tR} + m_3\, \overline \psi_{qR} \psi_{tL} + \mathrm{h.c.}\,.\hspace{1em}\nonumber
\label{eq:mix_st}
\end{eqnarray}
Notice that the peculiar structure of the mixing between the $\psi_q$ and $\psi_t$ fields,
as well as the structure of the elementary/composite mixing terms, is necessary to ensure
the finiteness of the fermion contribution to the Higgs potential at one loop.
If other mass mixing terms allowed by the symmetry structure are included,
such as $m \left(\overline \psi_{qR} U\right)_5 \left(U^T \psi_{tL}\right)_5 + \mathrm{h.c.}$,
$m \left(\overline \psi_{q} U\right)_5 \left(U^T \psi_{q}\right)_5$
or $m \left(\overline \psi_{t} U\right)_5 \left(U^T \psi_{t}\right)_5$,
the fermion contribution to the Higgs potential develops a logarithmic divergence. Notice that not including
these mass terms is an ad hoc assumption, because they are not protected by any symmetry.

The fermionic sector of the model is identical to the three-site construction.
This can be easily seen by the following identification of the fields
\begin{equation}
\psi_{qR} \rightarrow \psi_R\,, \quad \psi_{qL} \rightarrow \widetilde \psi_L\,, \quad
\psi_{tL} \rightarrow \psi_L\,, \quad \psi_{tR} \rightarrow \widetilde \psi_R\,.
\end{equation}
By working in the gauge in which the Goldstones are described only by the $\mathcal U$
matrix (and by the ${\mathcal U}_1$ in the three-site model), one can easily check that
the masses and mixing terms for the fermions can be identified with the ones given in
Eqs.~(\ref{eq:mix_Lagr_ferm_3-site}) and (\ref{eq:kin_Lagr_ferm_3-site})
if we allow for different mixings $\Delta$ for the left-
and right-handed fermion components by breaking parity.
Analogously, if we extend the construction of the present section to $n$ sites, the
fermionic sector will be similar to the one of a model with $2n-1$ sites built along the lines
of Sect.~\ref{sec:metter_sector}.
The only real difference between the three-site construction and the alternative models presented in this
section is due to the fact that the composite fermions are associated to different
global symmetries and, as a consequence, have different interactions with the composite gauge resonances.


\subsection{Locality in theory space}
\label{sec:non-local_operators}

To conclude the discussion about the construction of the multi-site models it is instructive to take a closer look at the
Lagrangian of the explicit realizations. The symmetry structure of the underlying non-linear $\sigma$-models
naturally led to a peculiar ``nearest-neighbor'' form for the Lagrangian.
This structure can be easily seen in the bosonic sector of the model.
The leading terms in the effective Lagrangian are the ones that respect the whole global
symmetry of the non-linear $\sigma$-models and include the kinetic
operators for the Goldstones (excluding the gauge interactions) and
for the gauge fields. All these operators are ``local'' in theory space, {\it{i.e.}}~they involve
fields belonging to a single site or to a single link. The other operators we introduced
in the Lagrangian do not respect the whole global symmetry and are suppressed
by powers of the couplings that induce the breaking. An example are the interactions
between the Goldstones and the gauge fields, which are weighted by the gauge couplings.
Notice that in the multi-site constructions the gauge couplings (including the ones of the composite
vector fields) are assumed to be perturbative.
Thus an insertion of a gauge coupling leads to an effective suppression
of the corresponding operator. Following this logic, the leading operators that break the global
symmetry are the ones that involve fields connected to the same link, or, in other words,
are of nearest-neighbor form. Operators that involve fields in non-near sites necessarily
break more symmetries, thus are suppressed by more insertions of the symmetry breaking
couplings.

It is easy to realize that the nearest-neighbor form is also tightly related to the collective breaking mechanism.
Indeed it guarantees that the various subgroups under which the Higgs transforms as
a Goldstone are broken independently of each other. As a consequence, insertions
of multiple symmetry-breaking couplings are necessary to generate operators which
depend on the Higgs VEV, such as the Higgs potential. If the locality assumption
is not respected the collective breaking mechanism is typically spoiled and the Higgs
observables become divergent.

The Lagrangian of the multi-site models does not naively follow the minimal One Scale One Coupling
(1S1S) power counting in Eq.~(\ref{POWCFull}), but it can be described by a simple extension of it.
The non-linear $\sigma$-models associated to the global
symmetry structure of the multi-site constructions are interpreted as an effective description of the underlying
strongly coupled dynamics. They are thus described by an effective Lagrangian that follows the 1S1S power
counting. The exact values of $m_*$ and $g_*$ characterizing this part of the Lagrangian do not have a
big impact on our constructions as long as $g_*$ is (much) larger than the gauge couplings of the vector
resonances.\footnote{For simplicity here we associate all the $\sigma$ models to a single $f$ decay constant and a single $m_*$.}
The elementary fields are external with respect to the composite dynamics and they enter
in the power-counting as external sources associated to a weak coupling which breaks the global invariance
of the composite sector (namely the global $\SO(5)^1_L$ symmetry at the left-most site).
There is however an important subtlety related to the vector and fermion states introduced at the composite sites.
Although these resonances are interpreted as composite states, they are assumed not be fully
strongly coupled so that their dynamics can be described perturbatively. This meas that from the point of view of the power-counting
we effectively split the composite sector into two parts: a fully strongly coupled sector encoded in the
non-linear $\sigma$-model and a semi-perturbative sector that includes the $\rho$ and $\psi$ states.
This assumption is incorporated in the power counting by treating the composite resonances as
weakly-coupled external sources with a coupling $\widetilde g_\rho$ that is smaller than the $\sigma$-model coupling $g_*$.

We will now briefly discuss how the multi-site models are modified by the introduction of non-nearest-neighbor operators.
As an example of non-minimal terms we can consider in the three-site model the non-nearest-neighbor interaction
\begin{equation}\label{eq:non-local_3-site}
{\mathcal L}_{non-local} = \frac{f'^2}{4} \Tr[(D_\mu ({\mathcal U}_1 {\mathcal U}_2))^T (D^\mu ({\mathcal U}_1 {\mathcal U}_2))]\,.
\end{equation}
This operator induces a breaking of the $\SO(5)_R^1$ and the $\SO(5)_L^2$ symmetries to their
diagonal subgroup and at the same time  feels the breaking of the $\SO(5)_L^1$
and $\SO(5)_R^2$ due to the gauging of the elementary fields $W$ and $B$ and the
composite vector resonances $\widetilde \rho$. As a consequence its coefficient is expected to
be subleading with respect to the one of the usual kinetic terms of the Goldstones.
It is easy to understand that the inclusion of ${\mathcal L}_{non-local}$ into the three-site Lagrangian reintroduces
a logarithmic divergence in the gauge contribution to the Higgs potential.
The operator in Eq.~(\ref{eq:non-local_3-site}) in fact is analogous to the
kinetic term for the Goldstones in the two-site model (Eq.~(\ref{eq:Goldstone_Lagr_2-site}))
as can be seen by the identification ${\mathcal U}_1 {\mathcal U}_2 \rightarrow {\mathcal U}$.
In exactly the same way the inclusion of the operator
\begin{equation}\label{eq:non-local_altern}
{\mathcal L}_{non-local} = \frac{f'^2}{4} \sum_i \left|\left(U^T {\mathcal U}^T D^\mu ({\mathcal U} U)\right)^i\right|^2
\end{equation}
in the alternative construction of Sect.~\ref{sec:alternative_constructions}
destroys the collective breaking mechanism and reintroduces a divergence
in the Higgs potential as in the non-linear $\sigma$-model case.
``Non-local'' operators such as (\ref{eq:non-local_altern}) have sometimes been considered in the
literature in connection to the multi-site constructions (see for example Ref.~\cite{DeCurtis:2011yx})
because they induce a tree-level correction to the $S$ parameter which can, in part,
compensate the sizable contribution coming from the heavy vector resonances (see Sect.~\ref{sec:ewpt_nonlocal}).
As we discussed before, however, this can be done only at the price of violating the power counting
and the collective-breaking mechanism.

Non-nearest neighbor interactions can also be constructed in the femionic sector.
An explicit example is the operator
\begin{equation}
{\mathcal L}_{non-local} = i \zeta \left(\overline {\widetilde \psi}\right)_i \gamma^\mu \left[({\mathcal U}_1 {\mathcal U}_2)^T
(D^\mu ({\mathcal U}_1 {\mathcal U}_2))\right]_{i5} \left(\widetilde \psi\right)_5 + \mathrm{h.c.}\,,
\label{eq:non-local_ferm_3-site}
\end{equation}
where $(\widetilde \psi)_i$ and $(\widetilde \psi)_5$ denote the $(\mathbf{2}, \mathbf{2})$ and
$(\mathbf{1}, \mathbf{1})$ $\SO(4)$ components inside the $\SO(5)$ fiveplet $\widetilde \psi$.
The above operator does not modify the fermion masses and contains only
derivative interactions with the Higgs. For this reason it does not contribute to the Higgs potential
at one loop and the violation of the collective breaking structure manifests itself only at higher order.
Operators like (\ref{eq:non-local_ferm_3-site})
correspond, after integrating out the heavy gauge resonances, to terms of the form
$i \zeta(\overline {\widetilde \psi})_i \gamma^\mu d_\mu^i (\widetilde \psi)_5 + \mathrm{h.c.}$,
where $d_\mu^i$ is the CCWZ $d$-symbol constructed with the physical Goldstone Higgs and the elementary
gauge fields. As we will show in Chap.~\ref{ch:EWPT} this kind of operators, if present with accidentally large coefficients,
can help in cancelling the contribution to $S$ coming from fermion loops.

\index{deconstruction}
To conclude the discussion, it is worth mentioning that the nearest-neighbor structure is also deeply related to
the holographic realizations of the composite Higgs scenario.
As we already mentioned, the multi-site constructions are analogous to a discretized or
deconstructed version of the five-dimensional models.
This can be clearly seen from
the schematic moose representations of the multi-site constructions: each site
can be interpreted as a lattice point along the fifth dimension. The replicas of
the $\SO(5)$ group at each site then correspond to the remnant of the
five-dimensional gauge invariance on the four-dimensional slices corresponding to each lattice point.

From this perspective, the nearest-neighbor assumption becomes equivalent
to the usual concept of locality along the fifth dimension. Moreover the collective-breaking
mechanism corresponds to the protection of the Higgs dynamics
in holographic models that is due to the five-dimensional gauge invariance.
Notice that, in this picture, a complete five-dimensional model coincides, at least formally, with
an effective theory with an infinite number of sites. This allows us to interpret the finiteness
of the Higgs potential in the holographic theories at any loop order as the result of a
collective breaking with an infinite number of independent symmetries.


\section{The Higgs potential}\label{sec:Higgs_pot_pheno}

The pNBG nature of the Higgs has a deep impact on the Higgs potential:
it can only be generated at the radiative level and is
tightly related to the collective breaking mechanism. One unavoidable, sizable source
of breaking of the Goldstone symmetry is the top quark Yukawa coupling. It is thus reasonable
to expect a tight relation between the Higgs mass and the fermionic sector
involved in the generation of the top mass. The general analysis presented in
Chap.~\ref{ch:SILH} confirms this expectation. In generic composite Higgs models a light mass scale
for the fermionic resonances associated with the top quark, the top partners,
is required to minimize the amount of fine tuning. Moreover in
a large class of minimal scenarios, as for instance the ones with
fermionic resonances in the fundamental representation of $\SO(5)$, a strict relation
exists between the Higgs mass and the mass of the lightest top partners.

The calculable phenomenological models described in the previous sections offer a privileged
framework to explicitly test the validity of the general results. The simplicity of the
multi-site constructions, moreover, allows us to derive explicit formulae for the Higgs
potential which can be used to refine the estimates presented in Chap.~\ref{ch:SILH}.
In the following we will analyze in details the structure of the Higgs potential in
two scenarios, following Refs.~\cite{Matsedonskyi:2012ym,Panico:2012uw}.
The first one is the minimal model we presented in the
previous sections in which the top partners belong to the fundamental representation
of $\SO(5)$. As a second scenario we consider a representative of the minimally tuned
models (see Chap.~\ref{ch:SILH}) in which the right-handed top component is fully composite.

\subsection[The $5+5$ model]{The $\mathbf{5+5}$ model}\label{sec:5+5_model}

The first scenario we consider is the three-site model presented in Sects.~\ref{sec:3-site_gauge} and \ref{sec:metter_sector}.
As we saw, the structure of the fermionic sector of the model is
in large part determined by the choice of the embedding of the elementary fermions
$q_L$ and $t_R$ in incomplete representations of the $\SO(5)$ group. In the model we consider both
elementary states are embedded in the fundamental representation, the ${\mathbf 5}$, thus we
will denote this scenarios as the ``$5 + 5$ model''.

In the general analysis of Chap.~\ref{ch:SILH}
we saw that this type of model belongs to the category of ``doubly-tuned'' scenarios
in which the amount of tuning needed to obtain a realistic configuration is significantly
higher than the naive expectation. The explicit computation of the Higgs potential
presented in this section will be also useful to explicitly verify the origin of the
additional tuning.

\subsubsection{Explicit form of the potential}

The most relevant contribution to the Higgs potential comes from the fer\-mio\-nic resonances
that are coupled to the top quark, namely the ones we included in our explicit multi-site constructions
in Sect.~\ref{sec:metter_sector}. The spectrum of the top partners in the three-site model contains two levels of resonances
with the same quantum numbers. The $\psi$ and $\widetilde \psi$ resonances transform in the fundamental
representation of $\SO(5)$ and have $\U(1)_X$ charge $2/3$. Under the unbroken $\SO(4) \times \U(1)_X$ group
they can be decomposed into a fouplet, in the $({\bf 2}, {\bf 2})_{2/3}$ representation,
and a singlet, in the $({\bf 1}, {\bf 1})_{2/3}$. Each fourplet contains two $\SU(2)_L$ doublets.
The one with $T^3_R = -1/2$ gives rise to a the $\{T, B\}$ multiplet with the same SM quantum numbers of the elementary $q_L$.
The remaining doublet, $\{X_{5/3}, X_{2/3}\}$, transforms in the
${\bf 2}_{7/6}$ representation under the $\SU(2)_L \times \U(1)_Y$ SM group and contains the exotic resonances
$X_{5/3}$ and $X_{2/3}$ with electric charges $5/3$ and $2/3$ respectively. Finally, the singlets
give rise to resonances with the same quantum numbers of the elementary $t_R$, which we denote by $\widetilde T$.
The spectrum of the resonances and their properties will be discussed in details in Chap.~\ref{ch:LHC_pheno}.

The Higgs potential can be
straightforwardly computed by using the standard Coleman--Weinberg formula
\begin{equation}\label{eq:Coleman-Weinberg}
V(H) = - \frac{2 N_c}{8 \pi^2} \int dp\, p^3 \log\left[\prod_i \left(p^2
+ m_i^2(H)\right)\right]\,,
\end{equation}
where $N_c = 3$ denotes the number of QCD colors and the product inside the logarithm argument
extends to all masses of the fermionic states $m_i(H)$ written as a function
of the Higgs VEV. A simple trick to compute the expression
inside the logarithm is to express it as a determinant of the complete mass matrix
of the fermions $M$
\begin{equation}
\prod_i \left(p^2 + m_i^2(H)\right) = \det\left[p^2 \mathbbm{1} + M^T M\right]\,.
\end{equation}
In order to remove the divergent Higgs-independent part one can subtract from the potential
in Eq.~(\ref{eq:Coleman-Weinberg}) its value at $H=0$.

The only relevant fermionic states that are coupled to the Higgs are the top and the resonances
of charge $2/3$. The contribution of these states to the potential has the form
\begin{eqnarray}
V(H) &=& - \frac{2 N_c}{8 \pi^2} \int dp\; p^3
\log\left[1 - \frac{F_1(p^2) + F_2(p^2)}{D(p^2)}\sin^2\left(\frac{H}{f}\right)\right.\nonumber\\
&& \hspace{8.5em}\left. +\; \frac{F_2(p^2)}{D(p^2)} \sin^4\left(\frac{H}{f}\right)\right]\,,\qquad
\label{eq:Higgs_pot_5+5}
\end{eqnarray}
where the $F_{1,2}(p^2)$ form factors are functions of the integration variable $p$ and
of the fermion mass parameters but not of the Higgs. The whole dependence on $h$ is encoded
in the trigonometric functions that appear inside the logarithm.
The factors $D(p^2)$ appearing in the denominator of the argument of the
logarithm is given by
\begin{equation}\label{eq:den_Higgs_pot_5+5}
D(p^2) = 2 p^2 \prod_{I = T, \widetilde T} (p^2 + m_{I_-}^2)
(p^2 + m_{I_+}^2)\,,
\end{equation}
where $m_{I\pm}$ denote the masses of the charge $2/3$ resonances before EWSB.
The $\pm$ sign refers to the two levels of composite resonances that are present in the
three-site set-up. Notice that all these masses include the shift due to the mixing with the
elementary states. The initial factor $p^2$ that appears in Eq.~(\ref{eq:den_Higgs_pot_5+5})
is due to the presence of the top which is massless before EWSB.
Finally the form factors $F_{1,2}(p^2)$ are given by
\begin{equation}\label{eq:F12_5+5}
\left\{
\begin{array}{l}
F_1(p^2) = (\widetilde m_Q - \widetilde m_T)\Delta^2 C_1(p^2)\left[(y_L^2 - 2 y_R^2)f^2 -
y_L^2 y_R^2 f^4 C_2(p^2)\right]\\
\rule{0pt}{1.8em}F_2(p^2) = - (\widetilde m_Q - \widetilde m_T)^2 \Delta^4 y_L^2 y_R^2 f^4
\end{array}
\right.\,.
\end{equation}
The $C_{1,2}(p^2)$ functions will not be needed for our analysis, however we report them here for completeness:
\begin{equation}
\left\{
\begin{array}{l}
C_1(p^2) = p^2 \left[(m + \widetilde m_T)(p^2 + \Delta^2 - m\, \widetilde m_Q)
+ (\widetilde m_Q \leftrightarrow \widetilde m_T)\right]\\
\rule{0pt}{1.75em}C_2(p^2) = (p^2 + \Delta^2 + \widetilde m_Q^2)/\left[(p^2 + m^2_{X_{2/3-}})
(p^2 + m^2_{X_{2/3+}})\right]
\end{array}
\right.\,.
\end{equation}

The potential can be approximated by expanding at leading order
the logarithm in Eq.~(\ref{eq:Higgs_pot_5+5}).
This approximation is sufficiently accurate if we are interested in phenomenologically
viable scenarios, in which a gap is needed between the Higgs VEV and the Godstone decay
constant: $\langle H \rangle/f \simeq v/f \ll 1$.\footnote{The expansion is not
valid in the limit $p \rightarrow 0$, in which the argument of the logarithm diverges. However in this
limit the factor $p^3$ in front of the logarithm compensate for the divergence and the
approximate integrand vanishes for $p \rightarrow 0$. The error introduced by this approximation
is thus small. As discussed in Sect.~\ref{pwcpot_sec}, the presence of the divergence is related to the
IR contribution to the Coleman--Weinberg potential coming from the top quark. A fully consistent computation
of the potential can be obtained by first isolating the top contribution and then expanding the remaining terms
which are regular for $p \rightarrow 0$.}
As will be clear in the following, in order to minimize the potential and compute the Higgs mass
we need to consider the terms of order $\sin^4(H/f)$ as well as the $\sin^2(H/f)$
ones. This means that, in principle we should also keep into account the second order of the
logarithm expansion in Eq.~(\ref{eq:Higgs_pot_5+5}). However, as we will see below, a sizable cancellation of the
$\sin^2(H/f)$ terms is needed in order to ensure $v/f \ll 1$, thus the second order
of the expansion is always negligible in realistic scenarios.

After the expansion and the integration, the potential takes the generic form
(compare Eq.~(\ref{pot_form}))
\begin{equation}\label{eq:Higgs_pot_5}
V(H) \simeq \alpha\, \sin^2(H/f) + \beta\, \sin^4(H/f)\,.
\end{equation}
By minimizing the potential we find the physical Higgs VEV
\begin{equation}
\xi = \sin^2 \left(\frac{\langle H \rangle}{f}\right)
= -\frac{\alpha}{2 \beta}\,,
\end{equation}
and the value of the Higgs mass
\begin{equation}\label{eq:mH_3-site}
m_H^2 = \frac{2 \beta}{f^2} \sin^2 \left(\frac{2 \langle H \rangle}{f}\right)\,.
\end{equation}
The requirement $v/f \ll 1$ implies that the $\alpha$ coefficient must be much smaller than $\beta$.
However, as can be seen from the explicit expressions of the form factors in Eq.~(\ref{eq:F12_5+5}),
the coefficient $\alpha$ is generated at leading order, ${\mathcal O}(y^2)$,
in the elementary/composite mixings, whereas $\beta$ is generated only at ${\mathcal O}(y^4)$.
This structure leads to the presence of an additional fine tuning in the
$5+5$ models with respect to the naive estimate $\xi = v^2/f^2$ in agreement with the results obtained
in the general analysis of Sect.~\ref{CHPOT} .
A more careful inspection of Eq.~(\ref{eq:F12_5+5}) shows that the leading contribution
to $\alpha$ is proportional to $y_L^2 - 2 y_R^2$. The cancellation thus leads to the
condition
\begin{equation}\label{eq:yL_yR_relation_3-site}
y_L \simeq \sqrt{2} y_R\,,
\end{equation}
{\it{i.e.}}~the left and right mixings of the top must be roughly of the same size.
This relation is very well satisfied numerically for realistic configurations~\cite{Panico:2011pw}.

The value of the coefficient $\beta$ can be easily computed analytically:
\begin{equation}\label{eq:beta_3-site}
\beta = \frac{N_c}{16 \pi^2} (\widetilde m_Q - \widetilde m_T)^2 \Delta^4 y_L^2 y_R^2 f^4
\!\!\!\!\sum_{\begin{array}{r@{\,}c@{\,}l}\scriptstyle I &\scriptstyle =&\scriptstyle T_-, T_+,\\
&&\scriptstyle \widetilde T_-, \widetilde T_+\end{array}}
\frac{\log(m_I^2/f^2)}{\prod_{I\neq J} (m_I^2 - m_J^2)}\,.
\end{equation}
In the limit in which the second level of resonances is much heavier that the first one,
we can use an expansion in the ratio of the heavy and light states masses and get a simple
approximate formula for $\beta$:
\begin{equation}\label{eq:beta_app_3-site}
\beta \simeq \frac{N_c}{16 \pi^2} (\widetilde m_Q - \widetilde m_T)^2 \Delta^4 y_L^2 y_R^2 f^4
\frac{\log\left(m_{T_-}^2/m_{\widetilde T_-}^2\right)}{\left(m_{T_-}^2 - m_{\widetilde T_-}^2\right)
m_{T_+}^2 m_{\widetilde T_+}^2}\,.
\end{equation}
As can be seen from the above formula, when one of the states $T_-$ and $\widetilde T_-$
is much lighter than the other, the contribution to $\beta$ coming from the first level
of resonances is enhanced by the logarithmic factor $\log\big(m_{T_-}^2/m_{\widetilde T_-}^2\big)$.
In this case the contribution from the light states completely dominates
and the corrections due to the second layer of resonances become negligible.
On the other hand, if the two light states have comparable masses, the second level of resonances,
in certain regions of the parameter space, can be relatively close in mass to the first one,
thus giving sizable corrections to the Higgs mass. The sign of these corrections is fixed
and they always determine a decrease of the Higgs mass. The size of the corrections is typically
below $50\%$.\footnote{Additional subleading corrections to the Higgs mass can also come from loops of vector partners. In particular the two-loop contribution due to gluon partners can be non-negligible in some regions of the parameter space~\cite{Barnard:2013hka}.}

\subsubsection{Light top partners for a light Higgs}

For a quantitative estimate of $m_H$ we need to determine the size of the $y_{L,R}$ mixings
that appear in the expression for $\beta$.
These mixings control the generation of the top Yukawa, thus we can relate them with
the top mass $m_t$.
An approximate expression for $m_t$ can be derived by using an expansion in $\sin^2(H/f)$.
The result can be conveniently rewritten in terms of the $y_{L,R}$ mixings and of the
masses of the $T$ and $\widetilde T$ resonances:
\begin{equation}
m_t^2 \simeq \frac{(\widetilde m_Q - \widetilde m_T)^2}{8} \frac{y_L^2 y_R^2 f^4 \Delta^4}
{m_{T_+}^2 m_{T_-}^2 m_{\widetilde T_+}^2 m_{\widetilde T_+}^2}
\sin^2 \left(\frac{2 H}{f}\right)\,.
\end{equation}
By comparing this expression with the approximate formula for the Higgs mass
in Eqs.~(\ref{eq:mH_3-site}) and (\ref{eq:beta_app_3-site}) we find a remarkable relation between
$m_H$ and the masses of the lightest $T$ and $\widetilde T$ resonances:
\begin{equation}\label{eq:mH_mt_3-site}
\frac{m_H^2}{m_t^2} \simeq \frac{N_c}{\pi^2} \frac{m_{T_-}^2 m_{\widetilde T_-}^2}{f^2}
\frac{\log\left(m_{T_-}^2/m_{\widetilde T_-}^2\right)}{m_{T_-}^2 - m_{\widetilde T_-}^2}\,.
\end{equation}

We can now compare the relation in Eq.~(\ref{eq:mH_mt_3-site}) with the general results
obtained in Chap.~\ref{ch:SILH}. As can be seen from Eq.~(\ref{mh5p5}), in the $5+5$ models
the Higgs mass can be estimated as
\begin{equation}
m_H^2 \simeq \frac{N_c}{2 \pi^2} \lambda_t^4 v^2\,,
\end{equation}
where $\lambda_t$ denotes any of the $y_{L,R}$. The $y_{L,R}$ mixings can be related to the top mass.
In particular if $y_L \sim y_R$ one can extract the relation
\begin{equation}\label{eq:est_ytop}
y_{top} \simeq y_L y_R\, m_\psi/f\,,
\end{equation}
where $m_\psi$ denotes the mass of the lightest state mixed with the top, $m_\psi = \min(m_{T_-}, m_{\widetilde T_-})$.
Inserting the result in Eq.~(\ref{eq:est_ytop}) into the estimate for the Higgs mass we get
\begin{equation}
\frac{m_H^2}{m_t^2} \simeq \frac{N_c}{\pi^2} \frac{m_\psi^2}{f^2}\,.
\end{equation}
This equation qualitatively reproduces the
relation between the Higgs mass and the masses of the lightest resonances $T$ and $\widetilde T$
found in Eq.~(\ref{eq:mH_mt_3-site}).
In the case $m_\psi^2 = m_{T_-}^2 = m_{\widetilde T_-}^2$ the two expressions exactly coincide,
while, when a large hierarchy between the two light states is present, they differ by a
factor of ${\mathcal O}(1)$. This shows that the general analysis of Chap.~\ref{ch:SILH}
correctly captures the main connection between the Higgs and the top partners masses,
both at a qualitative and a quantitative level.

\begin{figure}
\centering
\includegraphics[width=.47\textwidth]{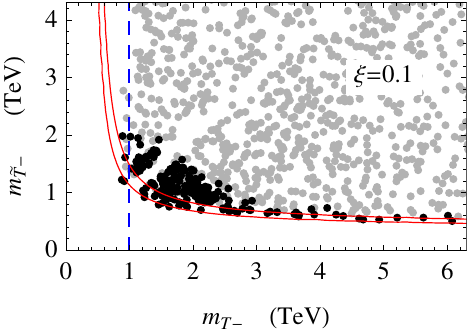}
\hfill
\raisebox{-.175em}{\includegraphics[width=.47\textwidth]{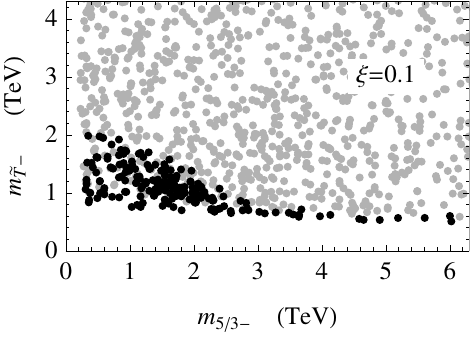}}
\caption{Scatter plot of the masses of the lightest $T$ and $\widetilde T$ resonances
(left panel) and of the $X_{5/3}$ and $\widetilde T$ resonances (right panel) in the
three-site model. The compositeness scale has been fixed to $\xi = 0.1$.
The black dots denote the points for which $115\ \mathrm{GeV} \leq m_H \leq 130\ \mathrm{GeV}$,
while the gray ones have $m_H > 130\ \mathrm{GeV}$. In the left plot the area between
the solid red lines represents the range obtained by applying the result
in Eq.~(\ref{eq:mH_mt_3-site}) for $115\ \mathrm{GeV} \leq m_H \leq 130\ \mathrm{GeV}$.
The dashed blue line corresponds to a lower bound on $m_{T_-}$ due to a saturation effect.
The plots are taken from Ref.~\cite{Matsedonskyi:2012ym}.}
\label{fig:mH_3-site_xi01}
\end{figure}

The scatter plot of the masses of the $T$ and $\widetilde T$ light resonances can be used
to check the validity of Eq.~(\ref{eq:mH_mt_3-site}). The numerical results obtained
in Ref.~\cite{Matsedonskyi:2012ym} for $\xi = 0.1$ are shown in the left panel
of Fig.~\ref{fig:mH_3-site_xi01}. As expected Eq.~(\ref{eq:mH_mt_3-site}) describes accurately
the relation between the Higgs and the resonances masses in the regions in which one state is
significantly lighter than the others. On the other hand, when the $T_-$ and $\widetilde T_-$
masses are comparable sizable deviations from Eq.~(\ref{eq:mH_mt_3-site}) can occur. These
are due to the possible presence of a relatively light second level of resonances, as already discussed.

The numerical results clearly show that resonances with a mass of the order
or below $1.5\ \mathrm{TeV}$ are needed in order to get a realistic Higgs mass in the case
$\xi = 0.1$. The upper bound becomes $1\ \mathrm{TeV}$ for $\xi = 0.2$. The prediction is
even sharper in the case in which only one state, the $\widetilde T_-$ is light. In this
region of the parameter space states with masses around $600\ \mathrm{GeV}$ are needed
for $\xi = 0.1$ and around $400\ \mathrm{GeV}$ for $\xi = 0.2$.

The result becomes even more stringent if we also take into account the masses of the
other composite resonances. As we discussed before, the first level of resonances contains, in addition
to $T_-$ and $\widetilde T_-$, three other states: a top-like state, the $X_{2/3-}$,
a bottom-like state, the $B_-$, and an exotic state with charge $5/3$, the $X_{5/3-}$.
the $X_{5/3-}$ can not mix with any other state, even after EWSB, and therefore it
remains always lighter than the other particles in the fourplet.\footnote{Some approximate
expressions for the masses of the resonances can be found in Ref.~\cite{Panico:2011pw}.
For a discussion of the details of the spectrum see also Chap.~\ref{ch:LHC_pheno}.}
The scatter plot for the masses of the $X_{5/3-}$ and $\widetilde T_-$ states
is shown in the right panel of Fig.~\ref{fig:mH_3-site_xi01}. In the parameter space regions
with realistic Higgs mass, the $X_{5/3-}$ resonance can be much lighter than the other states,
especially in the configurations in which the $T_-$ and $\widetilde T_-$ have comparable masses.
In these points the mass of the exotic state can be as low as $300\ \mathrm{GeV}$.

\subsubsection{Calculability in the two-site model}

We saw in Sect.~\ref{sec:multi-site} that the three-site model
provides an effective description of a composite Higgs in which the Higgs potential is
calculable at one-loop order. This property allowed us to decouple the UV physics and fully
characterize the model in terms of the parameters describing the elementary states and a
small set of composite resonances.

If we accept to give up a complete predictivity, however, the much simpler two-site construction
can be used to describe the low-energy dynamics of a composite Higgs and of just one level
of composite partners. As we saw, in this set-up the Higgs potential becomes logarithmically divergent at one loop.
There is however an interesting property which partially preserves predictivity.
For simplicity let us focus on the fermionic contribution from the top partners,
which dominates the Higgs potential. As shown in Sect.~\ref{sec:metter_sector} only the
leading terms in the expansion in powers of the elementary/composite mixings can develop a
logarithmic divergence. These terms have a fixed dependence on the Higgs VEV, namely they
only depend on $\sin^2(H/f)$ (see Eqs.~(\ref{eq:ferm_div_2-site_1})
and (\ref{eq:ferm_div_2-site_2})). As a consequence, to regulate the divergence only one counterterm is needed,
which corresponds to the renormalization of a single parameter. An interesting possibility is
to fix the value of the physical Higgs VEV, or more precisely the $v/f$ ratio,
as a renormalization condition, obtaining the Higgs mass as a prediction of the model.
In this sense $m_H$ is predictable also in the two-site set-up.

We will now explain a possible procedure to deal with the logarithmic divergence and compute
the Higgs potential in the two-site model. The simplest way to regulate the Coleman--Weinberg
potential in Eq.~(\ref{eq:Coleman-Weinberg}) is to introduce a hard cut-off $\Lambda$. With this
prescription we obtain the standard formula
\begin{equation}
V(H) = - \frac{N_c}{8 \pi^2} \Lambda^2 \sum_i m_i^2(H)
- \frac{N_c}{16 \pi^2} \sum_i m_i^4(H) \left[\log\left(\frac{m_i^2(H)}{\Lambda^2}\right)
-\frac{1}{2}\right]\,.
\end{equation}
In the two site model only a logarithmic divergence appears in the Higgs potential and
therefore the quadratically divergent term must be independent of the Higgs VEV.
This is ensured by the condition
\begin{equation}
\sum_i m_i^2(H) = \sum_i m_i^2(H = 0) =  \mathrm{const.}\,,
\end{equation}
which holds in the multi-site constructions as can be explicitly
verified.\footnote{If, as in the three-site case, the Higgs potential
is completely finite at one loop, an analogous condition holds for the logarithmic term,
i.e. $\sum_i m_i^4(H) = \sum_i m_i^4(H = 0) = \mathrm{const.}$.}
As discussed above, the logarithmic divergence is proportional to $\sin^2(h/f)$,
and this implies the relation 
\begin{equation}
\sum_i m_i^4(H) \propto \sin^2(H/f) + \mathrm{const.}\,.
\end{equation}
We can therefore cancel the divergence by introducing a single counterterm proportional to
\begin{equation}
\sum_i m_i^4(H) \left[\log\left(\frac{\Lambda^2}{\mu^2}\right) - \frac{1}{2}\right]\,,
\end{equation}
which depends on one free renormalization parameter, namely the scale $\mu$.
The renormalized potential thus takes the form
\begin{equation}
V(H) = -\frac{N_c}{16 \pi^2} \sum_i m_i^4(H) \log\left(\frac{m_i^2(H)}{\mu^2}\right)\,.
\end{equation}
To compute the Higgs potential we then need to choose a renormalization condition that fixes
the parameter $\mu$. A convenient choice is to fix the position of the minimum of the potential
to the required value of $v/f$. In this way the potential is completely determined and can be used to extract
a prediction for the Higgs mass.

It is interesting to notice that, as shown in Ref.~\cite{Matsedonskyi:2012ym}, the approximate
relation in Eq.~(\ref{eq:mH_mt_3-site}) between the Higgs mass and the masses of the top
partners is valid also in the two-site model. Of course in this case one needs to identify the
$T_-$ and $\widetilde T_-$ states with the only level of resonances included in the model.

\subsection[The $14+1$ model]{The ${\mathbf{14+1}}$ model}\label{sec:14+1_model}
\index{fully composite $t_R$}

The second scenario we consider is a representative of the class of minimally tuned models.
This class of models is characterized by the fact that the right-handed top component
is fully composite and is identified with one chiral resonance coming from
the composite dynamics. In order to implement this assumption we need to slightly modify
the fermionic sector of three-site construction as we will show in the following.

As an explicit example we will consider the case in which the elementary $q_L$ doublet is embedded
in the $\mathbf{14}$ representation of $\SO(5)$. The $t_R$, instead, is a total singlet. Under
the $\U(1)_X$ group the $q_L$ and $t_R$ fields as well as the composite top partners have charge $2/3$.\footnote{The reader is referred to Ref.~\cite{Pappadopulo:2013vca} for an implementation of the same setup in the framework of 5-d holographic models.}

\subsubsection{The structure of the model}

The global symmetry structure of the model is exactly equal to the one of the three-site construction
and the gauge sector coincides with the one described in Sect.~\ref{sec:3-site_gauge}.
The fermionic sector requires instead a few changes. First of all we embed the elementary
doublet $q_L$ in the $\mathbf{14}$ representation of $\SO(5)$. A suitable basis for this
representation is given by symmetric traceless $5 \times 5$ matrices.
Under the $\SO(4) \simeq \SU(2)_L \times \SU(2)_R$ subgroup they decompose as
$\mathbf{14} = \mathbf{9} \oplus \mathbf{4} \oplus \mathbf{1} = (\mathbf{3}, \mathbf{3}) \oplus
(\mathbf{2}, \mathbf{2}) \oplus (\mathbf{1}, \mathbf{1})$. The explicit form of the $\SO(4)$ multiplets is
\begin{equation}
\begin{array}{r@{\qquad}l}
(\mathbf{3}, \mathbf{3}): &
\left\{
\begin{array}{l@{\qquad}l}
M^{aa}_{ij} = \displaystyle \delta^a_i \delta^a_j + \delta^4_i \delta^4_j - \frac{1}{2} \delta^i_j\,, & a = 1,2,3\\
\rule{0pt}{1.75em}M^{ab}_{ij} = \displaystyle \frac{1}{\sqrt{2}}\left(\delta^a_i \delta^b_j + \delta^a_j \delta^b_i\right)\,,
& a < b\,, \quad a,b = 1, \ldots,4
\end{array}
\right.\\
\rule{0pt}{1.75em}(\mathbf{2}, \mathbf{2}): & \hspace{1.25em}  M^a_{ij} = \displaystyle \frac{1}{\sqrt{2}} \left(\delta^a_i \delta^5_j + \delta^a_j \delta^5_i\right)\,,
\qquad a = 1, \ldots, 4\\
\rule{0pt}{1.75em}(\mathbf{1}, \mathbf{1}): & \hspace{1.25em} M^0 = \displaystyle \frac{1}{2\sqrt{5}} \mathrm{diag} \left(1, 1, 1, 1, -4\right)\,,
\end{array}
\end{equation}
The elementary $q_L$ is embedded in the $(\mathbf{2}, \mathbf{2})$ multiplet, in particular in the
doublet with $T^3_R = -1/2$, analogously to what we did in the case of the representation
$\mathbf{5}$ (the explicit form of the embedding $Q_L$ is given in Eq.~(\ref{14source})).
The $q_L$ multiplet is associated to the leftmost site and formally transforms under the $SO(5)_L^1$ symmetry.

\begin{figure}
\centering
\includegraphics[width=.6\textwidth]{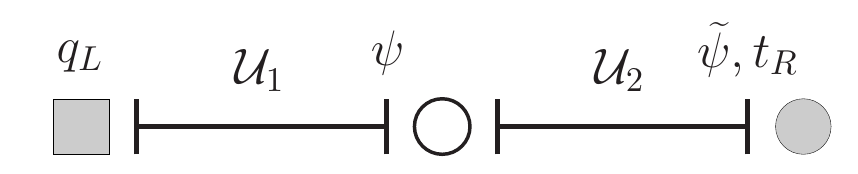}
\caption{Pictorial representation of the matter sector of the three-site $14+1$ model.}
\label{fig:3-site_14+1_moose}
\end{figure}

The composite sector now contains the $\psi$ and $\widetilde \psi$ fields in the $\mathbf{14}$ representation
and the $t_R$ as an $\SO(5)$ total singlet. The $\psi$ field is associated to the middle site and transforms under
the $\SO(5)_R^1$ symmetry. The $\widetilde \psi$ field belongs to the rightmost site and transforms
under the $\SO(5)_R^2$ subgroup. The $t_R$ field, being now composite, is associated to the rightmost
site as well. The moose representation of the fermionic sector of the model is shown in Fig.~\ref{fig:3-site_14+1_moose}.
The Lagrangian for the composite states reads
\begin{eqnarray}
{\mathcal L}_{cs} &=& \Tr\left[i \overline \psi \gamma^\mu D_\mu \psi - m \overline \psi \psi\right]\nonumber\\
&& +\; \Tr[i \overline {\widetilde \psi} \gamma^\mu D_\mu \widetilde \psi
- \widetilde m_\Phi \overline {\widetilde \Phi} \Psi - \widetilde m_Q \overline {\widetilde Q} \widetilde Q
- \widetilde m_T \overline {\widetilde T} \widetilde T]\nonumber\\
&& +\; i \overline t_R \gamma^\mu D_\mu t_R - \Tr[m_R \overline T_R \widetilde T_L + \mathrm{h.c.}]\nonumber\\
&& -\; \Delta \Tr[{\mathcal U}_2^T\, \overline \psi\, {\mathcal U}_2\, \widetilde \psi] + \mathrm{h.c.}\,,
\end{eqnarray}
where we denoted by $\widetilde \Phi$, $\widetilde Q$ and $\widetilde T$ respectively the
$\mathbf{9}$, $\mathbf{4}$ and $\mathbf{1}$ components of the $\widetilde \psi$ multiplet,
while $T_R$ denotes the embedding of the $t_R$ field into the singlet component of the
$\mathbf{14}$ representation.
The Lagrangian for the elementary states, including the elementary/composite mixing, is given by
\begin{equation}
{\mathcal L}_{el + mix}  = i \overline q_L \gamma^\mu D_\mu q_L - y_L f \Tr[{\mathcal U}_1^T\, \overline Q_L {\mathcal U}_1 \psi]
+ \mathrm{h.c.}\,.
\end{equation}

\subsubsection{Phenomenological properties}

To understand the phenomenological properties of the $14 + 1$ model it is useful to start from a comparison
with the $5+5$ set-up. As we discussed at length, in the latter model there is a tight connection between the Higgs mass
and the spectrum of the light top partners. In the previous sections we derived this result by a careful
analytic approach, there is however a simpler, although less rigorous, way to understand this connection
which can be easily extended to other models.

Let us start from the simplest description of the $5+5$ model, namely the non-linear $\sigma$-model
$\SO(5)/\SO(4)$ plus the elementary bosonic and fermionic sources. As shown in Sect.~\ref{CHPCH},
in this set-up only one independent operator contributes
to the Higgs potential at leading order in the elementary/composite mixings.
This operator is quadratically divergent at one loop, however it has a fixed dependence on $H/f$, namely $\sin^2(H/f)$,
thus it introduces a strong sensitivity to the UV dynamics only in the $\alpha$ coefficient in Eq.~(\ref{eq:Higgs_pot_5}).
On the contrary, the operators contributing to the $\sin^4(H/f)$ terms, which control the Higgs quartic coupling
and ultimately the Higgs mass, start to be generated at quartic order in the mixings and are only logarithmically divergent.
As a consequence the $\beta$ coefficient in the Higgs potential has a very mild dependence on the UV dynamics,
{\it{i.e.}}~on the details of the top partners spectrum. The correlation between the $\beta$ coefficient and the
mass of the ligtest top partners is not generated directly, instead it comes indirectly from the
size of the elementary/composite mixings which is fixed by the top mass.
At fixed $y_{top}$, indeed, the $y_{L,R}$ mixings are related to the mass of the lightest top partners as shown
in Eq.~(\ref{eq:est_ytop}) and this determines the tight connection between the Higgs mass and the top partners spectrum.
This peculiar structure explains the robustness of this connection
and its independence from the details of the explicit models.

The situation is totally different for the $14+1$ model. The $\mathbf{14}$
representation decomposes into three $\SO(4)$ multiplets and gives rise to two
independent invariants at the leading order in the elementary/composite mixings.
The two operators have a different dependence on $H/f$, thus their quadratic divergence at one loop
implies a strong dependence of the Higgs potential on the details of the top partners spectrum.
In particular the Higgs mass will not be any more determined only by the lightest top partners, but instead it
will depend on a larger set of resonances.

As shown in Chap.~\ref{ch:SILH}, the power counting estimates can be used to derive a relation (Eq.~(\ref{mh14p1})) between
the Higgs mass and the overall fermion mass scale $m_\psi = g_\psi f$
\begin{equation}\label{eq:mH_gpsi_14+1}
m_H \simeq \sqrt{\frac{N_c}{2 \pi^2}} y_t g_\psi v \simeq 500\ \mathrm{GeV}
\left(\frac{g_\psi}{5}\right)\,,
\end{equation}
where $y_t$ is the top Yukawa. It is important to stress that in the above formula $m_\psi$ does not denote
the mass of the lightest top parters, instead it must be interpreted as an ``average'' mass of the first
level of fermionic resonances.
Moreover, as clear from the above discussion, in the $14+1$ model large departures from the power-counting
estimates are possible. For instance, this can happen if accidental cancellations are present in the Higgs potential.

\begin{figure}
\centering
\includegraphics[width=.47\textwidth]{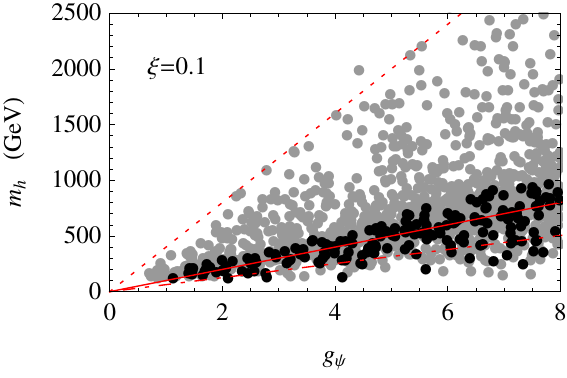}
\hfill
\raisebox{-.175em}{\includegraphics[width=.458\textwidth]{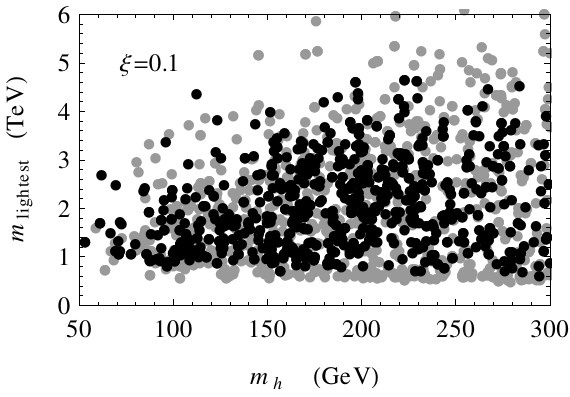}}
\caption{Left panel: scatter plot of the Higgs mass as a function of $g_\psi$ for $\xi = 0.1$
in the $14+1$ three-site model. Right panel: scatter plot of the mass of the lightest
fermionic resonance as a function of the Higgs mass. The red lines show the
estimate of the Higgs mass with $y_L = y_t$ (solid), $y_L = \sqrt{2/5} y_t$ (dot-dashed)
and $y_L = 4 y_t$ (dotted). The choice $y_L = 4 y_t$ corresponds to the largest value
for $y_L$ used in the scan.
The black dots correspond to the points with $y_L \leq y_t$, while the gray ones have $y_L > y_t$.
The plots are taken from Ref.~\cite{Panico:2012uw}.}
\label{fig:mH_3-site_14+1_xi01}
\end{figure}

The relation between the Higgs mass and the fermionic mass scale is reasonably well verified
in the explicit three-site model, although a significant amount of spread is present (see
left panel of Fig.~\ref{fig:mH_3-site_14+1_xi01})~\cite{Panico:2012uw}.\footnote{The mass scale
of the resonances $m_\psi$ has been identified in the scan with the quadratic
average of the fermion mass parameters present in the composite sector Lagrangian.}
The origin of the spread is mainly due to the relation between the top mass and the
elementary/composite mixing $y_L$. In a model with fully composite $t_R$ one generically expects
$y_L \simeq y_t$, however order-one corrections can be present. By using an approximate
analytic expression for the top mass one can find a lower bound on the value of $y_L$ needed to reproduce
the correct $m_t$
\begin{equation}
y_L \gtrsim \sqrt{\frac{2}{5}} y_t \simeq 0.6\,.
\end{equation}
Although the above inequality can be saturated, in a large part of the parameter space
some cancellation occurs and a value of $y_L$ significantly larger than the minimal one is required.
The spread on $y_L$ determines a corresponding spread in the relation between the Higgs mass
and the fermion scale $g_\psi$. If this effect is taken into account\footnote{For this purpose it
is sufficient to replace $y_t$ with $y_L$ in Eq.~(\ref{eq:mH_gpsi_14+1}).}
the agreement between the general estimate and the numerical results becomes quite good.

From the general estimates and the numerical results it is easy to see that there are only two
possibilities to get a realistic Higgs mass: considering the region of the parameter space in which
all the fermionic resonances are light ($g_\psi \lesssim 2$), or allow some extra tuning which
cancels the overall size of the effective potential. The amount of tuning required has been estimated
in Chap.~\ref{ch:SILH}.

As a final point we discuss the connection between the Higgs mass and the presence of light top partners.
As we discussed before, in the $14+1$ model a strict relation between the Higgs mass and the masses
of the lightest fermionic resonances does not exist. Given that many resonances determine the
Higgs potential it is possible to obtain a cancellation in the Higgs mass, by means of tuning,
even if all the resonances are heavy. This can be seen explicitly in the right panel of
Fig.~\ref{fig:mH_3-site_14+1_xi01}, where a scatter plot for the mass of the lightest
fermionic resonance is shown as a function of the Higgs mass. For a realistic Higgs mass
the resonances can be much heavier than the typical masses required in the $5+5$ model,
$m_{lightest} \lesssim 1.5\ \mathrm{TeV}$ for $\xi = 0.1$ (see Fig.~\ref{fig:mH_3-site_xi01}).
The plot shows nevetheless a preference for light states, obviously due to the lower
degree of tuning associated to a smaller value of $g_\psi$. The numerical result shows that,
at the price of tuning, resonances as heavy as $4\ \mathrm{TeV}$ can be obtained with a light Higgs.


\section{The Weinberg sum rules}\label{sec:Weinberg_sum_rules}

In the previous sections we addressed in a constructive manner the problem of finding
a calculable effective description of the composite Higgs scenarios. Our approach was to
add to the simple non-linear $\sigma$-model some extra symmetries which could protect
the Higgs potential thanks to a collective breaking mechanism. In this section we want to
consider the problem from a different perspective, namely we want to understand what
kind of information we can get about an effective description of the composite scenario
by requiring that the Higgs potential is calculable without introducing
by hand any additional symmetry. As we will see, the requirement of calculability
is strong enough to imply some stringent constraints of the structure of the effective theory.
In particular, at least in the simplest realization of the composite Higgs scenario,
a structure analogous to a multi-site model is always needed to ensure the finiteness
of the Higgs potential.

\subsection{The general effective Lagrangian}

For simplicity in our analysis we will only focus on the femionic part of the theory
and in particular on the states belonging to the top sector. For definiteness, we will also assume that
the elementary fermions $q_L$ and $t_R$ are embedded in (incomplete) multiplets in the
fundamental representation of $\SO(5)$. With this choice the elementary fermions can only
mix directly to composite resonances that transform as fourplets $\psi_4^i$ or singlets
$\psi_1^i$ under the unbroken $\SO(4)$ symmetry.
We can now construct the most general effective Lagrangian by using the CCWZ formalism.
The kinetic terms for the elementary and composite states are given by
\begin{eqnarray}
{\mathcal L}_{kin} &=& i \overline q_L \gamma^\mu D_\mu q_L + i \overline t_R \gamma^\mu D_\mu t_R\nonumber\\
&&+\; \sum_{i=1}^{N_S} \overline \psi_1^i (i \gamma^\mu D_\mu - m_{1i}) \psi_1^i
+ \sum_{i=1}^{N_Q} \overline \psi_4^i (i \gamma^\mu D_\mu - m_{4i}) \psi_4^i\,.\qquad
\end{eqnarray}
For simplicity, in writing the above formula, we chose to work in the basis in which the mass matrices for
the composite states are diagonal. $N_Q$ and $N_S$ denote the number of fourplets and singlets
included in the theory.
The most general elementary/composite mixing terms, following the partial compositeness
assumption, are given by
\begin{eqnarray}\label{eq:Wein_mix_Lagr}
{\mathcal L}_{mix} &=& \sum_{i=1}^{N_S} \left[y_{R1}^i f \left(\overline T_R U\right)_1 \psi_1^i
+ y_{R4}^i f \left(\overline T_R U\right)_4 \psi_4^i\right] + \mathrm{h.c.}\nonumber\\
&&+\; \sum_{i=1}^{N_Q} \left[y_{L1}^i f \left(\overline Q_L U\right)_1 \psi_1^i
+ y_{L4}^i f \left(\overline Q_L U\right)_4 \psi_4^i\right] + \mathrm{h.c.}\,,\quad
\end{eqnarray}
where $Q_L$ and $T_R$ denote the embedding of the elementary states
into incomplete $\SO(5)$ multiplets (the explicit expressions are given in Eq.~(\ref{eq:elem_ferm_embedding_5})) and $U$ is the Goldstone matrix.
The $4$ and $1$ subscripts in the expressions $(\overline Q_L U)_{4,1}$
and $(\overline T_R U)_{4,1}$ denote respectively the fourplet and singlet component.

\subsection{The Higgs potential and the Weinberg sum rules}
\index{Weinberg sum rules}

As a preliminary step for the computation of the fermion contribution to the Higgs potential
it is useful to derive the effective Lagrangian for the elementary top quark.
This can be done by integrating out the composite states $\psi_4^i$ and $\psi_1^i$.
The final result written in momentum space can be expressed as
\begin{equation}
{\mathcal L}_{eff} = \overline t_L \pslash\, \Pi_L t_L + \overline t_R \pslash\, \Pi_R t_R
- \left(\overline t_L \Pi_{LR} t_R + \mathrm{h.c.}\right)\,.
\end{equation}
The form factors are given by
\begin{equation}
\left\{
\begin{array}{l}
\Pi_L = \Pi_{L}^0 + \sin^2(H/f) \Pi_{L}^1\\
\rule{0pt}{1.25em}\Pi_R = \Pi_{R}^0 + \sin^2(H/f) \Pi_{R}^1\\
\rule{0pt}{1.25em}\Pi_{LR} = \sin(H/f) \cos(H/f) \Pi_{LR}^0
\end{array}
\right.\,,
\end{equation}
where
\begin{eqnarray}
&&\Pi_{L}^0 = 1 - f^2 \sum_{i=1}^{N_Q} \frac{|y^i_{L4}|^2}{p^2 - m_{4i}^2}\,,
\ \Pi_{L}^1 = -\frac{f^2}{2} \left(\sum_{i=1}^{N_S} \frac{|y^i_{L1}|^2}{p^2 - m_{1i}^2}
- \sum_{i=1}^{N_Q} \frac{|y^i_{L4}|^2}{p^2 - m_{4i}^2}\right)\,,\nonumber\\
&&\Pi_{R}^0 = 1 - f^2 \sum_{i=1}^{N_S} \frac{|y^i_{R1}|^2}{p^2 - m_{1i}^2}\,,
\ \Pi_{R}^1 = f^2 \left(\sum_{i=1}^{N_S} \frac{|y^i_{R1}|^2}{p^2 - m_{1i}^2}
- \sum_{i=1}^{N_Q} \frac{|y^i_{R4}|^2}{p^2 - m_{4i}^2}\right)\,,\nonumber\\
&&\Pi_{LR}^0 = \frac{f^2}{\sqrt{2}} \left(\sum_{i=1}^{N_S} y^{i*}_{R1} y_{L1}^i
\frac{m_{1i}}{p^2 - m_{1i}^2} - \sum_{i=1}^{N_Q} y^{i*}_{R4} y_{L4}^i
\frac{m_{4i}}{p^2 - m_{4i}^2}\right)\,.
\end{eqnarray}
The fermion contribution to the Higgs potential at one loop can be expressed in terms
of the form factors that appear in the top effective Lagrangian:
\begin{equation}
V(H) = -2 N_c \int \frac{d^4 p}{(2\pi)^4} \log\left(
p^2 \Pi_L(-p^2) \Pi_R(-p^2) + \left|\Pi_{LR}(-p^2)\right|^2\right)\,,
\end{equation}
where we rotated the integration variable in Euclidean space.

Similarly to what we did in the case of the multi-site models we can expand the
potential in a series in $\sin(H/f)$:
\begin{equation}
V(H) \simeq \alpha \sin^2(H/f) + \beta \sin^4(H/f)\,.
\end{equation}
The $\alpha$ and $\beta$ coefficients can be easily extracted from the expression of the
potential and read \footnote{The integral in the expression for $\beta$ has a spurious
IR divergence arising from the expansion of the potential. It can be cured by
inserting a small IR cut-off on the integration domain. Given that we are interested only
in the UV behavior we will ignore this subtlety.}
\begin{equation}\label{eq:exp_coeff_Weinberg_sum_rules}
\begin{array}{l}
\alpha = \displaystyle -\frac{N_c}{4 \pi^2} \int\! dp\; p^3 \left(\frac{\Pi^1_L}{\Pi^0_L}
+ \frac{\Pi^1_R}{\Pi^0_R} + \frac{(\Pi_{LR}^0)^2}{p^2 \Pi_L^0 \Pi_R^0}\right)\,,\\
\rule{0pt}{2.25em}\beta = \displaystyle \frac{N_c}{8 \pi^2} \int\! dp\; p^3 \left[
\left(\frac{\Pi^1_L}{\Pi^0_L}
+ \frac{\Pi^1_R}{\Pi^0_R} + \frac{(\Pi_{LR}^0)^2}{p^2 \Pi_L^0 \Pi_R^0}\right)^2
- 2\frac{p^2 \Pi_L^1 \Pi_R^1 - (\Pi_{LR}^0)^2}{p^2 \Pi_L^0 \Pi_R^0}\right]\,.
\end{array}
\end{equation}

We can now analyze the behavior of the integrands at large $p^2$ with the aim of determining
the UV divergent contributions. From the explicit expressions it is easy to see that
for large Euclidean momenta $\Pi_{L,R}^0 \sim p^0$, $\Pi_{L,R}^1 \sim p^{-2}$ and
$\Pi_{LR}^0 \sim p^{-2}$. It follows that the terms involving $\Pi_{LR}^0$ are all
finite. Analogously, all terms of order higher than four in the $\sin(H/f)$ expansion
are UV finite. The only divergent contributions come from the $\Pi_{L,R}^1$ form factors.
In particular the $\alpha$ coefficient is quadratically divergent, whereas
$\beta$ is only logarithmically divergent, in agreement with the results of our previous
analyses.

To ensure the calculability of the Higgs potential we can impose a set of sum rules on the integrands in
Eq.~(\ref{eq:exp_coeff_Weinberg_sum_rules})
demanding that $\alpha$ and $\beta$ are finite~\cite{Marzocca:2012zn,Pomarol:2012qf}.
These sum rules are analogous to the Weinberg sum rules in QCD~\cite{Weinberg:1967kj},
which constrain the behavior of the spectral functions of the axial and vector currents by interpreting them
as the result of the exchange of weakly coupled mesonic resonances.
The cancellation of the logarithmic divergence in $\beta$ requires
\begin{equation}\label{eq:Wein_sum_rule_I}
\begin{array}{rcl}
\displaystyle \lim_{p^2 \rightarrow \infty} \frac{-1}{f^2}p^2 \frac{\Pi_R^1(-p^2)}{\Pi_R^0(-p^2)} &=&
\displaystyle \sum_{i=1}^{N_S} \left|y_{R1}^i\right|^2
- \sum_{i=1}^{N_Q} \left|y_{R4}^i\right|^2 = 0\,,\\
\rule{0pt}{2.25em}\displaystyle \lim_{p^2 \rightarrow \infty} \frac{2}{f^2}p^2 \frac{\Pi_L^1(-p^2)}{\Pi_L^0(-p^2)} &=&
\displaystyle \sum_{i=1}^{N_S} \left|y_{L1}^i\right|^2
- \sum_{i=1}^{N_Q} \left|y_{L4}^i\right|^2 = 0\,.
\end{array}
\end{equation}
It is straightforward to check that, if the above conditions are satisfied, the quadratic
divergence in $\alpha$ is automatically canceled. A further condition is needed to cancel the
logarithmic divergence in $\alpha$:
\begin{eqnarray}\label{eq:Wein_sum_rule_II}
&&\lim_{p^2 \rightarrow \infty} \frac{2}{f^2} \left(\frac{\Pi_R^1(-p^2)}{\Pi_{R}^0(-p^2)}
+ \frac{\Pi_L^1(-p^2)}{\Pi_{L}^0(-p^2)}\right)\nonumber\\
&&\qquad = 2 \sum_{i=1}^{N_S} m_{1i}^2 \left(|y_{R1}^i|^2 - |y_{L1}^i|^2\right)
- \sum_{i=1}^{N_Q} m_{4i}^2 \left(|y_{R4}^i|^2 - |y_{L4}^i|^2\right) = 0\,.\qquad \qquad
\end{eqnarray}

The conditions in Eq.~(\ref{eq:Wein_sum_rule_I}) have a very simple interpretation.
The linearity of the mixing between the elementary and the composite states (see
Eq.~(\ref{eq:Wein_mix_Lagr})) implies that each elementary field is only mixed with
one linear combination of composite fourplets and one linear combination
of the singlets. By a field redefinition in the composite sector we can thus
go to a basis in which $y_{L,R}^i = 0$ for $i \geq 2$ and only the $y_{L,R}^1$
mixings are non-vanishing both for the fourplet and singlet components.
Of course in the new basis the mass matrix of the composite states is in general
non-diagonal. The sum rules in Eq.~(\ref{eq:Wein_sum_rule_I}) can now be rewritten as
\begin{equation}
|y_{R1}^1|^2 = |y_{R4}^1|^2
\qquad \textrm{and} \qquad |y_{L1}^1|^2 = |y_{L4}^1|^2\,.
\end{equation}
By a redefinition of the phases of the fields we can always choose
\begin{equation}\label{eq:Wein_cond_I}
y_{R1}^1 = y_{R4}^1 \equiv y_{R}
\qquad \textrm{and} \qquad y_{L1}^1 = y_{L4}^1 \equiv y_{L}\,.
\end{equation}
The Lagrangian with the elementary/composite mixing in the new basis reads
\begin{eqnarray}
{\mathcal L}_{mix} &=& y_{R} f \left(\overline T_R U\right)_1 \psi_{1L}^1
+ y_{R} f \left(\overline T_R U\right)_4 \psi_{4L}^1 + \textrm{h.c.}\nonumber\\
&& +\; y_{L} f \left(\overline Q_L U\right)_1 \psi_{1R}^1
+ y_{L} f \left(\overline Q_L U\right)_4 \psi_{4R}^1 + \textrm{h.c.}\nonumber\\
&=& y_{R} f \overline T_R U \psi_L^1 + y_{L} f \overline Q_L U \psi_R^1 + \textrm{h.c.}\,,
\label{eq:mix_Wein}
\end{eqnarray}
where $\psi_L^1$ is obtained by joining the fourplet and singlet fields
$\psi_{L1}^1$ and $\psi_{R1}^1$ to form a multiplet in the fundamental representation
of $\SO(5)$ and analogously for $\psi_R^1$.
The mixing Lagrangian in Eq.~(\ref{eq:mix_Wein}) exactly coincides with the one we considered
in the two-site model (see Eq.~(\ref{eq:Lagr_ferm_2-site})) once we identify $\psi^1$ with
$\widetilde \psi$ and $U$ with the $\mathcal U$ Goldstone matrix (in the gauge in which only
the Goldstone Higgs modes are present). This result shows that the
only extension of the basic non-linear $\sigma$-model in which the Higgs mass is calculable
necessarily has a two-site structure.

Let us now consider the additional condition in Eq.~(\ref{eq:Wein_sum_rule_II}) which ensure the
cancellation of the logarithmic divergences. Taking into account
the result in Eq.~(\ref{eq:Wein_cond_I}) we get
\begin{equation}\label{eq:Wein_cond_II}
2 \left[\big(M_1^\dagger M_1\big)_{11} - \big(M_4^\dagger M_4\big)_{11}\right] y_R^2
- \left[\big(M_1 M_1^\dagger\big)_{11} - \big(M_4 M_4^\dagger\big)_{11}\right] y_L^2 = 0\,,
\end{equation}
where $M_4$ and $M_1$ denote the mass matrices in the composite sector,
namely ${\mathcal L}_{mass} = \sum_{ij} \overline \psi_{4L}^i (M_4)_{ij} \psi_{4R}^j
+ \sum_{ij} \overline \psi_{1L}^i (M_1)_{ij} \psi_{1R}^j + \textrm{h.c.}$.
Similarly to what we did for the first Weinberg rum rules, we can simplify
the condition in Eq.~(\ref{eq:Wein_cond_II}) by a change of basis in the composite
sector. Given that the mass mixing terms between the $\psi^1_{4,1}$ fields and the
other resonances are linear in $\psi^1_{4,1}$, we can always redefine the $\psi^i_{4,1}$
fields with $i \geq 2$ in such a way that $\psi^1$ has mass mixing only with
$\psi^2_{4,1}$. In this basis the sum rule in Eq.~(\ref{eq:Wein_cond_II}) depends
only on the $\psi^1_{4,1}$ masses, $m_{4,1}^{(1)} \overline \psi_{L4,1}^1 \psi_{R4,1}^1$, and on the mass mixing terms
with $\psi^2_{4,1}$, $m_{4,1}^{(12)} \overline \psi_{L4,1}^1 \psi_{R4,1}^2 + \textrm{h.c.}$
and $m_{4,1}^{(21)} \overline \psi_{L4,1}^2 \psi_{R4,1}^1 + \textrm{h.c.}$.
In the new basis the sum rule in Eq.~(\ref{eq:Wein_cond_II}) becomes
\begin{eqnarray}
&&2 \left[\big|m_1^{(1)}\big|^2 + \big|m_1^{(21)}\big|^2
- \big|m_4^{(1)}\big|^2 - \big|m_4^{(21)}\big|^2\right] y_R^2\nonumber\\
&&- \left[\big|m_1^{(1)}\big|^2 + \big|m_1^{(12)}\big|^2
- \big|m_4^{(1)}\big|^2 - \big|m_4^{(12)}\big|^2\right] y_L^2 = 0\,,
\label{eq:Wein_cond_II_simpl}
\end{eqnarray}
Of course, given the large number of free parameters there is no unique
solution to the above equation. If we require Eq.~(\ref{eq:Wein_cond_II_simpl}) to
be satisfied for arbitrary values of the elementary/composite mixings $y_L$
and $y_R$ the possible solutions are limited. A natural way to satisfy the sum rule
is to assume that the relevant mass terms respect an $\SO(5)$ symmetry under which
$\psi^1$ and $\psi^2$ transform in the fundamental representation.
This structure ensures the relations $m_1^{(1)} = m_4^{(1)}$, $m_1^{(12)} = m_4^{(12)}$
and $m_1^{(21)} = m_4^{(21)}$ and provides a solution for Eq.~(\ref{eq:Wein_cond_II_simpl}).
Notice that the $\SO(5)$ assumption is automatically realized in the three-site construction.

\chapter{Collider phenomenology}
\label{ch:LHC_pheno}

In this chapter we focus on the phenomenology of the composite resonances.
Due to their ubiquitous presence and their tight connection with the Higgs and Electro-Weak (EW) dynamics,
these states are one of the primary targets to directly test the composite Higgs scenarios in collider
experiments.

Two main classes of composite states are generically present.
The first one includes the fermionic partners of the Standard Model (SM) matter fields, which
are responsible for generating the SM Yukawa couplings.
At the same time, these resonances give rise to the leading contributions to the Higgs effective
potential, thus triggering EW symmetry breaking.
The second class of composite resonances includes the vector states related to the global
symmetry of the composite dynamics, which can be considered as the partners of the
SM gauge fields.

For definiteness, in our discussion we will focus on the standard anarchic flavor scenario presented in Chap.~\ref{ch:flavor}.
In this set-up the most relevant collider signatures are due to the quark partners.
As we will see in Sect.~\ref{sec:lhc_fermions}, being the lightest composite states,
the quark partners have a particularly simple phenomenology which is almost completely fixed by their quantum numbers
under the global symmetry of the composite sector.
The vector resonances, on the other hand, are less directly involved in the tuning issue and are usually heavier than the
fermionic partners. Their collider phenomenology can be significantly affected by the details of the
composite dynamics and, in particular, by the presence of light fermionc states. We will discuss
this topic in the second part of the chapter (Sect.~\ref{sec:vector_resonances}).


\section{Fermionic resonances}\label{sec:lhc_fermions}

We start our discussion by analyzing the collider phenomenology of the fer\-mio\-nic resonances.
In particular we focus on the properties of the top partners, {\it{i.e.}}~the composite states responsible
for generating the top mass. As we explained in Chap.~\ref{ch:SILH}, these states
are unavoidably present in all minimal composite Higgs scenarios and
are required to be relatively light in all the models that aim to solve the hierarchy problem.
This feature makes the top partners the privileged channel to probe the
minimal composite Higgs frameworks in collider experiments.

It is important to stress that, from the point of view of the collider phenomenology, the concept
of top partner can be extended to any vector-like
fermionic resonance that has sizable mixing only with
the third-generation SM quarks, independently of any connection with
Naturalness or with the generation of the top mass. In fact, any such state
shares the same phenomenological properties of the ``canonical''
top partners and gives rise to analogous collider signatures.
Notice that, as we explained in Chap.~\ref{ch:flavor}, in the anarchic flavor scenario
all the quark partners have sizable couplings to the top sector, thus they can be considered as top partners.

As can be easily understood, in a large class of scenarios the main collider
signatures are determined by the lightest composite resonances.
The heavier states, due to the smaller production cross section, usually
play a marginal role and can be neglected in a first approximation.
This feature allows to study the collider phenomenology of the top partners by only focusing on a
small set of resonances, thus simplifying the analysis
and making it more model-independent.
It is important to mention that the top-partner dynamics can also be affected by the presence of composite
vector resonances, which can contribute to their production cross section.
We postpone a discussion of the interplay between vector and fermionic resonances to Sect.~\ref{sec:vector_resonances}.
In this section, instead, we focus exclusively on the fermionic states and
we assume that the vector resonances are heavy enough so that they have
a small impact on the collider phenomenology.

A possible way to parametrize the dynamics of the fermionic partners is provided by the multi-site
models discussed in Chap.~\ref{ch:Pheno_models}. This approach allows at the same time to
describe the phenomenology of the resonances and to relate their properties to the Higgs dynamics
and to the EW observables. The multi-site constructions thus offer a straightforward way to take
into account simultaneously the implications
of the direct searches and of the indirect experimental constraints, as for instance the EW precision tests.

In this chapter, however, we want to focus only on the collider phenomenology of the partners and we
prefer to adopt a simpler and more model-independent approach, which allows to avoid spurious effects
related to specific explicit constructions.
Following Refs.~\cite{DeSimone:2012fs,Grojean:2013qca,Matsedonskyi:in_prep}, we will thus parametrize the dynamics of the
to partners by the most generic effective Lagrangian compatible with the non-linearly realized
Goldstone symmetry. We will not require any extra assumptions, as for instance the full calculability
of the EW and Higgs observables, which is one of the main ingredients of the multi-site models.
General effective Lagrangians for the composite resonances
can be constructed by using the CCWZ formalism presented in Chap.~\ref{ch:Goldstone}. 
To simplify the analysis we will include in the effective models only a minimal set of
resonances, which should be interpreted as the lightest fermionic partners.
As a last ingredient, we will use the power-counting discussed in Chap.~\ref{ch:SILH} to define
a leading Lagrangian and estimate the relevance of additional higher-order operators.


\subsection{The effective parametrizations}\label{sec:effect_parametr}

As a first step we present the effective Lagrangians of the simplified models for the top partners.
As we did in the rest of these Notes, we restrict our attention to the
class of minimal composite Higgs realizations based on the symmetry pattern $\SO(5)/\SO(4)$.

In the following we consider three simplified models
that are representative of a large fraction of the explicit theories explored so far in the literature.
The first one is based on the standard partial compositeness set-up (see Chap.~\ref{ch:Goldstone}),
in which all the SM fermions have an elementary counterpart. In particular we focus on the
scenario in which the elementary fermions are mixed with composite operators in the fundamental
$\SO(5)$ representation. This choice reproduces the phenomenology of a large
class of ``minimal composite Higgs models'' (in particular the holographic MCHM$_5$ constructions~\cite{Contino:2006qr})
and is closely related to the $5+5$ phenomenological model presented in Sect.~\ref{sec:5+5_model}.
The other two simplified models are based on a slight modification of the classical partial compositeness
set-up in which the $t_R$ field is fully composite and is identified with a chiral
state coming from the strong dynamics.
Two models of this kind will be considered, in which the $q_L$ elementary doublet mixes with
operators in the fundamental $\SO(5)$ representation and in the $\mathbf{14}$ respectively.
The first choice is motivated by minimality, whereas the second is related to Naturalness considerations.
In fact, as we explained in Chap.~\ref{ch:SILH},
the models with a fully composite $t_R$ and resonances in the $\mathbf{14}$
representation minimize the amount of tuning needed to get a realistic Higgs mass.
The simplified model based on the $\mathbf{14}$ representation, moreover, describes the
collider phenomenology of the explicit $14+1$ construction presented in Sect.~\ref{sec:14+1_model}.


\subsubsection{The ${\mathbf{5+5}}$ model}\label{sec:5+5_effective}

The first model we consider is the one based on the usual partial compositeness
assumption, in which the SM multiplets, $q_L$ and $t_R$, are realized as elementary fields.
As we already anticipated, we assume that the composite operators that mix with the elementary fields transform
in the fundamental representation of $\SO(5)$, the $\mathbf{5}$.
Due to the connection with the phenomenological model of Sect.~\ref{sec:5+5_model}
we will denote this simplified scenario as the $5+5$ model.

Under the unbroken $\SO(4)$ symmetry, the representation $\mathbf{5}$
decomposes as $\mathbf{5} = \mathbf{4} \oplus \mathbf{1}$.
Thus, in this model, the top partners transform as fourplets and singlets under $\SO(4)$.
In our simplified description we will include only one layer of composite resonances,
namely one $\SO(4)$ fourplet and one singlet.

The leading effective Lagrangian can be written as a sum of three terms containing the dynamics
of the composite states, of the elementary ones and the mixing terms:
\begin{equation}
{\cal L} = {\cal L}_{comp} + {\cal L}_{elem} + {\cal L}_{mix}\,.
\end{equation}
The Lagrangian for the composite fermions contains the usual kinetic and mass terms
and an interaction term between the fourplet and the singlet that involves the
$d$-symbol operator:
\begin{eqnarray}\label{eq:Lagr_comp_5+5}
{\cal L}_{comp} &=& i\overline \psi_4 \slashed D \psi_4 + i \overline \psi_1 \slashed D \psi_1
- m_4 \overline \psi_4 \psi_4 - m_1 \overline \psi_1 \psi_1\nonumber\\
&&-\,\left(i\, c_L\, \overline \psi_{4L}^i \gamma^\mu d_\mu^i \psi_{1L} +
i\, c_R\, \overline \psi_{4R}^i \gamma^\mu d_\mu^i \psi_{1R}
+ {\mathrm{h.c.}}\right)\,,
\end{eqnarray}
where $\psi_{4,1}$ denote the composite fourplet and singlet.
The covariant derivatives for the composite fermions are given by
\begin{eqnarray}
D_\mu \psi_4 &=& \left(\partial_\mu - \frac{2}{3} i g' B_\mu - i e_\mu - i g_s G_\mu
\right) \psi_4\,,\\
D_\mu \psi_1 &=& \left(\partial_\mu - \frac{2}{3} i g' B_\mu - i g_s G_\mu\right) \psi_1\,,
\label{eq:cov_der_singlet}
\end{eqnarray}
where $G_\mu$ denotes the gluon field and $g_s$ is the $\SU(3)_c$ coupling.
In the above formulae $d_\mu$ and $e_\mu$ denote the CCWZ symbols defined in Eq.~(\ref{dedefgauge})
(their explicitly expressions are given in Appendix~\ref{so54}).
In terms of fields with definite $\SO(4) \simeq \SU(2)_L \times \SU(2)_R$ quantum numbers,
the fourplet $\psi_4$ decomposes as
\begin{equation}\label{eq:4-plet_structure}
\psi_4 = \frac{1}{\sqrt{2}}
\left[
\begin{array}{c}
-i B + i\, X_{5/3}\\
-B - X_{5/3}\\
-i\,T - i\,X_{2/3}\\
T - X_{2/3}
\end{array}
\right]\;.
\end{equation}
The four components of the multiplet
correspond to two $\SU(2)_L$ doublets, $(T, B)$ and $(X_{5/3}, X_{2/3})$, with
hypercharges $1/6$ and $7/6$ respectively. The first doublet has the same quantum numbers as the
elementary $q_L$ doublet, while the second one contains an exotic state, the $X_{5/3}$, 
with charge $5/3$ and a top-like state, the $X_{2/3}$, with charge $2/3$. The singlet $\psi_1$
has the same quantum numbers of the $t_R$ SM field. To make contact with the notation
of Sect.~\ref{sec:5+5_model}, we also denote this resonance by $\widetilde T$.

The Lagrangian for the elementary fermions is given by the usual kinetic terms
\begin{equation}
{\cal L}_{elem} = i\, \overline q_L \slashed D q_L + i\, \overline t_R \slashed D t_R\,
\end{equation}
where the covariant derivatives coincide with the SM ones
\begin{eqnarray}
D_\mu q_L &=& \left(\partial_\mu - i g W^i_\mu \frac{\sigma^i}{2}
- i\frac{1}{6} g' B_\mu - i g_S G_\mu\right) q_L\,,\\
D_\mu t_R &=& \left(\partial_\mu - i\frac{2}{3} g' B_\mu - i g_S G_\mu\right) t_R\,.
\end{eqnarray}

The mixing between the elementary and composite states is described by
\begin{eqnarray}
{\cal L}_{mix} &=& y_{L4} f \big(\overline q_L^{\mathbf{5}} U\big)_i \psi_4^i
+ y_{L1} f \big(\overline q_L^{\mathbf{5}} U\big)_5 \psi_1 + {\rm h.c.}\nonumber\\
&& +\, y_{R4} f \big(\overline t_R^{\mathbf{5}} U\big)_i \psi_4^i
+ y_{R1} f \big(\overline t_R^{\mathbf{5}} U\big)_5 \psi_1 + {\rm h.c.}\,,
\end{eqnarray}
where the index $i$ takes the values $i = 1, \ldots, 4$.
The embedding, $q_L^{\mathbf 5}$ and $t_R^{\mathbf 5}$, of the elementary states in incomplete
fundamental representations of $\SO(5)$ has been already given
in the previous chapters, we rewrite it here for
completeness \footnote{In order to avoid confusion with the notation used for the composite
states, we denote the embedding of the elementary fields in the fundamental $\SO(5)$ representation
by $q_L^{\mathbf 5}$ and $t_R^{\mathbf 5}$, and not by $Q_L$ and $T_R$ as in the previous chapters.
Later on we will adopt an analogous notation for the embedding in the $\mathbf{14}$.}
\begin{equation}\label{eq:embedding_5+5}
q_L^{\mathbf{5}} = \frac{1}{\sqrt{2}} \left[
\begin{array}{c}
-i\, b_L\\
-b_L\\
-i\, t_L\\
t_L\\
0
\end{array}
\right]\,,
\qquad \quad
t_R^{\mathbf{5}} = \left[
\begin{array}{c}
0\\
0\\
0\\
0\\
t_R
\end{array}
\right]\,.
\end{equation}
Notice that, as explained in Chap.~\ref{ch:Pheno_models},
in order to accommodate the correct hypercharges for the SM fermions an
extra $\mathrm{U}(1)_X$ subgroup must be included. Under this symmetry the
elementary fields $q_L$ and $t_R$ as well as all the composite multiplets
have charge $2/3$.

The complete effective Lagrangian contains $8$ free parameters, namely the $4$
elementary/composite mixings $y_{L4,1}$ and $y_{R4,1}$, the masses of the composite states,
$m_{4,1}$, and the coefficients of the $d$-symbol interactions $c_{L,R}$. All these coefficients are
in general complex. Five complex phases can be removed by suitable field redefinitions
but the remaining ones are physical and can not be eliminated. For simplicity, however,
we assume that the strong sector is invariant under CP, in this way all the parameters in the Lagrangian are real.

Let us now discuss the natural size of the parameters. The $d$-symbol term is a purely
strong sector interaction, thus its coefficient is expected to be of $\mathcal{O}(1)$.
The elementary/composite mixings, on the other hand, are external with respect to the
strong dynamics, thus their size is not fixed by the power counting. The partial compositeness
paradigm, however, implies that the mixing of an elementary states to the composite fourplet and singlet
are correlated. In the usual set-up, indeed, both terms are mainly generated from the mixing
of the elementary fields with only one composite operator in a complete $\SO(5)$ representation
(see Sect.~\ref{sec:PC}), thus we expect $y_{L4} \sim y_{L1}$
and $y_{R4} \sim y_{R1}$. The exact value of the mixings is determined by the requirement of
reproducing the correct top mass.

It is interesting to notice that the simplified model we presented so far almost coincides
with the fermionic sector of the two-site model discussed in Chap.~\ref{ch:Pheno_models}.
Indeed, it is straightforward to check that the Lagrangian obtained by integrating
out the vector resonances in the two-site $5+5$ model is equal to the simplified model
of this section for the following choice of the parameters:
\begin{equation}
y_{L4} = y_{L1} \equiv y_L\,, \qquad y_{R4} = y_{R1} \equiv y_R\,,
\qquad c_L = c_R = 0\,.
\end{equation}
In the two-site model the relation between the elementary/composite mixings is
dictated by the global symmetry structure and is essential to realize the collective
breaking mechanism which protects the Higgs potential.
In the two-site set-up the $d$-symbol term
in Eq.~(\ref{eq:Lagr_comp_5+5}) is only generated through non-nearest neighbor interactions,
it is thus suppressed with respect to the general power-counting estimate (see Sect.~\ref{sec:non-local_operators}).
In extended multi-site models, on the other hand, the $d$-symbol operator can arise as an effective interaction mediated
by heavy vector resonances associated to the $\SO(5)/\SO(4)$ coset and its expected size
follows the usual power-counting.

\paragraph{The mass spectrum}

We can now analyze the spectrum of the fermionic resonances.
The mass matrix of the charge $2/3$ states after EWSB has the simple form
\begin{equation}
\left[
\!\begin{array}{c}
\overline t_L\\
\rule{0pt}{1.em}\overline T_L\\
\overline X_{2/3L}\\
\rule{0pt}{1.5em}\overline {\widetilde T}_L
\end{array}
\!\!
\right]^{\!T}
\left[
\!\!
\begin{array}{c@{\hspace{.75em}}c@{\hspace{.75em}}c@{\hspace{.75em}}c}
0 & \displaystyle \frac{y_{L4} f}{2}(1 + c_{\varepsilon})
& \displaystyle \frac{y_{L4} f}{2} (1 - c_{\varepsilon})
& \displaystyle \frac{y_{L1} f}{\sqrt{2}} s_\varepsilon\\
\displaystyle \rule{0pt}{1.5em}-\frac{y_{R4} f}{\sqrt{2}} s_\varepsilon & -m_4 & 0 & 0\\
\displaystyle \rule{0pt}{1.75em}\frac{y_{R4} f}{\sqrt{2}} s_\varepsilon & 0 & -m_4 & 0\\
\rule{0pt}{1.25em}y_{R1} f c_\varepsilon & 0 & 0 & -m_1
\end{array}
\!
\right]
\left[
\!
\begin{array}{c}
t_R\\
\rule{0pt}{1.em}T_R\\
X_{2/3R}\\
\rule{0pt}{1.25em}{\widetilde T}_R
\end{array}
\!
\right]\,,
\label{eq:mass_5+5}
\end{equation}
where we defined $\varepsilon \equiv \langle h \rangle/f$, while $s_\varepsilon$
and $c_\varepsilon$ denote the sine and cosine of $\varepsilon$.
For completeness we also report here the relation between $\epsilon$ and the
EWSB scale $v$:
\begin{equation}
\xi = \frac{v^2}{f^2} = \sin^2 \epsilon\,.
\end{equation}
For small values of $\xi$ the approximate relation $\xi \simeq \epsilon^2$ is valid
with good accuracy.

An interesting feature of the mass matrix in Eq.~(\ref{eq:mass_5+5}) is the fact
that the dependence on the Higgs VEV appears only in the terms that mix the elementary
and the composite states. This shows that all the effects due to EWSB are necessarily
weighted by the elementary/composite mixings. As explained at length in Chap.~\ref{ch:Goldstone},
this feature is a consequence of the
Goldstone nature of the Higgs and, as we will see, implies a very peculiar structure
for the spectrum of the resonances.

The mass matrix for the charge $2/3$ states can not be exactly diagonalized
in closed analytic form.
However one can find some approximate formulae for the masses of the resonances
by using an expansion in $\xi$. The expression for the top mass is
\begin{equation}\label{eq:top_mass_5+5}
m_{top}^2 = \frac{\left( y_{L1} y_{R1} m_4 - y_{L4} y_{R4} m_1 \right)^2 f^4}
{(m_4^2 + y_{L4}^2 f^2)(m_1^2 + y_{R1}^2 f^2)}
\frac{\xi}{2} + \mathcal{O}\left(f^2 \xi^2\right)\,,
\end{equation}
while the masses of the heavy resonances are
\begin{eqnarray}
m_{X_{2/3}} &=& m_4\left[ 1 + \frac{y_{R4}^2 f^2}{4 m_4^2} \xi + \cdots\right]\,,\label{eq:mX23_elemtR}\\
m_{T} &=& \sqrt{m_4^2 + y_{L4}^2 f^2}\left[ 1 - \frac{(y_{L4}^2 - y_{R4}^2) f^2}
{4 m_4^2} \xi + \cdots\right]\,,\label{eq:mT_elemtR}\\
m_{\widetilde T} &=& \sqrt{m_1^2 + y_{R1}^2 f^2}\left[1 + \frac{(y_{L1}^2 - 2 y_{R1}^2) f^2}{4 m_1^2} \xi + \cdots\right]\,,\label{eq:mTtilde_elemtR}
\end{eqnarray}
where, inside the square brackets, we only kept the leading order terms in an expansion
in the elementary/composite mixings and in $\xi$.
In the above formulae we denoted each mass eigenstate by the same names of the
states with definite $\SO(4)$ quantum numbers. Of course this identification is valid
only as long as the elementary/composite mixings are smaller than the mass parameters in
the composite sector and the fourplet states are not close in mass to the singlet.
If these conditions are satisfied, in the limit of small $\xi$, each mass eigenstate
is approximately aligned with one field in the original basis, otherwise
the above expressions for the masses of the resonances are still approximately
valid but the mass eigenstates have sizable components along states with different
$\SO(4)$ quantum numbers.

Let us now consider the states with charge $-1/3$. For simplicity, in the effective model we do not
include a right-handed bottom component because its mixing with the composite dynamics is
typically small and it does not significantly modify the collider
phenomenology. Therefore the $b_L$ state remains in the spectrum
as a massless field. In addition to the $b_L$, the model also contains
a heavy $B$ whose mass is given by
\begin{equation}\label{eq:mB_elemtR}
m_B = \sqrt{m_4^2 + y_{L4}^2 f^2}\,.
\end{equation}
This formula is exact and does not receive corrections after EWSB. In fact the mass matrix
of the charge $-1/3$ states in the $5+5$ model does not depend on the Higgs VEV.

The heavy $B$ forms a nearly-degenerate $\SU(2)_L$ doublet with the
$T$ resonance. By comparing Eqs.~(\ref{eq:mT_elemtR}) and (\ref{eq:mB_elemtR})
one finds that the mass difference between the two states is of order
$\Delta m^2 \sim y^2 v^2$, where $y$ denotes the typical size of the $y_{L4}$ and
$y_{R4}$ mixings. Notice that the lightest state inside the doublet can be
either the $T$ or the $B$ depending on the values of the elementary/composite mixings.

The exotic state $X_{5/3}$ is the only field with electric charge $5/3$ in the model, thus
it can not mix with any other state. Its mass is simply given by the fourplet mass parameter
$m_{X_{5/3}} = m_4$. This state forms an approximate $\SU(2)_L$ doublet together with
the $X_{2/3}$ resonance. The spitting between the two states is of
order $\Delta m^2 \sim y_{R4}^2 v^2$ and the lightest state in the doublet is nearly always the
$X_{5/3}$ resonance.\footnote{The $X_{2/3}$ can be the lightest resonance inside the fourplet
due to level-repulsion effects if the singlet and fourplet are close in mass. In this case, however,
the lightest charge $2/3$ state is not purely the $X_{2/3}$, but contains a large admixture of the $\widetilde T$.}

\begin{figure}
\centering
\includegraphics[width=0.4\textwidth]{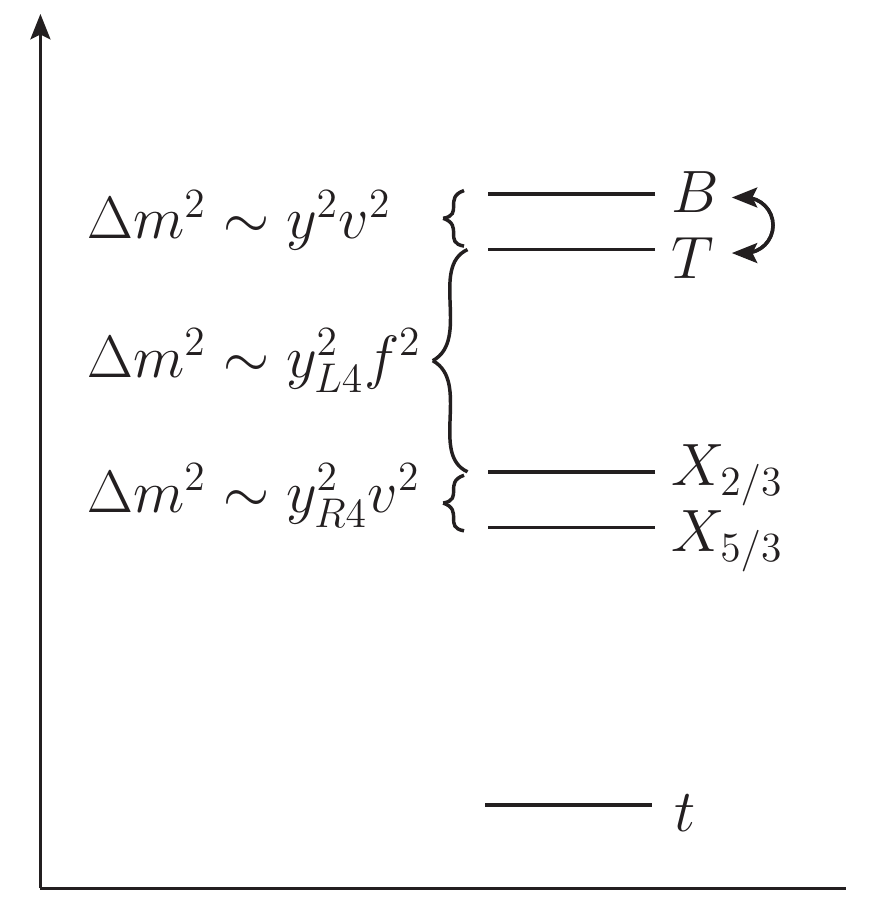}
\hspace{3.5em}
\includegraphics[width=0.4\textwidth]{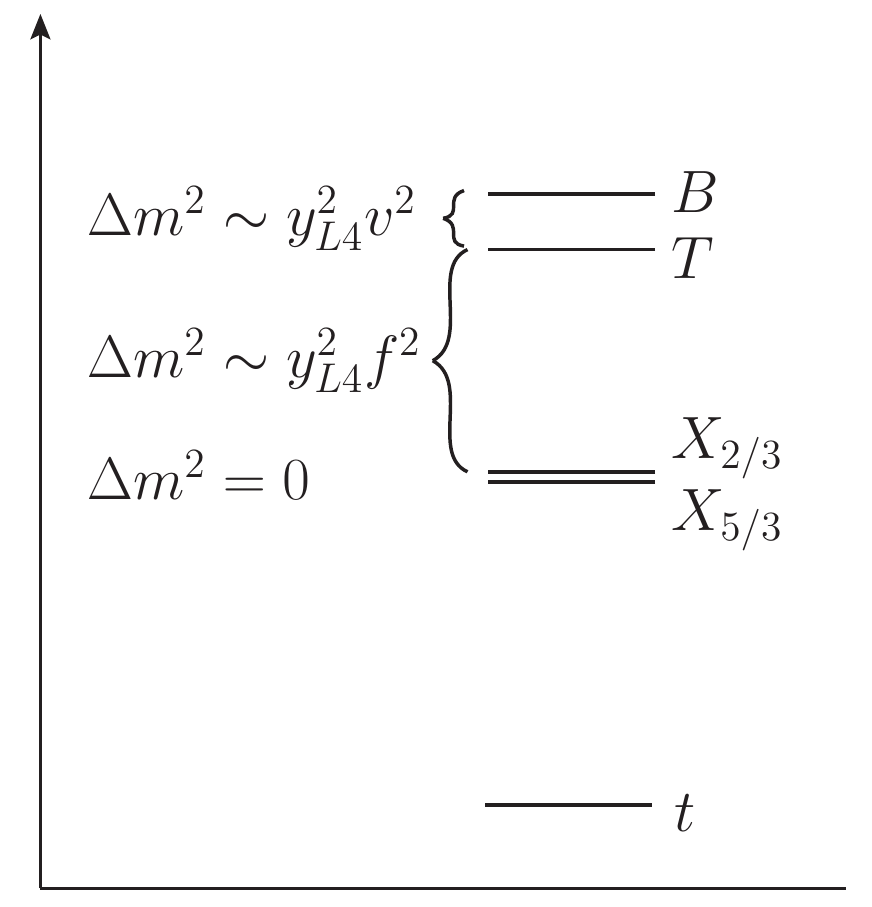}
\caption{Typical mass spectrum of the fourplet states. The left panel corresponds the the scenario
with an elementary $t_R$ (the $5+5$ model), while the right panel to the set-ups with a fully composite $t_R$
(the $5+1$ and $14+1$ models).}
\label{fig:mass_spectrum_elemtR}
\end{figure}

By comparing the masses of the two approximate $\SU(2)_L$ doublets we find that the splitting
inside each doublet is typically much smaller than the mass difference between the
two multiplets. As we saw before, the splitting inside the doublets is generated only after EWSB
and is thus proportional to the Higgs VEV. On the other hand, the mass difference between
the two doublets is induced by the mixing with the elementary $q_L$.
This effect is present even before EWSB and induces a split of order
$\Delta m^2 \sim y_{L4}^2 f^2$, which is enhanced by a factor $1/\xi \gg 1$ with respect to the
mass split inside the doublets.
The structure of the mass spectrum of the
resonances coming from the fourplet is schematically shown in the left panel of Fig.~\ref{fig:mass_spectrum_elemtR}.

Notice that the peculiar structure of the spectrum is a consequence
of the Goldstone nature of the Higgs. In a model in which the Higgs is a generic composite
resonance but not a Goldstone there is no particular structure in the mass spectrum and
the splitting among all the states in the fourplet is typically of the same order.


\subsubsection[The $5+1$ model]{The ${\mathbf{5+1}}$ model}
\index{fully composite $t_R$}

In the second simplified model we consider, the right-handed top component $t_R$ is identified
with a fully composite chiral state coming from the strong dynamics.
As in the previous model, the elementary doublet $q_L$ is assumed to mix with
composite operators in the fundamental representation of $\SO(5)$. In analogy
with the previous model, we denote the present set-up by $5+1$, where the two numbers
correspond to the $\SO(5)$ representations in which the SM fields are embedded.

The Lagrangian for the composite states, which now include also the $t_R$ identified with
a total $\SO(5)$ singlet, is given by
\begin{eqnarray}
{\cal L}_{comp} &=& i\overline \psi_4 \slashed D \psi_4 + i \overline \psi_1 \slashed D \psi_1
+ i\, \overline t_R \slashed D t_R
- m_4 \overline \psi_4 \psi_4 - m_1 \overline \psi_1 \psi_1\nonumber\\
&& -\, \left(i\, c_L \overline \psi_{4L}^i \gamma^\mu d_\mu^i \psi_{1L}
+ i\, c_R \overline \psi_{4R}^i \gamma^\mu d_\mu^i \psi_{1R}+ {\rm h.c.}\right)\nonumber\\
&&-\, \left(i\, c_t \overline \psi_{4R}^i \gamma^\mu d_\mu^i t_R + {\rm h.c.}\right)\,.
\label{eq:Lcomp_5+1}
\end{eqnarray}
In the above equation we did not include a possible mass mixing involving the $t_R$
and the $\psi_1$ fields. This term is allowed by the symmetry of the model,
however it can always be set to zero by a suitable field redefinition. Notice that the presence
of the $t_R$ field in the composite sector allows to write an additional $d$-symbol
interaction involving the fourplet $\psi_4$.

The Lagrangian for the elementary doublet $q_L$ contains only the usual
kinetic term: $\mathcal{L}_{elem} = i \overline q_L \slashed D q_L$.
The mixing terms between the elementary and composite states are given by
\begin{equation}\label{eq:Lmix_5+1}
{\cal L}_{mix} =
y_{Lt} f (\overline q_L^{\mathbf{5}} U)_5 t_R + y_{L4} f (\overline q_L^{\mathbf{5}} U)_i \psi_4^i
+ y_{L1} f (\overline q_L^{\mathbf{5}} U)_5 \psi_1 + {\rm h.c.}\,.
\end{equation}
An important difference with respect to the case with an elementary $t_R$ is the
fact that now the top field has a direct Yukawa term which comes from the $y_{Lt}$
mixing.

In the $5+1$ model there are $8$ free parameters, which can be forced to be real
by imposing CP invariance. For simplicity we will adopt this assumption in the following.
The $c_{L,R}$ and $c_t$ couplings correspond to purely strong sector interactions
and their coefficients are naturally of $\mathcal{O}(1)$ as can be inferred
from the power counting in Eq.~(\ref{POWCFull}).
Moreover, we expect the elementary/composite mixings to be of the same order,
$y_{Lt} \sim y_{L4} \sim y_{L1}$.

\paragraph{The mass spectrum.}

Let us now discuss the features of the spectrum. As a first
step we consider the charge $2/3$ fields, whose mass matrix is given by
\begin{equation}
\left[\!
\begin{array}{c}
\overline t_L\\
\rule{0pt}{1.em}\overline T_L\\
\overline X_{2/3L}\\
\rule{0pt}{1.5em}\overline {\widetilde T}_L
\end{array}\!\!
\right]^{\!T}
\left[\!
\begin{array}{c@{\hspace{.75em}}c@{\hspace{.75em}}c@{\hspace{.75em}}c}
\displaystyle \frac{y_{Lt} f}{\sqrt{2}} s_\varepsilon
& \displaystyle \frac{y_{L4} f}{2} (1 + c_\varepsilon)
& \displaystyle \frac{y_{L4} f}{2} (1 - c_\varepsilon)
& \displaystyle \frac{y_{L1} f}{\sqrt{2}} s_\varepsilon\\
\displaystyle \rule{0pt}{1.25em}0 & -m_4 & 0 & 0\\
\displaystyle \rule{0pt}{1.25em}0 & 0 & -m_4 & 0\\
\rule{0pt}{1.25em}0 & 0 & 0 & -m_1
\end{array}
\!\right]
\left[\!
\begin{array}{c}
t_R\\
\rule{0pt}{1.em}T_R\\
X_{2/3R}\\
\rule{0pt}{1.25em}{\widetilde T}_R
\end{array}
\!
\right]\,.
\label{eq:massm_5+1}
\end{equation}
It is interesting to notice that one combination of the $T$ and $X_{2/3}$ fields
does not mix with the other states, namely
\begin{equation}
X'_{2/3} = \frac{1}{\sqrt{2}\sqrt{1 + c^2_\varepsilon}} \left[
(1 + c_\varepsilon)\; X_{2/3} - (1 - c_\varepsilon)\; T\right]\,.
\end{equation}
Therefore the mass of the $X_{2/3}'$ state is just given by the composite mass $m_4$ and
is exactly degenerate with the exotic $X_{5/3}$ resonance. On the other hand, the
orthogonal combination of $T$ and $X_{2/3}$,
\begin{equation}
T' = \frac{1}{\sqrt{2}\sqrt{1 + c^2_\varepsilon}} \left[
(1 + c_\varepsilon)\;T + (1 - c_\varepsilon)\; X_{2/3}\right]\,,
\end{equation}
is mixed with the elementary fields and its mass acquires a shift controlled by the
$y_{L4}$ parameter, plus an additional corrections due to EWSB:\footnote{For simplicity
in the following we will drop the prime in front of the $T'$ and $X'_{2/3}$
resonances and we will denote them simply by $T$ and $X_{2/3}$.}
\begin{equation}\label{eq:mT_5+1}
m_T \simeq \sqrt{m_4^2 + y_{L4}^2 f^2}\left[1 - \frac{y_{L4}^2 f^2}{4 m_4^2}\xi + \cdots\right]\,.
\end{equation}
This state is close in mass to the charge $-1/3$ resonance coming from the
fourplet, the $B$, whose mass is $m_B = \sqrt{m_4^2 + y_{L4}^2 f^2}$. Apart from some
corner of the parameter space, the $T$ resonance is lighter than the $B$.

The spectrum of the fourplet resonances is schematically shown in the right panel of
Fig.~\ref{fig:mass_spectrum_elemtR}. As one can see, the structure of the spectrum is quite
similar to the one we found in the $5+5$ model. The only difference is the fact that
in the $5+1$ case the $X_{2/3}$ and the $X_{5/3}$  states are exactly degenerate.

The remaining charge-$2/3$ heavy state, the $\widetilde T$, which comes from the singlet,
has a mass given by
\begin{equation}\label{eq:mTtilde_5+1}
m_{\widetilde T} \simeq m_1 \left[1 + \frac{y_{L1}^2 f^2}{4 m_1^2}\xi + \cdots\right]\,.
\end{equation}

As we discussed before, in the models with a fully composite $t_R$ a direct mixing term
between the elementary $q_L$ and the $t_R$ is present in the effective Lagrangian.
The top mass is therefore mostly determined by the $y_{Lt}$ parameter and, at leading
order in the $v/f$ expansion, is given by
\begin{equation}\label{eq:m_top_5+1}
m_{top}^2 = \frac{1}{2} \frac{m_4^2}{m_4^2 + y_{L4}^2 f^2} y_{Lt}^2 f^2 \xi +
\mathcal{O}(f^2 \xi^2)\;.
\end{equation}


\subsubsection[The $14+1$ model]{The ${\mathbf{14+1}}$ model}
\index{fully composite $t_R$}

As in the $5+1$ model, in the third scenario we consider, the $14 + 1$ model,
the right-handed top component is a fully composite chiral state. Differently from the previous
case, however, the elementary doublet $q_L$ is mixed with some composite operators in the
symmetric (the $\mathbf{14}$) representation of $\SO(5)$. Under the unbroken $\SO(4)$ subgroup
the symmetric representation decomposes as $\mathbf{14} \simeq \mathbf{9} \oplus \mathbf{4} \oplus \mathbf{1}$.
This implies that, in addition to fourplets and singlets, the elementary states can also
mix with fermionic resonances that transform as nineplets under $\SO(4)$.

The Lagrangian for the composite states is analogous to the one we built for the $5+1$ model.
In particular the operators involving only the fourplet, $\psi_4$, and
the singlet, $\psi_1$, are exactly the same as
in the previous scenario and are given in Eq.~(\ref{eq:Lcomp_5+1}).

The nineplet, $\psi_9$, can be described by a $4 \times 4$ symmetric matrix. This
multiplet contains three $\SU(2)_L$ triplets with different $\SU(2)_R$ charges:
\begin{equation}
\begin{array}{r@{\hspace{1.9em}\textrm{with}\hspace{1.em}}l}
\left\{U_{8/3}\,,\; U_{5/3}\,,\; U_{2/3}\right\} & T_3^R = +1\,,\\
\rule{0pt}{1.35em}\left\{Y_{5/3}\,,\; Y_{2/3}\,,\; Y_{-1/3}\right\} & T_3^R = 0\,,\\
\rule{0pt}{1.35em}\left\{Z_{2/3}\,,\; Z_{-1/3}\,,\; Z_{-4/3}\right\} & T_3^R = -1\,,
\end{array}
\end{equation}
where the subscripts denote the electric charge. For shortness we do not give here
the embedding of the various components in the $4 \times 4$ matrix notation. It can be
found in Ref.~\cite{Matsedonskyi:2014lla}.

The dynamics of the nineplet, $\psi_9$, is described by additional kinetic and mass terms
in the composite sector Lagrangian:
\begin{equation}
\mathcal{L}_{comp}^{(9)} = i\, \Tr\left[\overline \psi_9 \slashed D \psi_9\right]
- m_9 \Tr\left[\overline \psi_9 \psi_9\right]\,,
\end{equation}
where the covariant derivative of the $\psi_9$ field is defined as
\begin{equation}
D_\mu \psi_9 = \left(\partial_\mu - \frac{2}{3} i g' B_\mu - i g_s G_\mu
\right) \psi_9 - i e_\mu \psi_9 + i \psi_9 e_\mu\,.
\end{equation}
The presence of a nineplet allows some additional interaction terms containing the $d$-symbol:
\begin{eqnarray}
{\cal L}_{comp} &=& - \left(i\, c_{L9} \overline \psi_{9L}^{ij} \gamma^\mu d_\mu^i \psi_{4L}^j
+ i\, c_{R9} \overline \psi_{9R}^{ij} \gamma^\mu d_\mu^i \psi_{4R}^j
+ {\rm h.c.}\right)\nonumber\\
&&-\, \left(i\, \frac{c_{t9}}{m_*} \overline \psi_{9L}^{ij} d^{\mu i} d_\mu^j t_R
+ {\rm h.c.}\right)\,.
\label{eq:Lcomp_dsymbol_14+1}
\end{eqnarray}
Notice that in the last line of Eq.~(\ref{eq:Lcomp_dsymbol_14+1}) we included a higher-dimensional
interaction involving two powers of $d_\mu$. The coefficient of this operator, following our
power counting, is expected to be suppressed by the cut-off of the effective theory, which
we identify with the mass scale of the heavy vector resonances, $m_*$. Although the interactions
coming from the higher-dimensional operator are suppressed with respect to the other ones,
they can be important for the collider phenomenology of the $U_{8/3}$ resonance, as we will
discuss in Sect.~\ref{sec:coll_phenomenology}.

The mixing of the elementary doublet $q_L$ with the composite states is described by the
following Lagrangian
\begin{eqnarray}\label{eq:Lmix_14+1}
{\cal L}_{mix} &=&
\frac{y_{Lt}}{2} f (U^t \overline q_L^{\mathbf{14}} U)_{55} t_R
+ y_{L9} f (U^t \overline q_L^{\mathbf{14}} U)_{ij} \psi_9^{ij}\nonumber\\
&&+\; y_{L4} f (U^t \overline q_L^{\mathbf{14}} U)_{i5} \psi_4^i
+ \frac{y_{L1}}{2} f (U^t \overline q_L^{\mathbf{14}} U)_{55} \psi_1
+ {\rm h.c.}\,,\hspace{2.em}
\end{eqnarray}
where $q_L^{\mathbf{14}}$ now denotes the embedding of the $q_L$ doublet into the $\mathbf{14}$
representation (compare Eq.~(\ref{14source})):
\begin{equation}\label{eq:embedding_14+1}
q_L^{\mathbf{14}} = \frac{1}{\sqrt{2}}
\left[
\begin{array}{ccccc}
0 & 0 & 0 & 0 & -i\,b_L\\
0 & 0 & 0 & 0 & -b_L\\
0 & 0 & 0 & 0 & -i\,t_L\\
0 & 0 & 0 & 0 & t_L\\
-i\, b_L & -b_L & -i\,t_L & t_L & 0
\end{array}
\right]\,.
\end{equation}
The normalization of the mixing terms in Eq.~(\ref{eq:Lmix_14+1}) has been chosen in such a way
that the Lagrangian for the fourplet and singlet states matches the one of the $5+1$ model
in Eq.~(\ref{eq:Lmix_5+1}) at leading order in the $v/f$ expansion.

\paragraph{The mass spectrum.}

As one can see from the mixing
terms in Eq.~(\ref{eq:Lmix_14+1}), analogously to the $5+1$ set-up, a direct Yukawa term for the top
quark is present and the top mass is mostly determined by the $y_{Lt}$ parameter.
At leading order in $\xi$ it can be approximated by
\begin{equation}
m_{top}^2 = \frac{1}{8} \frac{m_4^2}{m_4^2 + y_{L4}^2 f^2} y_{Lt}^2 f^2 \sin^2 2 \varepsilon
+ \mathcal{O}(f^2 \xi^2)\,.
\end{equation}

The spectrum of the fourplet and singlet resonances is quite similar to the one we found in the
$5+1$ model. In addition to these states now several other resonances coming from $\psi_9$ are present.
We do not report here the complete mass matrix and we only give some approximate
expressions for the resonance masses.

It is clear from Eq.~(\ref{eq:Lmix_14+1}) that the structure
of the mass matrix for the charge-$2/3$ states is analogous to the one in Eq.~(\ref{eq:massm_5+1}).
In particular the mass matrix is diagonal apart from the mixings of the composite states with the
$t_L$ components. This implies that the $t_L$ field is only mixed
with one combination of the $T$ and $X_{2/3}$ resonances in $\psi_4$ and with one combination
of the $U_{2/3}$, $Y_{2/3}$ and $Z_{2/3}$ states coming from $\psi_9$. The masses of the orthogonal
combinations, therefore, are not affected by the mixing and by EWSB and are simply given by
the composite mass parameters, $m_4$ for the fourplet states and $m_9$ for the states in the $\mathbf{9}$.
As in the $5+1$ model, the fourplet state which is mixed with the $t_L$ almost coincides with
the $T$, we will thus denote it with the same name, whereas we will denote the orthogonal
combination as the $X_{2/3}$ state. The mass of the $T$ resonances is approximately given by
\begin{equation}
m_T \simeq \sqrt{m_4^2 + y_{L4}^2 f^2} \left[1 - \frac{5 y_{L4}^2 f^2}{4 m_4^2}\xi + \cdots\right]\,,
\end{equation}
while the mass of the $X_{2/3}$ state is $m_{X_{2/3}} = m_4$.

Regarding the $\psi_9$ states, the resonance with the largest mixing with
the $t_L$ field is the $Z_{2/3}$. Its mixing is approximately twice as large as the one of the
$Y_{2/3}$, while the mixing of the $U_{2/3}$ arises at next order in $\xi$ and is thus negligible.
The mass of the $Z_{2/3}$ state can be well approximated by the formula \footnote{For shortness
we denote by $Z_{2/3}$ the combination of states that mixes with the top and by $Y_{2/3}$
and $U_{2/3}$ the orthogonal ones.}
\begin{equation}
m_{Z_{2/3}} \simeq m_9 \left[1 + \frac{5 y_{L9}^2 f^2}{8 m_9^2} \xi + \cdots\right]\,.
\end{equation}
The masses of the other states are instead simply $m_{U_{2/3}} = m_{Y_{2/3}} = m_9$ and do not
receive corrections after EWSB.
The last charge $2/3$ state, the $\widetilde T$ resonance contained in the singlet $\psi_1$, has a mass
\begin{equation}
m_{\widetilde T} \simeq m_1 \left[1 + \frac{y_{L1}^2 f^2}{4 m_1^2}\xi + \cdots\right]\,.
\end{equation}

Differently from the model based on the fundamental representation of $\SO(5)$, in the $14+1$ scenario
the resonances with charge $-1/3$ have some couplings with the Higgs and receive a mass shift after EWSB.
The mass of the $B$ state contained in the fourplet is given by
\begin{equation}
m_B \simeq \sqrt{m_4^2 + y_{L4}^2 f^2} \left[1 - \frac{y_{L4}^2 f^2}{2 (m_4^2 + y_{L4}^2 f^2)}\xi + \cdots\right]\,.
\end{equation}
Inside the $\mathbf{9}$ two additional states with charge $-1/3$ are present, namely the $Y_{-1/3}$
and the $Z_{-1/3}$. Similarly to the charge $2/3$ states,
only one combination, namely $B^-_{-1/3} = (Y_{-1/3} - Z_{-1/3})/\sqrt{2}$,
is coupled to the Higgs and receives a mass shift after EWSB:
\begin{equation}
m_{B^-_{-1/3}} \simeq m_9 \left[1 + \frac{y_{L9}^2 f^2}{2 m_9^2} \xi + \cdots\right]\,.
\end{equation}
The mass of the other field $B^+_{-1/3} = (Y_{-1/3} + Z_{-1/3})/\sqrt{2}$, instead, is
just given by $m_{B^+_{-1/3}} = m_9$.

The charge $5/3$ resonances, namely the $X_{5/3}$, the $U_{5/3}$ and the $Y_{5/3}$ are not coupled to
the Higgs and their masses are equal to the composite mass parameters, $m_{X_{5/3}} = m_4$ and
$m_{U_{5/3}} = m_{Y_{5/3}} = m_9$.
Finally the exotic resonances $U_{8/3}$ and $Z_{-4/3}$, being the only states with these electric charges
can not be mixed with any other field and have masses $m_{U_{8/3}} = m_{Z_{-4/3}} = m_9$.

To conclude the discussion of the resonances masses in the $14+1$ model we briefly summarize the
structure of the spectrum of each $\SO(4)$ multiplet. The spectrum of the fourplet states is
completely analogous to the one we found in the $5+1$ model and is schematically shown in
Fig.~\ref{fig:mass_spectrum_elemtR}.
In the spectrum of the resonances coming from the $\mathbf{9}$ representation
all the states are degenerate with mass $m_9$ apart from one charge $2/3$ state,
the $Z_{2/3}$, and one charge $-1/3$ state, the $B^-_{-1/3}$. The latter states are
heavier than the other ones and the mass split is of order $\Delta m^2 \sim y_{L9}^2 v^2$.


\subsection{General properties}\label{sec:partners_prop}

After the description of the effective Lagrangians and of the spectrum of the composite resonances,
we can now focus on the general properties that determine their collider phenomenology.
In particular we will analyze the structure of the couplings involving the composite partners and the
SM states and we will estimate their sizes. These couplings control the
relative importance of the various production channels as well as the decay branching ratios of the composite states.


\subsubsection{Production mechanisms}

\begin{figure}
\centering
\includegraphics[width=.32\textwidth]{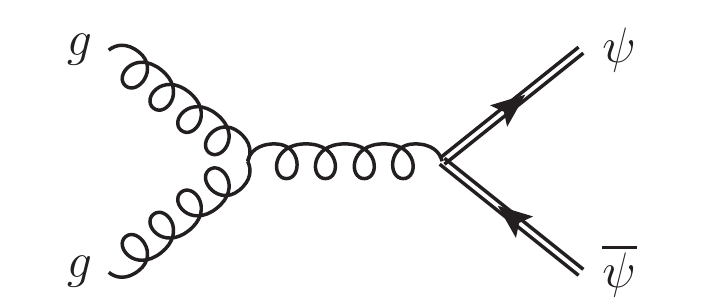}
\hspace{.0em}
\includegraphics[width=.32\textwidth]{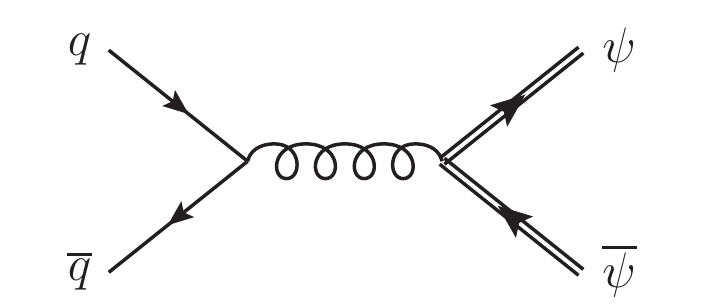}
\hspace{.0em}
\includegraphics[width=.31\textwidth]{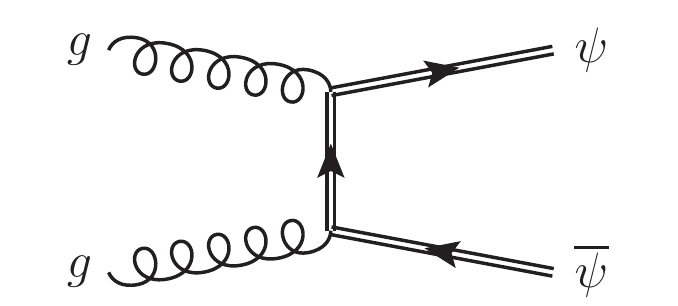}
\caption{Leading diagrams contributing to the pair production of composite top partners.}
\label{fig:pair_prod_diagram}
\end{figure}

As a first aspect we discuss the main production mechanisms of the top partners.
Being color triplets, they can be produced in pairs via QCD interactions through the
splitting of a virtual gluon (see diagrams in Fig.~\ref{fig:pair_prod_diagram}).
An important feature of this production channel is the fact that
its cross section is universal, that is, it depends exclusively on the partner mass $m_\psi$:
$\sigma_{pair} = \sigma_{pair}(m_\psi)$. QCD pair production is thus always present
for all top partners and has a substantial cross section due to the sizable QCD coupling.

Additional contributions to pair production are mediated by the EW gauge interactions.
All these contributions are typically suppressed with respect to the QCD ones due
to the smallness of the EW gauge couplings. Two classes of EW contributions
are present. The first class is due to diagrams with an s-channel exchange of an EW
gauge boson. The second class, instead, is generated by t-channel diagrams
containing ``flavor-changing'' gauge interactions that mix a SM quark and a fermionic resonance.
The first class of contributions is nearly universal and is determined by the EW quantum numbers
of the composite resonances (up to small effects of higher order in $v/f$). The second set of
contributions, instead, is sensitive to the details of the strong sector dynamics and
crucially depends on the parameters of the model that determine the size of the flavor-changing couplings.

\begin{figure}
\centering
\includegraphics[width=.8\textwidth]{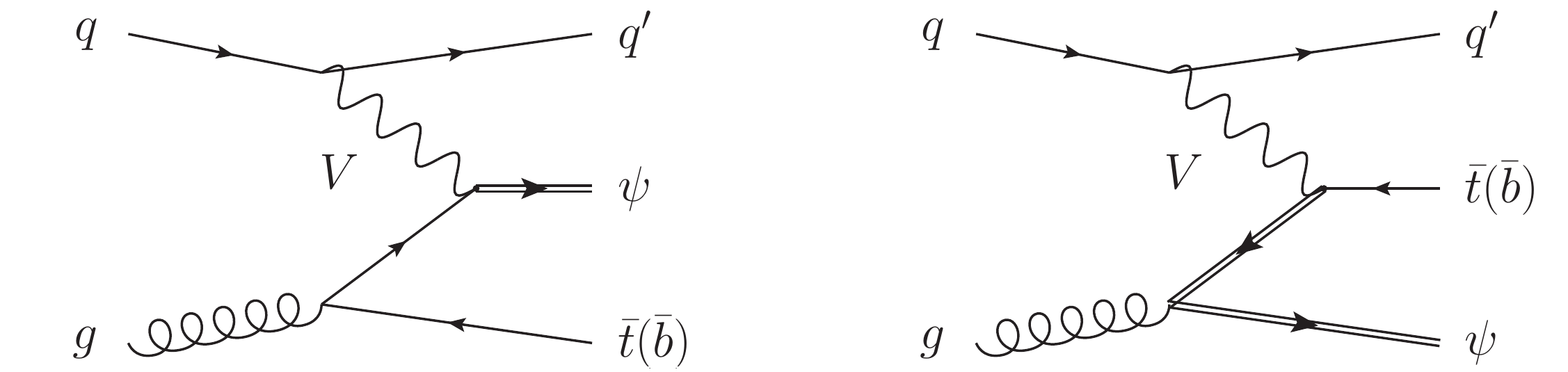}
\caption{Dominant diagrams contributing to the single production of
a composite top partner in association with a top or bottom quark.}
\label{fig:single_prod_diagram}
\end{figure}

Apart from pair production, the top partners can also be singly produced
in association with either a top or a bottom quark.
The structure of the diagrams that lead to the dominant contributions
is shown in Fig.~\ref{fig:single_prod_diagram}.
These processes are induced by a virtual EW boson $V = \{W^\pm, Z\}$ emitted from a light quark,
which then interacts with a top or a bottom quark produced by a gluon splitting. A distinctive
feature of the single production processes is the presence of a forward jet.
Differently from QCD pair production, single production is not universal and is determined
by the value of the flavor-changing EW couplings.

\begin{figure}
\centering
\includegraphics[width=.478\textwidth]{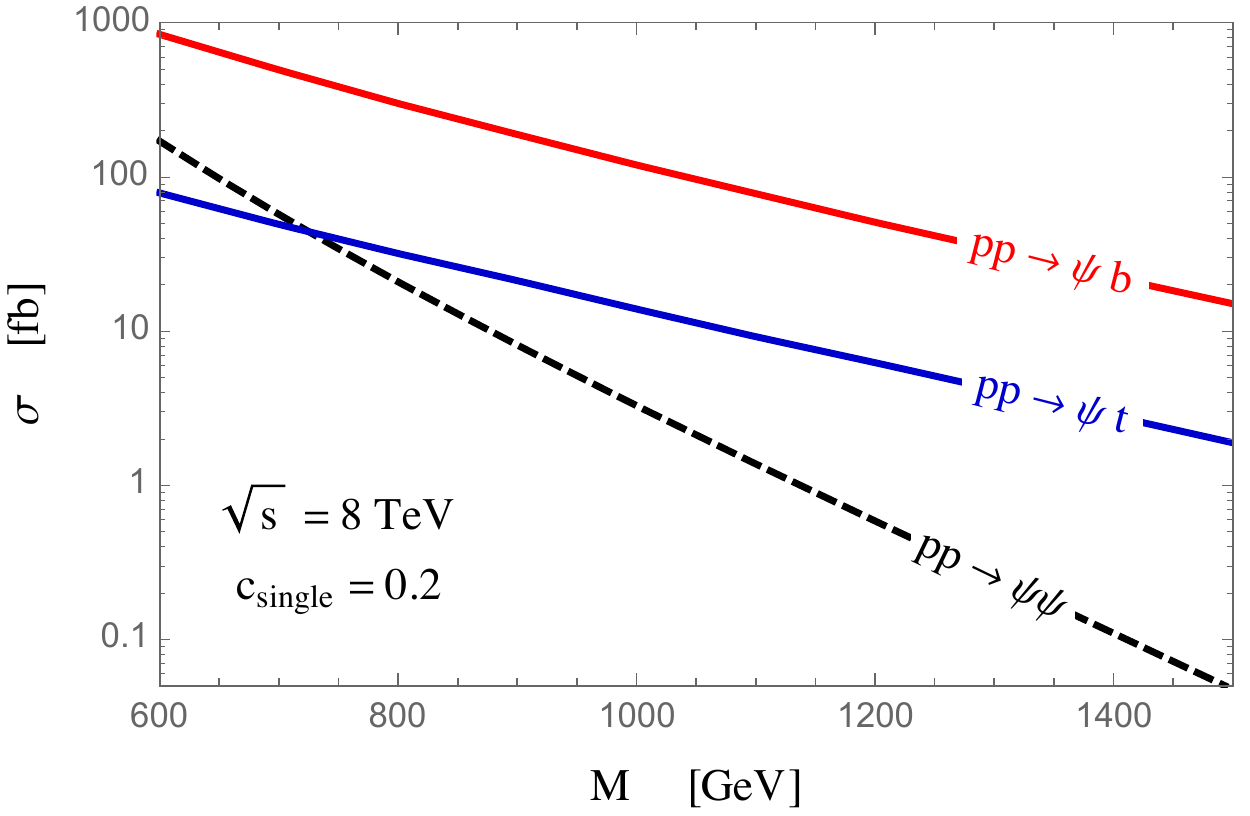}
\hfill
\includegraphics[width=.485\textwidth]{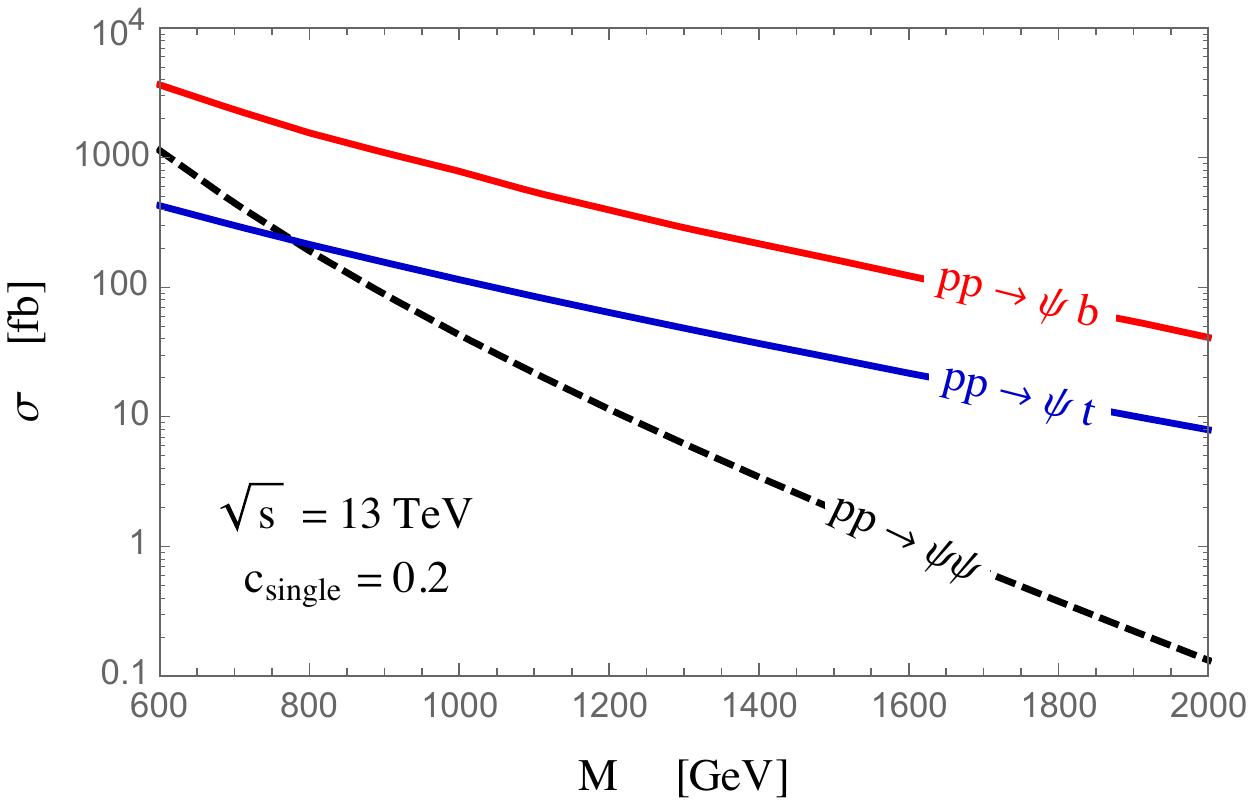}
\caption{Cross sections for the production of top partners at an hadronic collider with energy
$\sqrt s = 8\ \mathrm{TeV}$ (left panel) and $\sqrt s = 13\ \mathrm{TeV}$ (right panel).
The black dashed lines correspond to pair production, while the solid lines
correspond to $W$-mediated single production in association with a $b$ (red line) or a $t$ (blue line).
The single production couplings (in the unitary gauge) have been fixed to $c_{single} = 0.2$, which is a typical
value for $\xi = 0.1$.}
\label{fig:cross_sections}
\end{figure}

Due to the lower threshold, single production processes are favored over pair production
for higher top partner masses. The production cross section for typical values of the
single production couplings are shown
in Fig.~\ref{fig:cross_sections}.\footnote{The cross sections for pair and single production of
top partners at the $8$ TeV and $13$ TeV LHC can be found in Ref.~\cite{Matsedonskyi:2014mna}.}
As one can see from the plots, single production in association with a top quark becomes
comparable to pair production for $m_\psi \sim 1\ \mathrm{TeV}$.
On the other hand, the process in association with a bottom
quark can have a larger cross section than pair production even at low resonance masses.


\subsubsection{Couplings}

As we saw in the above discussion, the phenomenology of the light top partners
is relatively simple and is regulated by a few basic ingredients.
The production channels depend only on the QCD interactions and on the EW flavor-changing gauge couplings.
The latter couplings are also responsible for the top-partners decays.

Since we are typically interested in configurations in which the resonances are much heavier than the
EW bosons and the SM quarks, we can analyze the top-partner dynamics by using the Goldstone
boson equivalence theorem~\cite{Chanowitz:1985hj,Wulzer:2013mza}.
This theorem states that, at high energy, $E \gg m_W$, the longitudinal
components of the gauge bosons are described by the Goldstone fields. The transverse polarizations,
on the other hand, are well described by the vector fields in the absence of EWSB.
For the processes we are interested in, the transverse components typically give a negligible
contribution and the main effects come from the longitudinal modes.
This can be easily understood for the two-body decay processes into SM fermions, in which
the large mass difference between the composite resonance and the SM states implies that the decay products are very energetic.
A similar conclusion holds also for the single production processes. In this case, however,
the proof is not straightforward and the effective $W$ approximation~\cite{Kane:1984bb,Dawson:1984gx}
must be advocated to justify the result~\cite{DeSimone:2012fs}.

In order to derive the couplings to the Goldstone bosons it is convenient to work in an $R_\xi$ gauge,
in which the degrees of freedom associated to the unphysical Higgs components are retained.
The Higgs field can be parametrized as
\begin{equation}
H = \left[
\begin{array}{c}
\phi^+\\
\dfrac{1}{\sqrt{2}} \left(v + \rho + i \phi^0\right)
\end{array}
\right]\,,
\end{equation}
where $\rho$ denotes the physical Higgs boson and $\phi^{\pm,0}$ are the unphysical Goldstone modes.

The trilinear couplings of the Goldstones arise from three kind of terms: the mixings between the
elementary and the composite fermions, the $e$-symbol term contained in the covariant derivative
in the $\mathbf{4}$ and the $\mathbf{9}$ representations and the $d$-symbol terms. The last two types of operators
generate trilinear interactions involving a SM fermion only after the rotation to the mass eigenstate basis.
An important feature of the $e_\mu$ and $d_\mu$ operators is the fact that they give rise to interactions that
contain the derivative of the Goldstones and whose strength crucially depends on the energy of the
process. To simplify the analysis it is convenient to integrate by parts these interaction terms and use
a field redefinition to get rid of the derivatives. If we neglect the top and bottom
masses, the effect of the redefinition is just to replace the derivative acting on a heavy resonance
with the mass of the resonance itself:
\begin{equation}
i \gamma^\mu \partial_\mu \psi \rightarrow m_\psi \psi\,,
\qquad
i \partial_\mu \overline\psi \gamma^\mu \rightarrow -m_\psi \overline\psi\,.
\end{equation}

The quantum numbers of the resonances determine at which order in the $v/f$ expansion each trilinear coupling
is generated. It is easy to check that only the couplings involving fermions in different $\SO(4)$ representations
can be generated at leading order. On the contrary, couplings between fields in the same representation are necessarily
suppressed by powers of $v/f$ and are typically subleading. The order at which the various couplings between the
heavy resonances and the SM fields are generated is summarized in Table~\ref{tab:coupling order}.

\begin{table}[t!]
\centering
\tabcolsep=3pt
\begin{tabular}{c | c c c | c c c | c c c}
\rule[-6pt]{0pt}{1.5em}& $W t_R$ & $Z\hspace{.5pt}t_R$ & $h\hspace{1.pt}t_R$ & $W t_L$ & $Z\hspace{.5pt}t_L$ & $h\hspace{1.pt}t_L$
& $W b_L$ & $Z\hspace{.5pt}b_L$ & $h\hspace{1.pt}b_L$\\
\hline
\hline
\rule[-6pt]{0pt}{1.75em}$X_{5/3}$ & $\checkmark$ & --- & --- & \small $\varepsilon$ & --- & --- & --- & --- & ---\\
\rule[-6pt]{0pt}{1.5em}$X_{2/3}$ & --- & $\checkmark$ & $\checkmark$ & --- & \small $\varepsilon$ & \small $\varepsilon$
&\, \small $\varepsilon$\ {\footnotesize (not $5+1$)} & --- & ---\\
\rule[-6pt]{0pt}{1.5em}$T$ & --- & $\checkmark$ & $\checkmark$ & --- & \small $\varepsilon$ & \small $\varepsilon$
& \small $\varepsilon$ & --- & ---\\
\rule[-6pt]{0pt}{1.5em}$B$ & $\checkmark$ & --- & --- & \small $\varepsilon$ & --- & --- & --- & ---
&\, \small $\varepsilon$\ {\footnotesize (only $14+1$)}\\
\hline
\rule[-6pt]{0pt}{1.75em}$\widetilde T$ & --- & \small $\varepsilon$ & \small $\varepsilon$ & ---
& $\checkmark$ & $\checkmark$ & $\checkmark$ & --- & ---\\
\hline
\rule[-6pt]{0pt}{1.75em}$Y_{5/3}$ & $\varepsilon$ & --- & --- & $\checkmark$ & --- & --- & --- & --- & ---\\
\rule[-6pt]{0pt}{1.5em}$Y_{2/3}$ & --- & $\varepsilon$ & $\varepsilon$ & --- & $\checkmark$ & $\checkmark$ & $\checkmark$ & --- & ---\\
\rule[-6pt]{0pt}{1.5em}$Z_{2/3}$ & --- & $\varepsilon$ & $\varepsilon$ & --- & $\checkmark$ & $\checkmark$ & $\checkmark$ & --- & ---\\
\rule[-6pt]{0pt}{1.5em}$B^+_{-1/3}$ & $\varepsilon$ & --- & --- & $\checkmark$ & --- & --- & --- & $\checkmark$ & ---\\
\rule[-6pt]{0pt}{1.5em}$B^-_{-1/3}$ & $\varepsilon$ & --- & --- & $\checkmark$ & --- & --- & --- & --- & $\checkmark$\\
\rule[-6pt]{0pt}{1.5em}$Z_{-4/3}$ & --- & --- & --- & --- & --- & --- & $\checkmark$ & --- & ---
\end{tabular}
\caption{Couplings of the composite resonances with the third generation SM
quarks mediated by the gauge fields or the Higgs.
The couplings that are present at leading order are denoted by a $\checkmark$ mark,
while the ones generated only at subleading order in $v/f$ are denoted
by $\varepsilon$.
}
\label{tab:coupling order}
\end{table}

A few peculiar things regarding the couplings of the charge $-1/3$ partners are worth mentioning.
It is straightforward to check that the composite sector is automatically invariant under the $P_{LR}$
discrete symmetry, which arises as an accidental invariance of the leading effective
Lagrangian~\cite{Mrazek:2011iu}.\footnote{See Appendices~\ref{discrete} and \ref{sec:custodial_symmetries}
for a detailed discussion of the $P_{LR}$ symmetry and its implications.}
This symmetry is also preserved by the mixing of the elementary $b_L$ and $t_R$ fields, whereas it is
broken by the mixing of the $t_L$.
An important implication of the $P_{LR}$ invariance is the fact that
the $Z$ coupling to a $P_{LR}$ eigenstate is canonical and is not modified even after EWSB.
Moreover the $Z$ boson can not mediate
flavor-changing currents between two eigenstates with the same $P_{LR}$ parity. Corrections to the $Z$ couplings can only arise
through the couplings that break the $P_{LR}$ symmetry. As far as we are only interested in tree-level effects,
however, the breaking induced by the $t_L$ mixing can not affect the charge $-1/3$ fermionic sector, which is thus
completely invariant under $P_{LR}$.\footnote{Other
corrections can arise from finite-mass effects due to the $Z$ boson. These effects however are suppressed by
$m_Z^2/m_\rho^2$ and are negligible.}
From the embedding of the elementary and composite fields into $\SO(4)$ multiplets, one can check that
all the charge $-1/3$ states are odd under $P_{LR}$, with the exception
of the $B^+_{-1/3}$ field coming from the $\mathbf{9}$, which instead is even.
As a consequence, the only resonance that can be mixed to the $b_L$
through the $Z$ is the $B^+_{-1/3}$ state, whereas the $B$ and $B^-_{-1/3}$ do not have such couplings.

Another peculiar result on the $b_L$ couplings is obtained in the $5+1$ model. In this set-up
for accidental reasons the mass eigenstate $X_{2/3}$ is exactly orthogonal to the state that
is coupled to the $b_L$ through the $W$ boson. This means that the only state that is coupled
to the $b_L$ is the $T$ and not the $X_{2/3}$. Notice that this accidental alignment is not
present in other models, for instance in the $14+1$ set-up the coupling $W X_{2/3} b_L$ exists,
although, as expected, it is suppressed by a $v/f$ factor.

Finally, the trilinear coupling of the $b_L$ and $B$ fields involving the physical Higgs
is present only in the $14+1$ model and arises at order $v/f$. In the $5+5$ and $5+1$
models, instead, the charge $-1/3$ fields are not coupled at all with the physical Higgs.

Differently from the $b_L$ couplings, the interactions involving the top field follow
a much more regular pattern. At leading order in $v/f$, the fourplet states are
coupled only to the $t_R$ field which is realized as an $\SO(4)$ singlet in all the scenarios
we considered. On the contrary the leading interactions of the $\widetilde T$ resonance
and of the $\bf 9$ involve the left-handed top component. It is worth mentioning that, in the limit
in which the fourplet is completely decoupled, the $Z$ boson coupling
of the $\widetilde T$ resonance with the $t_R$ exactly vanishes at all orders in $v/f$.

\begin{table}[t]
\centering
\tabcolsep=3pt
\begin{tabular}{@{}c | c | c@{}}
&\rule[-6pt]{0pt}{1.25em}$t_R$ elementary & $t_R$ composite\\
\hline
\hline
\rule[-6pt]{0pt}{2.em}$\phi^+ \overline X_{5/3L} \,t_R$
& $- y_{R4} \,c_{\phi_R}\! +\! \sqrt{2} \, c_R \,s_{\phi_R} m_{X_{5/3}}/f$
& $ \sqrt{2} \, c_t  \, m_{X_{5/3}}/f$ \\
\rule[-6pt]{0pt}{2.em}$(\rho+i \phi^{0})\overline X_{2/3L} \,t_R $
& $- \frac{y_{R4}}{\sqrt 2} \,c_{\phi_R} + c_R \, s_{\phi_R} \, m_{X_{2/3}}/f $
& $ c_t  \, m_{X_{2/3}}/f$  \\
\rule[-6pt]{0pt}{2.85em}$(\rho-i \phi^{0})\overline{T}_L \,t_R $
&
$
\begin{array}{c}
\frac{y_{L1}}{\sqrt 2} s_{\phi_L} s_{\phi_R} + \frac{y_{R4}}{\sqrt{2}} c_{\phi_L} c_{\phi_R}\\
\rule{0pt}{.85em} -c_R \, s_{\phi_R} \, m_{T}/f
\end{array}
$
& $\frac{y_{Lt}}{\sqrt 2} \,s_{\phi_L} - c_t  \, m_{T}/f $\\
\rule[-14pt]{0pt}{3.35em}$\phi^- \overline{B}_L \,t_R$
&
$
\begin{array}{c}
-  y_{L1} s_{\phi_L} s_{\phi_R} - y_{R4} c_{\phi_L} c_{\phi_R}\\
\rule{0pt}{1em} + \sqrt 2 \, c_R \, s_{\phi_R} \, m_{B}/f
\end{array}
$
& $- y_{Lt}\, s_{\phi_L}  + \sqrt 2 \, c_t  \, m_{B}/f$  \\
 \hline
\rule[-14pt]{0pt}{3.25em}$(\rho+i \phi^{0}) \overline {\widetilde T}_R \,t_L $
&
$
\begin{array}{c}
- \frac{y_{L1}}{\sqrt 2} c_{\phi_L} c_{\phi_R} - \frac{y_{R4}}{\sqrt{2}} s_{\phi_L} s_{\phi_R}\\
\rule{0pt}{.85em} + c_L \, s_{\phi_L} \, m_{\widetilde T}/f
\end{array}
$
&  $- \frac{y_{L1}}{\sqrt 2} \,c_{\phi_L}\! + c_L \, s_{\phi_L} m_{\widetilde T}/f $   \\
 \hline
\rule[-12pt]{0pt}{3.em}$ \phi^+ \overline{\widetilde{T}}_R \,b_L$
&
$
\begin{array}{c}
y_{L1} c_{\phi_L} c_{\phi_R} + y_{R4} s_{\phi_L} s_{\phi_R}\\
\rule{0pt}{1.15em} - \sqrt 2 \, c_L \, s_{\phi_L} \, m_{\widetilde T}/f
\end{array}
$
&  $y_{L1}\, c_{\phi_L}\! - \sqrt 2 \, c_L \, s_{\phi_L} m_{\widetilde T}/f $   \\
\end{tabular}
\caption{Leading-order trilinear couplings of the Goldstone bosons involving one heavy resonance
and one third-generation SM fermion. On the second column ``$t_R$ elementary'' we report the
results for the $5+5$ model. On the third column ``$t_R$ composite'' we list the results
for the $5+1$ and $14+1$ models that give rise to the same coupling at leading order
in the $v/f$ expansion. We denoted by $s_{\phi_{L,R}}$ and $c_{\phi_{L,R}}$ the sine and cosine of the
mixing angles $\phi_{L,R}$.}
\label{tab:leading_couplings}
\end{table}
The explicit expressions for the leading order couplings of the fourplet and singlet states
are reported in Table~\ref{tab:leading_couplings}.\footnote{For shortness we do not report the
explicit couplings of the resonances in the $\bf 9$ multiplet. The explicit expressions for the leading
couplings of these states can be found in Ref.~\cite{Matsedonskyi:2014lla}.}
To simplify the results we introduced the two angles $\phi_L$ and $\phi_R$ defined by
\begin{equation}
\tan \phi_L = \frac{y_{L4} f}{m_4}\,,
\qquad \quad
\tan \phi_R = \frac{y_{R1} f}{m_1}\,,
\end{equation}
which parametrize the amount of compositeness of the $q_L$ and $t_R$ SM fields or,
in other words, correspond to the rotation angles that connect the gauge eigenstate
basis with the mass eigenstate one before EWSB.

The leading order couplings in the case of the $5+1$ and $14+1$ models are exactly
the same. Moreover they coincide with the ones derived in the $5+5$ model in the limit
of large $t_R$ compositeness, $\sin \phi_R = 1$, as can be seen by a trivial
identification of the $d$-symbol term coefficients. It is important to stress that, although the interactions
of the resonances in the fully composite $t_R$ models can be obtained as a limiting case
of the scenarios with an elementary $t_R$, the two set-ups do not fully coincide in this limit.
Indeed in the $5+5$ model the mixing of the elementary $t_R$ breaks the global $\SO(5)$ invariance
(for instance it generates a contribution to the Higgs potential), whereas in the $5+1$ and $14+1$
models the $t_R$ field and its interactions are completely invariant under $\SO(5)$.

Another interesting result is the fact that
the interactions of the $X_{5/3}$ and $T$ resonances coincide with the
ones for the $X_{2/3}$ and $B$ states apart from an overall $\sqrt{2}$ factor.
This is a consequence of the $\SU(2)_L$ invariance that is respected by the leading-order couplings.

As can be seen from the explicit results in Table~\ref{tab:leading_couplings},
the strength of the $d$-symbol interactions is controlled by the mass of the heavy resonance and
thus can significantly enhance the trilinear couplings. This is especially true in the scenarios
with a fully composite $t_R$, where the terms proportional to $c_t$ control the
interactions between the fourplet and the $t_R$. However, also in the other cases, the
contributions from the strong sector interactions can be sizable. For instance, in the
$5+5$ model, although the $d$-symbol term does not directly generate a coupling
involving the elementary fermions, it can significantly modify the trilinear interactions
in Table~\ref{tab:leading_couplings} through the elementary/composite mixings.
Of course this contribution is only present if both the fourplet and the singlet
resonances are present in the effective Lagrangian. The decoupling of this effect is
governed by the mixing angles $\sin \phi_{L,R}$, thus even a relatively heavy multiplet
can affect the couplings of the lighter resonances provided that its mixing angle is not
too small.

It is also important to stress that, from the point of view of our effective approach,
the $d$-symbol interactions are controlled by free parameters, $c_t$ and $c_{L,R}$,
and the values of the trilinear couplings relevant for the single-production
and the decays are, to a large extent, arbitrary.
This has to be contrasted with what happens in renormalizable theories with additional vector-like
quarks. In that case the trilinear interactions involving a SM field and a resonance arise only from the
usual gauge interactions after rotating the fields from the gauge-eigenstate basis to the mass-eigenstate one.
This means that the value of these couplings is tightly related to the mixings between the elementary
and the composite states and their maximal size is determined by the SM gauge couplings and
by the EW charges of the resonances. In composite Higgs models, on the contrary, this constraint is not there and these
couplings can be significantly larger.

To conclude the discussion it is useful to comment on the connections between the Goldstone
couplings and the gauge couplings in the unitary gauge which is commonly used
in the explicit computations and for parametrizing the production cross sections
(see Refs.~\cite{DeSimone:2012fs,Matsedonskyi:2014mna}).
The gauge bosons couplings in the unitary gauge at the leading order in $v/f$
can be easily derived from the leading Goldstone couplings listed in
Table~\ref{tab:leading_couplings}. The connection is given by the relation
\begin{equation}
g_{W\!Xt_{L,R}} = \frac{g v}{2 m_X} g^{Gold}_{\phi^\pm \!Xt_{L,R}}\,,
\qquad \quad
g_{Z\!Xt_{L,R}} = \frac{g v}{2 c_w m_X} g^{Gold}_{\phi^0 \!Xt_{L,R}}\,,
\end{equation}
where $g$ is the $\SU(2)_L$ gauge coupling and $c_w$ is the cosine of the weak mixing angle.
Analogous relations are valid for the leading $b_L$ couplings.

In general the size of the single production couplings can vary considerably in different models
depending on the resonance species and on the detailed implementation of partial compositeness.
It is however possible to derive a simple generic estimate which can be used to get a rough idea of
the expected coupling size. The single production interactions in the unitary gauge
are always proportional to the EW symmetry breaking scale $v$. This can be easily understood
by noticing that the gauge interactions are flavor diagonal if the EW symmetry is unbroken.
In the composite Higgs scenarios any $v$ insertion is accompanied by a factor $1/f$, therefore
the couplings are proportional to the universal factor
\begin{equation}\label{eq:coupl_est}
c_{single} \sim g \frac{v}{f} = g \sqrt{\xi}\,,
\end{equation}
where the EW coupling factor $g$ is due to the fact that the single production couplings
are EW gauge interactions. Given that, in reasonably Natural and viable scenarios, $\xi \sim 0.1$,
the above estimate suggests a typical value $c_{single} \sim 0.2$, even though considerable numerical
enhancements are possible in explicit models.


\subsection{Collider phenomenology}\label{sec:coll_phenomenology}

In this subsection we briefly discuss the collider phenomenology of the top partners.
For simplicity we focus on simplified scenarios in which all the relevant light fermionic resonances
belong to a single $\SO(4)$ multiplet. Although more generic situations are possible, this assumption is not
a very restrictive one. From he point of view of collider phenomenology, a mass difference of a few
hundred $\mathrm{GeV}$ is typically enough to suppress the role of the heavier states due
to a reduced production cross section. In this case considering only the lightest
multiplet is typically a very good approximation. Exceptions to this rule are possible in non-generic cases in
which the heavier states have exceptionally large couplings to the SM fields or lead to final states which are
particularly easy to identify. In these situations considering only the lightest multiplet can lead to an
underestimation of the signal and to milder exclusion bounds.

\subsubsection{The fourplet}

As a first case we concentrate on the scenario with only a light fourplet.
The lightest state in the spectrum is given by the exotic $X_{5/3}$ resonance.
Due to its electric charge its only gauge interaction with the SM fermions involves the top
quark, hence its only non-negligible decay channel is into a top and a $W$ boson,
$\mathrm{BR}(X_{5/3} \rightarrow W^+ t) \simeq 1$.
The size of the coupling to the top has a different parametric dependence in the scenarios with a
fully composite and an elementary $t_R$. In the former case, the coupling strength follows the general estimate
in Eq.~(\ref{eq:coupl_est}); in the unitary gauge one gets
\begin{equation}
g_{X_{5/3}t_R} = \frac{g}{\sqrt{2}} c_t \frac{v}{f}\,.
\end{equation}
On the other hand, in the generic scenarios with an elementary $t_R$ the coupling tends to be suppressed at high
resonance masses:\footnote{Significant deviations from this estimate can appear
if a light singlet is present or if the $t_R$ mixing is much larger than the $t_L$ one
($y_R \gg y_L$). In these cases the coupling follows the general estimate
in Eq.~(\ref{eq:coupl_est}).}
\begin{equation}
g_{X_{5/3}t_R} \simeq \frac{g}{2} \frac{v}{f} \sqrt{y_{top}\frac{f}{m_{X_{5/3}}}}\,.
\end{equation}

Apart from pair production, which typically dominates
at small masses ($m_{X_{5/3}} \lesssim 700\ \mathrm{GeV}$ at the $8\ \mathrm{TeV}$ LHC), the $X_{5/3}$
resonance can also be singly produced in association with a top.
Both production channels give rise to final states containing same-sign
leptons, which can be efficiently detected at the LHC~\cite{Contino:2008hi,Mrazek:2009yu,Dissertori:2010ug}.

Contributions to the same final states also come from the $B$ resonance. This state is always heavier than the
$X_{5/3}$, thus its contribution to the signal is relevant only if the mass split is not too large
($\Delta m \ll 500\ \mathrm{GeV}$)~\cite{Mrazek:2009yu,Matsedonskyi:2014mna}.
The dominant coupling of the $B$ involves the top quark, thus it can be singly produced in association with a top
and its most relevant decay channel is $B \rightarrow Wt$.

In all the scenarios we considered the $X_{5/3}$ resonance forms a nearly-degenerate doublet with the $X_{2/3}$.
At leading order, the $X_{2/3}$ couples to the top through the $Z$ boson and the Higgs. These two couplings have
similar strength and determine the dominant branching ratios
$\mathrm{BR}(X_{2/3} \rightarrow Z t) \approx \mathrm{BR}(X_{2/3} \rightarrow h t) \approx 1/2$.

The last state inside the fourplet, the charge-$2/3$ $T$, is close in mass to the $B$. Its phenomenology
is similar to the one of the $X_{2/3}$ resonance and the two states have similar production and decay channels
($\mathrm{BR}(T \rightarrow Z t) \approx \mathrm{BR}(T \rightarrow h t) \approx 1/2$).
Given the sizable mass gap between the $T$ and the lightest resonances inside the fourplet, the chain decays
$T \rightarrow Z X_{2/3}$, $T \rightarrow h X_{2/3}$ and $T \rightarrow W X_{5/3}$ are usually kinematically
allowed. The corresponding couplings, however, arise at subleading order on $v/f$, thus the direct decays
into SM states are favored.

To conclude we summarize the constraints coming from the $8$ TeV LHC searches.
The experimental collaborations performed several searches for pair-produced fermionic
resonances. The most sensitive ones exploit the same-sign leptons final states to look for
exotic $X_{5/3}$ resonances. They provide the bound $m_{X_{5/3}} \gtrsim 770\ \mathrm{GeV}$~\cite{Chatrchyan:2013wfa,TheATLAScollaboration:2013jha}.
Several searches for charge-$2/3$ resonances have also been presented.
The strongest bounds on these states are given by $m_{X_{2/3}} \gtrsim 700\ \mathrm{GeV}$ and
have been obtained by the CMS collaboration by combining the three decay channels $bW$, $tZ$ and $tH$~\cite{Chatrchyan:2013uxa}.
Other searches for charge-$2/3$ resonances performed by the ATLAS
collaboration are available in the literature. They include, in particular, searches for resonances decaying
into a single channel ($Zt$~\cite{TheATLAScollaboration:2013oha} and $ht$~\cite{ATLAS:2013ima}),
as well as searches for resonances giving rise to final states with two same-sign
leptons~\cite{TheATLAScollaboration:2013jha}.

\begin{figure}
\centering
\includegraphics[height=.32\textwidth]{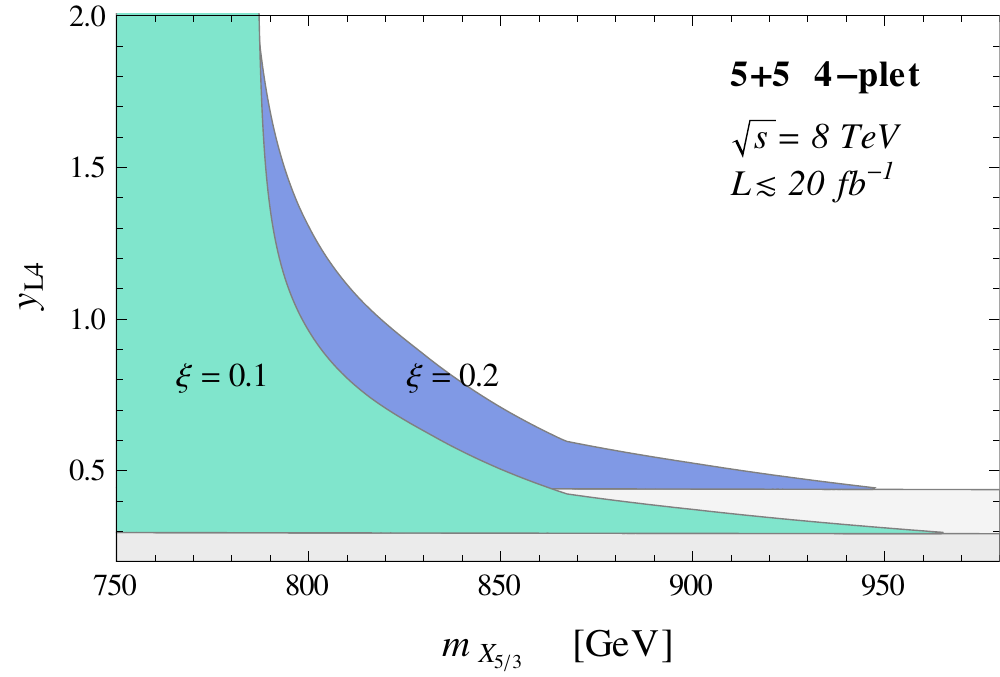}
\hfill
\includegraphics[height=.32\textwidth]{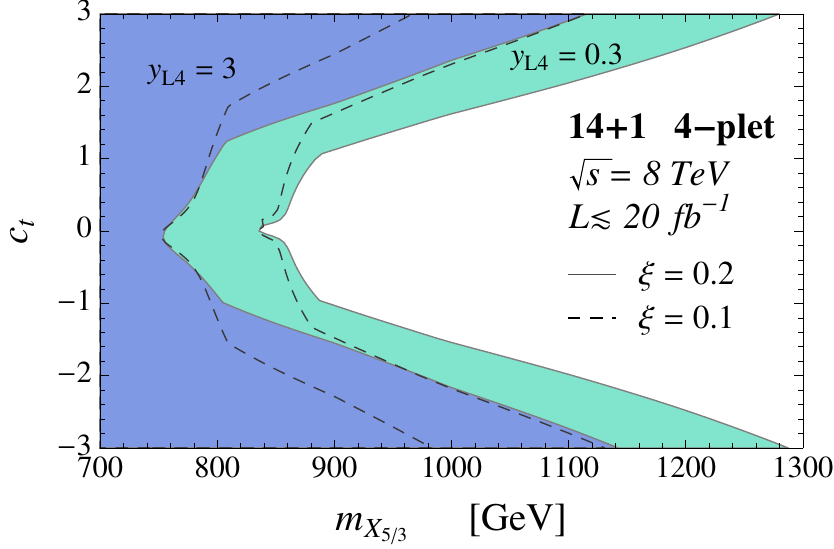}
\caption{Exclusions in the scenarios with only a light fourplet
for the $8\ \mathrm{TeV}$ LHC data.
On the left panel: $5+5$ model for $\xi = 0.1$ (green region) and $\xi = 0.2$ (blue region). The gray areas
are theoretically excluded.
On the right panel: exclusions for the $14+1$ model in the $(m_{X_{5/3}}, c_t)$ plane.
The blue (green) region shows the excluded points for $y_{L4} = 3$ ($y_{L4} = 0.3$) for $\xi = 0.2$.
The exclusions for $\xi = 0.1$ are denoted by the dashed contours.
}
\label{fig:4plet_8TeV}
\end{figure}

Although the single production channels have not been taken into account in the experimental analyses
at the $8$ TeV LHC, their impact on the exclusions can be non-negligible.
Estimates of the constraints on the parameter space of the $5+5$ and $14+1$ scenarios are shown in
Fig.~\ref{fig:4plet_8TeV}.\footnote{One free parameter, namaly $y_{L4}$ in
the $5+5$ model and $y_{Lt}$ in the $14+1$, has been fixed by requiring the correct value of the top mass.}
These results include the same-sign lepton signal coming
from single production of the $X_{5/3}$ resonance and the additional contributions due to the $B$ resonance.
One can see that these effects can significantly increase the bound on $m_{X_{5/3}}$ and push
it above $1\ \mathrm{TeV}$ in a sizable part of the parameter space.
The exclusions for the $5+1$ scenario only slightly differ from
the ones for the $14+1$ case, the differences in the bounds being of order $20\ \mathrm{GeV}$.
The full LHC program should be able to extend the reach for top partners
to the $2-3\ \mathrm{TeV}$ range.

\subsubsection{The singlet}

The second scenario we focus on is the one with only a light composite singlet. The phenomenology of
this scenario is quite simple because the low-energy spectrum contains only one light resonance, the $\widetilde T$.
As shown in Table~\ref{tab:coupling order}, the $\widetilde T$ resonance has three leading-order
couplings, which involve the left-handed top and bottom quarks.
The $\SU(2)_L$ invariance fixes the relative size of these couplings (see Table~\ref{tab:leading_couplings})
and determines the dominant decay branching ratios: $\mathrm{BR}(\widetilde T \rightarrow Wb) \approx 1/2$ and
$\mathrm{BR}(\widetilde T \rightarrow Zt) \approx \mathrm{BR}(\widetilde T \rightarrow ht) \approx 1/4$.
Notice that the decay channel into the bottom quark is further enhanced by
the larger phase space, although this effect is relevant only for small $\widetilde T$ mass.

The presence of large $W$ and $Z$ couplings to the top and bottom quarks
implies a sizable cross section for the single production channels.
In particular single production in association
with a bottom quark is the dominant production mechanism in a large part of the parameter space.
The leading-order couplings of the $\widetilde T$ resonance can be easily estimated.
In the case with an elementary $t_R$ one finds
\begin{equation}
g_{\widetilde T b_L} = -\sqrt{2} c_w g_{\widetilde T t_L} \simeq \frac{g}{2}\frac{v}{f} \sqrt{y_{top} \frac{f}{m_{\widetilde T}}}\,,
\end{equation}
which is suppressed at large resonance masses with respect to the general estimate in Eq.~(\ref{eq:coupl_est}).
In the models with a fully composite $t_R$ the estimate of the coupling strength becomes
\begin{equation}\label{eq:gTbl_composite}
g_{\widetilde T b_L} = -\sqrt{2} c_w g_{\widetilde T t_L} = \frac{g}{2} y_{L1} \frac{v}{m_{\widetilde T}}\,.
\end{equation}
It is interesting to notice that the expression for $g_{\widetilde T b_L}$ in
Eq.~(\ref{eq:gTbl_composite}) depends only on the $y_{L1}$ mixing and on the $\widetilde T$
resonance mass, but not on the compositeness scale $f$. This implies that the exclusion
bounds, when expressed as a function of $y_{L1}$ are almost independent of $\xi$.

In the scenarios with only a light singlet a strong correlation exists between the
coupling responsible for $b$-associated single production, $g_{\widetilde T b_L}$, and the
$V_{tb}$ element of the CKM matrix. Before EWSB, the only $W$-mediated coupling involving
the $b$ quark is the usual $\overline t_L \slashed W b_L$ vertex, included in the
elementary-fields Lagrangian. After EWSB, due to the mixing between the top and the
L-handed component of the $\widetilde T$ resonance, the $g_{\widetilde T b_L}$ coupling is
generated together with some corrections to the $\overline t_L \slashed W b_L$ coupling.
It is straightforward to see that the following relation holds
\begin{equation}\label{eq:gTb-Vtb}
g_{\widetilde T b_L}^2 + g_{tb}^2 = g^2/2 \qquad \Rightarrow
\qquad g_{\widetilde T b_L} = g \sqrt{\delta V_{tb} - \delta V_{tb}^2/2}\,,
\end{equation}
where $g_{tb}$ denotes the $\overline t_L \slashed W b_L$ coupling and $\delta V_{tb} = 1 - |V_{tb}|$.
The requirement that the correction to $V_{tb}$ should be small implies an upper
bound on the $g_{\widetilde T b_L}$ coupling and on the single production cross section.
The current measurements of the $V_{tb}$ matrix element give the value
$|V_{tb}| = 1.021 \pm 0.032$~\cite{Agashe:2014kda}. Taking into account the fact that in our scenario
$|V_{tb}| \leq 1$, the experimental bound implies $g_{\widetilde T b_L} \leq 0.21\, g$
at the $2\sigma$ level.
Obviously, if additional relatively light resonances are present, the relation in
Eq.~(\ref{eq:gTb-Vtb}) may be modified and larger values of $g_{\widetilde T b_L}$ could be
compatible with sufficiently small deviations in $V_{tb}$. This however would probably require
a certain degree of additional tuning.

The $8\ \mathrm{TeV}$ LHC bounds on this scenario come from the searches for pair produced charge-$2/3$ states.
The strongest ones ($m_{\widetilde T} \gtrsim 700\ \mathrm{GeV}$)
have been obtained by the CMS collaboration~\cite{Chatrchyan:2013uxa}.
Slightly milder bounds have been derived by the ATLAS
collaboration by looking into single channels ($Wb$~\cite{TheATLAScollaboration:2013sha},
$Zt$~\cite{TheATLAScollaboration:2013oha} and $ht$~\cite{ATLAS:2013ima}))
or looking for final states with two same-sign leptons~\cite{TheATLAScollaboration:2013jha}.

\begin{figure}
\centering
\includegraphics[height=.315\textwidth]{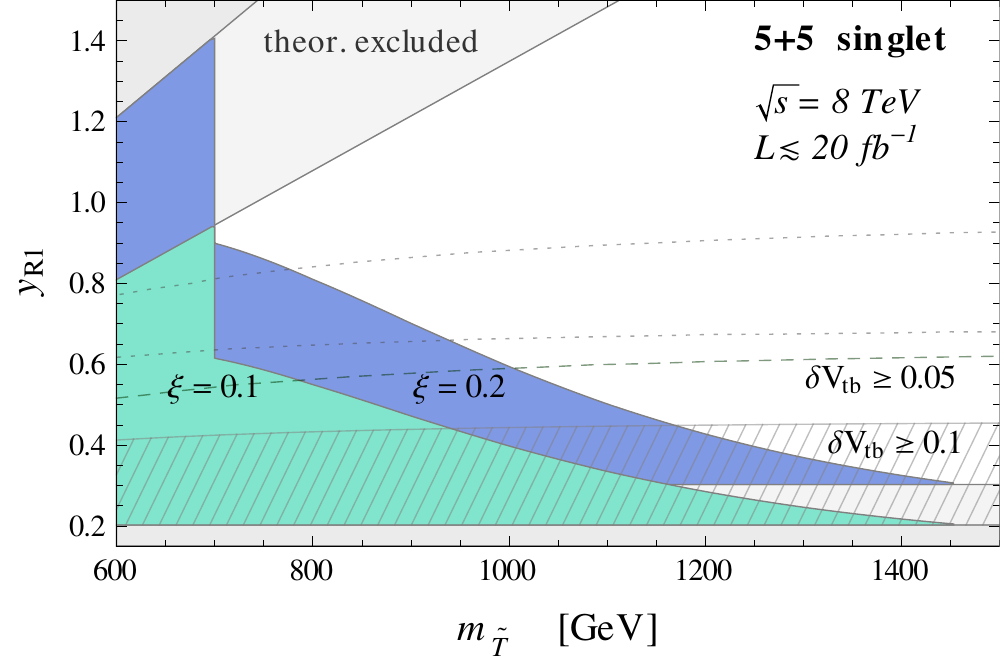}
\hfill
\includegraphics[height=.321\textwidth]{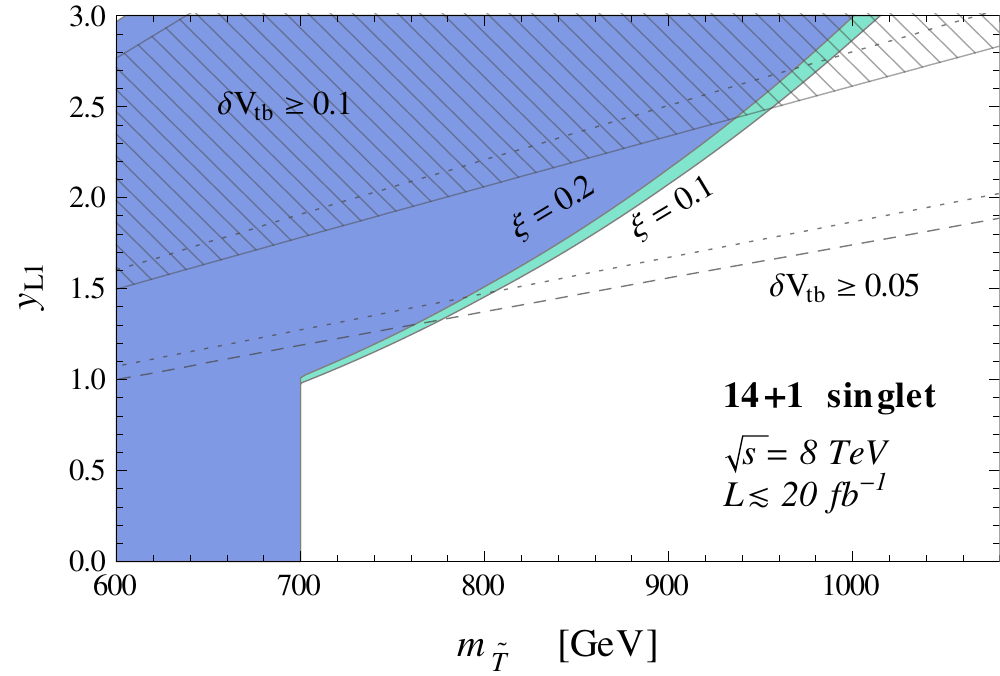}
\caption{Exclusions in the $5+5$ (left panel) and $14+1$ scenarios (right panel) with only a composite singlet
for the $8\ \mathrm{TeV}$ LHC data.
The blue (green) region shows the excluded points for $\xi = 0.2$ ($\xi = 0.1$).
The striped regions correspond to the points
with $\delta V_{tb} \geq 0.1$ for $\xi = 0.1$, while the boundary of the region with $\delta V_{tb} \geq 0.05$
is denoted by the dashed gray lines. The dotted gray lines show how these regions change for $\xi = 0.2$.}
\label{fig:singl_8TeV}
\end{figure}

Due to the large single production cross section in association with a bottom quark, it is conceivable
that the bounds can be significantly increased by a modified search strategy which could be
sensitive to this additional channel. Possible search strategies have been
proposed in the literature~\cite{Ortiz:2014iza}. The exclusion bounds in the $5+5$ and $14+1$ scenarios for
the $8\ \mathrm{TeV}$ LHC data are shown in Fig.~\ref{fig:singl_8TeV}.
The results shown in the plots include the bounds from the experimental analyses and the estimate
of the constraints coming from single production searches. The bounds for the $5+1$ scenario
are very close to the ones for the $14+1$ model.

\subsubsection{The nineplet}

As a last case we discuss the scenario with only a light nineplet. As we saw in Sect.~\ref{sec:effect_parametr}
in this model all the resonances are almost degenerate, thus they can all be pair produced with similar rates via
QCD interactions. Moreover direct decays into SM states, if allowed, are favored with respect to chain decays, which
should necessarily proceed off-shell. The phenomenology of the states with charge $2/3$,
$-1/3$ and $5/3$ is quite similar to the one we discussed in the two previous scenarios.\footnote{The interested reader can find a more
detailed discussion in Ref.~\cite{Matsedonskyi:2014lla}.}
In particular the bounds coming from the $8\ \mathrm{TeV}$ LHC experimental searches apply to the nineplet resonances as well.
A small difference with respect to the previous scenarios is the fact that, due to the presence of multiple almost-degenerate states,
each final state benefits from an enhanced signal.
By taking into account this effect slightly stronger bounds than in the fourplet and singlet cases can be derived.

In addition to the above mentioned resonances, the nineplet also contains two exotic-charge states,
the $U_{8/3}$ and the $Z_{-4/3}$. The latter has a simple phenomenology since it can couple only to the bottom quark.
This fixes its decay channel, $Z_{-4/3} \rightarrow W^- b$, and implies that it can be singly produced only in association with a $b$.
On the other hand, the phenomenology of the $U_{8/3}$ resonance is more peculiar. Due to its charge, it can not
simply decay into a gauge boson and a SM quark. Instead it must decay into a three-body final state $U_{8/3} \rightarrow W^+ W^+ t$.
This decay can be mediated by an off-shell charge-$5/3$ resonance (the $X_{5/3}$ or the $Y_{5/3}$) or by
a contact interaction coming from the higher-order operator
\begin{equation}
{\cal L}_{UV} = i \frac{c_{t9}}{m_*} \overline \psi_9^{ij} d_\mu^i d^{\mu, j} t_R + \mathrm{h.c.}\,,
\end{equation}
where $c_{t9}$ is expected to be of order one.
It turns out that, if $m_*$ is in the multi-TeV range, the contribution due to the contact interaction is subleading
with respect to the one mediated by an off-shell resonance. For smaller values of $m_*$, instead, the
contact interaction can play a significant role and can modify the angular and invariant mass distributions of the decay products.

The $U_{8/3}$ state can be easily probed
in final states with two or three same-sign leptons. The searches for charge-$5/3$ states, which exploit the former channel,
can thus be reinterpreted for the $U_{8/3}$ resonance. Since several $W$ bosons
are present in the final state, the probability of obtaining two or more same sign leptons is quite high.
As a consequence, the efficiency for the two same-sign lepton searches is enhanced (it is nearly an
order of magnitude higher than for a charge-$5/3$ resonance), making the $U_{8/3}$ the best target to probe a
light nineplet. The $8\ \mathrm{TeV}$ LHC searches can be recast to get the bound $m_9 \gtrsim 990\ \mathrm{GeV}$~\cite{Matsedonskyi:2014lla},
which is much stronger than the ones obtained from pair production in the light fourplet and light singlet scenarios.

\subsection{Other fermionic partners}

To conclude the discussion about the fermionic partners, it is worth mentioning that
other classes of fermionic resonances are usually present in the composite Higgs scenarios.
As we saw in Chaps.~\ref{ch:Goldstone} and \ref{ch:flavor}, in several models,
most noticeably the ones in which the composite states belong to the
fundamental representation of $\SO(5)$, additional fermionic partners specifically connected to the
bottom sector are present. These states are responsible for generating the bottom mass and
are mainly coupled to the $b_R$ component and to the $q_L = (t_L, b_L)$ doublet.
Being dominantly coupled to third-generation quarks, the bottom partners
have the same collider phenomenology of top partners. The only peculiarity of these resonances
is the fact that their mixing with the elementary states, being controlled by the bottom mass,
is typically much smaller than the one of the top partners. The bottom partners can be easily parametrized
by a generalization of the formalism we used in this chapter. Their relevance for
collider experiments has been analyzed in Refs.~\cite{Kumar:2010vx,Alvarez:2013qwa,Gillioz:2013pba}.

As already mentioned and explained in details in Sect.~\ref{sec:anarchic_pc},
in the anarchic flavor scenarios all the quark partners have sizable
mixing with the third-generation quarks, thus they unavoidably belong to the wide category of top (or bottom) partners. 
The situation can be quite different in models based on flavor symmetries. In this case the light-generation
quarks can be associated to specific sets of fermionic partners whose mixing with the heavy SM quarks
is quite small. These states can be copiously produced at hadron colliders and their cross section can be significantly
enhanced by the fact that the light SM quarks can have a sizable amount of compositeness.
The decay channels of the light-generation partners
are quite different from the ones of the top partners. These states mainly decay into light quarks (i.e.~jets)
via interactions mediated by the SM gauge bosons or the Higgs. This kind of decays leads to very distinctive collider signatures.
We refer the interested reader to the original
literature~\cite{Delaunay:2010dw,Redi:2011zi,Delaunay:2011vv,DaRold:2012sz,Delaunay:2013iia,Redi:2013eaa,Delaunay:2013pwa}.

Finally, in addition to the quark partners, composite Higgs models can also predict lepton resonances.
For instance, this happens if the lepton sector follows the usual partial compositeness structure
(see Sect.~\ref{sec:leptonic_sector} for a discussion on this topic). Obviously the collider phenomenology of the
lepton partners differs significantly from the one of the quark partners. A fundamental difference is the fact that
the lepton partners are neutral under QCD, thus they can not be easily produced at a hadron collider.
Only a limited attention has been devoted so far to the phenomenology of these resonances. The main
results can be found in Refs.~\cite{delAguila:2010es,Falkowski:2013jya}, which focus on scenarios characterized
by a large tau compositeness.


\section{Vector resonances}\label{sec:vector_resonances}

In the previous section we focused on the phenomenology of the fermionic resonances.
In particular we considered them ``in isolation'', that is, we assumed them to be the only light composite states
coming from the new strong dynamics and we neglected effects due to other possible resonances.
However, as we saw in the general discussions in Chaps.~\ref{ch:Goldstone} and \ref{ch:SILH},
the composite dynamics typically gives rise to a much richer set of relatively light resonances.
For instance in One-Scale-One-Coupling theories (see Chap.~\ref{ch:SILH}) we expect the presence of
composite vector states with a mass
roughly comparable to the one of the fermionic resonances. In such a scenario a non-trivial interplay in the
phenomenology of the fermionic and vector resonances is expected.

Going beyond the One-Scale-One-Coupling scenarios and taking into account Naturalness considerations,
a slightly more refined picture emerges. The fermionic states, in particular the top partners, are usually required
to be quite light ($m_\psi \sim f$) in order to obtain a light Higgs and minimize the amount of fine-tuning.
On the other hand, the Naturalness ``pressure'' on the vector resonances is less severe,
allowing these states to be considerably heavier than the fermions without worsening the amount of tuning.
A high mass for the vector resonances is also preferred by the indirect constraints coming from
flavor physics (see Chap.~\ref{ch:flavor}) and EW precision measurements (see Chap.~\ref{ch:EWPT}).
The former constraints force the QCD vector resonances to be quite heavy, while the latter impose a
lower bound of a few TeV on the mass of $\SU(2)_L$-charged states. Of course these bounds, as
usual for indirect constraints, should be interpreted as rough estimates
and not as strict limits. Cancellations in the indirect bounds are always possible, although they usually
require some additional amount of tuning.

Analogously to what we did for the fermionic resonances in the previous section, in the following
we will discuss the phenomenology of the vector resonances by using a model-independent approach.
Guided by the above considerations, however, we will not describe the vector states alone, instead
we will write some more complete simplified models that also include the dynamics of the composite fermions.
Given that in many scenarios the top partners tend to be the lightest fermionic states, we will only
include them in the effective parametrization and we will assume that other possible fermionic states
are heavier and are not relevant for collider phenomenology.
For most of our discussion we will follow Refs.~\cite{Pappadopulo:2014qza,Greco:2014aza}, 
related studies can be found in Refs.~\cite{Contino:2011np,Bellazzini:2012tv}.

\subsection{The effective parametrizations}

The effective Lagrangians describing the vector resonances can be straightforwardly built by using the
CCWZ formalism.
The vector resonances are classified in terms of their quantum numbers
under the unbroken $\SO(4) \simeq \SU(2)_L \times \SU(2)_R$ group. In the following we will consider
three relevant scenarios, with resonances in the $({\bf 3}, {\bf 1})$, $({\bf 1}, {\bf 3})$ and $({\bf 1}, {\bf 1})$
representations.

In addition to the $\SO(5) \times \U(1)_X$ group, the global symmetry of the composite dynamics
also contains the usual $\SU(3)_c$ QCD group.
This implies that vector resonances with QCD quantum numbers can also be present. These resonances
are usually called ``heavy gluons'' or, adopting the language of extra-dimensional models, ``Kaluza-Klein gluons''.
As we saw in Chap.~\ref{ch:flavor}, these states are important in the context of flavor physics because they can mediate
flavor changing neutral currents, leading to strong bounds in generic composite Higgs scenarios.
From the point of view of the effective CCWZ description, the heavy gluons are singlets under $\SO(4)$ and transform
as an octet under QCD.

In the effective Lagrangians we present in the following we only include the kinetic and mass terms for the
vector resonances and the relevant interactions with the top partners. For simplicity we only consider top partners
that transform as fourplets and singlets under $\SO(4)$. The extension of the formalism to other representations
can be straightforwardly worked out. The Lagrangians for the top partners and for the SM third-generation
quarks coincide with the ones discussed in Sect.~\ref{sec:effect_parametr}. The elementary gauge fields
as well as the light quarks and the leptons are described by the usual SM Lagrangian.

Notice that the effective Lagrangians for the vector resonances can be used both in the standard partial compositeness
scenarios and in the scenarios with a totally composite $t_R$. When necessary we will explicitly comment on the few differences between the two cases.

\subsubsection[$\SU(2)_L$ triplets]{$\SU{\mathbf{(2)_L}}$ triplets}

The first scenario we consider is the one with vector resonances, $\rho^L_\mu$,
belonging to the $({\bf 3}, {\bf 1})$ representation.
We assume that these states transform non-homogeneously under $\SO(4)$:
\begin{equation}
\rho^L_\mu \equiv \rho^{a_L}_\mu t^{a_L}_L \rightarrow
h\left[\Pi; g\right] \rho^L_\mu\, h\left[\Pi; g\right]^T
+ i \left(h\left[\Pi; g\right] \partial_\mu h\left[\Pi; g\right]^T\right)_{L}\,,
\end{equation}
where $t^{a_L}_L$ are the $\SU(2)_L$ generators in $\SO(4)$ and $h\left[\Pi; g\right]$ is the non-linear
$\SO(4)$ transformation corresponding to an $\SO(5)$ element $g$
(the exact definition can be found in Chap.~\ref{ch:Goldstone}, Eq.~(\ref{hmatrix})).
The subscript $L$ in the second term on the right-hand side means that
only the $(\mathbf{3}, \mathbf{1})$ components must be taken (see Appendix~\ref{so54}).
The $\rho^L_\mu$ resonance is also assumed to be neutral under the $\U(1)_X$ symmetry and under QCD.

Due to its quantum numbers, the $\rho^L_\mu$ resonance can be directly coupled to top partners, $\psi_4$, in the
fundamental representation of $\SO(4)$. On the other hand, no interaction
with $\SO(4)$ singlets arises at leading order. The effective Lagrangian reads
\begin{equation}\label{eq:vect_L_lagr}
{\cal L}_{\rho_L} = -\frac{1}{4 g_{\rho_L}^2} \rho^{a_L}_{\mu\nu} \rho^{a_L \mu\nu}
+ \frac{m_{\rho_L}^2}{2 g_{\rho_L}^2} \left(\rho^{a_L}_{\mu} - e^{a_L}_{\mu}\right)^2
+ \kappa\, \overline \psi_4^i \gamma^\mu \left(\rho^{a_L}_{\mu} - e^{a_L}_{\mu}\right) (t^{a_L}_L)_{ij} \psi_4^j\,,
\end{equation}
where $\kappa$ is a free parameter (expected to be of order one) and the $\rho^L_\mu$ field strength is defined as
$\rho^{a_L}_{\mu\nu} = \partial_\mu \rho^{a_L}_\nu - \partial_\nu \rho^{a_L}_\mu
+ \epsilon^{a_L b_L c_L} \rho^{b_L}_\mu \rho^{c_L}_\nu$.
Notice that, due to the non-homogeneous transformation of $\rho^L_\mu$ under $\SO(4)$,
we needed to use the $e_\mu$ CCWZ symbol in order to write suitable mass and interaction terms.
The $\rho_\mu^L - e_\mu^L$ combination, indeed, transforms homogeneously under $\SO(4)$ and
can be used to build the invariant operators in the effective Lagrangian.

The mass term in Eq.~(\ref{eq:vect_L_lagr}) induces a mixing between the vector resonance and the elementary
$\SU(2)_L$ gauge fields. This means that the SM gauge fields do not coincide exactly with the elementary states
but rather acquire a (small) component along the composite vectors. This also implies that
the SM $\SU(2)_L$ gauge coupling, $g$, differs from the elementary coupling $g_0$ and is given by
the combination
\begin{equation}
\frac{1}{g^2} = \frac{1}{g_0^2} + \frac{1}{g_{\rho_L}^2}\,.
\end{equation}
The $\U(1)_Y$ and QCD couplings, instead,
coincide with the elementary ones, $g' = g'_0$ and $g_s = g_{s0}$.

The masses of the SM gauge bosons, at the linear order in $\xi$, are given by the expressions
\begin{equation}\label{eq:sm_bosons_mass}
m_W^2 = \frac{g^2}{4} f^2 \xi\,,
\qquad
m_Z^2 = \frac{g^2 + g'^2}{4} f^2 \xi\,.
\end{equation}
The mass spectrum of the vector resonances is quite simple. The masses of all the resonances coincide at order $\xi$:
\begin{equation}
M^2_{\rho_L^{\pm}} = M^2_{\rho_L^{0}} \simeq \frac{g_{\rho_L}^2}{g_{\rho_L}^2 - g^2} m_{\rho_L}^2
- \frac{\xi}{4}\frac{g^2}{g_{\rho_L}^2 - g^2}\left(2 m_{\rho_L}^2 - f^2 g^2\right)\,.
\end{equation}
Due to the custodial invariance a mass split can only arise at order $\xi^2$ and must be weighted by the
hypercharge coupling $g'$:
\begin{equation}
M^2_{\rho_L^{0}} - M^2_{\rho_L^{\pm}} \simeq \frac{\xi^2}{16} \frac{{g'}^2}{g_{\rho_L}^2} m_{\rho_L}^2
\left(1 - \frac{f^2 g^2}{m_{\rho_L}^2}\right)\,.
\end{equation}
As a consequence the charged and neutral vector resonances have a mass split typically below
$0.1\%$ and can be considered degenerate from the point of view of collider phenomenology.

\subsubsection[$\SU(2)_R$ triplets]{$\SU{\mathbf{(2)_R}}$ triplets}

The effective description of vector resonances in the $({\bf 1}, {\bf 3})$ representation, which we denote by $\rho^R_\mu$,
is quite similar to the one of the $\rho_\mu^L$ states. Under the $\SO(4)$ subgroup  we assume that the $\SU(2)_R$ triplet
transforms as
\begin{equation}
\rho^R_\mu \equiv \rho^{a_R}_\mu t^{a_R}_R \rightarrow
h\left[\Pi; g\right] \rho^R_\mu h\left[\Pi; g\right]^T
+ i \left(h\left[\Pi; g\right] \partial_\mu h\left[\Pi; g\right]^T\right)_{R}\,,
\end{equation}
where $t^{a_R}_R$ correspond to the $SU(2)_R$ generators in $\SO(4)$. As denoted by the subscript $R$,
only the $(\mathbf{1}, \mathbf{3})$ components of the shift term must be considered.
Analogously to the $\SU(2)_L$ triplet, the
$\rho^R_\mu$ resonances can be directly coupled to top partners in the fundamental representation of $\SO(4)$,
while no interactions with the singlets arise at leading order. The effective Lagrangian describing the $\rho^R_\mu$
resonances can be written as
\begin{equation}
{\cal L}_{\rho_R} = -\frac{1}{4 g_{\rho_R}^2} \rho^{a_R}_{\mu\nu} \rho^{a_R \mu\nu}
+ \frac{m_{\rho_R}^2}{2 g_{\rho_R}^2} \left(\rho^{a_R}_{\mu} - e^{a_R}_{\mu}\right)^2
+ \kappa\, \overline \psi_4^i \gamma^\mu \left(\rho^{a_R}_{\mu} - e^{a_R}_{\mu}\right) (t^{a_R}_R)_{ij} \psi_4^j\,,
\end{equation}
where $\rho^{a_R}_{\mu\nu} = \partial_\mu \rho^{a_R}_\nu - \partial_\nu \rho^{a_R}_\mu
+ \epsilon^{a_R b_R c_R} \rho^{b_R}_\mu \rho^{c_R}_\nu$ is the $\rho^R_\mu$ field strength.

Due to the mass term in the effective Lagrangian, the neutral resonance in the $SU(2)_R$ triplet is mixed to the
elementary hypercharge gauge boson $B_\mu$. The SM gauge $\U(1)_Y$ gauge coupling is then given by
\begin{equation}
\frac{1}{{g'^2}} = \frac{1}{g'^2_0} + \frac{1}{g_{\rho_R}^2}\,,
\end{equation}
while the $SU(2)_L$ and QCD couplings coincide with the elementary ones, $g = g_0$ and $g_s = g_{s0}$.
The breaking of the custodial symmetry due to the hypercharge gauging determines a small split of the
neutral and the charged $\rho^R_\mu$ resonances. The masses, up to corrections of order $\xi^2$
are given by
\begin{equation}
M^2_{\rho^\pm_R} = m^2_{\rho_R}\,,
\qquad
M^2_{\rho^0_R} = \frac{g_{\rho_R}^2}{g_{\rho_R}^2 - g'^2} m^2_{\rho_R}
- \frac{\xi}{4} \frac{g'^2}{g_{\rho_R}^2 - g'^2} \left(2 m_{\rho_R}^2 - f^2 g'^2\right)\,.
\end{equation}
The neutral resonance is always slightly heavier than the charged ones.
The mass difference, however, is sizable only for quite small values of the $g_{\rho_R}$ coupling ($g_{\rho_R} \lesssim 1$).
The SM gauge boson masses at leading order in $\xi$ coincide with the expressions given in Eq.~(\ref{eq:sm_bosons_mass}).

\subsubsection{Singlets}

The next set-up we consider is the scenario in which the vector resonance, denoted by $\rho^X_\mu$,
is an $\SO(4)$ and QCD singlet. For definiteness we consider the case in which
the vector state is associated to the $\U(1)_X$ symmetry, that is we assume it to transform as a gauge
field under the Abelian $\U(1)_X$ subgroup:\footnote{Notice that the assumption that $\rho^X_\mu$ transforms
as a gauge field does not imply any real constraint on its properties. In full generality one can define a shifted version
of the $\rho^X_\mu$ field, namely $\rho'^X_\mu \equiv \rho^X_\mu - g'_0 B_\mu$, that is invariant under $\U(1)_X$
and rewrite the effective Lagrangian in terms of the new field.}
\begin{equation}
\rho^X_\mu \rightarrow \rho^X_\mu + \partial_\mu \alpha^X\,.
\end{equation}
The $\rho^X_\mu$ fields is thus naturally mixed with the elementary hypercharge boson $B_\mu$
that gauges the diagonal subgroup of $\SU(2)_R \subset \SO(4)$ and $\U(1)_X$.

The general structure of the effective Lagrangian for the singlet vector resonance is
\begin{equation}\label{eq:Lagr_rhoX}
{\cal L}_{\rho_X} = -\frac{1}{4 g_{\rho_X}^2} \rho^{X}_{\mu\nu} \rho^{X \mu\nu}
+ \frac{m_{\rho_X}^2}{2 g_{\rho_X}^2} \left(\rho^{X}_{\mu} - g'_0 B_\mu\right)^2
+\,\left(\rho^{X}_{\mu} - g'_0 B_\mu\right) J^\mu_{\rho_X}\,,
\end{equation}
where the $\rho^X_{\mu\nu}$ field strength is given
by $\rho^X_{\mu\nu} = \partial_\mu \rho^X_\nu - \partial_\nu \rho^X_\mu$ and
we denoted collectively by $J^\mu_{\rho_X}$ any singlet current constructed from the composite fermion
states. Notice that, in general, the $\rho^X_\mu$ resonance is coupled to all the fermionic resonances that can give rise
to a singlet current. For instance it can couple to the currents obtained from the fourplet fields,
$J^\mu_{\rho_X} \ni \kappa_4\, \overline \psi_4^i \gamma^\mu \psi_4^i$, as well as from the singlets,
$J^\mu_{\rho_X} \ni \kappa_1\, \overline \psi_1 \gamma^\mu \psi_1$.\footnote{This situation is not uncommon in explicit models.
For instance in the minimal scenarios all the top partners are charged under the $\U(1)_X$ subgroup, thus we expect
them to be coupled to vector fields with the quantum numbers of the $\rho^X_\mu$ resonance.}
It is interesting to notice that, in the scenarios with a fully composite right-handed top,
the $\rho^X_\mu$ field can be directly coupled to the $t_R$ field. Mixed couplings involving the $t_R$ and the singlet
resonance $\psi_1$ can also be present:
\begin{equation}
J^\mu_{\rho_X} \ni \kappa_t\, \overline t_R \gamma^\mu t_R\,,
\; \kappa_{t1}\, \overline t_R \gamma^\mu \psi_{1R} + \mathrm{h.c.}\,.
\end{equation}

Analogously to the scenario with an $\SU(2)_R$ vector triplet, the hypercharge gauge coupling is given by a combination
of the elementary $\U(1)_Y$ coupling $g'_0$ and of the vector resonance coupling $g_{\rho_X}$, namely
\begin{equation}
\frac{1}{{g'^2}} = \frac{1}{g'^2_0} + \frac{1}{g_{\rho_X}^2}\,.
\end{equation}
The $\SU(2)_L$ and QCD couplings, on the other hand, coincide with the corresponding elementary couplings.
The mass of the vector resonance is given (up to corrections of order $\xi^2$) by the following expression
\begin{equation}
M^2_{\rho_X} = \frac{g_{\rho_X}^2}{g_{\rho_X}^2 - g'^2} m_{\rho_X}^2 + 
\frac{g'^4}{g_{\rho_X}^2 - g'^2}\frac{f^2 \xi}{4}\,.
\end{equation}
The SM gauge boson masses at leading order in $\xi$ coincide with the expressions given in Eq.~(\ref{eq:sm_bosons_mass}).

\subsubsection{Heavy gluons}

The last class vector states we consider are the heavy gluons, $\rho^g_\mu$, or, in other words, vector resonances
that transform as an octet under QCD:
\begin{equation}
\rho^g_\mu \rightarrow g_{SU(3)} \rho^g_\mu g_{SU(3)}^\dagger + i g_{SU(3)} \partial_\mu g_{SU(3)}^\dagger\,,
\end{equation}
where $g_{SU(3)}$ denotes an $\SU(3)_c$ transformation. The phenomenology of the heavy gluons is quite similar
to the one for the singlet vector resonances. The Lagrangian, in fact, is analogous to the
one we wrote for the $\rho^X_\mu$ field (see Eq.~(\ref{eq:Lagr_rhoX})). There are only two obvious differences:
the heavy vectors are now mixed with the elementary gluons, and the current $J^\mu_g$ transforms in the adjoint
representation of $\SU(3)_c$. The Lagrangian is thus given by
\begin{equation}
{\cal L}_{\rho_g} = -\frac{1}{4 g_{\rho_g}^2} \rho^{g}_{\mu\nu} \rho^{g \mu\nu}
+ \frac{m_{\rho_g}^2}{2 g_{\rho_g}^2} \left(\rho^{g}_{\mu} - g_{s0} G_\mu\right)^2
+ \left(\rho^{g}_{\mu} - g_{s0} G_\mu\right) J^\mu_g\,.
\end{equation}
The heavy gluons are directly coupled with the top partners and can have a direct coupling to the right-handed
top if it is a fully composite state.

Given that the QCD bosonic sector is not affected by EWSB, the mass of the heavy gluons is simply given by
\begin{equation}
M^2_{\rho_g} = \frac{g_{\rho_g}^2}{g_{\rho_g}^2 - g_s^2} m_{\rho_g}^2\,,
\end{equation}
and the SM QCD gauge coupling is related to the elementary coupling $g_{s0}$ by
\begin{equation}
\frac{1}{g_s^2} = \frac{1}{g_{s0}^2} + \frac{1}{g_{\rho_g}^2}\,.
\end{equation}

\subsection{Collider phenomenology}

We can now discuss the collider phenomenology of the vector resonances. As a preliminary step we will analyze
the structure of the couplings with the SM fields and with the top partners. These couplings are of fundamental importance
for collider phenomenology because they determine
the main production and decay channels of the vector resonances.
After this preliminary discussion we will present in details the most promising channels for the direct detection
of the vector states at hadronic colliders.

\subsubsection{Couplings}

The couplings of the vector resonances that are important for collider phenomenology belong to three
main classes: coupling to the SM gauge fields and the Higgs, couplings to light SM fermions (namely the leptons and the
quarks from the first two generations) and couplings to heavy SM quarks and top partners.
Each class has specific features that depend on the way the couplings are generated.
Moreover each class influences the collider phenomenology of the vector resonances in a different way
leading to specific collider signatures.

The couplings to the SM bosons we are mainly interested in are the ones that involve two
SM states and one composite resonance. These couplings are relevant for the decay of the vector states
and, at the same time, control the Vector Boson Fusion (VBF) production channel.
As we did for the top partners couplings in Sect.~\ref{sec:partners_prop}, we can simplify the analysis
by relying on the Goldstone boson equivalence theorem,
according to which the longitudinal components of the $W$ and $Z$ gauge bosons
are well described by the corresponding Goldstone bosons in the high-energy limit.
Due to the large mass gap between the vector resonances and the SM gauge fields, the
equivalence theorem is expected to provide an excellent approximation. Remembering that the Goldstone bosons
transform in the fundamental representation of the unbroken $\SO(4)$ group, we can easily link the
structure of the vector resonances couplings to their quantum numbers under $\SO(4) \simeq \SU(2)_L \times \SU(2)_R$.
The vector resonances in the $({\bf 3}, {\bf 1})$ and $({\bf 1}, {\bf 3})$ representations can be directly coupled to the
Goldstone multiplet and the corresponding couplings can be estimated to scale as
\begin{equation}\label{eq:coupling_rho_VV}
g_{\rho_{L,R} \phi \phi} \sim \left(\frac{m_{\rho_{L,R}}}{g_{\rho_{L,R}} f} \right)^2 g_{\rho_{L,R}}
\sim g_{\rho_{L,R}}\,,
\end{equation}
where $\phi$ denotes any of the Goldstone bosons, including the physical Higgs.
To obtain the last estimate in the above equation
we applied to usual One-Scale-One-Coupling power-counting, according to which
$m_{\rho_{L,R}}/g_{\rho_{L,R}} \sim f$. As one can see from
Eq.~(\ref{eq:coupling_rho_VV}), the coupling of the vector triplets to the SM gauge bosons
is of the order of the resonances coupling, hence it is usually sizable.
This result is not unexpected since the $g_{\rho_{L,R} \phi \phi}$ coupling parametrizes an interaction involving only
composite states.

On the other hand, the singlet vector states and the heavy gluons, which are not charged under $\SO(4)$,
have strongly suppressed couplings to the SM gauge fields since they can not be directly coupled to the Godstones.
The singlets, indeed, interact mostly with the transverse
gauge bosons and the corresponding coupling is generated only after EWSB through the mixing between
the vector resonances and the $B_\mu$ boson. This coupling is very small, of order
\begin{equation}
g_{\rho_X WW} \sim g_{\rho_X ZH} \sim \frac{g'^2}{g_{\rho_X}} \xi\,.
\end{equation}
In the case of the heavy gluons, instead, a trilinear coupling involving two SM gluons is altogether absent
due to the unbroken $\SU(3)_c$ gauge invariance.

\begin{figure}
\centering
\includegraphics[width=.32\textwidth]{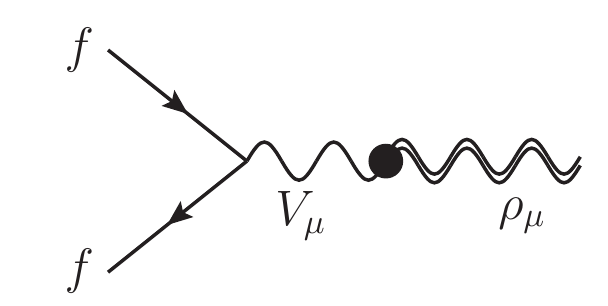}
\caption{Structure of the leading diagrams giving rise to the coupling of the composite vector resonances
with the light SM fermions. The black dot denotes the mass mixing between the composite states and the elementary gauge fields.}
\label{fig:rho_coupling_lightf}
\end{figure}

The second class of couplings, which involves the light SM fermions, is mainly generated through the
mixing of the vector resonances with the elementary gauge fields. This effect is ``universal'' and depends
only on the quantum numbers of the fermion species, which determine the couplings to the elementary gauge fields.
The structure of the diagrams giving rise to these couplings are shown in Fig.~\ref{fig:rho_coupling_lightf}.
They arise from the exchange of an
elementary gauge field which is mixed with the composite vector states through a mass term.
The strength of the effective interactions depends on the coupling of the ``mediator'' gauge boson
and on the $\xi$ order at which the mixing is generated. We will discuss the various cases in the following.

For the $\SU(2)_L$ triplet $\rho_L$ the main couplings are due to the mixing with the $W$ boson. This mixing
is already present before EWSB, thus the effective coupling is not suppressed by powers of $\xi$.
These interactions obviously involve the left-handed fermion components and their strength can be estimated as
\begin{equation}
g_{\rho_L f_L f_L} \simeq c_L \frac{g^2}{g_{\rho_L}}\,,
\end{equation}
where $c_L$ denotes the $SU(2)_L$ charge of the fermions. Another class of contributions to the $\rho_L$ couplings
comes from the $\rho^0_L$ mixing with the $B_\mu$ field.
This mixing, however, arises only after EWSB, thus the induced couplings scale
as $g'^2/g_{\rho_L} \xi$ and can typically be neglected.

The couplings of the $\SU(2)_R$ triplet $\rho_R$ and of the singlet $\rho_X$ are mainly due to the
mixing with the hypercharge gauge boson. They can be estimated as
\begin{equation}
g_{\rho^0_R f\! f} \simeq c_Y \frac{g'^2}{g_{\rho_R}}\,,
\qquad
g_{\rho_X f\! f} \simeq c_Y \frac{g'^2}{g_{\rho_X}}\,,
\end{equation}
where $c_Y$ denotes the hypercharge of the fermions. Notice that, differently from the couplings of the neutral
components of the $\SU(2)_R$ triplet, the couplings of the $\rho_R^\pm$ resonances to the light SM fermions
are generated only after EWSB (through the mixing with the $W$ bosons)
and are thus strongly suppressed: $g_{\rho^\pm_R f\! f} \sim g^2/g_{\rho_R} \xi$.

Finally, the heavy gluons are coupled to the light SM quarks thanks to the mixing with the elementary gluons.
The corresponding couplings are of order
\begin{equation}
g_{\rho_g f\! f} \simeq \frac{g_s^2}{g_{\rho_g}}\,.
\end{equation}

We can now discuss the couplings that involve the third generation quarks and the top partners.
The main source of these interactions is the coupling of the vector resonances with the composite
fermions, which is generated directly by the strong dynamics. As can be easily understood, the strength of these
couplings is equal to the vector resonance coupling and is weighted by the amount of compositeness
of the fermions.

To be more explicit, any interaction involving two fully composite fermions has a strength
\begin{equation}
g_{\bar \rho X X'} \sim g_{\bar \rho}\,,
\end{equation}
where $\bar \rho$ collectively denote any vector resonance, while
$X$ and $X'$ correspond to any fermion resonance whose mixing with the elementary states is small. This estimate
applies, for instance, the to the $X_{5/3}$ state, which is purely composite, or to the $X_{2/3}$ field, which is mixed
to the elementary states only after EWSB. The same estimate is valid for the couplings involving the right-handed
top component in the scenarios where it is fully composite.

On the other hand, an interaction involving a partially composite SM quark or a top partner with a large mixing
to the elementary sector has a strength suppressed by the fermions compositeness angle. For example,
the $t_L$ and $b_L$ fields acquire a degree of compositeness due to the mixing with the composite fourplet $\psi_4$
and the corresponding mixing angle is (see Sect.~\ref{sec:lhc_fermions})
\begin{equation}
\tan \phi_L = \frac{y_{L4} f}{m_4}\,.
\end{equation}
An interaction involving the $t_L$ field is thus weighted by a factor $\sin \phi_L$ for each $t_L$ field:
\begin{equation}
g_{\rho_L t_L t_L} \sim g_{\rho_L} \sin^2 \phi_L\,,
\quad
g_{\rho_R t_L X_{5/3L}} \sim g_{\rho_R} \sin \phi_L\,,
\quad
\ldots
\end{equation}
Analogously, the $T_L$ and $B_L$ partners inside the fourplet are always accompanied by a compositeness
factor $\cos \phi_L$:
\begin{equation}
g_{\rho_L T_L T_L} \sim g_{\rho_L} \cos^2 \phi_L\,,
\quad
g_{\rho_L t_L T_{L}} \sim g_{\rho_L} \sin \phi_L \cos \phi_L\,,
\quad
\ldots
\end{equation}
The same happens for the partially composite $t_R$ and for the right-handed component
of the singlet partner $\widetilde T_R$.
The corresponding compositeness angle is
\begin{equation}
\tan \phi_R = \frac{y_{R1} f}{m_1}\,.
\end{equation}

Similarly to the light-SM-fermions case, the interactions with the heavy SM quarks and the top partners
receive an additional contribution from
the mixing between the vector resonances and the SM gauge fields. This contribution scales as
$\bar g^2/g_{\bar \rho}$, where $\bar g$ denotes the SM gauge coupling involved in generating the interaction,
and is typically subleading with respect to the direct strong-dynamics effects.

\subsubsection{Main production and decay channels}

At a hadronic machine the two main production mechanisms for the vector resonances are
Drell-Yan (DY) processes (mediated by the interactions with the light SM quarks)
and VBF processes (mediated by the couplings with the EW gauge bosons).
As we discussed above, the strength of the interactions with the light SM fermions scales like $1/g_\rho$,
thus the DY production cross section is reduced if the strong-sector coupling is large. On the contrary,
the coupling to the SM gauge bosons usually scales like $g_\rho$, leading to an enhancement of VBF
for large couplings.

To determine the relative importance of the two production channels, however,
we also need to take into account the initial parton luminosities.
The parton luminosity for the DY processes is much larger than the VBF
one, since the latter is suppressed by an additional factor $\alpha_{EW}^2$ coming from the
emission of the virtual EW gauge bosons. This large suppression makes the VBF channel usually subleading
with respect to DY. This result is valid at the LHC for all the resonances we considered above.
The only exception are the charged components of the $SU(2)_R$ triplet, $\rho^\pm_R$, whose
coupling to the light SM fermions is suppressed by a factor $\xi$. In this case the DY and VBF channels
have comparable cross section, though both of them are quite small.
Notice that for the singlet resonance $\rho_X$ and for the heavy gluons the couplings to the SM gauge bosons
are highly suppressed and the VBF channel is totally negligible.

It is important to stress that, although all the couplings to the light SM fermions scale in the same way,
their absolute size can significantly vary depending on the quantum numbers of the vector resonances.
For instance the couplings mediated by the $B$ gauge boson are smaller with respect to the
one mediated by the $W$ by a factor $(g'/g)^2 \simeq 0.3$, which implies a one order of magnitude suppression
in the production cross section. The heavy gluon couplings, on the other hand, are enhanced with respect
to the EW ones, so that their production is roughly two orders of magnitude bigger than for an $\SU(2)_L$ triplet.
The parton luminosities and the typical size of the cross sections for the various vector resonances
can be found in Refs.~\cite{Bini:2011zb,Pappadopulo:2014qza,Greco:2014aza}.

We can now discuss the typical decay channels. Due to the sizable coupling, when kinematically allowed,
the vector resonances preferentially decay into pairs of composite fermions. This is basically true for all vector resonances
unless somewhat extreme corners of the parameter space are considered.
Notice that, when the decay into pairs of composite fermions is allowed, the vector resonances tend to be
quite broad. This is a consequence of the large couplings as well as of the fact that multiple decay channels are
usually present due to the existence of various top partners species.\footnote{See for instance Ref.~\cite{Azatov:2015xqa} for a collider study of heavy gluons decaying to top partners.}

Below the threshold for the decay into two partners, the direct decays into SM states usually become dominant.
The vector resonances charged under $\SO(4)$, namely the $\rho_L$ and $\rho_R$ triplets, have large
branching ratios into pairs of SM gauge bosons or a gauge boson and the Higgs. These decays are mediated by the
${\mathcal O}(g_\rho)$ couplings involving the Goldstone bosons.
If the top and bottom quarks have a significant amount of compositeness, another important channel is the direct
decay into third generation quarks. This is true for almost all the
vector resonances,
the only exception being the charged $\rho_R^\pm$ resonances, whose couplings to the SM quarks arise
only after EWSB.

Heavy-light decays involving one top partner and one third-generation quark can also be relevant when kinematically
allowed. This typically happens only in a relatively small mass window, given that, above the $m_\rho > 2 m_\psi$
threshold the decay into two top partners usually dominates.

In addition to the mentioned channels, the vector resonances can also decay directly into light SM fermions.
A particularly interesting channel is the one involving a pair of leptons, which is available for
the $\rho_L$, $\rho_R^0$ and $\rho_X$ resonances. Although these decays usually have only a small branching ratio,
they can be important for collider searches due to the easily detectable final states.

The main decay channels of the vector resonances are summarized in Table~\ref{tab:vector_decays}.

\begin{table}
{\small
\begin{tabular}{c|ccccc}
\rule[-6pt]{0pt}{1.em}& gauge & light SM & 3$^{rd}$ gen. & heavy-light & top partners\\
\hline
\hline
\rule[-4pt]{0pt}{2.2em}$\rho_L^0$ & $WW$, $Zh$ & $l^+ l^-$, $u \bar u$ & $t \bar t$, $b \bar b$ & $T \bar t$, $B \bar b$
&
$
\begin{array}{c}
X_{5/3} \overline X_{5/3}, X_{2/3} \overline X_{2/3},\\
(T \overline T, B \overline B)
\end{array}
$\\
\rule[-6pt]{0pt}{1.9em}$\rho_L^\pm$ & $WZ$, $Wh$ & $l \nu$, $u \bar d$ & $t \bar b$ & $B \bar t$, $T \bar b$ & $X_{5/3} \overline X_{2/3}$, $T \overline B$\\
\hline
\rule[-4pt]{0pt}{2.2em}$\rho_R^0$ & $WW$, $Zh$ & $l^+ l^-$, $u \bar u$ & $t \bar t$, $b \bar b$ & $T \bar t$, $B \bar b$ &
$
\begin{array}{c}
X_{5/3} \overline X_{5/3}, X_{2/3} \overline X_{2/3},\\
(T \overline T, B \overline B)
\end{array}
$\\
\rule[-6pt]{0pt}{1.9em}$\rho_R^\pm$ & $WZ$, $Wh$ &  &  & $X_{5/3} \bar t$, $X_{2/3} \bar b$ & $X_{5/3} T$, $X_{2/3} B$\\
\hline
\rule[-4pt]{0pt}{2.3em}$\rho_X$ & & $l^+ l^-$, $u \bar u$ & $t \bar t$, $b \bar b$ & $T \bar t$, $B \bar b$, $\widetilde T \bar t$ &
$
\begin{array}{c}
X_{5/3} \overline X_{5/3}, X_{2/3} \overline X_{2/3},\\
\widetilde T \overline {\widetilde T}, (T \overline T, B \overline B)
\end{array}
$\\
\hline
\rule[-4pt]{0pt}{2.3em}$\rho_g$ & & $u \bar u$ & $t \bar t$, $b \bar b$ & $T \bar t$, $B \bar b$, $\widetilde T \bar t$ &
$
\begin{array}{c}
X_{5/3} \overline X_{5/3}, X_{2/3} \overline X_{2/3},\\
\widetilde T \overline {\widetilde T}, (T \overline T, B \overline B)
\end{array}
$
\end{tabular}
}
\caption{Main decay channels of the composite vector resonances. In the column for the
decays into light SM fermions, $u \bar u$ also includes the $d \bar d$, $s \bar s$ and $c \bar c$ channels,
and similarly $u \bar d$ also includes the $c \bar s$ channel. In the ``top partners'' column
the parenthesis enclosing $T \overline T$ and $B \overline B$ indicate that these channels are
always suppressed with respect to $X_{5/3} \overline X_{5/3}$ and $X_{2/3} \overline X_{2/3}$ because the
$T$ and $B$ resonances are always heavier than the $X_{5/3}$ and $X_{2/3}$.}
\label{tab:vector_decays}
\end{table}

\subsubsection{Collider searches}

Several decay channels can be exploited to search the vector resonances at hadronic colliders.
At the $8\ \mathrm{TeV}$ LHC the experimental analyses focused on the direct decays into SM states.
The main final states that have been considered for the charged resonances are
\begin{itemize}
\item heavy SM quarks, $\rho^+ \rightarrow t \bar b$~\cite{TheATLAScollaboration:2013iha,CMS:pwa},
\item leptons, $\rho^+ \rightarrow l \nu$~\cite{ATLAS:2014lnu,Khachatryan:2014tva},
\item gauge bosons with fully hadronic final state, $\rho^+ \rightarrow WZ \rightarrow jj$~\cite{CMS:2013fea,Khachatryan:2014hpa},
\item gauge bosons with fully leptonic final state, $\rho^+ \rightarrow WZ \rightarrow 3l \nu$~\cite{ATLAS:20143lnu,CMS:2013vda}.
\end{itemize}
For the neutral states the following searches have been performed
\begin{itemize}
\item top quarks, $\rho^0 \rightarrow t \bar t$~\cite{TheATLAScollaboration:2013kha,CMS:2013vca},
\item leptons, $\rho^0 \rightarrow l^+ l^-$~\cite{Aad:2014cka,CMS:2013qca},
\item $\tau$ leptons, $\rho^0 \rightarrow \tau \tau$~\cite{TheATLAScollaboration:2013yha},
\item gauge bosons with fully hadronic final state, $\rho^0 \rightarrow WW \rightarrow jj$~\cite{CMS:2013fea},
\item gauge bosons with semi-leptonic final state, $\rho^0 \rightarrow WW \rightarrow l \nu jj$~\cite{CMS:2013xea}.
\end{itemize}

The constraints coming from the $8$ TeV LHC data have been discussed in Refs.~\cite{Pappadopulo:2014qza,Greco:2014aza}.
We will briefly summarize them in the following.
The searches for resonances decaying directly into leptons are relevant for the ``EW'' vector resonances
(namely the $\rho_{L,R}$ and the $\rho_X$) for relatively small values of the $g_\rho$ coupling
($g_\rho \sim 1$).
In the case of the $\rho_L$, these searches can exclude resonances with a mass $m_{\rho_L} \lesssim 2 - 3\ \mathrm{TeV}$.
The bound quickly deteriorates for larger $g_{\rho_L}$ and disappears for $g_{\rho_L} \gtrsim 2 - 3$ due to the suppressed
branching fraction into leptons. In the case of the $\rho_R$ and $\rho_X$ resonances, the bounds are usually
$\sim 500\ \mathrm{GeV}$ weaker than the ones for the $\rho_L$ due to the smaller production cross section.

Another relevant decay channel is the one into SM gauge bosons, whose importance
increases at larger $g_\rho$. This channel allows to exclude a $\rho_L$ resonance
with a mass $m_{\rho_L} \lesssim 1.5 - 2\ \mathrm{TeV}$ for $g_{\rho_L} \lesssim 4 - 5$.
This decay channel can also be relevant for the $\rho_R$ resonances although, similarly to the lepton channel,
the bounds are less stringent ($m_\rho \lesssim 1 - 1.5\ \mathrm{TeV}$ for $g_\rho \lesssim 2$) than for the $\rho_L$.

Finally, the $t \bar t$ decay channel can also be used to set exclusions. In the case of the ``EW'' vector resonances
these bounds are usually weaker than the ones coming from the channels involving leptons
and gauge bosons.
The situation is different for the heavy gluons, which can only decay into colored states. In this case the
$t \bar t$ channel is the best one to set bounds on $m_{\rho_g}$.

Notice that, if the decay into a pair of top partners or the heavy-light channel are kinematically allowed,
the branching ratio for the direct decays into SM states is usually small and
the above mentioned searches drastically lose effectiveness. Although the top partners decay into third-generation quarks
and gauge bosons, the presence of the intermediate fermion resonances in the decay chain significantly affects the
kinematics making the searches for $t \bar t$, $t \bar b$ and $b \bar b$ final states less efficient.
It has been shown that dedicated search strategies focused on the decays to composite fermions could significantly improve the
sensitivity to these channels~\cite{Bini:2011zb,Chala:2014mma}.

\chapter{EW precision tests}
\label{ch:EWPT}

The composite Higgs dynamics gives rise to a rich set of new-physics effects that can be
used to probe this scenario through a comparison with the experimental data.
One of the most distinctive phenomena is the presence of composite resonances
around the TeV scale, which can be straightforwardly tested in collider experiments. A second
important signature is the peculiar pattern of distortions of the Higgs
couplings, which constitutes a direct manifestation of the non-linear Nambu--Goldstone structure.
In addition to these features, the composite dynamics gives also rise to many indirect effects.
These noticeably include a set of corrections to the Electro-Weak (EW) observables that describe the physics of the
light Standard Model (SM) fermions and of the gauge fields. The importance of these observables comes from the
fact that they can be measured in high-precision experiments and thus can be used to test even tiny
corrections coming from a new-physics dynamics.
Obvious examples are the $Z$-pole observables measured at the LEP experiment
and the properties of the bottom quark easily accessible at $b$-factories.
All these measurements agree with the SM and have been extensively used to set stringent constraints on beyond the Standard Model (BSM) scenarios. They constitute the so-called EW Precision Tests (EWPT) of the SM.

It is important to stress that the corrections to the precision observables ultimately come from exactly the same features
that can be tested in direct searches, namely the composite resonances and
the Goldstone structure. Thus they provide a complementary approach to test the
main properties of the composite dynamics. The virtue of the precision measurements is the
fact that, in principle, they can probe new physics at scales much higher than the actual energy of the experiment. 
As we will see in the following,
a typical example are the $Z$-pole LEP data, which can be sensitive to the presence of composite
resonances with masses much above the TeV scale. This high sensitivity, however, comes with a price.
Differently from a direct discovery, which usually point towards a specific type of new physics,
a deviation in a precision observable can in principle be due to very different sources.
Similar ambiguities are present in the case of no discovery, when the experimental data are used to set
bounds on the BSM dynamics. As we saw in Chap.~\ref{ch:LHC_pheno}, direct searches of new states
can probe and constrain new physics in a virtually model-independent way. Indirect effects, instead,
can receive contributions from several sources including, in specific cases, sizable UV effects, and thus
their interpretation depends much more on the exact details of the new-physics model.

In this chapter we will discuss the main corrections to the precision observables that arise in composite Higgs scenarios.
Our primary aim will be to provide an overview of the broad class of effects that characterize the composite Higgs models,
we will thus proceed similarly to what we did in the previous chapters and follow a general effective
field theory approach. Interestingly, an important set of the EW precision observables
are mainly determined by the low-energy dynamics of the composite sector.
In these cases the most relevant contributions
come from the lightest composite resonances belonging either to the quark sector (the usual top partners) or to the
gauge sector. These effects can be reliably computed within an effective approach and thus provide a set
of robust predictions which can be compared with the EWPT.


\section{The oblique parameters}\label{sec:oblique_parameters}
\index{oblique parameters}

Composite Higgs models are, to a good approximation, ``universal'' theories of EW symmetry breaking and thus their corrections to the EW precision observables can be conveniently encapsulated in the so-called ``oblique parameters'' defined in Ref.~\cite{Barbieri:2004qk}. The approach of Ref.~\cite{Barbieri:2004qk}, which we will now briefly summarize, extends the original prescription by Peskin and Takeuchi~\cite{Peskin-Takeuchi} to models containing new heavy vector bosons. Universal theories of EW symmetry breaking are those in which the only interactions of the leptons and of the light quark fields are of the form
\beq
{\mathcal{L}}_{\textrm{int}}=\overline{\Psi}\gamma^\mu(T^\alpha\hat{W}_\mu^\alpha+Y\hat{B}_\mu)\Psi\,,
\label{univ}
\eeq
where $T^\alpha$ denotes the \mbox{SU$(2)_L$} SM generators and $Y$ is the hypercharge. The QCD interactions are of course also present for the quarks, but they play no role in the discussion and can be safely ignored. Only the light quarks and the leptons are considered in the definition of universal theories because these are the particles which enter more directly in the EWPT. Deviations from universality, for instance in the bottom sector, can be systematically taken into account as we will discuss in Sect.~\ref{sec:Zbb}. Notice that a field redefinition might be needed in specific theories to put the fermion interactions in the form of Eq.~(\ref{univ}). We call universal theories those in which a field basis exists where Eq.~(\ref{univ}) is satisfied and we consider that basis for the study of EW precision physics.

The SM, in the limit in which we neglect the Yukawa operators, is the simplest example of a universal theory. Fermions are coupled by the habitual gauge interactions like in Eq.~(\ref{univ}), where $\hat{W}=g W$ and $\hat{B}=g' B$ are the rescaled \mbox{SU$(2)_L\times$U$(1)_Y$} gauge fields. Composite Higgs models are universal theories as well since the light quarks and leptons mixings with the composite sector are typically very small and negligible.\footnote{Considerable departures from universality are possible in \mbox{U$(3)^3$} flavor-symmetric scenario, as we briefly discussed in Sect.~\ref{U33mod}.} The only relevant interactions originate from the elementary sector Lagrangian and are dictated by gauge symmetry. Namely they take once again the form of Eq.~(\ref{univ}) but with $\hat{W}=g_0 W_0$ and $\hat{B}=g'_0 B_0$ where $g_0$, $g'_0$ and $W_0$, $B_0$ denote, respectively, the elementary gauge couplings and the elementary gauge fields. Notice that $\hat{W}$ and $\hat{B}$ do not coincide with the light SM vector boson mass eigenstates and are not even proportional to them. In generic composite Higgs models, such as the ones we studied in the previous chapters, the $\hat{W}$ and $\hat{B}$ fields are linear combinations of the vector bosons and of the heavy vector resonance fields. With a terminology borrowed from extra-dimensional models of EW Symmetry Breaking (EWSB), $\hat{W}$ and $\hat{B}$ are sometimes called ``holographic'' fields.

Since they couple with the light fermions, the holographic fields are the right objects to discuss EW precision physics. In particular, precision EW processes involve four-fermion amplitudes and thus they are only affected by new physics modifications of the holographic fields propagators with respect to the SM predictions. The oblique parameters are thus defined in terms of the new physics corrections to the transverse vacuum polarization amplitudes $\Pi_{VV'}$ where $V,\,V'=\{\hat W^\alpha, \hat B\}$, or better in terms of their low-momentum expansion which is a good approximation since the EW precision processes take place much below the new physics scale. The oblique parameters that are relevant for the composite Higgs phenomenology are $\widehat S$ and $\widehat T$, defined as~\footnote{The parameters $g$ and $v$ are defined, following Ref.~\cite{Barbieri:2004qk}, in terms of the $\Pi_{W^+ W^-}$ correlator,
namely $1/g^2 = \Pi'_{W^+ W^-}(p^2 = 0)$ and $v^2 = -4\,\Pi_{W^+ W^-}(p^2 = 0) \simeq (246\ \mathrm{GeV})^2$.}
\bea
&&\widehat S = g^2 \left.\frac{d \Pi_{W^3B}}{d p^2}\right|_{p^2 = 0}\,,\label{eq:def_S}\\
&&\widehat T = \frac{4}{v^2}\left.\left(\Pi_{W^3 W^3} -\Pi_{W^1 W^1}\right)\right|_{p^2 = 0}\label{eq:def_T}\,.
\eea
We stress once again, in view of some confusion that emerges in the literature, that the usage of the holographic fields $\hat{W}$ and $\hat{B}$ is compulsory for a proper discussion of the EWPT and no alternative can be considered. In particular, using the SM vector bosons mass eigenstate basis to define the oblique parameters makes absolutely no sense because in this basis the heavy resonance fields also couple with the fermions and contribute to the EW precision physics. The corrections from heavy vectors exchange are of exactly the same order as the ones from the modified couplings with the physical EW bosons and thus they can not be ignored.

In the composite Higgs scenario the corrections to the oblique parameters come from
two main sources. The first one is the intrinsic non-linear dynamics associated to the
Goldstone nature of the Higgs. The non-renormalizable Lagrangian describing the Higgs
field contains distortions of the gauge bosons and Goldstone couplings that induce
a modification of the SM contributions to the vacuum polarization amplitudes. The second class of
corrections to $\widehat S$ and $\widehat T$ comes from the presence of composite vector and fermionic resonances.

\begin{figure}
\centering
\includegraphics[width=0.45\textwidth]{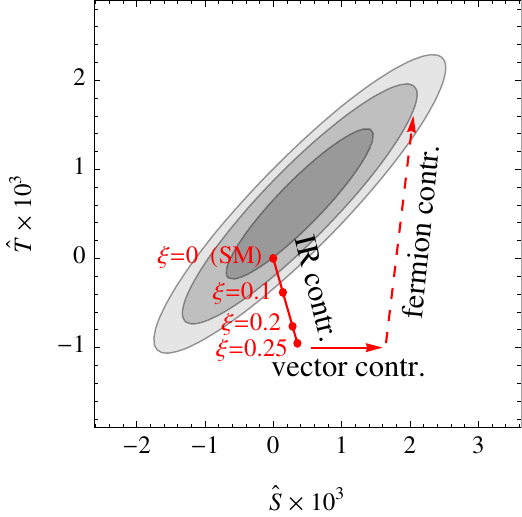}
\caption{Constraints on the oblique EW parameters $\widehat S$ and
$\widehat T$~\cite{Baak:2012kk}.
The gray ellipses show the exclusion contours at $68\%$, $95\%$ and $99.7\%$ confidence
level for $m_h = 126\ \mathrm{GeV}$.
The red lines schematically show the contributions that arise in composite Higgs models
as explained in the main text. The IR contribution corresponds to the corrections due
to the non-linear Higgs dynamics, approximately given in Eqs.~(\ref{eq:S_IR})
and (\ref{eq:T_IR}), and is obtained by fixing $m_\rho = 3\ \mathrm{TeV}$.
The plot is taken from Ref.~\cite{Grojean:2013qca}.}
\label{fig:ST_plane}
\end{figure}

The various contributions to the oblique parameters are shown schematically in
Fig.~\ref{fig:ST_plane} together with the current experimental constraints.
It can be clearly seen that, unless a very high compositeness scale $f$ is chosen,
a certain amount of compensation between the different contributions is needed
to respect the bounds. In particular a sizable positive contribution
to $\widehat T$ coming from the fermions is usually essential.

In the following we will analyze in details each class of contributions. For definiteness, we will
focus on the minimal composite Higgs scenarios based on the $\SO(5)/\SO(4)$ symmetry pattern.

\subsection{IR corrections}
\label{sec:oblique_pars_universal}

The first class of corrections to the oblique EW parameters is due the non-linear
Higgs dynamics which induces a modification of the Higgs couplings with the EW gauge bosons.
This distortion is present in any composite Higgs scenario and is fully determined
by the symmetry breaking pattern that gives rise to the Goldstones. In particular the
leading logarithmically-enhanced contributions are ``universal'' and are completely fixed
by the IR dynamics of the theory~\cite{oai:arXiv.org:0706.0432}. For this reason they can be
simply computed in the $\SO(5)/\SO(4)$ non-linear $\sigma$-model that describes the Higgs and the EW gauge bosons, introduced in Chap.~\ref{ch:Goldstone}. In this case, since no extra massive vector bosons are present, the holographic fields coincide with the EW bosons like in the SM.

\begin{figure}
\centering
\includegraphics[width=0.35\textwidth]{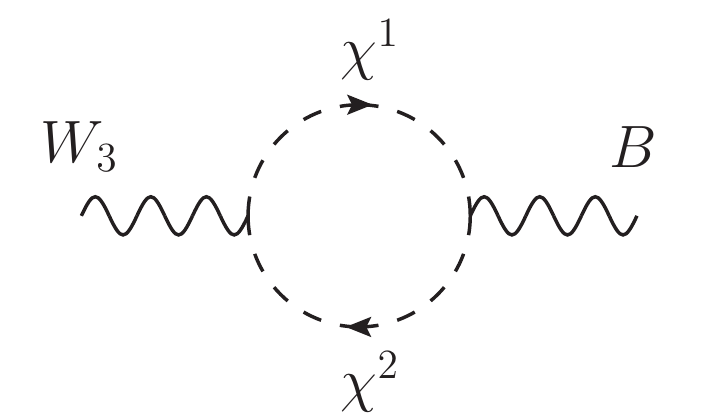}
\hspace{.5em}
\includegraphics[width=0.35\textwidth]{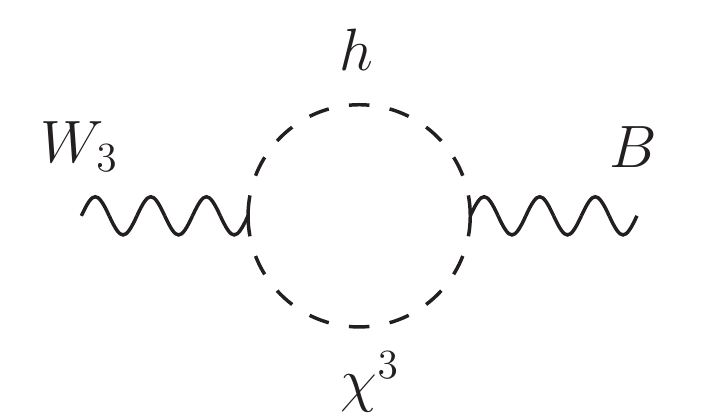}\\
\vspace{.75em}
\includegraphics[width=0.35\textwidth]{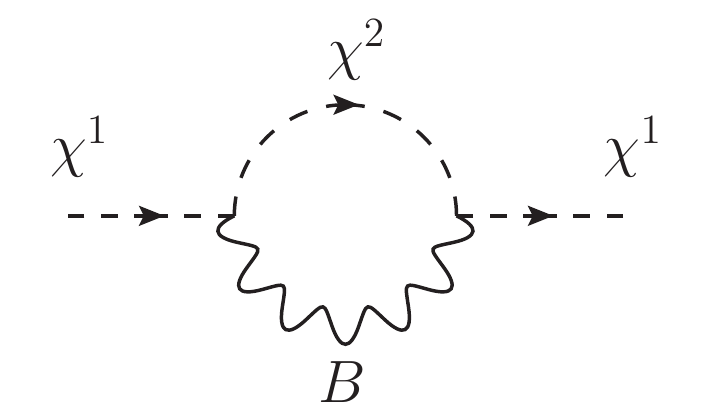}
\hspace{.5em}
\includegraphics[width=0.35\textwidth]{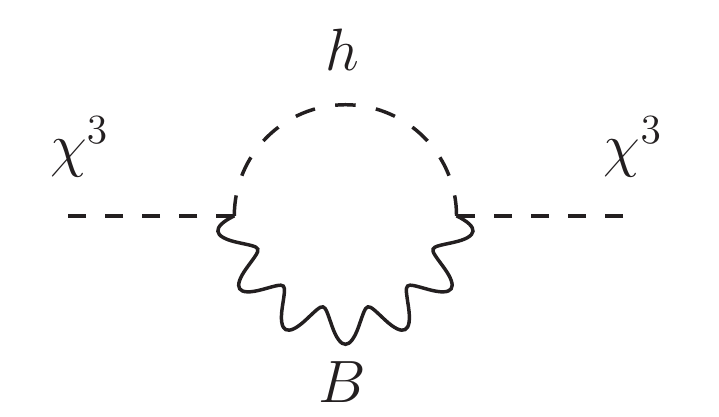}
\caption{Logarithmically divergent one-loop contributions to $\widehat S$ (upper row)
and $\widehat T$ (lower row).
Although the oblique parameters are defined in terms of the correlators of the transverse holographic gauge fields,
they can be more efficiently computed in a gauge in which the unphysical Goldstone modes $\chi$ inside the Higgs multiplet
are retained. In this gauge the $\widehat T$ parameter can be extracted from the two-point function of the $\chi^{1,3}$
fields.
}
\label{fig:ST_IR_diagrams}
\end{figure}

The relevant contributions to the $\widehat S$ and $\widehat T$ parameters come from
diagrams that involve one loop of the Higgs doublet fields as shown in
Fig.~\ref{fig:ST_IR_diagrams}. Such diagrams are present in the SM as well and they have been computed in Ref.~\cite{Peskin-Takeuchi}.
In this case, however, the renormalizability of the theory ensures that
the logarithmic divergence coming from the loops of the would-be Goldstone bosons $\chi^a$
is exactly canceled by the physical Higgs contributions.
In a non-renormalizable theory, such as the composite Higgs scenario, the argument on
the finiteness of the oblique parameters is no more valid. In particular the modifications of the physical Higgs couplings to the EW bosons imply that the Higgs contribution does not exactly cancel the logarithmic divergence coming from the $\chi$ fields. Notice that the interactions involving two $\chi$ fields and the EW gauge bosons coincide with the ones of the SM as can be checked explicitly by expanding the first term in Eq.~(\ref{CHlag0}) and using the relation between $V$ and $v$ in Eq.~(\ref{xval}). This mismatch leads to
a contribution to $\widehat S$ and $\widehat T$ given by
\begin{equation}\label{eq:S_IR}
\Delta \widehat S = \frac{g^2}{192 \pi^2} \xi \log \left(\frac{m_\rho^2}{m_H^2}\right)
\simeq 1.4 \times 10^{-3}\, \xi
\end{equation}
and
\begin{equation}\label{eq:T_IR}
\Delta \widehat T = -\frac{3{g'}^2}{64 \pi^2} \xi \log \left(\frac{m_\rho^2}{m_H^2}\right)
\simeq -3.8 \times 10^{-3}\, \xi\,.
\end{equation}
In the above equations we identified the UV cut-off at which the logarithmic divergence
is regulated with the mass scale of the vector resonances $m_\rho$ (we fixed $m_\rho = 3\ \mathrm{TeV}$ to derive the numerical estimates). This is indeed what happens in 
calculable implementations of the composite Higgs scenario, as we will discuss
in Sect.~\ref{sec:ferm_EW_parameters}.

It is interesting to notice that the sign of the corrections to the oblique parameters in Eqs.~(\ref{eq:S_IR}) and (\ref{eq:T_IR})
is fixed, namely $\widehat S$ receives a positive contribution while $\widehat T$ a negative one.
In the absence of further corrections, the current
experimental bounds imply a severe constraint on the strong sector scale $f$.
As can be seen from Fig.~\ref{fig:ST_plane}, the constraints come essentially from the
shift in the $\widehat T$ parameter, whereas the correction to $\widehat S$ has a small
impact. At the $2\sigma$ level the bound is $\xi \lesssim 0.05$, corresponding to
$f \gtrsim 1.1\ \mathrm{TeV}$, and is relaxed to $\xi \lesssim 0.08$
($f \gtrsim 870\ \mathrm{GeV}$) at $3\sigma$.

In generic composite Higgs models, as we will see in the following, additional contributions
to the EW parameters are present, which can improve the compatibility with the data allowing 
larger values of $\xi$. The above analysis, however, shows that a sizable value of $\xi$
can be viable only at the price of some tuning between the unavoidable ``universal'' IR
corrections to $\widehat S$ and $\widehat T$ and the extra contributions coming from the
heavy resonances.

To conclude the discussion it is useful to analyze the divergence structure of the corrections to the
oblique parameters. As shown in Eq.~(\ref{eq:def_S}), the $\widehat S$ parameter is defined as the momentum-derivative
of the two-point correlator of the $W^3_\mu$ and $B_\mu$ holographic fields.
In the non-linear $\sigma$-model description this operator is associated to a logarithmic degree of divergence at one loop,
as can be inferred from the counting in Eq.~(\ref{eq:divergence_counting}). The result in Eq.~(\ref{eq:S_IR}) indeed confirms
this expectation.

Let us now consider the $\widehat T$ parameter, which is defined in terms of the zero-momentum two-point
correlators of the elementary $W^a_\mu$ fields as in Eq.~(\ref{eq:def_T}).
By naively following the divergence counting, we would expect $\widehat T$ to develop a quadratic divergence
at one loop. This expectation, however, is not verified because of the presence of the $\SO(3)_c$ custodial invariance,
which, as we will show explicitly in Appendix~\ref{sec:custodial_symmetries}, forbids corrections to $\widehat T$.
This symmetry coincides with the diagonal combination of the $\SU(2)_L$
and $\SU(2)_R$ groups inside $\SO(4)$. It is thus automatically respected by the composite dynamics of the
$\SO(5)/\SO(4)$ models. Moreover it is left unbroken after EWSB because the Higgs VEV is invariant under $\SO(3)_c$.
The only breaking of the custodial symmetry comes from the mixing of the elementary sector with the composite dynamics.
In particular it is broken by $\U(1)_Y$ hypercharge gauging, whereas it is preserved by the $\SU(2)_L$ gauging.
Additional sources of breaking come from the mixing of the elementary fermions, as we will discuss later on.
In order to generate a radiative correction to $\widehat T$ from loops of Goldstones and gauge bosons it is necessary
to insert at least two powers of the hypercharge coupling $g'$. As a consequence the one-loop degree of divergence is
reduced from quadratic to logarithmic. This explains the structure of the result in Eq.~(\ref{eq:T_IR}) and
the fact that the corrections to $\widehat T$ are weighted by a $g'^2$ factor.


\subsection{The vector resonances contribution}
\label{sec:ST_vector}

A second class of corrections to the EW oblique parameters comes from the presence
of composite vector states.

We start our analysis by considering the $\widehat T$ parameter. As a consequence of the custodial invariance, $\widehat T$
does not receive corrections at tree level. Indeed, as we explained before, the only breaking of $\SO(3)_c$ in the gauge sector is
due to the elementary hypercharge gauge field $B_\mu$, which can not contribute at tree-level to the
two point correlator of the elementary $W_\mu$ bosons.\footnote{In computing the oblique
parameters at tree-level, the holographic gauge fields must be treated as external sources and must not
be used as virtual fields propagating in the diagrams. In particular the mass mixing between the $W^3_\mu$ component
and $B_\mu$ induced after EWSB does not induce a tree-level contribution to $\widehat T$.
The holographic bosons can be used as virtual fields only at the radiative level, as we did in the
computation of the IR corrections in Sect.~\ref{sec:oblique_pars_universal}.}
The only contributions of the vector resonances to the $\widehat T$ parameter are induced at the radiative level.
These contributions however
are subleading with respect to the IR effects in Eq.~(\ref{eq:T_IR}) because they lack the $\log(m_\rho^2/m_H^2)$
enhancement that characterizes the IR corrections. The contributions to $\widehat T$ from the vector resonances
can thus be usually neglected~\cite{Contino:2015mha}.

\begin{figure}
\centering
\includegraphics[width=0.75\textwidth]{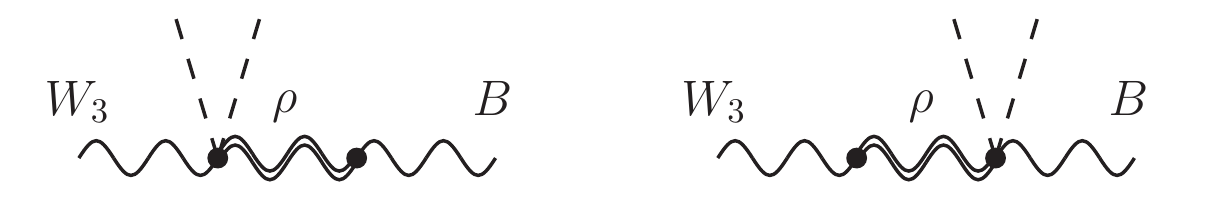}
\caption{Tree-level diagrams contributing to the $\widehat S$ parameter through the
exchange of a massive vector resonance. The dashed lines represent insertions of the Higgs
VEV.}
\label{fig:S_tree}
\end{figure}

The $\widehat S$ parameter, on the other hand, is not protected and thus can be generated at tree-level
through the exchange of composite vector states. The origin of these contributions is the mass mixing
between the vector resonances and the elementary gauge bosons. The structure of the
corresponding diagrams is shown in Fig.~\ref{fig:S_tree}.
To understand the origin of these corrections it is
useful to start from a simple explicit example: the two-site model introduced
in Chap.~\ref{ch:Pheno_models}. In this set-up the kinetic term for the Goldstone fields
(see Eq.~(\ref{eq:Goldstone_Lagr_2-site})) contains a mass mixing between the elementary
gauge fields, $W_\mu$ and $B_\mu$, and the composite vectors, $\widetilde \rho_\mu$.
Obviously before EWSB the mixing is diagonal and links the elementary fields to the vector states with
the same $\SU(2)_L \times \U(1)_Y$ charges. When the Higgs acquires a VEV, however, off-diagonal
mixing terms are generated which give rise to a contribution
to the two-point function of the $W^3_\mu$ and $B_\mu$ bosons. 

The explicit expression of the correction to the $\widehat S$ parameter in the two-site
model is given by
\begin{equation}\label{eq:S_tree}
\Delta \widehat S = \frac{g_0^2}{2 \widetilde g_\rho^2} \xi
\simeq \frac{m_W^2}{m_\rho^2}\,,
\end{equation}
where $m_\rho$ denotes the mass of the composite vector resonances,
$m_\rho^2 \simeq \widetilde g_\rho^2 f^2/2$ (see Eq.~(\ref{eq:gauge_res_masses_2-site})).
It is interesting to notice that the tree-level shift in Eq.~(\ref{eq:S_tree}) is strictly positive.
When this additional contribution is taken into account together with the IR corrections
in Eqs.~(\ref{eq:S_IR}) and (\ref{eq:T_IR}), the overall fit of the EW data worsens and the
bound on $\xi$ becomes stronger.

The constraints on $\widehat S$ can be also used to derive a robust lower bound on the mass of the
$\SO(5)$ composite vector resonances. By marginalizing over $\widehat T$, one finds an absolute
upper bound on $\widehat S$, namely $\widehat S \lesssim 2.5 \times 10^{-3}$ (see Fig.~\ref{fig:ST_plane}).
By assuming that the correction in Eq.~(\ref{eq:S_tree}) is the dominant contribution
to $\widehat S$ (or at least that no strong accidental cancellation happens), this constraint can be translated
into a lower bound on the mass of the vector resonances $m_\rho \gtrsim 2\ \mathrm{TeV}$.

The result we obtained in the two-site set-up is in agreement with the general
estimate of the tree-level corrections to $\widehat S$ derived in Chap.~\ref{ch:SILH} (see Eq.~(\ref{Shatest})) \begin{equation}\label{eq:S_tree_est}
\Delta \widehat S \sim \frac{m_W^2}{m_*^2}\,,
\end{equation}
where $m_*$ denotes the mass scale of the lightest vector resonances. If a larger
set of resonances is present, additional contributions to $\widehat S$ are
generated. For instance, if some ``axial'' resonances in the $\SO(5)/\SO(4)$ coset are present,
the tree-level correction becomes~\cite{Contino:2010rs}
\begin{equation}\label{eq:S_altern}
\Delta \widehat S = m_W^2 \left(\frac{1}{m_\rho^2} + \frac{1}{m_a^2}\right)\,,
\end{equation}
where $m_a$ denotes the mass of the axial resonances. This result coincides with the one
obtained in the alternative two-site construction presented
in Sect.~\ref{sec:alternative_constructions}. Notice that in the minimal constructions
$m_\rho < m_a$, thus the correction due to the ``axial'' resonances is subleading.
Similar considerations apply to possible additional heavy vector resonances, whose contribution
to $\widehat S$ is suppressed by the larger mass and can be usually neglected with respect to the one coming
from the lightest states.

The vector resonances can also induce radiative corrections to $\widehat S$. As in the case of the $\widehat T$
parameter, these corrections are subdominant with respect to the IR effects in Eq.~(\ref{eq:S_IR}) and can thus be
usually neglected.\footnote{A more accurate computation of the corrections to $\widehat S$
can be obtained through the use of a dispersion relation approach, which allows to partially take into
account additional UV contributions~\cite{Orgogozo:2012ct}. These effects, however, turn out to be numerically
small so that the full results is very well approximated by the sum of the tree-level contributions and of the
IR corrections.}

\subsubsection{Contributions from ``non-local'' operators}\label{sec:ewpt_nonlocal}

Before concluding the discussion of the vector resonances corrections to the oblique parameters,
we briefly discuss how the above results are modified
in the presence of non-nearest neighbor (or ``non-local'') interactions.
The role of this class of interactions in the multi-site constructions has been
discussed in Sect.~\ref{sec:non-local_operators}, where we showed that they can spoil
the collective breaking protection and reintroduce a divergence in the Higgs potential.

For definiteness in the following we will focus on the alternative two-site construction
presented in Sect.~\ref{sec:alternative_constructions}, similar results, however, are obtained
in the any other model. The leading ``non-local'' interaction in this set-up
is given in Eq.~(\ref{eq:non-local_altern}) and its main effect is to introduce an additional
contribution $f'^2$ to the Goldstone decay constant, $f$, which is modified as
\begin{equation}
f^2 = f'^2 + \frac{f_1^2 f_2^2}{f_1^2 + f_2^2}\,.
\end{equation}
The modification of the relation between $f$ and the two $\sigma$-model decay constants
$f_{1,2}$ leads to a change in the tree-level corrections to the $\widehat S$
parameter given in Eq.~(\ref{eq:S_altern}), which now becomes~\cite{Contino:2010rs,DeCurtis:2011yx}
\begin{equation}
\Delta \widehat S = m_W^2 \frac{f^2 - f'^2}{f^2} \left(\frac{1}{m_\rho^2} + \frac{1}{m_a^2}\right)\,.
\end{equation}
From this expression one can see that a positive value of $f'^2$ determines a reduction
of the corrections to $\widehat S$. The tree-level correction can even become negative
if $f' > f$. For this to happen without leading to negative squared masses for the
vector resonances, one needs to choose $f_2^2 < 0$. In these configurations the vector resonances
in the adjoint of $\SO(4)$ are heavier than the axial ones in the coset $\SO(5)/\SO(4)$,
as can be seen from the explicit expressions
\begin{equation}
m_\rho^2 = \frac{1}{2} g_\rho^2 f_1^2\,,
\qquad \quad
m_a^2 = \frac{1}{2} g_\rho^2 (f_1^2 + f_2^2)\,,
\end{equation}
where $g_\rho$ is the gauge resonances coupling.
If the ``non-local'' operators are not present, instead, one necessarily gets $m_\rho < m_a$
and a positive tree-level correction to the $\widehat S$ parameter.


\subsection{The fermion contribution}\label{sec:ferm_EW_parameters}

Another important set of corrections to the oblique EW parameters is generated at the radiative
level due to loops of fermionic states. As we explained before and summarized in Fig.~\ref{fig:ST_plane},
these contributions can be crucial to reconcile the composite Higgs scenarios with the EW data.
This is especially true in the scenarios favored by Naturalness, which requires a relatively large value for $\xi$.
In this section we analyze in details the fermion contributions to the oblique
parameters and derive some estimates of their size.

\subsubsection{The ${\mathbf{\widehat S}}$ parameter}

We start the analysis by discussing the corrections to the $\widehat S$ parameter.
As a preliminary step it is interesting to determine the degree of divergence associated to these corrections,
which will tell us if they are ``calculable'' or not in the various models.

We saw in Sect.~\ref{sec:oblique_pars_universal} that in the simplest and more general description of a Goldstone Higgs,
the non-linear $\sigma$-model, the one-loop corrections to $\widehat S$ are logarithmically divergent. In that case we could
only estimate the radiative effects by retaining the dominant IR running correction which is enhanced with
respect to the UV contributions by the large logarithm. Obviously the minimal non-linear $\sigma$-model does not
contain a description of the fermionic states, so for the present discussion we need to consider a more complete
implementation of the composite Higgs scenarios.

One possibility is to consider the multi-site constructions described in Chap.~\ref{ch:Pheno_models}.
The divergence structure in these models is considerably different
from the one in the $\sigma$-model description. Indeed in the multi-site constructions all the
effects related to EWSB are protected by a collective breaking mechanism and the degree
of divergence associated to the corresponding operators is automatically lowered.
For instance, in the $5+5$ two-site model discussed in Sect.~\ref{sec:metter_sector},
the leading fermionic contribution to $\widehat S$ corresponds to an effective operator with the structure
\begin{eqnarray}
\frac{c_S}{(16 \pi^2)^2 f^2} {\mathcal O}_S^f &=& \frac{c_S}{(16 \pi^2)^2 f^2}
\Tr\left[A_{\mu\nu}\, {\mathcal U}\, \widetilde m^2\, {\mathcal U}^t A^{\mu\nu}\right]\nonumber\\
& \supset & \frac{1}{2} \frac{c_S}{16 \pi^2} \frac{\widetilde m_Q^2 - \widetilde m_T^2}{16 \pi^2 f^2}
\sin^2 \left(\frac{H}{f}\right) g_0 g_0' W^3_{\mu\nu} B^{\mu\nu}\,.\hspace{2em}
\label{eq:S_ferm_op}
\end{eqnarray}
As required by the collective-breaking mechanism, this operator contains two insertions
of the $\widetilde m$ mass parameter which are needed to break the
global $\SO(5)_R$ invariance associated to the composite resonances.
Notice that the contribution to $\widehat S$ vanishes if the masses of the $\SO(4)$
fourplet and singlet components of $\widetilde \psi$ are equal, $\widetilde m_Q = \widetilde m_T$.
In this limit, indeed, the global $\SO(5)_R$ invariance is restored
and the collective-breaking structure ensures that no EWSB effect can be generated.
According to the counting in Eq.~(\ref{eq:divergence_counting_ferm}), the insertions of the $\widetilde m$ parameter
lower the degree of divergence associated to the ${\mathcal O}_S^f$ operator making it finite at one loop.

Similar considerations apply to the radiative contributions coming from the vector resonances,
which turn out to be calculable at one loop in the multi-site models. This feature explains why
in the IR contributions in Eqs.~(\ref{eq:S_IR}) and (\ref{eq:T_IR}) we identified the scale at which the
logarithmic divergence is regulated with the mass of the vector resonances.

There is, however, an important subtlety that needs to be mentioned. What we showed by the previous
analysis is that in the two-site model no counterterm needs to be introduced at one loop for the operator
corresponding to the $\widehat S$ parameter.
In the language of Feynman diagrams, this means that the 1PI (one particle irreducible) contributions
to $\widehat S$ are finite. However additional corrections coming from non-1PI diagrams are in general also present.
These diagrams can include some sub-divergences which must be removed by a renormalization of the
parameters of the model. In other words, the corrections to the $\widehat S$ parameter are
finite only when expressed in terms of renormalized parameters.
It is only in this sense that we can consider $\widehat S$ ``calculable'' in the multi-site constructions.

\begin{figure}
\centering
\includegraphics[width=0.925\textwidth]{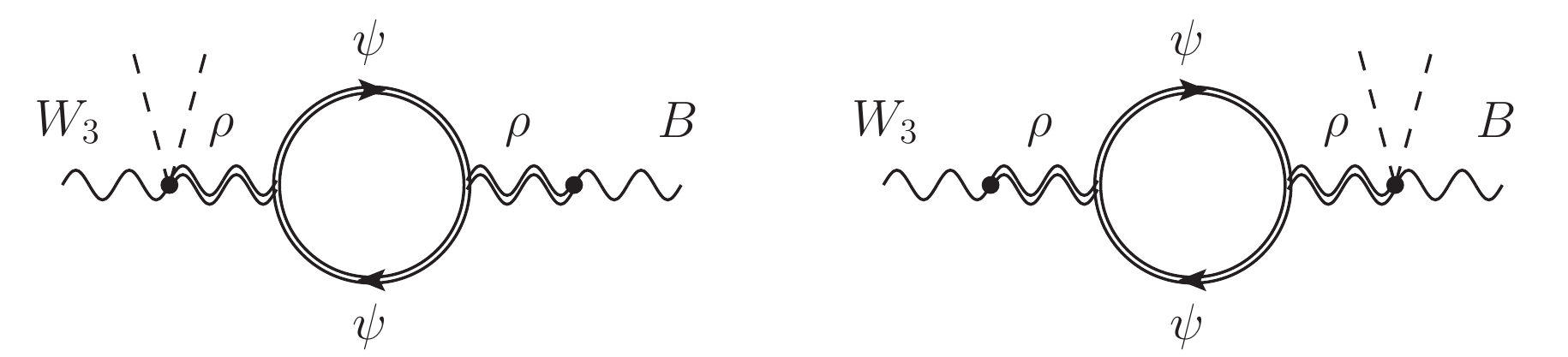}
\caption{Radiative contribution to the $\widehat S$ parameter generated by a
composite fermion loop. The dashed lines represent insertions of the Higgs VEV.}
\label{fig:S_ferm}
\end{figure}

A simple example of sub-divergent contributions to $\widehat S$ in the two-site model
comes from the mixing of the $\widehat S$ operator with the two-point function of the vector resonances.
We already saw that this mixing, which is induced by the mass terms that link the elementary fields to the composite
vector states, is responsible for generating the tree-level corrections to $\widehat S$.
At one loop the correlator of the vector fields acquires a logarithmic divergence, which leads to
a sub-divergence in $\widehat S$. The sub-divergence is obviously regulated by the renormalization of
the vector resonances coupling $\widetilde g_\rho$.
Typical diagrams giving rise to this effect are shown in Fig.~\ref{fig:S_ferm}.
Notice that they include a loop containing only composite fermions and not elementary states.
The elementary fermions, indeed, are not directly coupled to the composite vectors and can
enter in the loop only through the mass mixing with the composite fermions.
Insertions of the mass mixings introduce additional fermion propagators making the diagram finite,
therefore the elementary fermions do not contribute to the divergence.

Although the multi-site constructions allow a full determination of the corrections to $\widehat S$,
the actual computation is substantially  cumbersome. Moreover the result crucially depends
on the details of the model and can not be discussed in full generality. Fortunately in many cases a simpler and more general
approach can be used to extract the leading corrections to the oblique parameters.
This can be done by noticing that the fermionic top partner resonances are preferentially lighter than the vectors
for reasons related with the generation of the Higgs mass that we discussed in the previous chapters.
In this situation we can use a simplified description of the
composite Higgs theories that only includes the Higgs dynamics, the elementary fields and the light fermionic partners,
where the heavy vector states are integrated out.
This minimal description is provided by the CCWZ constructions presented in Chap.~\ref{ch:LHC_pheno}, in which
the Higgs (and the elementary gauge fields) is described by the $\sigma$-model
Lagrangian discussed in Sect.~\ref{sec:pheno_models_sigma-model},
whereas the elementary SM fermions and their composite partners are described by
the most general effective Lagrangian compatible with the Goldstone symmetry.
Obviously, these effective models must be interpreted as a valid descrption of the composite Higgs scenarios up to
a cut-off that coincides with the mass of the vector resonances, $m_\rho$.
In the following, we will use these effective descriptions to quantitatively analyze the corrections to the oblique parameters.

Given that no collective-breaking structure is present in the effective Lagrangian,
it is easy to understand that the divergence counting in the general CCWZ constructions coincides with the one of the usual
non-linear $\sigma$-model, which is encoded in Eq.~(\ref{eq:divergence_counting}).
The one loop corrections to the $\widehat S$ parameter are thus expected to be logarithmically divergent.
This results is valid irrespectively of the fields that circulate in the loop, thus it applies to the gauge contributions
(as we explicitly verified when we computed the IR effects) as well as to the fermion ones.

The leading logarithmically enhanced contributions to $\widehat S$ coming from fermion loops can be
easily computed in the effective theory. As an explicit example we consider the $5+5$ model, whose general effective
Lagrangian has been presented in Sect.~\ref{sec:5+5_effective}. Similar results are however valid in the other models.
The logarithmically enhanced corrections to $\widehat S$ coming from the top partners
are given by~\cite{Grojean:2013qca,Azatov:2013ura}
\begin{equation}\label{eq:S_log_ferm}
\Delta \widehat S = \frac{g^2 N_c}{24 \pi^2} (1 - c_L^2 - c_R^2)\, \xi \log \left(\frac{m_\rho^2}{m_4^2}\right)\,,
\end{equation}
where $N_c$ is the number of QCD colors. It is important to notice that this class of divergent contributions to $\widehat S$
is generated only if composite multiplets in non-trivial representations of $\SO(4)$ are present in the effective theory.
In the $5+5$ model, for instance, the correction in Eq.~(\ref{eq:S_log_ferm}) is due to the light $\SO(4)$ fourplet.
This explains why the argument of the logarithm is written in terms of the fourplet mass $m_4$.
On the other hand, loops containing only $\SO(4)$ singlets can not generate a divergence. This follows from the fact
that the gauge interactions of an $\SO(4)$ singlet with the SM gauge fields are trivially given by the standard covariant
derivative and are not modified after EWSB (this can be explicitly seen from Eq.~(\ref{eq:cov_der_singlet})).

The fact that the correction in Eq.~(\ref{eq:S_log_ferm}) is independent of the elementary/composite mixings $y_{L,R}$
is quite remarkable. In particular it implies that any light non-trivial $\SO(4)$ multiplet contributes to $\widehat S$ with
a sizable shift.\footnote{Similar contributions to $\widehat S$ are also present in technicolor
models~\cite{Golden:1990ig,Barbieri:2008zt}.}
In models in which many light fermionic resonances are present, as for instance
if all the SM fermion generations have light partners, the natural size of the corrections
to $\widehat S$ can become very large and some amount of cancellation may be required to pass
the experimental bounds.

\begin{figure}
\centering
\includegraphics[width=0.5\textwidth]{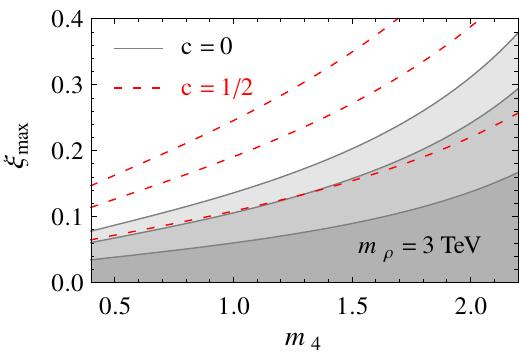}
\caption{Upper bound on $\xi$ in the $5 + 5$ model as a function of the fourplet mass $m_4$
for $m_\rho = 3\ \mathrm{TeV}$ and different values of $c \equiv c_L = c_R$.
The results have been obtained by considering the contributions
to $\widehat S$ given in Eqs.~(\ref{eq:S_IR}), (\ref{eq:S_tree}) and (\ref{eq:S_log_ferm}) and
by marginalizing on $\widehat T$. The shaded regions correspond to the points compatible
with the constraints at the $68\%$, $95\%$ and $99\%$ confidence level for $c = 0$.
The dashed red curves show how the bounds are modified for $c = 1/2$.}
\label{fig:xi_bound_S}
\end{figure}

The logarithmically enhanced contribution to $\widehat S$ in Eq.~(\ref{eq:S_log_ferm}) is
sizable if $c_L^2 + c_R^2$ is not too close to $1$ and is typically much larger than the  IR
effects in Eq.~(\ref{eq:S_IR}). Even if we assume that only
one multiplet of partners is light, as in the minimal two-site $5+5$ model,
the correction due to fermion loops can be comparable
with the tree-level contribution in Eq.~(\ref{eq:S_tree}) if the strong coupling $\widetilde g_\rho$
is large, $\widetilde g_\rho \gtrsim 5$.
From an effective theory point of view, the coefficients
$c_{L,R}$ are just free parameter, thus in principle the fermion contribution to $\widehat S$ can
have an arbitrary sign. In particular for $c_L^2 + c_R^2 > 1$ a sizable negative shift is induced,
which could improve the agreement with the EW precision measurements (see Fig.~\ref{fig:ST_plane}).
On the other hand, if no cancellation is present a large positive shift in $\widehat S$
is unavoidable in the presence of relatively light composite multiplets. For example
for $m_4 \simeq 700\ \mathrm{GeV}$ and $m_\rho \simeq 3\ \mathrm{TeV}$, independently of the value of $\widehat T$,
a tight upper bound, $\xi \lesssim 0.1$, is obtained for $c_L = c_R = 0$, which corresponds to the two-site set-up.
The limits on the compositeness scale for $c \equiv c_L = c_R = 0$ and $c \equiv c_L = c_R = 1/2$ as a function of the
fourplet mass are shown in Fig.~\ref{fig:xi_bound_S}.

Before concluding this discussion it is interesting to notice that, from the effective theory
point of view, the logarithmically enhanced correction in Eq.~(\ref{eq:S_log_ferm})
can be interpreted as a running of the two dimension-six operators which contribute
to $\widehat S$
\begin{equation}
{\mathcal O}_W = i \left(H^\dagger \sigma^i \dblarrow{D}_\mu H\right)(D^\nu W_{\mu\nu})^i\,,
\quad
{\mathcal O}_B = i \left(H^\dagger \dblarrow{D}_\mu H\right)(D^\nu B_{\mu\nu})\,,
\end{equation}
where $H^\dagger \dblarrow D_\mu H$ is the derivative $H^\dagger (D_\mu H) - (D_\mu H)^\dagger H$.
An essential ingredient for the generation of the running is the presence
of the non-renormalizable gauge interactions
due to the non-linear Higgs dynamics. Interactions of this kind are contained
in the $e_\mu$ term in the covariant derivative of the composite fermions and in the
$d$-symbol term (see the explicit expressions in Appendix~\ref{so54}).
The logarithmically enhanced contribution is instead absent
in a theory with only renormalizable interactions~\cite{Carena:2006bn,Grojean:2013qca}.

\subsubsection{The ${\mathbf{\widehat T}}$ parameter}
\label{THATLOOP}

We can now analyze the corrections to the $\widehat T$ parameter.
Similarly to what happens for $\widehat S$, the $\widehat T$ parameter is calculable in the multi-site set-ups.
The custodial protection, however, provides an additional reduction of the degree of divergence,
so that, as we will discuss in the following, $\widehat T$ is in many cases calculable even in a general
effective theory independently of a collective-breaking mechanism.

As a preliminary step to exploit the implications of the custodial symmetry, we need
to determine the $\SO(4) \simeq \SU(2)_L \times\SU(2)_R$ quantum numbers of the spurions that control
the breaking of $\SO(3)_c$ induced by the mixing with the elementary states.
For this purpose, a useful choice is to assume that the elementary fields are only charged under $\SU(2)_L$
with the usual SM quantum numbers, whereas they are neutral under $\SU(2)_R$.
The Higgs, on the other hand, belongs to the $(\mathbf{2}, \mathbf{2})$ representation. In order to formally
restore the $\SO(4)$ invariance we then need to associate
the elementary/composite mixings and the elementary gauge couplings
to a set of spurions. It is easy to see that the
spurions must be neutral under the $\SU(2)_L$ group, while they have non-trivial transformation properties under
$\SU(2)_R$.  In Table~\ref{tab:spurions} we list the quantum numbers of the spurions for different choices of the representations
of the composite operators involved in the mixings.
\index{spurions}
\begin{table}
\centering
\setlength\tabcolsep{10pt}
\setlength\doublerulesep{1.5pt}
\begin{tabular}{c | c  c  c  c}
\rule[-6pt]{0pt}{10pt}field & composite op. & spurion & $\SO(5)$ reps. & $\widehat T$ contrib.\\
\hline
\hline
\raisebox{-.85em}[0pt][-.85em]{$q_L$} & \rule[-6pt]{0pt}{1.7em} $({\mathbf 2}, {\mathbf 1})$ & $y_L \in ({\mathbf 1}, {\mathbf 1})$
& $\mathbf{4}$ & ---\\
& \rule[-6pt]{0pt}{1.7em} $({\mathbf 2}, {\mathbf 2})$ & $y_L \in ({\mathbf 1}, {\mathbf 2})$
& $\mathbf{5}$, $\mathbf{10}$, $\mathbf{14}$ & $4$ insertions \\
\hline
& \rule[-6pt]{0pt}{1.7em} $({\mathbf 1}, {\mathbf 1})$ & $y_R \in ({\mathbf 1}, {\mathbf 1})$
& $\mathbf{5}$, $\mathbf{14}$ & ---\\
$t_R$ & \rule[-6pt]{0pt}{1.7em} $({\mathbf 1}, {\mathbf 2})$ & $y_R \in ({\mathbf 1}, {\mathbf 2})$
& $\mathbf{4}$ & $4$ insertions\\
& \rule[-6pt]{0pt}{1.7em} $({\mathbf 1}, {\mathbf 3})$ &$y_R \in ({\mathbf 1}, {\mathbf 3})$
& $\mathbf{10}$ & $2$ insertions\\
\hline
$W_\mu^a$ & \rule[-6pt]{0pt}{1.7em} $({\mathbf 3}, {\mathbf 1})$ & $g_0 \in ({\mathbf 1}, {\mathbf 1})$
& $\mathbf{10}$ & ---\\
\hline
$B_\mu$ & \rule[-6pt]{0pt}{1.7em} $({\mathbf 1}, {\mathbf 3})$ & $g'_0 \in ({\mathbf 1}, {\mathbf 3})$
& $\mathbf{10}$ & $2$ insertions
\end{tabular}
\caption{Quantum numbers of the spurions associated to the elementary/composite mixings
for some of the most common $\SO(4) \simeq \SU(2)_L \times \SU(2)_R$ representations of the composite operators.
The fourth column shows the $\SO(5)$ representations that contain each $\SO(4)$ multiplet.
In the fifth column we report the minimal number of insertions needed
to generate a contribution to the $\widehat T$ parameter (if no number is given the
corresponding spurion does not break the custodial invariance). The $\widehat T$ parameter
is finite in the effective theory if more than two insertions of the spurions are needed,
whereas it is logarthmically divergent if only two insertions are enough.}
\label{tab:spurions}
\end{table}

As we will show in Appendix~\ref{sec:custodial_symmetries}, the $\widehat T$ operator belongs to the
$(\mathbf{5}, \mathbf{1})$ representation of $\SU(2)_L \times \SU(2)_R$.
Given that all the spurions are neutral under $\SU(2)_L$, a contribution to $\widehat T$
necessarily requires at least four Higgs insertions:
$H^4 \simeq (\mathbf{5}, \mathbf{5}) \oplus (\mathbf{3}, \mathbf{3}) \oplus (\mathbf{1}, \mathbf{1})$.
To build a suitable operator we also need to insert some powers of the spurions
that break the custodial invariance. Each spurion insertion lowers the
degree of divergence by one power. If two insertions are sufficient the one-loop contribution to $\widehat T$
can be logarithmically divergent, while more than two insertions ensure that $\widehat T$ is finite.
Notice that in the latter case the fermion contributions to $\widehat T$ is fully finite and does not contain
sub-divergences coming from non-1PI diagrams. Indeed the insertions of the spurions are associated to the
fermionic lines in the diagrams and must necessarily appear in the loop.

As can be seen from Table~\ref{tab:spurions}, in many
models the minimal number of $y_{L,R}$ insertions is four, implying that $\widehat T$ is calculable
also in the effective theory. It is interesting to notice that this happens when the elementary fermions
are mixed with strong sector operators in the $\mathbf{5}$ and $\mathbf{14}$ representations
of $\SO(5)$. As we will see in Sect.~\ref{sec:Zbb}, in these cases a custodial protection for the
$b_L$ couplings to the $Z$ boson is also present, which improves the compatibility of the
model with the experimental data.

\begin{figure}
\centering
\includegraphics[width=0.375\textwidth]{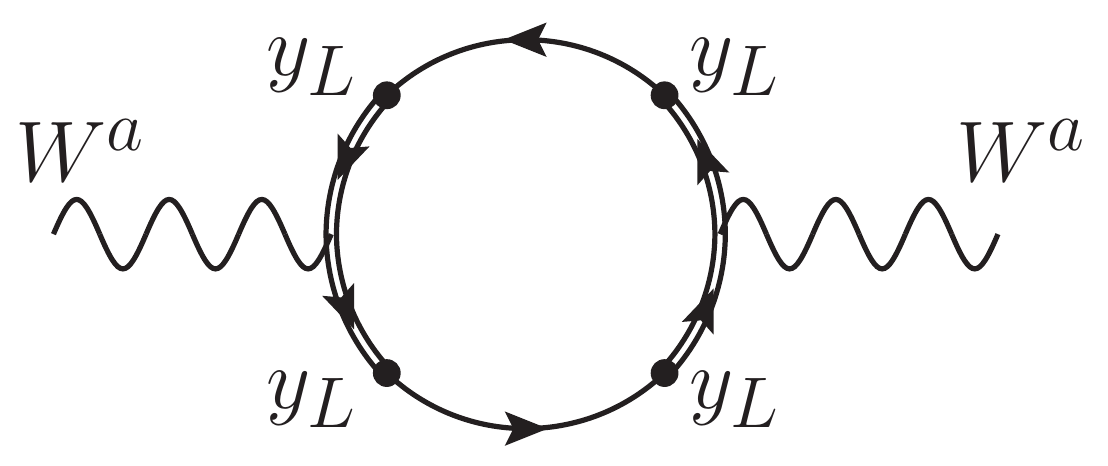}
\caption{Schematic structure of the radiative contribution to the $\widehat T$ parameter
generated by a composite fermion loop at leading order in the $y$ expansion.}
\label{fig:T_ferm}
\end{figure}

We can now derive some estimates of the size of the one-loop corrections to $\widehat T$ induced by fermion loops.
For definiteness, in the following we will focus on the scenarios in which the $\widehat T$ parameter is finite
in the effective theory. In this case a typical diagram contributing at leading order in the $y$
expansion is shown in Fig.~\ref{fig:T_ferm}.
It is straightforward to estimate the leading corrections to $\widehat T$ coming from fermion loops:
\begin{equation}\label{eq:T_loop_ferm}
\Delta \widehat T \simeq \frac{N_c}{16 \pi^2} \frac{y_L^4 f^2}{m^2} \xi\,,
\end{equation}
where $m$ is the mass of the lightest top partner and $y_L$ collectively denotes the elementary/composite mixing
of the third-generation $q_L$ doublet. To get a quantitative estimate we can extract the value of the $y_L$ mixing from the
top mass. If we assume that the elementary/composite mixings have comparable sizes,
$y_{L} \simeq y_{R} \simeq y$, the top Yukawa can be
estimated as $y_t \simeq y^2 f/m$. By using this expression we get the estimate
\begin{equation}\label{eq:T_ferm}
\Delta \widehat T \simeq \frac{N_c}{16 \pi^2} y_t^2 \xi \simeq 2 \times 10^{-2} \xi\,.
\end{equation}
Notice that this contribution is typically dominant with respect to the IR
correction in Eq.~(\ref{eq:T_IR}). Moreover, the sign of the fermion contribution
can be positive, thus compensating the negative shift in Eq.~(\ref{eq:T_IR}).

Additional contributions to $\widehat T$ coming from heavier resonances can also be present. They are however
suppressed with respect to the corrections in Eq.~(\ref{eq:T_loop_ferm}) by the larger resonance mass.


\subsection{Explicit results}

In the general analysis presented in the previous sections we found that, in a large class
of composite Higgs models, the corrections to the oblique EW parameters can be computed in
a reliable way from the low-energy dynamics of the theory. In particular the leading corrections
can be extracted within an effective theory which only includes the SM fields and the
lightest fermionic resonances.

Given the importance of the fermionic contributions in making the composite scenarios compatible with the
EW precision data, it is worth analyzing them carefully in explicit models.
In the following we present such analysis in two classes of minimal models in which the composite resonances
belong to the fundamental representation of $\SO(5)$, namely the $5+5$ and $5+1$ scenarios described
in Sect.~\ref{sec:effect_parametr}. The explicit results will also allow us to get
a solid quantitative determination of the constraints coming from the EW precision measurements.

\subsubsection{The ${\mathbf{5+5}}$ model}

As a first example we will focus on the $5+5$ model.
This scenario follows the usual partial compositeness structure, in which the $q_L$ and $t_R$ fields are realized
as elementary states. The composite operators that mix with the elementary fermions transform
in the fundamental representation of $\SO(5)$, thus the fermionic resonances belong to the fourplet
and singlet representations of $\SO(4)$.

Before discussing the results in the complete model, it is interesting to consider two simple
limits, in which only one $\SO(4)$ multiplet of composite fermions is present in the low-energy
spectrum.

\begin{figure}[t!]
\centering
\includegraphics[width=0.45\textwidth]{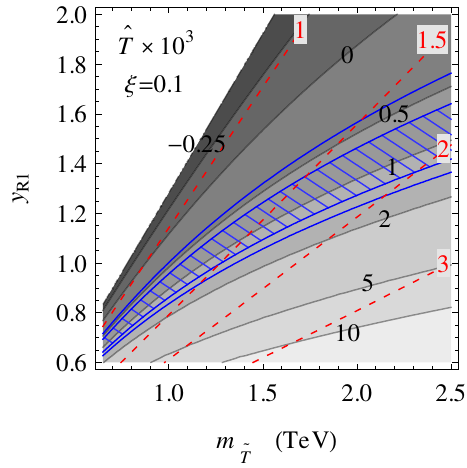}
\caption{Corrections to the $\widehat T$ parameter in the $5+5$ model as a function of the singlet mass $m_{\widetilde T}$
and the $y_{R1}$ mixing for $\xi = 0.1$. The results correspond to the limit with only a light singlet and includes
the exact fermion one-loop corrections and the IR contribution in Eq.~(\ref{eq:T_IR}).
The dashed red lines correspond to the contours with fixed $y_{L1}$.
The solid blue contours give the regions that are compatible with the constraints on the
oblique parameters at the $68\%$ and $95\%$ confidence level.}
\label{fig:T_singlet}
\end{figure}

In the case in which only the singlet is light, the fermionic contributions to $\widehat S$
are finite and are negligible with respect to the tree-level corrections in Eq.~(\ref{eq:S_tree}).
The correction to $\widehat S$ is thus positive and is fixed by the mass of the
heavy vector resonances.
The $\widehat T$ parameter, on the other hand, receives a sizable positive contribution from the fermion loops.
At leading order in $v/f$ the explicit result reads
\begin{equation}\label{eq:T_singlet}
\Delta \widehat T = \frac{3 \xi}{64 \pi^2} \frac{y_{L1}^4 m_1^2 f^2}{(m_1^2 + y_{R1}^2 f^2)^3}
\left\{m_1^2 + 2 y_{R1}^2 f^2 \left[
\log\left(\frac{2 (m_1^2 + y_{R1}^2 f^2)^2}{v^2 y_{L1}^2 y_{R1}^2 f^2}\right)
- 1\right]
\right\}\,.
\end{equation}
In a large part of the parameter space this correction can compensate the negative shift that comes from
the IR contribution in Eq.~(\ref{eq:T_IR}). The total shift in $\widehat T$ is shown in
Fig.~\ref{fig:T_singlet} for the reference value $\xi = 0.1$ corresponding to $f = 780\ \textrm{GeV}$.
The results are shown as a function of the singlet mass $m_{\widetilde T} \simeq m_1^2 + y_{R1}^2 f^2$
and the $y_{R1}$ mixing.\footnote{The top mass, as in all the numerical results we show in this chapter, has been
fixed to the value $m_t = m_t^{\overline{MS}}(2\ \textrm{TeV}) = 150\ \textrm{GeV}$, which corresponds to
a pole mass $m_t^{pole} = 173\ \textrm{GeV}$.} To derive the value of $\widehat S$ the value
$m_\rho = 3\ \textrm{TeV}$ has been used for the vector resonances mass scale.
It can be seen that sizable positive values of $\widehat T$ can easily be obtained for reasonable values of
the singlet mass and of the elementary/composite mixings.

As a second simplified limit we consider the case in which the resonance spectrum contains only
a light fourplet. In this case the dominant contribution to the $\widehat S$ parameter
comes from the logarithmically enhanced corrections due to the loops of fermionic resonances.
The explicit result coincides with the one in Eq.~(\ref{eq:S_log_ferm}) with $c_L=c_R=0$:
\begin{equation}\label{eq:S_4plet}
\Delta \widehat S = \frac{g^2}{8\pi^2} \xi \log\left(\frac{m_\rho^2}{m_4^2}\right)
\simeq 1.6 \times 10^{-2}\, \xi\,,
\end{equation}
where the numerical estimate has been obtained by setting $m_4 \simeq 700\ \mathrm{GeV}$
and $m_\rho \simeq 3\ \mathrm{TeV}$. If the gauge resonances are heavy $m_\rho/f \simeq g_\rho \gtrsim 4$,
the correction in Eq.~(\ref{eq:S_4plet}) is comparable to the tree-level one.
The sizable positive contribution to $\widehat S$ implies a quite stringent bound on the
compositeness scale, $\xi \lesssim 0.1$, irrespectively of the value of $\widehat T$.
An even stronger constraint is obtained if one also
considers the corrections to the $\widehat T$ parameter. The fermion contribution at the leading
order in the $y$ expansion is given by
\begin{equation}
\Delta \widehat T = - \frac{\xi}{32 \pi^2} \frac{y_{L4}^4 f^2}{m_4^2}\,.
\end{equation}
The approximate result shows that the shift in $\widehat T$ tends to be negative.
A numerical computation confirms that this result is valid in a large part of the
parameter space of the model.

The fact that the shift in $\widehat T$ is necessarily negative makes the constraints coming
from the oblique parameters extremely severe. Using the fit in Fig.~\ref{fig:ST_plane}
an upper bound $\xi \lesssim 0.02$ at the $99\%$ confidence level is obtained, which corresponds
to a lower bound $f \gtrsim 1.7\ \mathrm{TeV}$.

\begin{figure}[t!]
\centering
\includegraphics[width=0.45\textwidth]{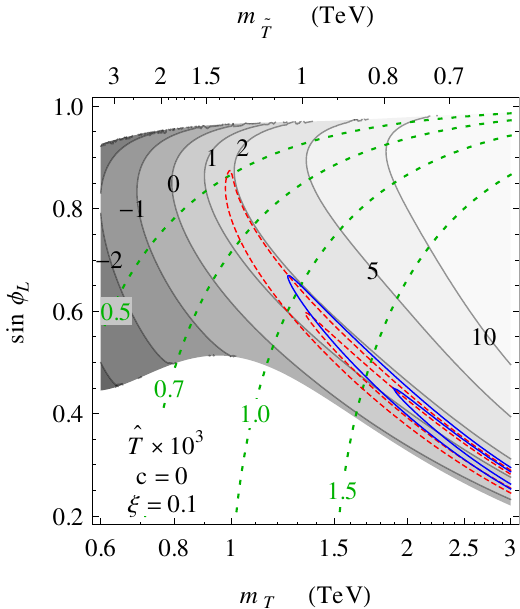}
\hfill
\includegraphics[width=0.45\textwidth]{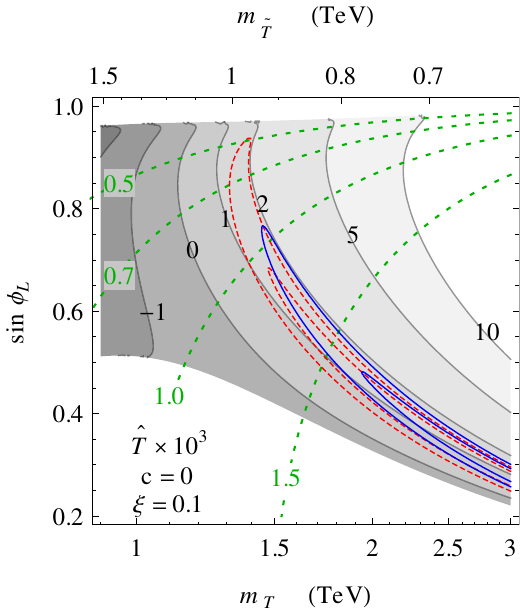}
\caption{Corrections to the $\widehat T$ parameter as a function of the mass of the top partners
and of the $q_L$ compositeness in the $5+5$ model with $c=0$ for $\xi = 0.1$.
The two plots correspond to the two choices of $y_{R}$ that allow to obtain the correct Higgs and
top masses at fixed $m_T$ and $\phi_L$. In the white regions the Higgs and top masses can not
be reproduced. The dashed green contours show the mass (in $\mathrm{TeV}$) of the exotic composite state
$X_{5/3}$. The solid blue contours give the regions that are compatible with the constraints
on the oblique parameters at the $68\%$ and $95\%$ confidence level, while the dashed red lines
show how the bounds are modified if we assume a $25\%$ reduction in $\widehat S$.
The plots are taken from Ref.~\cite{Grojean:2013qca}.}
\label{fig:T_2site}
\end{figure}

\begin{figure}
\centering
\includegraphics[width=0.45\textwidth]{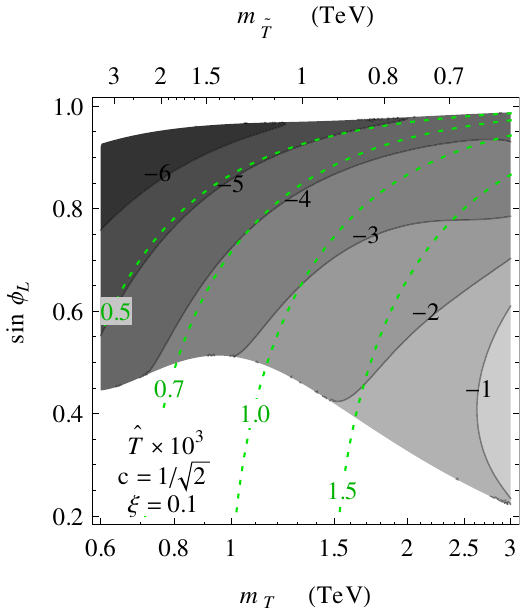}
\hfill
\includegraphics[width=0.45\textwidth]{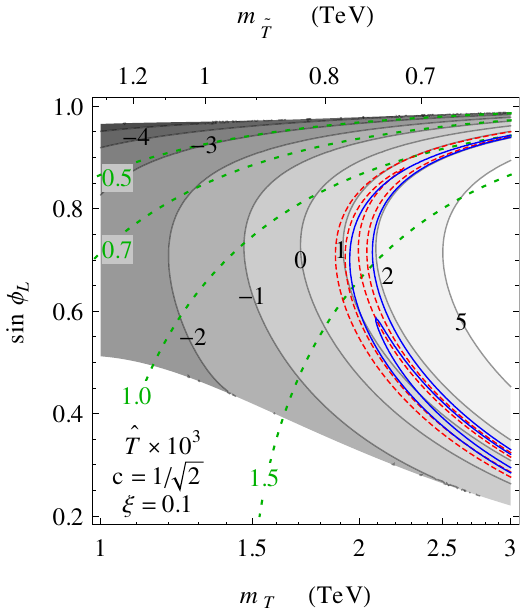}
\caption{Corrections to the $\widehat T$ parameter as a function of the mass of the top partners
and of the $q_L$ compositeness in the $5+5$ model with $c=1/\sqrt{2}$ for $\xi = 0.1$.
For a description of the symbols used see caption of Fig.~\ref{fig:T_2site}.
The plots are taken from Ref.~\cite{Grojean:2013qca}.}
\label{fig:T_c}
\end{figure}

As a final case we consider two more complete models that include both a fourplet and a singlet.
In order to reduce the number of free parameters it is useful to choose a common value for the
left and right elementary mixings, $y_{L4} = y_{L1} = y_L$ and $y_{R4} = y_{R1} = y_R$,
and for the coefficients of the $d$-symbol operators, $c_L = c_R = c$. With this
choice the effective Lagrangian (excluding the interactions with the gauge fields)
reproduces the one of the two-site $5+5$ model discussed in Sect.~\ref{sec:5+5_model}. This means
that the Higgs mass is calculable and the result in Eq.~(\ref{eq:mH_mt_3-site}) can be used
to relate it to the masses of the lightest top partners.
By imposing this relation and requiring the correct value for the top mass, we can describe the parameter space of the
model in terms of three free parameters, which we can choose to be the mass of the $T$ resonance
inside the fourplet ($m_T \simeq m_4^2 + y_L^2 f^2$), the $q_L$ compositeness angle $\phi_L$,
defined as
\begin{equation}
\sin \phi_L \equiv \frac{y_L}{\sqrt{m_4^2 + y_L^2 f^2}}\,,
\end{equation}
and the coefficient of the $d$-symbol term, $c$.
With the above choice the right mixing $y_R$ is determined up to a twofold ambiguity. In the
numerical results (in Figs.~\ref{fig:T_2site} and \ref{fig:T_c}) we show two plots that correspond
to the two choices of $y_R$.

For definiteness we focus on two explicit models obtained for particular choices of $c$. The first
one is the case $c=0$, which corresponds to the minimal $5+5$ two-site model of Sect.~\ref{sec:5+5_model}.
The second case corresponds to the choice $c = 1/\sqrt{2}$ for which the logarithmically enhanced
fermion contribution to $\widehat S$ vanishes (see Eq.~(\ref{eq:S_log_ferm})).

The numerical results for $\widehat T$ for the case $c=0$ are shown in Fig.~\ref{fig:T_2site}
for $\xi = 0.1$, that roughly corresponds to the maximal value allowed by the bounds
on the $\widehat S$ parameter. In the numerical results the Higgs mass has been fixed to
the value $m_h = 126\ \mathrm{GeV}$. As expected from the results discussed in the simplified cases
with only one light multiplet, in the region in which the fourplet is light the corrections to
$\widehat T$ are negative, whereas a light singlet typically implies a positive shift.
The fit of the oblique parameters puts strong bounds on the parameter space of the model. In the plots
the allowed regions for $68\%$ and $95\%$ confidence level are shown.

As can be seen from Fig.~\ref{fig:T_2site}, the oblique parameters can be used to set some lower bounds
on the masses of the resonances coming from the composite fourplet. At the $95\%$ confidence level,
one finds $m_{X_{2/3}} \simeq m_{X_{5/3}} \gtrsim 950\ \mathrm{GeV}$ for the mass of the
exotic doublet $\mathbf{2}_{7/6}$ and $m_T \simeq m_B \gtrsim 1.2\ \mathrm{TeV}$ for the
$\mathbf{2}_{1/6}$ states. If one assumes a $25\%$ cancellation in the corrections to $\widehat S$,
the bounds are significantly relaxed: $m_{X_{2/3}} \simeq m_{X_{5/3}} \gtrsim 500\ \mathrm{GeV}$
and $\mathbf{2}_{7/6}$ and $m_T \simeq m_B \gtrsim 1\ \mathrm{TeV}$. Notice that these bounds
are competitive with the ones obtained from direct searches (see Chap.~\ref{ch:LHC_pheno}).

Finally the results for the case $c = 1/\sqrt{2}$ are given in Fig.~\ref{fig:T_c} for $\xi = 0.1$.
Notice that in this case the main corrections to $\widehat S$ come from the tree-level effects, thus,
in principle, higher values of $\xi$ could be allowed. In particular for the value $m_\rho = 3\ \mathrm{TeV}$
used in the plots, the corrections to $\widehat S$ are well below the absolute upper bound given by
the current data.
By comparing the corresponding plots one can see that the results for $c= 1/\sqrt{2}$ significantly
differ from the ones for $c = 0$. In the case $c = 1/\sqrt{2}$ the corrections to $\widehat T$ tend to be
more negative and a much lighter singlet (and consequently a heavier fourplet) is needed to pass
the constraints on the oblique parameters: $m_{\widetilde T} \lesssim 800\ \mathrm{GeV}$
and $m_T \gtrsim 2\ \mathrm{TeV}$.

\subsubsection{The case of a fully composite ${\mathbf{t_R}}$}

The second scenario we consider, the $5+1$ model, is a modification of the partial compositeness structure
in which only the left-handed doublet $q_L$ is elementary, while the $t_R$ is a fully composite state.

The IR contributions to the oblique parameters and the tree-level corrections to
$\widehat S$ due to the gauge resonances do not depend on the assumptions on the fermion compositeness.
The presence of a fourplet of composite fermions induces a logarithmically enhanced contribution to
$\widehat S$, which is given by
\begin{equation}
\Delta \widehat S = \frac{g^2}{8 \pi^2} (1 - c_L^2 - c_R^2 - c_t^2) \xi \log\left(\frac{m_\rho^2}{m_4^2}\right)\,.
\end{equation}
The main difference with respect to the $5+5$ case is the presence of the
additional contribution depending on $c_t$, which comes from the
$d$-symbol operator involving the $t_R$ and the composite fourplet (see Eq.~(\ref{eq:Lcomp_5+1}))
This term can lead to a cancellation of the leading contributions even if
no light singlet is present so that $c_L = c_R = 0$.

\begin{figure}
\centering
\includegraphics[width=0.45\textwidth]{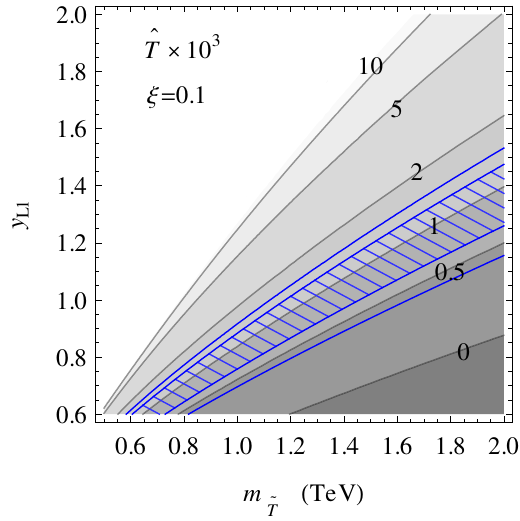}
\hfill
\includegraphics[width=0.45\textwidth]{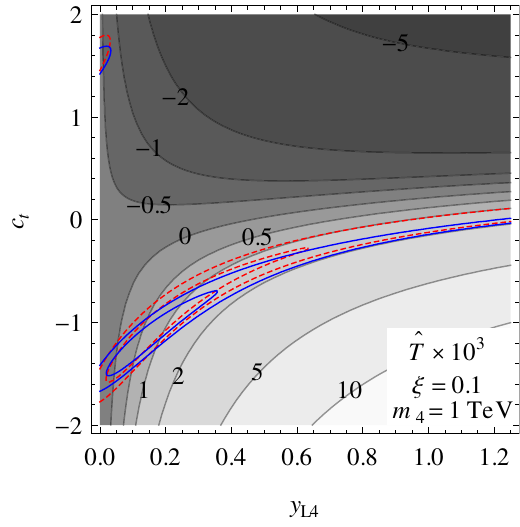}
\caption{Corrections to the $\widehat T$ parameter in the model with a fully composite $t_R$.
On the left panel the case with only a light singlet is shown as a function of the
singlet mass $m_{\widetilde T}$ and of the $q_L$ mixing. On the right panel we plot
the results as a function of the $q_L$ mixing and of $c_t$ in
the case with only a fourplet of mass $m_4 = 1\ \mathrm{TeV}$.
Both plots are obtained for $\xi = 0.1$.
The solid blue contours give the regions that are compatible with the constraints
on the oblique parameters at the $68\%$ and $95\%$ confidence level, while the dashed red lines
in the right plot show how the bounds are modified if we assume a $25\%$ reduction in $\widehat S$.
The plots are taken from Ref.~\cite{Grojean:2013qca}.}
\label{fig:T_comptR}
\end{figure}

In the following we focus on the simplified cases in which only one light $\SO(4)$ multiplet
of massive composite fermions is present.
As a first scenario we consider the case in which only a fourplet is present.
The contributions to the $\widehat S$ parameter are dominated by the tree-level corrections. The fermion
contribution to $\widehat T$ can be sizable and is typically positive. At leading order in $v/f$ it reads
\begin{equation}
\Delta \widehat T = \frac{3}{64 \pi^2}\xi \frac{y_{L1}^2 f^2}{m_1^2}
\left\{y_{L1}^2 + 2 y_{Lt}^2 \left[\log\left(\frac{2 m_1^2}{v^2 y_{Lt}^2}\right) - 1\right]\right\}\,.
\end{equation}
The total correction to $\widehat T$ is shown in the left panel of Fig.~\ref{fig:T_comptR}
for $\xi = 0.1$.
One can notice that in this set-up the overall results for the
corrections to the oblique parameters are similar to the ones we found in the case of a partially composite $t_R$.

The second scenario we consider is the one with only a light fourplet. Once we fix the top mass,
the parameter space of the model can be described by three free parameters:
the elementary composite mixing, $y_{L4}$, the fourplet mass, $m_4$ and the coefficient of the
$d$-symbol operator $c_t$.
The contributions to the $\widehat T$ parameter coming from fermion loops at leading order
in $v/f$ are given by
\begin{eqnarray}
\Delta \widehat T &=& - \frac{\xi}{32 \pi^2} \frac{y_{L4} f^2}{m_4^2}
\Bigg\{3 c_t^2 y_{L4} \left(y_{L4}^2 - 4 y_{Lt}^2\right) + y_{L4}^2 \left(y_{L4} - 3 \sqrt{2} c_t y_{Lt}\right)
\nonumber\\
&& -\; 3 y_{Lt}^2 \left(y_{L4} - 4 \sqrt{2} c_t y_{Lt}\right) \left[
\log\left(\frac{2 m_4^2}{v^2 y_{Lt}^2}\right) - 1\right]
\Bigg\}\,.
\end{eqnarray}
The terms related to the $d$-symbol operator come with accidentally large
coefficients, thus the value of $\Delta \widehat T$ has a strong dependence on $c_t$.
In the right panel of Fig.~\ref{fig:T_comptR} we show the total correction to $\widehat T$
as a function of $y_{L4}$ and $c_t$ for a fixed value of the fourplet mass, $m_4 = 1\ \mathrm{TeV}$.
The parameter space regions with better agreement with the EW data are the ones
with $c_t \sim -1$, in which the logarithmically enhanced shift in $\widehat S$ is
partially canceled.


\section[The $Z b_L \overline b_L$ coupling]{The ${\mathbf{Z b_L \overline b_L}}$ coupling}\label{sec:Zbb}

In the previous section we focused our attention on the oblique parameters, $\widehat{S}$ and $\widehat{T}$,
which encode the universal new physics effects that affect in the same way
all the fermion generations.
The oblique parameters are enough to capture all the relevant corrections
in the limit of small fermion compositeness.
This is usually an excellent approximation for the
light quark generations, but not for the third one, whose
mixing with the strong dynamics needs to be sizable to generate the large top mass.
As a consequence, the observables related to the third quark generation can receive
non-universal shifts that need to be taken into account separately from the universal effects.
Among such observables the coupling of the $Z$ boson to the bottom field plays a
prominent role in constraining the BSM dynamics. The current experiments, indeed,
tested this coupling at the $few \times 10^{-3}$ level. In this section we will consider in datails
how the $Z \overline b b$ coupling is modified in the composite Higgs scenario.

Before starting the actual analysis it is useful to fix our notation.
We parametrize the $Z$ interactions with the bottom quark by the Lagrangian
\begin{equation}\label{eq:Lagr_Z_b}
{\mathcal L}^Z = \frac{g}{\cos \theta_w} Z_\mu \overline b \gamma^\mu \left[
(g^{\textrm{SM}}_{b_L} + \delta g_{b_L}) P_L + (g^{\textrm{SM}}_{b_R} + \delta g_{b_R}) P_R\right] b\,,
\end{equation}
where $g^{\textrm{SM}}$ denotes the SM couplings (including the loop corrections), $\delta g$
corresponds to the corrections due to new physics and $P_{L,R}$ are the left and
right projectors. The weak mixing angle is denoted by $\theta_w$.
The tree-level values for the SM couplings are
\begin{equation}
g_{b_L}^{\textrm{SM,tree}} = -\frac{1}{2} + \frac{1}{3} \sin^2 \theta_w\,,
\qquad \quad
g_{b_R}^{\textrm{SM,tree}} = \frac{1}{3} \sin^2 \theta_w\,,
\end{equation}
while the one loop corrections (computed in the limit $g \rightarrow 0$) are
\begin{equation}
g_{b_L}^{\textrm{SM,loop}} = \frac{m_t^2}{16 \pi^2 v^2}\,,
\qquad \quad
g_{b_R}^{\textrm{SM,loop}} = 0\,.
\end{equation}

\begin{figure}
\centering
\includegraphics[width=0.5\textwidth]{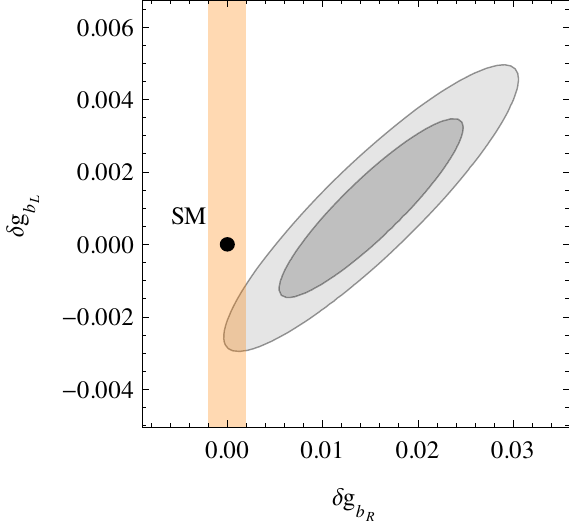}
\caption{Constraints on the new physics corrections to the $Z$ boson couplings with
the bottom quark. The ellipses show the exclusion contours at $68\%$ and $95\%$ confidence
level~\cite{Batell:2012ca,Guadagnoli:2013mru}.
The vertical band shows the expected size of the corrections to the $g_{b_R}$
coupling in minimal models. The plot is taken from Ref.~\cite{Grojean:2013qca}.}
\label{fig:deltag_plane}
\end{figure}

The current bounds on the deviations of the $Z$ couplings to the bottom are shown in
Fig.~\ref{fig:deltag_plane}. The deviation to the left-handed bottom coupling are constrained
to be at the level of $3 \times 10^{-3}$, while the bounds on the coupling with the
right-handed bottom are one order of magnitude less stringent. Notice however that a
strong correlation exists between the bounds on $\delta g_{b_L}$ and $\delta g_{b_R}$.
In many minimal composite Higgs scenarios the corrections to the $g_{b_R}$ coupling are
typically small, at most of the same order of the deviations in $g_{b_L}$. If we impose the
constraint $|\delta g_{b_R}| \lesssim few \times 10^{-3}$, a negative value for
$\delta g_{b_L}$, of order $-2 \times 10^{-3}$ is preferred, whereas a positive shift
worsens the fit with respect to the SM. The region favored by the current fit in the case
of small $\delta g_{b_R}$ is shown in Fig.~\ref{fig:deltag_plane} and corresponds to the
intersection of the gray ellipses with the vertical band.


\subsection[Tree-level corrections and the $P_{LR}$ symmetry]{Tree-level corrections and the ${\mathbf{P_{LR}}}$ symmetry}\label{sec:P_LR_symmetry}

\begin{figure}
\centering
\raisebox{.9em}{\includegraphics[width=0.35\textwidth]{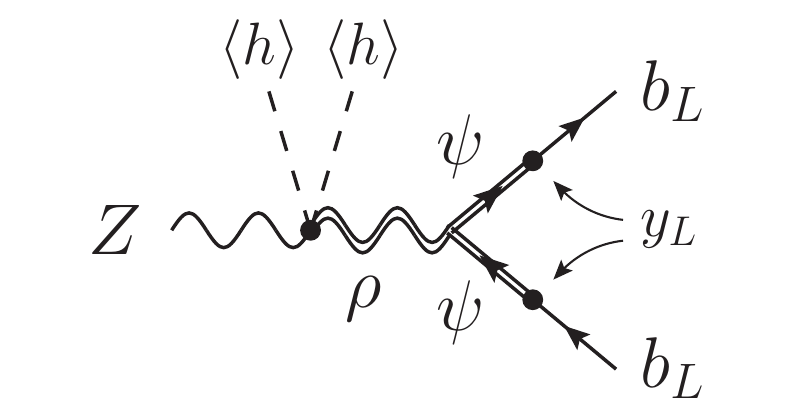}}
\hspace{.5em}
\includegraphics[width=0.35\textwidth]{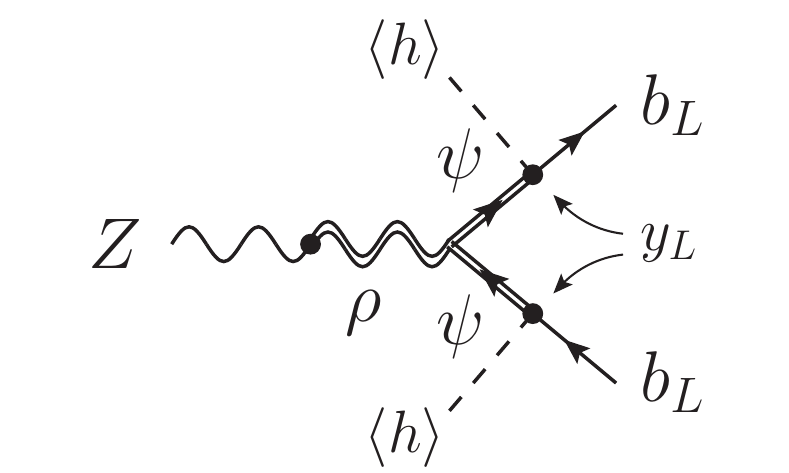}
\caption{Example of diagrams contributing to the deviations of the $Z b_L \overline b_L$
vertex in the absence of custodial protection.}
\label{fig:Zbb_tree}
\end{figure}

As a first step in our analysis, we discuss the tree-level corrections to the left-handed
bottom coupling. In generic composite Higgs models the main contributions come from the mixing
of the $b_L$ with the partners related to the top sector. These corrections are potentially
sizable because the $q_L$ doublet must be strongly coupled to the top partners
in order to generate the top mass. The corrections to $g_{b_L}$ can be estimated
as~\cite{Giudice:2007fh}
\begin{equation}\label{eq:deltagb_est_naive}
\delta g_{b_L} \sim \frac{y_L^2}{g_\psi^2} \xi \gtrsim \frac{y_t^2}{g_\psi^2} \xi\,,
\end{equation}
where the last inequality follows from the absolute lower bound on the left-handed
mixing $y_L \gtrsim y_t$, which is obtained when the $t_R$ is fully composite.
The schematic structure of the diagrams that generate the tree-level corrections
to the $g_{b_L}$ coupling is shown in Fig.~\ref{fig:Zbb_tree}.
By comparing the above estimate with the current data
a stringent upper bound on $\xi$ is found. This is especially true if the top partners
are relatively light, $g_\psi = m_\psi/f \lesssim 2$, as expected from naturalness considerations.

\index{P@$P_{LR}$ symmetry}
It is important to stress that Eq.~(\ref{eq:deltagb_est_naive}) is only a naive estimate.
To derive it we assumed that the corrections to the $Z b_L \overline b_L$ coupling are generated
at the lowest possible order and no protection mechanism is present. However,
as we will discuss in the following, in many composite Higgs scenarios the corrections
to $g_{b_L}$ are naturally suppressed thanks to a custodial protection~\cite{Agashe:2006at}.
This happens when the composite sector is invariant
under an $\mathrm{O}(4)$ symmetry and not just under $\SO(4)$.
The extended symmetry includes, in addition to the
usual $\SO(4) \simeq \SU(2)_L \times \SU(2)_R$ group, a discrete parity, $P_{LR}$, defined
as the interchange between the generators of the $\SU(2)_L$ and $\SU(2)_R$ subgroups.
The $P_{LR}$ symmetry can be represented by the $\mathrm{O}(4)$ element
$P_{LR} = \mathrm{diag}(-1,-1,-1,1)$.\footnote{The
$P_{LR}$ symmetry can be also seen as a transformation of $\mathrm{O}(5)$ corresponding to the
element $P_{LR} = \mathrm{diag}(-1,-1,-1,1,1)$. For more details on the $P_{LR}$ symmetry see
Appendix~\ref{discrete}.}
The fourth component of the Higgs multiplet is invariant under this
transformation, therefore the $P_{LR}$ symmetry is not broken by the Higgs VEV.

We can now analyze the implications of the $P_{LR}$ symmetry on the $Z$ couplings, extending the
discussion of Sect.~\ref{pfer}.
Let us consider a fermion $\psi$ that is an eigenstate of $P_{LR}$.
It can be shown that its coupling to the $Z$ boson is protected and does not receive corrections after EWSB.
Moreover, as in the SM, the coupling is completely determined by the electric
charge and by the hypercharge (or, equivalently, by the $\U(1)_X$ charge) of the fermion $\psi$.
A proof of this result is postponed to Appendix~\ref{sec:custodial_symmetries}.
If $\psi$ has definite quantum numbers under $\SU(2)_L \times \SU(2)_R$, the condition to be
an eigenstate of $P_{LR}$ is equivalent to demanding that it has the same charges
under $\SU(2)_L$ and $\SU(2)_R$, namely
\begin{equation}\label{eq:PLR_conditions}
T_L = T_R
\qquad \mathrm{and} \qquad
T^3_L = T^3_R\,.
\end{equation}

Let us now apply these results to the bottom field.
To protect the $b_L$ coupling to the $Z$ boson
we need to embed the $q_L$ doublet in an $\SO(4)$ multiplet is such a way that
Eq.~(\ref{eq:PLR_conditions}) is verified for the bottom component. The only way to do this
is to embed $q_L$ into the $\mathbf{4} = (\mathbf{2}, \mathbf{2})$ representation with
$\U(1)_X$ charge $2/3$. With this choice the $b_L$ quantum numbers are
\begin{equation}
T_L = T_R = 1/2
\qquad \mathrm{and} \qquad
T^3_L = T^3_R = -1/2\,,
\end{equation}
and the $b_L$ field is odd under the $P_{LR}$ symmetry.

It is easy to understand that the choice of the $q_L$ embedding
also determines the $t_R$ quantum numbers under
$\SO(4) \times \U(1)_X$. In order to be able to write the top Yukawa only two charge
assignments are possible:
\begin{equation}\label{eq:tR_embedding}
t_R \in \mathbf{1}_{2/3} = (\mathbf{1}, \mathbf{1})_{2/3}
\qquad \mathrm{or} \qquad
t_R \in \mathbf{6}_{2/3} = (\mathbf{1}, \mathbf{3})_{2/3} \oplus (\mathbf{3}, \mathbf{1})_{2/3}\,.
\end{equation}

To complete our construction we must take into account
the fact that the Higgs is a NGB of an extended $\SO(5)$ symmetry. This implies that the
elementary fields must be embedded in representations of the whole global group and
not only of the unbroken $\SO(4)$ subgroup. In order to obtain a custodial protection for the
$b_L$ coupling we need to embed the $q_L$ doublet into an $\SO(5)$ representation that contains,
in addition to a $(\mathbf{2}, \mathbf{2})$ multiplet, another multiplet that can be coupled to the
$t_R$. For the first choice in Eq.~(\ref{eq:tR_embedding}) we can use the $\mathbf{5}_{2/3}$
representation of the global group that decomposes as
\begin{equation}
\mathbf{5}_{2/3} = (\mathbf{2}, \mathbf{2})_{2/3} \oplus (\mathbf{1}, \mathbf{1})_{2/3}\,.
\end{equation}
Another simple choice is the $\mathbf{14}_{2/3}$ representation whose decomposition is
\begin{equation}
\mathbf{14}_{2/3} = (\mathbf{2}, \mathbf{2})_{2/3} \oplus
(\mathbf{3}, \mathbf{3})_{2/3} \oplus (\mathbf{1}, \mathbf{1})_{2/3}\,.
\end{equation}
Explicit examples of models with these two embeddings have been presented in Chap.~\ref{ch:Pheno_models}.
For the second embedding of the $t_R$ in Eq.~(\ref{eq:tR_embedding}) a minimal choice is
the $\mathbf{10}_{2/3}$ multiplet:
\begin{equation}
\mathbf{10}_{2/3} = (\mathbf{2}, \mathbf{2})_{2/3} \oplus
(\mathbf{1}, \mathbf{3})_{2/3} \oplus (\mathbf{3}, \mathbf{1})_{2/3}\,.
\end{equation}

It is interesting to remark that, when the $q_L$ doublet is embedded in the $\mathbf{5}$ or
$\mathbf{14}$ representation, the $P_{LR}$ invariance arises as an accidental
symmetry of the lowest order interactions in the effective theory.
This means that no additional condition
needs to be imposed to obtain the custodial protection~\cite{Mrazek:2011iu}.
This is not the case for the representation $\mathbf{10}$. To respect the
$P_{LR}$ symmetry for this embedding it is necessary to impose
suitable conditions on the operators involving
the $(\mathbf{1}, \mathbf{3})$ and $(\mathbf{3}, \mathbf{1})$ multiplets.

Before concluding the discussion, it is important to mention that
the $P_{LR}$ symmetry can only protect the coupling of the $Z$ boson to the fermions at zero momentum.
The effective Lagrangian for the SM fields indeed contains also higher dimensional
interactions that include powers of the momenta, and, in general, these additional operators are
not protected by the custodial symmetry. The physical value of the couplings
coincide with the interactions computed with on-shell states and not with the ``zero momentum'' one,
so that the higher-dimensional operators can induce some corrections.

Obviously the additional corrections can come from terms proportional to the $Z$ mass
or to the mass of the fermions.
In the case of the $b_L$ field the corrections due to the non-zero bottom mass are clearly
negligible, but they can be important if we are interested in the couplings of the top
quark as we will see in Sect.~\ref{sec:top_couplings}.

\subsubsection{The ${\mathbf{b_L}}$ coupling in the presence of custodial protection}

We can now analyze the size of the corrections to the coupling of the bottom quark with the
$Z$ boson. In the cases with $P_{LR}$ symmetry the only relevant corrections are the ones
induced at non-zero momentum. For instance they are due to operators of the form
$D_\mu F^{\mu\nu} \overline q_L \gamma_\nu q_L$. The estimate of their effects is
\begin{equation}\label{eq:Zbb_tree_corr_top}
\delta g_{b_L} \sim \frac{y_L^2 f^2}{m_\psi^2} \frac{m_Z^2}{m_\rho^2}
\simeq 8 \times 10^{-4} \frac{f}{m_\psi} \left(\frac{4 \pi}{g_\rho}\right)^2 \xi\,,
\end{equation}
where $m_\psi$ is the mass scale of the composite fields mixed with the bottom.
To obtain the numerical estimate we assumed the relation $y_L \sim y_R \sim \sqrt{y_t g_\psi}$.

Notice that the embedding of the $q_L$ doublet in the representation $\mathbf{5}_{2/3}$ (or
$\mathbf{14}_{2/3}$) does not allow to generate a bottom mass. Indeed the $\SO(5)$ multiplet
does not contain any component with the quantum numbers of the $b_R$ field. Therefore to introduce a bottom
mass the $q_L$ doublet must be also embedded in an additional multiplet with
suitable quantum numbers. This additional embedding could determine a breaking of the
$P_{LR}$ protection for the $b_L$ field and generate corrections to the $g_{b_L}$ coupling.
For instance this happens if we add an extra multiplet in the fundamental representation,
$\mathbf{5}_{-1/3}$ (or in the $\mathbf{14}_{-1/3}$). With this embedding the
$b_L$ has charges $T^3_L = - T^3_R = -1/2$ and does not respect the conditions in
Eq.~(\ref{eq:PLR_conditions}). The contribution to the $Z b_L \overline b_L$ vertex coming from
the bottom partners can be estimated as
\begin{equation}\label{eq:Zbb_tree_corr_mB}
\delta g_{b_L} \sim \frac{(y_L^b f)^2}{m_B^2} \xi\,,
\end{equation}
where we denoted by $y_L^b$ the mixing of the $q_L$ to the new multiplets and by $m_B$ the typical
mass scale of the new bottom partners. We can relate $y_L^b$ to the bottom Yukawa by assuming
that $y_L^b \sim y_R^b$, which implies $y_L^b \sim y_R^b \sim \sqrt{y_b m_B/f}$.
The estimate in Eq.~(\ref{eq:Zbb_tree_corr_mB}) becomes
\begin{equation}
\delta g_{b_L} \sim y_b \frac{f}{m_B} \xi \simeq 2 \times 10^{-2} \frac{f}{m_B} \xi\,.
\end{equation}
This correction can easily have a size comparable with the current bounds if the
bottom partners are relatively light. Of course, if we relax the assumption $y_L^b \sim y_R^b$
or if we chose $m_B \gg f$, the contribution in Eq.~(\ref{eq:Zbb_tree_corr_mB}) can be
sufficiently suppressed.

\subsubsection{The ${\mathbf{b_R}}$ coupling}

Let us now consider the corrections to the $b_R$ coupling with the $Z$ boson.
As for the $t_R$ the quantum numbers of the embedding of the $b_R$ field are determined
by the embedding of $q_L$. If we use the $\mathbf{5}_{-1/3}$ or $\mathbf{14}_{-1/3}$
multiplets to generate the bottom mass the $b_R$ field must be embedded in the
$(\mathbf{1}, \mathbf{1})_{-1/3}$ representation of $\SO(4) \times \U(1)_X$.
It is easy to see that the $b_R$ quantum numbers satisfy the conditions in
Eq.~(\ref{eq:PLR_conditions}), thus the $g_{b_R}$ coupling is protected by the
custodial symmetry. The tree-level corrections to the $b_R$ coupling are then
tiny and completely negligible with respect to the current bounds.

Non-negligible corrections to the $g_{b_R}$ coupling can be obtained if we use a different
embedding of $q_L$ to give a mass to the bottom quark. Several possibilities have
been discussed in Ref.~\cite{Agashe:2006at}. In the following for simplicity we will focus
on the case in which the correction to the $b_R$ coupling are small.


\subsection{Loop corrections}

\index{P@$P_{LR}$ symmetry}
We can now consider the one-loop contributions to the $Z b_L \overline b_L$ vertex.
As a first step we analyze the associated degree of divergence.
By using the counting rule in Eq.~(\ref{eq:divergence_counting_ferm}) it is
straightforward to check that the $Z b_L \overline b_L$ operator is
naively associated to a quadratic degree of divergence. In the scenarios in which
the $b_L$ is invariant under the $P_{LR}$ custodial symmetry, however, the degree of divergence
is automatically reduced. This is an obvious consequence of the fact that a new physics
contribution to the $g_{b_L}$ coupling can be generated only through the insertion
of the couplings that break the $P_{LR}$ symmetry. In the set-up in which the
elementary fermions are coupled to operators in the fundamental representation of $\SO(5)$
the main breaking of this symmetry is due to the $y_L$ elementary/composite
mixing\footnote{We are neglecting here the contribution from the bottom partners.
Due to the small mixings $y_{L,R}^b$ needed to generate the bottom mass,
these states do not play a significant role in the one loop corrections.}
and it is easy to see that at least two insertions of the corresponding spurion
(see Chap.~\ref{ch:Pheno_models}) are needed to generate a contribution to the $Z b_L \overline b_L$ vertex.
The degree of divergence at one loop is therefore reduced to logarithmic.

By an explicit analysis of the operators that correct the $g_{b_L}$ coupling we can get
another interesting insight on the nature of the divergent
contributions~\cite{Grojean:2013qca}. For definiteness we will consider the $5+5$ model, whose effective
description has been discussed in Sect.~\ref{sec:5+5_effective}.
For this analysis it is convenient to work in the
basis of elementary and composite fields. In this basis the physical $b_L^{SM}$ field
corresponds to a linear combination of the elementary $b_L$ and of the
composite $B_L$ contained inside the fourplet $\psi_4$.
The operators that induce
a distortion of the $g_{b_L}$ coupling can be immediately related to the ones that correct the
couplings of the $Z$ boson to the elementary and the composite states by using the transformation
\begin{equation}\label{eq:b_LSM_embed}
b_L \rightarrow \frac{m_4}{\sqrt{m_4^2 + y_{L4}^2 f^2}} b_L^{SM}\,,
\qquad
B_L \rightarrow \frac{y_{L4} f}{\sqrt{m_4^2 + y_{L4}^2 f^2}} b_L^{SM}\,.
\end{equation}

\index{spurions}
By means of a spurion analysis it can be shown that the leading operator involving the elementary $b_L$ field is given by
\begin{equation}\label{eq:bL_elem_op}
{\mathcal O} = i \left(\overline q_L \Lambda_L^\dagger \gamma^\mu \Lambda_L q_L\right)
\left(U^t_{5I} \Lambda_L^{\alpha I} (\Lambda^\dagger_L)^{\alpha J} U_{Ji} d_\mu^i\right) + \mathrm{h.c.}\,,
\end{equation}
where the $\Lambda_L^{\alpha I}$ spurion corresponds to the mixing $y_{L4}$ of the $q_L$ doublet with the composite states
and is defined analogously to the one we introduced in Sect.~\ref{CHPCH} (see Eq.~\ref{eq:spurion_55_pot}).
In particular the index $\alpha$ transforms as a doublet under an elementary $\SU(2)_L$ group, while $I$
transforms in the fundamental representation of $\SO(5)$.
This operator contains four powers of $y_L$, thus it corresponds to
a finite one loop contribution.
The operator related to the composite resonance, on the other hand, contains only
two spurion insertions and is associated to a logarithmic divergence:
\begin{equation}\label{eq:BL_comp_op}
{\mathcal O} = i \left(\overline \psi_4 \gamma^\mu \psi_4\right)
\left(U^t_{5I} \Lambda_L^{\alpha I} (\Lambda^\dagger_L)^{\alpha J} U_{Ji} d_\mu^i\right) + \mathrm{h.c.}\,.
\end{equation}
As can be seen from Eq.~(\ref{eq:b_LSM_embed}), the contribution to the $b_L^{SM}$ coupling
due to the above operator contains two additional powers of $y_L$ coming from the
rotation angle between the composite states and the SM ones. Therefore, also in this case,
the correction to $g_{b_L}$ is proportional to the fourth power of $y_L$. However, differently
from the case of the operator in Eq.~(\ref{eq:bL_elem_op}) the two additional powers of $y_L$
correspond to insertions on the external legs of the diagrams and do not reduce the degree
of divergence.

It is important to stress that we built the operator in Eq.~(\ref{eq:BL_comp_op})
by using the composite resonances $\psi_4$ in the fourplet
representation of $\SO(4)$. This choice is essential because a contribution to the $g_{b_L}$ interaction
can only come from multiplets that contain a resonance with the same quantum numbers
of the $b_L$ field. For example, if the composite sector contains only $\SO(4)$ singlets,
the analogous of the operator in Eq.~(\ref{eq:BL_comp_op}) does not contribute to the
$Z b_L \overline b_L$ vertex and the leading corrections to $g_{b_L}$ are finite at one loop.

Another way to understand why four powers of $y_L$, instead of two, are needed to generate
a correction to the $g_{b_L}$ coupling is the following. Due to the partial compositeness
assumption, each external $b_L$ field is necessarily associated to a power of $y_L$.
However, being the $b_L$ fields invariant under
the $P_{LR}$ symmetry, these insertions do not lead to a breaking of the custodial
protection. As a consequence at least two additional insertions of $y_L$ inside the loop
are required to generate a non-vanishing contribution.

An interesting outcome of the above analysis is the fact that the divergence in the $Z b_L \overline b_L$ vertex
is not directly due to the elementary states, but instead it is related to a corresponding divergence
in the couplings of the composite fields.
Only one counterterm and one renormalization condition is thus needed to fix both divergent corrections.
For instance, if we choose to express everything in terms of the renormalized couplings
of the composite resonances, the one-loop corrections to the $Z b_L \overline b_L$ vertex become
completely finite and calculable.

Before concluding the general analysis of the degree of divergence of the $Z b_L \overline b_L$ vertex
it is interesting to discuss how the above results are modified if we embed our
theory into a two-site model. In this case the collective
breaking mechanism ensures that no counterterm exists for the $Z$ couplings to the
$b_L$ and $B_L$ fields, which are finite at one loop.
Analogously to what happens for the oblique parameters, however, the non-$1$PI diagrams
contributing to the $Z b_L \overline b_L$ vertex
acquire a logarithmic divergence related to the running of the strong sector couplings.
The radiative corrections to $g_{b_L}$ are thus finite only
when expressed in terms of the renormalized strong sector parameters.
Accessing these couplings experimentally, however, would be a difficult
task and the poor knowledge of their value would in any case
forbid a full prediction of the radiative corrections to the $Z b_L \overline b_L$ vertex.
As we did for the oblique parameters, it is then more convenient to adopt a more practical
point of view and compute only the leading corrections coming from the lightest fermionic resonances.
In this way we encode our ignorance on the values of
the strong sector couplings in the unknown UV contributions.

\subsubsection{A closer look at the divergent contributions}

After the general discussion we presented so far, we want to analyze in a more
detailed way the structure of the contributions to the $Z b_L \overline b_L$ vertex.
As we will see, a large class of diagrams are automatically finite thanks to the presence
of a selection rule and only a very special subset can generate a logarithmic divergence.
As explained before, we will work in an effective theory in which the
vector resonances are integrated out and only the lightest fermionic resonances
are retained.

\begin{figure}
\centering
\includegraphics[width=.65\textwidth]{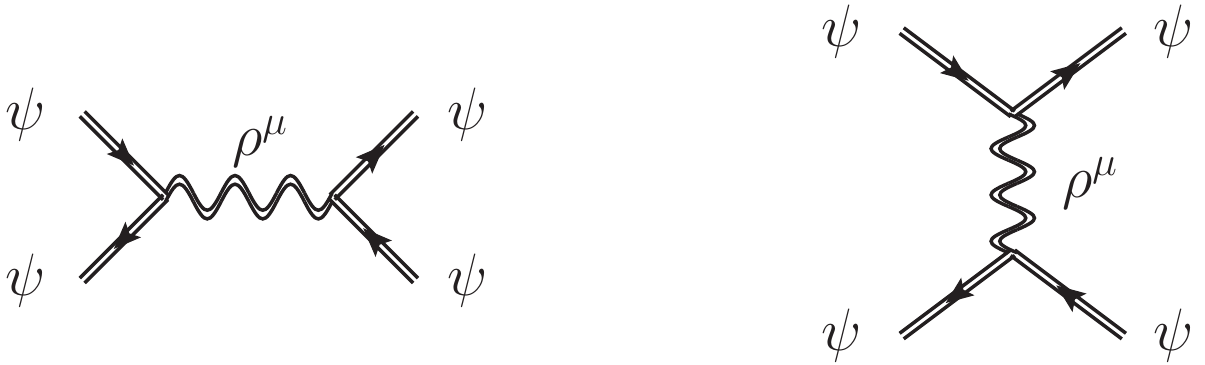}
\caption{Structure of the Feynman diagrams that generate $4$-fermion contact
operators among the composite fermions through the exchange of heavy gauge resonances.}
\label{fig:4-ferm}
\end{figure}

\index{four-fermion interactions}
In Sect.~\ref{sec:oblique_parameters} we saw that the one-loop contributions to the oblique
parameters come only from the leading terms in the effective $\sigma$-model Lagrangian.
The situation is different for $g_{b_L}$ coupling, whose leading corrections also come
from some dimension-six operators, namely some contact interactions involving
four composite fermions. The schematic structure of four-fermion contact operators is
\begin{equation}
{\mathcal O}_{\textrm{4-ferm}} \sim \frac{1}{f^2} (\overline \psi \gamma^\mu \psi)
(\overline \psi \gamma_\mu \psi)\,.
\end{equation}
In spite of having dimension six, these operators are not suppressed by powers of
the cut-off, instead their natural coefficient is of order $1/f^2$ (see
the power-counting  in Eq.~(\ref{POWCFull})). Operators of this kind are typically generated by the
strong sector dynamics. For instance they can derive from the exchange of heavy vector
resonances as shown in the diagrams in Fig.~\ref{fig:4-ferm}. In these diagrams
the suppression due to the propagator of the
heavy boson is exactly compensated by the large coupling, $g_\rho \simeq m_\rho / f$, thus
explaining the order $1/f^2$ coefficient.

\begin{figure}
\centering
\includegraphics[height=.275\textwidth]{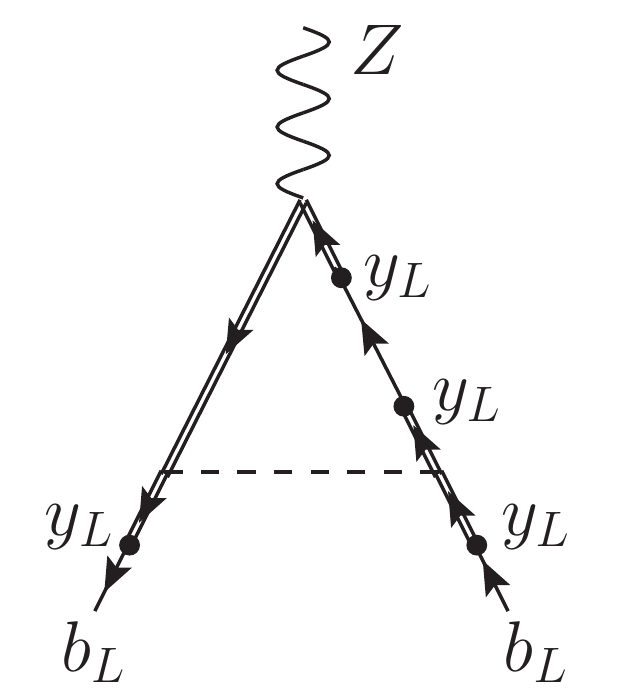}
\hspace{4.em}
\includegraphics[height=.275\textwidth]{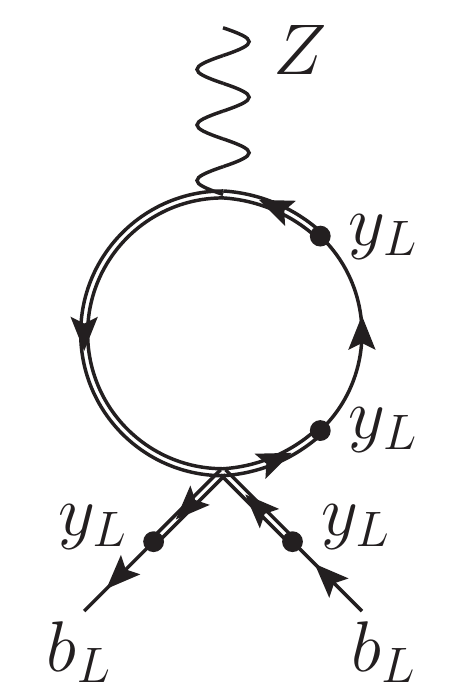}
\caption{Schematic structure of fermion one-loop loop diagrams contributing to the $Z b_L \overline b_L$
vertex at leading order in the $y_L$ expansion. Double lines denote the composite fields,
while single lines correspond to the elementary ones. The dashed line denotes a Goldstone propagator.}
\label{fig:Zbb_diagrams}
\end{figure}

The corrections to the $Z b_L \overline b_L$ vertex come from two classes of contributions.
The first one is generated from diagrams containing only interactions coming from the
leading terms in the effective Lagrangian. An example of such diagrams are the
``triangle''-type ones schematically shown on the left of Fig.~\ref{fig:Zbb_diagrams}.
The second class of diagrams are the ``bubble''-type ones, shown on the right of
Fig.~\ref{fig:Zbb_diagrams}, that include a vertex coming from the four-fermion contact
operators. As we will see the structure of the elementary/composite mixings implies the
presence of a selection rule that forbids logarithmically divergent corrections coming
from a large class of diagrams~\cite{Grojean:2013qca}.
In particular the only diagrams that lead to a divergent contribution
to $g_{b_L}$ are a subset of the ``bubble''-type ones, while the others are finite.

The origin of the selection rule can be easily understood by analyzing the ``triangle''-type
diagrams in an expansion in $y_L$. In particular, as explained before, we are interested
in the diagrams that contain two $y_L$ insertions on the external legs (see Fig.~\ref{fig:Zbb_diagrams}).
The external $b_L$ fields are both mixed with the $B_L$ state contained in the fourplet $\psi_4$.
In order to generate a divergence, the vertices with a Goldstone boson must also contain a
power of the momentum, {\it{i.e.}}~they must come from an interaction of the form
$\partial_\mu \phi \overline \psi_L \gamma^\mu \psi_L$, where we generically denoted by $\phi$ the
Goldstone field and by $\psi$ the composite fermions.\footnote{In the effective Lagrangian vertices
of this kind are generated by the $d$-symbol terms.} The chirality structure of the vertex
implies that the composite fermions that enter in the loop are necessarily left-handed.
But, in the leading order Lagrangian, the left-handed composite fermions mix with the elementary
states only through $y_R$. As a consequence, in order to generate a non-vanishing contribution
to $g_{b_L}$ some $y_R$ or some composite mass insertions are needed in addition to the four
powers of the $y_L$ mixing. This lowers the degree of divergence making the diagrams finite.

The only diagrams that can give rise to a logarithmic divergence are the ``bubble''-type ones.
They of course crucially depend on the presence of four-fermion operators in the effective Lagrangian.
Two types of contact interactions can generate a contribution to $g_{b_L}$. The first type
has the form
\begin{equation}\label{eq:4-ferm_L}
{\mathcal O}_L^{\mathrm{4-ferm}} = \frac{e_L}{f^2} \left(\overline B_L \gamma^\mu B_L\right)
\left(\overline {\mathcal T}_L \gamma_\mu {\mathcal T}_L\right)\,,
\end{equation}
where $\mathcal T$ denotes any composite state with charge $2/3$ contained in the fourplets.
For shortness in Eq.~(\ref{eq:4-ferm_L}) we did not specify the color structure, which is not relevant
for the present discussion. By adapting the previous analysis, it is straightforward to show that
the ``bubble''-type diagrams with the ${\mathcal O}_L^{\mathrm{4-ferm}}$ vertex are protected by
the selection rule and are finite. The second type of contact interactions is of the form
\begin{equation}\label{eq:4-ferm_R}
{\mathcal O}_R^{\mathrm{4-ferm}} = \frac{e_R}{f^2} \left(\overline B_L \gamma^\mu B_L\right)
\left(\overline {\mathcal T}_R \gamma_\mu {\mathcal T}_R\right)\,.
\end{equation}
In this case the selection rule is violated because the ${\mathcal T}_R$ fields can mix with the $q_L$
doublet through $y_L$. This class of vertices gives rise to a logarithmically divergent contribution
to the $Z b_L \overline b_L$ vertex.

\subsubsection{Estimates}

We can now derive some estimates for the contributions to the $g_{b_L}$ vertex coming
from fermion loops. The leading contributions are of course the ones related to the
logarithmically divergent diagrams, which can be estimated as
\begin{equation}\label{eq:Zbb_loop_log}
\delta g_{b_L} \simeq \frac{y_L^2}{16 \pi^2} \frac{y_{L4}^2 f^2}{m_4^2 + y_{L4}^2 f^2}
\xi \log\left(\frac{m_\rho^2}{m_4^2}\right)\,,
\end{equation}
where $y_L$ collectively denotes any of the mixing parameters of the $q_L$ elementary doublet.
Notice that in the above estimate we explicitly included a factor
$y_{L4}^2 f^2/(m_4^2 + y_{L4}^2 f^2)$, which corresponds to the mixing between
the elementary $b_L$ and the composite $B_L$ that
appears in the external legs of the logarithmically divergent diagrams.
If we assume that $y_L \simeq y_R$ then we can use the relation between the mixings and the
top Yukawa $y_L \sim y_R \sim \sqrt{y_t g_\psi}$ to get the result
\begin{equation}
\delta g_{b_L} \simeq \frac{y_t^2}{16 \pi^2} \xi \log\left(\frac{m_\rho^2}{m_4^2}\right)
\simeq 2 \times 10^{-2}\,\xi\,,
\end{equation}
where for the numerical estimate we set $m_\rho \simeq 3\ \mathrm{TeV}$ and
$m_4 \simeq 700\ \mathrm{GeV}$.

The presence of a logarithmic divergence implies that the $Z b_L \overline b_L$
coupling can also receive an unsuppressed contribution from the UV dynamics.
This unknown UV contribution can be estimated as
\begin{equation}\label{eq:Zbb_loop_uv}
\delta g_{b_L} \simeq \frac{y_L^2}{16 \pi^2} \frac{y_{L4}^2 f^2}{m_4^2 + y_{L4}^2 f^2} \xi \simeq
\frac{y_t^2}{16 \pi^2} \xi
\simeq 6 \times 10^{-3} \,\xi\,.
\end{equation}
To derive this estimate we assumed that the only sources of breaking of the
$P_{LR}$ symmetry are the $y_L$ elementary/composite mixings that we included
in our effective Lagrangian. As in the estimate of the logarithmically enhanced
terms, we included a factor $y_{L4}^2 f^2/(m_4^2 + y_{L4}^2 f^2)$. This takes into account
the fact that an unsuppressed UV contribution can only come from operators analogous to
the one in Eq.~(\ref{eq:BL_comp_op}) and thus is always accompanied two powers of
the $b_L$ compositeness angle.

The UV contribution is typically of the same order of the finite corrections coming from
the diagrams protected by the selection rule, whose estimate reads
\begin{equation}\label{eq:Zbb_loop_fin}
\delta g_{b_L} \simeq \frac{y_L^2}{16 \pi^2} \frac{y_L^2 f^2}{m_\psi^2} \xi
\simeq \frac{y_t^2}{16 \pi^2} \xi
\simeq 6 \times 10^{-3} \,\xi\,,
\end{equation}
where $m_\psi$ is the mass scale of the top partners. Clearly in a generic model
only the logarithmically enhanced contributions, which are parametrically dominant,
can be reliably computed and not the additional finite corrections. An exception to this rule is the case in which
the mixing angle of the $b_L$ with composite fourplets is negligible, as for instance
when only $\SO(4)$ composite singlets are present in the spectrum of the low energy theory.
As we saw, in this case the corrections to $g_{b_L}$ are finite and are dominated
by the loops of the lightest resonances, whereas the UV contributions are suppressed
by powers of the cut-off.

Before concluding, it is useful to compare the loop corrections with the tree-level effects.
The corrections in Eqs.~(\ref{eq:Zbb_loop_log}) and (\ref{eq:Zbb_loop_fin})
are usually larger than the tree-level contribution given in Eq.~(\ref{eq:Zbb_tree_corr_top}).
This is especially true if the mass of the resonances is not too small $m_\psi \gtrsim f$
and the vector resonances coupling is large $g_\rho \gtrsim 5$. The corrections due to the
bottom partners, estimated in Eq.~(\ref{eq:Zbb_tree_corr_mB}), can in principle be comparable
to the ones coming from the resonances loops if the bottom partners are light $m_B \sim f$.
These corrections crucially depend on the quantum numbers of the bottom partners.
In minimal scenarios (bottom partners in the fundamental representation of $\SO(5)$) they are
positive and some cancellation is required to pass the present bounds.


\subsection{Explicit results}

After the general analysis presented in the previous section, we now analyze the corrections to the $Z b_L \overline b_L$
coupling in some explicit models. As we did in the case of the oblique parameters, in the following we will consider
two classes of minimal models in which the composite resonances
belong to the fundamental representation of $\SO(5)$, namely the $5+5$ and $5+1$ scenarios described
in Sect.~\ref{sec:effect_parametr}.
In order to be as model independent as possible
we will compute only the leading contributions calculable within the effective models.
Moreover we will work in the limit of vanishing SM gauge couplings, {\it{i.e.}}~we include only the
effects due to the Goldstone fields. This approximation is however known to be a very good one in the SM.
As we explained before, in the classes of models we consider the tree-level corrections to the $Z b_L \overline b_L$
vertex come from the bottom partners, whose mass is not required to be light by Naturalness considerations.
For simplicity we will neglect these effects in our analysis.

\subsubsection{Results in the ${\mathbf{5+5}}$ model}

As a first scenario we consider the $5+5$ model.
An interesting limit of this model is the case in which only a light composite $\SO(4)$
singlet is present in the spectrum. In this case the corrections to the $g_{b_L}$ coupling
are finite and fully calculable within the effective theory.
The absence of fourplets in the effective theory also implies that additional contributions
coming from four-fermion contact interactions and from the UV dynamics are suppressed by the
cut-off scale and can be expected to be subleading.
At leading order in $v/f$ the shift in $g_{b_L}$ is given by
\begin{equation}
\delta g_{b_L} = \frac{\xi}{64 \pi^2} \frac{y_{L1}^2 m_1^2 f^2}{(m_1^2 + y_{L1}^2 f^2)^3}
\left\{m_1^2 + 2 y_{R1}^2 f^2 \left[\log\left(\frac{2 (m_1^2 + y_{R1}^2 f^2)^2}{v^2 y_{L1}^2 y_{R1}^2 f^2}\right)
- 1\right]\right\}\,.
\end{equation}
It is interesting to compare this result with the one-loop fermion contribution to
$\widehat T$ in the same set-up given in Eq.~(\ref{eq:T_singlet}). One can notice that a strict
relation exists between the two quantities: $\Delta \widehat T = 3\, \delta g_{b_L}$~\cite{oai:arXiv.org:0706.0432,Gillioz:2008hs,Grojean:2013qca}.
In particular the positive correction to $\widehat T$ is related to a corresponding positive
shift in $g_{b_L}$. For the typical size of the fermion contribution to $\widehat T$
needed to satisfy the experimental bounds, $1 \times 10^{-3} < \Delta \widehat T < 2 \times 10^{-3}$,
a moderate contribution to $\delta g_{b_L}$ is found: $0.33 \times 10^{-3} < \delta g_{b_L} < 0.66 \times 10^{-3}$. As we discussed before (see Fig.~\ref{fig:deltag_plane}), the experimental
measurements disfavor a positive contribution to the $Z b_L \overline b_L$ coupling. Thus the
scenario with only a light singlet tends to be in worse agreement with the EW precision
data than the SM.

In the more general scenario in which composite fermions in the fourplet representation of $\SO(4)$
are present, the only contributions that can be reliably computed in the effective theory are
the logarithmically divergent ones due to the presence of four-fermion contact interactions.
As an example we report here the leading correction to $g_{b_L}$ induced by the operator
\begin{equation}\label{eq:4-ferm_example}
{\mathcal O} = \frac{e_R}{2 f^2} \left(\overline \psi_4^a \gamma^\mu \psi_4^a\right)
\left(\overline \psi_4^b \gamma_\mu \psi_4^b\right)\,,
\end{equation}
where $a$ and $b$ are color indices.\footnote{Different color structures lead to results
that only differ by group theory factors.}
This operator contains a vertex of the form
\begin{equation}
\frac{e_{R}}{f^2} \left(\overline B_L^a \gamma^\mu B_L^a\right)
\left(\overline T^b_R \gamma_\mu T_R + \overline X^b_{2/3R} \gamma_\mu X^b_{2/3R}\right)\,,
\end{equation}
which gives rise to a logarithmically divergent correction to $g_{b_L}$:
\begin{equation}
\delta g_{b_L}^{\mathrm{4-ferm}} = \frac{3 e_{R}}{32 \pi^2}\xi \frac{y_{L4}^2 f^2}{m_4^2 + y_{L4}^2 f^2}
y_{L4}^2 \log\left(\frac{m_\rho^2}{m_4^2}\right)\,.
\end{equation}
Notice that the sign of this contribution crucially depends on the sign of the $e_R$ coefficient.
In the effective theory $e_R$ is a completely free parameter, thus its sign is not fixed.
From the UV perspective, instead, some constraints on the size and the sign of the
four-fermion interactions could be present.

Apart from the analysis performed in Ref.~\cite{Grojean:2013qca}, some additional studies of the
corrections to the $Z b_L \overline b_L$ vertex due to fermion loops in some specific effective
models have been presented in Refs.~\cite{Lodone:2008yy,Gillioz:2008hs,Anastasiou:2009rv},
these works however do not include the effects related to the four-fermion contact interactions.
A computation of the one-loop corrections to the $Z b_L \overline b_L$ vertex including the
effects due to the SM gauge couplings is presented in Ref.~\cite{Anastasiou:2009rv}.


\subsubsection{The case of a fully composite ${\mathbf{t_R}}$}
\index{fully composite $t_R$}

As a second explicit model we consider the $5+1$ scenario.
Analogously to what happens in the $5+5$ model,
in the case in which the low-energy spectrum contains only a composite $\SO(4)$ singlet
the one-loop contributions to $Z b_L \overline b_L$ are finite and dominated by the IR
contributions. The leading order corrections are given by
\begin{equation}
\delta g_{b_L} = \frac{1}{64 \pi^2} \xi \frac{y_{L1}^2 f^2}{m_1^2} \left\{
y_{L1}^2 + 2 y_{Lt}^2 \left[\log\left(\frac{2 m_1^2}{v^2 y_{Lt}^2}\right) - 1\right]
\right\}\,.
\end{equation}
Also in this model the corrections to $g_{b_L}$ and the ones to the $\widehat T$ parameters
are related to each other by the relation $\Delta \widehat T = 3\, \delta g_{b_L}$.
The values of $\widehat T$ compatible with the experimental bounds
($0 \lesssim \widehat T \lesssim 2 \times 10^{-3}$) imply a moderate positive shift in $g_{b_L}$.
This slightly worsens the agreement with the EW data with respect to the SM.

In the presence of an $\SO(4)$ fourplet in the effective theory, the corrections to the
$Z b_L \overline b_L$ vertex are divergent. The leading contribution comes from
logarithmically divergent diagrams containing four-fermion contact interactions.
As an example we report the contribution induced by the operator given in Eq.~(\ref{eq:4-ferm_example}). In this case the following contribution arises:
\begin{equation}
\delta g_{b_L} = \frac{e_{R}}{32 \pi^2}\xi \frac{y_{L4}^2 f^2}{m_4^2 + y_{L4}^2 f^2} y_{L4}
\left(y_{L4} - \sqrt{2} c_t y_{Lt}\right)\log\left(\frac{m_\rho^2}{m_4^2}\right)\,.
\end{equation}
The correction in the above formula depends on the parameter $e_R$ that fixes the coefficient
of the four-fermion interaction, but also on the coefficient of the $d_\mu$-symbol term $c_t$
(see the Lagrangian in Eq.~(\ref{eq:Lcomp_5+1})). Depending on the size and sign of the various
parameters the correction to $g_{b_L}$ can become negative and satisfy the current bounds.


\section{The top couplings}\label{sec:top_couplings}

So far we devoted our attention to a set of observables related to the physics
of the light fermion generations (the oblique EW parameters) and to the bottom quark.
The tight experimental bounds on these observables do not allow for large
deviations from the SM predictions and lead to stringent bounds on the new physics
effects. Another class of observables, in particular the ones related to the top quark,
are instead less constrained by the present data so that sizable deviations from the SM are still allowed.
Large corrections to the top couplings are naturally predicted in the scenarios with
partial or full compositeness due to the strong mixing of the third generation with
the composite dynamics.

As discussed in Sect.~\ref{sec:P_LR_symmetry} in the scenarios in which the $Z$ coupling
to the $b_L$ are protected by the $P_{LR}$ symmetry, the $t_L$ can not have a custodial
protection. The right-handed top component, on the other hand, being necessarily embedded
in a component with quantum numbers $T_L^3 = T_R^3 = 0$ is protected by a discrete subgroup $P_C$
of the custodial symmetry $\SO(3)_c$~\cite{Agashe:2006at}
(see Sect.~\ref{sec:DeltaF=1} for the definition of $P_C$).\footnote{When the
$t_R$ is embedded in the $(\mathbf{1}, \mathbf{1})_{2/3}$ representation its coupling to the
$Z$ is also protected by the usual $P_{LR}$.}
As we explained in Sect.~\ref{sec:P_LR_symmetry} (see also Appendix~\ref{sec:custodial_symmetries}),
the custodial protection ensures
that at zero momentum no distortion of the $Z$ coupling can arise.
Additional corrections, however, can be generated when we consider the top and the $Z$ boson on-shell.
In particular the corrections due to the top mass $m_t$ are only suppressed by
powers of the ratio between $m_t$ and the top partners mass.
This means that they can be sizable if the top partners are relatively light.

The top couplings to the Z boson are described by the following Lagrangian
\begin{equation}
{\mathcal L}^Z = \frac{g}{\cos \theta_w} Z_\mu \overline t \gamma^\mu
\left[(g_{t_L}^{\mathrm{SM}} + \delta g_{t_L}) P_L + (g_{t_R}^{\mathrm{SM}} + \delta g_{t_R}) P_R\right] t\,,
\end{equation}
where we used a notation similar to the one for the bottom couplings (see Eq.~(\ref{eq:Lagr_Z_b})).
The tree-level values of the SM couplings are given by
\begin{equation}
g_{t_L}^{\mathrm{SM}} = \frac{1}{2} - \frac{2}{3} \sin^2 \theta_w\,,
\qquad \quad
g_{t_R}^{\mathrm{SM}} = - \frac{2}{3} \sin^2 \theta_w\,.
\end{equation}
The couplings of the left-handed top component with the charged W boson are related
to the $V_{tb}$ element of the CKM matrix. We will parametrize the new physics
contributions as $V_{tb} = 1 - \delta V_{tb}$.

The current LHC results already constrain the new physics
contribution to $V_{tb}$ at the $10\%$ level: $V_{tb} = 1.020 \pm 0.046\, \mathrm{(meas.)}
\pm 0.017\, \mathrm{(theor.)}$~\cite{Chatrchyan:2012ep}.
The bounds on the models coming from this measurement are
usually weaker than the ones coming from the EW precision data and become competitive only
in some corners of the parameter space.


\subsection[A relation between $\delta g_{t_L}$ and $\delta V_{tb}$]{A relation between \boldmath${\mathbf{\delta g_{t_L}}}$ and \boldmath${\mathbf{\delta V_{tb}}}$}

Before discussing the results in some explicit models, we present a general
relation that links the deviations in the $Z t_L \overline t_L$ vertex to the
corrections to $V_{tb}$~\cite{delAguila:2000aa,delAguila:2000rc,Aguilar-Saavedra:2013pxa}.
In the effective Lagrangian describing the
Higgs doublet $H$ and the SM fermions only two dimension-six operators contribute
to the corrections to the $t_L$ couplings~\cite{delAguila:2000aa,Giudice:2007fh,Grzadkowski:2010es,Contino:2013kra}:
\begin{equation}\label{eq:dim-6_oper_tL}
{\mathcal L} = i \frac{c_{H_q}}{f^2} (\overline q_L \gamma^\mu q_L)
\left(H^\dagger \dblarrow D_\mu H\right)
+ i \frac{c'_{H_q}}{f^2} (\overline q_L \sigma^i \gamma^\mu q_L)
\left(H^\dagger \sigma^i \dblarrow D_\mu H\right)\,,
\end{equation}
where $H^\dagger \dblarrow D_\mu H$ is the derivative
$H^\dagger (D_\mu H) - (D_\mu H)^\dagger H$.
One combination of the two operators in Eq.~(\ref{eq:dim-6_oper_tL}) controls the corrections to the
$Z b_L \overline b_L$ vertex and is thus tightly constrained by the experimental data.
Given that the precision on the bottom couplings is much higher than on the top ones,
for the present analysis we can assume that the corrections to $g_{b_L}$ vanish,
which implies the relation $c'_{H_q} = - c_{H_q}$~\cite{Agashe:2006at,AguilarSaavedra:2012vh,Grojean:2013qca}.
By imposing this condition we find that
the operators in Eq.~(\ref{eq:dim-6_oper_tL}) give rise to the following contributions to the interactions of
the top quark with the EW gauge bosons
\begin{equation}
{\mathcal L} \supset 2 c_{H_q} v^2 \left[\frac{g}{\cos \theta_w} \overline t_L Z_\mu \gamma^\mu t_L
+ \frac{g}{2} \left( \overline t_L \left(W_\mu^1 - i W_\mu^2\right) \gamma^\mu b_L + \mathrm{h.c.}\right)\right]\,.
\end{equation}
From this equation one can easily see that the leading corrections to the $Z t_L \overline t_L$
vertex and to the $V_{tb}$ matrix element satisfy the relation
\begin{equation}\label{eq:ztt_vtb}
\delta g_{t_L} = -\delta V_{tb}\,.
\end{equation}
It is important to stress that the above result holds only at order $v^2/f^2$. The subleading terms, as for instance
the ones coming from dimension-eight operators, can generate independent corrections to $g_{t_L}$ and $V_{tb}$.

Notice that the above analysis is valid in a generic BSM scenario and is not
related to the Goldstone boson nature of the Higgs.


\subsection{Explicit results}

To conclude the discussion about the corrections to the top couplings we report in the
following the explicit results for the leading order corrections to $g_{t_L}$,
$V_{tb}$ and $g_{t_R}$ in the composite models with elementary fermions
embedded in the fundamental representation of $\SO(5)$~\cite{Grojean:2013qca}. As we did for the
oblique parameters and the $Z b_L \overline b_L$ vertex, we consider two
general CCWZ parametrizations with only one level of resonances (for the
details of the models see Sect.~\ref{sec:effect_parametr}).

\subsubsection{The case of an elementary ${\mathbf{t_R}}$}

As a first case we consider the scenarios with an elementary $t_R$.
The corrections to the $t_L$ couplings at leading order in $v/f$ are given by
\begin{eqnarray}
\delta g_{t_L} = - \delta V_{tb} &=& - \frac{\xi}{4} \frac{f^2}{m_4^2 + y_{L4}^2 f^2}
\Bigg[\left(\frac{m_4 m_1 y_{L1} + y_{L4} y_{R4} y_{R1} f^2}{m_1^2 + y_{R1}^2 f^2}
- \sqrt{2} c_L\, y_{L4}\right)^2\nonumber\\
&& \hspace{7.5em}+\; (1-2 c_L^2)y_{L4}^2\Bigg]\,.
\end{eqnarray}
The coupling of the $t_R$ field with the $Z$ boson receives the following leading
corrections
\begin{eqnarray}
\delta g_{t_R} &=& \frac{\xi}{4} \frac{(m_4 y_{L1} y_{R1} - m_1 y_{L4} y_{R4})f^2}
{(m_4^2 + y_{L4}^2 f^2)(m_1^2 + y_{R1}^2 f^2)}\Bigg[
\frac{y_{L4}^2 (m_4 y_{L1} y_{R1} - m_1 y_{L4} y_{R4}) f^4}{m_4^2 (m_4^2 + y_{L4}^2 f^2)}\nonumber\\
&& +\; 2 \frac{m_1}{m_4^2} y_{L4} y_{R4} f^2 - 2 \sqrt{2} \frac{c_R y_{L4} y_{R1} f^2}{m_4} \Bigg]\,.
\end{eqnarray}
In the above equation the factor in front of the brackets is proportional to the
top mass, hence the correction to the $g_{t_R}$ coupling vanishes in the limit of
zero $m_t$ as a consequence of the custodial protection.

The above equations can be directly applied to the two-site $5+5$ model. In this case
one finds that, in the regions of the parameter space favored by the EW data, the corrections
to the $V_{tb}$ matrix element are typically below $10\%$ ($-0.12 \lesssim \delta V_{tb} \lesssim 0$).
The current bounds can already exclude a corner in the parameter space where $|\delta V_{tb}| \gtrsim 0.10$.
The deviations to the $g_{t_R}$ coupling, on the other hand, are always quite small, $\delta g_{t_R} \lesssim 0.01$.

For completeness we also give the simplified results in the cases in which only one
$\SO(4)$ multiplet is light. In the limit with only a light singlet one gets
\begin{equation}
\delta g_{t_L} = -\delta V_{tb} = -\frac{\xi}{4} \frac{m_1^2 y_{L1}^2 f^2}{(m_1^2 + y_{R1}^2 f^2)^2}\,,
\qquad \quad
\delta g_{t_R} = 0\, .
\end{equation}
The above formula shows that the corrections to the $t_L$ couplings are suppressed in the parameter space
region with a sizable $t_R$ compositeness ($y_{R1} f > m_1$ and $y_{R1} > y_{L1}$).

In the case with only a light fourplet one gets the following results
\begin{equation}
\delta g_{t_L} = - \delta V_{tb} = -\frac{\xi}{4} \frac{y_{L4}^2 f^2}{m_4^2 + y_{L4}^2 f^2}\,,
\end{equation}
and
\begin{equation}
\delta g_{t_R} = -\frac{\xi}{4} \frac{y_{L4}^2 y_{R4}^2 f^2}{m_4^2 + y_{L4}^2 f^2}
\left(\frac{f^2}{m_4^2} + \frac{f^2}{m_4^2 + y_{L4}^2 f^2}\right)\,.
\end{equation}
In this case the experimental bounds on $V_{tb}$ can be directly translated into an upper bound
on the $t_L$ compositeness.

\subsubsection{The case of a fully composite ${\mathbf{t_R}}$}

As a second example we consider the scenario with a fully composite $t_R$.
The leading corrections to the $V_{tb}$ matrix element and to the top couplings
to the $Z$ boson are given by
\begin{equation}
\delta g_{t_L} = - \delta V_{tb} = -\frac{\xi}{4} \frac{f^2}{m_4^2 + y_{L4}^2 f^2}
\Bigg[\left(\frac{m_4 y_{L1}}{m_1} - \sqrt{2} c_L\, y_{L4}\right)^2 + (1- 2 c_L^2) y_{L4}^2\Bigg]
\end{equation}
and
\begin{equation}
\delta g_{t_R} = \frac{\xi}{4} \frac{y_{L4} y_{Lt} f^2}{(m_4^2 + y_{L4}^2 f^2)^2}
\left[y_{L4} y_{Lt} f^2 - 2 \sqrt{2} c_t (m_4^2 + y_{L4}^2 f^2)\right]\,.
\end{equation}
As in the case with an elementary $t_R$, also in the present setup the initial factor
in the expression for $\delta g_{t_R}$ vanishes if the top mass is zero.

In the limits with only one light multiplet the above results can be drastically simplified.
If only a light singlet is present in the effective theory  one gets
\begin{equation}
\delta g_{t_L} = - \delta V_{tb} = - \frac{\xi}{4} \frac{y_{L1}^2 f^2}{m_1^2}\,,
\qquad \quad \delta g_{t_R} = 0\,.
\end{equation}
In this case the corrections to the $Z t_L \overline t_L$ coupling can become
sizable if the composite singlet is light.

In the limit with only a light composite fourplet the corrections to the
top couplings become
\begin{equation}
\delta g_{t_L} = - \delta V_{tb} = -\frac{\xi}{4} \frac{y_{L4}^2 f^2}{m_4^2 + y_{L4}^2 f^2}
\end{equation}
and
\begin{equation}
\delta g_{t_R} = \frac{\xi}{4} \frac{y_{L4} y_{Lt} f^2}{(m_4^2 + y_{L4}^2 f^2)^2}
\left[y_{L4} y_{Lt} f^2 - 2\sqrt{2} c_t (m_4^2 + y_{L4}^2 f^2)\right]\,.
\end{equation}
Analogously to the case with an elementary $t_R$, the corrections to the $V_{tb}$
matrix element can be used to put an upper bound on the degree of compositeness of the
$q_L$ doublet.

\begin{subappendices}

\section{The custodial symmetries}\label{sec:custodial_symmetries}

In the analysis of the constraints from EWPT we encountered two ``custodial'' symmetries
that are of fundamental importance in keeping under control the corrections to the EW parameters.
The first symmetry is the standard custodial group $\SO(3)_c$, which
forbids corrections to the $\widehat{T}$ parameter. The second one is the discrete $P_{LR}$ invariance which protects the couplings
of the $Z$ boson to the SM fields and, in particular, to the bottom quark. These symmetries are also responsible for a number of peculiar properties and selection rules on the mass-spectrum and on the couplings of the composite sector resonances, some of which we encountered in the main text. The way in which these protections work is explained below.

\subsection[The $\SO(3)_c$ symmetry]{The $\SO{\mathbf{(3)_c}}$ symmetry}

We start from assuming an $\SO(4)$ global symmetry of the sector responsible for EWSB, under which the four real Higgs field components $\Pi^i$ form a fourplet or, equivalently, a $(\mathbf{2},\mathbf{2})$ pseudo-real matrix (see Eq.~(\ref{hmat}))
\begin{equation}\label{eq:H_matrix}
{\Sigma} = \frac{1}{\sqrt{2}}\left(
i \sigma_\alpha \Pi^\alpha + \Id_2 \Pi^4
\right)\,.
\end{equation}
Under $\SO(4) \simeq \SU(2)_L \times \SU(2)_R$, $\Sigma$ transforms as
\begin{equation}\label{eq:H_transf}
{\Sigma} \rightarrow g_L {\Sigma} g_R^\dagger\,.
\end{equation}
In composite Higgs models, this  $\SO(4)$ symmetry is part of (or coincides with, in the minimal coset $\SO(5)/\SO(4)$) the unbroken subgroup $\mathpzc{H}$ of the $\mathpzc{G}/\mathpzc{H}$ coset and it is by assumption an exact symmetry of the composite sector. In the SM, $\SO(4)$ is instead an accidental symmetry of the Higgs doublet Lagrangian, if considered in isolation. Both in composite Higgs and in the SM, the $\SO(4)$ symmetry is broken by the gauge fields and fermions couplings.

The Higgs VEV, $\langle \Pi^i \rangle = v\, \delta_{i4}$, breaks $\SO(4)$ spontaneously to an $\SO(3)$ subgroup, realizing the symmetry breaking pattern
\beq
\SO(4)\rightarrow \SO(3)\,.
\eeq
The unbroken $\SO(3)$ is what we call the custodial $\SO(3)_c$. Its action can be either viewed as rotations of the first three $\ve\Pi$ vector components (with the physical Higgs in $\Pi^4$ being a scalar) or, equivalently, as the vector subgroup of $\SU(2)_L \times \SU(2)_R$, $\SU(2)_V$, defined by equal left and right transformations $g_L=g_R=g_V$. Indeed the Higgs VEV in the matrix notation
\beq
\langle\Sigma\rangle = \frac{v}{\sqrt{2}} \Id_2\,,
\eeq
is invariant under the vector transformations.

Both in the SM and in composite Higgs the $W_\mu^\alpha$ and $B_\mu$ fields weakly gauge the $\SU(2)_L\times\U(1)_Y$ subgroup of $\SO(4)$. Actually when dealing with fermions (see Sect.~\ref{sec:HCsigmaM}) an extra unbroken $\U(1)_X$ group needs to be introduced, but this will play no role in what follows. The $W^\alpha$ fields, which fully gauge $\SU(2)_L$, preserve $\SO(4)$ provided we assign them to the $(\mathbf{3},\mathbf{1})$ representation of the group. The effect of the hypercharge gauging, which instead breaks $\SO(4)$, will be discussed later on. 

The cancellation of the $\widehat{T}$ parameter immediately follows from this symmetry structure. Indeed $\widehat{T}$ is defined (see Eq.~(\ref{eq:def_T})) in terms of the amputated two-point $W$ field correlators at zero transferred momentum, and thus it should correspond to a non-derivative mass-term operator in the effective action. However the only such term which is compatible with the unbroken $\SO(3)_c$ (and also happens to respect the full $\SO(4)$) is \footnote{The normalization of the operator that follows is chosen to match the SM $W$ mass term.}
\beq
\label{massterm}
{\mathcal{L}}_{\textrm{mass}}=\frac{g^2v^2}8 W_{\mu}^\alpha W^{\mu}_\alpha\,.
\eeq
This term contributes in the same way to $\Pi_{W^3W^3}$ and to $\Pi_{W^1W^1}$, thus it does not contribute to $\widehat{T}$, which is proportional to the difference between the two. The custodial $\SO(3)_c$ symmetry thus implies $\widehat{T}=0$. 

A non-vanishing $\widehat{T}$ would correspond to the presence, in the effective Lagrangian, of an operator of the form
\beq
\frac{v^2}8 {\mathcal{S}}_{\alpha\beta}W_{\mu}^\alpha W^{\beta,\mu}\,,
\eeq
where ${\mathcal{S}}$ is a symmetric traceless (since the trace component does not contribute) tensor in the $(\mathbf{5},\mathbf{1})$ representation of $\SU(2)_L\times\SU(2)_R$. In terms of ${\mathcal{S}}$, $\widehat{T}$ can be expressed as $\widehat{T}={\mathcal{S}}_{33}-{\mathcal{S}}_{11}$. In a perfectly invariant theory no parameter exists with non-trivial $\SO(4)$ transformation properties. A non-vanishing ${\mathcal{S}}$, and thus in turn a non-vanishing $\widehat{T}$, can only be constructed in terms of spurions, whose presence signals the explicit breaking of the symmetry.

Both in the SM and in composite Higgs, explicit $\SO(4)$ breaking emerges from the hypercharge gauging and from the coupling to fermions. The contributions to $\widehat{T}$ from the latter are described extensively in Sect.~\ref{THATLOOP}, here we briefly discuss the effects of the former breaking. The breaking appears because only one of the three $\SU(2)_R$ generators, the third one, is gauged by the hypercharge field $B_\mu$,
{\it{i.e.}}~only one of the three $\SU(2)_R$ gauge sources $W_R$ is a truly dynamical field while the other components are set to zero. This breaking corresponds to a spurion ${\mathcal{G}}'$ in the $(\mathbf{1},\mathbf{3})$, which can be inserted in the relation between the $W_R$ source and the physical field $B$, namely
\beq
W_{R,\mu}^\alpha={{\mathcal{G}}'}^\alpha B_\mu\,.
\eeq
By two powers of this spurions, plus four powers of the Higgs VEV, which transforms in the $(\mathbf{2},\mathbf{2})$, a non-vanishing ${\mathcal{S}}$ tensor can be constructed and a contribution to $\widehat{T}$ is generated. Notice however that the lack of symmetry in the hypercharge gauging becomes visible only at the loop level because it is only in the presence of at least one $B_\mu$ field propagator that we can distinguish the case in which all the three $W_R$ fields are dynamical, and the symmetry is preserved, from the one in which only $B_\mu$ is dynamical and the symmetry is broken. Therefore $\widehat{T}$ remains zero at tree-level and Eq.~(\ref{massterm}) gets generalized in the only possible way compatible with the unbroken electromagnetic $\U(1)$ symmetry, namely
\beq
{\mathcal{L}}_{\textrm{mass}}=\frac{v^2}8\left[(g\,W^1)^2+ (g\,W^2)^2 + (g\,W^3-g'B)^2 \right]\,.
\eeq
The term above, which just coincides with the habitual SM one, gives masses to the $W$ and to the $Z$ that obey the $\rho=1$ relation.

\subsection[The $P_{LR}$ symmetry]{The ${\mathbf{P_{LR}}}$ symmetry}

We now turn to the $P_{LR}$ symmetry and describe how it can protect the coupling of the $Z$ boson to fermions. The right starting point, even before introducing the $P_{LR}$ symmetry itself, is to remind ourselves how the SM gauge fields are introduced in the theory. The $W_\mu^\alpha$ fields gauge the $\SU(2)_L$ group while $B_\mu$ gauges the $\U(1)_Y$ hypercharge generator, defined in Sect.~\ref{sec:HCsigmaM} as the sum of $t_R^3$ in $\SU(2)_R$ and the $X$ charge of the $\U(1)_X$ group, namely
\beq
\label{uai}
Y=t_R^3+X\,.
\eeq
The gauging is conveniently described, as we saw in Sect.~\ref{sources_and_local}, by introducing external sources associated to all the group generators and identifying part of them as dynamical fields only at a late stage of the calculation. We thus consider three $\SU(2)_L$, three $\SU(2)_R$ and one  $\U(1)_X$ sources, namely
\bea
\label{gaugedef}
&&W_{L,\mu}=W_{L,\mu}^\alpha t_L^\alpha=g\, W_\mu^\alpha t_L^\alpha\,,\nonumber\\
&&W_{R,\mu}=W_{R,\mu}^\alpha t_R^\alpha=g'B_\mu t_R^3\,,\nonumber\\
&&{\mathcal{X}}_\mu=g' B_\mu\,,
\eea
where $t_{L,R}^\alpha$ are the $\SU(2)_L\times\SU(2)_R$ generators, for which a normalized explicit representation is provided in Eq.~(\ref{eq:SO4_gen}). The physical value of the source fields, in terms of three $W$'s and $B$, is also reported in the equation above. 

The full $\SU(2)_L\times\SU(2)_R\times\U(1)_X$ global group can be formally promoted to a local symmetry by regarding the $W_L$, $W_R$ and ${\mathcal{X}}$ sources as the gauge fields associate to the three semi-simple factors. However only the unbroken (or linearly-realized) subgroup $\SO(3)_c\times\U(1)_X$ will be relevant in what follows. This subgroup acts on the sources as
\begin{eqnarray}
\label{trcust}
&&W_{L,\mu} \rightarrow g_V\cdot\left( W_{L,\mu} + i \partial_\mu \right)\cdot g_V^\dagger\,,\nonumber\\
\rule{0pt}{1.25em}
&&W_{R,\mu} \rightarrow g_V\cdot\left( W_{R,\mu} + i \partial_\mu \right)\cdot g_V^\dagger\,,\nonumber\\
\rule{0pt}{1.25em}
&&{\mathcal{X}}_\mu \rightarrow {\mathcal{X}}_\mu+\partial_\mu\alpha_X\,,
\end{eqnarray}
where $\alpha_X$ denotes the $\U(1)_X$ transformation parameter. What is peculiar in the expression above is that $W^\mu_{L}$ and $W^\mu_{R}$, in spite of being two distinct fields, both transform as if they were gauge connections associated to the $\SO(3)_c$ local group. These local symmetry transformations will be very effective in constraining the fermion couplings.

We now introduce $P_{LR}$, which is defined in Appendix~\ref{discrete} as the discrete $\Zdouble_2$ transformation that interchanges the
$\SU(2)_L$ and $\SU(2)_R$ generators inside $\SO(4)$. Therefore it acts on the gauge sources as
\beq
W_{L,\mu}^\alpha \leftrightarrow W_{R,\mu}^\alpha\,, 
\eeq
while it leaves ${\mathcal{X}}_\mu$ invariant. On the Higgs fourplet, $P_{LR}$ acts like a parity reflection of the first three components
\beq
\label{plrfund2}
{{P}}_{LR}^{\mathbf{4}}=\textrm{diag}(-1,-1,-1,+1)\,,
\eeq 
therefore the Higgs VEV is even and $P_{LR}$ survives as an unbroken symmetry after EWSB. In the presence of $P_{LR}$, the unbroken group $\SO(3)_c\times \U(1)_X$ is enlarged to ${\textrm{O}}(3)_c\times \U(1)_X$.

Let us now analyze the implications of the gauge symmetry transformations in Eq.~(\ref{trcust}) on the zero-momentum, {\it{i.e.}}~non-derivative, couplings of the gauge fields to the fermions. Since we are interested in the $Z$ boson couplings, we restrict our attention to the interactions of the neutral sources $W_L^3$ and $W_R^3$ to one charge-eigenstate chiral fermion $\psi$. The only interactions allowed by the symmetries are 
\bea
\label{intterm}
&& i\,\overline\psi\gamma^\mu\left(\partial_\mu-i\,t_L^3W_{L,\mu}^3-i\,t_R^3W_{R,\mu}^3-i\,X {\mathcal{X}}_{\mu}\right)\psi\nonumber\\
&&+\;c\,(W_{L,\mu}^3-W_{R,\mu}^3)\overline\psi\gamma^\mu\psi\,.
\eea
In particular, the terms on the first line are enforced by the covariant derivative structure of the kinetic term and thus their coefficient is uniquely determined by the $t_{L,R}^3$ and $X$ eigenvalue of $\psi$. The one on the second line is instead separately gauge-invariant given that the shift term in the local transformation of Eq.~(\ref{trcust}) cancels when we take the difference $W_L-W_R$. Therefore it has an arbitrary coefficient ``$c$''. Not surprisingly, since they are rigidly fixed by the gauge symmetries, the interactions on the first line reproduce the SM $W_3$ and $B$ vertices, which after the weak angle rotation reduce to the standard photon and $Z$ boson couplings. This is immediately verified by substituting the explicit value of the sources in Eq.~(\ref{gaugedef}), obtaining
\beq
t_L^3W_{L,\mu}^3+t_R^3W_{R,\mu}^3+X {\mathcal{X}}_{\mu}=g\,t_L^3W_{\mu}^3+g'(t_R^3+X)B_\mu\,,
\eeq
and noticing that $t_R^3+X=Y$ as in Eq.~(\ref{uai}). Since the ones on the first line match with the SM, the only deviation comes from the term on the second line. Using the explicit value of the sources and performing the weak angle rotation to the $Z$ and photon field basis one immediately finds that \mbox{$W_{L}^3-W_{R}^3=g/\cos\theta_w Z$}. No corrections to the zero-momentum photon coupling are thus generated, as an obvious consequence of the unbroken electromagnetic gauge group, while the $Z$ boson interaction can be distorted by an amount
\beq
\delta g_\psi = c\,,
\eeq
having adopted the standard convention of normalizing the coupling deviation by the $g/\cos\theta_w$ factor (see Eq.~(\ref{eq:Lagr_Z_b})).

However $W_{L}^3-W_{R}^3$ is odd under $P_{LR}$, therefore if $\psi$ is a $P_{LR}$ eigenstate, no matter if even or odd,  $P_{LR}$ enforces $c=0$ and no corrections to the $Z$ couplings can occur.
This result can be easily extended to multiple fermionic fields with definite $P_{LR}$ parity. In particular the $Z$ boson couplings to a set of eigenstates with the same $P_{LR}$ parity are necessarily canonical and flavor-diagonal.
Flavor-violating $Z$ interactions, indeed, can only involve eigenstates with opposite parity.

When we consider the SM fermions, in order for the $Z$ couplings to be protected, $P_{LR}$ must be a symmetry not only of the composite sector (possibly an accidental one as we saw happening in some cases in the main text), but also of the partial compositeness mixing of the elementary SM field we are interested in.
The left-handed bottom quark coupling to the $Z$ is particularly relevant, let us thus discuss under which condition it benefits of the $P_{LR}$ protection. We start from the case in which the $q_L$ doublet mixes with a fundamental of $\SO(5)$, {\it{i.e.}}~with the $(\mathbf{2},\mathbf{2})$ fourplet $\SO(4)$ representation inside the fundamental. The embedding of the doublet, provided in Eq.~(\ref{qlemb}), immediately shows that the $Z$ coupling to $b_L$ is protected in this case, compatibly with what we found in Sects.~\ref{pfer} and \ref{sec:Zbb}. Indeed the $b_L$ field only appears in the first and in the second component of the fourplet, which are both $P_{LR}$-odd according to Eq.~(\ref{plrfund2}). Therefore $P_{LR}$ is preserved by the mixing, provided $b_L$ is regarded as an odd field, and the $b_L$ coupling is protected. We also see from the same equation that the $t_L$ fields appear instead both in the third component of the multiplet, which is odd, and in the fourth one which is even. The $t_L$ mixing thus breaks $P_{LR}$ and no protection is present for its coupling with the $Z$. This was for the first mixing of the $q_L$, the one with the $\mathbf{5}_{2/3}$ multiplet that participate in the generation of the top quark mass. The situation is reversed for the mixing with the $\mathbf{5}_{-1/3}$, for which the embedding is reported in Eq.~(\ref{qlbemb}). The $t_L$ coupling is protected in that case, while the $b_L$ one is not, given that the $b_L$ mixes with both an even and an odd component. The modifications of the $Z b_L{\overline{b}}_L$ coupling is thus induced only by the second mixing parameter, $\lambda_{b_L}$, and not by $\lambda_{t_L}$. It should be rather obvious, at this point, that no protection is instead present when the $q_L$ mixes with operators in the spinorial representation. Indeed the mixing occurs in this case with a $(\mathbf{2},\mathbf{1})$ representation of $\SO(4)$ and the $P_{LR}$ symmetries interchanges the $(\mathbf{2},\mathbf{1})$ and the $(\mathbf{1},\mathbf{2})$ components of the spinorial. Since the mixing occurs with the former and not with the latter, it breaks $P_{LR}$ and no protection is found. This is in accordance with the results of Sect.~\ref{pfer}.

\end{subappendices}

\backmatter

\phantomsection
\addcontentsline{toc}{chapter}{Bibliography}
\providecommand{\href}[2]{#2}


\end{document}